\documentclass{aa} 

\usepackage{latexsym, graphicx, natbib, longtable}
\usepackage{longtable}
\usepackage{lscape}
\usepackage{txfonts}
\def\HI{\hbox{\rm H\,{\sc i}}}

\begin{document}
   \title{Diffuse neutral hydrogen in the {\HI} Parkes All Sky Survey}

   %\subtitle{}

   \author{A. Popping
          \inst{1} \inst{2} 
          \and
          R. Braun\inst{3}
          }

   \offprints{A. Popping \email{attila.popping@icrar.org} }

   \institute{Kapteyn Astronomical Institute, P.O. Box 800, 9700 AV Groningen, the Netherlands
   \and
   International Centre for Radio Astronomy Research, The University of Western Australia, 35 Stirling Hwy, Crawley, WA 6009, Australia
    \and
    CSIRO Astronomy and Space Science, P.O. Box 76, Epping, NSW 1710, Australia}

   \date{}

% \abstract{}{}{}{}{} 
% 5 {} token are mandatory
 
  \abstract
  {Observations of neutral hydrogen can provide a wealth of
    information about the distribution and kinematics of
    galaxies. To learn more about large scale structures and accretion
    processes, the extended environment of galaxies must also be
    observed. Numerical simulations predict a cosmic web of extended
    structures and gaseous filaments.}
   {To detect {\HI} beyond the ionisation edge of galaxy disks, column
  density sensitivities have to be achieved that probe the regime of Lyman limit
  systems. Typically {\HI} observations are limited to a brightness
  sensitivity of $N_{HI} \sim 10^{19}$ cm$^{-2}$ but this has to be
  improved by at least an order of magnitude.}
{In this paper, reprocessed data is presented that was originally
  observed for the {\HI} Parkes All Sky Survey (HIPASS). HIPASS
  provides complete coverage of the region that has been observed for
  the Westerbork Virgo Filament {\HI} Survey (WVFS), presented in
  accompanying papers, and thus is an excellent product for data
  comparison. The region of interest extends from 8 to 17
    hours in right ascension and from $-$1 to 10 degrees in
    declination.  Although the original HIPASS product already has
  good flux sensitivity, the sensitivity and noise characteristics can
  be significantly improved with a different processing method.}
   {The newly processed data has an $1\sigma$ RMS
     flux sensitivity of $\sim 10$ mJy beam$^{-1}$ over 26 km
     s$^{-1}$, corresponding to a column density sensitivity of $\sim 3
     \cdot 10^{17}$ cm$^{-2}$. While the RMS sensitivity is improved
     by only a modest 20\%, the more substantial benefit is in the
     reduction of spectral artefacts near bright sources by more than
     an order of magnitude. In the reprocessed region we confirm all
     previously catalogued HIPASS sources and have identified 29
     additional sources of which 14 are completely new {\HI}
     detections. We derived spectra and moment maps for all detections
     together with total fluxes determined both by integrating the
     spectrum and by integrating the flux in the moment maps within
     the source radius. Extended emission or companions were sought in
     the nearby environment of each discrete detection. Ten
     extra-galactic filaments are marginally detected within the
     moment maps. }
   {With the improved sensitivity after reprocessing and its large sky
     coverage, the HIPASS data is a valuable resource for detection of
     faint {\HI} emission. This faint emission can correspond to
       extended halos, dwarf galaxies, tidal remnant and potentially
       diffuse filaments that represent the trace neutral fraction of
       the Cosmic Web.}


   \maketitle
%
%________________________________________________________________

\section{Introduction}
Current cosmological models ascribe about 4\% of the density to
baryons \citep{2007ApJS..170..377S}. At low redshift most of these
baryons do not reside in galaxies, but are expected to be hidden in
extended gaseous web-like filaments. (e.g. \cite{1999ApJ...511..521D},
\cite{2001ApJ...552..473D}, \cite{1999ApJ...514....1C}). Calculations
suggets that in the current epoch baryons are almost equally
distributed amongst three components: (1) galactic concentrations, (2)
a warm-hot intergalactic medium (WHIM) and (3) a diffuse intergalactic
medium (seen as the Ly$\alpha$ forest). These simulations predict that
the three components are each coupled to a decreasing range of
baryonic overdensity $\log(\rho / \bar{\rho}_H) > 3.5$, 1 - 3.5, and
$< 3.5$.

Direct detection of the inter-galactic gas, or the WHIM, is very
difficult in the EUV and X-ray bands \citep{1999ApJ...514....1C}. In
this and accompanying papers we make an effort to detect traces of the
inter and circum-galactic medium in neutral hydrogen. Due to the moderately high
temperatures in the intergalactic medium (above $10^4$ Kelvin), most
of the gas in the Cosmic Web is highly ionised. To detect the trace
neutral fraction in the Lyman Limit Systems using the 21-cm
line of neutral hydrogen, a column density sensitivity of $N_{HI} \sim
10^{17-18}$ is required. A more detailed background and introduction
to this topic is given in \cite{2010PhDT..APOPPING} and
\cite{2011A&A...527A..90P}.

A first example of detection in {\HI} emission of the likely
counterpart of a Lyman Limit absorption System is shown in
\cite{2004A&A...417..421B}, where a very diffuse {\HI} structure is
seen with a peak column density of only $N_{HI} \sim 10^{18}$
cm$^{-2}$, connecting M31 and M33. To be able to detect a large number
of diffuse {\HI} features, extended blind surveys are required with an
excellent brightness sensitivity. One of the first such efforts is
presented in \cite{2011A&A...527A..90P}, where the Westerbork
Synthesis Radio Telescope (WSRT) is used to undertake a deep
fully-sampled survey of the galaxy filament joining the Local Group to
the Virgo Cluster (Westerbork Virgo Filament {\HI} Survey) extending
from 8 to 17 hours in RA and from $-$1 to +10 degrees in
Declination. Data products were created from both the cross-, as well
as the auto-correlation data, to achieve a very high brightness
sensitivity at a variety of spatial resolutions.  The total-power
product of the WVFS is presented in \cite{2011A&A...527A..90P},
  while the interferometric data is presented in
  \cite{2011A&A...528A..28P}. In these papers new detections of
neutral hydrogen are reported. Although these detections are very
interesting, they are difficult to interpret or to confirm, as no
comparison data is currently available at a comparable sensitivity.

In this paper we use reprocessed data of the {\HI} Parkes All Sky
Survey to complement the WVFS observation.  The {\HI} Parkes All Sky
Survey (HIPASS) \citep{2001MNRAS.322..486B} includes the complete
Southern sky and the Northern sky up to +25.5 degrees in
Declination. The Northern part of the survey is described in
\cite{2006MNRAS.371.1855W}. This survey currently has the best
available {\HI} brightness sensitivity yet published. As the Northern
part of the survey completely covers the region that has been observed
for the Westerbork Virgo Filament Survey, HIPASS is an excellent
product for data comparison.

Although neither the flux sensitivity, nor the brightness sensitivity
of HIPASS is equivalent to that of the WVFS total power data, we can
still learn more about faint {\HI} detections in the WVFS, by taking
into account the limitations of both surveys. The low column densities
of some new {\HI} detections in the WVFS might be confirmed,
indicating that the gas is indeed very diffuse. Conversely, if column
densities measured in the HIPASS data are significantly higher than in
the WVFS, this would imply that the gas is more condensed than it
appeared, with the emission diluted by the large beam of the WVFS.

Although the HIPASS data is completely reduced and the processed cubes
are publicly available, for our purpose we have begun anew with the
raw, unprocessed observational data. Increased computing capacity, and
different calibration algorithms allow significant
improvements to be achieved over the original HIPASS products.

In the observations and data reduction sections we will explain in
detail the processing employed and the improvements achieved. A new
list of objects detected in the region of interest is given in the
results section. Although the improved data reduction method can be
applied to the complete HIPASS survey area, we emphasise that we have
only applied it to the region of overlap with the WVFS both spatially
and spectrally.

In section 2 of this
paper we will briefly summarise the observations that have been used,
followed by the data reduction strategy in section 3. In section 4 the
results will be presented; the general properties of each detected
object are given, but new {\HI} detections in the HIPASS data are
discussed in more detail. We close with a short discussion and
conclusion in sections 5 and 6. Detailed analysis and comparison of the
data, together with a discussion of the nature of the {\HI} emission
will be presented in a future paper. In that paper we will compare
the results of the HIPASS data together with the auto- and
cross-correlation products of the WVFS.

\section{Observations}
In our search for diffuse {\HI} emission, we have employed data that was
originally acquired for the {\HI} Parkes All Sky Survey (HIPASS). The
data is described in detail in \cite{2001MNRAS.322..486B} and we will
only summarise the relevant properties. All data has been obtained
using the Parkes 21-cm Multibeam system, containing a cooled 13 beam
receiver and digital correlator. The Multibeam correlator has an
instantaneous bandwidth of 64 MHz divided into 1024 channels. For the
HIPASS observations the receivers were tuned to a central frequency of
1394.5 MHz, offering a velocity range of $-$1280 to 12700 km s$^{-1}$
with a mean channel separation of 13.4 km s$^{-1}$. The central beam
FWHM is 14.0 arcmin, and the 13 beams are separated from one another
by about 30 arcmin. Data acquisition was started in 1997 February and
completed in 2000 March. Observations were obtained by scanning the
Telescope in Declination strips of 8 degrees length. The multibeam
receiver is rotated relative to the scan direction, to get
approximately uniformly spaced sampling of the sky over a strip of
$\sim 1.7$ degrees width. Each Declination scan maps approximately
$8\times 1.7$ degrees. To obtain full coverage of the sky at full
sensitivity, subsequent scans are displaced by 7 arcmin in RA, which
means that each of the 13 beams maps the sky with Nyquist
sampling. The scan rate of each strip is 1 degree min$^{-1}$. Using
all 13 beams, the total integration time of the HIPASS survey results
in $7\times10^3$ s deg$^{-2}$, or 450 s beam$^{-1}$. The typical
sensitivity of the original HIPASS product is 13.3 mJy beam$^{-1}$
over 18 km s$^{-1}$.

For the original HIPASS product, cubes were created of $8\times8$
degrees in size, centered at Declinations between $-$90 and +24
degrees. For our purpose we have selected all original HIPASS scans
centered at a Declination of $-$2, 6 and 14 degrees, and
between 8 and 17 hours in Right Ascension. With these data we achieve
the best possible coverage and sensitivity in the region between $-$1
and 10 degrees in Declination. This region was selected to exactly
overlap with the region observed in the Westerbork Virgo Filament
Survey (WVFS). The WVFS is an unbiased survey of $\sim1500$ squared degrees,
undertaken with the Westerbork Synthesis Radio Telescope (WSRT),
directed at the galaxy filament connecting the Local Group with the Virgo
Cluster.  The WVFS total power data has an effective beam size of
$\sim49$ arcmin with a sensitivity of 16 mJy beam$^{-1}$ over 16 km
s$^{-1}$. The HIPASS data has a slightly superior flux sensitivity,
but because of the smaller beam size the column density sensitivity is
about an order of magnitude worse. Nevertheless the HIPASS data is an
excellent product to use for comparison with the WVFS data.

\section{Data Reduction}

\subsection{Bandpass removal in the original HIPASS product}
Although the original unprocessed HIPASS data have been used, the
reduction method is slightly different, to obtain an improved end
product. The most challenging aspect of calibrating an observed total
power spectrum is the accurate estimation of the system bandpass
shape. Bandpass calibration of a single dish telescope is
traditionally accomplished by observing in {\it signal/reference}
mode. The telescope alternately tracks the target position and a
suitable nearby reference region for the same amount of time. The
reference position is used to estimate the bandpass shape and is
divided out of the signal spectrum. For HIPASS the telescope was
scanning the sky continuously, so the straight-forward {\it
  signal/reference} method could not be employed. The method that has
been used, was to estimate the bandpass shape of each spectrum, by
using a combination of earlier and later spectra, observed by the same
feed of the multibeam receiver. The bandpass was estimated by taking a
channel-by-channel {\it median} of the earlier and later spectra. The
{\it median} reference spectrum was preferred above the {\it mean}
reference spectrum, as the median statistic is more robust to outlying
data points, and is independent of the magnitude of deviation of 
outlying points.

The strategy that has been used in reducing the raw HIPASS data works
well in the absence of line emission, but breaks down in the vicinity of
bright detections. Some {\HI} sources are sufficiently bright that
bandpass estimates just prior and after the target spectrum are
elevated by the source itself. This results in negative
artefacts, or {\it bandpass sidelobes} that appear as depressions in
the spectra north and south of strong {\HI} sources.

\subsection{Bandpass removal in reprocessed HIPASS data}
All unprocessed HIPASS spectra are archived and can be reprocessed
using different methods. Techniques can be developed to improve the
bandpass-sidelobes and preserve spatially extended emission. An
example of such an approach is given by \cite{2003ApJ...586..170P} where a
different processing algorithm has been employed to image HVCs and the
Magellanic Stream. We have tested many different algorithms, including
the original HIPASS processing pipeline and the method used by
\cite{2003ApJ...586..170P}, and achieved the lowest residual RMS
fluctuation level with the approach outlined below.

The data were bandpass-corrected, calibrated and Doppler-tracked using
the {\it aips++} program {\it LiveData} \citep{2001MNRAS.322..486B} in
the following manner. 

\begin{itemize}

\item{The spectra were hanning smoothed over three channels to a
    velocity resolution of $\sim 26$ km s$^{-1}$.}

\item{In estimating the shape of the bandpass, a complete 8$^\circ$ scan
    is used instead of just a few time steps before and after the
    target spectrum. By using a complete scan instead of a subset the
    statistics are improved, making the bandpass estimate more
    robust.}

\item{A third order polynomial has been fit to the data in the time
    domain. Data points outside 2 times the standard deviation were
    excluded from the polynomial fit. This process is iterated three
    times, to get the best possible outlier rejection.}

\item{After fitting and correcting the data in the time domain, a
    second order polynomial was fit in the frequency domain. Higher
    order polynomials in frequency were tested but did not improve the
    result.}

\end{itemize}

All the processed scans were gridded with {\it Gridzilla}
\citep{2001MNRAS.322..486B} using a pixel size of 4'. Cubes were
created with a size of 24 degrees in Declination, ranging from $-$6 to
18 degrees and typical width of 1 hour in Right Ascension with an
overlap of one degree between the adjacent cubes.

All overlapping scans were averaged using the system temperature
weighted median of relevant data points. The value of the median is
strongly dependent on the form of the weighting function {\it w(r)}
and the radius $r_{max}$ out to which spectra are included. The
weighting procedure for HIPASS is described in detail in
\cite{2001MNRAS.322..486B}. A Gaussian beam-shape is assumed, so the
weighting function has the functional form:

\begin{equation}
w(r) = \left\{
\begin{array}{l l}
\exp \Big[ - \big( \frac{r}{\sigma} \big) ^2 /2 \Big] & \quad \mbox{for $r \leq r_{max}$}\\
0 & \quad \mbox{for $r > r_{max}$}\\
\end{array} \right.
\end{equation} 

For gridding the HIPASS data, a value of $r_{max} = 6$ arcmin. has
been adopted. The estimated flux at a given pixel is determined by
the weighted median of all spectra contributing to that pixel and has
the form:

\begin{equation}
F_e = \frac{\textrm{median}(F)}{\textrm{median}(w)}
\end{equation}

For a random distribution of data points or observations, the median
of the weights [median($w$)] is determined by the weighting function
where the radius divides the smoothing area in two equivalent parts,
i.e. $w(r_{max}/\sqrt{2})$. For the adopted $r_{max}$ of 6 arcmin,
median($w$)=1.28, which has been taken into account when gridding the
data. Tests during the gridding of HIPASS data have shown that the
input spectra are very nearly randomly distributed on the sky
\citep{2001MNRAS.322..486B}.

A top-hat kernel of 12' has been used to smooth data spatially. The
final beam size of the gridded data cubes is approximately 15.5' FWHM,
although as discussed at some length in \cite{2001MNRAS.322..486B},
the effective beam-size is dependent on the signal-to-noise ratio of a
detection.

After the processed cubes were formed, sub-cubes were created using
the inner 14 degrees in Declination with the highest uniform
sensitivity overlapping with the WVFS data. A velocity range was
selected from 200 to 1700 km s$^{-1}$, again to match with the
velocity coverage of the WVFS data.

In the reprocessed data, we typically achieve an RMS flux sensitivity
of $\sim10$ mJy beam$^{-1}$ over 26 km s$^{-1}$. This is an
improvement on the original HIPASS processing, although we have
degraded the velocity resolution to 26 km s$^{-1}$ compared to 18 km
s$^{-1}$, due to the Hanning smoothing that has been used instead of
the Tukey smoothing. When scaling both noise values to the same
velocity resolution, the achieved sensitivity is a significant
improvement on the sensitivity of the first HIPASS product. The
typical sensitivity of the first HIPASS product is $\sim 13$ mJy
beam$^{-1}$ over 18 km s$^{-1}$. For the northern Declinations we are
concentrating on, the sensitivity is slightly worse at $\sim 14$ mJy
beam$^{-1}$ over 18 km s$^{-1}$ \citep{2006MNRAS.371.1855W}
corresponding to $\sim 12$ mJy beam$^{-1}$ over 26 km s$^{-1}$. The
previous reprocessing of the HIPASS data \citep{2003ApJ...586..170P}
resulted in a similar rms value as we are achieving, however was not
able to correct for the negative artefacts.  Rather than an improved
rms value, the more important benefit of reprocessing the data is that
the negative artefacts in the vicinity of bright sources are
suppressed by more than an order of magnitude.

\begin{figure*}[t]
  \includegraphics[width=0.5\textwidth]{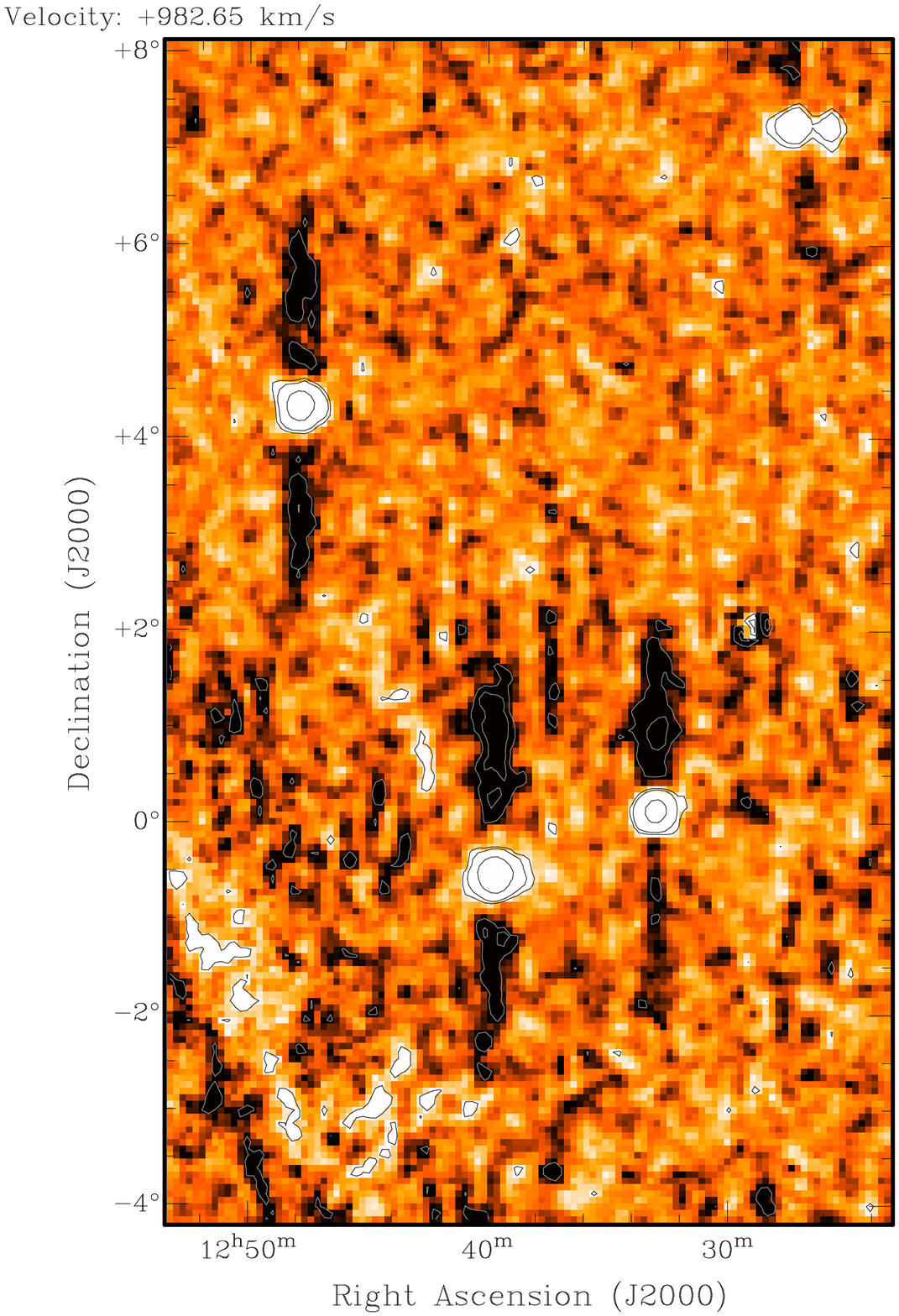}
  \includegraphics[width=0.5\textwidth]{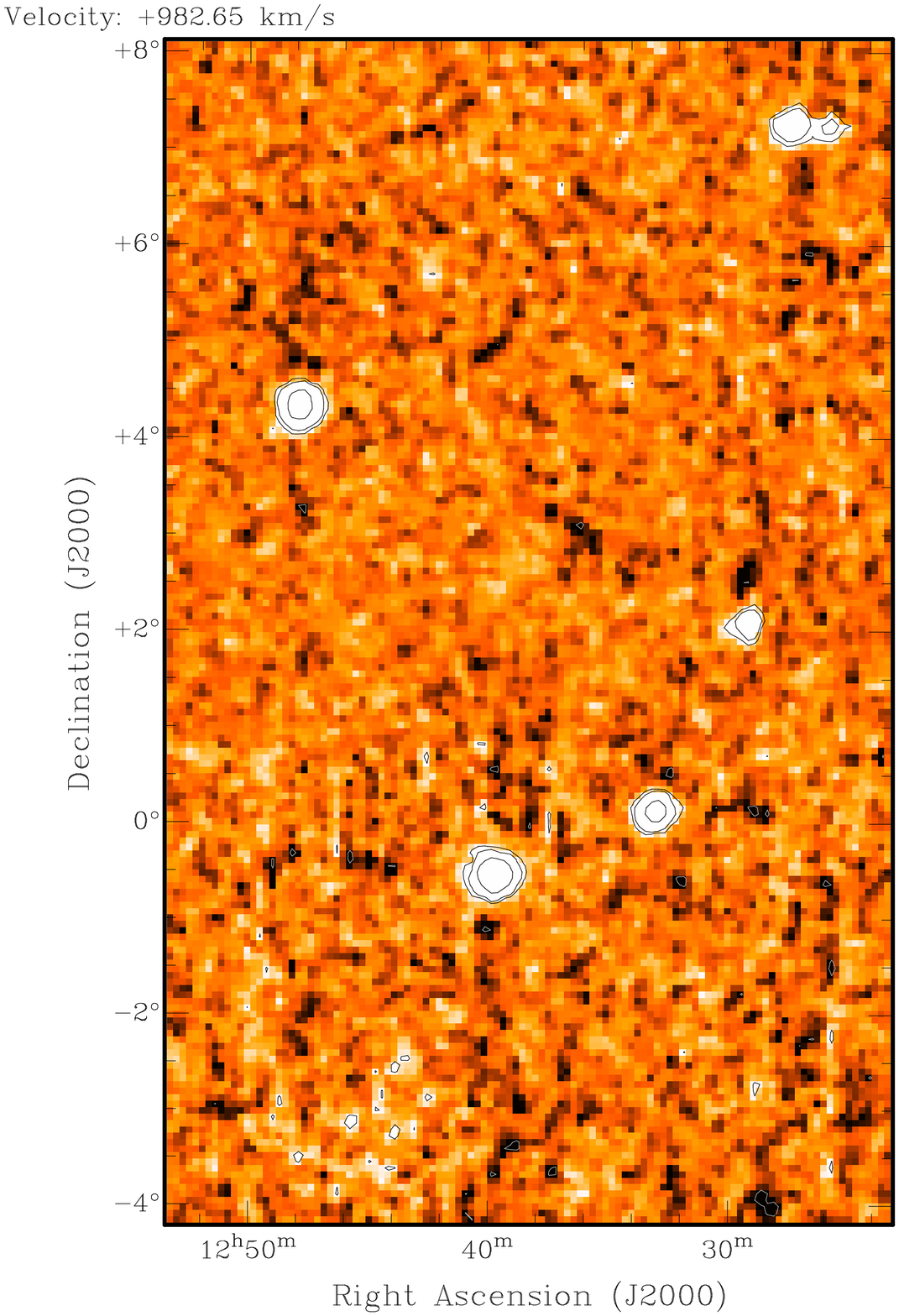}
  \caption{Example of the artefacts in the original HIPASS
  pipeline ({\it left panel}) and the reprocessed data ({\it right
  panel}). Both panels show the same region of the sky at the same
  velocity, the intensity ranges from -40 to 40 mJy
  beam$^{-1}$. Contours are plotted at -60, -30, 30, 60 and 300 mJy
  beam$^{-1}$, negative contours are coloured white}
\label{lobes}
\end{figure*}

An example of the reduced data and its artefacts is given in
Fig.~\ref{lobes}. The left panel shows a region of the sky processed
with the original HIPASS pipeline. There are strong negative artefacts
north and south of strong {\HI} sources. The right panel of
Fig.~\ref{lobes} shows the same region processed with the improved
reduction pipeline. Although the spectral artefacts are still visible,
there is a dramatic improvement. The reprocessed data will be much
more sensitive to diffuse {\HI} emission, especially in the direct
vicinity of bright {\HI} objects. Using this follow-up HIPASS product,
better flux estimates can be determined for discrete
sources. Moreover, it also allows investigation of the nearby
environment of discrete bright objects, which has previously been
impossible.

\section{Results}
A significant sky area of more than 1500 deg$^2$ has been reprocessed
from the original HIPASS data. We achieve an RMS sensitivity that is
improved by $\sim20$\% compared to the published HIPASS product. More
important than this improvement in noise, is the very effective
suppression of the negative artefacts surrounding all bright {\HI}
detections. We expect slightly different flux values, especially for
the bright {\HI} objects, as the newly processed data is more
sensitive to extended emission. Furthermore, we expect to have more
{\HI} detections of diffuse or companion galaxies due to the improved
sensitivity and artefact suppression.

For the detection of sources we have used two methods. Firstly an
automated source finder was employed to identify sources. Developing a
fully automated and reliably source finder is extremely difficult,
especially when looking for faint objects, therefore all the cubes
were also inspected visually to identify features that might be missed
by the automated source finder.

\subsection{Automatic source detection}

We use the {\it Duchamp} \citep{2008glv..book..343W} three dimensional
source finding algorithm to identify candidate detections. Features
are sought with a peak flux exceeding 5 times the noise
value. Detections are {\it grown down} from the peak to a cutoff value
of three times the noise, to be more sensitive to extended or diffuse
emission. The requirement for candidate sources to be accepted is to
have a size of at least 10 pixels (after growing) in one channel and
to contain pixels from at least 2 adjacent channels. As a result the
minimum velocity width of a detection is 26 km s$^{-1}$ which is
sufficient to detect most galaxies, but low mass dwarf galaxies or
companions might be missed. The search criteria
were chosen to eliminate isolated noise peaks and make the detections
more reliable. 

The original reduced data has a velocity resolution of 26 km
  s$^{-1}$. Hanning smoothed versions of the data cubes have been
  created at a velocity resolution of 52 and 104 km s$^{-1}$. These
  smoothed cubes are more sensitive to structures that are extended in
  the velocity domain.  {\it Duchamp} was applied in a similar way as
  to the original data cubes, however this did not result in additional
  detections.

Spectra and moment maps of all the candidate detections have
been inspected visually for reliability. Strong artefacts due to
e.g. solar interference can be easily identified in the integrated
maps and these features were eliminated from the source list. The
spectra can be used to identify false detections that are caused by
ripples in the bandpass seen toward continuum sources.

\subsection{Visual source detection}

Although automatic source finders are an excellent tool to search for
candidate detections they have their limitations. Detecting point
sources is relatively easy, as they are clearly defined in both the
spatial and velocity domain. We are especially interested in diffuse
features that have a peak brightness of only a few times the
noise. These sources can only be found when source finders are "tuned"
to the actual properties, the spatial and spectral extent, of these
features. Since we do not know the appearance or existence of the
sources beforehand, it is difficult to employ dedicated source
finders. Another complication is that the noise is not completely
uniform. The noise can be elevated, for example, in the vicinity of continuum
sources or in the case of solar interference. Several efforts have
been made to look for diffuse but very extended features, both
spatially and in velocity. Unfortunately these efforts have not been
very successful and resulted in previously known or unreliable
detections.
  
Visual inspection of the data was necessary, to have a better
understanding of the quality and features in the data, but also to
look for features that have been missed by the automated source finder. All
cubes were inspected in the spatial domain as well as in the two
velocity domains (RA versus velocity and Dec versus velocity) to
search for objects that had been missed. 

\subsection{Detected sources}

For each candidate detection an optical counterpart was sought in the
NASA Extragalactic Database (NED) \footnote{The NASA/IPAC
  Extragalactic Database (NED) is operated by the Jet Propulsion
  Laboratory, California Institute of Technology, under contract with
  the National Aeronautics and Space Administration.} within a search
radius of 14 arcmin. This radius is approximately the diameter of the
HIPASS beam and only objects within this radius can contribute to the
measured flux. Another requirement for potential optical counterparts
is to have a known radial velocity that is comparable to the radial velocity
of the {\HI} detection. Candidate detections with a clearly identified
optical counterpart are accepted. For each candidate detection the
noise is determined based on the line-width at 20\% of the peak
($W_{20}$). For line-widths smaller than 250 km s$^{-1}$ the noise is
given as:

\begin{equation}
n=\sqrt{\frac{1.5 W_{20}}{v_{res}}}\cdot \textrm{RMS} \cdot v_{res} 
\end{equation} 
where $n$ is in units of [Jy km s$^{-1}$], RMS is the sensitivity of
the cubes and $v_{res}$ is the velocity resolution. For broader
profiles the term $(1.5 W_{20})$ is replaced by ($W_{20} + 50$
km~s$^{-1}$) to account for the line wings.

Using this noise measurement, a signal-to-noise ratio can be
determined for each candidate detection. Candidate detections with an
integrated signal-to-noise larger than 5 are accepted as detections.

Finally, a list of 203 detections was obtained of which 31 were
detected by visual inspection; all detections and their observed
properties are listed in table.~\ref{sources}. The first column gives
the catalogue name, composed of the three characters "{\it HIR}", an
abbreviation of "{\it HI}PASS {\it R}eprocessed", followed by two sets
of four digits indicating the Right Ascension and Declination of the
detection. The second column gives the optical identification of the
detection if any is known and the third column gives the original
HIPASS ID if available.  Completely new {\HI} detections are
  indicated in the third column by {\it "new''}. In the case a
  detections is not mentioned in the HIPASS catalogue however is detected
  by ALFALFA, this is indicated with {\it "$\alpha\alpha$"}. The fourth,
fifth and sixth column give the spatial position and heliocentric
velocity of the source. This is followed by velocity width of the
object, at 20\% of the peak flux. The seventh and eighth columns give
the integrated flux (3-D) and the integrated line-strength (1-D)
respectively. We will discuss the differences between the two
different flux measurements in the following section.  The last column
indicates whether the object is detected by the automatic source
finder (A) or by visual inspection (V). In some cases the objects
are completely at the edge of the processed bandwidth. Although a
detection here can be a solid detection, the estimated flux and
line-width values are underestimated as a part of the spectrum is
missing.  For objects at the edge of the bandwidth, this is indicated
in the table with a letter $''(u)''$

%%%%%%%%%%%%%%%%%%
%%%%%%%%%%%%%%%%%%%
\onecolumn

\begin{landscape}
\begin{center}
\begin{longtable}{lllrrccccc}

\caption{Observed properties of {\HI} detections in reprocessed HIPASS
  data within the Westerbork Virgo Filament Survey region\\
{\it (a)}: Units of $V_{Hel}$ are given in [km s$^{-1}$].\\
{\it (b)}: Units of $W_{20}$ are given in [km s$^{-1}$].\\
{\it (c)}: Units of $S_{int}$ are given in [Jy km s$^{-1}$].\\
{\it (d)}: Units of $S_{line}$ are given in [Jy km s$^{-1}$].\\
{\it (u)}: The flux and line-width values are underestimated as the
object is at the edge of the processed bandwidth..\\
} 
\label{sources}\\

%This is the header for the first page of the table...

\hline
\hline
Name           &  Optical ID.              &   HIPASS ID &   RA [hh:mm:ss] & Dec [dd:mm:ss] & $V_{Hel}$$^{(a)}$  & $W_{20}$$^{(b)}$ & $S_{int}$$^{(c)}$  & $S_{line}$$^{(d)}$ & det. \\
\hline   
\endfirsthead

%This is the header for the remaining page(s) of the table...
\hline
\hline
Name           &  Optical ID.              &   HIPASS ID  &  RA [hh:mm:ss] & Dec [dd:mm:ss] & $V_{Hel}$$^{(a)}$  & $W_{20}$$^{(b)}$ & $S_{int}$$^{(c)}$ & $S_{line}$$^{(d)}$ & det. \\
\hline
\endhead

%This is the footer for all pages except the last page of the table...
\hline
  \multicolumn{8}{c}{{Continued on Next Page\ldots}} \\
\endfoot

%This is the footer for the last page of the table...
\hline \hline
\endlastfoot

%%%%%%%%%%%%%%%%%%%%%%%%%%%%%%%
%%%%%%%%%%%%%%%%%%%%%%%%%%%%%%

HIR0821-0025 &  UGC 04358              &   HIPASS J0821-00     &    8:21:41   &  -0:25:00    &     1775$^u$  &   50$^u$    &      4.1$^u$    &     3.5$^u$   &   A \\   
HIR0859+1109 &  UGC 04712             &    new    &    8:59:28   &  11:09:23    &     1643  &   80    &      1.9    &     3.0   &   A   \\   
HIR0906+0618 &  UGC 04781              &  HIPASS J0906+06     &    9:06:36   &   6:18:27    &     1433  &   168   &     19.7    &    16.0   &   A  \\   
HIR0908+0555 &  UGC 04797               &   HIPASS J0908+05a    &    9:08:12   &   5:55:54    &     1326  &   102   &      5.6    &     5.3    &   A \\   
HIR0908+0517 &  SDSS J090836.54+051726.8 &   HIPASS J0908+05b   &    9:08:43   &   5:17:49    &     600   &   57    &      1.3    &     2.2    &   A \\   
HIR0910+0711 &  NGC 2777                  &  HIPASSJ0910+07  &    9:10:34   &   7:11:43    &     1500  &   139   &     11.6    &     9.9     &   A\\   
HIR0911+0024 &  No object found.	&   new   &    9:11:21   &   0:24:59    &     1286  &   83    &      2.8    &     2.7     &   A\\   
HIR0921+0725 &  No object found.	&   new  &    9:21:07   &   7:25:32    &     1369  &   110   &      3.7    &     4.5     &   A\\   
HIR0944-0038 &  UGC 05205               &  HIPASS J0944-00a     &    9:44:06   &  -0:38:42    &     1485  &   178   &     15.0    &    13.1     &   A\\   
HIR0944+0937 &  IC 0559                 &   HIPASS J0944+09    &    9:44:31   &   9:37:06    &     522   &   132   &      5.6    &     4.8     &   A\\   
HIR0944-0040 &  SDSS J094446.23-004118.2  &  HIPASS J0944-00b   &    9:44:43   &  -0:40:37    &     1222  &   161   &      7.0    &     8.5   &   A  \\   
HIR0946+0141 &  SDSS J094602.54+014019.4  &  new  &    9:46:00   &   1:41:07    &     1763$^u$  &   53$^u$    &      1.0$^u$    &     2.3$^u$    &   A \\   
HIR0946+0031 &  UGC 05238                 & HIPASS J0946+00    &    9:46:55   &   0:31:09    &     1697$^u$  &   120$^u$   &      7.0$^u$    &     7.7$^u$     &   A\\   
HIR0947+0241 &  UGC 05249                &  HIPASS J0947+02    &    9:47:44   &   2:41:15    &     1776  &   25    &      1.3    &     1.6     &   A\\   
HIR0951+0750 &  UGC 05288               &   HIPASS J0951+07    &    9:51:16   &   7:50:12    &     555   &   112   &     33.0    &     27.8    &   A\\   
HIR0953+0135 &  NGC 3044                &    HIPASS J0953+01   &    9:53:42   &   1:35:06    &     1292  &   341   &     47.3    &    49.4     &   A\\   
HIR0954+0915 &  NGC 3049                 &   HIPASS J0954+09   &    9:54:41   &   9:15:53    &     1471  &   223   &      6.1    &     9.1     &   A\\   
HIR1005+0139 &  2dFGRS N421Z115      &    new     &   10:05:20   &   1:39:48    &     1260  &   95    &      2.0    &     2.4    &   V \\   
HIR1007+1022 &  UGC 05456               &  HIPASS J1007+10    &   10:07:18   &  10:22:18    &     547   &   90    &      4.8    &     7.2     &   A\\   
HIR1013+0702 &  UGC 05522               &   HIPASS J1013+07    &   10:13:57   &   7:02:36    &     1219  &   232   &     44.7    &    42.8     &   A\\   
HIR1014+0329 &  NGC 3169                &  HIPASS J1014+03     &   10:14:17   &   3:29:28    &     1246  &   457   &    141.4    &    98.7     &   A\\   
HIR1015+0242 &  UGC 05539              &   HIPASS J1015+02     &   10:15:52   &   2:42:11    &     1275  &   157   &     11.4    &    13.7     &   A\\   
HIR1017+0421 &  UGC 05551              &    HIPASS J1017+04   &   10:17:12   &   4:21:37    &     1340  &   90    &      9.4    &     6.0     &   A\\   
HIR1028+0335 &  UGC 05677              &  HIPASS J1028+03      &   10:28:33   &   3:35:28    &     1146  &   139   &      3.2    &     5.4     &   A\\   
HIR1031+0428 &  UGC 05708              &   HIPASS J1031+04     &   10:31:15   &   4:28:27    &     1169  &   198   &     33.9    &    33.9     &   A\\   
HIR1038+1024 &  CGCG 065-074          &    HIPASS J1038+10     &   10:38:14   &  10:24:26    &     1167  &   192   &      8.1    &     6.9     &   A\\   
HIR1039+0145 &  UGC 05797               &   HIPASS J1039+01    &   10:39:28   &   1:45:15    &     706   &   71    &      3.9    &     3.0     &   A\\   
HIR1044+1134 &  MESSIER 095             &   HIPASS J1044+11   &   10:44:02   &  11:34:18    &     798   &   280   &     12.2    &    32.8     &   A\\   
HIR1046+0149 &  NGC 3365                 &   HIPASS J1046+01   &   10:46:11   &   1:49:32    &     985   &   248   &     36.1    &    43.1     &   A\\   
HIR1051+0550 &  NGC 3423                &  HIPASS J1051+05     &   10:51:17   &   5:50:24    &     1010  &   186   &     42.4    &    45.8     &   A\\   
HIR1051+0327 &  HIPASS J1051+03        &   HIPASS J1051+03    &   10:51:31   &   3:27:20    &     1062  &   85    &     13.0    &    13.3     &   A\\   
HIR1051+0435 &  UGC 05974                &  HIPASS J1051+04    &   10:51:34   &   4:35:53    &     1050  &   175   &     11.1    &    15.3     &   A\\   
HIR1052+0002 &  MGC 0013223             &    new  &   10:52:49   &   0:02:35    &     1776$^u$  &   30$^u$    &      1.3$^u$    &     1.2$^u$     &   A\\   
HIR1053+0232 &  LSBC L1-137              &   HIPASS J1053+02   &   10:53:08   &   2:32:41    &     1035  &   89    &      6.4    &     6.4     &   A\\   
HIR1055+0511 &  No object found.        &   new   &   10:55:27   &   5:11:57    &     982   &   142   &      6.4    &     3.8     &   V\\   
HIR1101+0338 &  NGC 3495                 &  HIPASS J1101+03    &   11:01:15   &   3:38:01    &     1119  &   324   &     27.4    &    30.5     &   A\\   
HIR1105-0002 &  NGC 3521                &  HIPASS J1105-00     &   11:05:48   &  -0:02:05    &     793   &   444   &    298.8    &   199.0     &   A\\   
HIR1107+0710 &  NGC 3526                 &  HIPASS J1106+07    &   11:07:00   &   7:10:54    &     1422  &   194   &     10.6    &     6.6     &   A\\   
HIR1110+0107 &  CGCG 011-018           &   HIPASS J1110+01     &   11:10:56   &   1:07:45    &     983   &   96    &      3.4    &     5.1     &   A\\   
HIR1112+1014 &  UGC 06248                &  HIPASS J1112+10   &   11:12:45   &  10:14:03    &     1286  &   75    &      1.8    &     2.7     &   A\\   
HIR1117+0434 &  NGC 3604                 &  HIPASS J1117+04    &   11:17:30   &   4:34:25    &     1511  &   209   &      7.9    &    10.8     &   A\\   
HIR1119+0939 &  SDSS J111928.10+093544.2 &  HIPASS J1119+09    &   11:19:46   &   9:39:12    &     1075  &   298   &      4.6    &     3.6    &   V \\   
HIR1120+0232 &  UGC 06345                &  HIPASS J1120+02    &   11:20:11   &   2:32:48    &     1604  &   141   &     19.7    &    22.0     &   A\\   
HIR1124+0318 &  NGC 3664                  &  HIPASS J1124+03   &   11:24:25   &   3:18:44    &     1379  &   140   &     20.5    &    22.7     &   A\\   
HIR1124+1121 &  NGC 3666                 &  HIPASS J1124+11    &   11:24:25   &  11:21:16    &     1056  &   278   &     41.0    &    45.9     &   A\\   
HIR1125+0958 &  IC 0692                   &  HIPASS J1126+10   &   11:25:53   &   9:58:09    &     1154  &   102   &      3.1    &     3.5     &   A\\   
HIR1127+0846 &  IC 2828                   &  HIPASS J1127+08   &   11:27:03   &   8:46:06    &     1048  &   79    &      1.0    &     1.6     &   V\\   
HIR1127-0058 &  UGC 06457              &    HIPASS J1127-00    &   11:27:10   &  -0:58:30    &     956   &   106   &      7.2    &     7.9     &   A\\   
HIR1128+0923 &  NGC 3692                &   HIPASS J1128+09a    &   11:28:19   &   9:23:35    &     1617  &   236   &     11.7    &     9.4     &   A\\   
HIR1130+0917 &  NGC 3705                &   HIPASS J1130+09    &   11:30:02   &   9:17:38    &     1018  &   353   &     29.6    &    34.4     &   A\\   
HIR1136+0049 &  UGC 06578              &    HIPASS J1136+00b    &   11:36:32   &   0:49:04    &     1099  &   95    &      5.5    &     5.7     &   A\\   
HIR1140+1128 &  NGC 3810                &   HIPASS J1140+11    &   11:40:59   &  11:28:07    &     995   &   268   &     33.6    &    38.5     &   A\\   
HIR1144+0210 &  SDSS J114454.28+020946.8 &  HIPASS J1145+02    &   11:44:52   &   2:10:15    &     1009  &   52    &      4.1    &     3.5     &   A\\   
HIR1158-0127 &  UGC 06970                &  HIPASS J1158-01    &   11:58:41   &  -1:27:28    &     1471  &   166   &     46.4    &     4.4     &   A\\   
HIR1200-0105 &  NGC 4030                 &   HIPASS J1200-01   &   12:00:26   &  -1:05:52    &     1465  &   346   &     64.0    &    50.8     &   A\\   
HIR1204-0131 &  UGC 07053               &  HIPASS J1204-01     &   12:04:17   &  -1:31:37    &     1469  &   123   &      6.5    &     8.2     &   A\\   
HIR1207+0249 &  HIPASS J1208+02        &   HIPASS J1208+02    &   12:07:57   &   2:49:05    &     1318  &   220   &     84.7    &    55.7     &   A\\   
HIR1211+0201 &  UGC 07178                &   HIPASS J1211+02a   &   12:11:04   &   2:01:37    &     1334  &   95    &      6.2    &     8.0     &   A\\   
HIR1211+0256 &  UGC 07185                &   HIPASS J1211+02b   &   12:11:26   &   2:56:53    &     1300  &   110   &      5.8    &     7.6     &   A\\   
HIR1212+0248 &  LEDA 135791             &   new   &   12:12:27   &   2:48:29    &     877   &   117   &      3.1    &     3.9     &   A\\   
HIR1212+1054 &  NGC 4178                 &   HIPASS J1212+10   &   12:12:51   &  10:54:29    &     476$^u$   &   56$^u$    &     10.0$^u$    &     8.7$^u$     &   A\\   
HIR1214+0747 &  UGC 07239               &  HIPASS J1214+07     &   12:14:12   &   7:47:05    &     1220  &   148   &      5.3    &     5.3     &   A\\   
HIR1214+0911 &  VCC 0117                 &  HIPASS J1214+09    &   12:14:48   &   9:11:33    &     1776$^u$  &   64$^u$    &      1.4$^u$    &     1.8$^u$     &   A\\   
HIR1215+0935 &  NGC 4207                 &   HIPASS J1215+09a   &   12:15:26   &   9:35:37    &     587   &   247   &      3.7    &     4.2     &   V\\   
HIR1217+1001 &  UGC 07307               &   HIPASS J1216+10    &   12:17:05   &  10:01:41    &     1178  &   73    &      6.0    &     5.6     &   A\\   
HIR1217+0027 &  UGC 07332               &   HIPASS J1217+00    &   12:17:57   &   0:27:01    &     932   &   81    &     15.4    &    18.7     &   A\\   
HIR1218+0640 &  NGC 4241                 &  HIPASS J1218+06    &   12:18:00   &   6:40:59    &     720   &   126   &      5.4    &     7.3     &   A\\   
HIR1219+0639 &  VCC 0381                 &   HIPASS J1219+06b  &   12:19:50   &   6:39:46    &     481$^u$   &   76$^u$    &      2.0$^u$    &     2.6$^u$     &   A\\   
HIR1220+0019 &  CGCG 014-010            &  HIPASS J1220+00     &   12:20:10   &   0:19:35    &     884   &   83    &      2.3    &     2.5     &   A\\   
HIR1220+0126 &  UGC 07394               &   HIPASS J1220+01    &   12:20:34   &   1:26:39    &     1617  &   226   &      5.5    &     4.7     &   V\\   
HIR1221+0429 &  MESSIER 061             &   HIPASS J1221+04    &   12:21:51   &   4:29:23    &     1562  &   180   &     85.1    &    87.2     &   A\\   
HIR1222+0434 &  NGC 4301                &                          &  12:22:28   &   4:34:36    &     1267  &   135   &     14.9    &    17.1     &   A\\   
HIR1222+0814 &  VCC 0566                 &         $\alpha \alpha$        &   12:22:40   &   8:14:39    &     1392  &   106   &      4.1    &     2.9     &   V\\   
HIR1222+1118 &  NGC 4330                &         $\alpha \alpha$         &   12:22:56   &  11:18:37    &     1590  &   273   &      5.6    &     7.6     &   A\\   
HIR1223+0922 &  NGC 4316                &   HIPASS J1222+09    &   12:23:01   &   9:22:46    &     1233  &   321   &      7.3    &     3.8     &   V\\   
HIR1223+0517 &  NGC 4324                &   HIPASS J1223+05    &   12:23:15   &   5:17:18    &     1540  &   131   &      2.9    &     4.1     &   A\\   
HIR1224+0636 &  IC 3268                  &                         &   12:24:08   &   6:36:20    &     720   &   129   &      5.9    &     5.3     &   A\\   
HIR1224+0359 &  VCC 0737               &   HIPASS J1224+03a     &   12:24:37   &   3:59:05    &     1720$^u$  &   137$^u$   &      4.2$^u$    &     3.4$^u$     &   A\\   
HIR1224+0319 &  VCC 0739               &    HIPASS J1224+03b    &   12:24:41   &   3:19:30    &     921   &   68    &     13.8    &    11.4     &   A\\   
HIR1225+0545 &  NGC 4376               &    HIPASS J1225+05    &   12:25:19   &   5:45:19    &     1141  &   123   &      3.0    &     3.8      &   A\\   
HIR1225+0714 &  IC 3322A                 &    HIPASS J1225+07  &   12:25:36   &   7:14:11    &     996   &   298   &     24.6    &    21.0     &   V\\   
HIR1225+0210 &  UGC 07512               &   HIPASS J1225+02    &   12:25:42   &   2:10:01    &     1498  &   78    &      4.5    &     6.2     &   A\\   
HIR1225+0548 &  VCC 0848                &   HIPASS J1226+05    &   12:25:50   &   5:48:50    &     1531  &   207   &      5.2    &     6.7     &   A\\   
HIR1226+1026 &  NGC 4390                &  HIPASS J1225+10     &   12:26:05   &  10:26:35    &     1101  &   152   &      4.3    &     2.4     &   V\\   
HIR1226+0853 &  NGC 4411                &       $\alpha\alpha$           &   12:26:40   &   8:53:28    &     1269  &   108   &     24.2    &    21.2     &   A\\   
HIR1226+1131 &  IC 3356                  &  HIPASS J1226+11    &   12:26:52   &  11:31:08    &     1097  &   95    &     11.3    &    17.3     &   A\\   
HIR1226+0230 &  NGC 4409              &   HIPASS J1226+02      &   12:26:54   &   2:30:21    &     1680$^u$  &   193$^u$   &      9.7$^u$    &    10.6$^u$     &   A\\   
HIR1226+0800 &  NGC 4416              &      $\alpha\alpha$        &   12:26:56   &   8:00:26    &     1392  &   125   &      2.5    &     2.6     &   A\\   
HIR1227+0553 &  NGC 4423              &    HIPASS J1227+05     &   12:27:10   &   5:53:45    &     1128  &   193   &     10.8    &    10.8     &   A\\   
HIR1227+1052 &  IC 3371                 &   HIPASS J1227+10    &   12:27:12   &  10:52:42    &     930   &   188   &      7.7    &     9.2     &   A\\   
HIR1227+0713 &  UGC 07557             &    HIPASS J1227+07     &   12:27:13   &   7:13:56    &     930   &   172   &     23.2    &    21.2     &   V\\   
HIR1227+0615 &  NGC 4430               &    HIPASS J1227+06    &   12:27:20   &   6:15:58    &     1415  &   143   &      3.6    &     5.0     &   A\\   
HIR1227+0132 &  HI 1225+01            &     HIPASS J1227+01    &   12:27:27   &   1:32:48    &     1286  &   85    &     35.1    &    25.1     &   A\\   
HIR1228+0843 &  UGC 07590              &   HIPASS J1228+08     &   12:28:21   &   8:43:54    &     1115  &   179   &      8.0    &     8.6     &   A\\   
HIR1228+0334 &  NGC 4457                &   HIPASS J1228+03    &   12:28:44   &   3:34:38    &     890   &   138   &      3.6    &     3.1     &   A\\   
HIR1229+0243 &  UGC 07612              &            &   12:29:11   &   2:43:53    &     1630  &   62    &      4.1    &     4.0     &   A\\   
HIR1229+0644 &  IC 3414                  &  HIPASS J1229+06    &   12:29:26   &   6:44:45    &     550$^u$   &   185$^u$   &      7.0$^u$    &     4.6$^u$     &   A\\   
HIR1230+0013 &  No object found.       &     new        &   12:30:29   &   0:13:01    &     1524  &   138   &      2.1    &     3.4     &   V\\   
HIR1230+0929 &  HIPASS J1230+09       &    HIPASS J1230+09    &   12:30:41   &   9:29:04    &     495$^u$   &   71$^u$    &      2.8$^u$    &     2.4$^u$     &   A\\   
HIR1231+0145 &  CGCG 014-054           &     new        &   12:31:02   &   1:45:09    &     1101  &   52    &      1.6    &     1.2     &   V\\   
HIR1231+0357 &  NGC 4496A               &   HIPASS J1231+03    &   12:31:36   &   3:57:15    &     1736$^u$  &   145$^u$   &     34.2$^u$    &    43.7$^u$     &   A\\   
HIR1232+0024 &  NGC 4517A              &    HIPASS J1232+00a    &   12:32:31   &   0:24:28    &     1520  &   182   &     36.9    &    35.7     &   A\\   
HIR1232+0007 &  NGC 4517                &   HIPASS J1232+00b    &   12:32:43   &   0:07:30    &     1133  &   325   &    108.9    &   113.5     &   A\\   
HIR1233+0436 &  VCC 1468                &           &   12:33:01   &   4:36:54    &     1220  &   62    &      0.8    &     1.4     &   V\\   
HIR1233-0032 &  HIPASS J1233-00       &    HIPASS J1233-00    &   12:33:11   &  -0:32:50    &     719   &   97    &      1.8    &     2.6     &   A\\   
HIR1233+0840 &  NGC 4519                 &   HIPASS J1233+08   &   12:33:26   &   8:40:24    &     1207  &   219   &     48.9    &    49.6     &   A\\   
HIR1234+0236 &  NGC 4527                &   HIPASS J1234+02a    &   12:34:02   &   2:36:40    &     1723$^u$  &   241$^u$   &     55.1$^u$    &    56.2$^u$     &   V\\   
HIR1234+0332 &  UGC 07715              &            &   12:34:03   &   3:32:42    &     1101  &   114   &      0.7    &     1.8     &   V\\   
HIR1234+0212 &  NGC 4536               &   HIPASS J1234+02b     &   12:34:34   &   2:12:32    &     1776$^u$  &   153$^u$   &     29.6$^u$    &    34.6$^u$     &   V\\   
HIR1236+0638 &  IC 3576                 &   HIPASS J1236+06    &   12:36:35   &   6:38:14    &     1070  &   71    &     13.0    &    13.4     &   A\\   
HIR1236+0306 &  UGC 07780             &    HIPASS J1236+03     &   12:36:40   &   3:06:19    &     1445  &   141   &      2.9    &     4.1     &   A\\   
HIR1237+0655 &  IC 3591                 &   HIPASS J1237+06    &   12:37:00   &   6:55:52    &     1626  &   132   &      7.1    &     7.9     &   A\\   
HIR1239-0031 &  NGC 4592              &    HIPASS J1239-00     &   12:39:17   &  -0:31:19    &     1070  &   172   &    162.7    &    98.1    &   A \\   
HIR1241+0124 &  UGC 07841             &    HIPASS J1241+01     &   12:41:15   &   1:24:40    &     1670$^u$  &   158$^u$   &      9.4$^u$    &     7.0$^u$     &   A\\   
HIR1242+0547 &  VCC 1918              &    HIPASS J1242+05     &   12:42:18   &   5:47:07    &     983   &   112   &      1.4    &     1.8     &   A\\   
HIR1242-0120 &  NGC 4629              &    HIPASS J1242-01     &   12:42:28   &  -1:20:26    &     1109  &   172   &     22.7    &    22.8     &   A\\   
HIR1242+0358 &  NGC 4630              &    HIPASS J1242+03     &   12:42:32   &   3:58:17    &     719   &   151   &      4.0    &     5.6     &   A\\   
HIR1242-0004 &  NGC 4632               &   HIPASS J1242-00     &   12:42:33   &  -0:04:57    &     1696$^u$  &   185$^u$   &     36.6$^u$    &    34.1$^u$     &   A\\   
HIR1243+0739 &  VCC 1952               &   HIPASS J1243+07    &   12:43:11   &   7:39:24    &     1311  &   86    &      1.9    &     2.4     &   A\\   
HIR1243+1132 &  NGC 4647               &     $\alpha\alpha$      &   12:43:28   &  11:32:37    &     1394  &   201   &      2.4    &     4.3     &   A\\   
HIR1244+0028 &  UGC 07911             &    HIPASS J1244+00     &   12:44:23   &   0:28:19    &     1178  &   124   &     12.1    &    10.4     &   A\\   
HIR1245-0027 &  NGC 4666               &    HIPASS J1245-00    &   12:45:11   &  -0:27:43    &     1524  &   402   &     75.3    &    78.0     &   A\\   
HIR1246+0557 &  UGC 07943             &    HIPASS J1246+05     &   12:46:45   &   5:57:37    &     830   &   141   &      6.2    &     7.6     &   A\\   
HIR1247+0420 &  NGC 4688               &    HIPASS J1247+04    &   12:47:46   &   4:20:33    &     982   &   73    &     31.1    &    31.8     &   A\\   
HIR1247+1058 &  VCC 2062                &     $\alpha\alpha$     &   12:47:56   &  10:58:09    &     1140  &   131   &      6.7    &     9.1     &   A\\   
HIR1248+0826 &  NGC 4698                &   HIPASS J1248+08    &   12:48:28   &   8:26:22    &     995   &   430   &     19.9    &    17.9     &   A\\   
HIR1249+0325 &  NGC 4701                &   HIPASS J1249+03    &   12:49:11   &   3:25:10    &     722   &   178   &     58.2    &    55.7     &   A\\   
HIR1249+0519 &  NGC 4713                &   HIPASS J1250+05    &   12:49:57   &   5:19:33    &     649   &   189   &     47.5    &    46.9     &   A\\   
HIR1253+0428 &  NGC 4765                &   HIPASS J1253+04    &   12:53:11   &   4:28:09    &     719   &   118   &     13.4    &    19.2     &   A\\   
HIR1253+0115 &  NGC 4771                &   HIPASS J1253+01    &   12:53:24   &   1:15:08    &     1128  &   304   &      9.3    &     9.9     &   A\\   
HIR1253+0212 &  NGC 4772                &   HIPASS J1253+02    &   12:53:29   &   2:12:49    &     1035  &    43   &      6.9    &     6.8     &   A\\   
HIR1254+0240 &  NGC 4809               &           &   12:54:51   &   2:40:16    &     918   &   158   &     10.9    &    11.7     &   A\\   
HIR1255+0008 &  UGC 08041             &   HIPASS J1255+00      &   12:55:12   &   0:08:12    &     1311  &   205   &     17.0    &    17.9     &   A\\   
HIR1255+0414 &  NGC 4808               &   HIPASS J1255+04b     &   12:55:50   &   4:14:24    &     741   &   270   &     81.4    &    60.5     &   A\\   
HIR1257+0242 &  UGC 08074             &    HIPASS J1257+02     &   12:57:49   &   2:42:26    &     918   &   113   &      3.0    &     3.9     &   A\\   
HIR1300+0230 &  NGC 4900               &    HIPASS J1300+02b    &   13:00:37   &   2:30:28    &     943   &   133   &     16.8    &    17.2     &   A\\   
HIR1300-0000 &  NGC 4904                &   HIPASS J1300-00    &   13:00:53   &  -0:00:55    &     1167  &   198   &      8.0    &     9.0     &   A\\   
HIR1306+1027 &  CGCG 071-109        &   HIPASS J1306+10       &   13:06:26   &  10:27:07    &     928   &   65    &      3.1    &     4.4     &   A\\   
HIR1311+0530 &  UGC 08276              &    HIPASS J1312+05   &   13:11:56   &   5:30:11    &     908   &   113   &      2.6    &     3.5     &   A\\   
HIR1312+0711 &  UGC 08285              &   HIPASS J1312+07     &   13:12:32   &   7:11:02    &     890   &   134   &      3.8    &     4.9     &   A\\   
HIR1313+1012 &  UGC 08298              &   HIPASS J1313+10     &   13:13:19   &  10:12:16    &     1152  &   96    &     12.9    &    12.3     &   A\\   
HIR1317-0100 &  UM 559                  &   HIPASS J1317-00    &   13:17:43   &  -1:00:31    &     1207  &   111   &      3.9    &     3.4     &   A\\   
HIR1320+0524 &  UGC 08382             &    HIPASS J1320+05    &   13:20:36   &   5:24:36    &     956   &   119   &      4.6    &     4.3     &   A\\   
HIR1320+0947 &  UGC 08385             &    HIPASS J1320+09     &   13:20:38   &   9:47:51    &     1115  &   162   &     11.2    &    14.1     &   A\\   
HIR1326+0206 &  NGC 5147               &    HIPASS J1326+02    &   13:26:19   &   2:06:48    &     1087  &   171   &     13.1    &    17.3     &   A\\   
HIR1326+0229 &  SDSS J132615.73+022729.5 &   HIPASS J1326+02B   &   13:26:29   &   2:29:57    &     1035  &   110   &      0.8    &     1.0    &   V \\   
HIR1327+1003 &  UGC 08450                &  HIPASS J1327+10    &   13:27:08   &  10:03:44    &     1049  &   103   &      4.4    &     3.2     &   A\\   
HIR1328+0219 &  HIPASS J1328+02       &    HIPASS J1328+02    &   13:28:05   &   2:19:03    &     1022  &   73    &      7.4    &     3.1     &   A\\   
HIR1337+0853 &  NGC 5248               &    HIPASS J1337+08   &   13:37:17   &   8:53:53    &     1158  &   288   &    100.1    &    78.0     &   A\\   
HIR1337+0739 &  UGC 08614             &   HIPASS J1337+07      &   13:37:27   &   7:39:15    &     1040  &   175   &     16.0    &    22.9     &   A\\   
HIR1338+0826 &  UGC 08629              &    HIPASS J1338+08    &   13:38:40   &   8:26:50    &     1022  &   143   &      0.9    &     3.2     &   V\\   
HIR1348+0356 &  NGC 5300               &   HIPASS J1348+03     &   13:48:10   &   3:56:31    &     1170  &   221   &     11.0    &    15.2     &   A\\   
HIR1352-0105 &  NGC 5334               &   HIPASS J1352-01     &   13:52:53   &  -1:05:30    &     1387  &   228   &     24.8    &    26.7     &   A\\   
HIR1355+0504 &  NGC 5364               &    HIPASS J1356+05    &   13:55:41   &   5:04:48    &     1260  &   291   &     49.4    &    48.8     &   A\\   
HIR1401+0247 &  No object found.       &    new      &   14:01:04   &   2:47:58    &     1040  &   111   &      3.4    &     2.8     &   V\\   
HIR1404+0848 &  UGC 08995             &   HIPASS J1404+08b      &   14:04:49   &   8:48:43    &     1234  &   182   &      9.2    &     7.5     &   A\\   
HIR1411-0109 &  NGC 5496               &   HIPASS J1411-01     &   14:11:38   &  -1:09:18    &     1541  &   267   &     67.7    &    66.2     &   A\\   
HIR1416+0350 &  HIPASS J1416+03       &    HIPASS J1416+03    &   14:16:59   &   3:50:21    &     1470  &   116   &      7.0    &     8.8     &   A\\   
HIR1417-0130 &  2dFGRS N275Z229       &    HIPASS J1417-01     &   14:17:14   &  -1:30:07    &     1551  &   88    &      2.2    &     2.4     &   A\\   
HIR1417+0651 &  No object found         &   new      &   14:17:45   &   6:51:08    &     1167  &   92    &      2.3    &     2.4     &   V\\   
HIR1419+0922 &  UGC 09169               &   HIPASS J1419+09    &   14:19:44   &   9:22:36    &     1280  &   165   &     20.3    &    23.1     &   A\\   
HIR1420+0358 &  NGC 5569                &           &   14:20:29   &   3:58:41    &     1750$^u$  &   135$^u$   &      7.8$^u$    &     9.0$^u$     &   A\\   
HIR1420+0834 &  SDSS J142044.53+083735.8 &  HIPASS J1420+08    &   14:20:48   &   8:34:55    &     1286  &   130   &      2.5    &     3.2    &   V \\   
HIR1421+0326 &  NGC 5577                &   HIPASS J1421+03    &   14:21:28   &   3:26:51    &     1485  &   275   &      9.0    &     8.4     &   A\\   
HIR1422-0022 &  NGC 5584               &    HIPASS J1422-00    &   14:22:25   &  -0:22:57    &     1657  &   223   &     30.8    &    31.0     &   A\\   
HIR1423+0143 &  UGC 09215             &     HIPASS J1423+01    &   14:23:29   &   1:43:21    &     1383  &   239   &     20.7    &    22.1     &   A\\   
HIR1424+0820 &  UGC 09225              &   HIPASS J1424+08     &   14:24:22   &   8:20:17    &     1247  &   130   &      4.9    &     5.6     &   A\\   
HIR1427+0842 &  UGC 09249             &    HIPASS J1427+08     &   14:27:02   &   8:42:16    &     1365  &   148   &      9.8    &     9.7     &   A\\   
HIR1429-0000 &  UGC 09299             &    HIPASS J1429-00     &   14:29:34   &  -0:00:24    &     1535  &   213   &     50.8    &    46.3     &   A\\   
HIR1430+0717 &  NGC 5645              &    HIPASS J1430+07     &   14:30:39   &   7:17:03    &     1365  &   195   &     19.3    &    17.3     &   A\\   
HIR1431+0257 &  IC 1024                 &   HIPASS J1431+03    &   14:31:21   &   2:57:38    &     1445  &   215   &      8.6    &    10.9     &   V\\   
HIR1432+0954 &  NGC 5669              &    HIPASS J1432+09     &   14:32:41   &   9:54:10    &     1365  &   211   &     31.8    &    41.7     &   A\\   
HIR1432+0257 &  CGCG 047-085         &            &   14:32:42   &   2:57:51    &     1537  &   146   &      4.1    &     5.4     &   V\\   
HIR1432+0016 &  UGC 09348              &           &   14:32:55   &   0:16:36    &     1657  &   198   &      5.5    &     5.2     &   V\\   
HIR1433+0426 &  NGC 5668                &    HIPASS J1433+04   &   14:33:29   &   4:26:57    &     1577  &   124   &     50.7    &    50.1     &   A\\   
HIR1435+0517 &  UGC 09385              &    HIPASS J1435+05    &   14:35:22   &   5:17:30    &     1635  &   106   &      7.4    &     7.8     &   A\\   
HIR1437+0217 &  NGC 5690                &    HIPASS J1437+02   &   14:37:40   &   2:17:30    &     1710$^u$  &   193$^u$   &     10.2$^u$    &    13.0$^u$     &   A\\   
HIR1439+0259 &  UGC 09432              &   HIPASS J1439+02     &   14:39:06   &   2:59:08    &     1564  &   110   &      7.4    &     6.4     &   A\\   
HIR1439+0521 &  NGC 5701                &   HIPASS J1439+05    &   14:39:10   &   5:21:44    &     1499  &   139   &     77.2    &    57.4     &   A\\   
HIR1439-0040 &  NGC 5705                &   HIPASS J1439-00    &   14:39:45   &  -0:40:20    &     1736$^u$  &   136$^u$   &     18.2$^u$    &    19.2$^u$     &   A\\   
HIR1440-0026 &  NGC 5719                &   HIPASS J1440-00    &   14:40:43   &  -0:26:53    &     1643  &   247   &     36.9    &    31.2     &   A\\   
HIR1440+0210 &  NGC 5725               &    HIPASS J1440+02    &   14:40:46   &   2:10:53    &     1617  &   179   &      1.8    &     3.4     &   V\\   
HIR1443+0451 &  IC 1048                 &   HIPASS J1443+04    &   14:43:02   &   4:51:25    &     1670$^u$  &   307$^u$   &     19.0$^u$    &    18.2$^u$     &   V\\   
HIR1444+0142 &  NGC 5740               &    HIPASS J1444+01    &   14:44:25   &   1:42:53    &     1556  &   334   &     30.4    &    23.8     &   A\\   
HIR1445+0751 &  UGC 09500              &   HIPASS J1445+07    &   14:45:18   &   7:51:59    &     1682  &   56    &      9.7    &     9.4     &   A\\   
HIR1446+1011 &  No object found         &    new     &   14:46:27   &  10:11:55    &     969   &   154   &      3.5    &     3.9     &   V\\   
HIR1453+0333 &  NGC 5774                &  HIPASS J1453+03     &   14:53:45   &   3:33:13    &     1723$^u$  &   308$^u$   &     89.4$^u$    &    84.9$^u$     &   A\\   
HIR1458-0104 &  NGC 5792                 &   HIPASS J1458-01   &   14:58:30   &  -1:04:54    &     1740$^u$  &   92$^u$    &     16.4$^u$    &    16.9$^u$     &   A\\   
HIR1458+0646 &  KUG 1456+069           &   HIPASS J1458+06     &   14:58:35   &   6:46:30    &     1681  &   138   &      6.2    &     4.6     &   A\\   
HIR1500+0155 &  NGC 5806                 &  HIPASS J1500+01    &   15:00:00   &   1:55:55    &     1352  &   322   &      9.1    &     7.4     &   V\\   
HIR1504-0052 &  UGC 09682                &   HIPASS J1504-00   &   15:04:20   &  -0:52:29    &     1776$^u$  &   101$^u$   &      3.9$^u$    &     3.5$^u$     &   A\\   
HIR1515+0603 &  No object found        &    new      &   15:15:02   &   6:03:22    &     1696  &   103   &      2.2    &     3.1     &   V\\   
HIR1521+0505 &  NGC 5921                &   HIPASS J1521+05    &   15:21:56   &   5:05:01    &     1472  &   189   &     35.2    &    31.9     &   A\\   
HIR1537+0558 &  NGC 5964                &   HIPASS J1537+05    &   15:37:35   &   5:58:48    &     1446  &   212   &     36.5    &    38.6     &   A\\   
HIR1546+0654 &  UGC 10023               &    HIPASS J1546+06   &   15:46:05   &   6:54:55    &     1407  &   170   &      5.5    &     5.2     &   A\\   
HIR1605+0832 &  CGCG 079-046            &   	HIPASS J1606+08    &   16:05:40   &   8:32:03    &     1379  &   153   &      4.3    &     4.7     &   V\\   
HIR1608+0733 &  IC 1197                  &   HIPASS J1608+07   &   16:08:15   &   7:33:15    &     1360  &   221   &     13.5    &    11.4     &   A\\   
HIR1609-0006 &  UGC 10229              &   	HIPASS J1609-00    &   16:09:42   &  -0:06:08    &     1491  &   129   &      9.1    &     8.1     &   A\\   
HIR1618+0725 &  NGC 6106               &    HIPASS J1618+07    &   16:18:45   &   7:25:10    &     1445  &   253   &     19.7    &    20.7     &   A\\   
HIR1619+0142 &  CGCG 024-001          &    HIPASS J1619+01a     &   16:19:19   &   1:42:02    &     1497  &   165   &      6.8    &     9.5     &   A\\   
HIR1656+0800 &  HIPASS J1656+08       &    HIPASS J1656+08     &   16:56:39   &   8:00:53    &     1471  &   113   &      3.0    &     4.3     &   A\\   
HIR1728+0725 &  UGC 10862              &   HIPASS J1728+07     &   17:28:07   &   7:25:35    &     1685  &   150   &     13.1    &    15.7     &   A\\   
HIR1732+0705 &  NGC 6384               &   HIPASS J1732+07     &   17:32:26   &   7:05:45    &     1683$^u$  &   313$^u$   &     42.6$^u$    &    50.5$^u$     &   A\\

%%%%%%%%%%%%%%%%%%%%%%%%%%%%%%%
%%%%%%%%%%%%%%%%%%%%%%%%%%%%%%%

\end{longtable}
\end{center}

\end{landscape}

\twocolumn

%%%%%%%%%%%%%%%%%%%
%%%%%%%%%%%%%%%%%%

\subsection{New {\HI} detections}
All sources in the HIPASS catalogue have been re-detected in the
reprocessed HIPASS product. Apart from these previously known {\HI}
detections, 29 detections have been made that were not listed in the
HIPASS catalogue, of which 14 are completely new {\HI}
detections. Thirteen of the detections that were not listed in the
  original HIPASS catalogue were recovered through visual inspection. 
Amongst the detections that are not listed in the HIPASS
  catalogue, 6 objects have since been detected in the ALFALFA
  survey and are listed in the third data catalog
  \citep{2008AJ....136..713K} covering right ascensions from 11:36 to
  13:52 hours and declinations form +8 to +12 degrees.  In fact, all the
  newly detected HIPASS objects within this area can be confirmed
  with the ALFALFA catalogue.

\begin{figure*}
\begin{center}
 
  \includegraphics[width=0.45\textwidth]{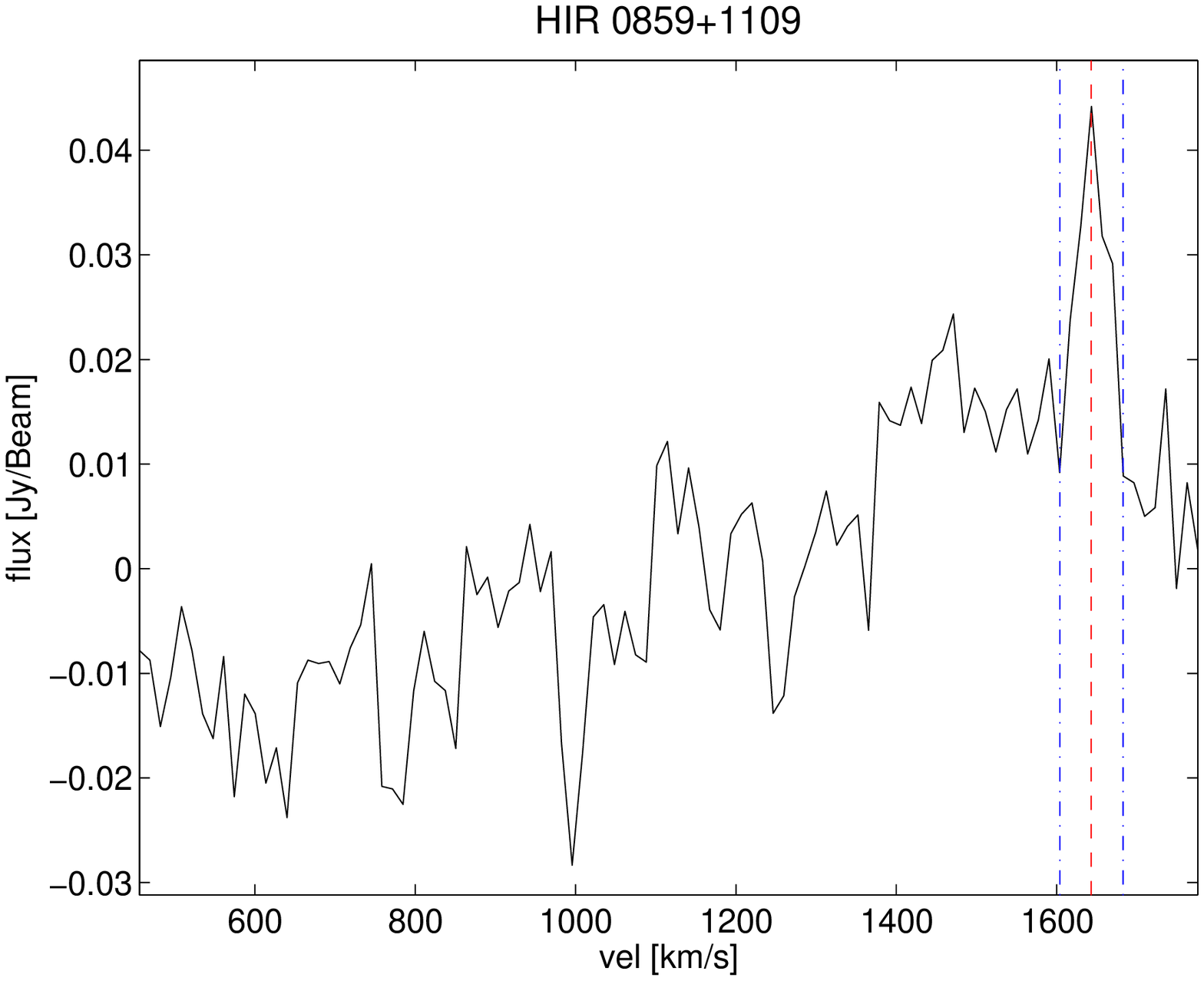}
  \includegraphics[width=0.45\textwidth]{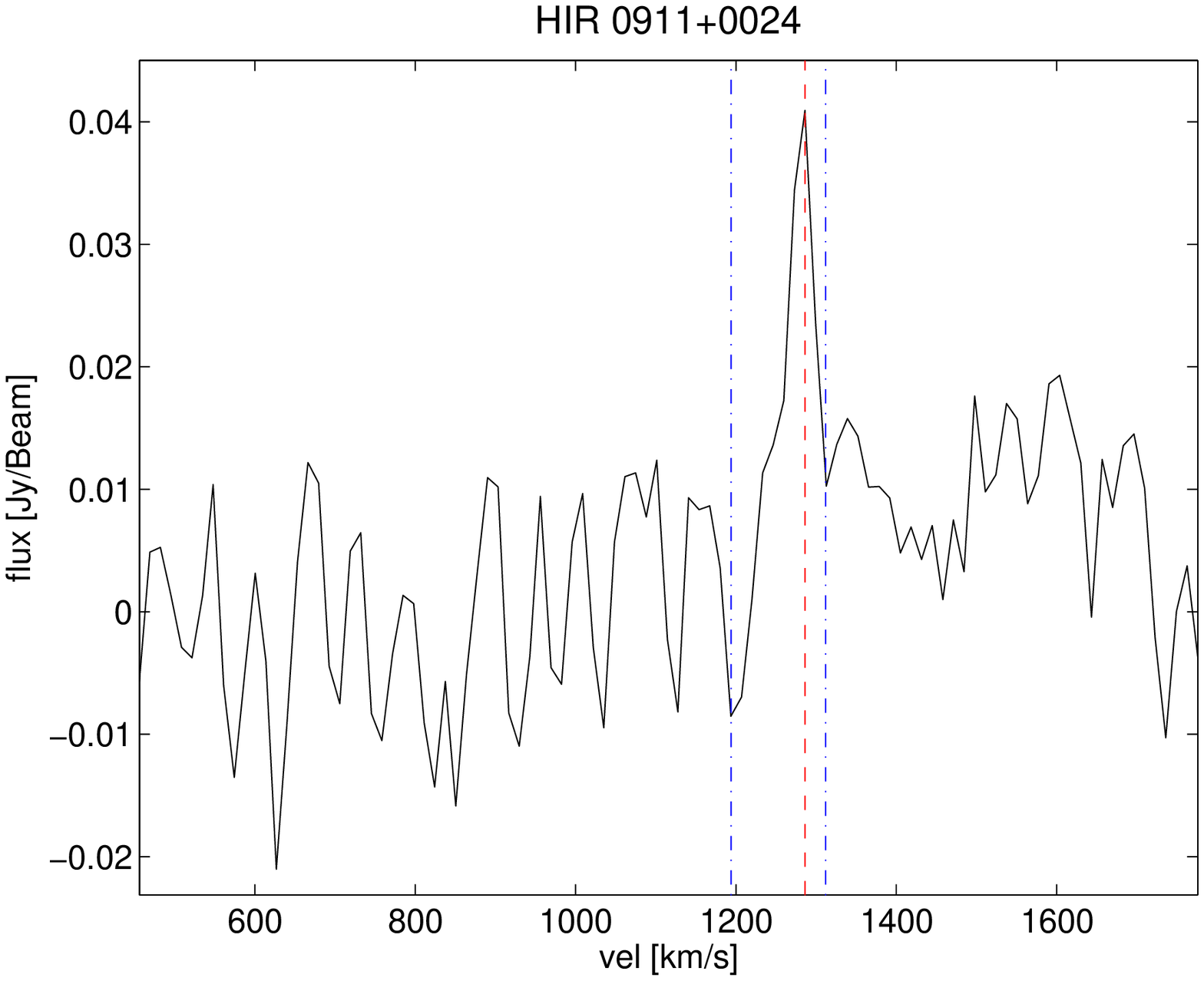}
  \includegraphics[width=0.45\textwidth]{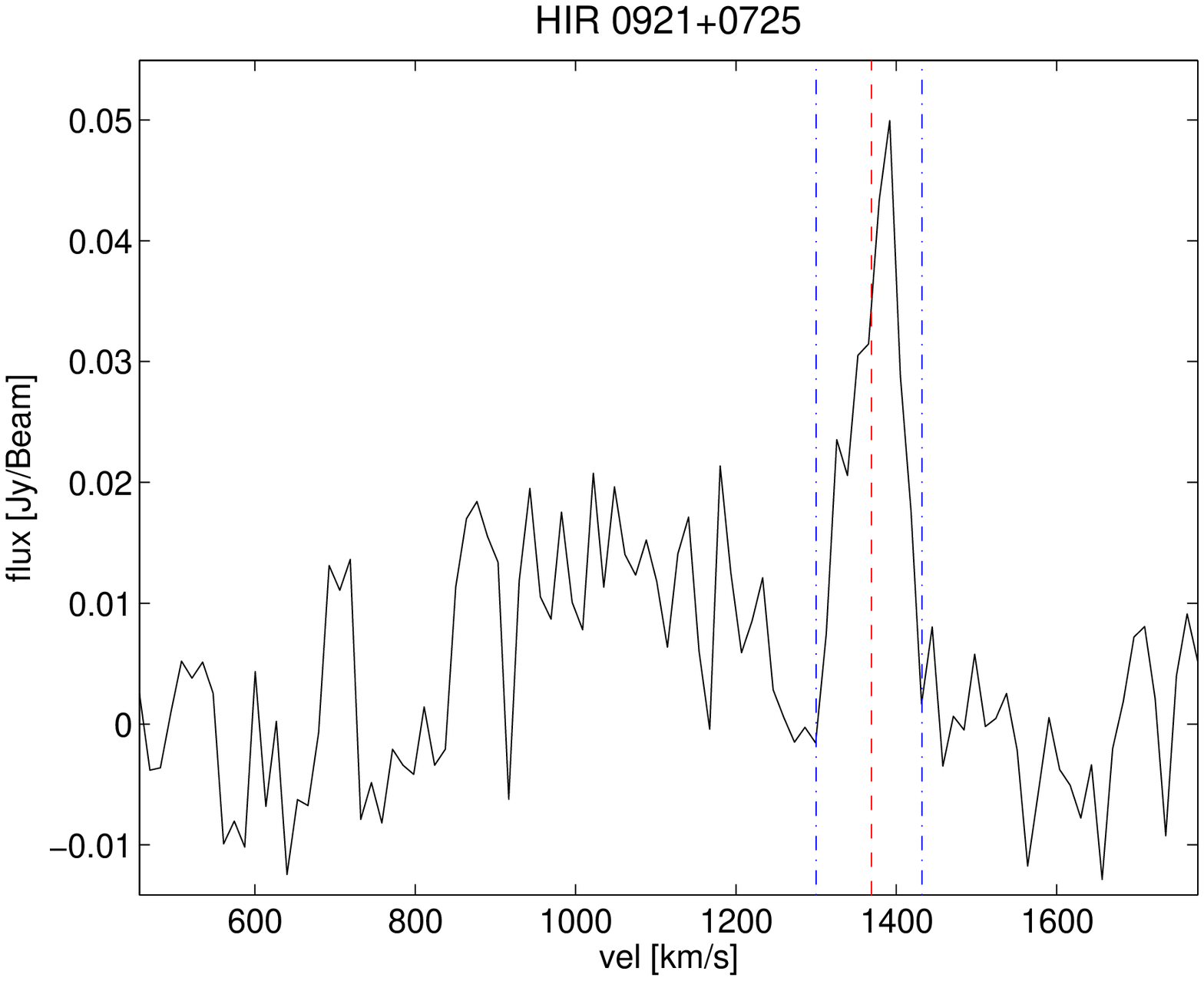}
  \includegraphics[width=0.45\textwidth]{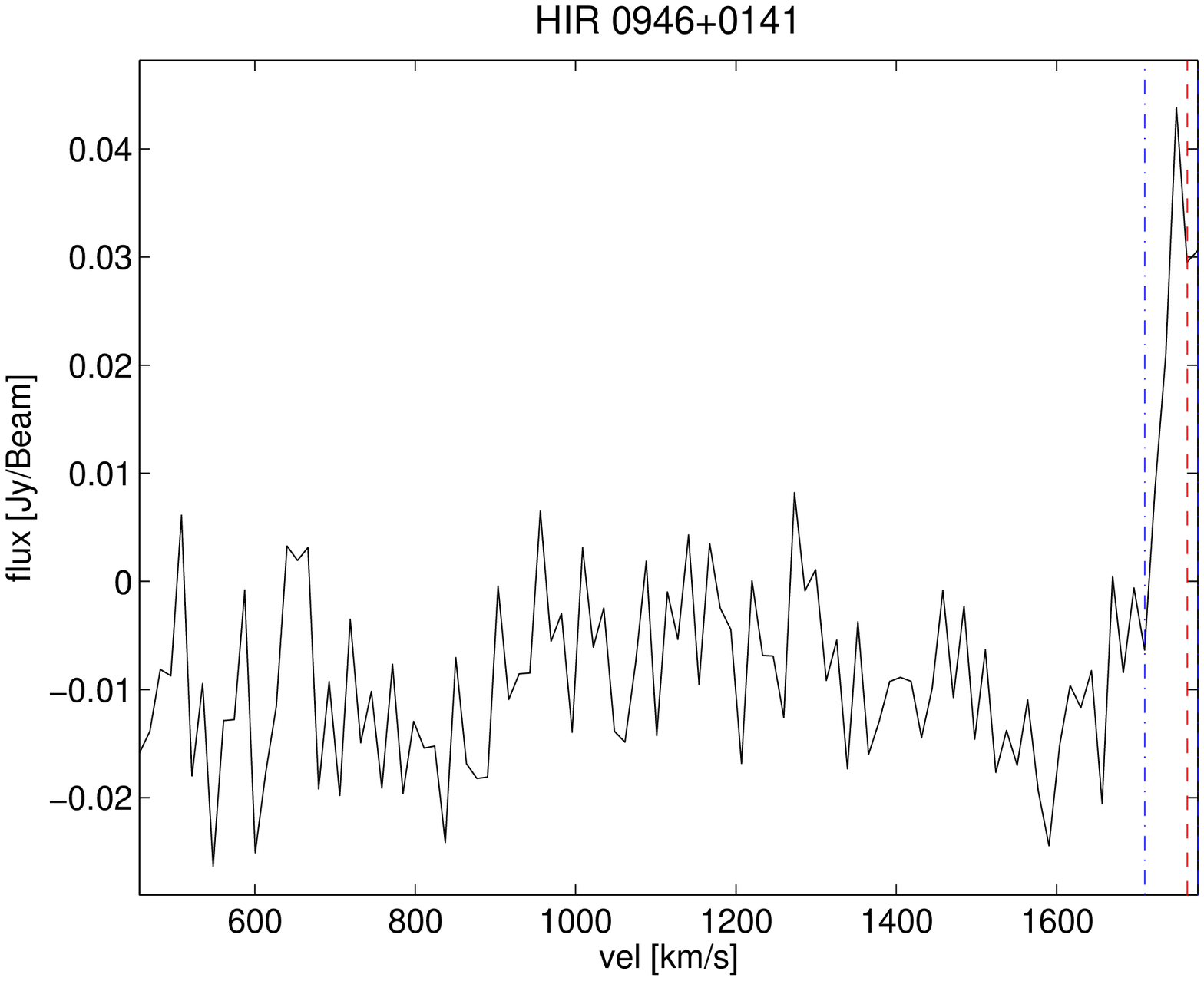}
  \includegraphics[width=0.45\textwidth]{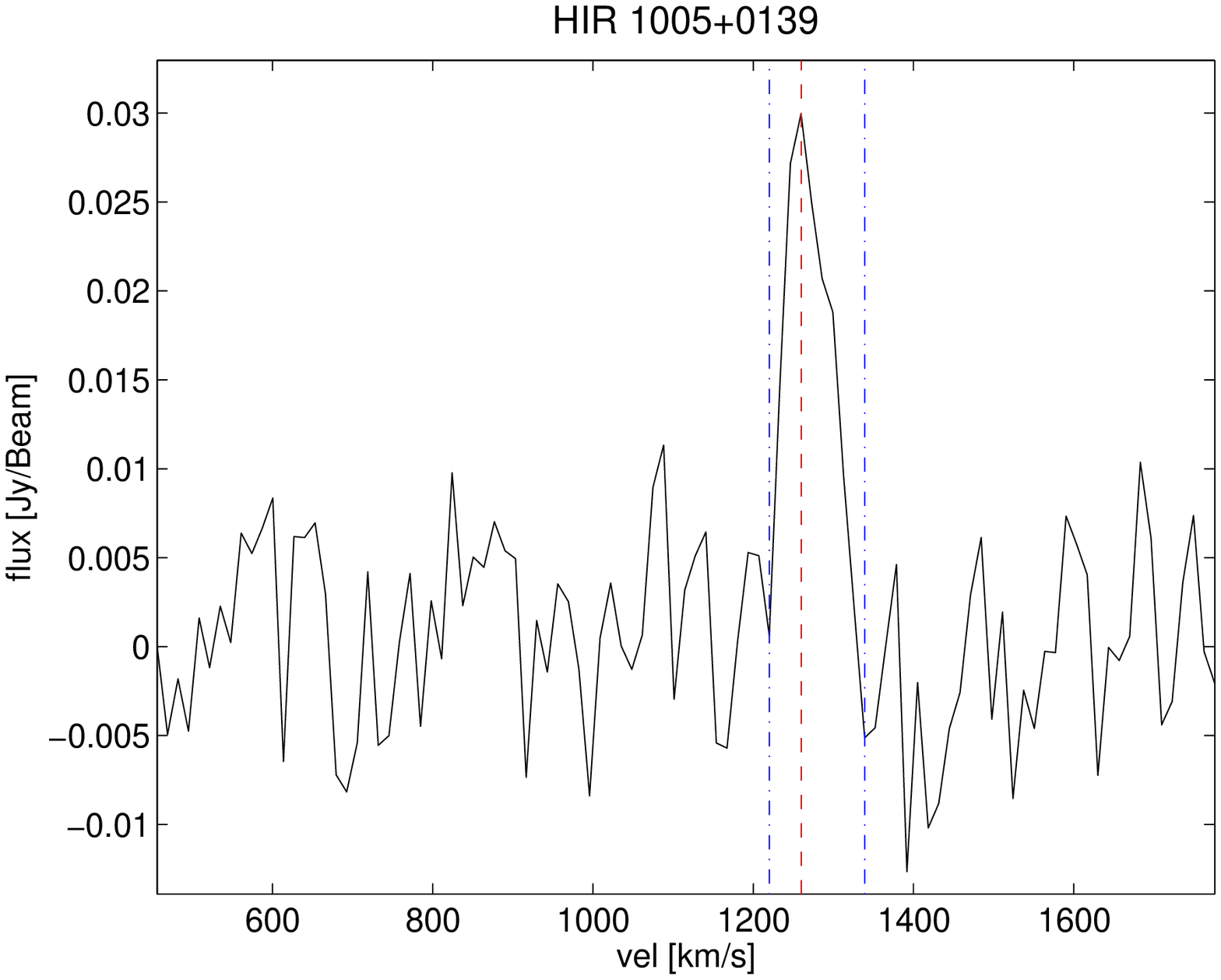}
  \includegraphics[width=0.45\textwidth]{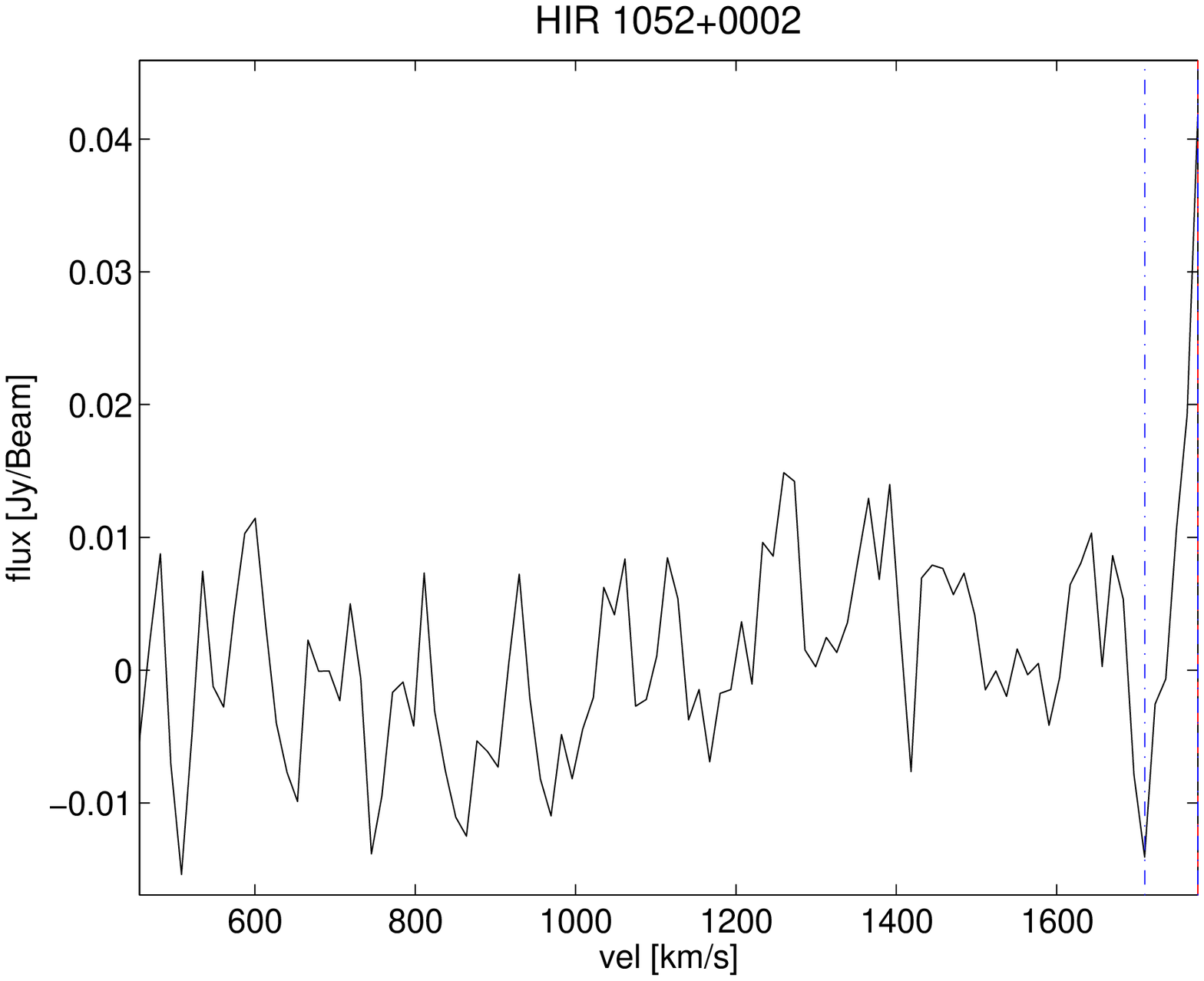}
  
\end{center}
  \caption{Spectra of new {\HI} detections obtained from reprocessed
    HIPASS data. The radial velocity of each object is indicated by
    the red dashed vertical, the velocity width is indicated by the blue dash-dotted line.}
  \label{newspectra}
\end{figure*}

  Spectra of completely new {\HI} detections are shown in
  Fig.~\ref{newspectra}. The spectra of the new detections do not
    show some particular feature and they are both found by visual (8
    detections) and automated (6 detections) inspection. 
Only the velocity interval that overlaps with the velocity
  coverage of the WVFS survey, from $\sim400$ to $\sim1800$ km
  s$^{-1}$, has been reprocessed. As the bandwidth of the reprocessed
  data is relatively small compared to the full HIPASS frequency
  coverage, detections at the edge of the reprocessed data appear at
  the very edge in Fig.~\ref{newspectra} .

%9-64
{\it HIR 0859+1109}: This is a new {\HI} detection for which no optical
galaxy is known at the relevant radial velocity. At a radial velocity
of 1988 $\pm$40 km s$^{-1}$ and with an offset of 1.6 arcmin is UGC
4712, which is more than 300 km s$^{-1}$ higher than the radial
velocity of HIR0859+1109. It is possible that HIR~0859+1109 is the
{\HI} counterpart of UGC~4712.

%9-48
{\it HIR 0911+0024}: When looking at the spectrum of this detection,
there is one narrow peak that looks significant. There is no optical
galaxy known at this redshift.

%9-54
{\it HIR 0921+0725}: The nature of this {\HI} detection is not clear. The
DSS image shows a diffuse feature at a location of RA=09:21:26.3 and
DEC=07:21:57, but many SDSS objects are listed at this location, all
without any distance information. Based on the appearance of the
optical feature, we expect that all these are at a higher redshift and not
related to HIR~0921+0725.

%10_73
{\it HIR 0946+0141}: This {\HI} feature is very likely the 
counterpart of SDSS~J094602.54+014019.4, a spiral galaxy at a radial
velocity of 1753 km s$^{-1}$. Both the radial velocity and the DSS
image are well-matched to the {\HI} detection. 

%10-204
{\it HIR 1005+0139}: At a spatial separation of only 1.5 arcmin is
2dFGRS~N421Z115 with a radial velocity within 30 km s$^{-1}$ of
HIR~1005+0139. This detection is a completely new {\HI} detection and
is the {\HI} counterpart of 2dFGRS~N421Z115 with high certainty.

{\it HIR 1052+0002}: When inspecting the DSS image, a small galaxy can
be identified at the peak of the {\HI} contours. This is the irregular
galaxy MGC 0013223 at a radial velocity of 1772 km s$^{-1}$, which is
very similar to the observed {\HI} radial velocity. Although {\HI} has not been
observed before in this galaxy, HIR~1052+0002 is very likely the
neutral counterpart of MGC~0013223.

%11-201
{\it HIR 1055+0511}: Although the peak of this detection is not very
bright, the line is broad enough to make it significant. The DSS image
shows a diffuse object at RA=10:55:16.3 and DEC=05:12:19.5 that might
be relevant for this candidate detection. However, this is an SDSS object at a
cataloged radial velocity of almost 6000 km s$^{-1}$. If correct, any relation with
HIR~1055+0511 is highly unlikely.

%12-23
{\it HIR 1212+0248}: Less than half an arcmin separated and at exactly
the same radial velocity is LEDA~135791. HIR~1212+0248 is the first {\HI}
detection of this dwarf irregular galaxy.

%12-205
{\it HIR 1230+0013}: This feature is about 30 arcmin separated from NGC
4517A. Although there is no sign of any optical counterpart in the DSS
images, HIR~1230+0013 is possibly associated with NGC~4517A, as the
radial velocity is very similar.

%12-207
{\it HIR 1231+0145}: At an angular offset of less than 5 arcmin and a
similar radial velocity is the irregular galaxy CGCG 014-054. Although
this galaxy has no reported {\HI}, HIR~1231+0145 is most likely the {\HI}
component of CGCG~014-054 because of the good correspondence in
position and velocity.

%14-213
{\it HIR 1401+0247}: There is no known galaxy at the relevant
redshift, however when looking at the DSS image, there is the massive
galaxy cluster Abell 1835 at redshift 0.253 centered at RA=14:01:02.0
and DEC=02:51:32, coincident with the peak of the apparent {\HI}
contours. An association with an {\HI} signal at the cluster redshift
is clearly out of the question, while applying the cluster redshift to
the detected spectral feature would imply a rest frequency of 1773.6
MHz, where no known transition occurs. For comparison, some known
radio frequency transitions are OH 1720.53 MHz, H$_2$CO 4829.66 MHz,
and CH$_3$OH 6668.52 MHz. The cluster is known to act as a
gravitational lens \citep{2005ApJ...627...32S}, although current
attempts to determine the redshift of lensed features have not been
successful. While unlikely, the detected feature might correspond to
H$_2$CO at z~=~2.4, or CH$_3$OH at z~=~3.7.

%14-214
{\it HIR 1417+0651}: When looking at the DSS image at the location of
this object, two small galaxies can be recognised at the peak of the
contours. One is CGCG~046-087 which is irrelevant because of the
radial velocity of 7559 km s$^{-1}$. The other object is a GALEX
source at RA=14:17:50.7 and DEC=06:50:22 without any redshift
information.

%15-216
{\it HIR 1446+1011}: The DSS image shows a galaxy at the peak of the {\HI}
contours. However, this is CGCG~076-029 at almost 16,000 km
s$^{-1}$. Beyond this, there is no sign of an optical source that can
be easily linked to HIR~1446+0011.

%15-215
{\it HIR 1515+0603}: There is a very small and diffuse SDSS object at
RA=15:14:57.23 and DEC=06:06:03.10. Although there is no redshift
information about this object, based on the visual appearance a
connection with HIR~1515+0603 is possible, but not very likely.

%%%%%%%%%%%%%%
%%%%%%%%%%%%%

\begin{figure*}
\begin{center}
 \includegraphics[width=0.45\textwidth]{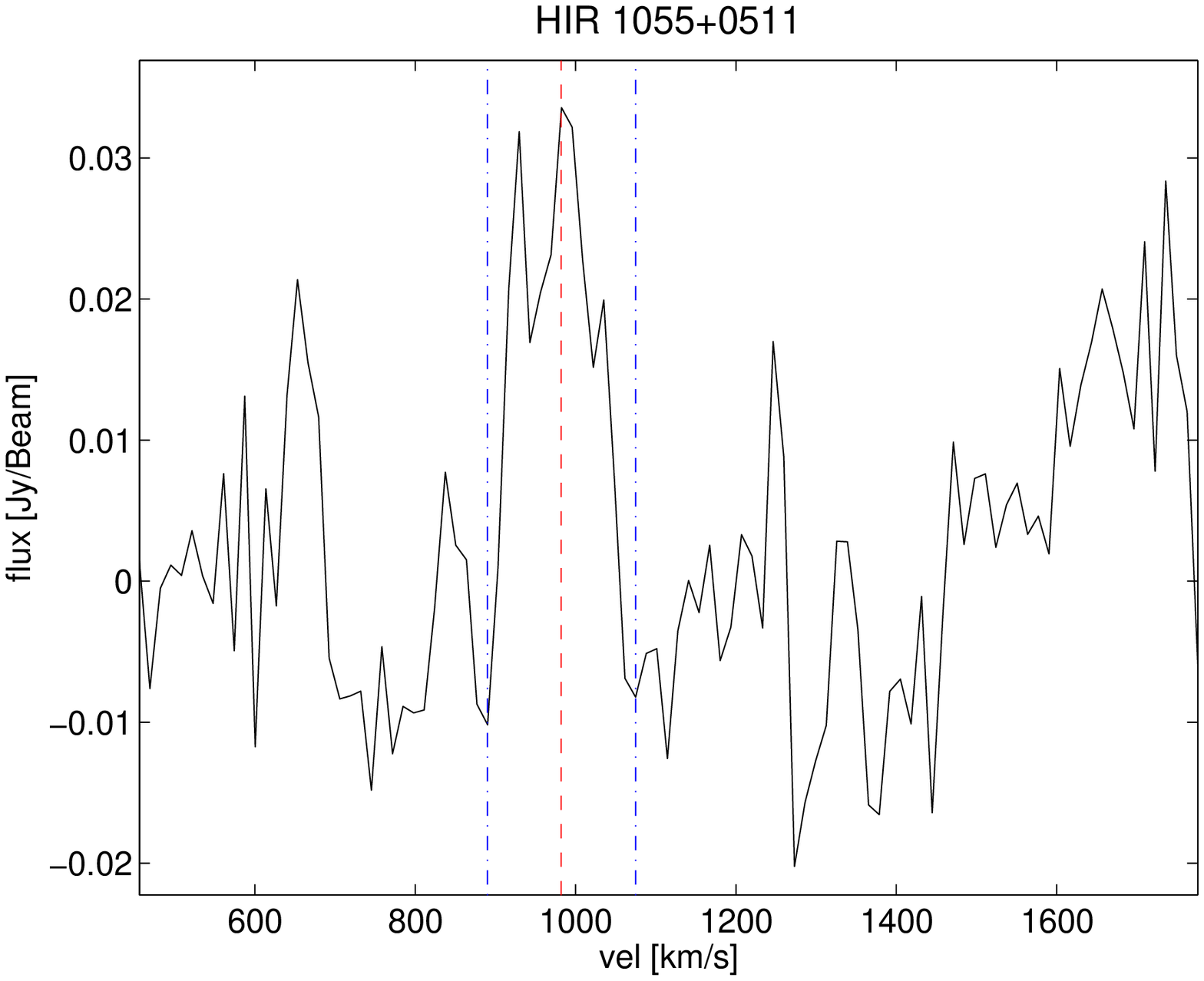}
 \includegraphics[width=0.45\textwidth]{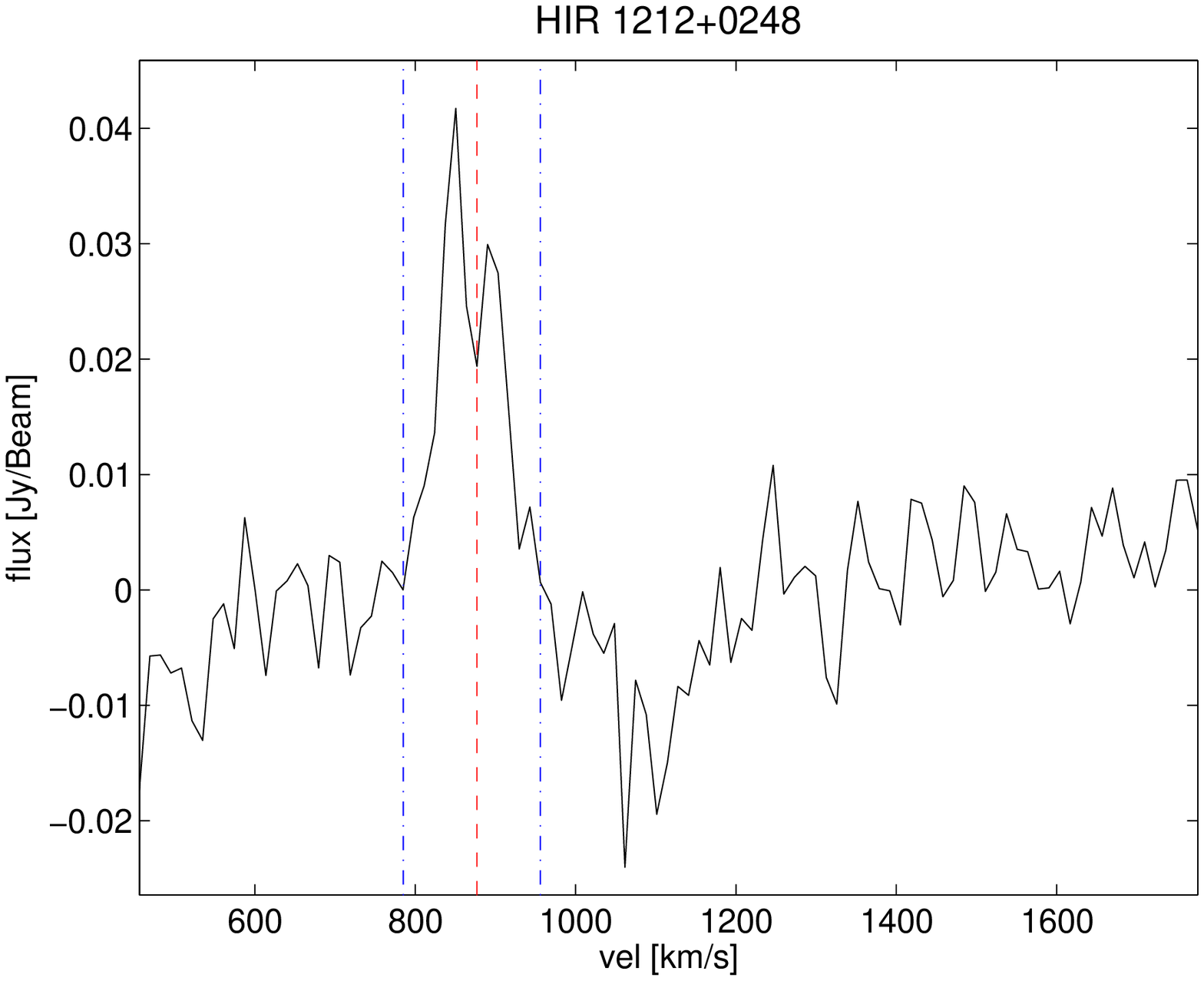}
  \includegraphics[width=0.45\textwidth]{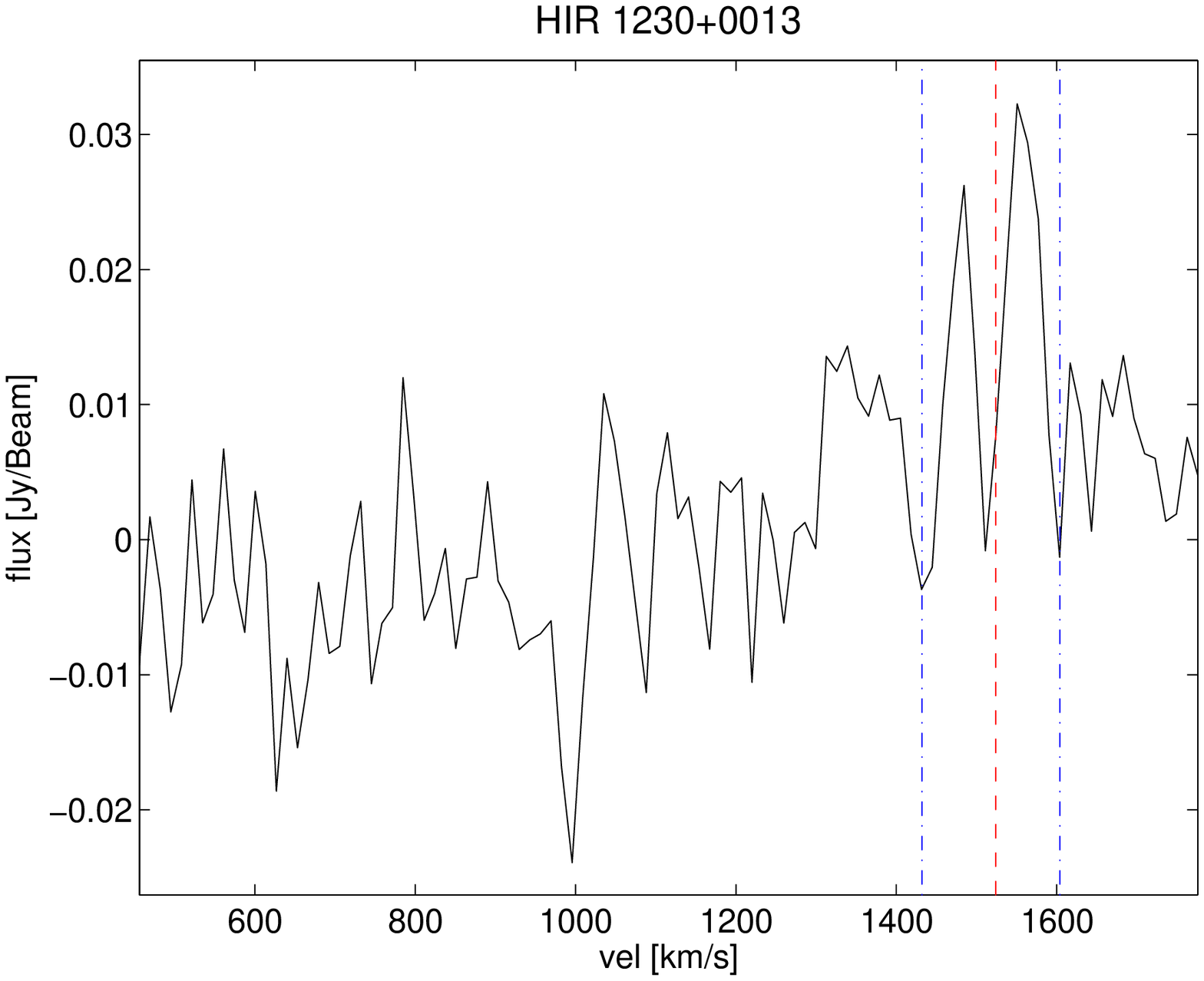}
  \includegraphics[width=0.45\textwidth]{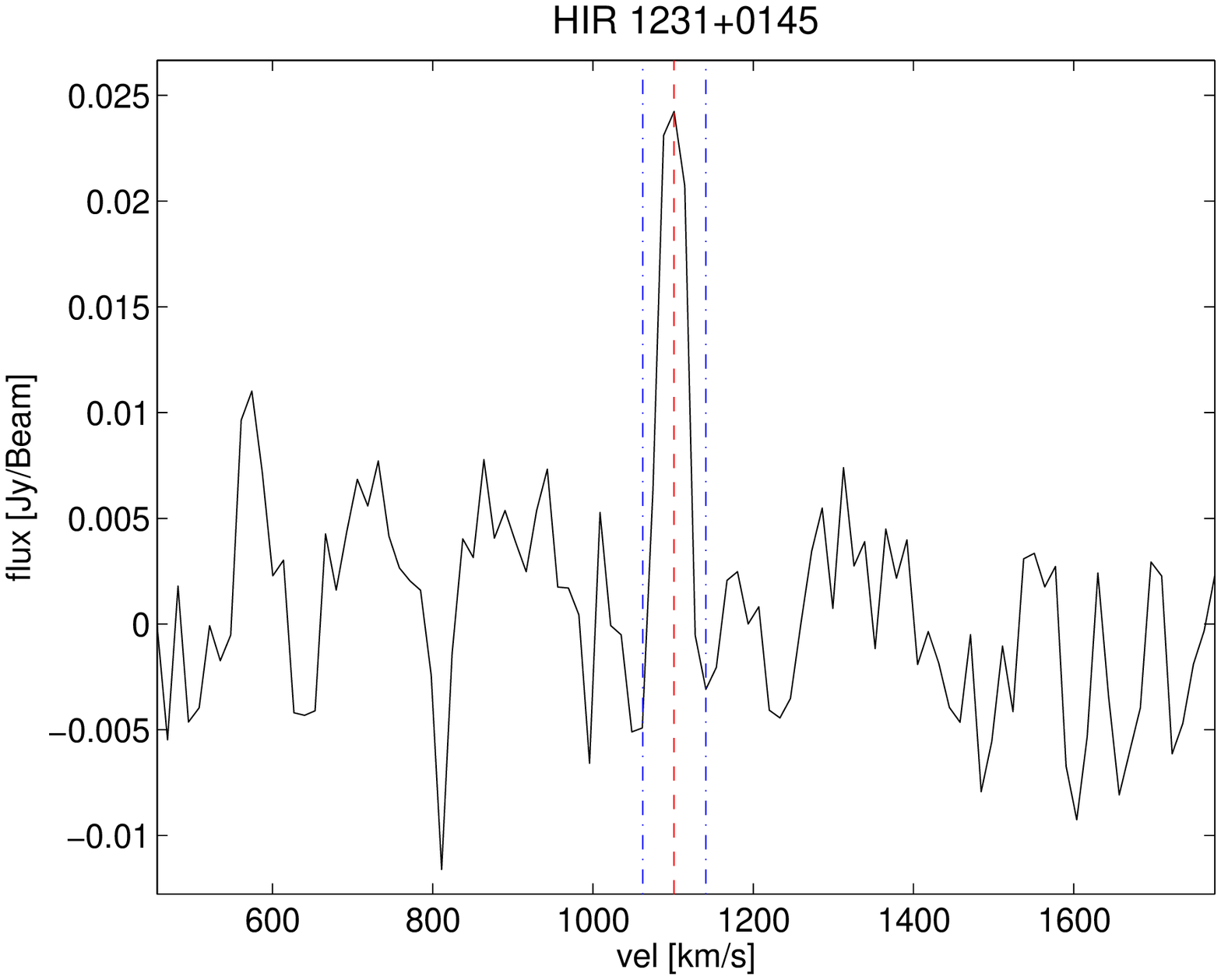}
  \includegraphics[width=0.45\textwidth]{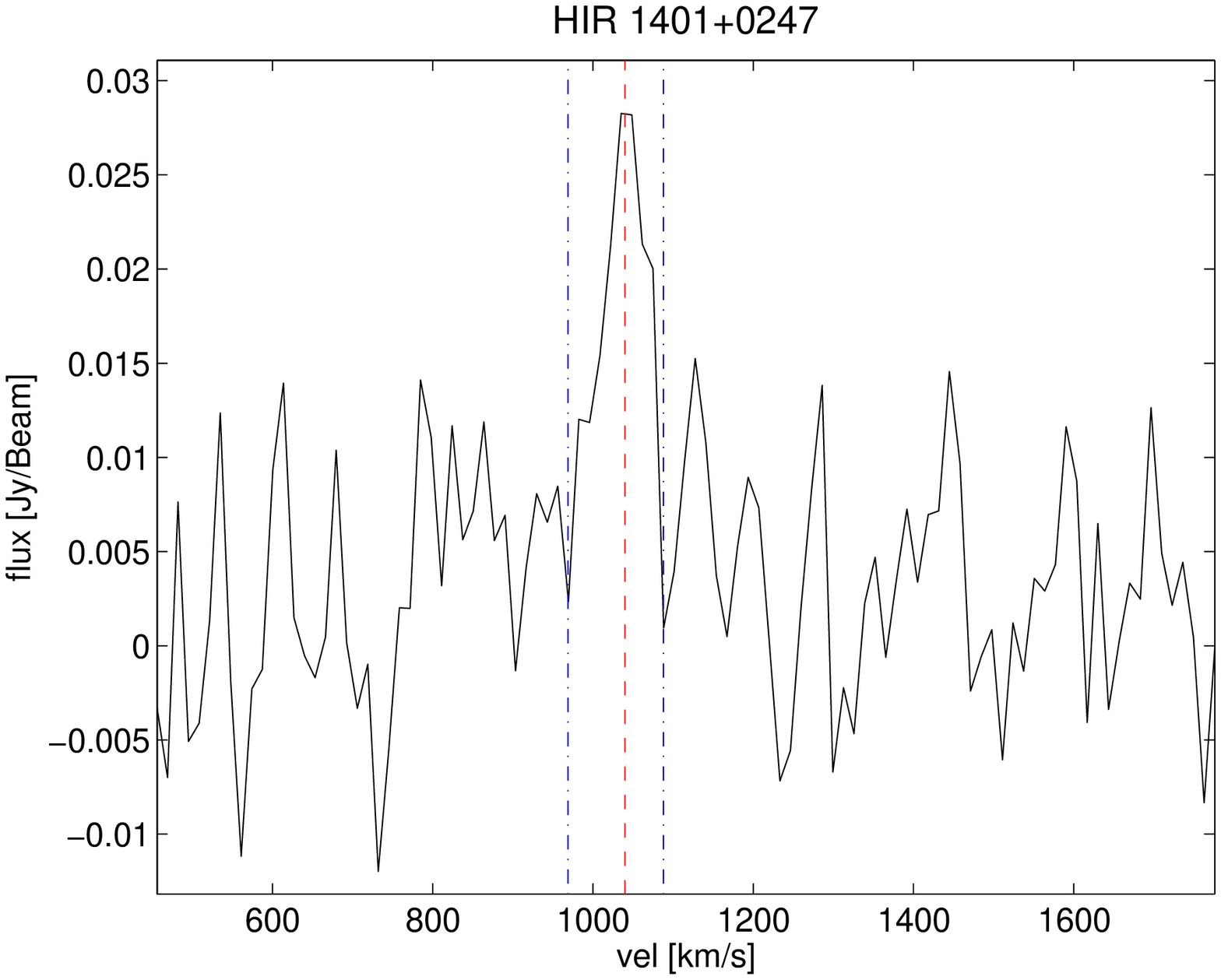}
  \includegraphics[width=0.45\textwidth]{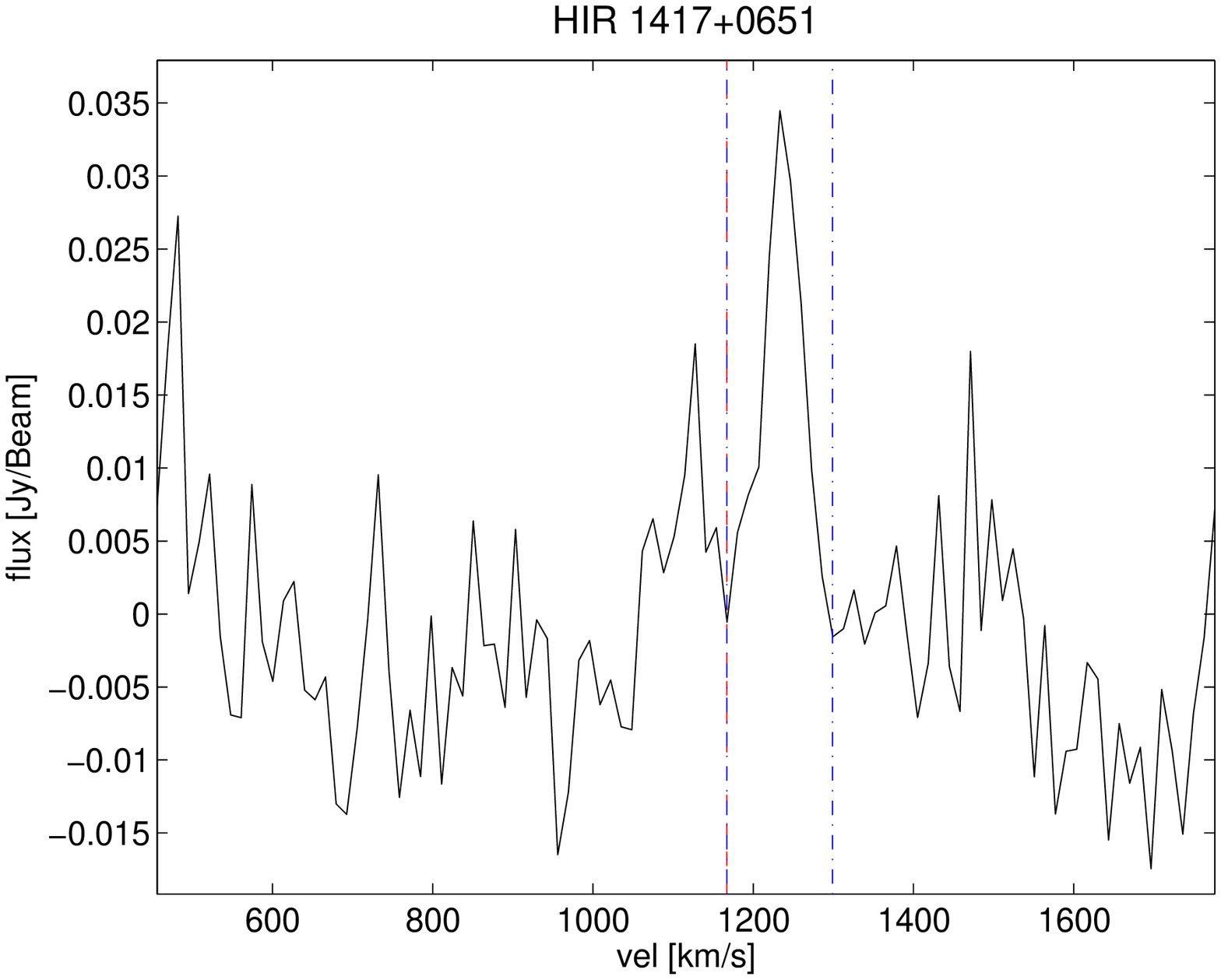}
  
\end{center}
  {\bf Fig~\ref{newspectra}.} (continued)

\end{figure*}

\begin{figure*}
\begin{center}
 
 \includegraphics[width=0.45\textwidth]{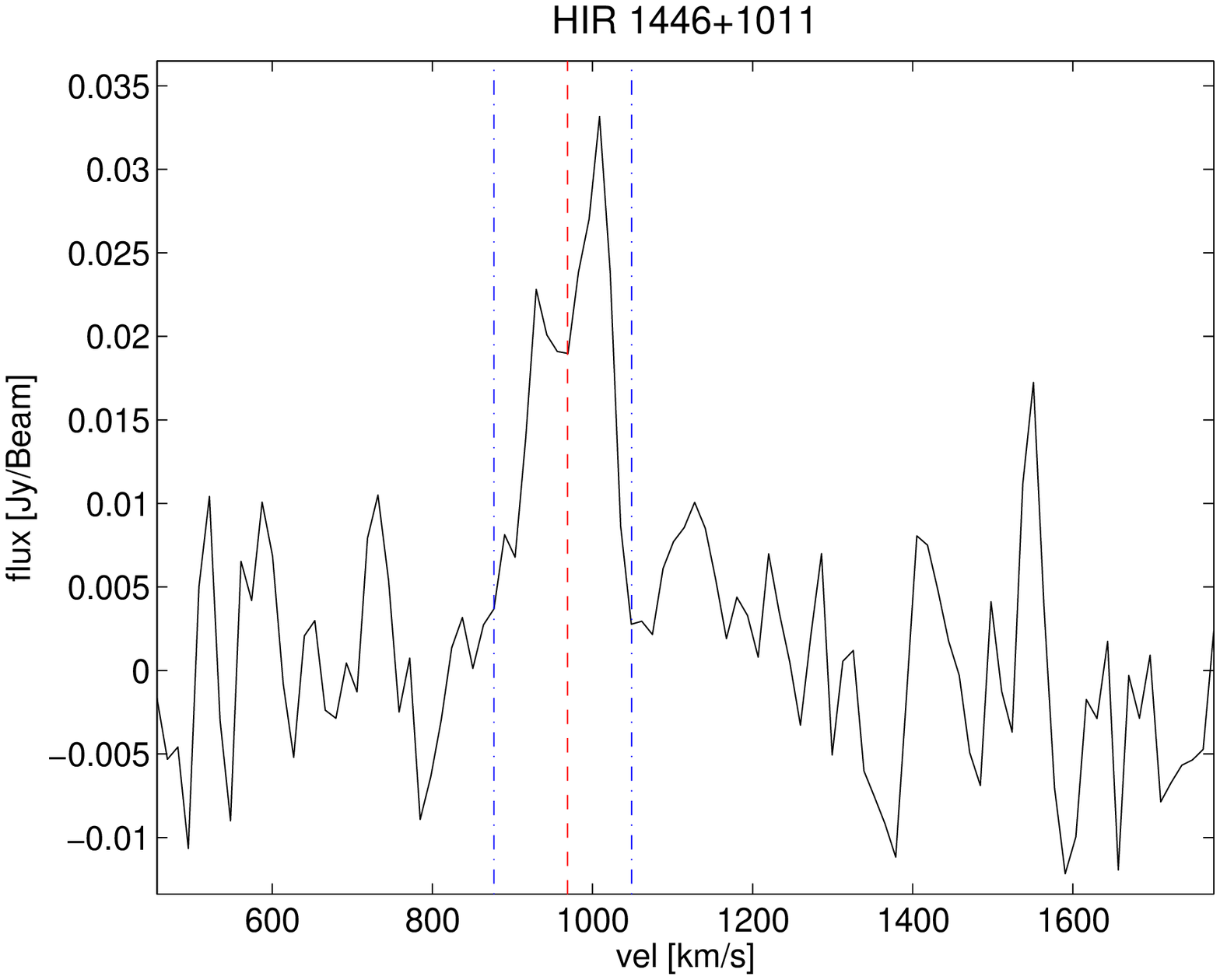}
  \includegraphics[width=0.45\textwidth]{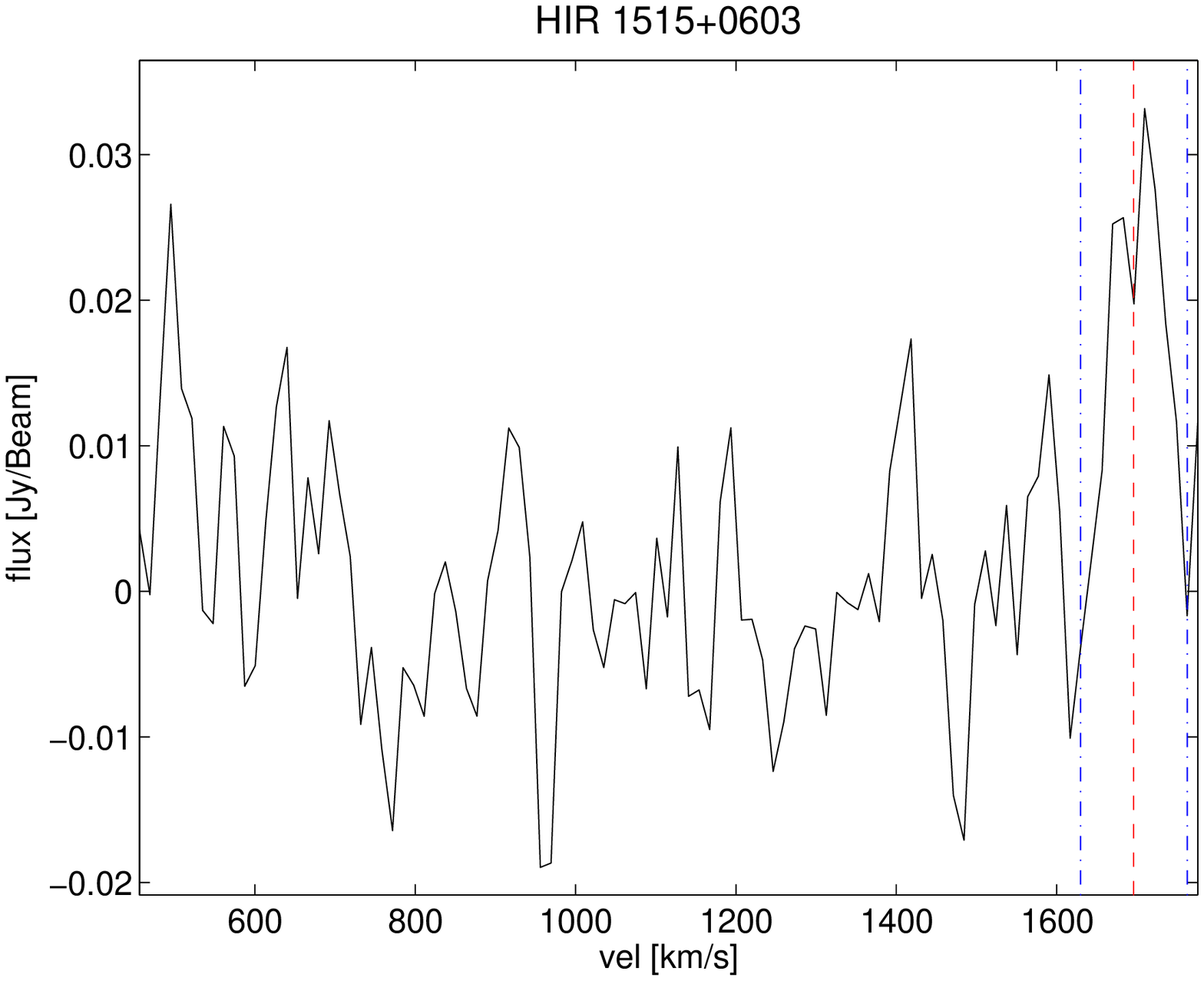}

\end{center}
  {\bf Fig~\ref{newspectra}.} (continued)

\end{figure*}

\subsection{Flux densities}
The total flux densities have been determined for all detections using
two independent methods. The line-widths and the integrated
line-strengths have first been determined by integrating the brightness
over the full velocity width of an object along the single spectrum
that contains the overall peak brightness. We assume that all sources
are unresolved and fully contained within the 15.5 arcmin beam. In a
second approach the total flux has been determined from the integrated
moment maps. Moment maps are created by collapsing the cube in the
direction of the velocity axis over the full line width. The
visualisation package KARMA \citep{1996ASPC..101...80G} has been used
to integrate the flux. The radial profile of an object can be plotted,
including a fit to the data points within a user defined circle. This
circle was chosen to completely enclose the object, including any
possible extended emission. The integrated flux densities have to be
corrected for the beam integral to convert from Jy Beam$^{-1}$ km
s$^{-1}$ to Jy km s$^{-1}$. An integrated flux density is determined
by simply adding the pixels values, and a fitted flux density is
determined by fitting a Gaussian to the radial profile. Since we aim
to be sensitive to extended features that do not necessarily have a
Gaussian profile, we use the pixel integral for our flux density
determinations rather than a Gaussian fit.

\begin{figure*}[th]
  \includegraphics[width=0.5\textwidth]{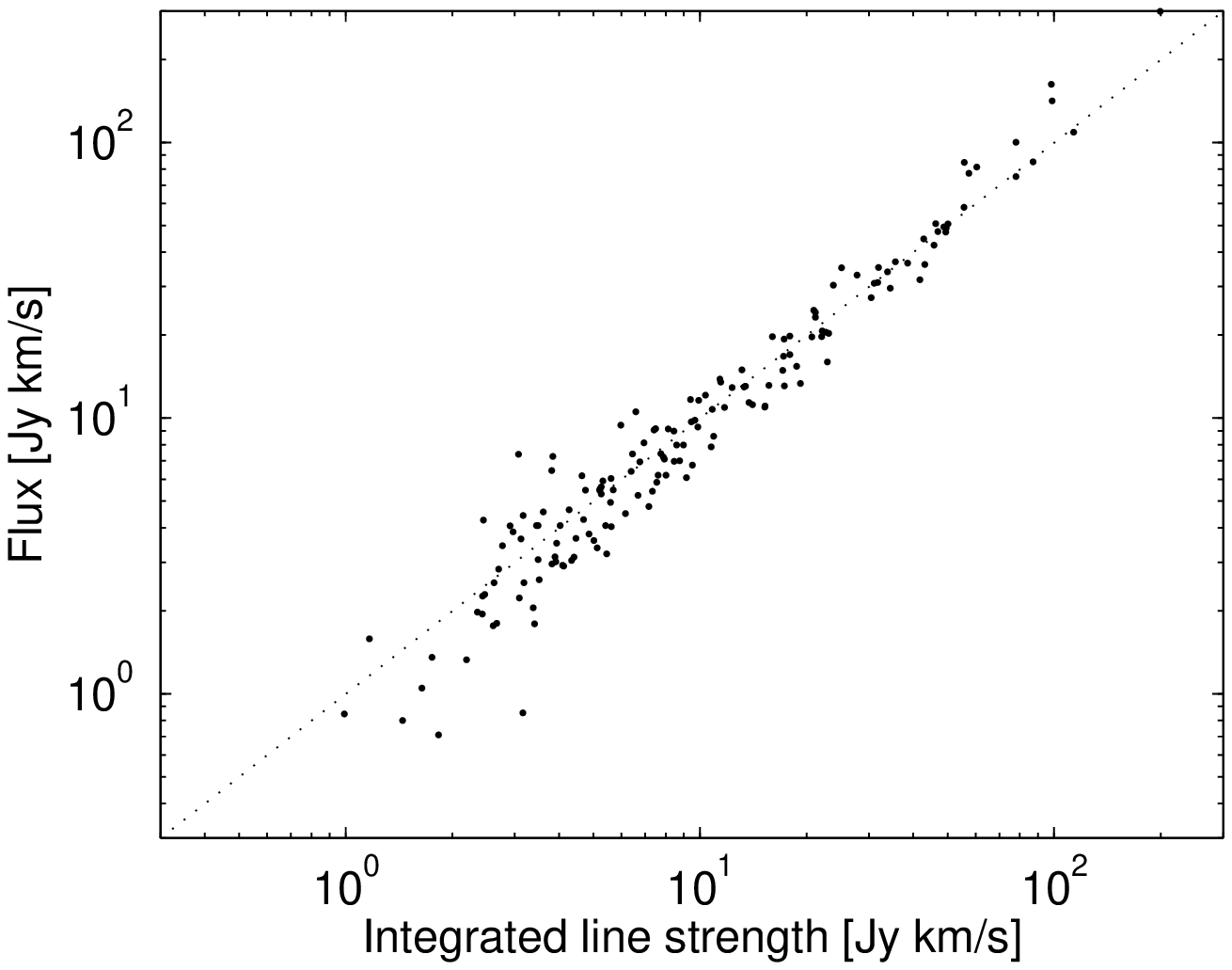}
  \includegraphics[width=0.5\textwidth]{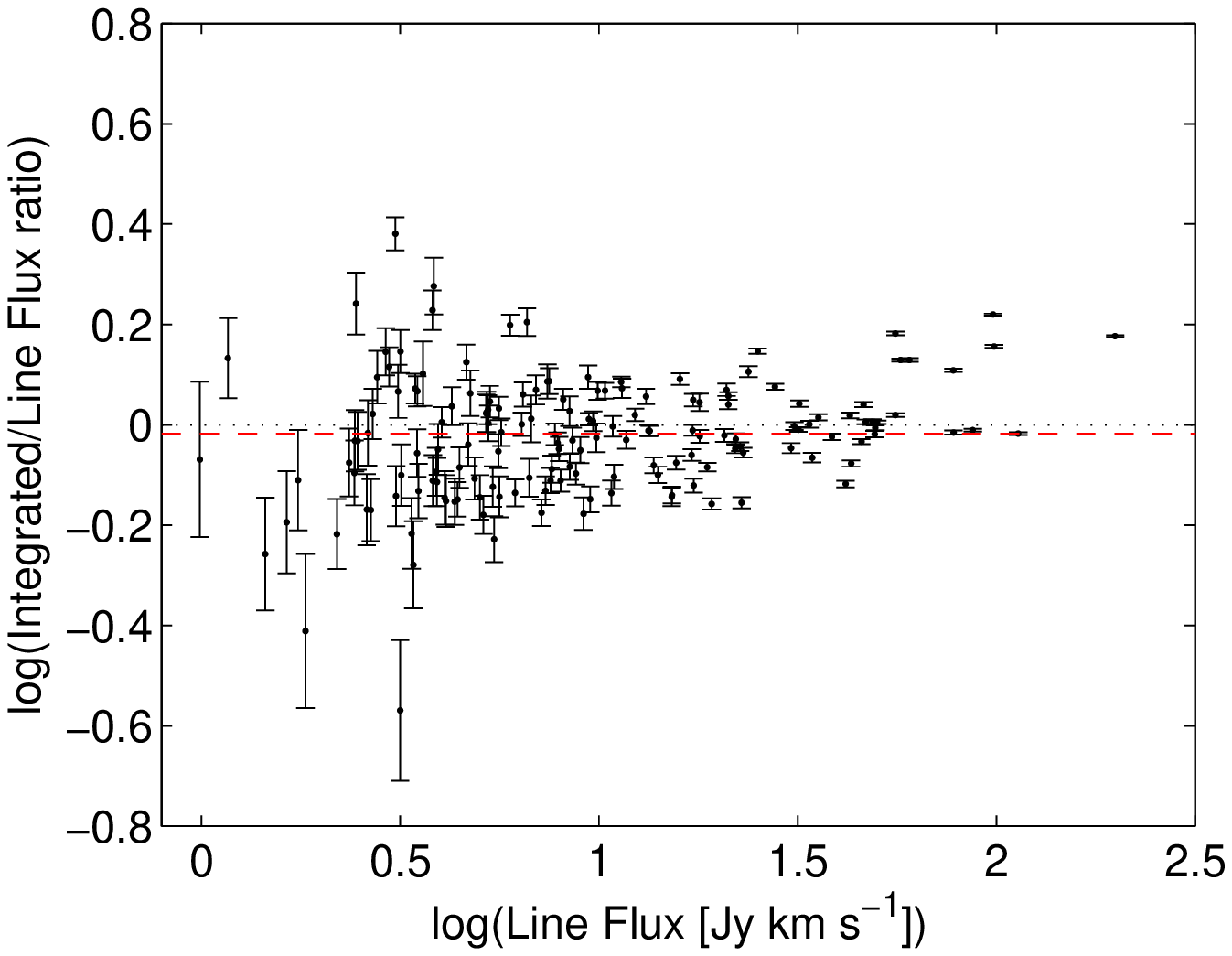}
  \caption{Left panel: Flux obtained from the moment maps is plotted
    as function of the line strength, the dotted line indicates where
    the two flux measurements are equivalent. Right panel: ratio of
    integrated flux and line-strength is plotted as function of
    line-strength on a logarithmic scale. The dotted line indicates
    where the ratio is one and the fluxes are equivalent. The dashed
    line shows the median of the ratios.}
  \label{flux_comp}
\end{figure*}

When integrating the line strength of a single spectrum, not all the
flux is measured if the source is resolved by the beam, or if there
are extended emission features like filaments. The flux densities are
plotted and compared in Fig.~\ref{flux_comp} where in the left panel
the total flux obtained from the integrated pixel values is plotted as
function of the single pixel integrated line strength. The dotted line
goes through the origin and indicates where the fluxes are equal. Only
those sources are plotted that are completely covered by our data
cubes in both the spectral and spatial directions. The measured fluxes
match the dotted line very well, meaning that there is typically no
large discrepancy between the two different methods. The ratio of the
fluxes is plotted on a logarithmic scale in the right panel of
Fig.~\ref{flux_comp}, the dotted line indicating again where the
fluxes are equal. The mean of the ratios is 0.99, with a standard
deviation of 0.28. The dashed line in the right panel of
Fig.~\ref{flux_comp} represents the median of the flux ratios, which
is 0.96. As both the mean and median values are close to one, there is
generally very good agreement between the fluxes.  For large flux
values above $\sim 50$ Jy km s$^{-1}$ the fluxes derived by
integrating the individual pixel values are typically larger by 20 to
50\%. This is due to the fact that these are typically large and
extended sources that are resolved by the beam.  At low flux levels,
there are a number of sources for which the integrated line-strength
exceeds the integrated flux of the moment map by more than a factor of
two. Since the line-strength of the peak spectrum provides a {\it
  lower limit} to the true integral, there must be a residual
  artefact in either the spectrum or the moment map. The spectra and
  moment maps of these sources were inspected and either the spectral
  bandpass appears slightly elevated, or there are negative residuals
  in the moment maps which influence the flux estimates.

\begin{figure*}[th]
  \includegraphics[width=0.5\textwidth]{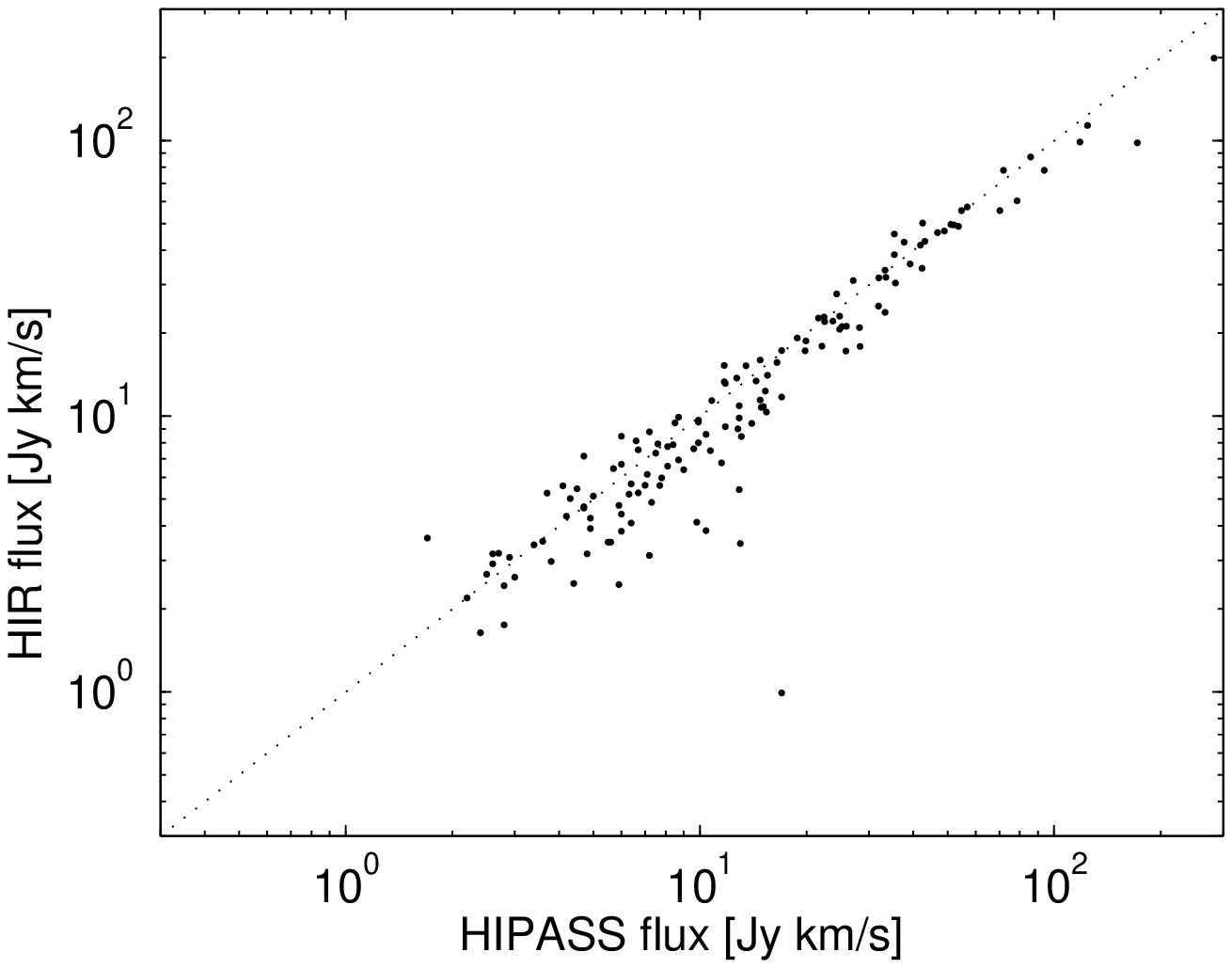}
  \includegraphics[width=0.5\textwidth]{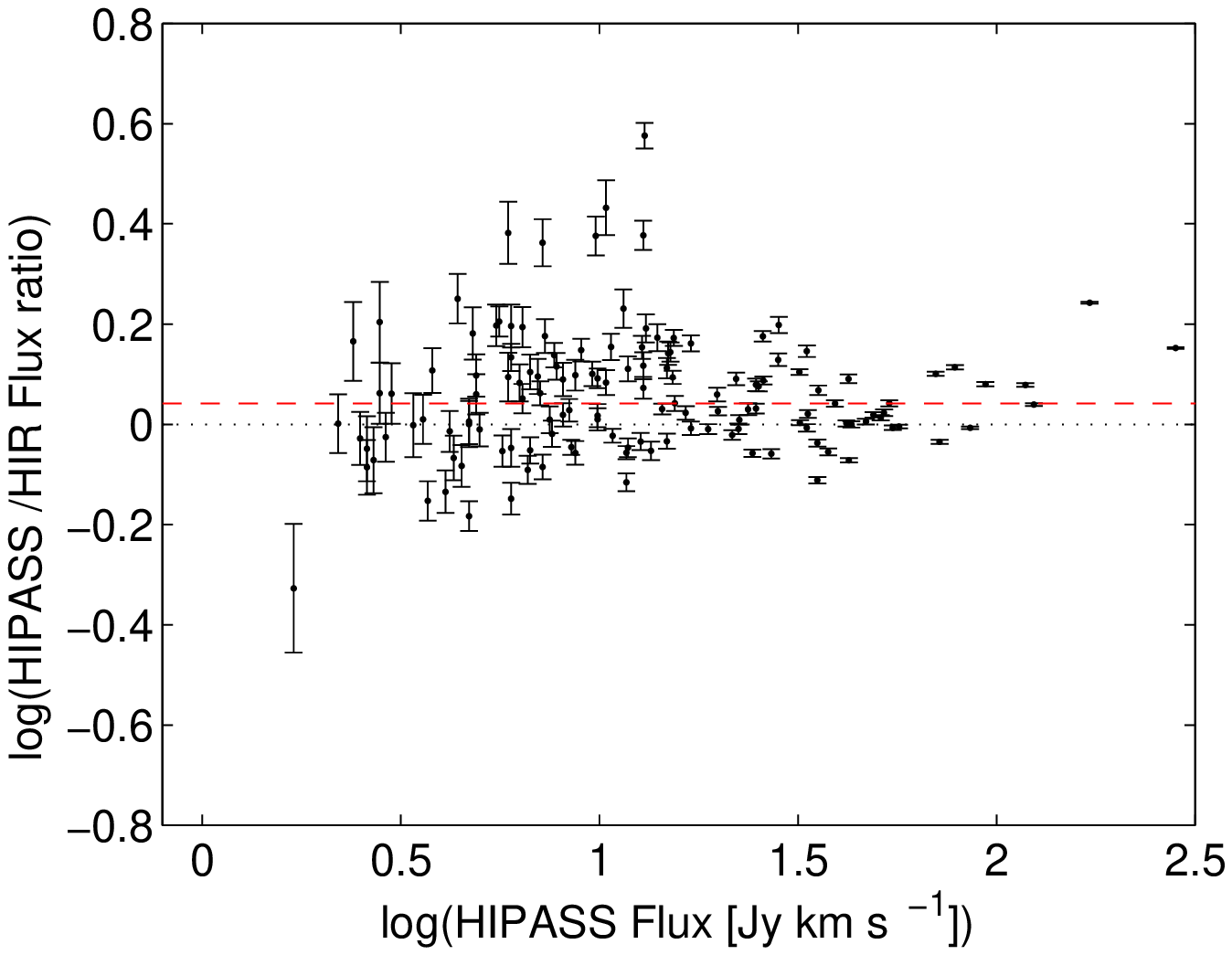}
  \caption{Left panel: The integrated line strength of the
    reprocessed HIPASS data (HIR) is plotted as function of the original
    HIPASS fluxes. Right panel: ratio of the first and reprocessed HIPASS
    product is plotted against the original HIPASS flux. The dashed
    line indicates where the fluxes are equal, while the dashed line
    represents the median of the ratios. On average there is a small
    excess in the flux values of the original HIPASS product.}
  \label{hipass_comp}
\end{figure*}

Another important comparison to make is of our measured fluxes with
those obtained in the first HIPASS product. In Fig.~\ref{hipass_comp}
fluxes derived from the reprocessed HIPASS data are compared with
fluxes in the HIPASS catalogue \citep{2004MNRAS.350.1195M,
  2006MNRAS.371.1855W}. Again, only those sources are plotted that are
completely covered by the data cubes. The left panel shows the
reprocessed fluxes as function of original fluxes, while the right
panel shows the ratio. The dashed line indicates the median of the
flux ratios which is 1.10. The mean of the ratios is 1.31 with a
standard deviation of 1.41. In this case the median estimate gives a
better representation of the general trend instead of the mean, as the
effect of strong outliers is suppressed.

The $\sim 10$\% excess in flux in the published HIPASS product over
that in our reprocessed result may reflect the variation in effective
beam size with signal-to-noise ratio that is a consequence of median
gridding, as discussed at some length in \cite{2001MNRAS.322..486B}.
Exactly the same data has been used in both processing methods, so
another effect that may contribute to the difference in flux is how
the bandpass is determined. The gridding of the data has been
  done in a similar fashion as for the original HIPASS product, so any
  differences are caused in the pre-gridding.  For several sources the
individual spectra in both the original and the re-reduced HIPASS
product have been compared. Despite the difference in flux, the
spectra look very similar. A difference in the fit to the bandpass can
enhance the whole spectrum slightly, without significantly affecting
the shape.

\subsection{Companions}

Moment maps have been generated for all detected objects by integrating
over their velocity widths. Initially, maps of 2 by 2 degrees in size
are generated, to completely cover the detection itself, including the
nearby environment. All moment maps were inspected by eye for diffuse
emission features. For objects that showed tentative signs of
filaments or companions, another moment map was generated of 5 by 5
degrees in size. These moment maps were inspected in detail, by
searching for local peaks in the spatial domain as well as possible
line features in the spectrum at the relevant velocity.

\begin{figure*}[t]
  \includegraphics[width=0.5\textwidth]{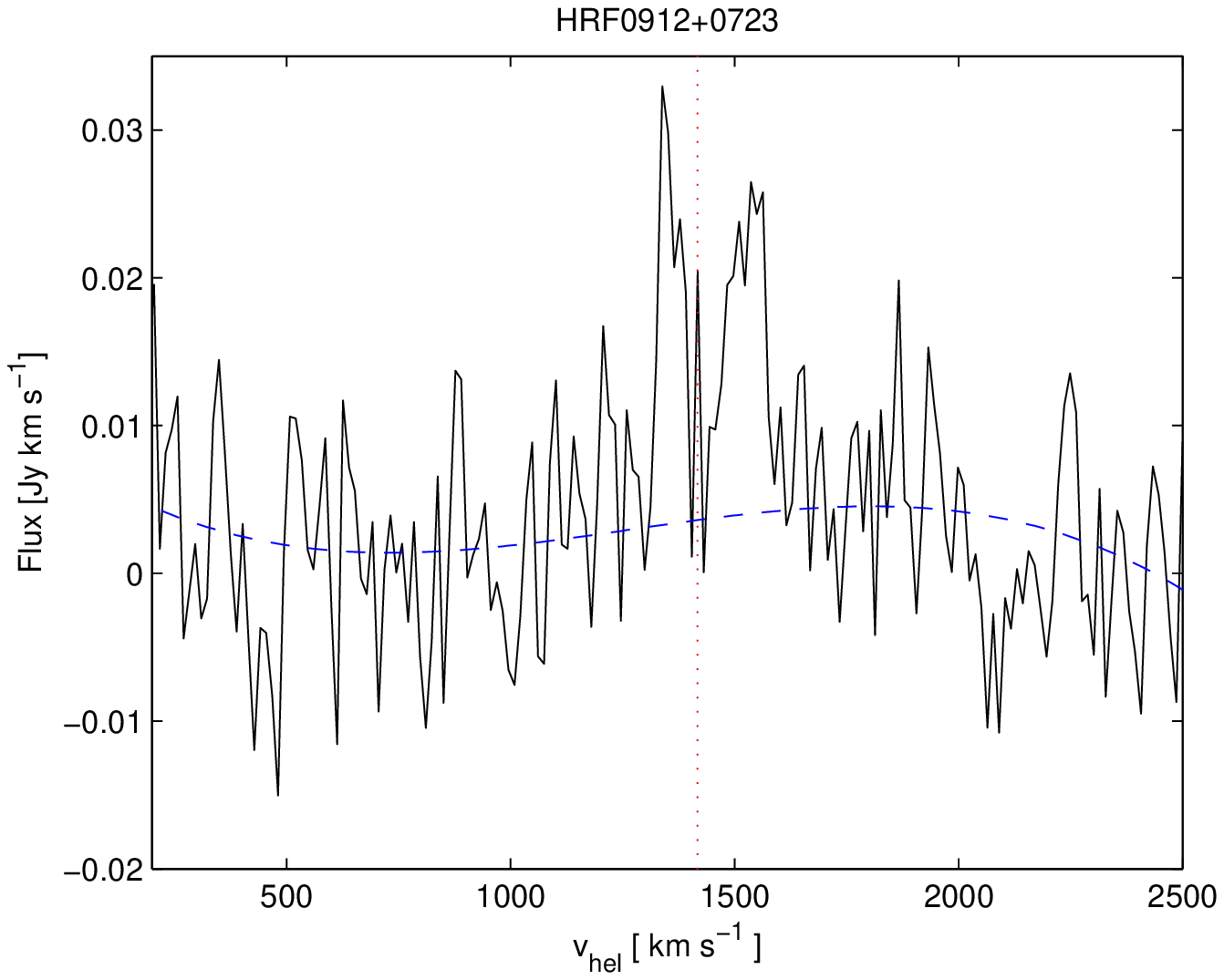}
  \includegraphics[width=0.5\textwidth]{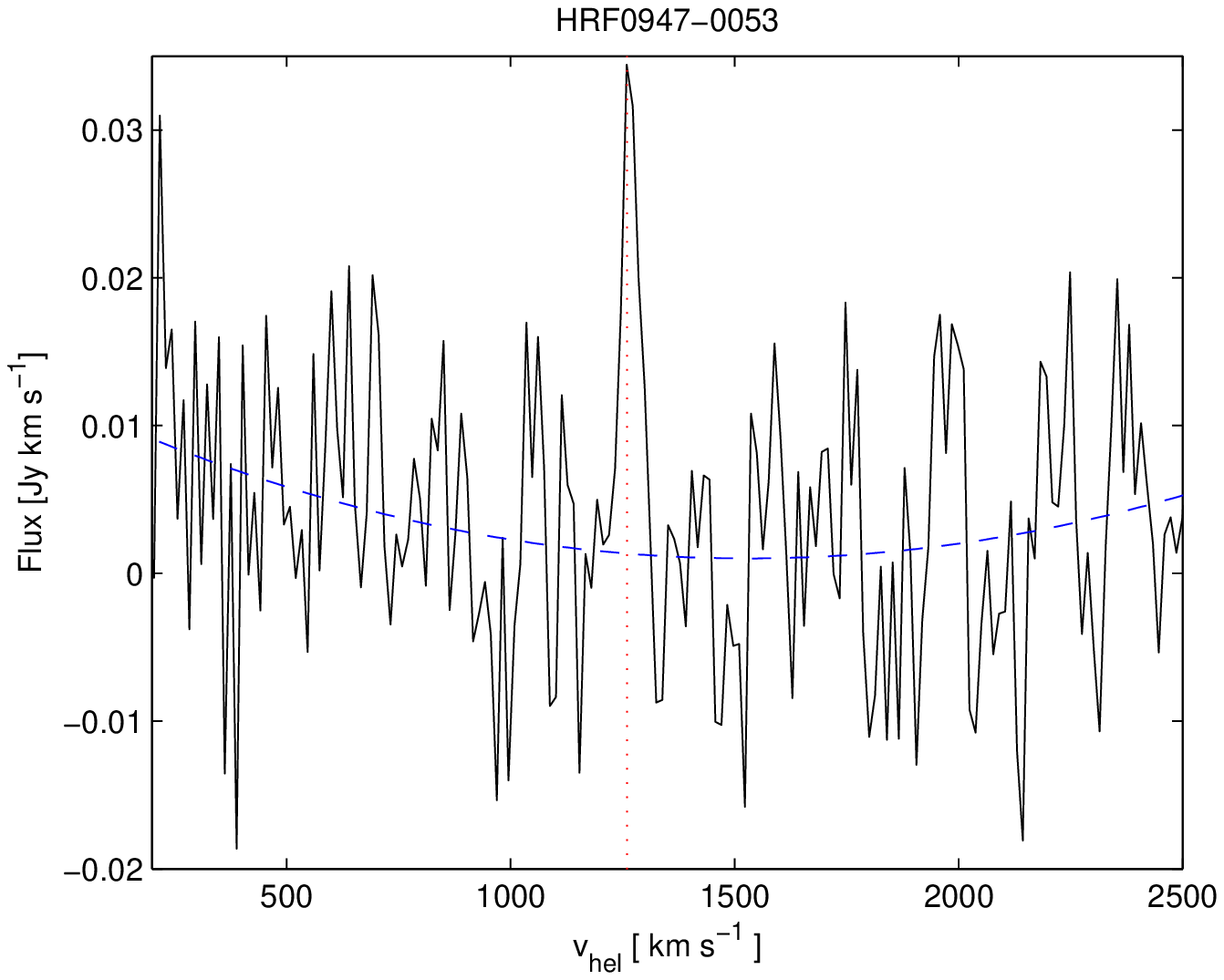}
  \includegraphics[width=0.5\textwidth]{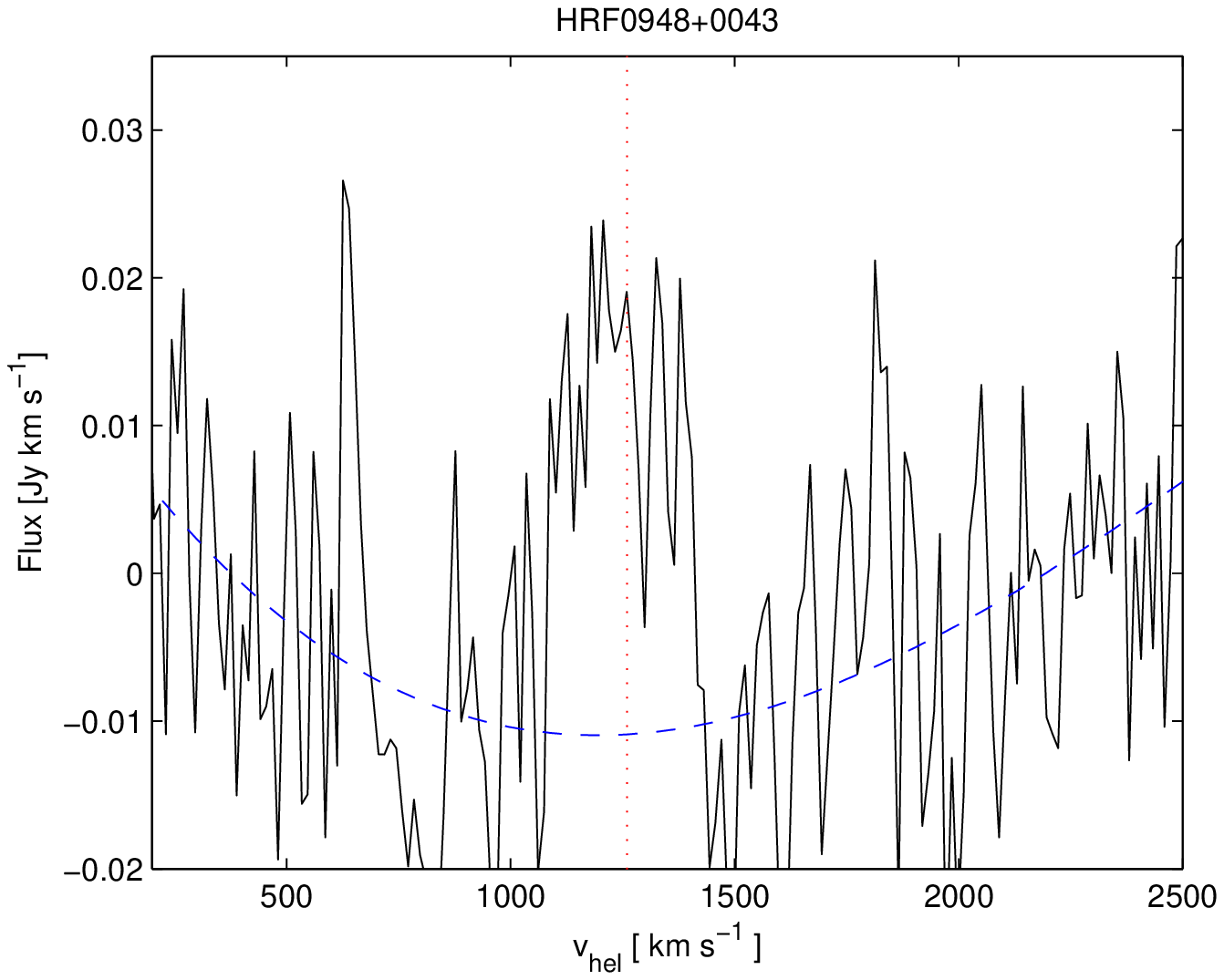}   
  \includegraphics[width=0.5\textwidth]{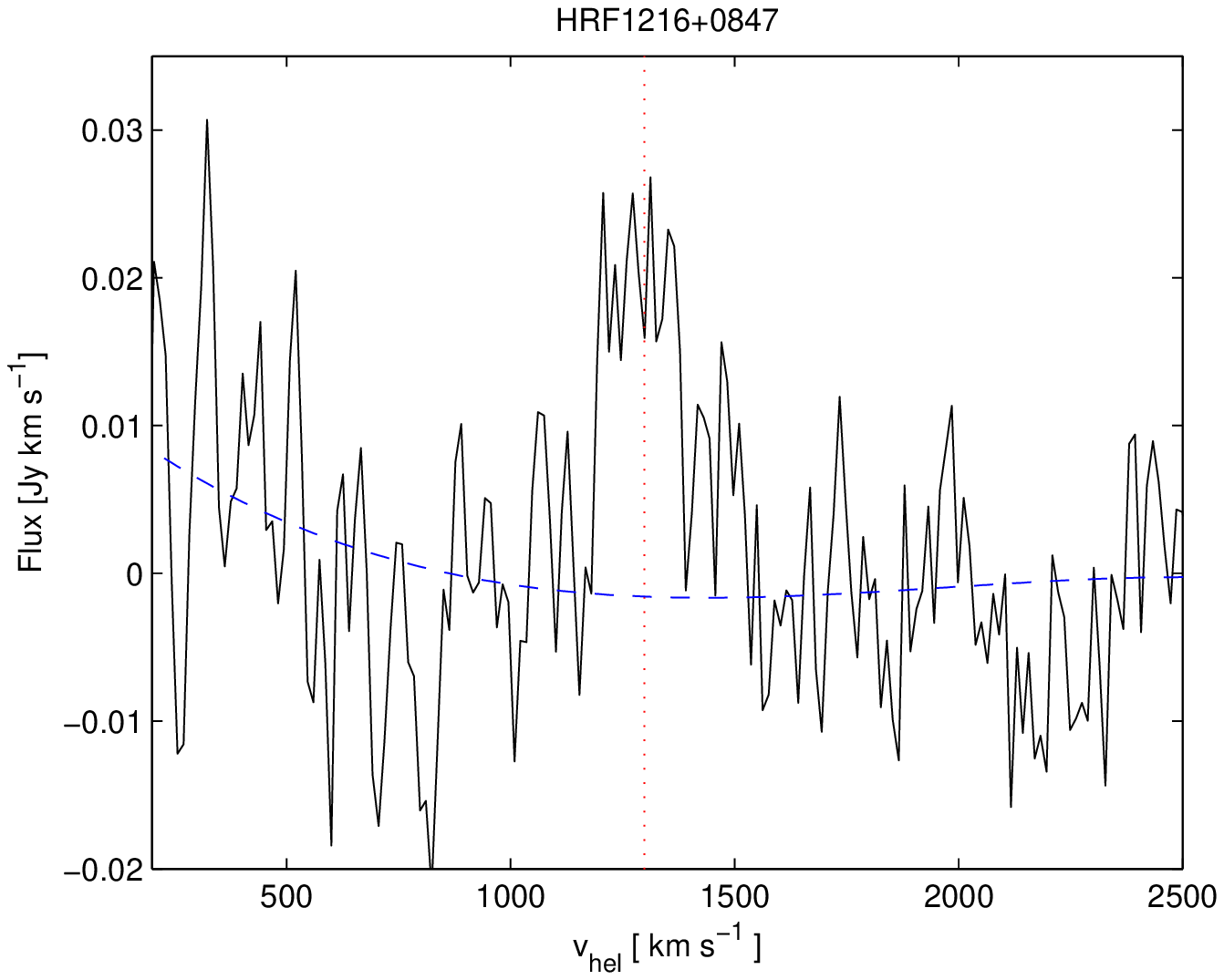}
  \includegraphics[width=0.5\textwidth]{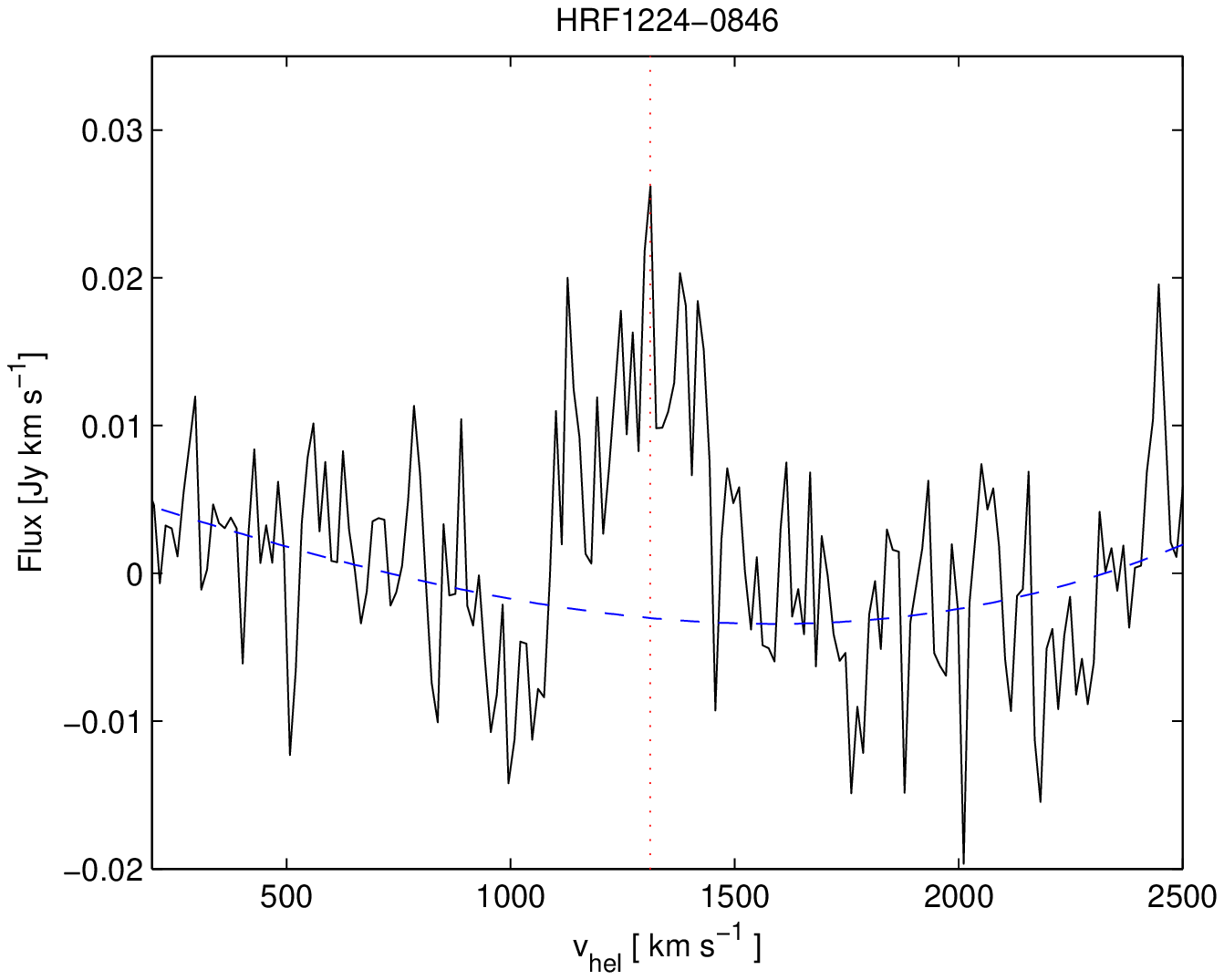}
  \includegraphics[width=0.5\textwidth]{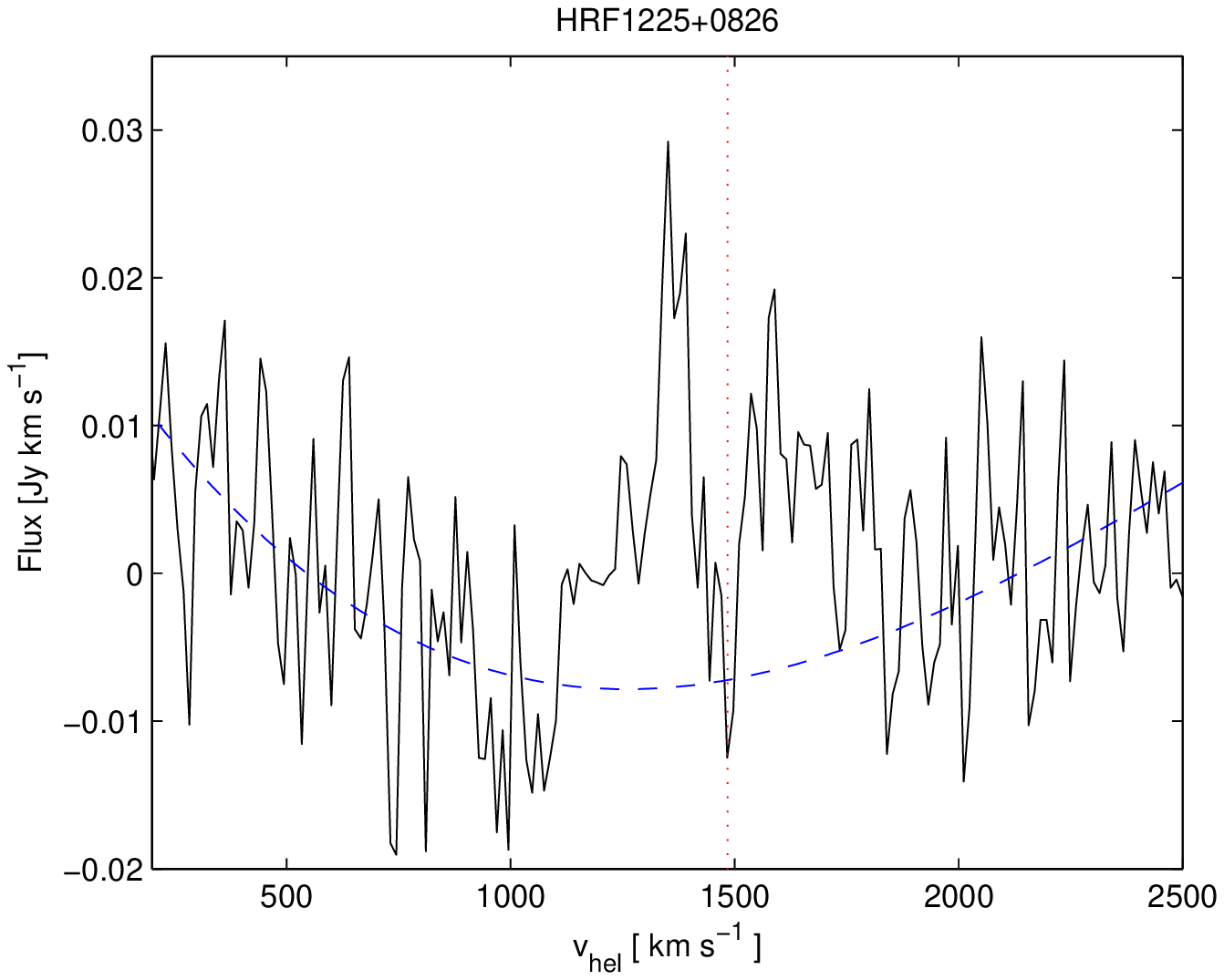}
 
  \caption{Spectra of {\HI} features that have been detected in the
    vicinity of other objects, by inspecting the moment maps. A third
    order polynomial is fit to the bandpass and is indicated by
    the dashed line. The vertical dotted line represents the radial
    velocity of the detected feature.}
  \label{filaments}
\end{figure*}

Several faint features have been marginally detected, that have
not emerged from our earlier source finding procedure. These
  detections appear very interesting but would need further
  confirmation to make them robust. They are usually very faint, but
some of them have very broad line-widths.  None of the features have
an optical counterpart, therefore the origin of the features is not
straightforward. 
All of these tentative detections were found by visually
inspecting the moment maps around bright sources. As none of them
passed the criteria of previous source finding algorithms, it is very
likely that a significant number of comparable features are still
present in the data. We will discuss the detailed properties of all
features below. The relevant spectra are shown in
Fig.~\ref{filaments}. A third order polynomial was fit to the
spectrum, excluding the line itself and galactic emission, to correct
for bandpass instabilities at these very low flux values. We will
leave statements about the possible origin of these features to the
discussion.

\begin{figure*}[t]
  \includegraphics[width=0.5\textwidth]{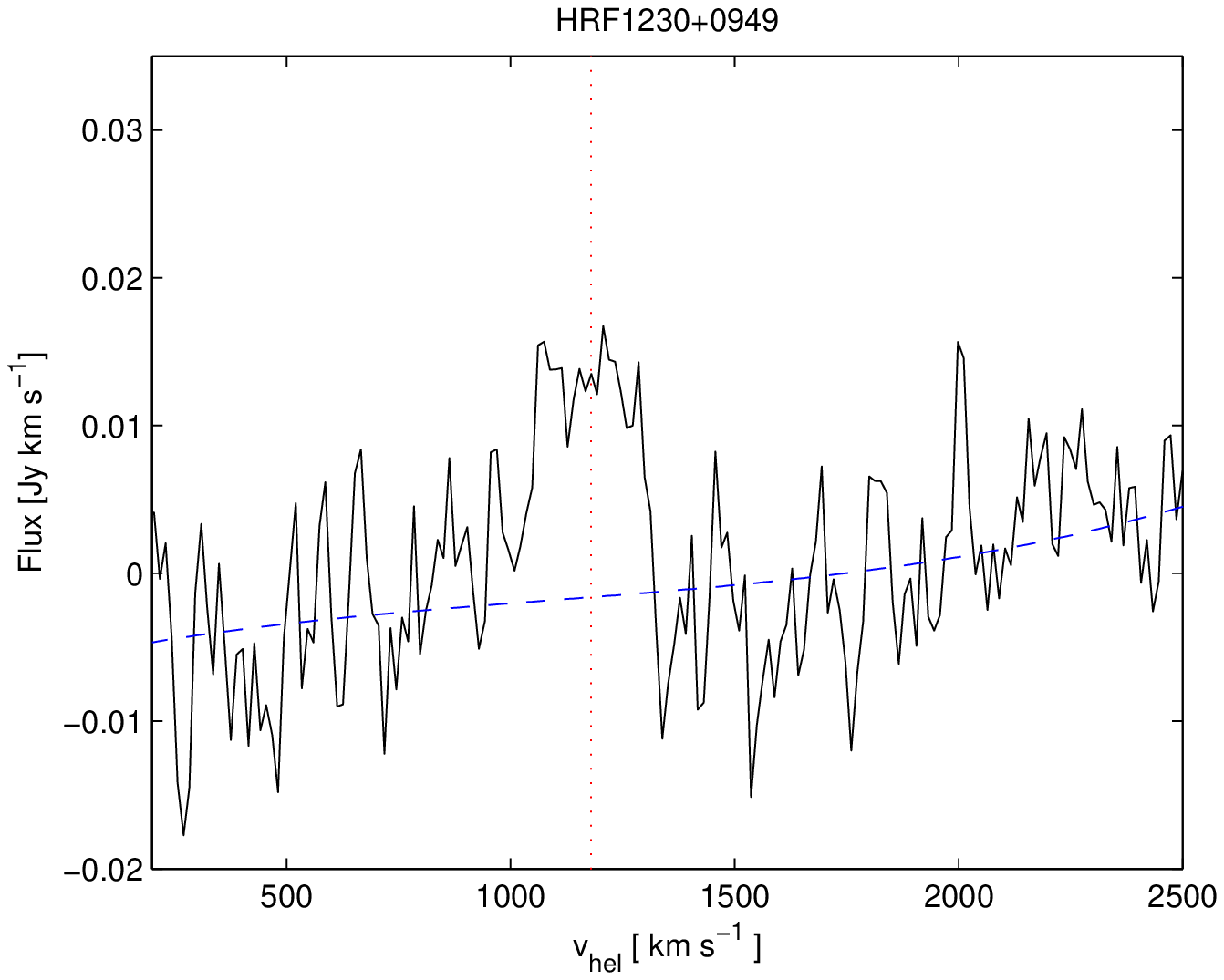}
  \includegraphics[width=0.5\textwidth]{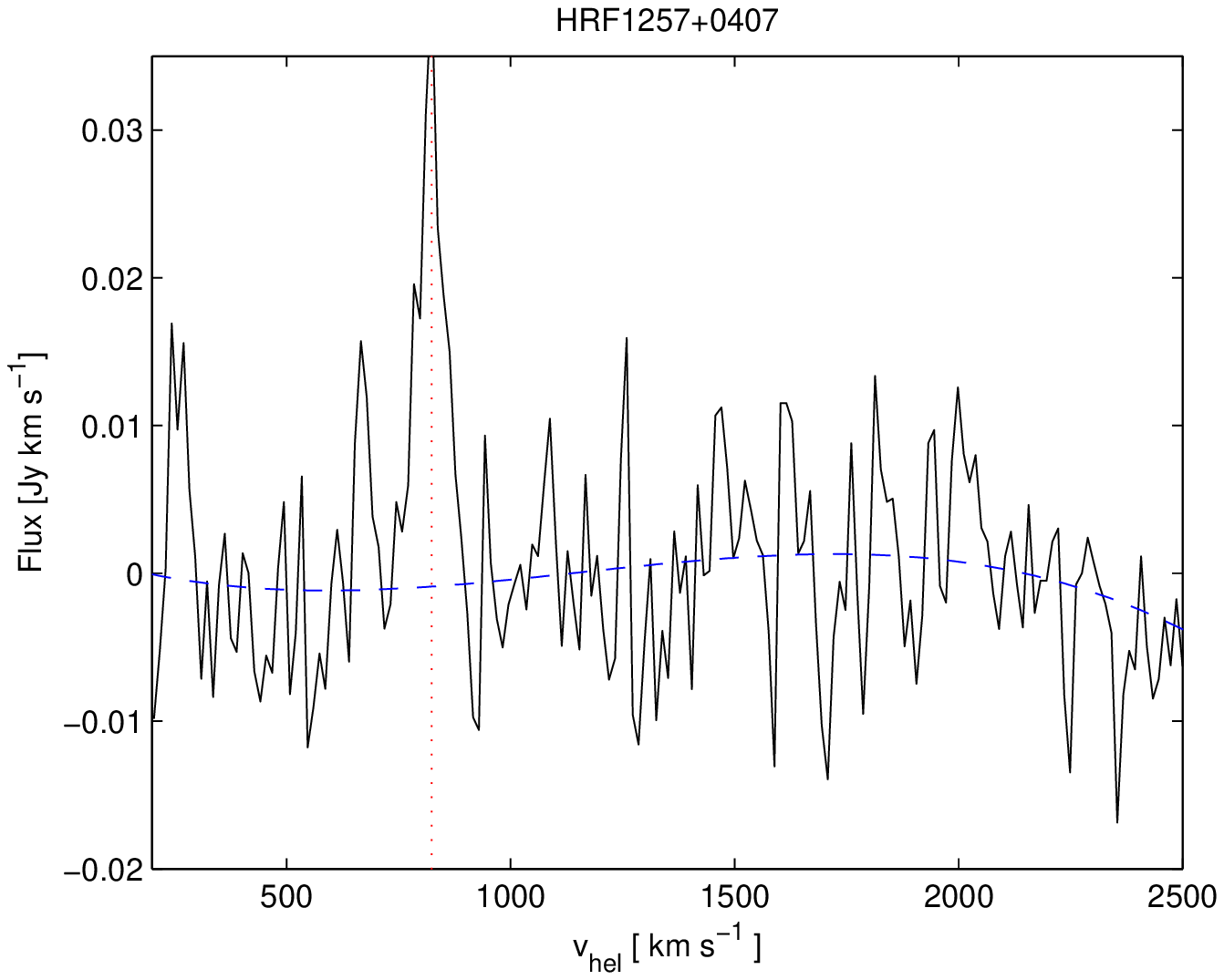}
  \includegraphics[width=0.5\textwidth]{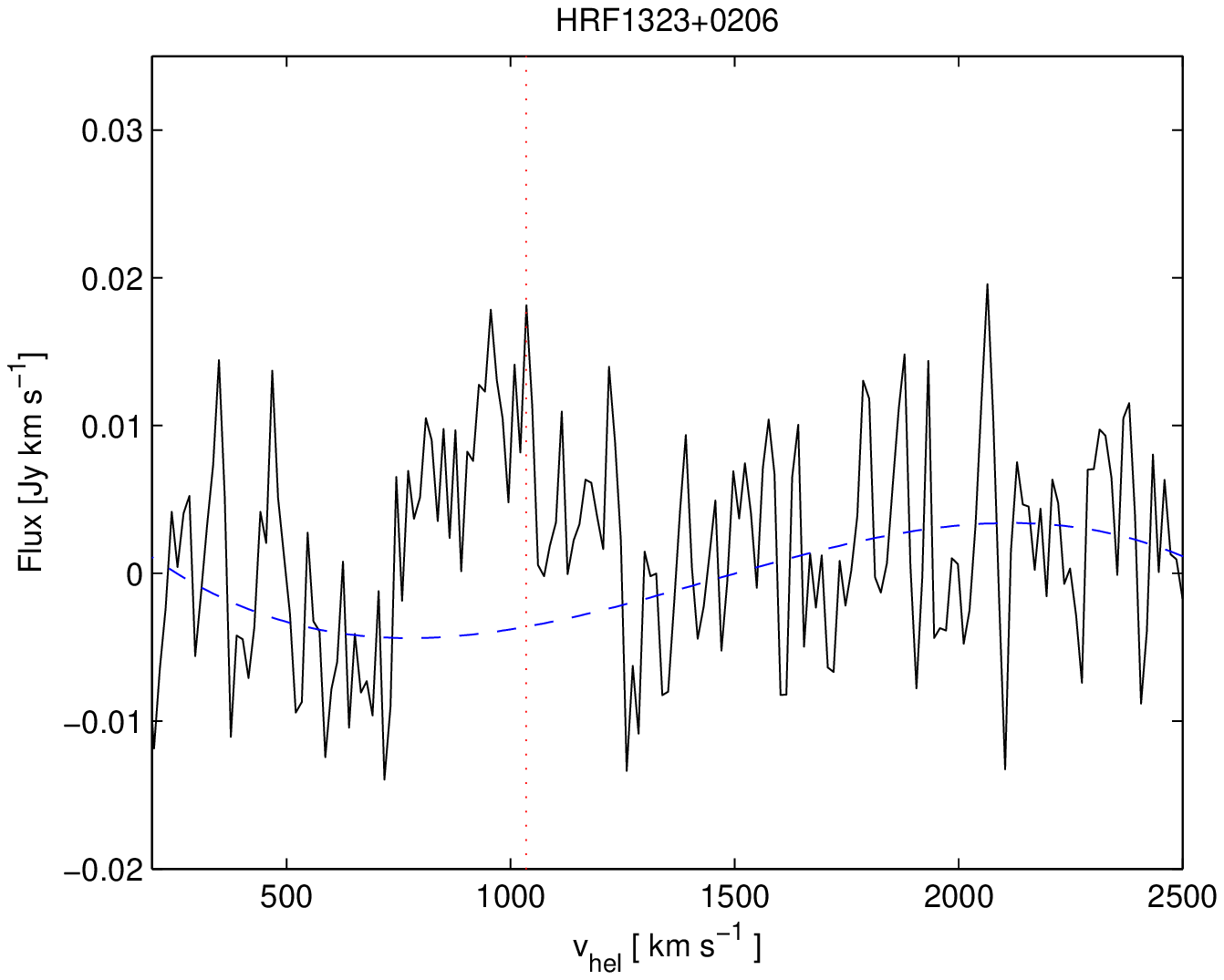}
  \includegraphics[width=0.5\textwidth]{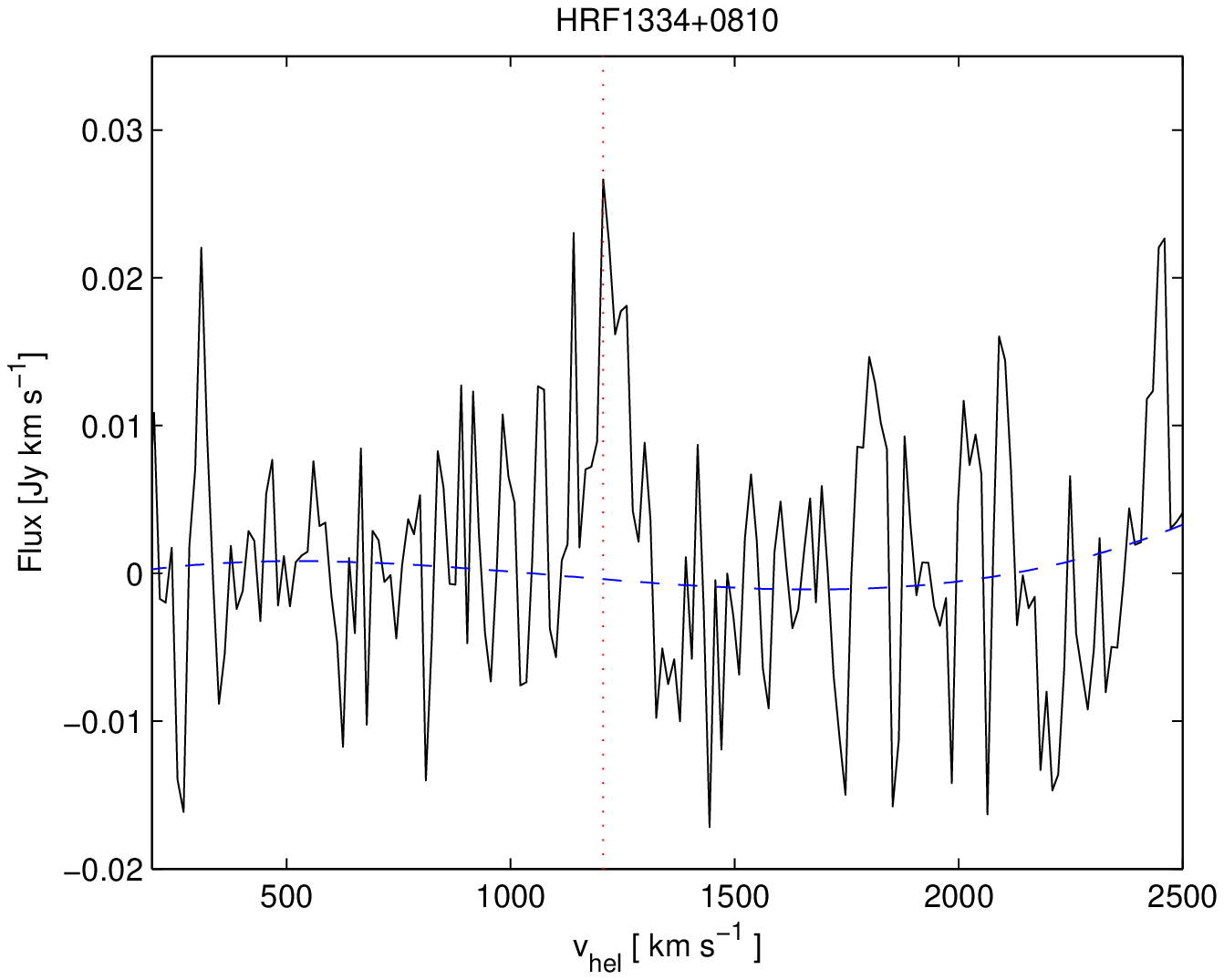}
  {\bf Fig~\ref{filaments}.} (continued)                                        
\end{figure*}

{\bf HIRF~0912+0723}: The position of this feature, at RA=09:12:55 and
DEC=07:23:34, is close to UGC~4781 and NGC~2777 and it has a
comparable radial velocity to these galaxies of 1458 km s$^{-1}$. The
spectrum has two peaks, that are separated by 140 km s$^{-1}$. The two
peaks are reminiscent of a double horned profile, although they may
simply be due to two unrelated structures within the telescope beam. A
direct relation to either of the two cataloged galaxies is not
obvious, as the line is quite broad and has a very different
character. The total line integral at the indicated position is 3.9 Jy
km s$^{-1}$, after correction of the spectral baseline. With an integrated
signal-to-noise of $\sim 7$, this detection has moderately high
significance.

{\bf HIRF~0947-0053} : This feature with RA=09:47:11 and DEC=-00:53:05
is near the optical source SDSS~J094446.23-004118.2, in the same field
as the previous detection. In contrast to HIRF~0948+0043, this
detection is relatively narrow, with a $W_{20}$ value of 77 km
s$^{-1}$ and it has one peak with a maximum brightness of $\sim 33$
mJy beam$^{-1}$. The integrated line-strength is only 1.3 Jy km
s$^{-1}$, yielding a marginally significant signal-to-noise of 5.

\begin{figure}[t]
\begin{center}
  \includegraphics[width=0.5\textwidth]{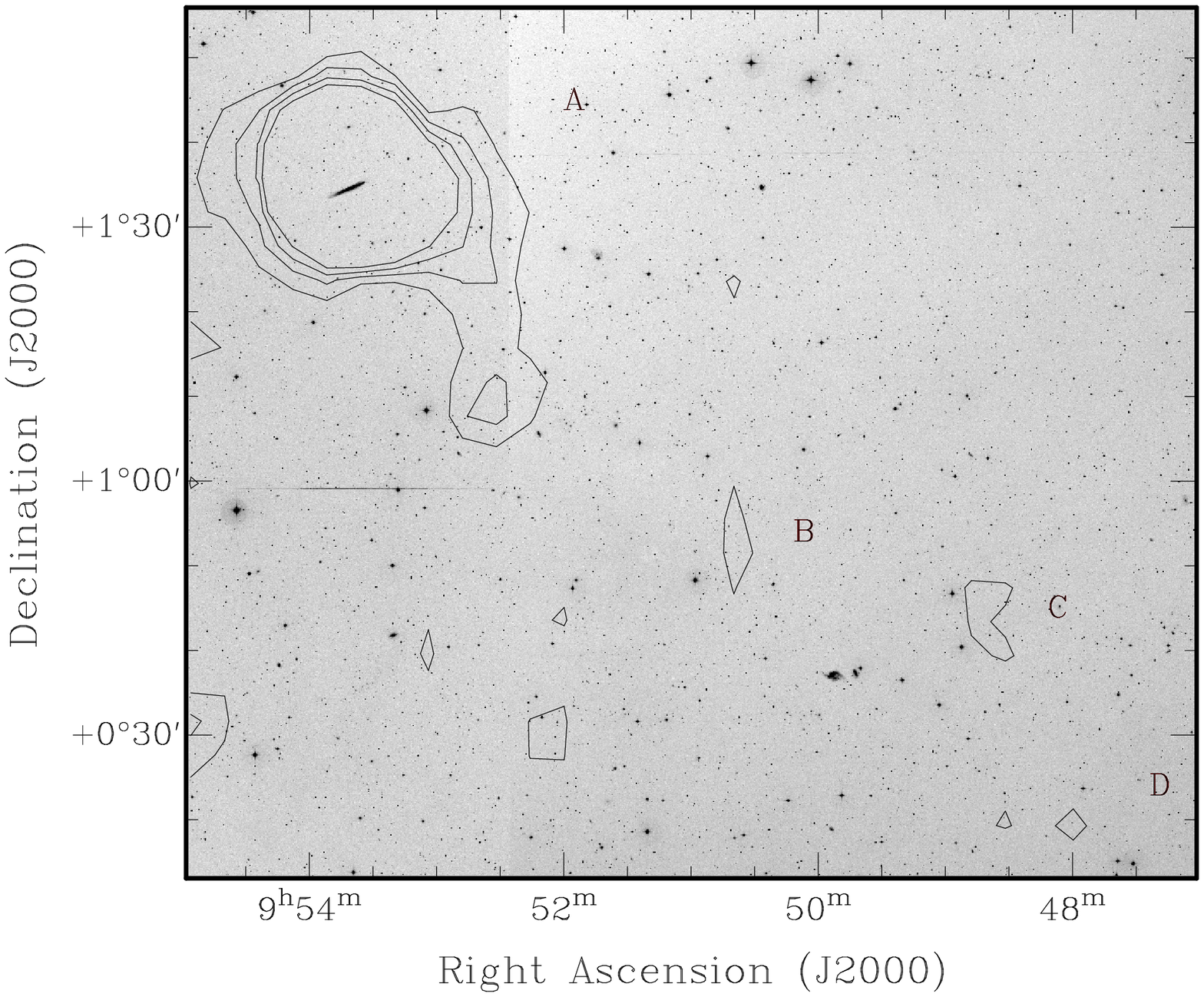}
\end{center}  
  \caption{Second generation DSS image with {\HI} contours at 2, 3, 4 and
    5 Jy beam$^{-1}$ km s$^{-1}$. The large galaxy is NGC~3044, three
    companions are identified with very similar line-widths. The line
    profile of each object is shown in Fig.~\ref{10_43_spect}.}
  \label{HIRF0948+0043}                                       

\end{figure}

\begin{figure*}[t]
\begin{center}
  \includegraphics[width=1.0\textwidth]{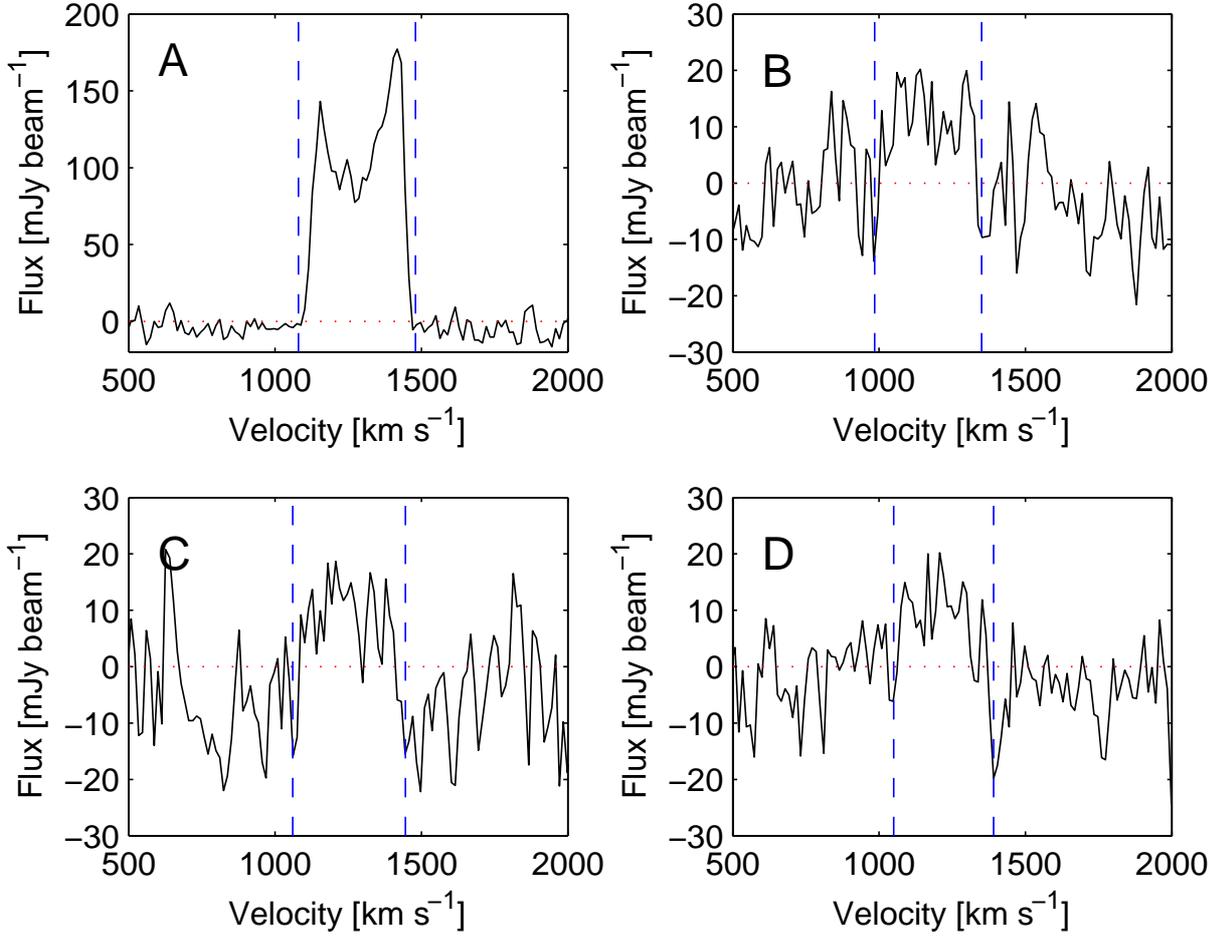}
\end{center}  
  \caption{Line profiles of NGC~3044 (A) and three faint detection (B,
    C, D) in the extended environment of the galaxy.}
  \label{10_43_spect}                                       

\end{figure*}

{\bf HIRF~0948+0043}: This feature is located at RA= 09:48:32 and
DEC=00:43:23, offset by several degrees from NGC~3044 which has a
consistent radial velocity of 1289 km s$^{-1}$.  This is a very broad
profile with a $W_{20}$ value of $\sim 320$ km s$^{-1}$. Although the
brightness in each channel is only about 10 mJy Beam$^{-1}$, this
brightness is present over 25 channels, which yields a high
significance. The line-integral without any further smoothing is 7.6
Jy km s$^{-1}$, which corresponds to a signal to noise of
13. Intriguingly, there appear to be additional filamentary features
in the field of this object. Similar broad line profiles can be
recognised at several positions along the filament, albeit with low
signal-to-noise. A DSS image of the region around HIRF~0948+0043 is
shown in Fig.~\ref{HIRF0948+0043}, which is indicated by the letter C
in this image. The main galaxy in the top left of the plot is NGC~3044
which has extended emission. Letters B, C and D are assigned to
regions in the environment showing emission. The line profiles are
shown in Fig.~\ref{10_43_spect}, although the fluxes in all the
companions are very low, they all have a line-width that is very
comparable to NGC~3044. It appears as if these denser regions form a
more extended filament.  In \cite{1997ApJ...490..247L} a high
resolution {\HI} image of NGC3044 obtained with the VLA is presented.
Although the brightness sensitivity of these observations is much
lower than the Parkes data, at column density levels of a few times
$N_{HI} \sim 10^{20}$ cm$^{-2}$ small companions can be seen in the
same direction as the companions we detect. 

{\bf HIRF~1216+847}: At RA=12:16:45 and DEC=08:47:22 this detection has
a radial velocity of 1299 km s$^{-1}$ and $W_{20}$=199 km
s$^{-1}$. Although the peak brightness is not very substantial; $\sim
20$ mJy beam$^{-1}$ over the full line width, the integrated flux of
4.3 Jy km s$^{-1}$ corresponds to a signal-to-noise of 9.

{\bf HIRF~1224+0846}: With an RA of 12:24:52 and DEC of 08:46:25, this
detection is within half a degree of NGC~4316. The radial velocity of
1312 km s$^{-1}$ matches that of NGC~4316 fairly well. Of particular
note is the very broad line-width of approximately 400 km
s$^{-1}$. While the peak brightness is modest; only 24 mJy beam$^{-1}$
it extends quite uniformly across the entire line-width. This yields an
integrated flux of 5.3 Jy km s$^{-1}$ corresponding to a signal-to-noise
of 9.

{\bf HRIF~1225+0826}: Located at RA=12:25:40 and DEC=08:26:16, this
intriguing feature's properties are very sensitive to the method of
baseline definition. The overlaid baseline fit results in a central
radial velocity of 1484 km s$^{-1}$, an integrated flux of 8.4 Jy km
s$^{-1}$, an extremely broad line-width of about 700 km s$^{-1}$ at
20\% of the peak flux and a signal-to-noise of 10. However, other
spectral baselines would severely diminish the apparent line-width,
flux and overall significance of this feature. A slightly
  enhanced baseline fit would also cut the source in half, therefore
  potentially this detection could also consist of two different
  sources with a different spectral position within the telescope
  beam.   Confirming observations will be necessary to establish it's
reliability.

{\bf HRIF~1230+0949}: This is a very similar detection to several of
those discussed previously, with a low peak brightness, but a very
broad line-width. This objects is located at RA=12:30:24 and
DEC=09:49:45 with a radial velocity of 1180 km s$^{-1}$. The
integrated flux of 3.7 Jy km s$^{-1}$ over a line-width $W2_{20}$ of
280 km s$^{-1}$, yield a signal-to-noise of 7.

{\bf HIRF~1257+0407}: This detection at RA=12:57:20 and DEC=04:07:59 is
in the direct environment of NGC~4804 with a radial velocity of
824 km s$^{-1}$ that is comparable to this galaxy. The peak flux of 35 mJy
beam s$^{-1}$ is relatively strong, compared to the other detections
listed here. The line-width at 20\% of the peak is 102 km s$^{-1}$ and
the integrated flux of 2.4 Jy km s$^{-1}$ has a signal-to-noise of 7.

{\bf HIRF~1323+0206}: This detection is at RA= 13:23:17 and
DEC=02:06:06, with a radial velocity of 980 km s$^{-1}$. This is
another example of a very faint source with a very broad line width of
more than 500 km s$^{-1}$. The brightness in each individual channel
barely exceeds the $1\sigma$ level. The line-width at 20\% of the peak
is 505 km s$^{-1}$ and the integrated line strength is 5.2 Jy km
s$^{-1}$, which corresponds to a signal-to-noise of 7.

{\bf HIRF~1334+0810}: This detection at RA=13:34:21 and DEC=08:10:14
is only marginal. The peak brightness is $\sim 25$ mJy beam$^{-1}$ and
the integrated flux is 2.2 Jy km s$^{-1}$ and has a signal to noise of
only 6.5 when taking into account the $W_{20}$ of 110 km s$^{-1}$.
Interesting however is that this feature is about half a degree south
of an extended chain of galaxies that is connected to NGC 5248 in the
data. Although these galaxies individually could not be resolved in
the HIPASS data, optical images from DSS can reveal UGC~8575 and
CGCG~073-036. Both these galaxies and HIRF~1324+0810 have a similar
radial velocity of $\sim$1200 km s$^{-1}$ and a narrow line profile
that is completely embedded in the profile of NGC~5248. This diffuse
detection seems to be connected to the filament of galaxies, although
the connecting bridge is very faint.

\subsection{Completeness and Robustness}

Our source catalogue has been constructed from all sources which have
a peak brightness exceeding our 5$\sigma$ limit of $\sim$50 mJy
beam$^{-1}$ at ``full'' velocity resolution of 26~km~s$^{-1}$ and
correspondingly fainter brightnesses after velocity smoothing to 52
and 104~km~s$^{-1}$. Simulations involving the injection of artificial
sources into similar total power {\HI} data-cubes by
\cite{2002ApJ...567..247R} have shown that an asymptotic completeness
of about 90\% is reached at a signal-to-noise ratio of 8, while the
completeness at a signal-to-noise ratio of 5 is only likely to be
about 30\%. 

We have used the {\it Duchamp} \citep{2008glv..book..343W}
source finding tool to find candidate sources in the data cubes at the
different velocity resolutions. The source finder was found to be
robust down to this 5$\sigma$ limit on peak brightness. All candidate
sources were subsequently inspected visually. Artefacts from solar
interference or due to bright continuum sources could easily be
rejected. A lower threshold for the source finder resulted in a
much larger proportion of candidates that were deemed unreliable after
visual inspection. Moment maps of all candidate detections have been
further inspected and analysed interactively resulting in some
additional detections of interesting features. As these features were
sought preferentially in the direct vicinity of other galaxies, they
have not been cataloged to the same level of completeness throughout
the data volume. Although many of these features appear highly
significant, their derived properties are often quite sensitive to the
form of the spectral baseline that is subtracted.

We have found two classes of objects. The first ones are relatively
narrow lines, with a total line width of $\sim 100$ km s$^{-1}$ or a
FWHM of $\sim 50$ km s$^{-1}$ and a peak brightness per 26~km~s$^{-1}$
velocity channel that only exceeds the local noise by a factor three
or four.  The second class of objects has very broad line widths of up
to $\sim 500$ km s$^{-1}$, however the measured brightness per line
channel only exceeds the local noise by a factor two, or not at
  all. Because of the broad line widths, the integrated detections
have a high significance, but such features are particularly difficult
to detect with an automatic source finder.  Although the detections
seem significant, they need further confirmation as the nature and
broad line-width of these objects is unexpected. Nevertheless it seems
unlikely that these detections are an artefact of the reduction
pipeline or the method of bandpass estimation. Only a second order
polynomial has been fit to the bandpass in each spectrum. For most of
the broad lines, a range of channels is systematically elevated above
the rest of the spectrum and the transition is quite sharp.  Higher
order polynomials would be needed to artificially create such
features.

\section{Discussion}
In this paper we describe how a significant part of the raw HIPASS
observations have been reprocessed. The reprocessed region covers the
right ascension range from 8 to 17 hours and Declinations from -1 to
10 degrees. A source catalogue and {\HI} features without optical
counterparts have been presented. Although the original HIPASS product
is an excellent one in its own right, by improving the reduction and
processing pipeline the quality and number of detections can be
significantly improved. The main purpose of reprocessing this
particular region is because of the overlap with the WVFS survey as
described in \cite{2011A&A...527A..90P} and \cite{2011A&A...528A..28P}
and the HIPASS data complements the two data products of the
WVFS survey.  The first WVFS product has a worse flux sensitivity
  than the HIPASS data, but a better column density sensitivity due to
  the very low resolution of this dataset. On the opposite the second
  WVFS product has a better flux sensitivity than the HIPASS data, but
  a worse column density sensitivity as the resolution of this data is
  higher than the resolution of the Parkes telescope.

We will leave detailed analysis and discussion to a
later paper, when the three independent datasets will be compared. A
few comments will be made that are relevant to this dataset and these
detections.

\subsection{Red-shifted OH}
There are several detections, both in the source list, as well as
identified as possible filaments, that do not have an optical
counterpart. Because there is no optical counterpart, the origin of
these features is not straightforward. All detections were considered
to be {\HI} detections that could reside in the vicinity of other
objects as tidal remnants or Cosmic Web features. 
Another scenario that has not been explored is red-shifted OH emission
from sources at a redshift of $z\sim0.15$. The 1665.401/1667.358 MHz
doublet of an OH megamaser emitted at this redshift would have an
observed frequency of $\sim 1415$ MHz. Although confirmed detections
of these red-shifted OH megamasers have not yet been reported, they
are predicted to be found in blind {\HI} surveys . ALFALFA
\citep{2005AJ....130.2598G} expects to find several dozen OHMs in the
redshift interval 0.16-0.25. Although the area covered by ALFALFA is
significantly larger, based on these numbers we can expect to detect a
few OHMs in our survey volume. To detect the OH doublet, two similar
peaks should be identified with a separation of $\sim 350$ km
s$^{-1}$. When looking at the profiles of known OHMs in
e.g. \cite{2002AJ....124..100D}, the doublet is not always clearly
apparent and so this requirement might be somewhat relaxed. All the
documented OHMs do have a broad line-width, typically larger than 300
km s$^{-1}$. This consideration rules out most of our detections
without optical counterparts as candidate OHMs, as the line-widths are
much narrower. There are however a few cases, where this may be a
possible scenario, namely: HRF 0948+0043, HRF 1224+0846, HRF
1225+0826 and HRF 1323+0206. The other prediction of
this scenario is that a suitable ULIRG at $z\sim0.15$ should be
coincident with the $\sim$1415 MHz line detection. We have sought
for objects at the appropriate redshift that coincide with the spatial
positions of these detections, but did not find any sources that
could cause redshifted OH emission. 

\subsection{Gas accretion modes}
An interesting question regarding structure formation is how the
intergalactic medium fuels the galaxies; ie. how gas is accreted. The
two most discussed scenarios are hot mode and cold mode accretion
\citep{2005MNRAS.363....2K}. The line-width of a detection can be used
to estimate the upper limit of the kinetic temperature of the gas and
is given by:

\begin{equation}
T_{kin} \leq \frac{m_H\Delta V^2}{8k\ln 2}
\end{equation}

where $m_H$ is the mass of an hydrogen atom, $k$ is the Boltzmann
constant and $\Delta V$ is the {\HI} line-width at FWHM. This equation
gives an upper limit to kinetic temperature, as internal
turbulence or rotation can also increase the line-width of an object.

In the case of cold mode accretion with temperatures of the order of
$T < 10^5$ K, the line-widths of the gas are relatively narrow, up to
$\sim100$ km s$^{-1}$. The conditions to observe such gas are
relatively easily satisfied, the neutral fraction in cold gas is
  still significant so that the {\HI} column density is still high. 
Although it is difficult to distinguish tidal remnants from pristine
gas that is fuelling the galaxies, gas accretion is a very plausible
scenario. The details of our detections will be discussed later when
other data products are included. There are however a handful of
detections in the direct vicinity of other galaxies that are not
bright, but have line-widths up to $>500$ km s$^{-1}$. One possible
scenario is that this is red-shifted OH emission as is discussed in
the previous subsection, but this is very unlikely. These
line-widths are however also the line-widths that are expected in the
case of hot-mode-accretion. This is gas that is gradually shock
  heated during structure formation to virial temperatures and than
  rapidly cools down to accrete onto the galaxies, for a more extended
  explanation see \cite{2005MNRAS.363....2K}. Because this gas is
highly ionised, the neutral fraction is very low, so the {\HI}
component of such gas is extremely small and very unlikely.
 
\section{Conclusion}
Original data of the {\HI} Parkes All Sky Survey has been reprocessed,
that overlaps in sky coverage with the Westerbork Virgo Filament
Survey. This region was selected to complement the WVFS and use HIPASS
to confirm candidate detections. Furthermore, HIPASS is an excellent
dataset, to search for diffuse {\HI} features that can be related to the
neutral component of the Cosmic Web.

By using an improved reduction strategy, we achieved a reduced RMS
value and lower artefact level, compared to the original
HIPASS product. In the reprocessed data, we achieve a noise value of
$\sim 10$ mJy beam$^{-1}$ over 26 km s$^{-1}$, which is a $\sim 20$\%
improvement over the original HIPASS product. The data has a
{\bf $1\sigma$} brightness sensitivity of $\sim 3.5 \cdot 10^{17}$ cm$^{-2}$, which
allows direct detection in {\HI} emission of some of the higher column
density Lyman Limit Systems seen in QSO absorption line studies.

The major difference with respect to the original HIPASS
cubes is that negative artefacts in the bandpass in the vicinity of
bright sources are almost completely eliminated. This allows us to search
for diffuse and extended {\HI} emission, which has not been possible before.

In total we have detected 203 objects in the reprocessed region, of
which 29 had not been catalogued in the original HIPASS source
catalogue. Fourteen of these detections are completely new {\HI}
detections, of which many do not have an optical counterpart. Although
these detections are briefly mentioned, detailed discussion and
possible confirmation from other data sets will be presented in
a subsequent paper in this series.

In this work only a relatively small part of the HIPASS data has been
reprocessed, as that survey covers the complete Southern sky and the
Northern sky up to +24 degrees in Declination. Apart from this work,
steps have been undertaken to further improve the reduction pipeline
and reprocess the complete survey area. Improved data cubes, together
with improved object-searching algorithms will permit detection of
significantly more sources.  This will provide improved statistics on
the distribution of {\HI} in the local universe. The reprocessed cubes
will also permit an unbiased search for diffuse neutral hydrogen and
Cosmic Web filaments, although currently there is no other all-sky {\HI}
survey with an appropriate brightness sensitivity and distinct angular
resolution to complement this data.

\begin{acknowledgements}
We would like to thank several members of the HIPASS team and
especially Mark Calabretta for help and useful discussions, that
helped in improving the quality of the reduced HIPASS data. The  Parkes telescope is part of the Australia Telescope which is funded by the Commonwealth of Australia for operation as a National Facility managed by CSIRO.
\end{acknowledgements}

%\bibliographystyle{aa}
%\bibliography{names,bibliography}

\newpage

\onecolumn
\begin{appendix}
\section{Spectra {\HI} detections in the reprocessed {\HI} Parkes All Sky Survey data.}

%%%%%%
\begin{figure*}[!h]
  \begin{center}

 \includegraphics[width=0.3\textwidth]{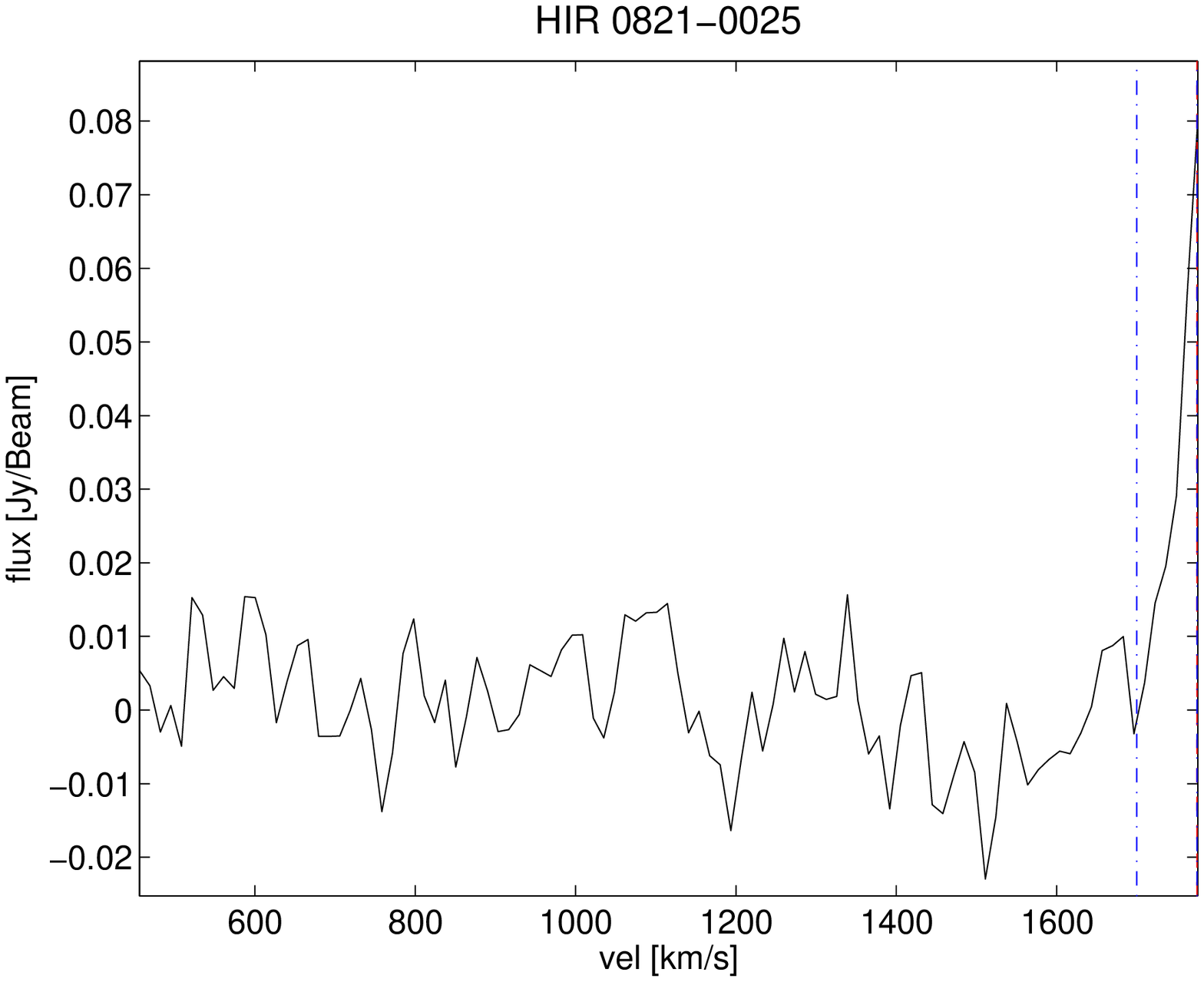}
 \includegraphics[width=0.3\textwidth]{0859+1109_spec.eps}
 \includegraphics[width=0.3\textwidth]{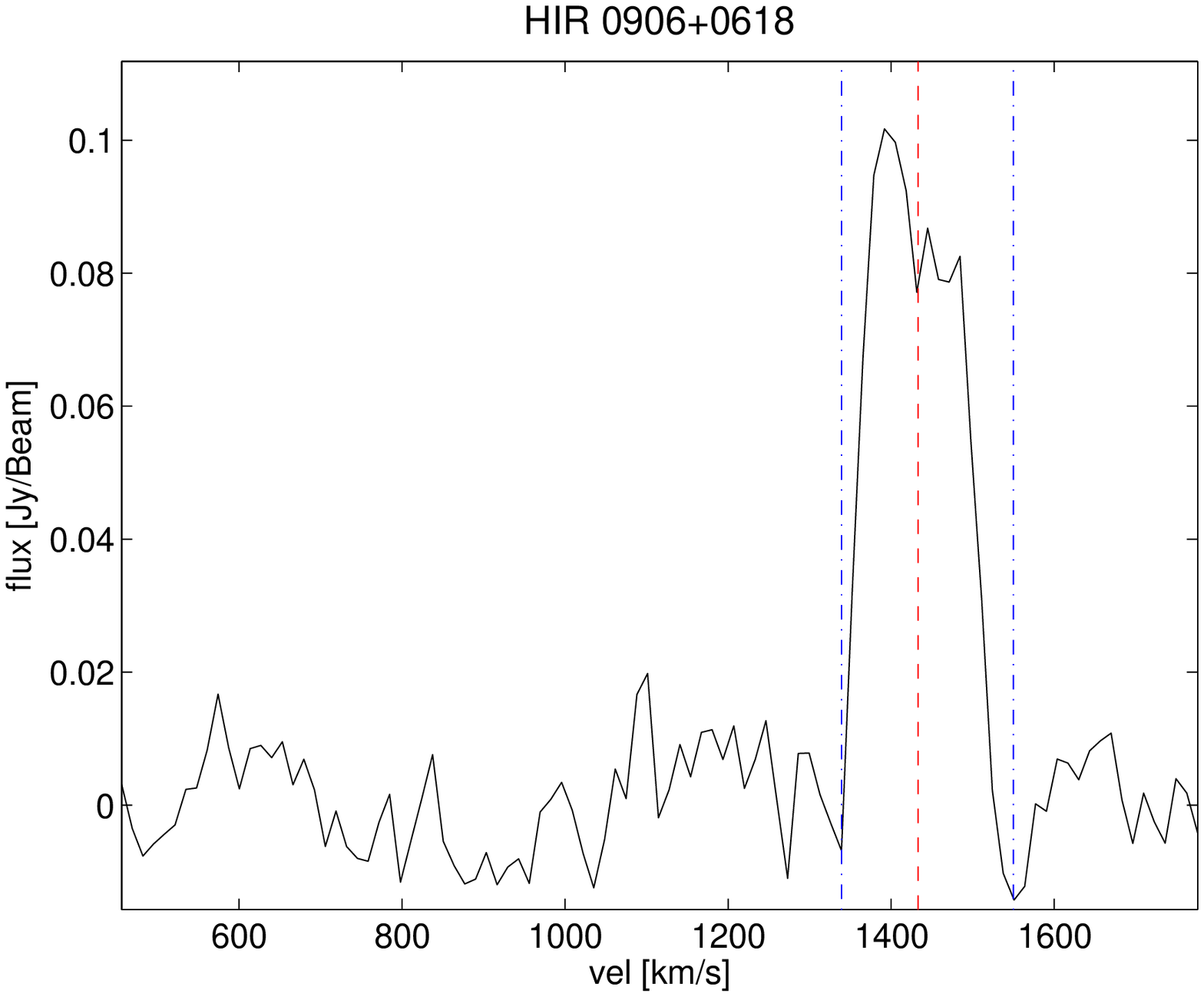}
 \includegraphics[width=0.3\textwidth]{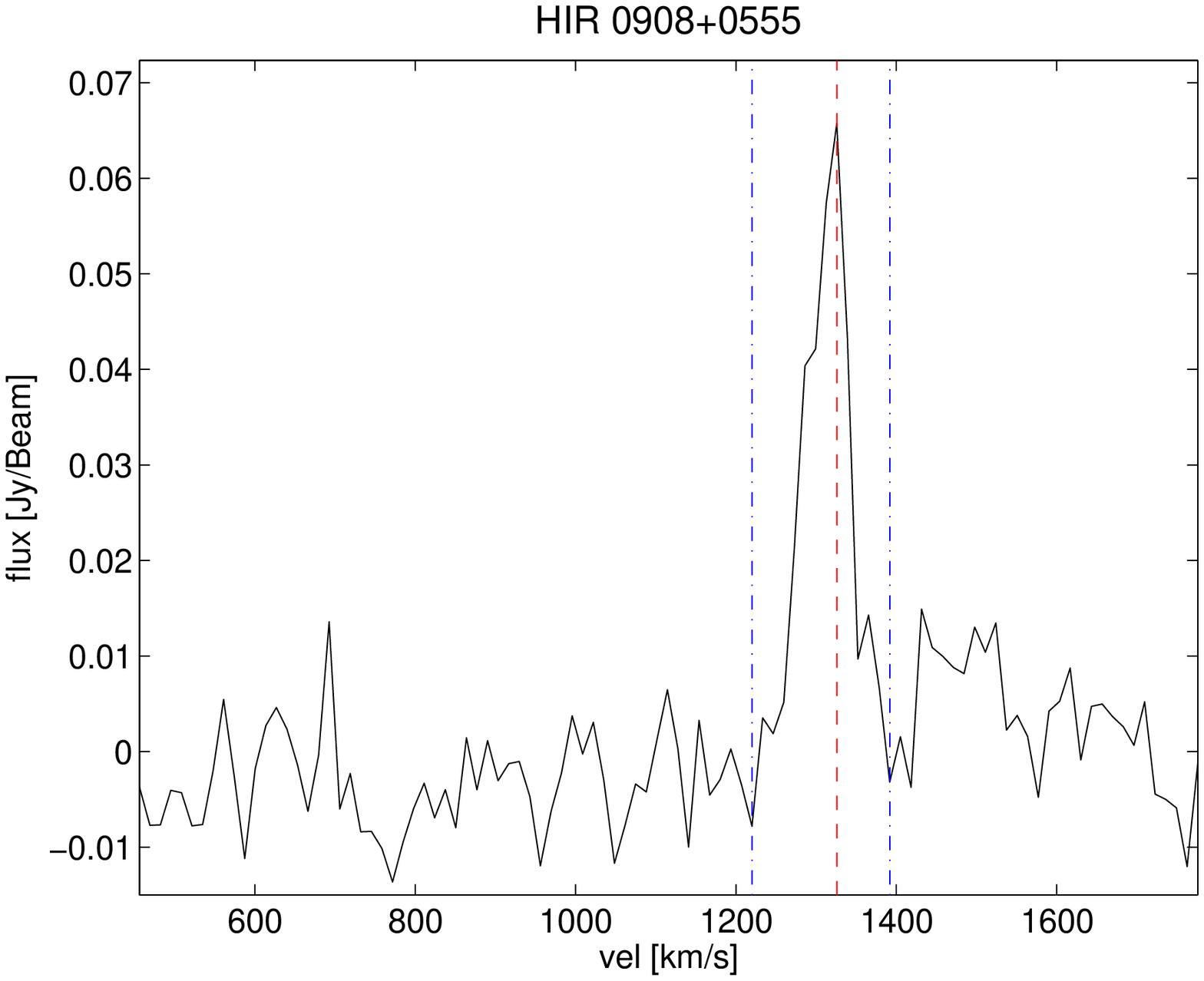}
 \includegraphics[width=0.3\textwidth]{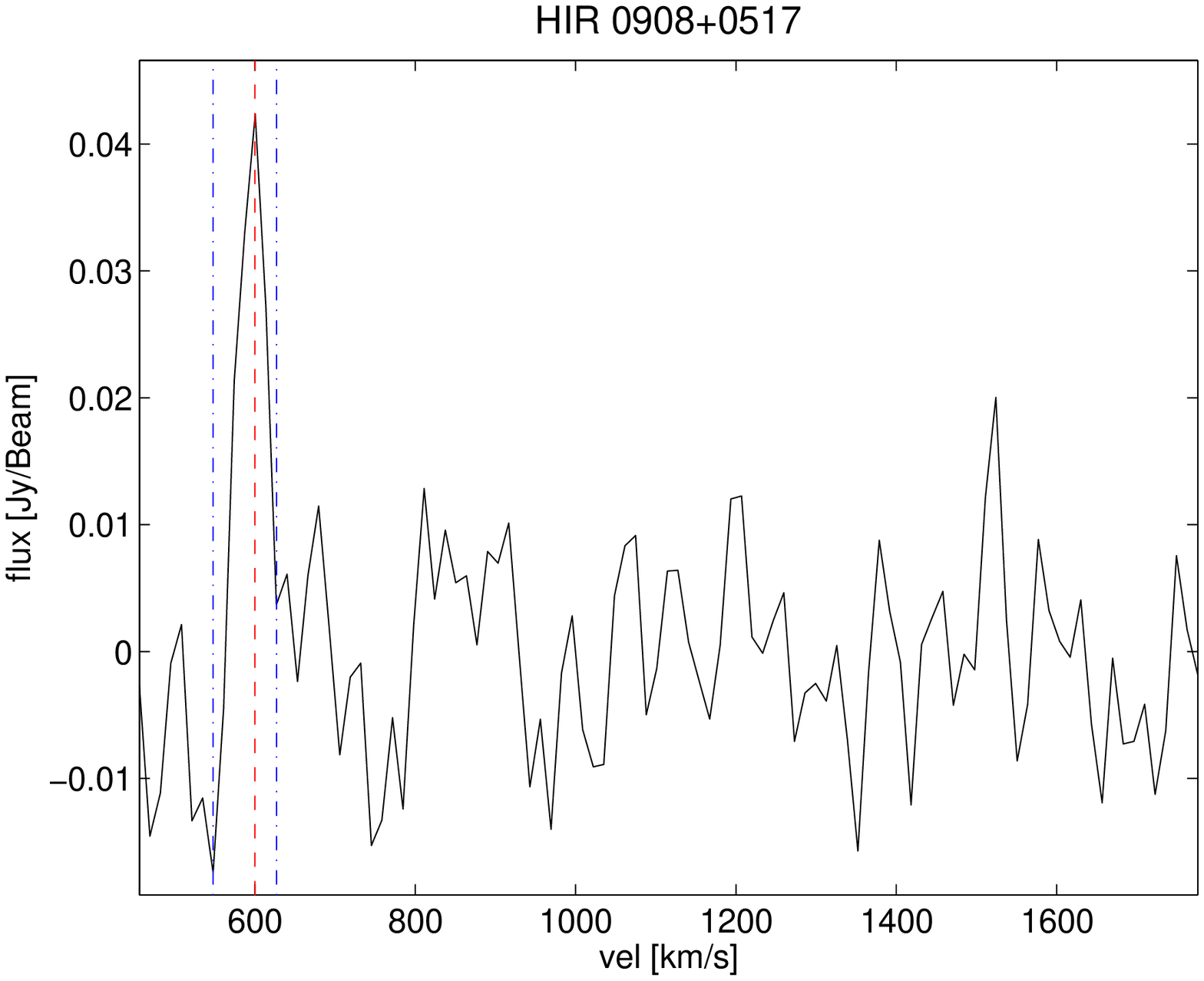}
 \includegraphics[width=0.3\textwidth]{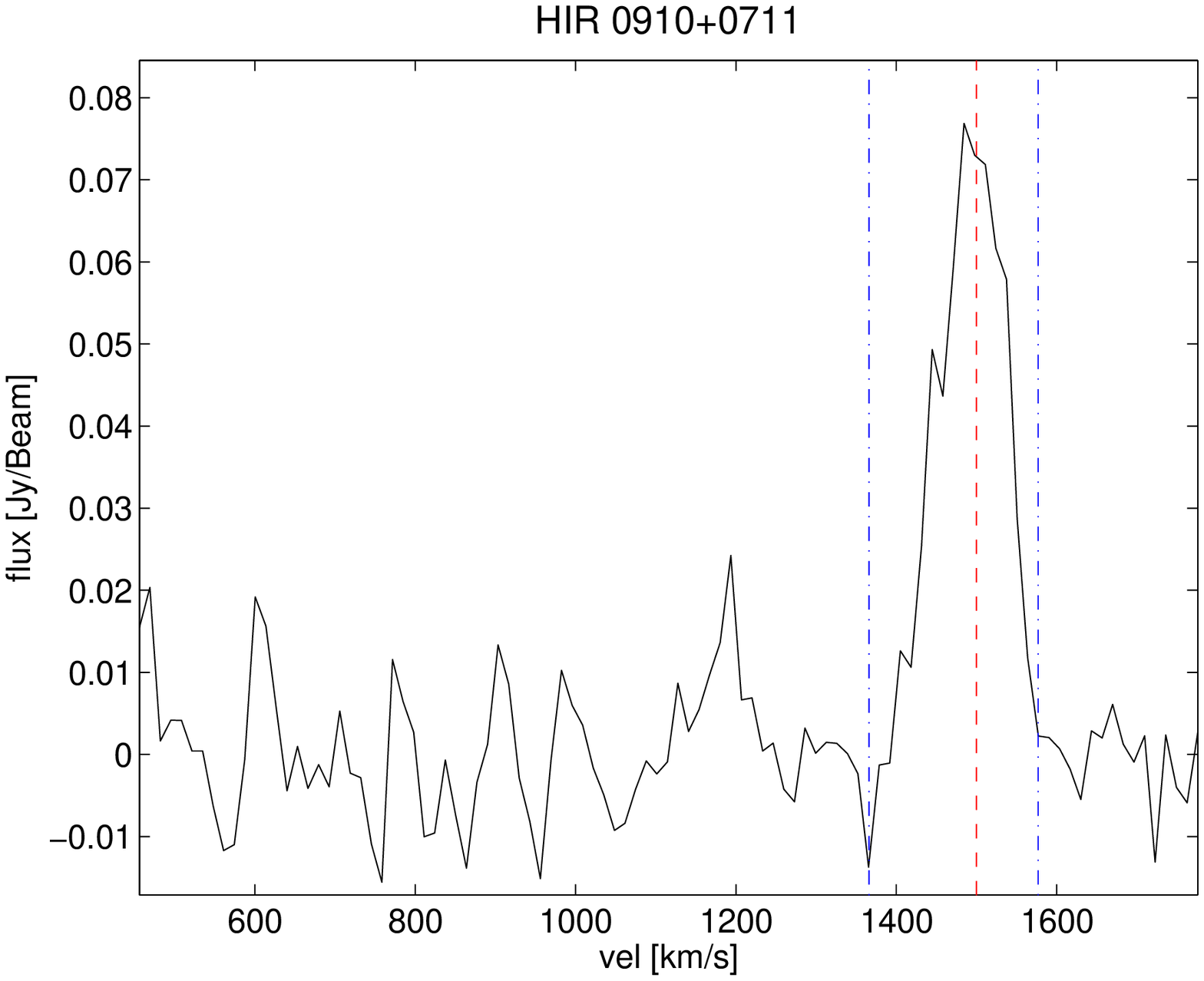}
 \includegraphics[width=0.3\textwidth]{0911+0024_spec.eps}
 \includegraphics[width=0.3\textwidth]{0921+0725_spec.eps}
 \includegraphics[width=0.3\textwidth]{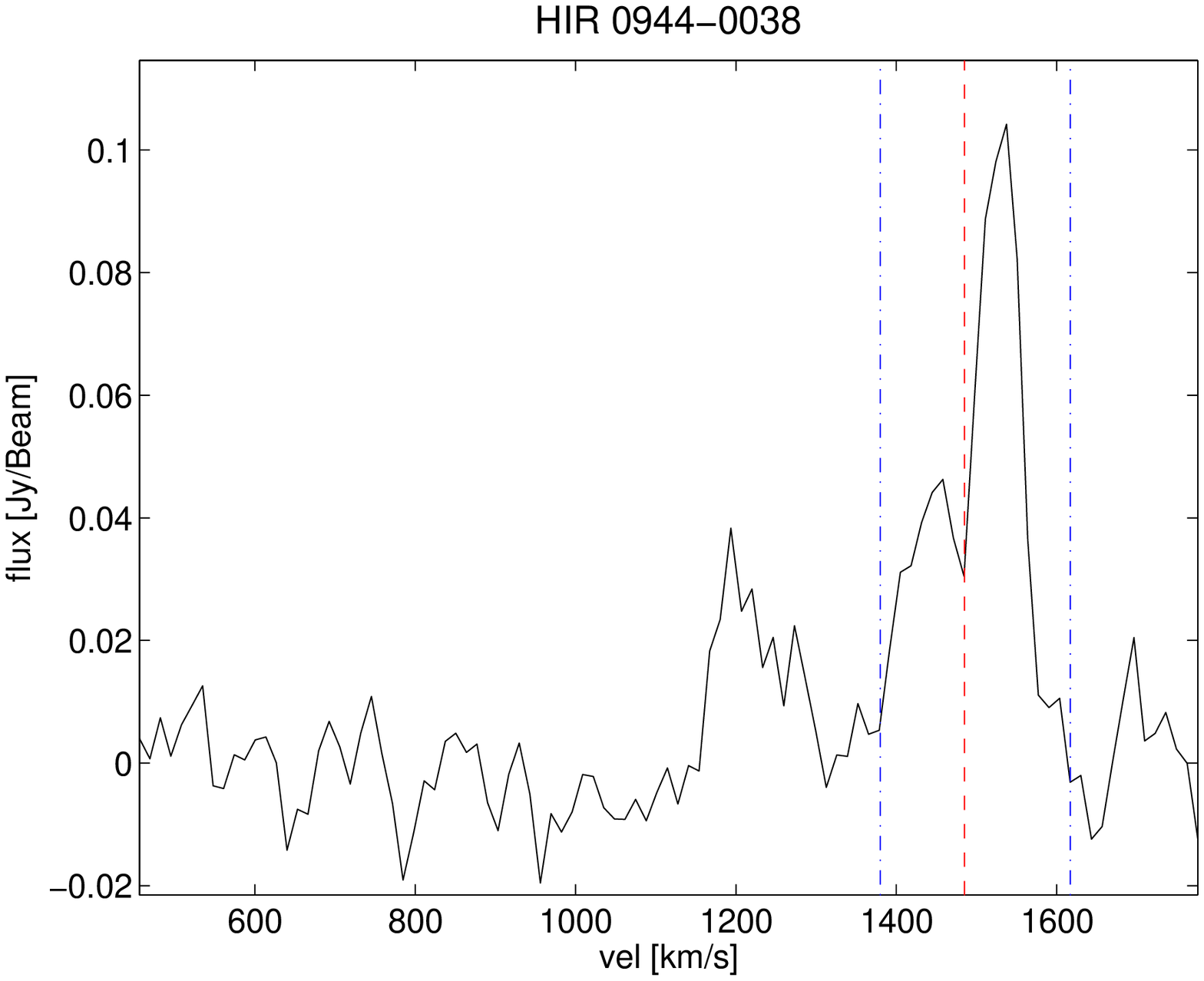}
 \includegraphics[width=0.3\textwidth]{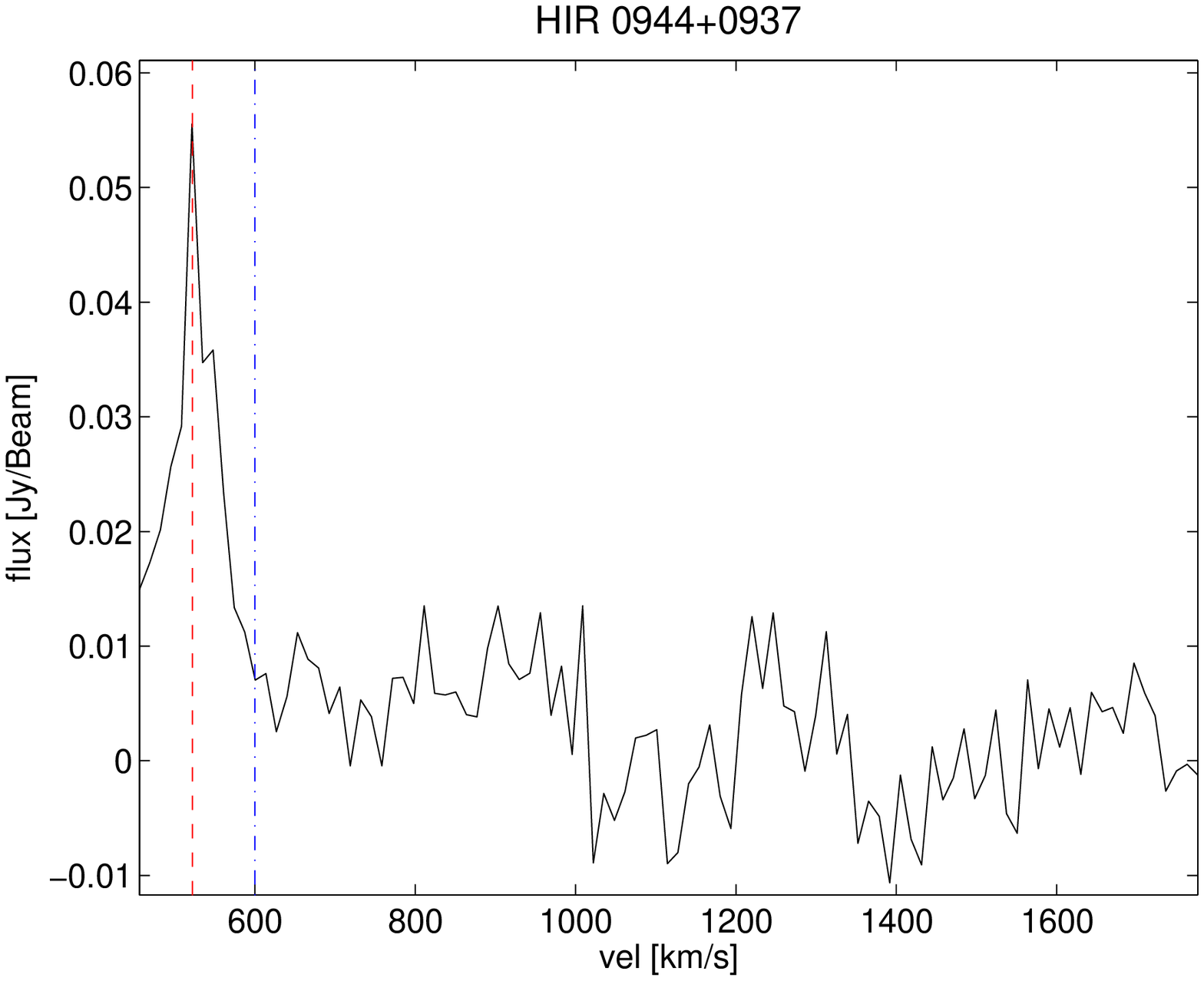}
 \includegraphics[width=0.3\textwidth]{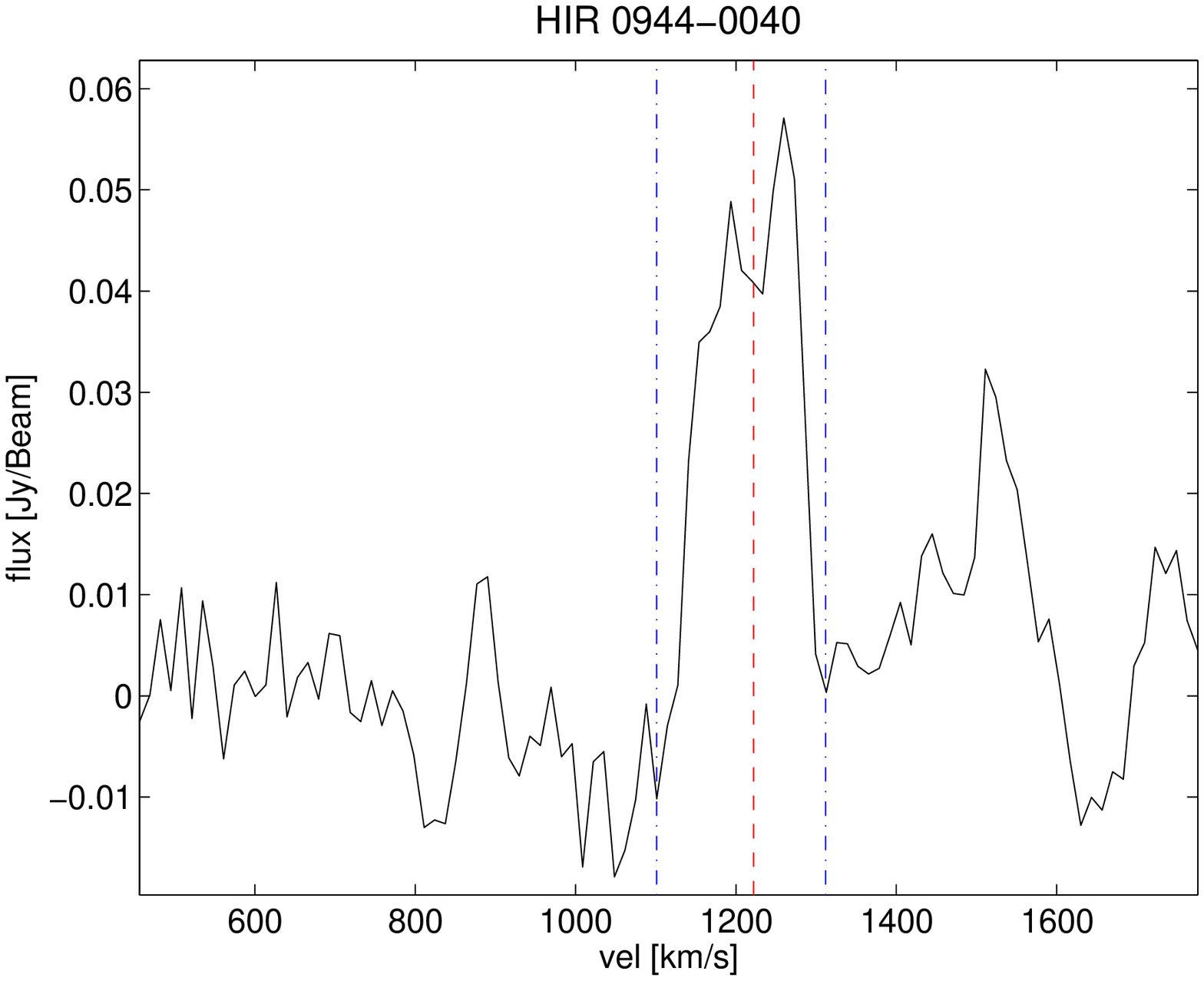}
 \includegraphics[width=0.3\textwidth]{0946+0141_spec.eps}

 \end{center}                                            
  \caption{Spectra of all detections of neutral hydrogen in the
    reprocessed region of the {\HI} Parkes All Sky Survey. The
    velocity width of each object is indicated by the two blue
    dash-dotted lines, while the central velocity is shown by the red
    dashed line.}
  \label{all_spectra}                                    
\end{figure*} 

%%%%%%%%%

\begin{figure*}
  \begin{center}
 \includegraphics[width=0.3\textwidth]{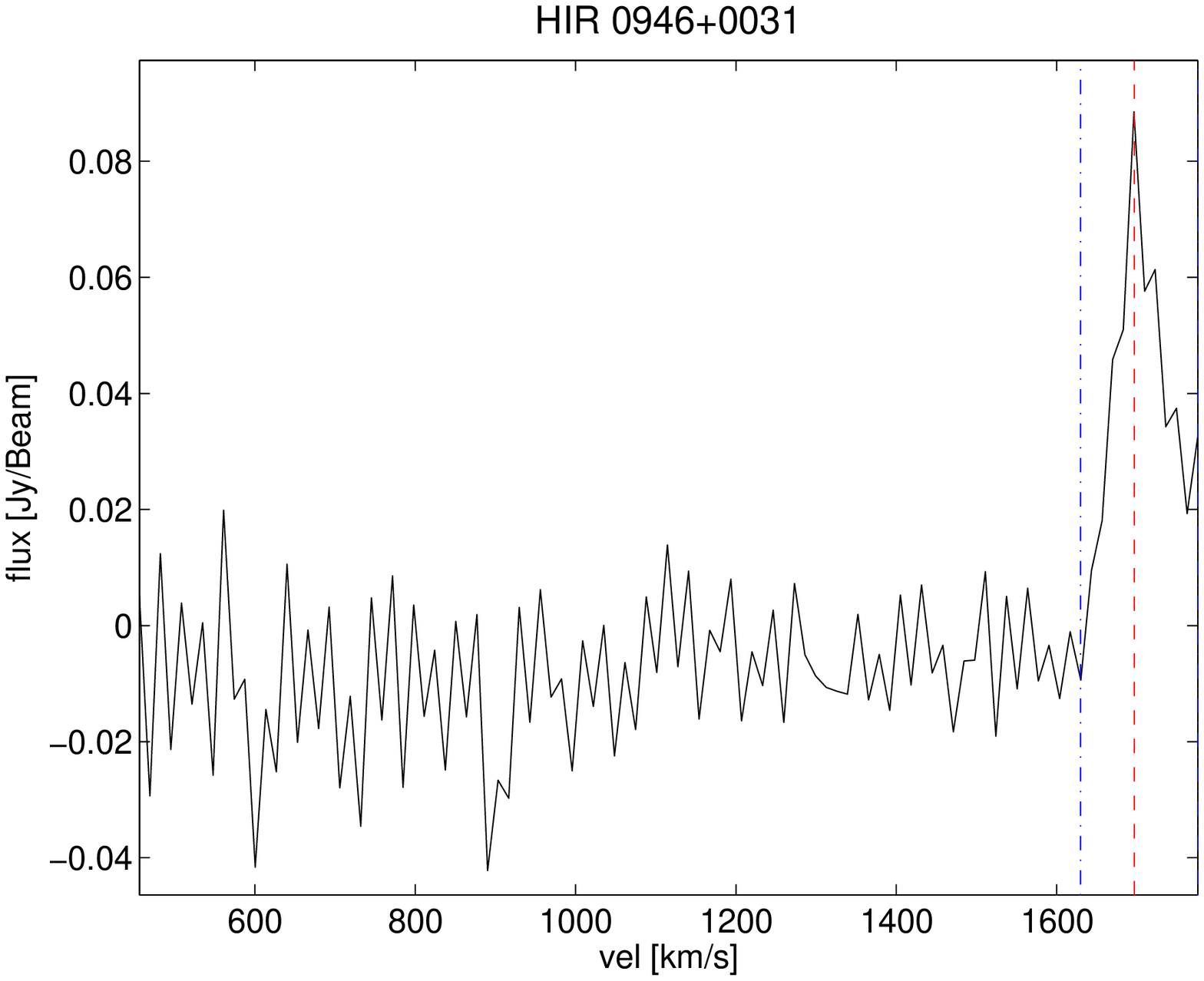}
 \includegraphics[width=0.3\textwidth]{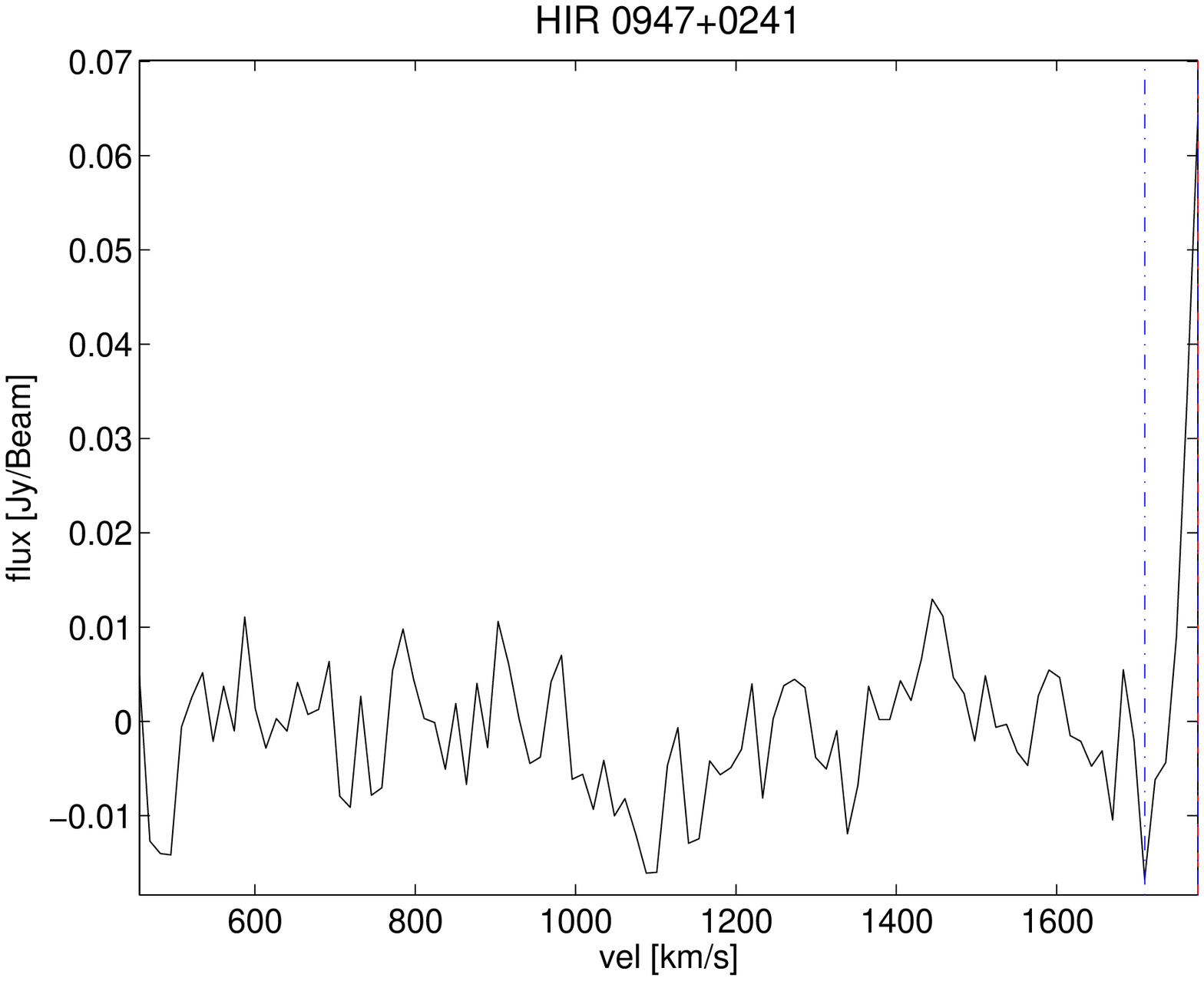}
 \includegraphics[width=0.3\textwidth]{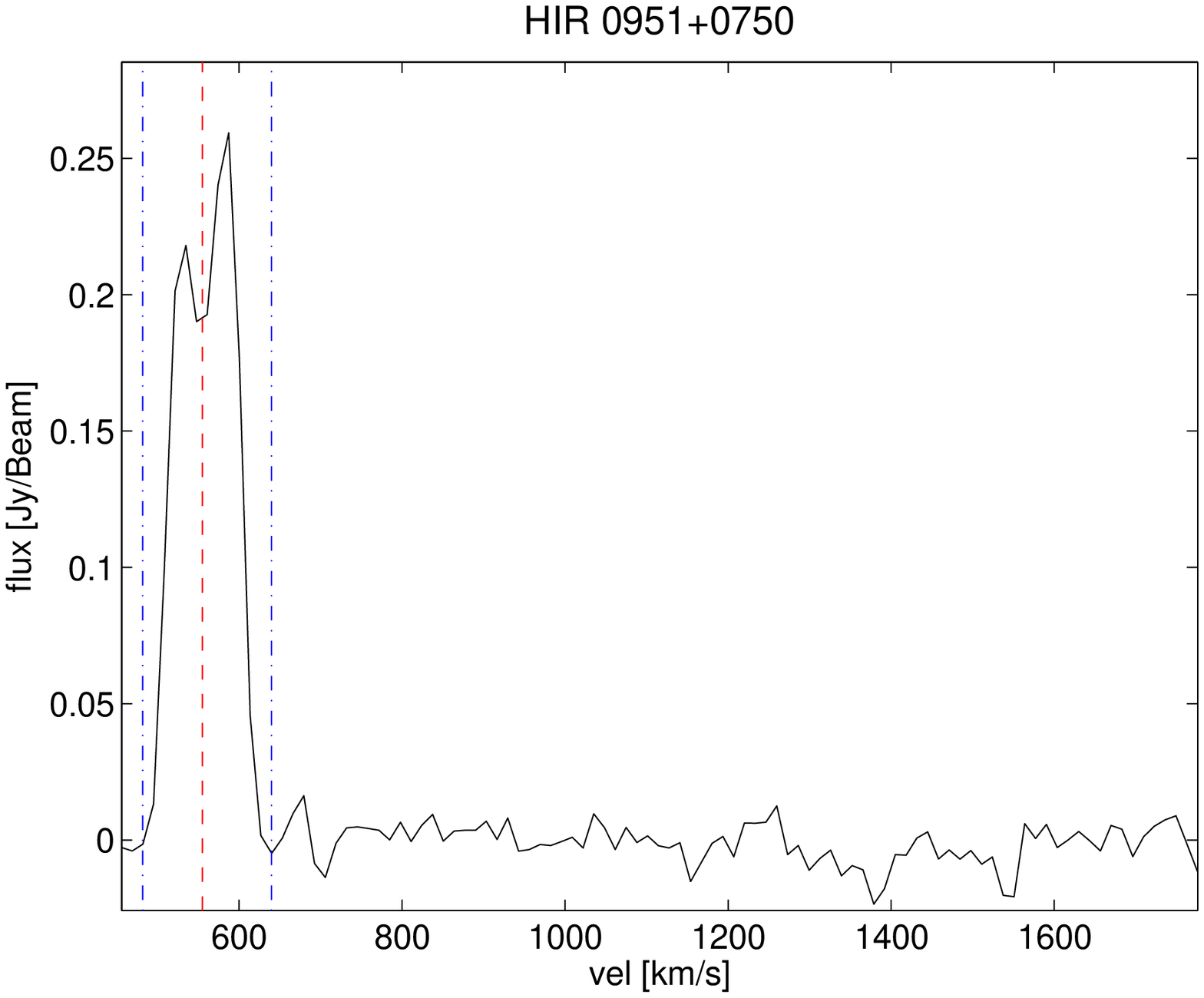}
 \includegraphics[width=0.3\textwidth]{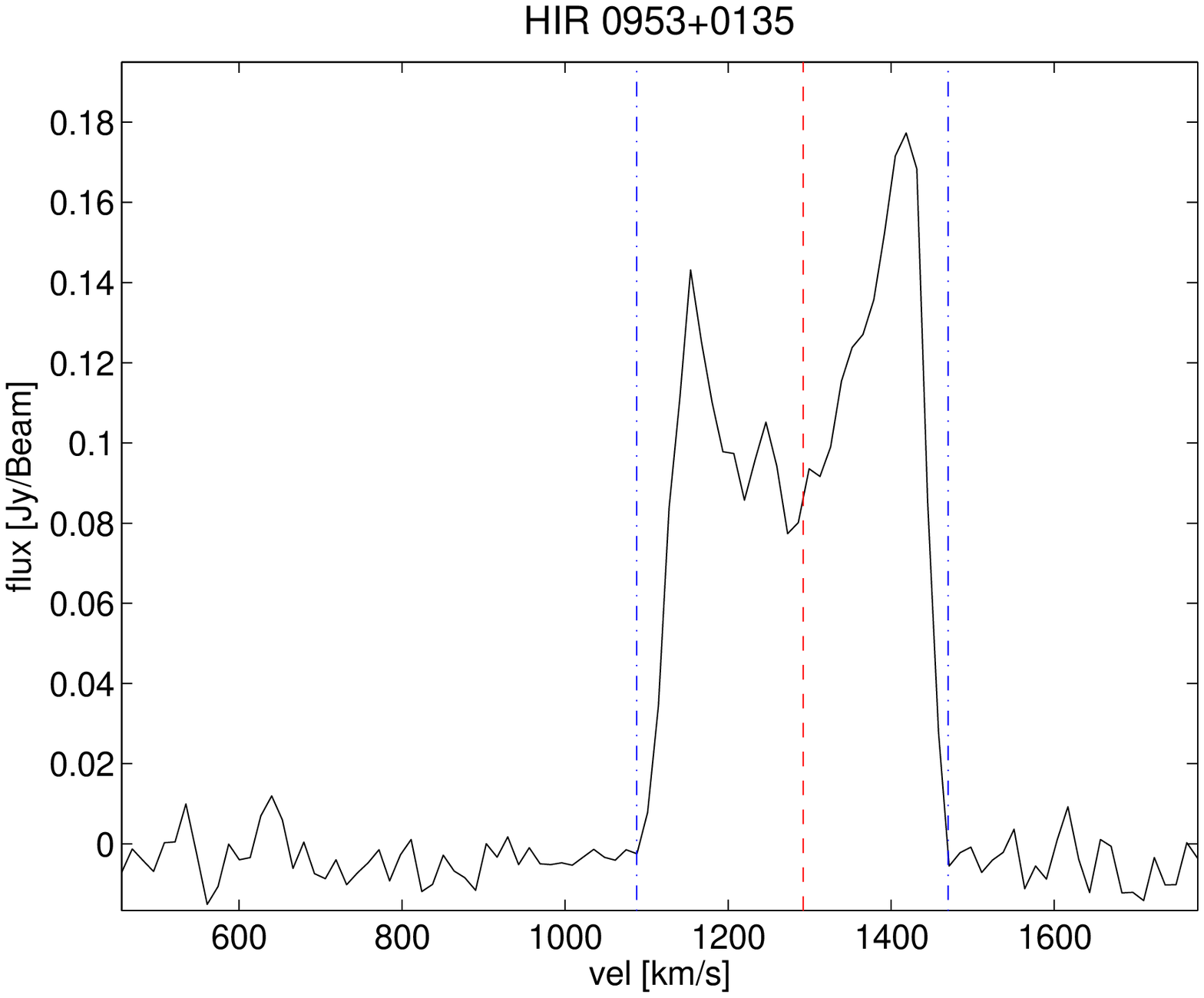}
 \includegraphics[width=0.3\textwidth]{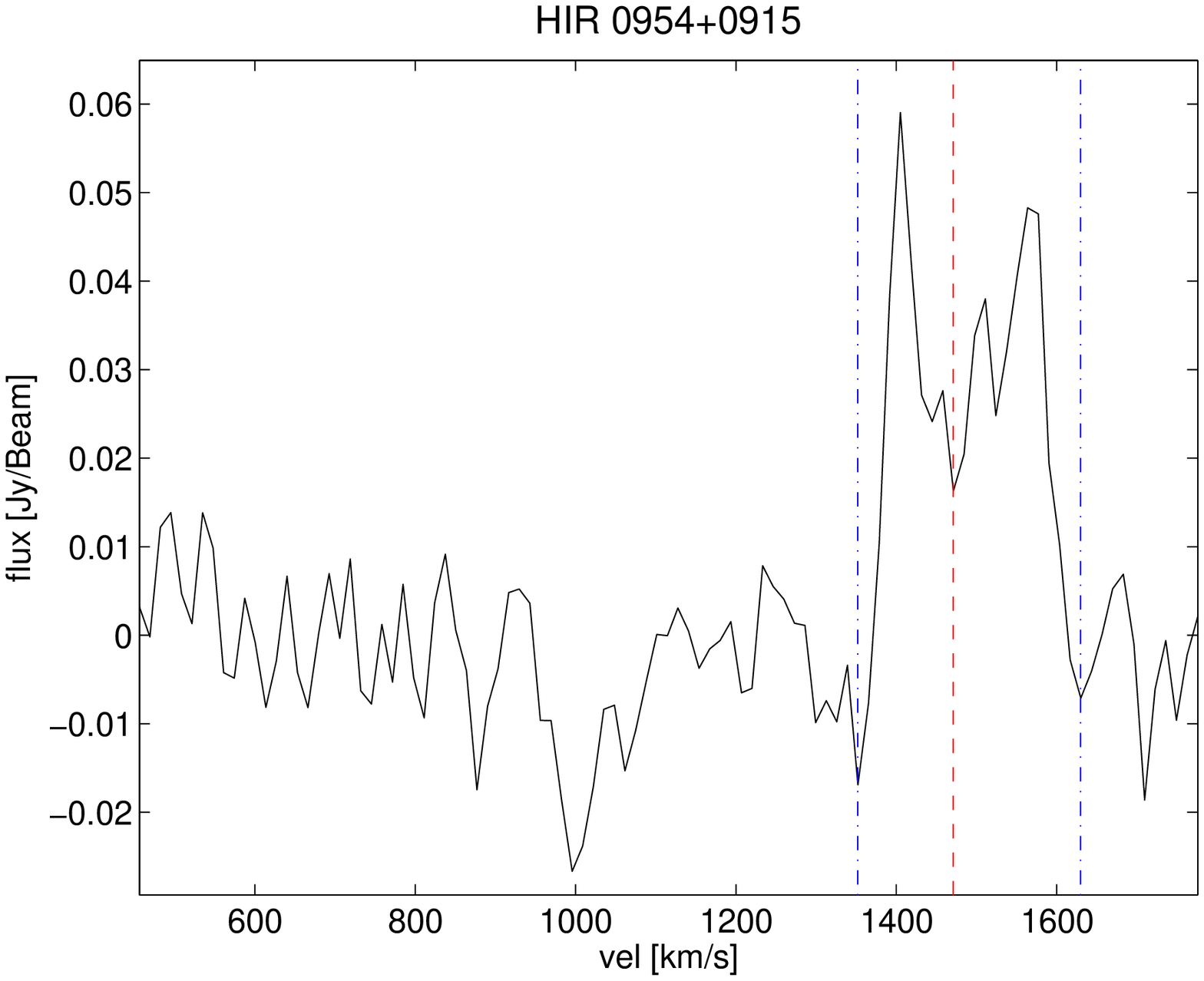}
 \includegraphics[width=0.3\textwidth]{1005+0139_spec.eps}
 \includegraphics[width=0.3\textwidth]{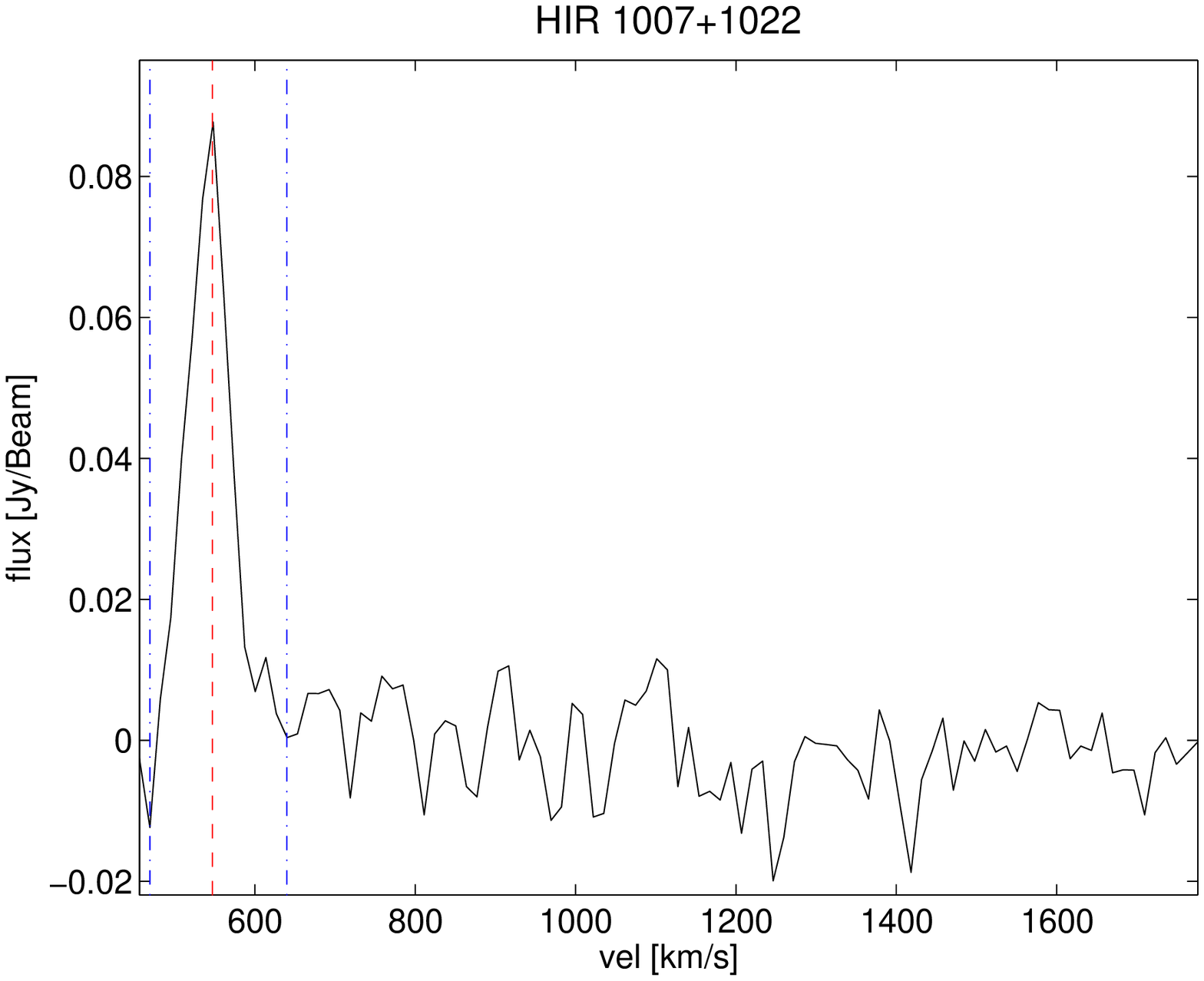}
 \includegraphics[width=0.3\textwidth]{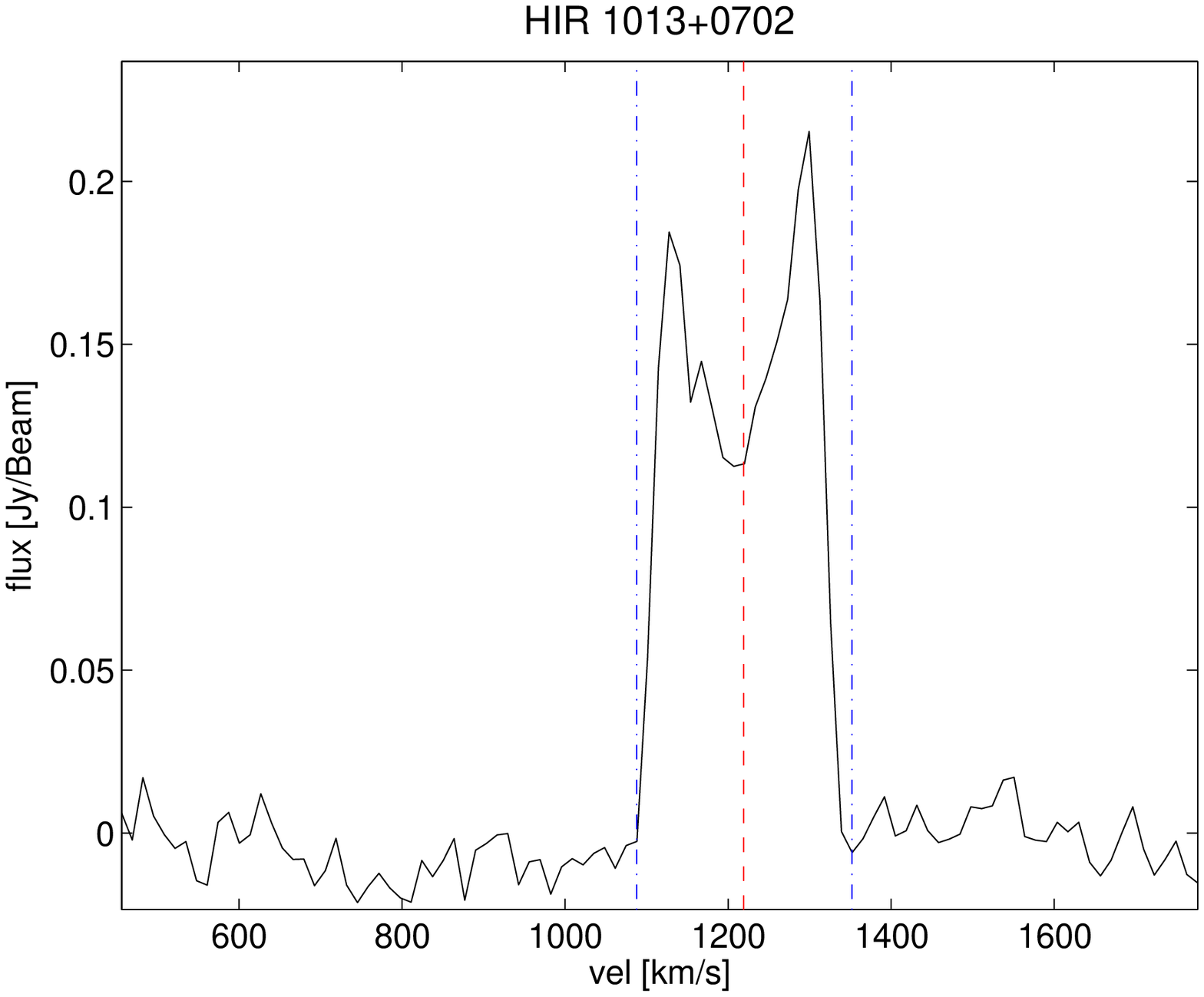}
 \includegraphics[width=0.3\textwidth]{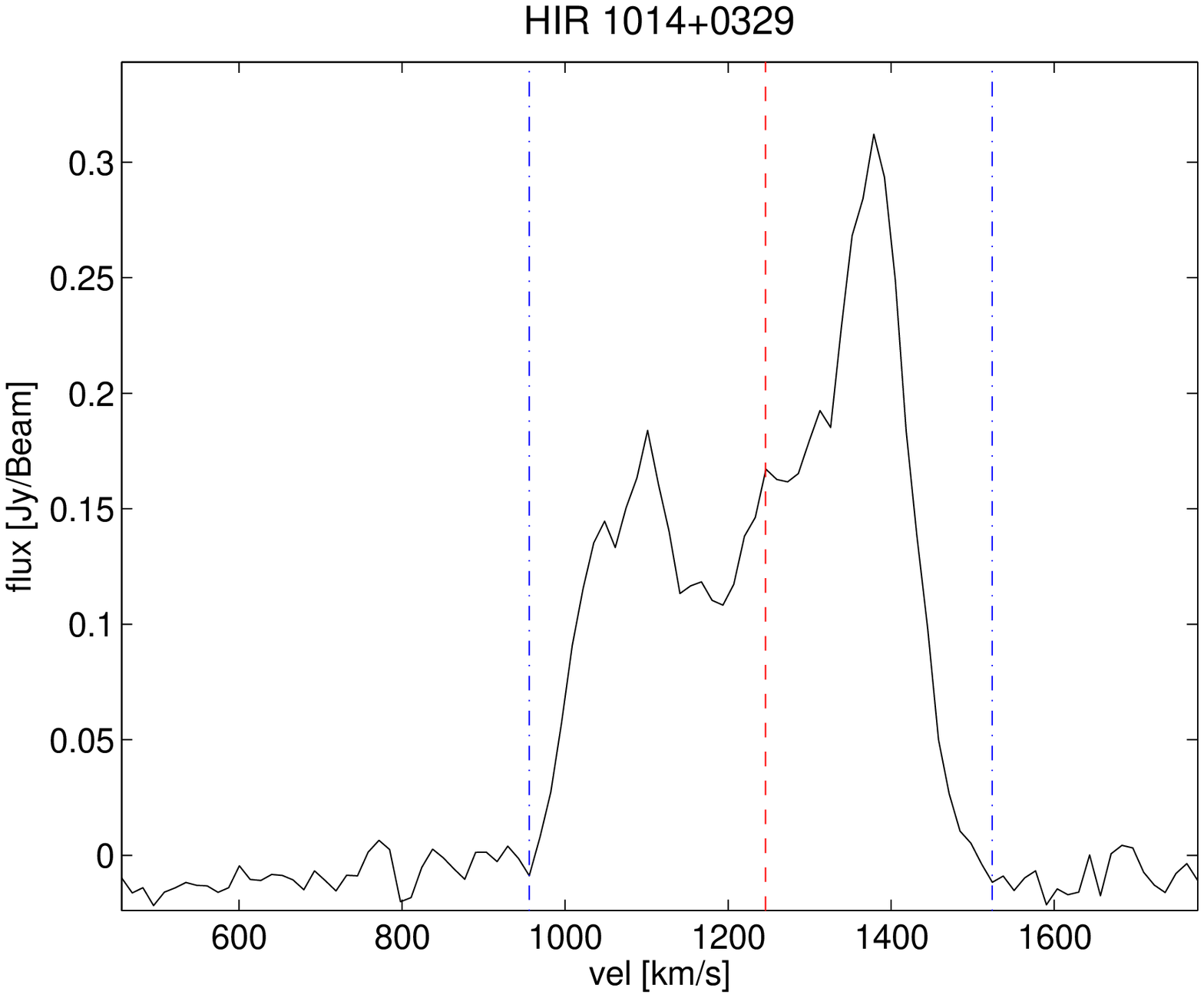}
 \includegraphics[width=0.3\textwidth]{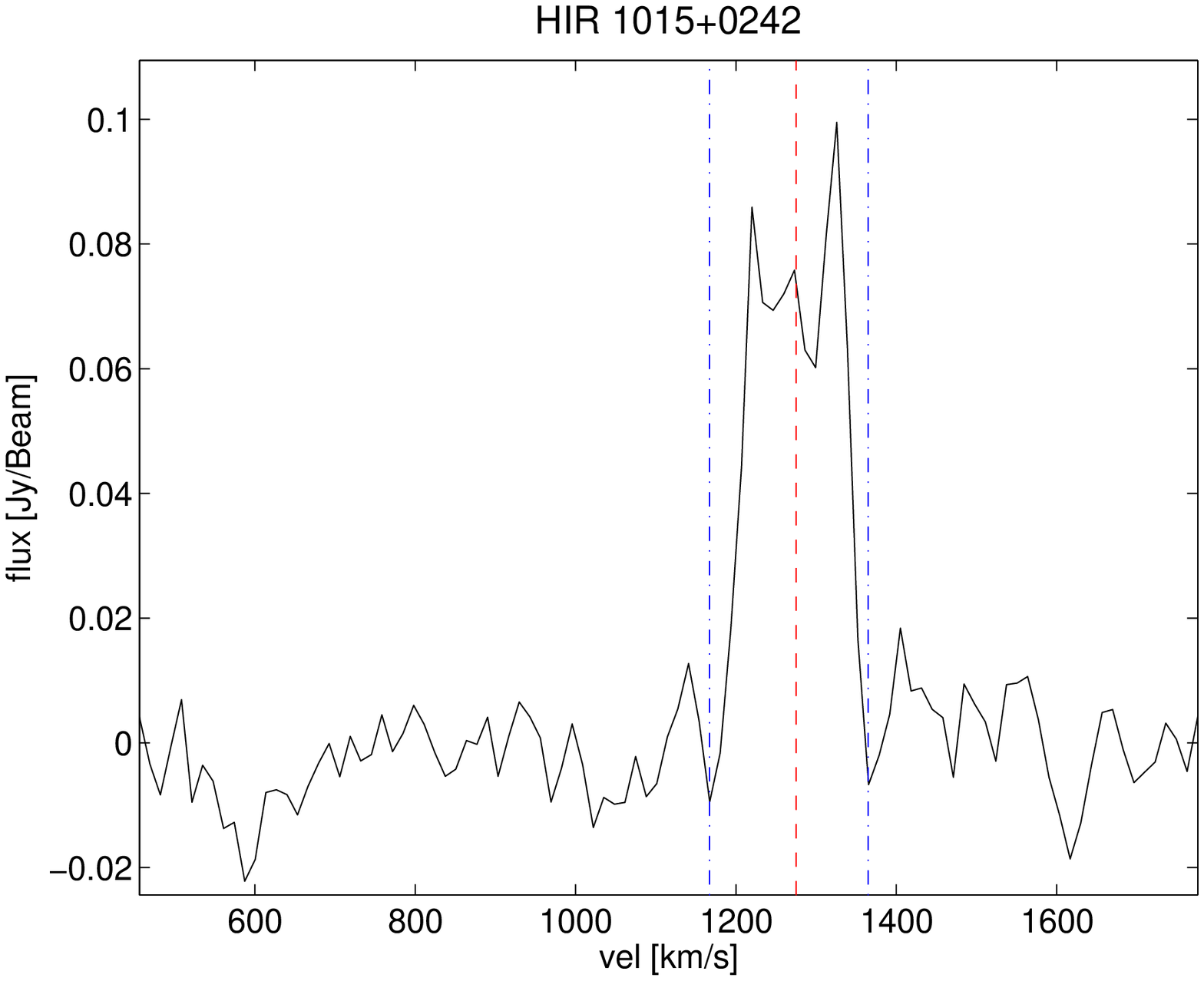}
 \includegraphics[width=0.3\textwidth]{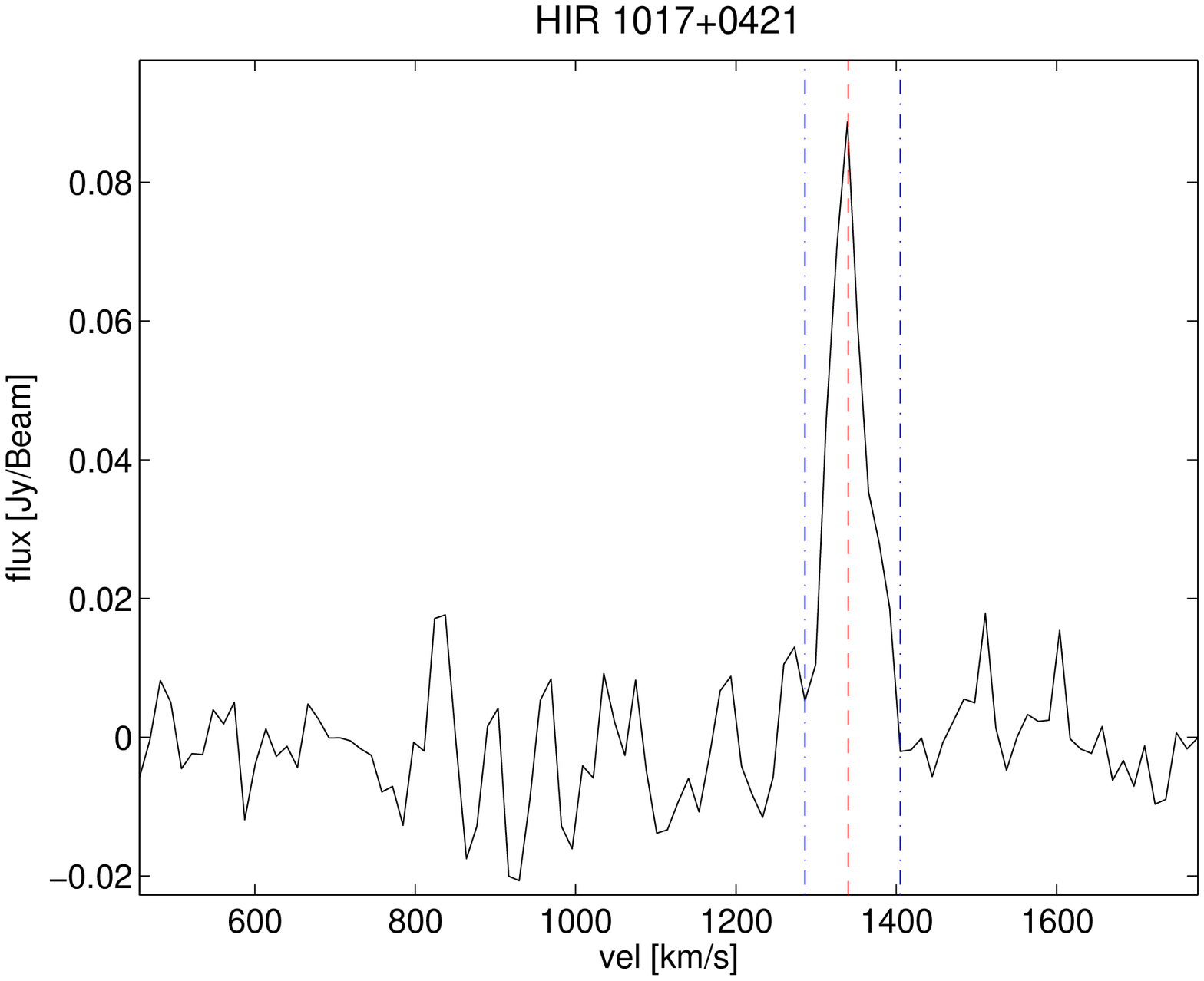}
 \includegraphics[width=0.3\textwidth]{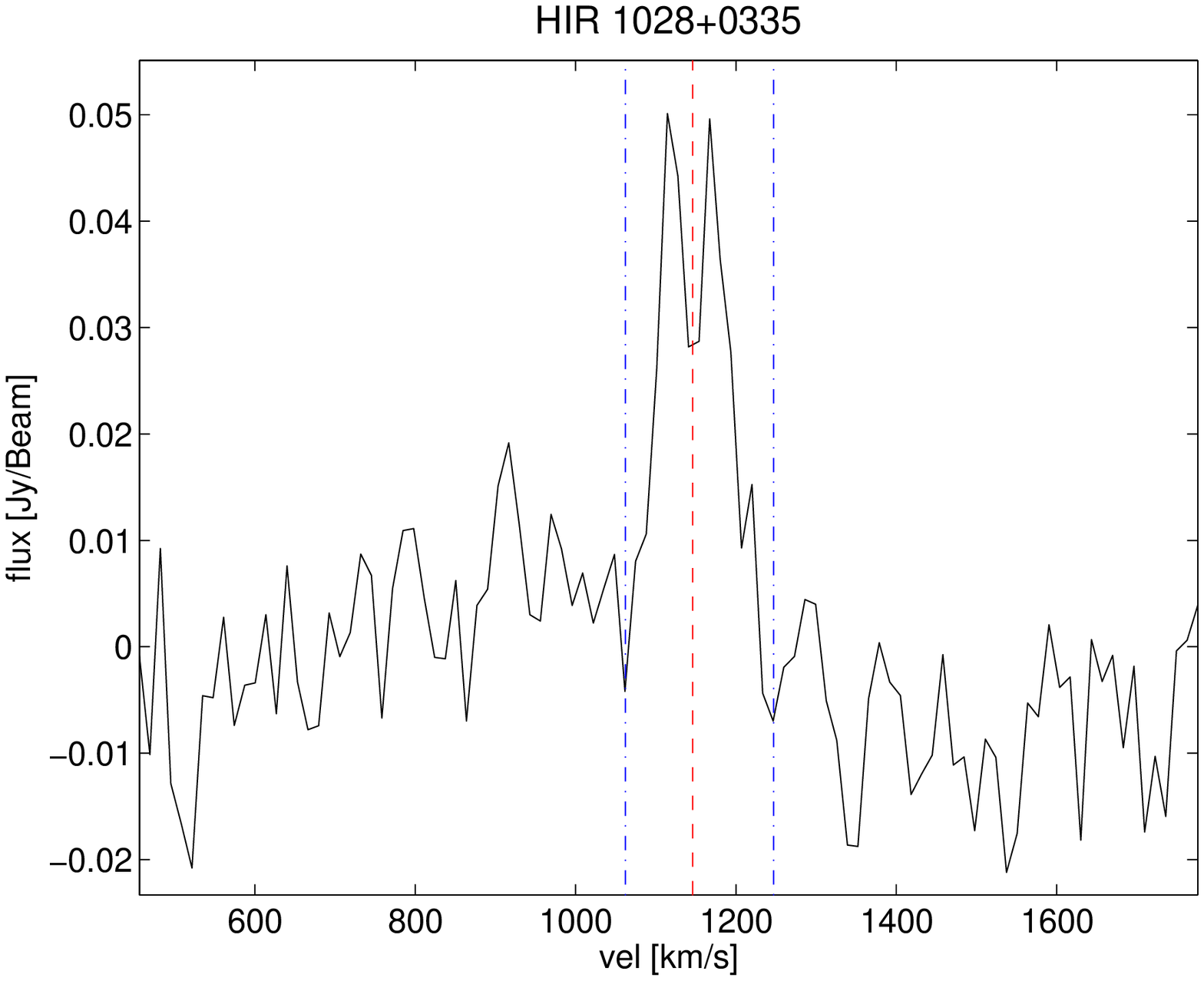}
 \includegraphics[width=0.3\textwidth]{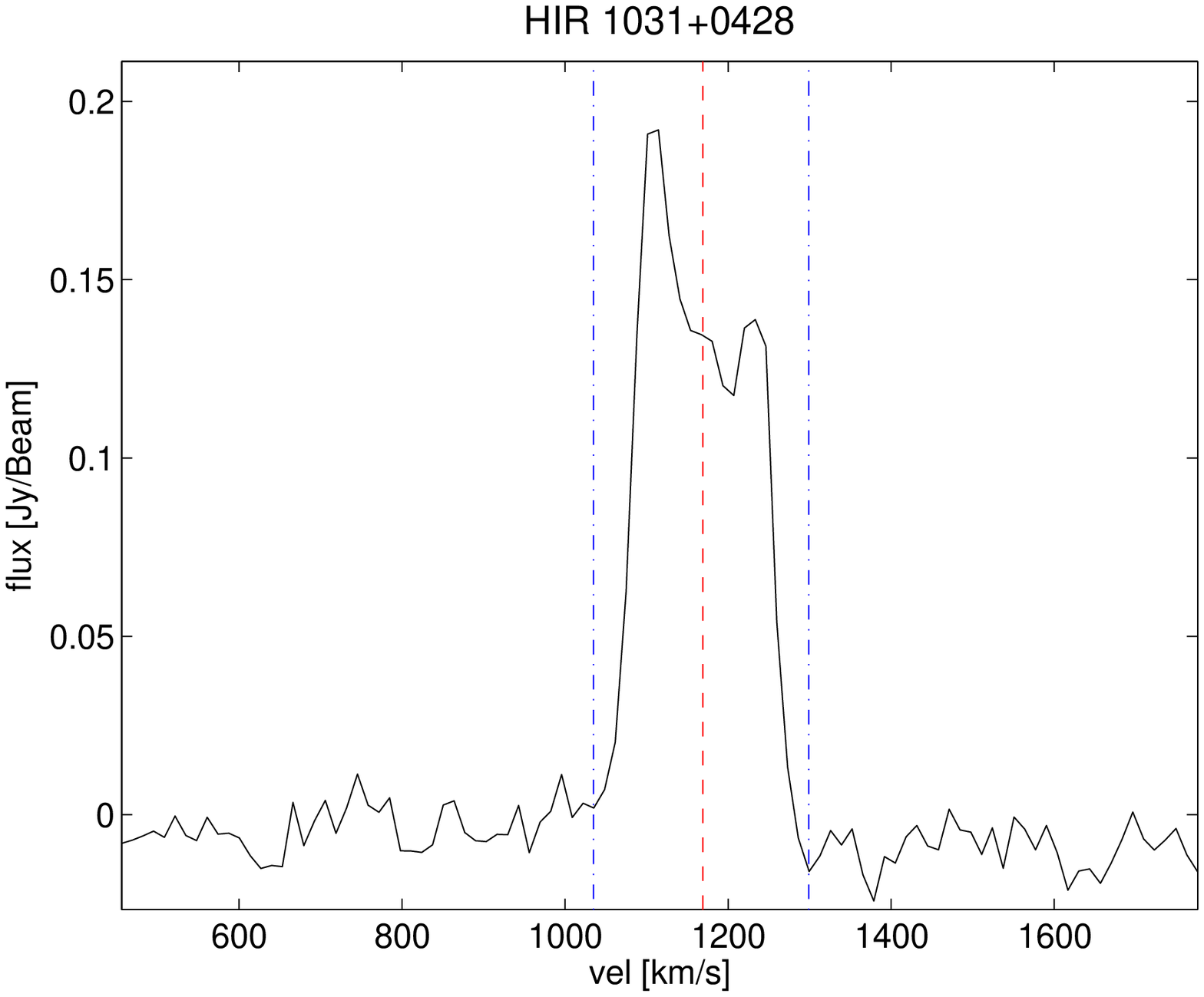}
 \includegraphics[width=0.3\textwidth]{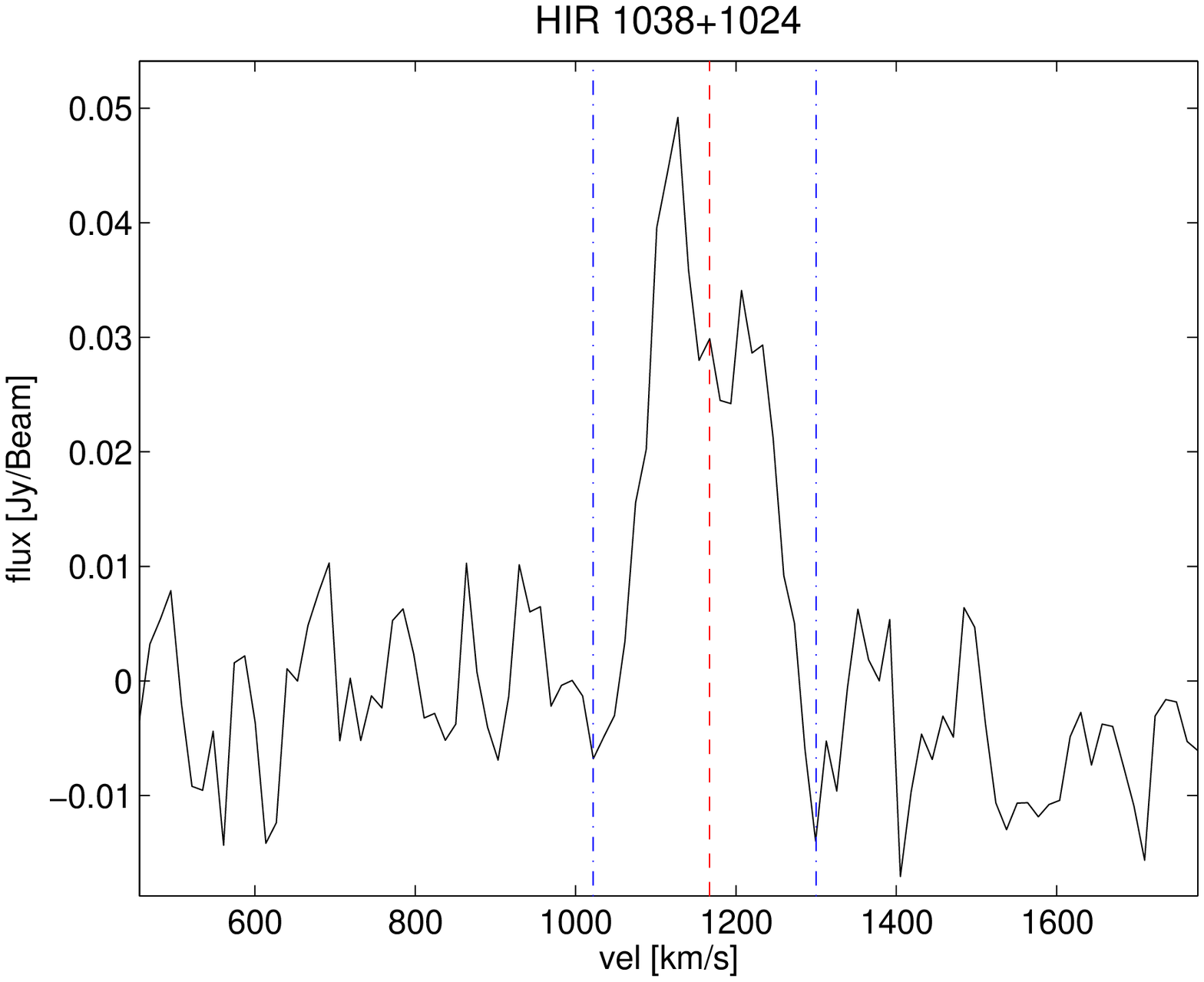}
 \includegraphics[width=0.3\textwidth]{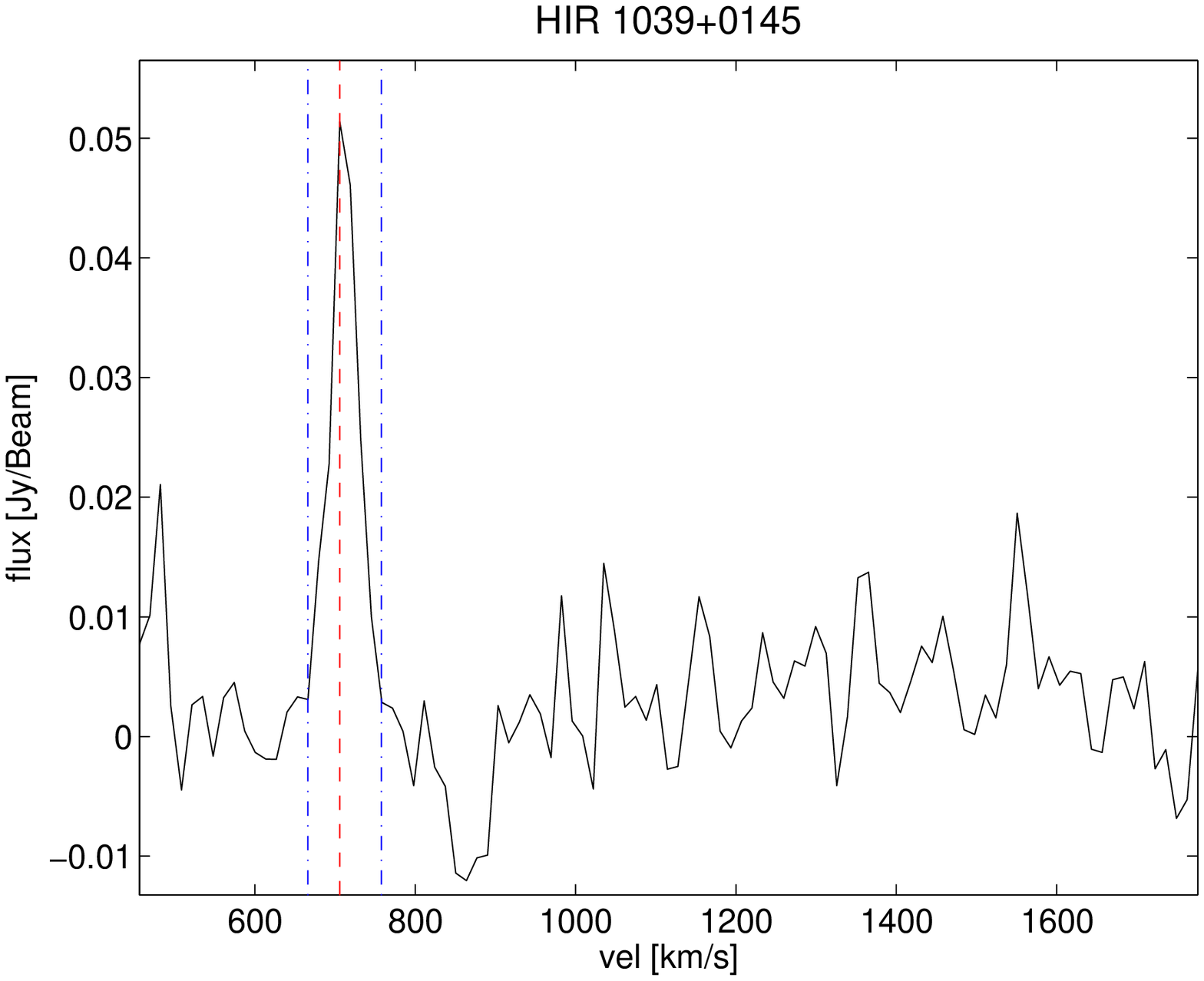}

 \end{center}                                            
{\bf Fig~\ref{all_spectra}.} (continued)                                        
 
\end{figure*}                                            

\begin{figure*}
  \begin{center}

  \includegraphics[width=0.3\textwidth]{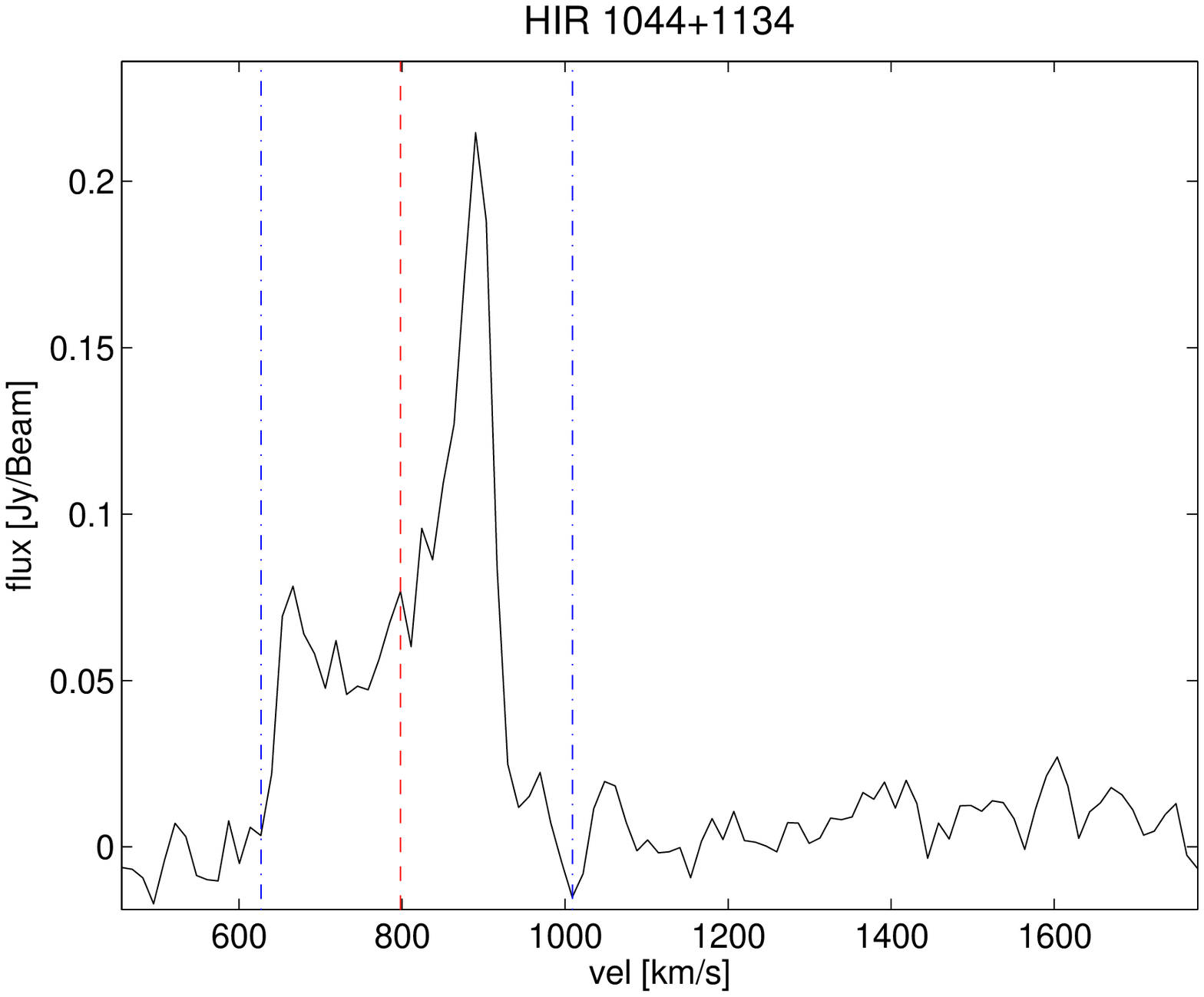}
 \includegraphics[width=0.3\textwidth]{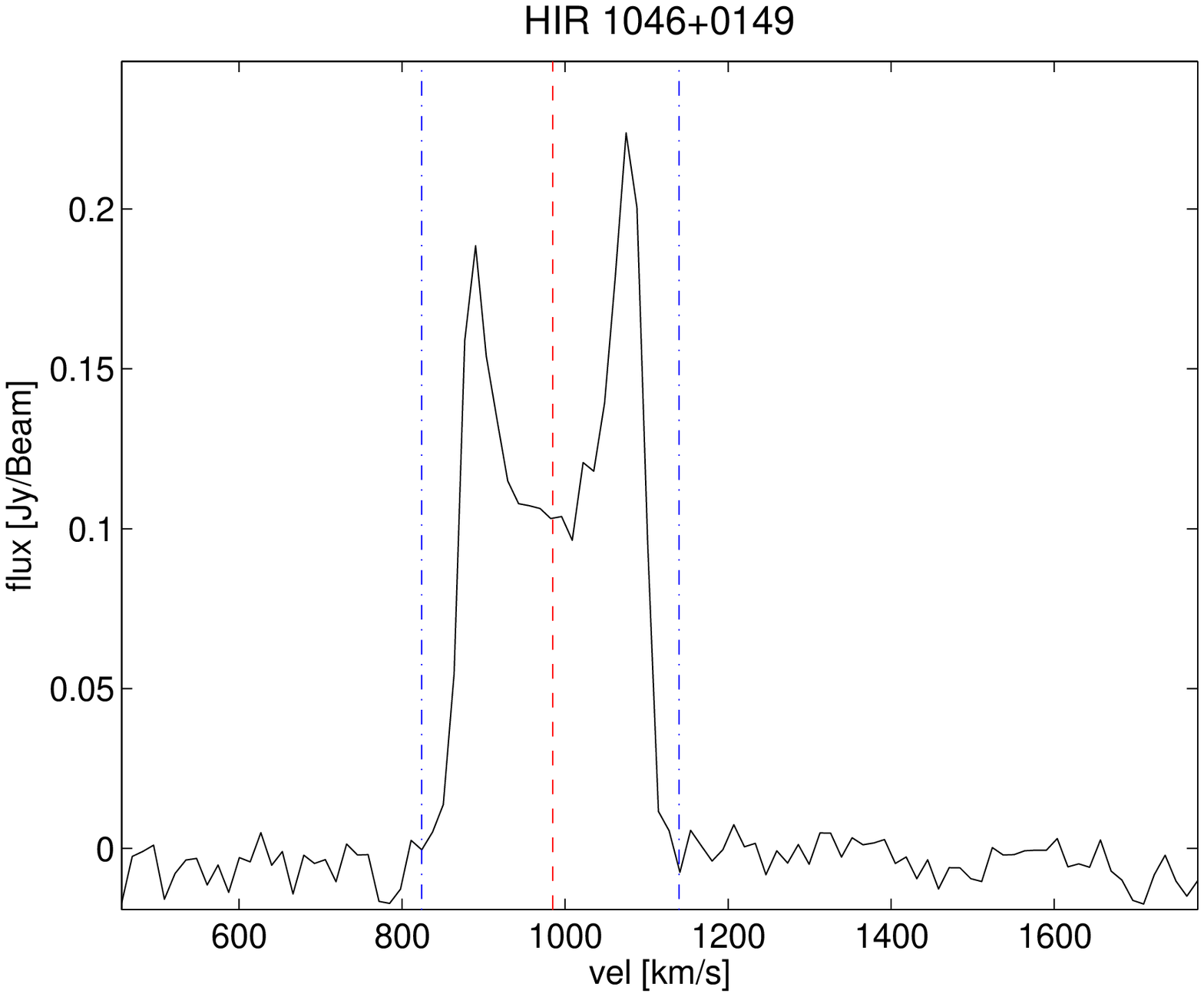}
 \includegraphics[width=0.3\textwidth]{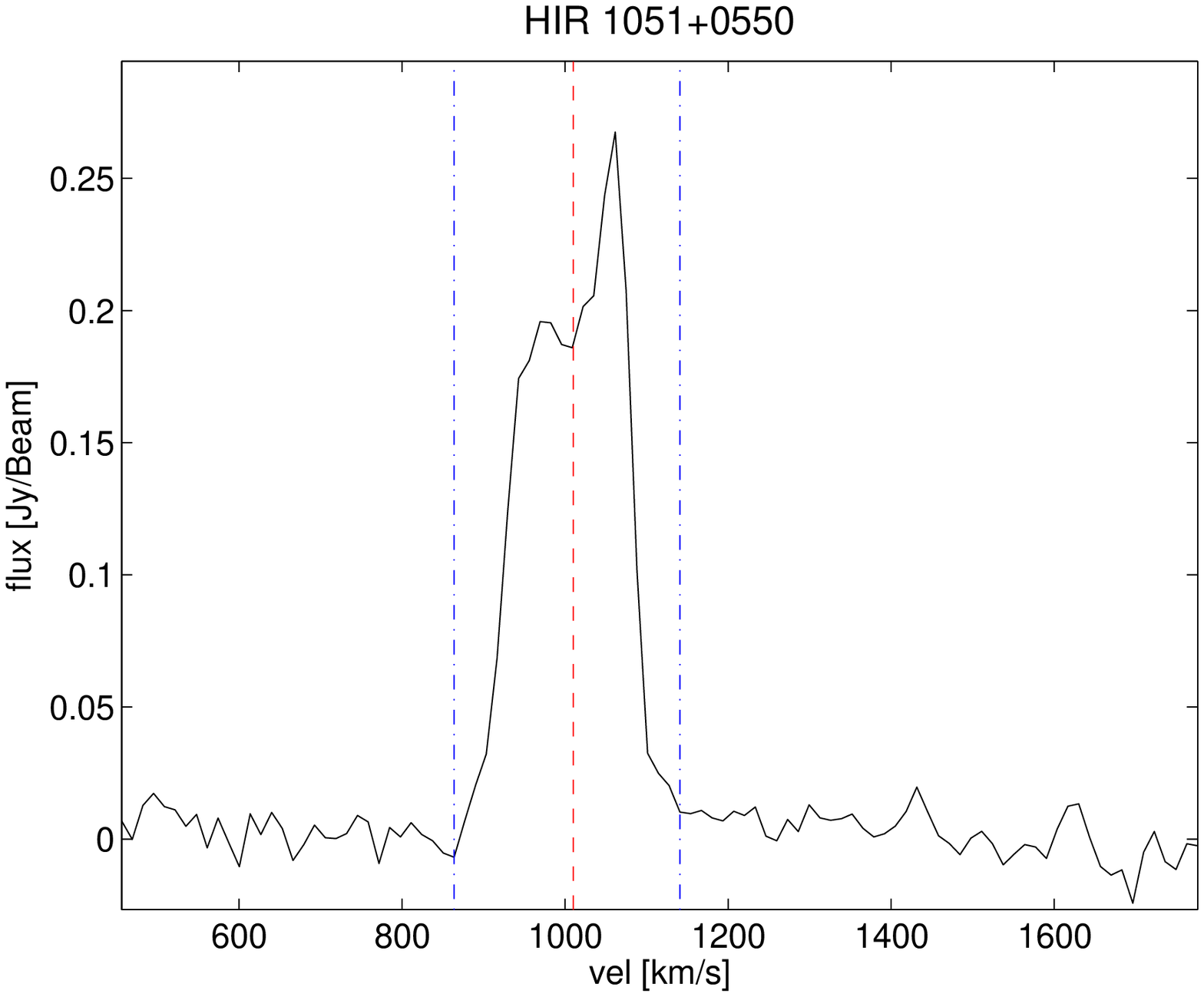}
 \includegraphics[width=0.3\textwidth]{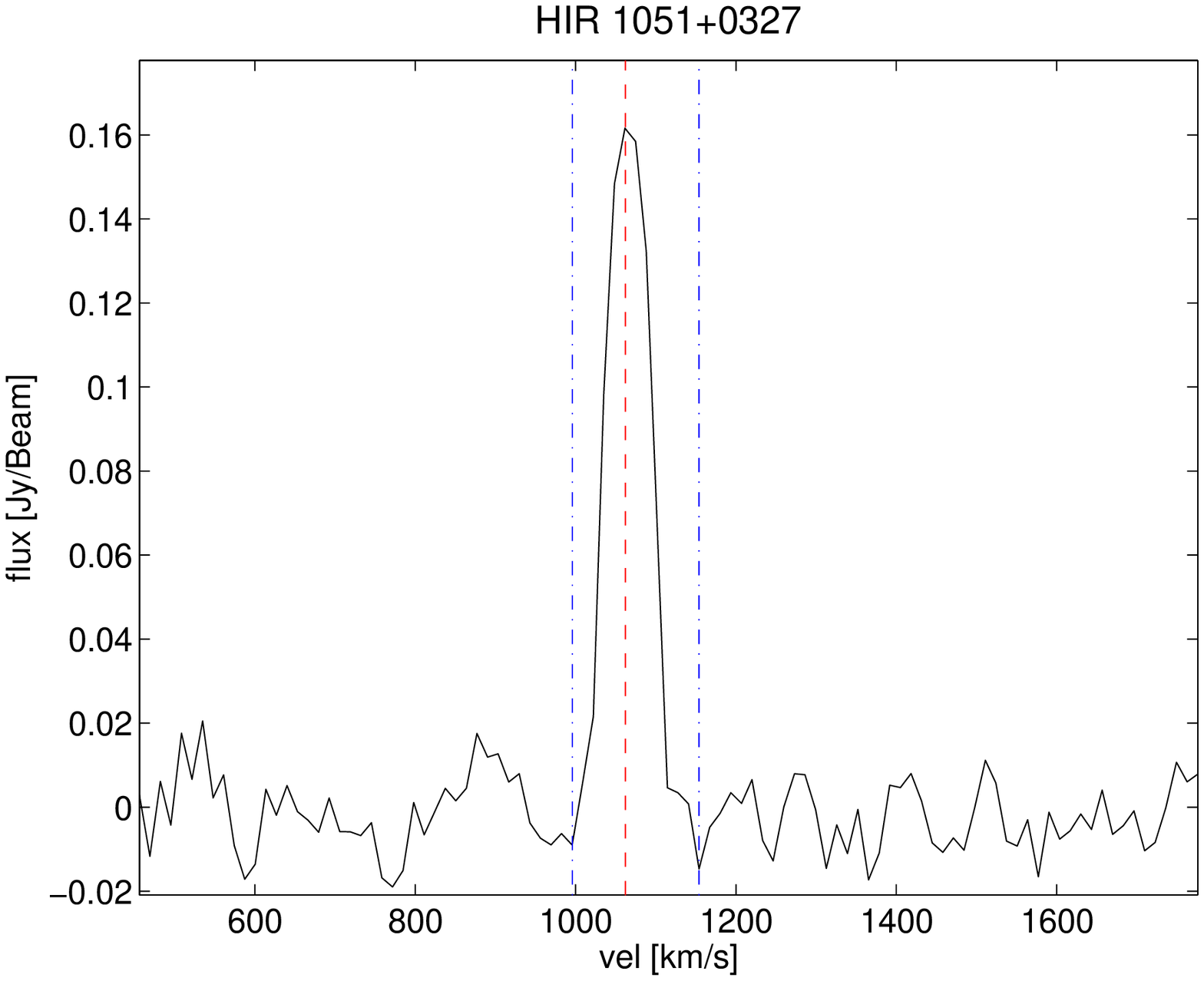}
 \includegraphics[width=0.3\textwidth]{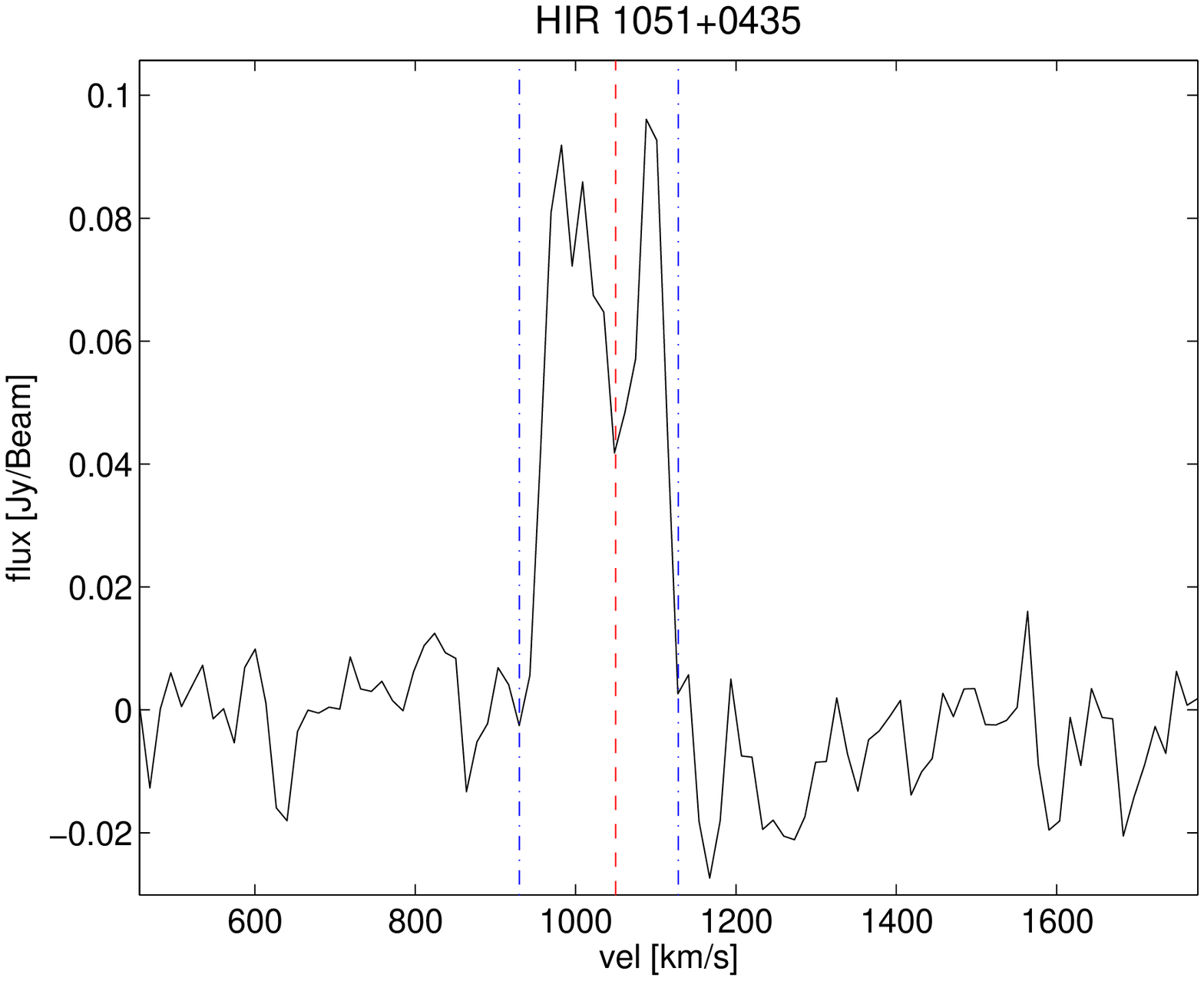}
 \includegraphics[width=0.3\textwidth]{1052+0002_spec.eps}
 \includegraphics[width=0.3\textwidth]{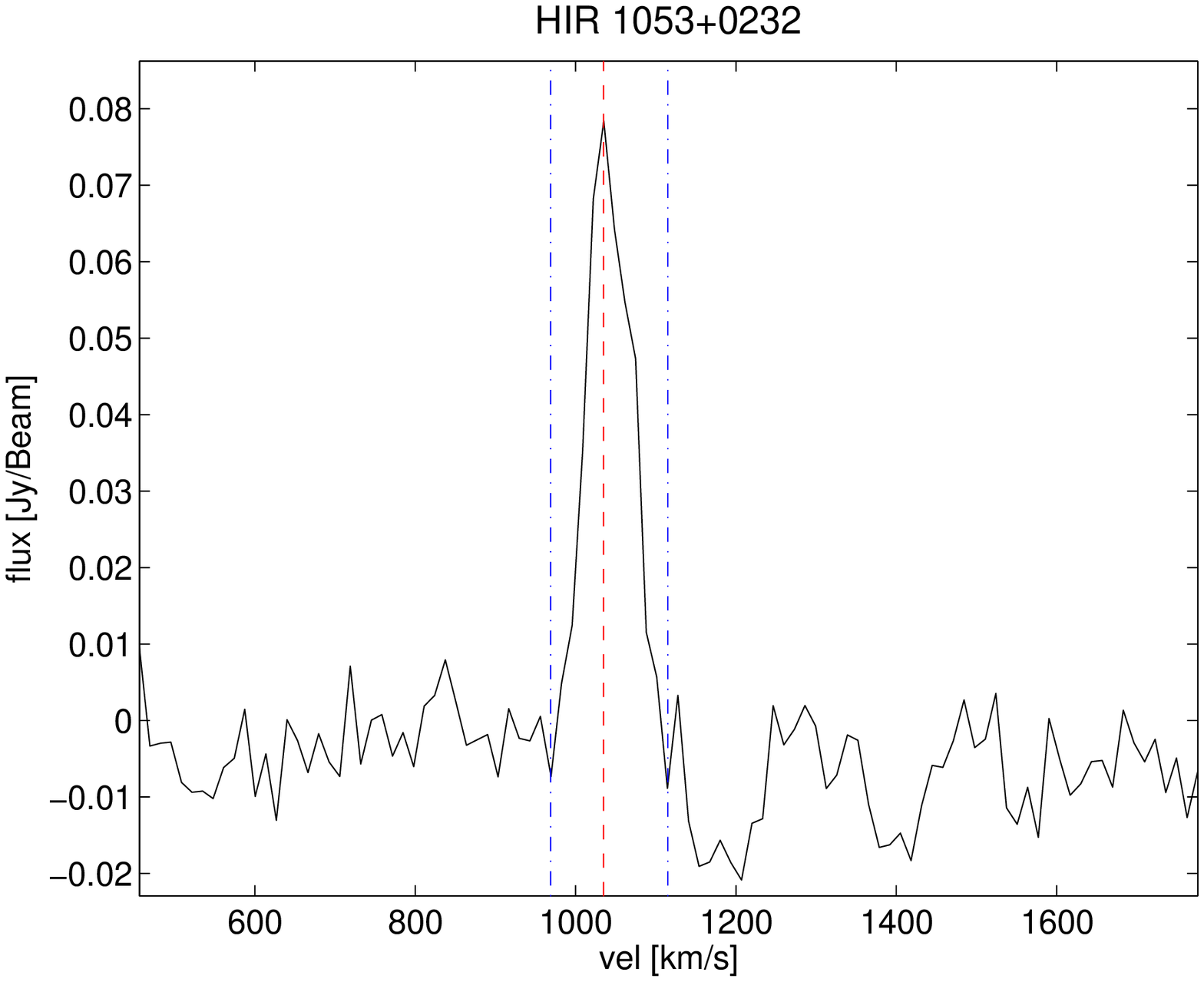}
 \includegraphics[width=0.3\textwidth]{1055+0511_spec.eps}
 \includegraphics[width=0.3\textwidth]{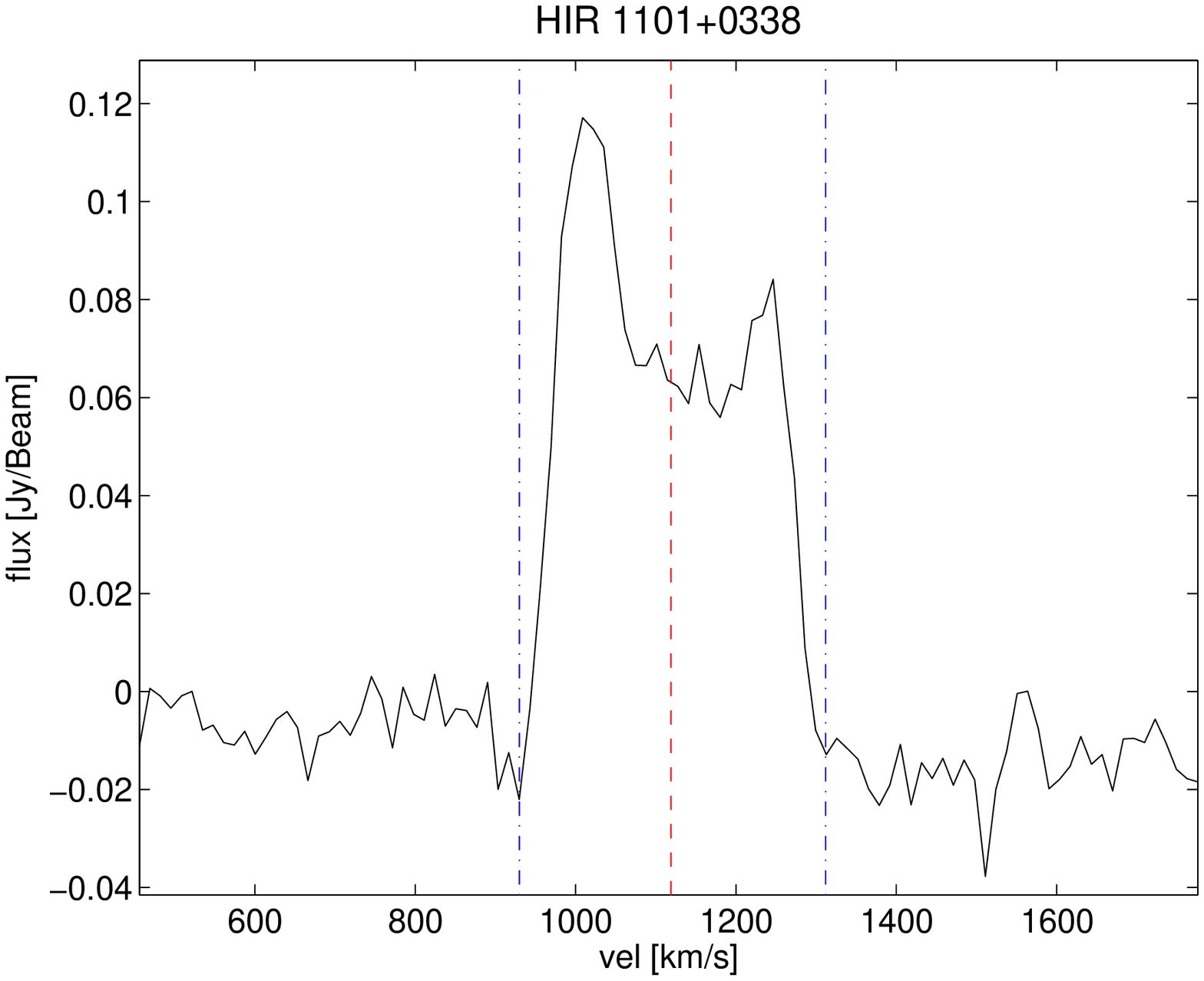}
 \includegraphics[width=0.3\textwidth]{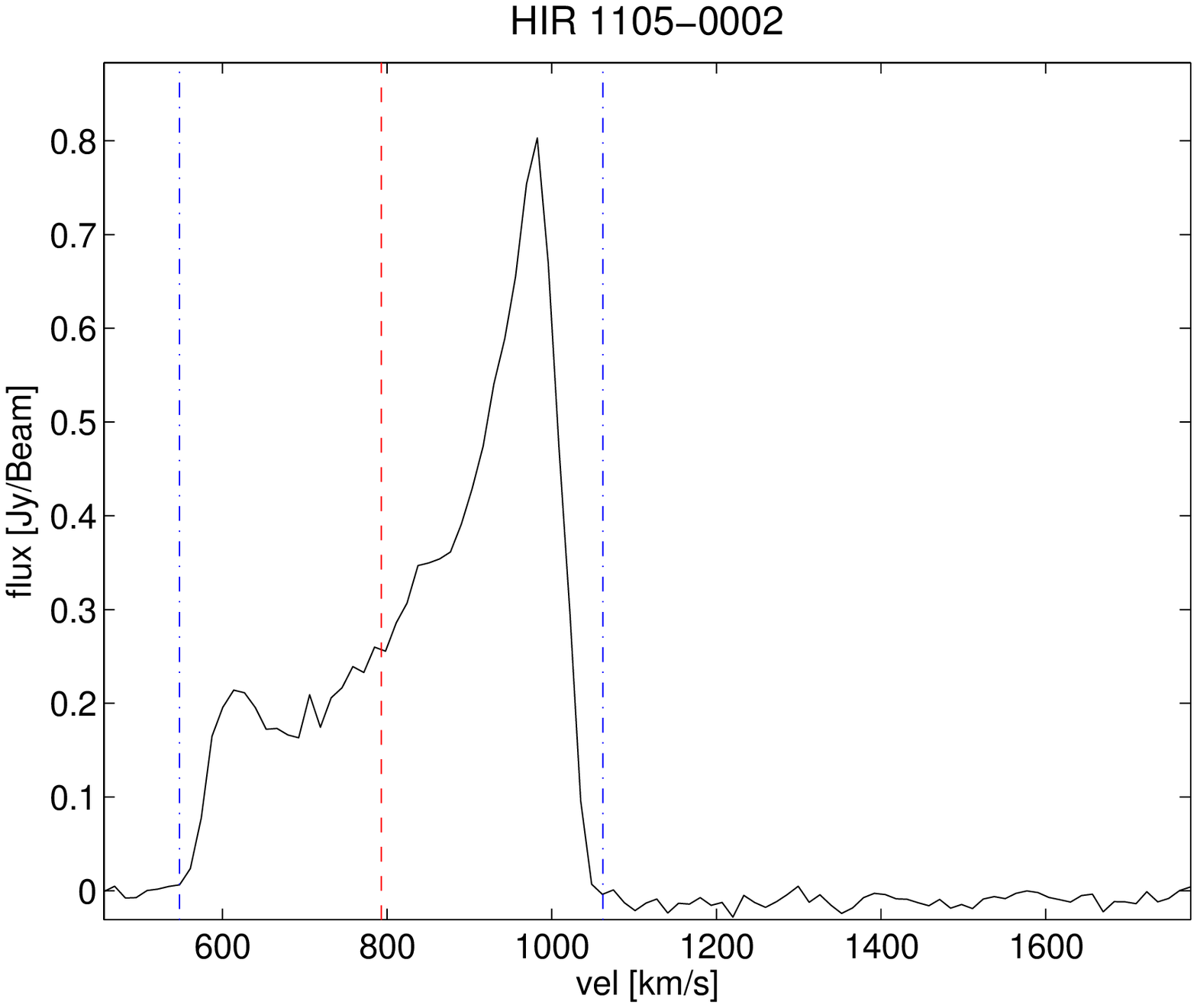}
 \includegraphics[width=0.3\textwidth]{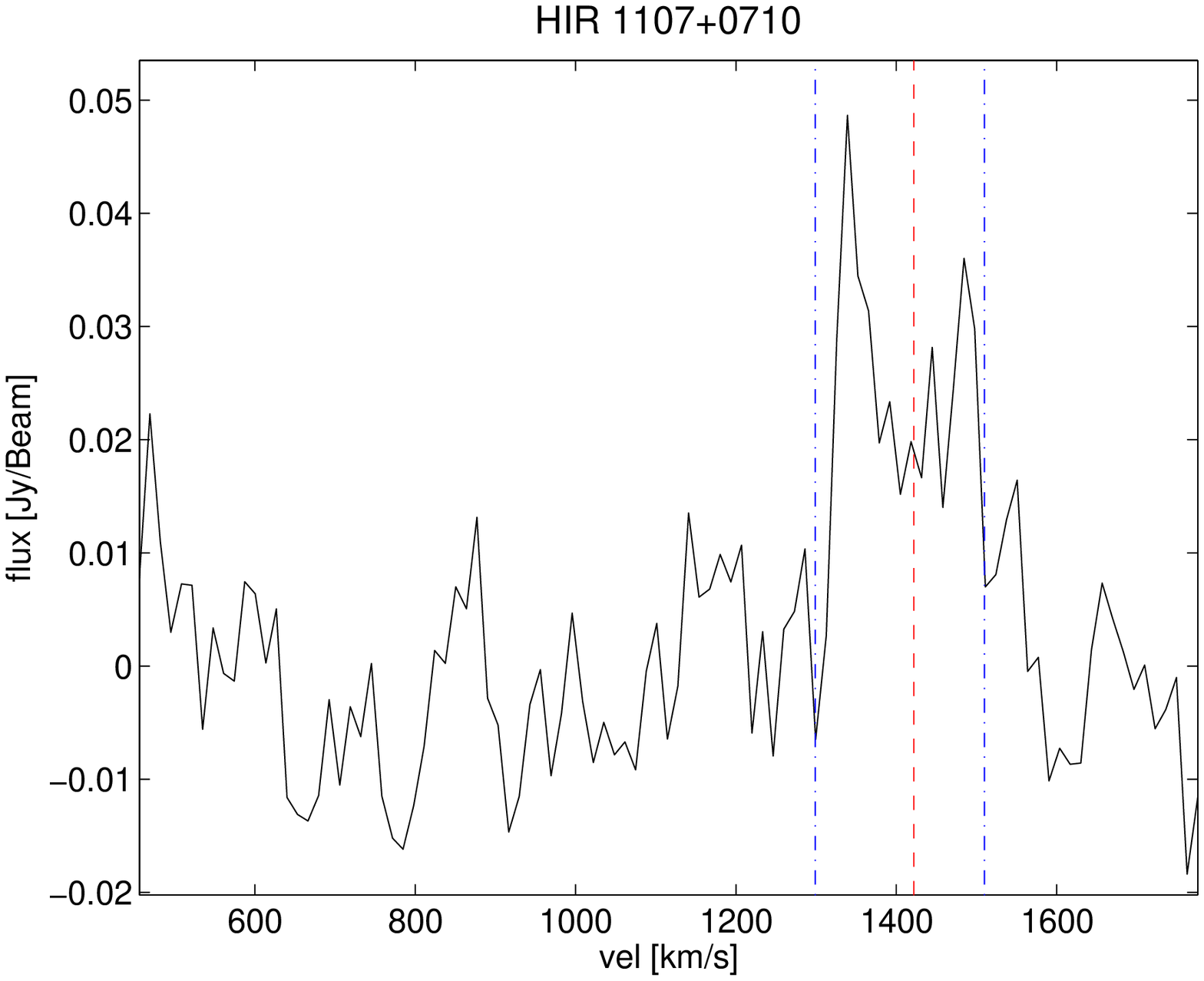}
 \includegraphics[width=0.3\textwidth]{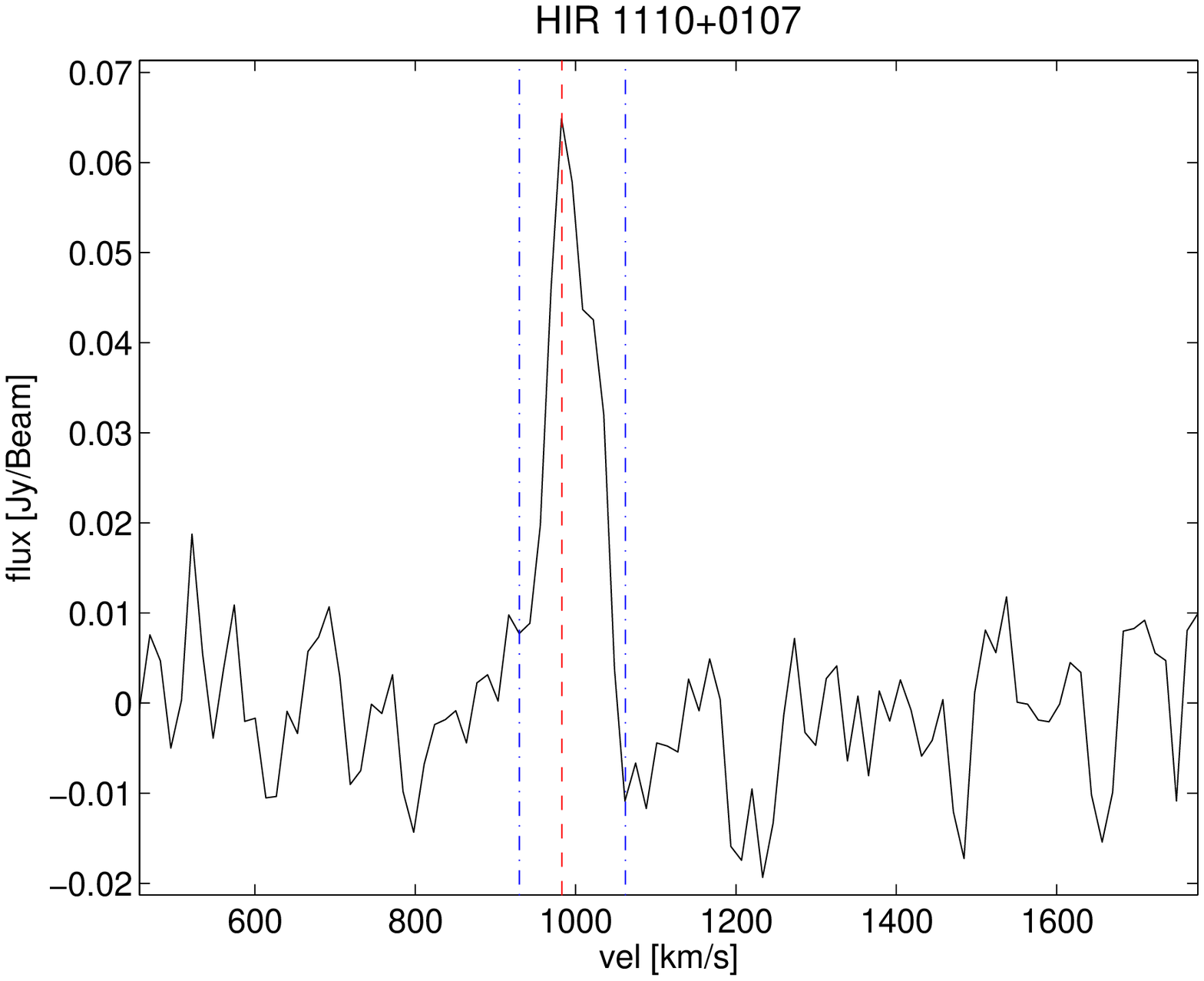}
 \includegraphics[width=0.3\textwidth]{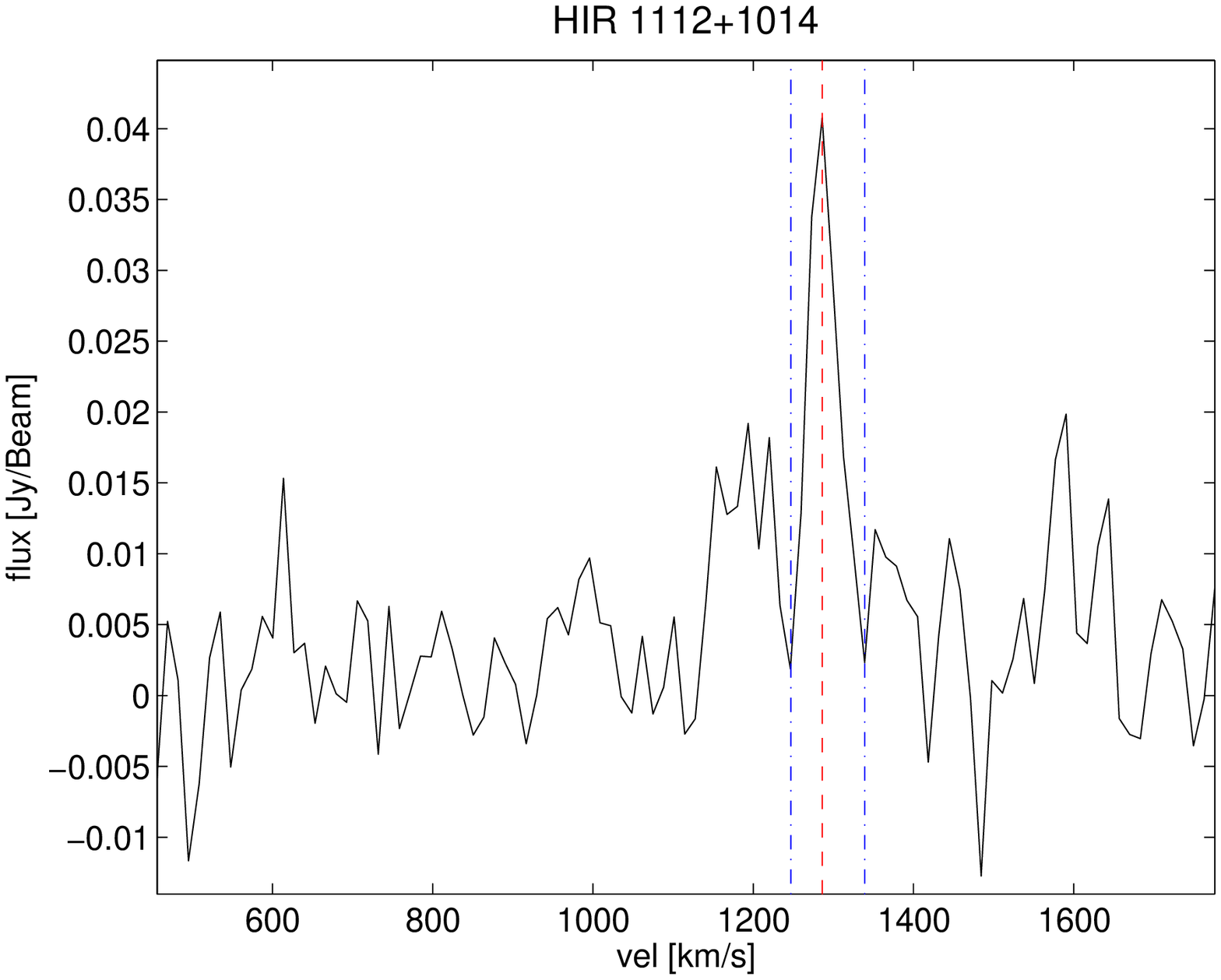}                                                        
 \includegraphics[width=0.3\textwidth]{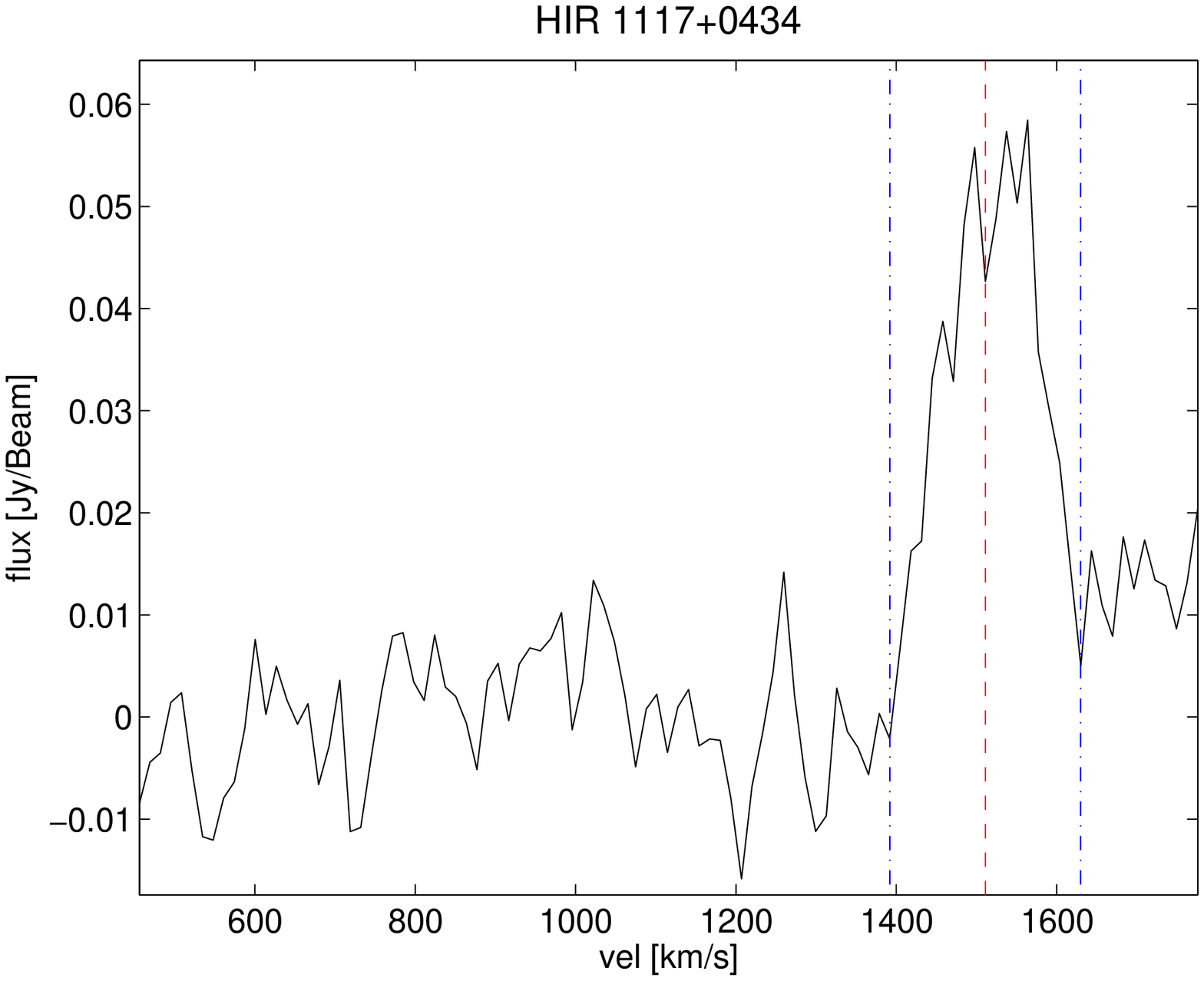}
 \includegraphics[width=0.3\textwidth]{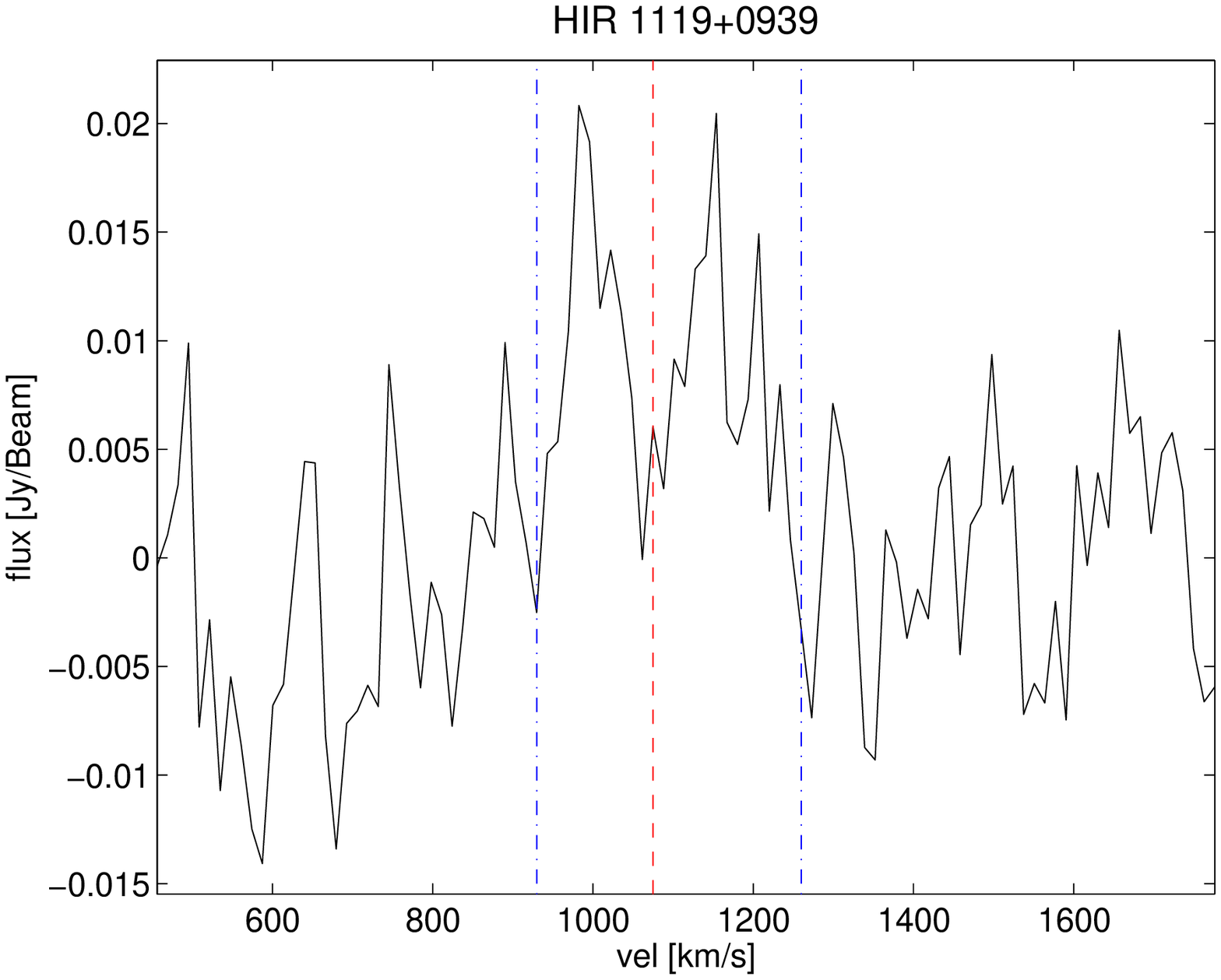}
                                                        
 \end{center}                                            
{\bf Fig~\ref{all_spectra}.} (continued)                                        
 
\end{figure*}

\begin{figure*}
  \begin{center}

 \includegraphics[width=0.3\textwidth]{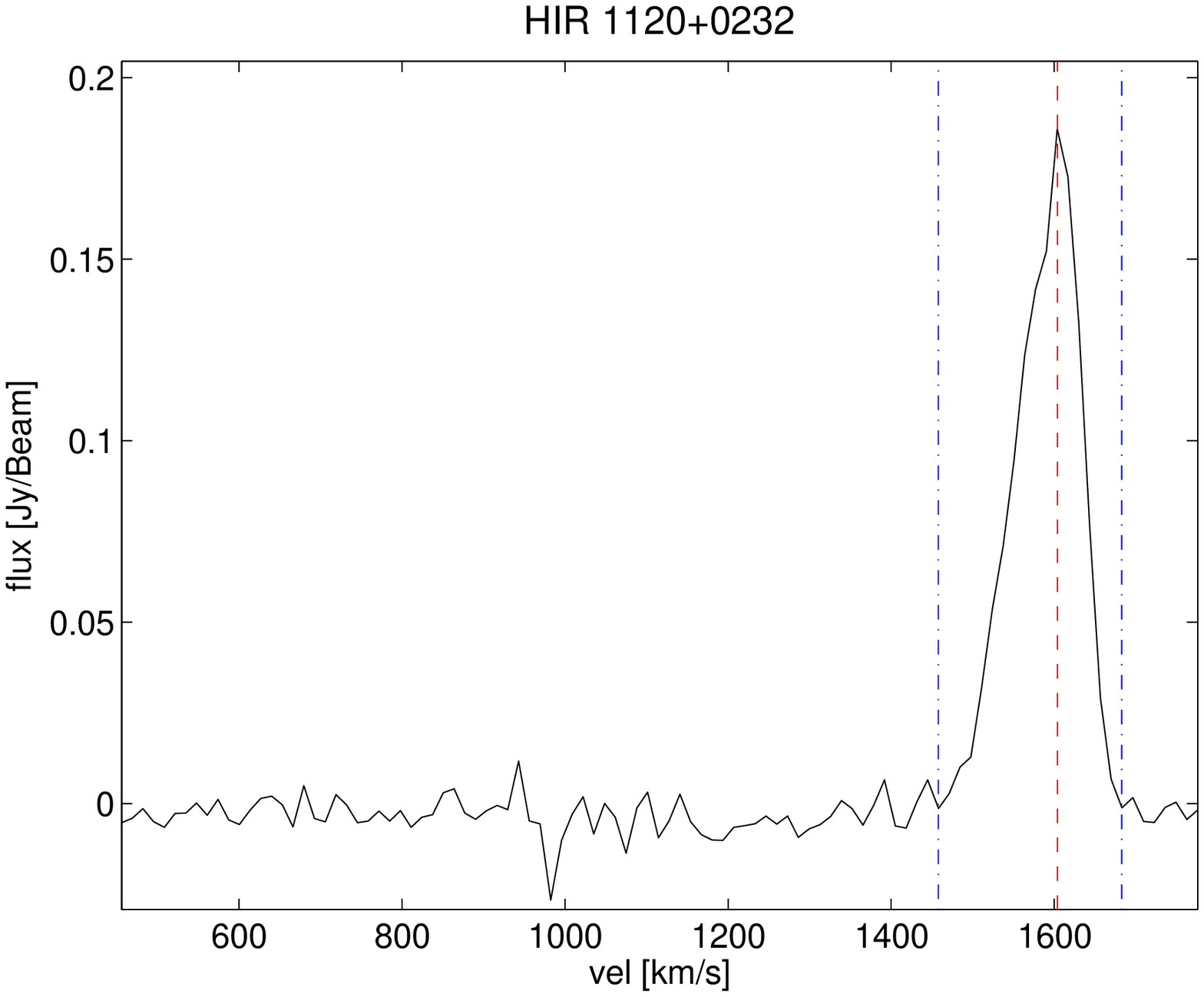}
 \includegraphics[width=0.3\textwidth]{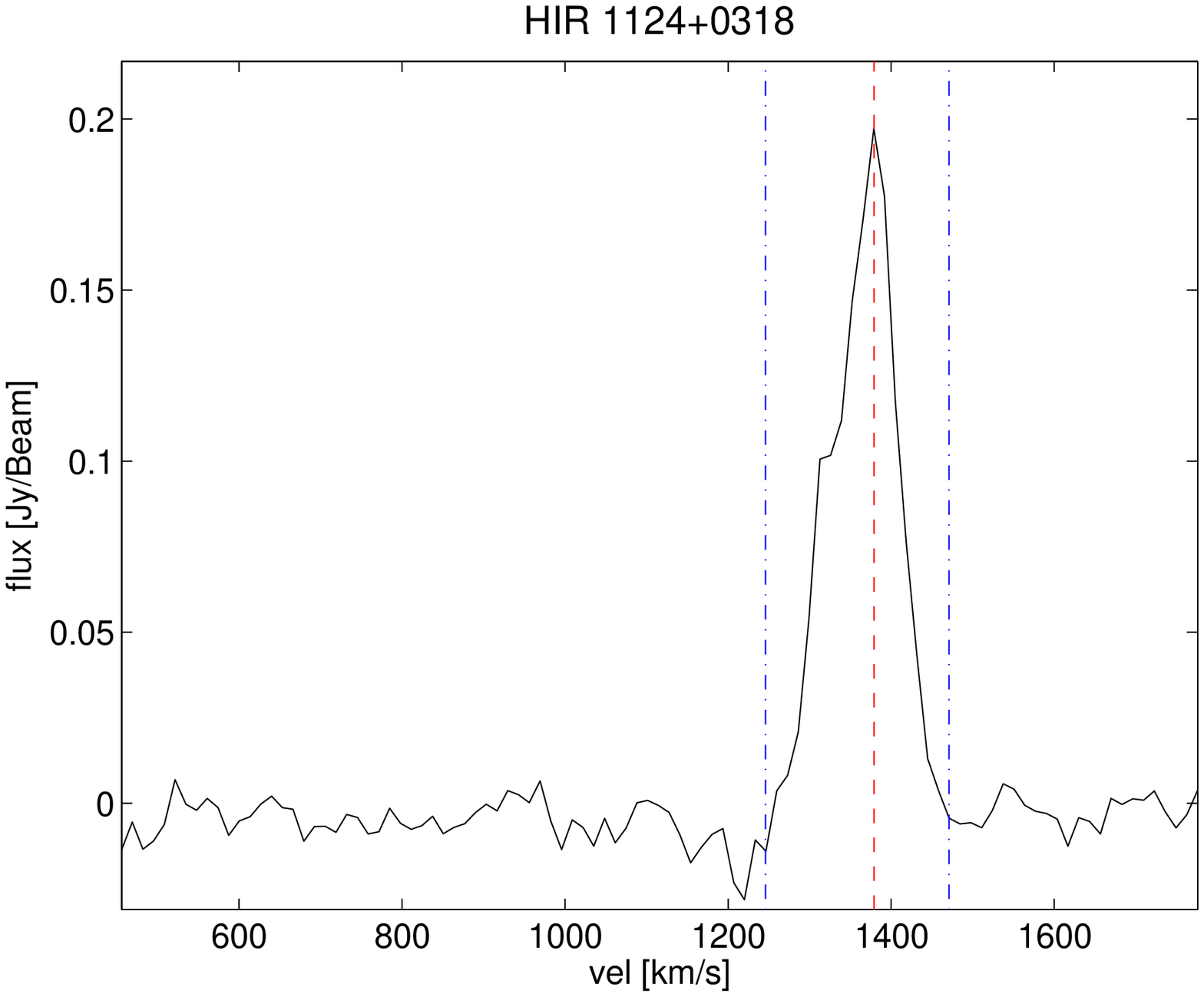}
  \includegraphics[width=0.3\textwidth]{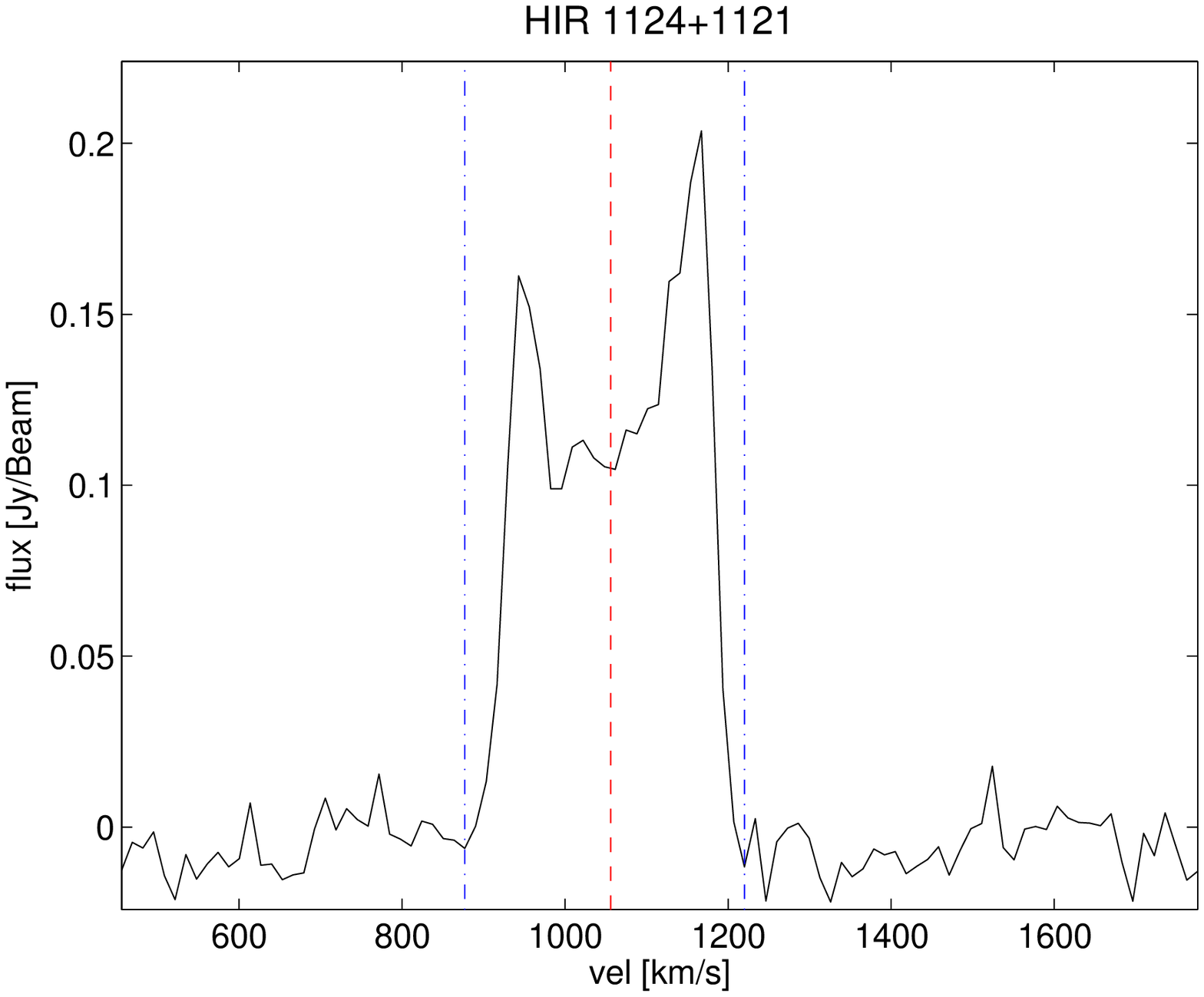}
 \includegraphics[width=0.3\textwidth]{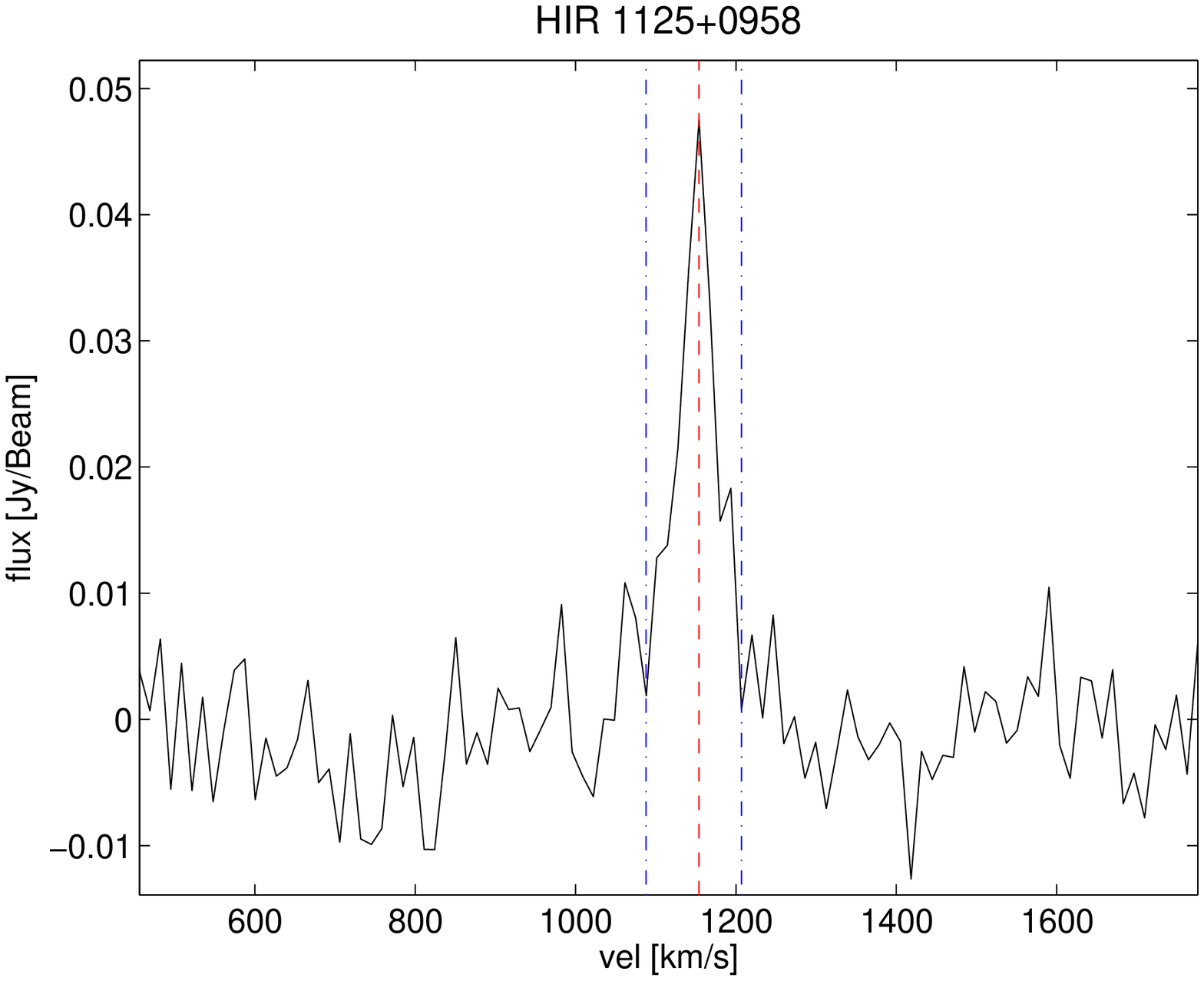}
 \includegraphics[width=0.3\textwidth]{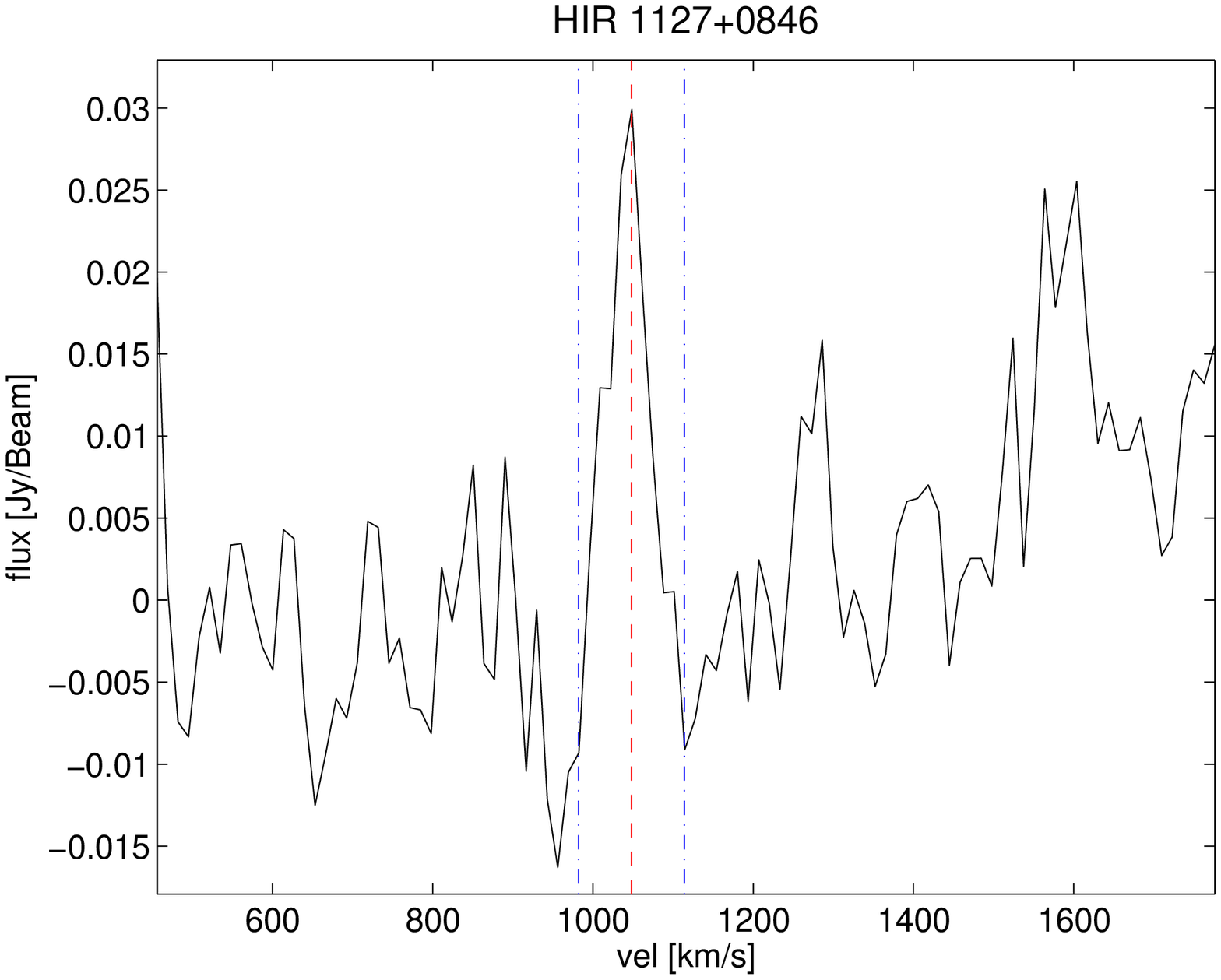}
 \includegraphics[width=0.3\textwidth]{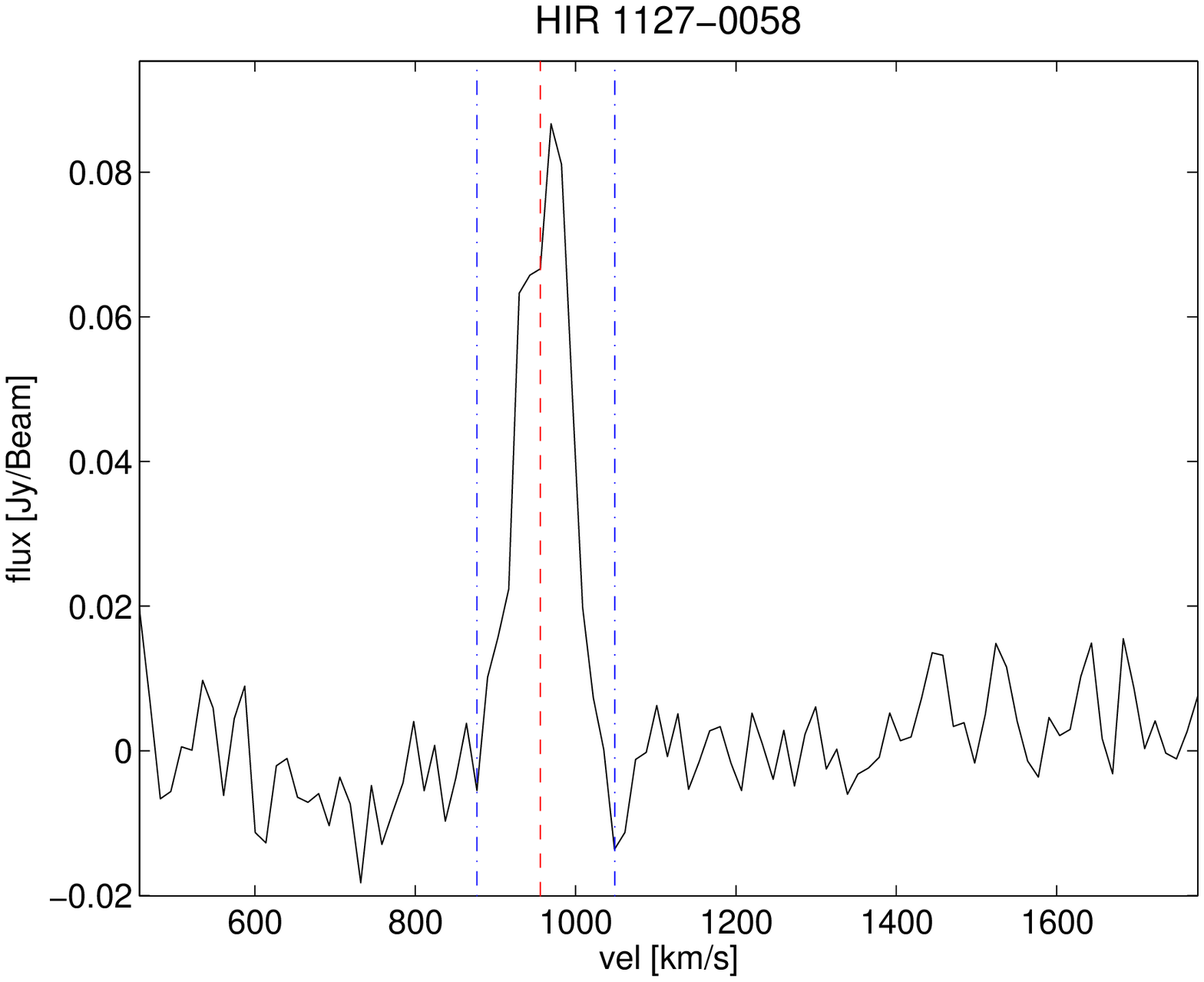}
 \includegraphics[width=0.3\textwidth]{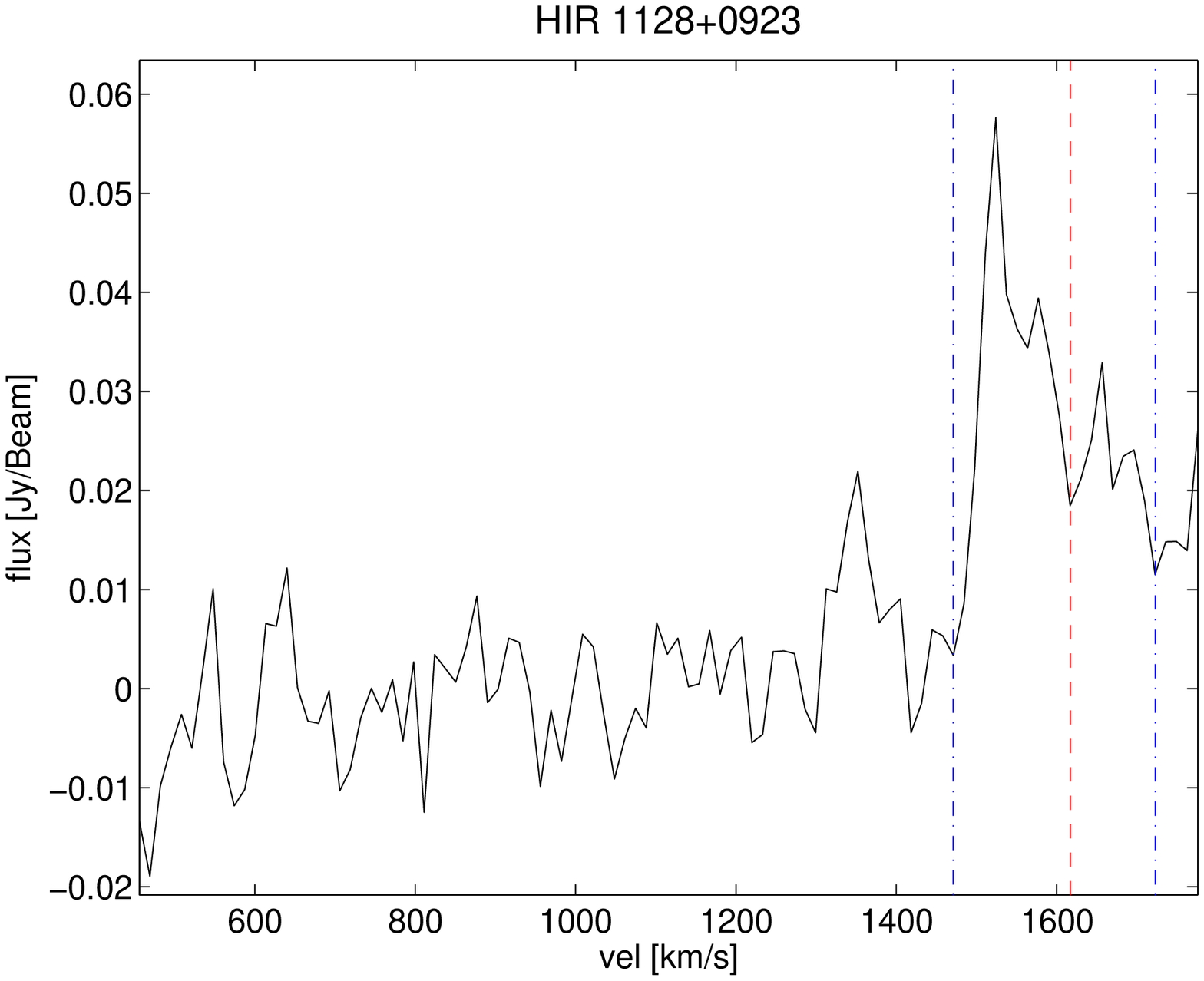}
 \includegraphics[width=0.3\textwidth]{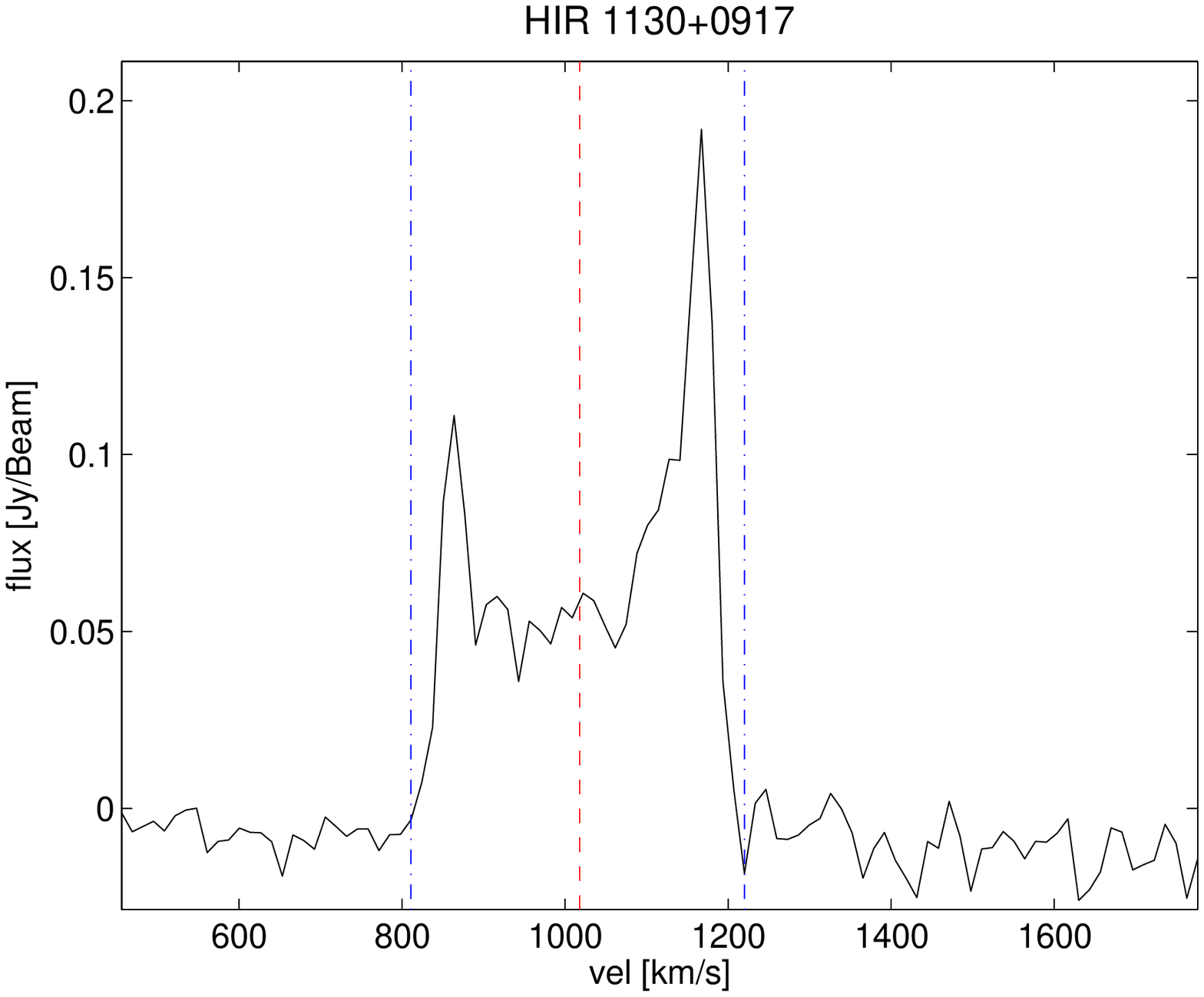}
 \includegraphics[width=0.3\textwidth]{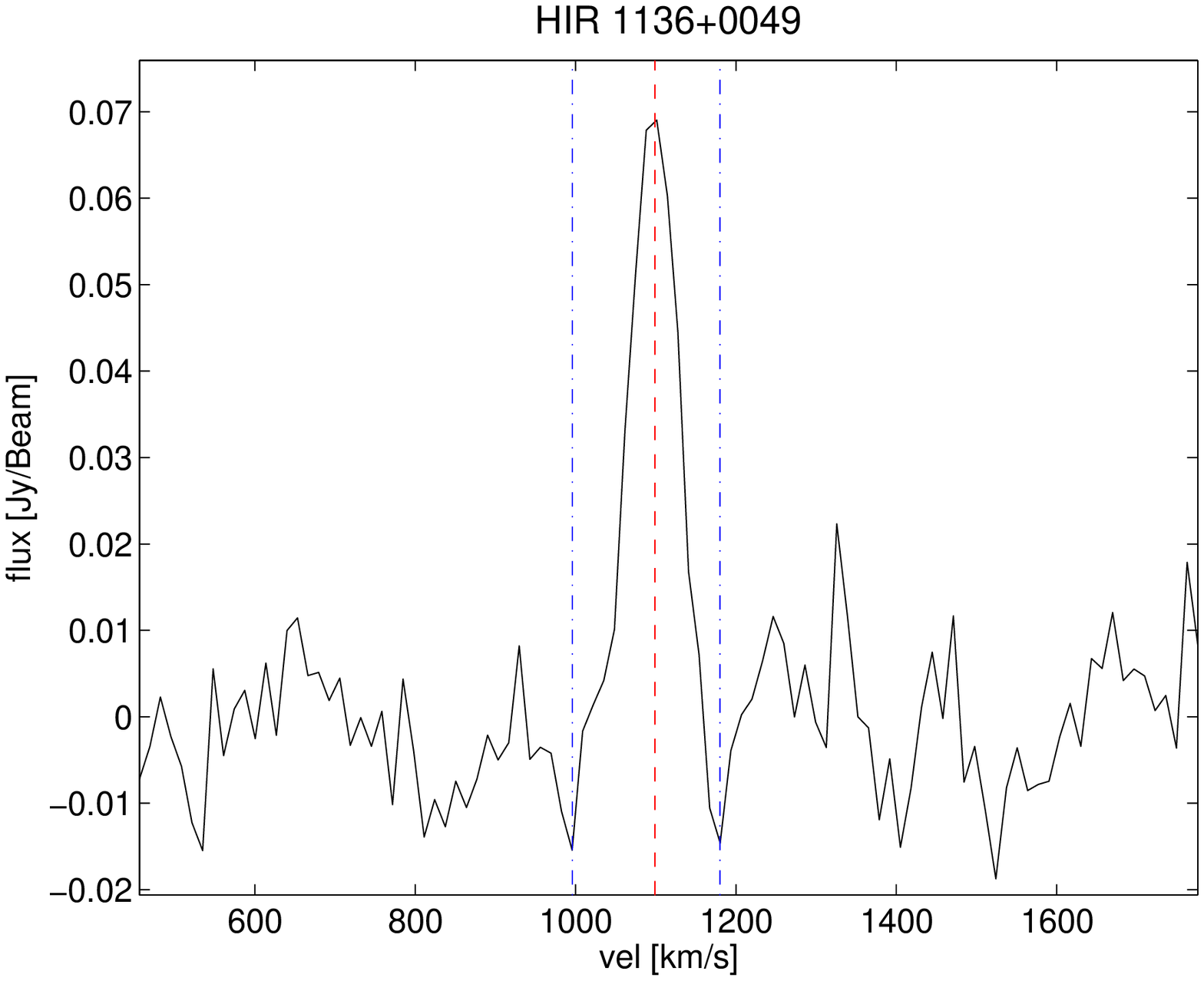}
 \includegraphics[width=0.3\textwidth]{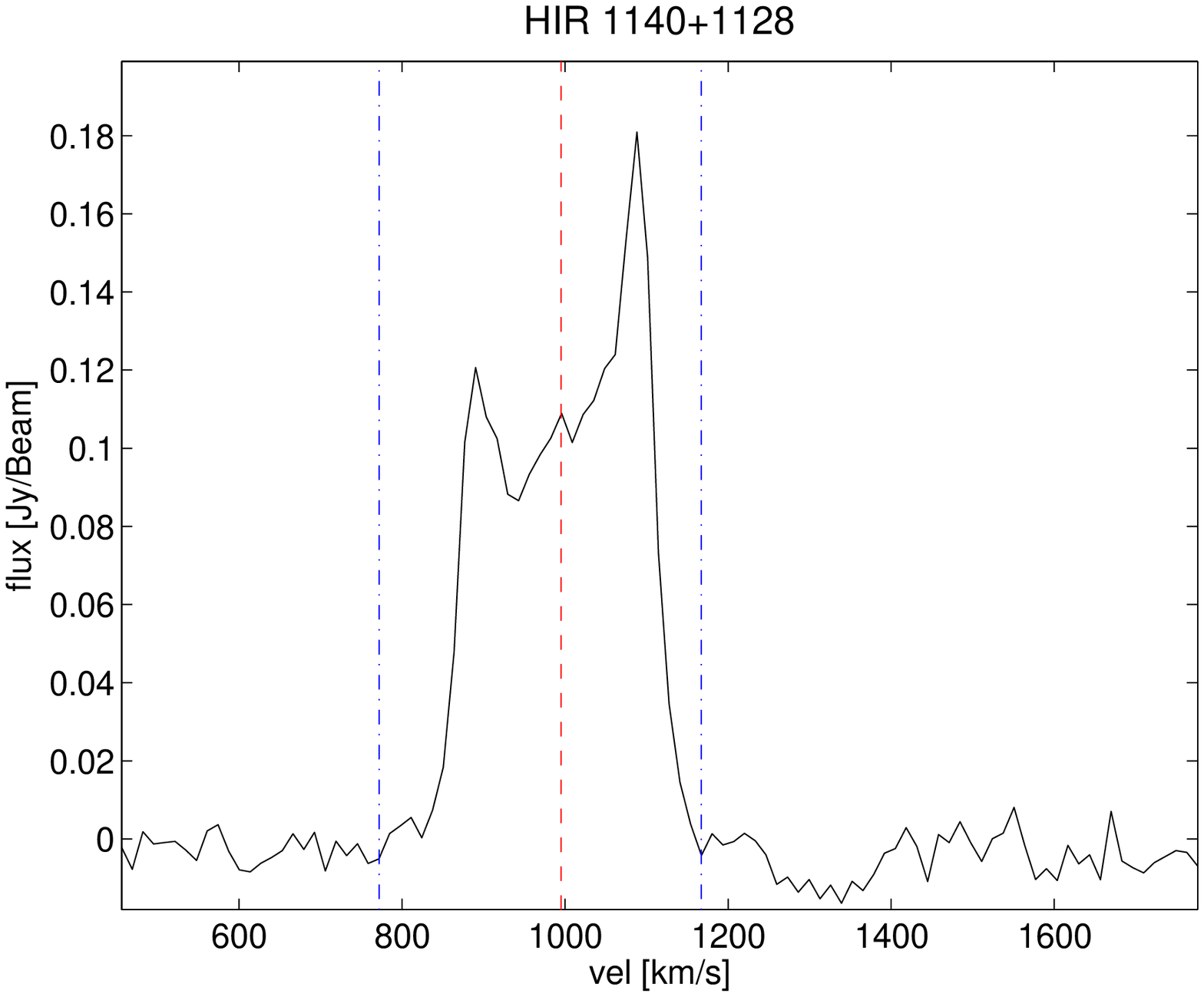}
 \includegraphics[width=0.3\textwidth]{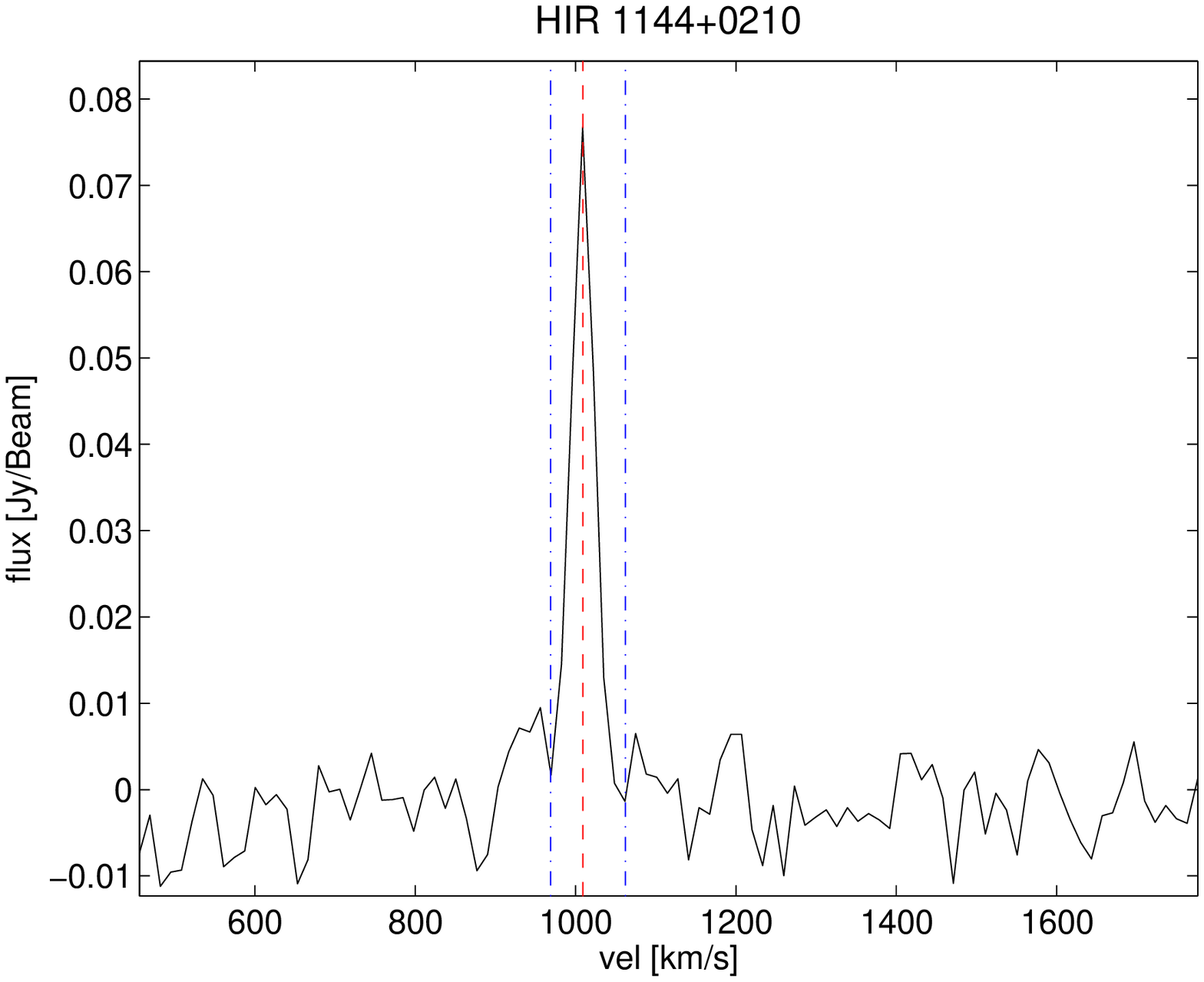}
 \includegraphics[width=0.3\textwidth]{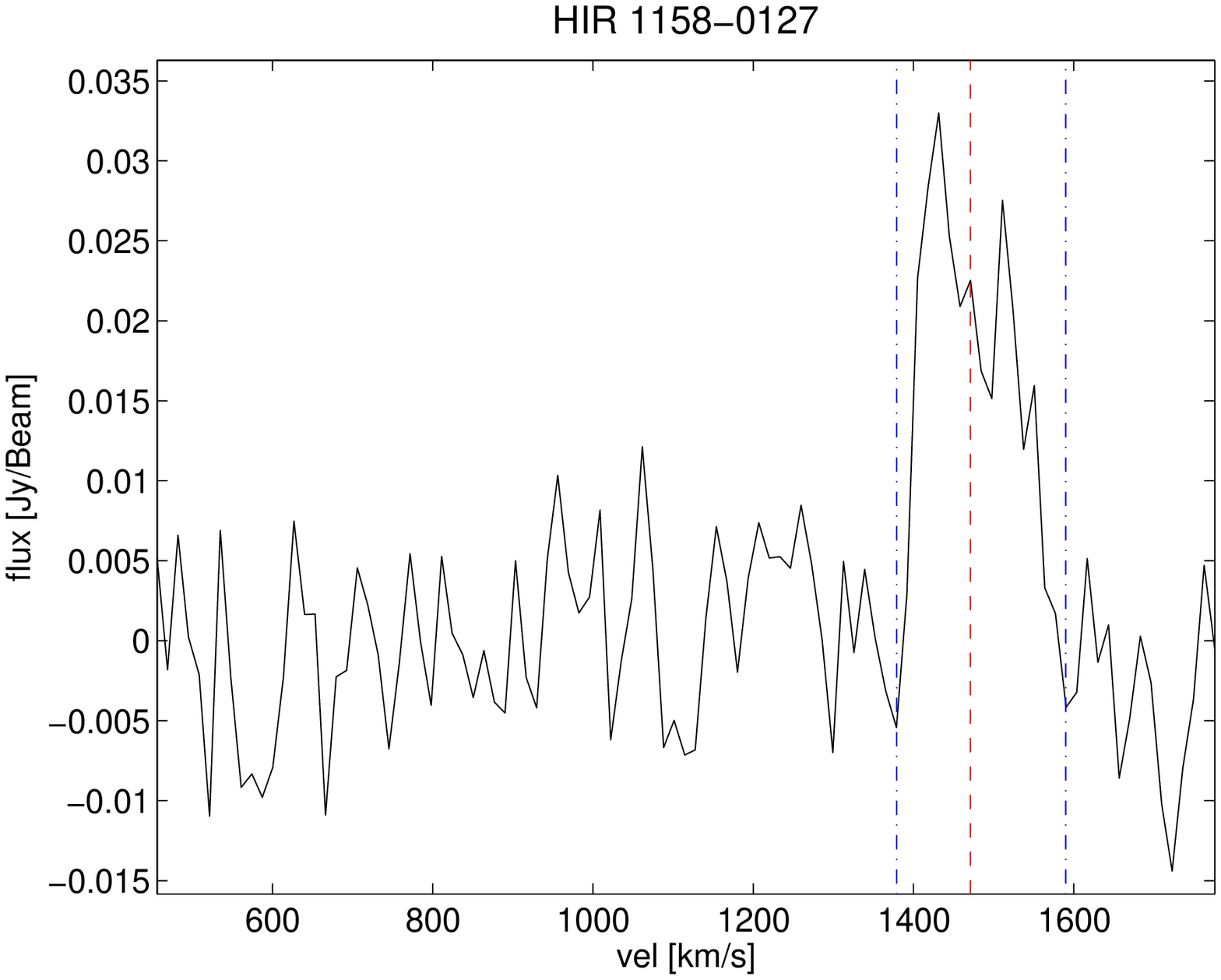}
 \includegraphics[width=0.3\textwidth]{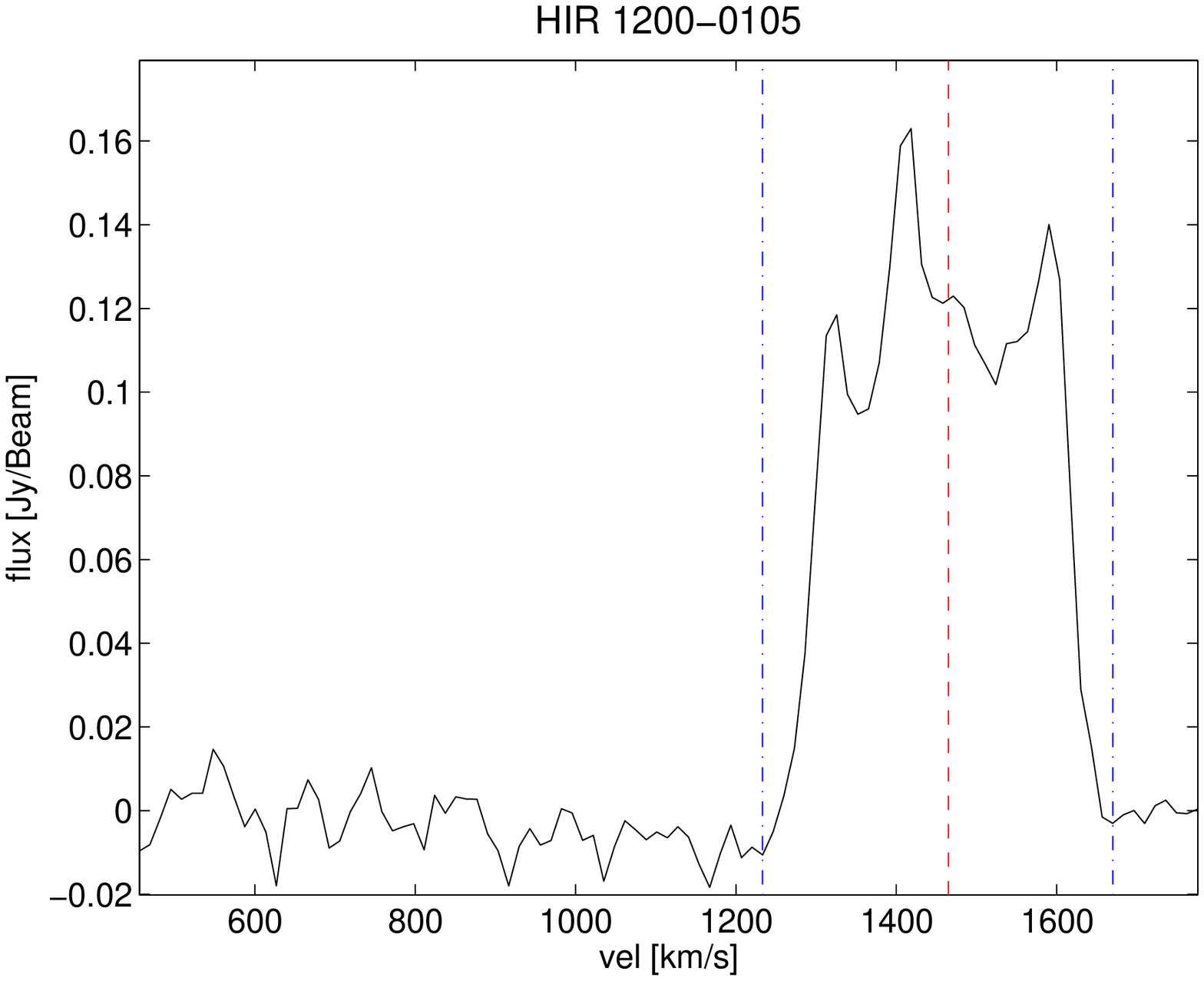}
 \includegraphics[width=0.3\textwidth]{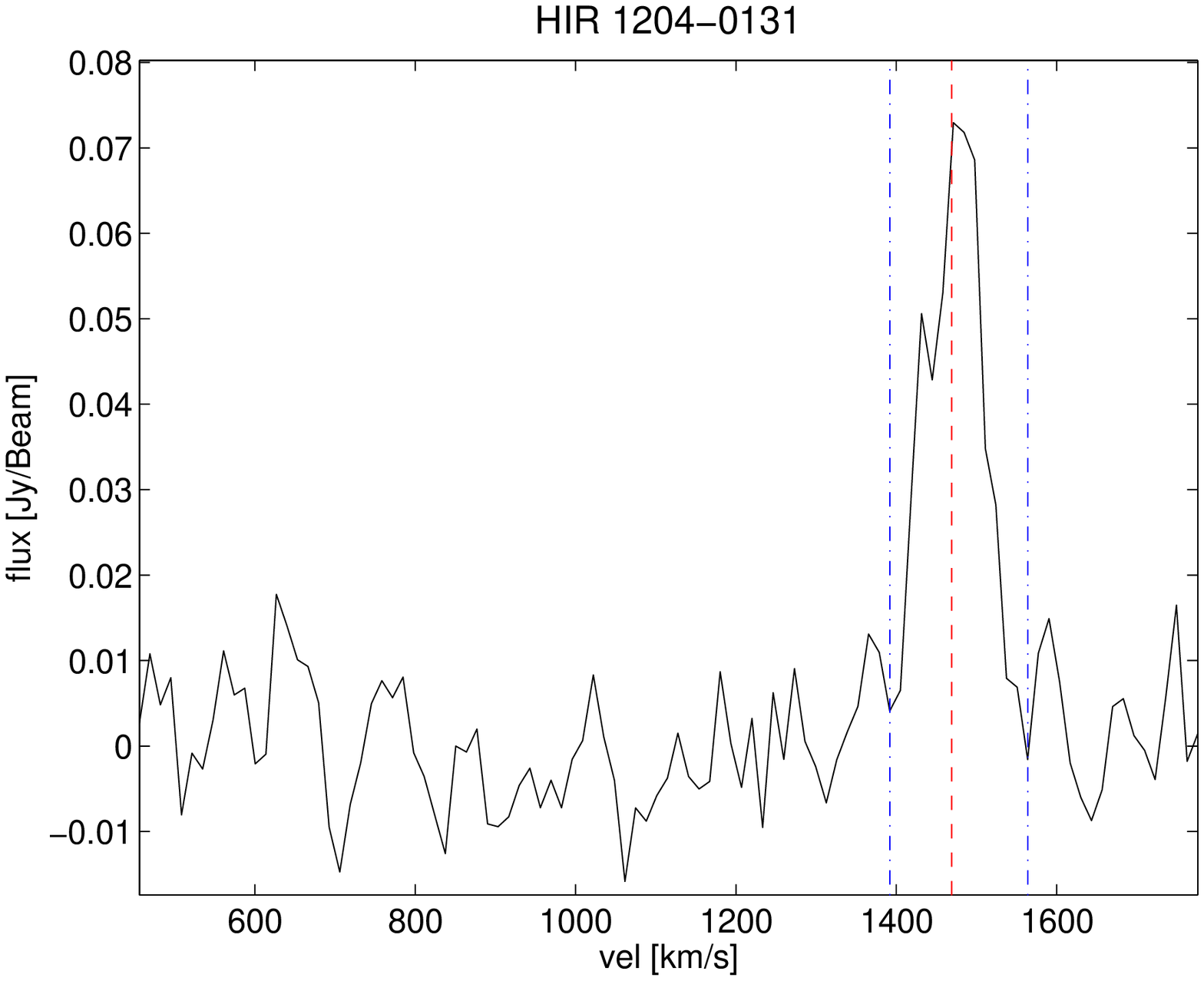}
 \includegraphics[width=0.3\textwidth]{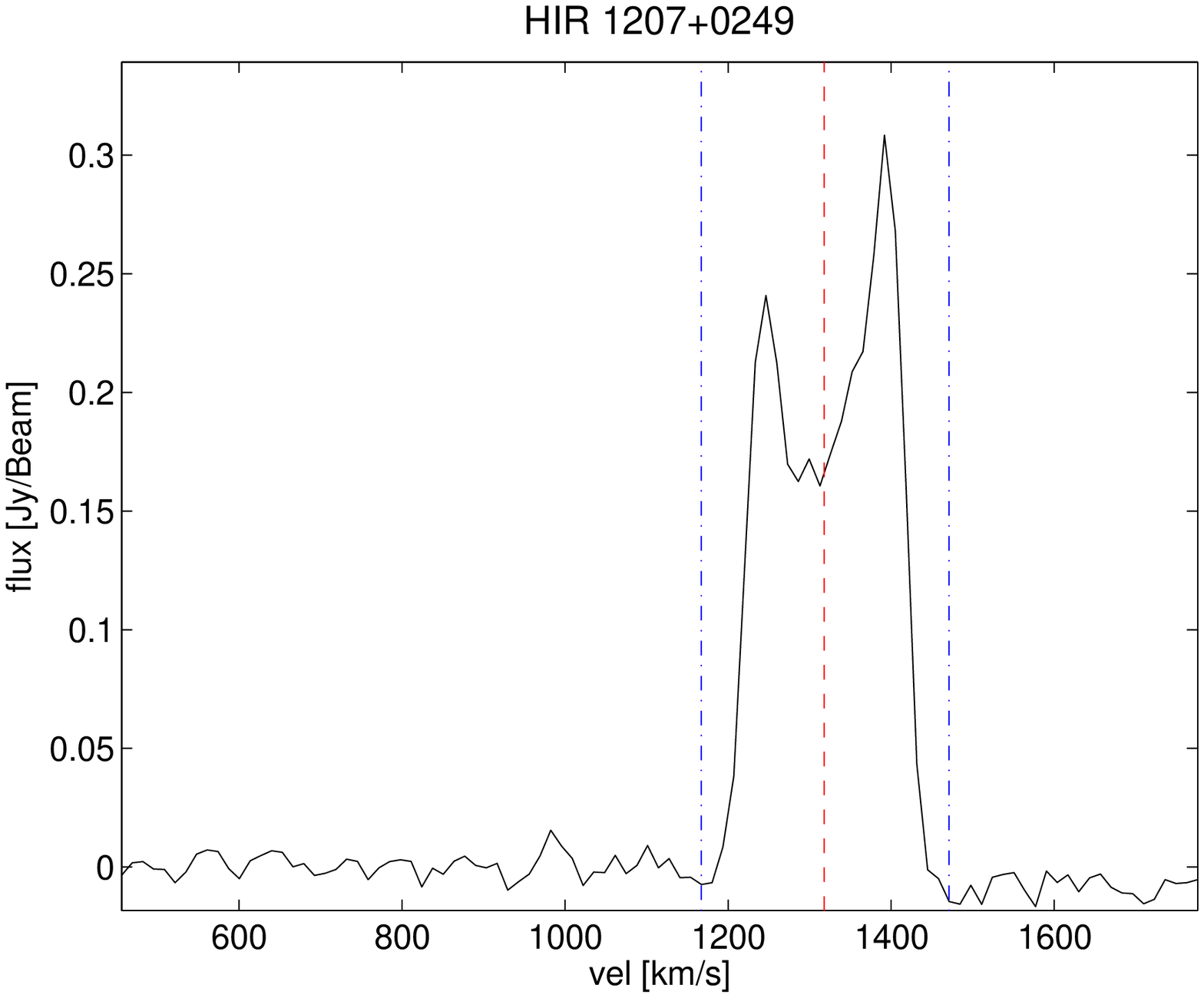}

 \end{center}                                            
{\bf Fig~\ref{all_spectra}.} (continued)                                        
 
\end{figure*}

\begin{figure*}
  \begin{center}

 \includegraphics[width=0.3\textwidth]{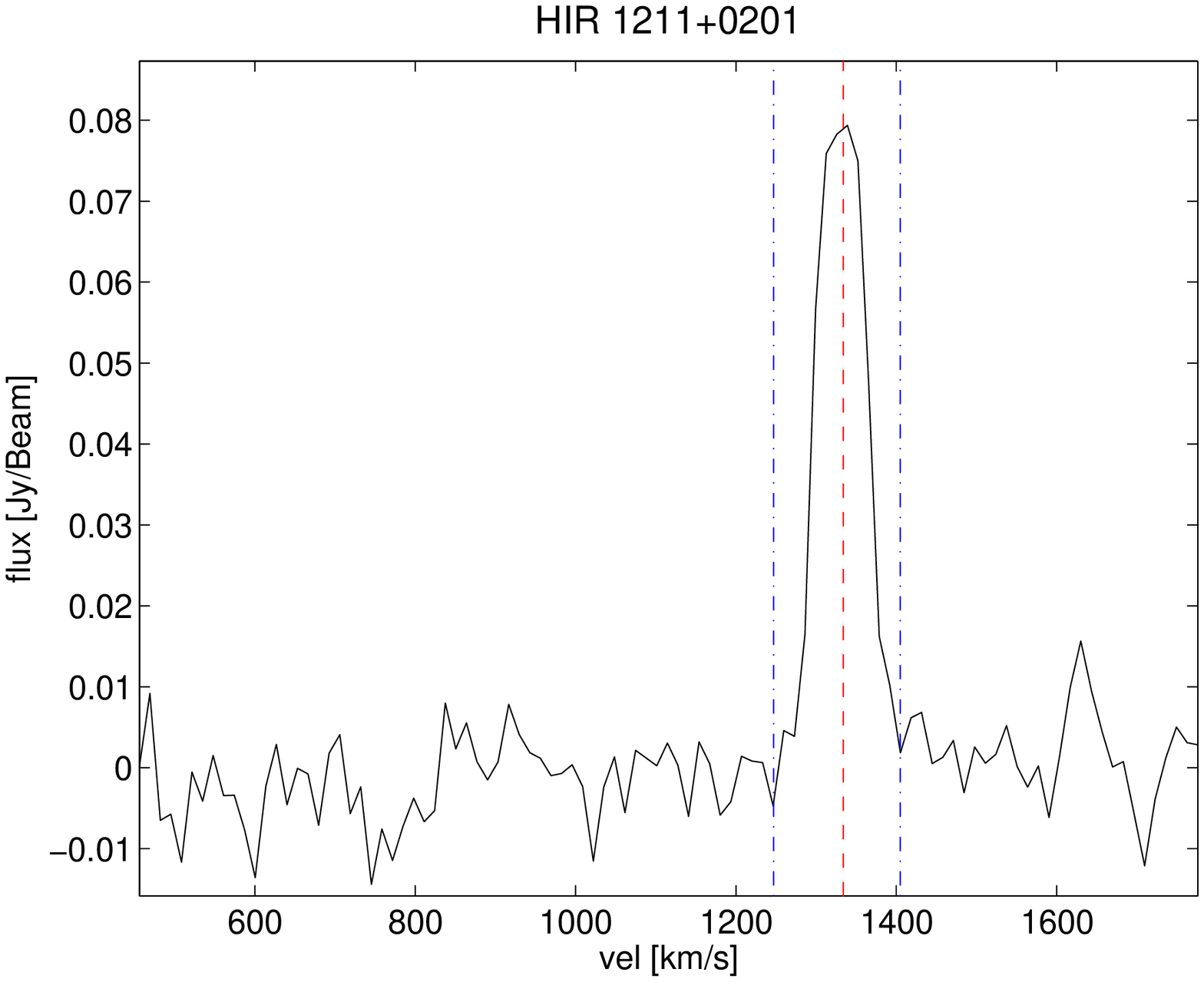}
 \includegraphics[width=0.3\textwidth]{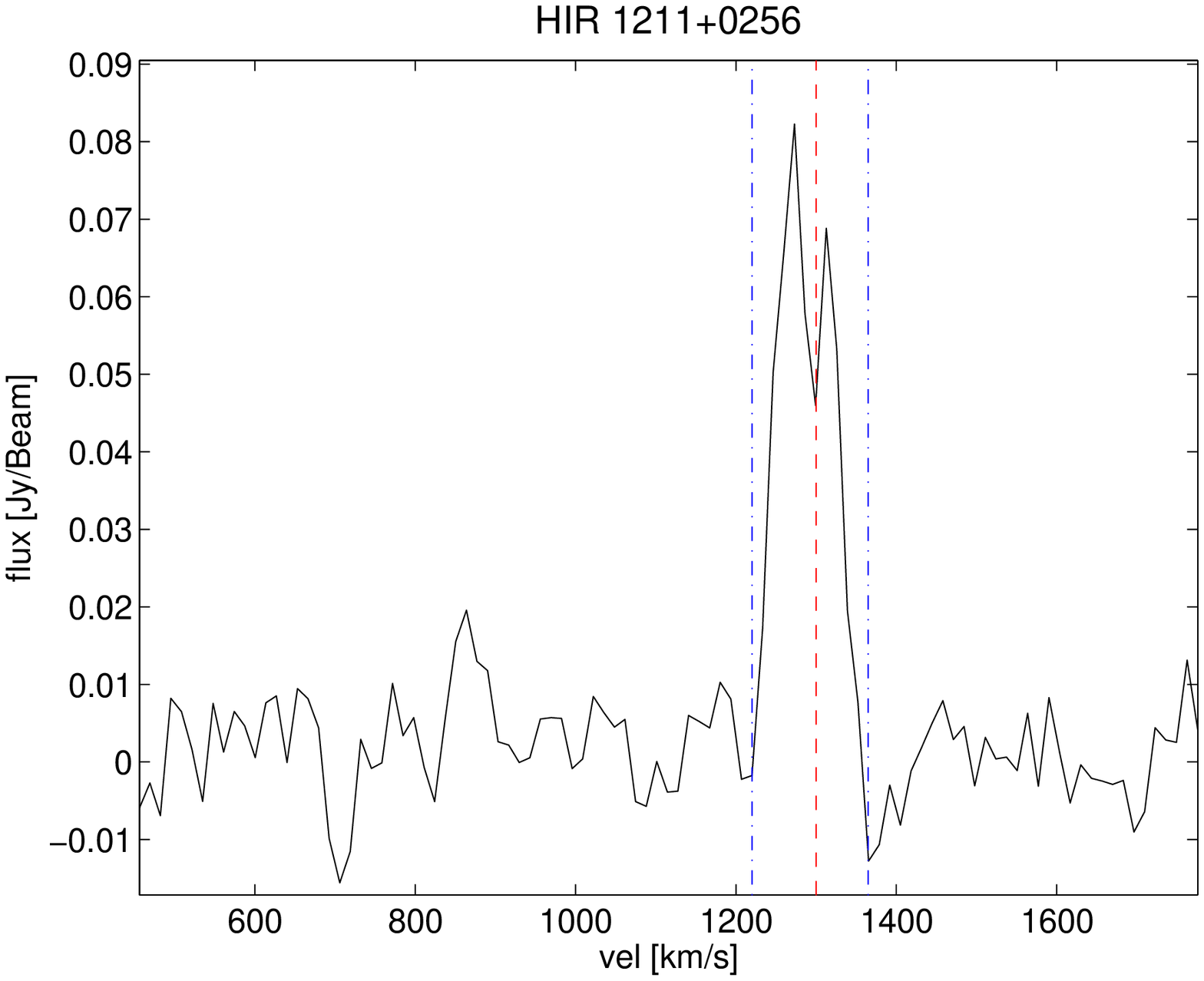}
 \includegraphics[width=0.3\textwidth]{1212+0248_spec.eps}
 \includegraphics[width=0.3\textwidth]{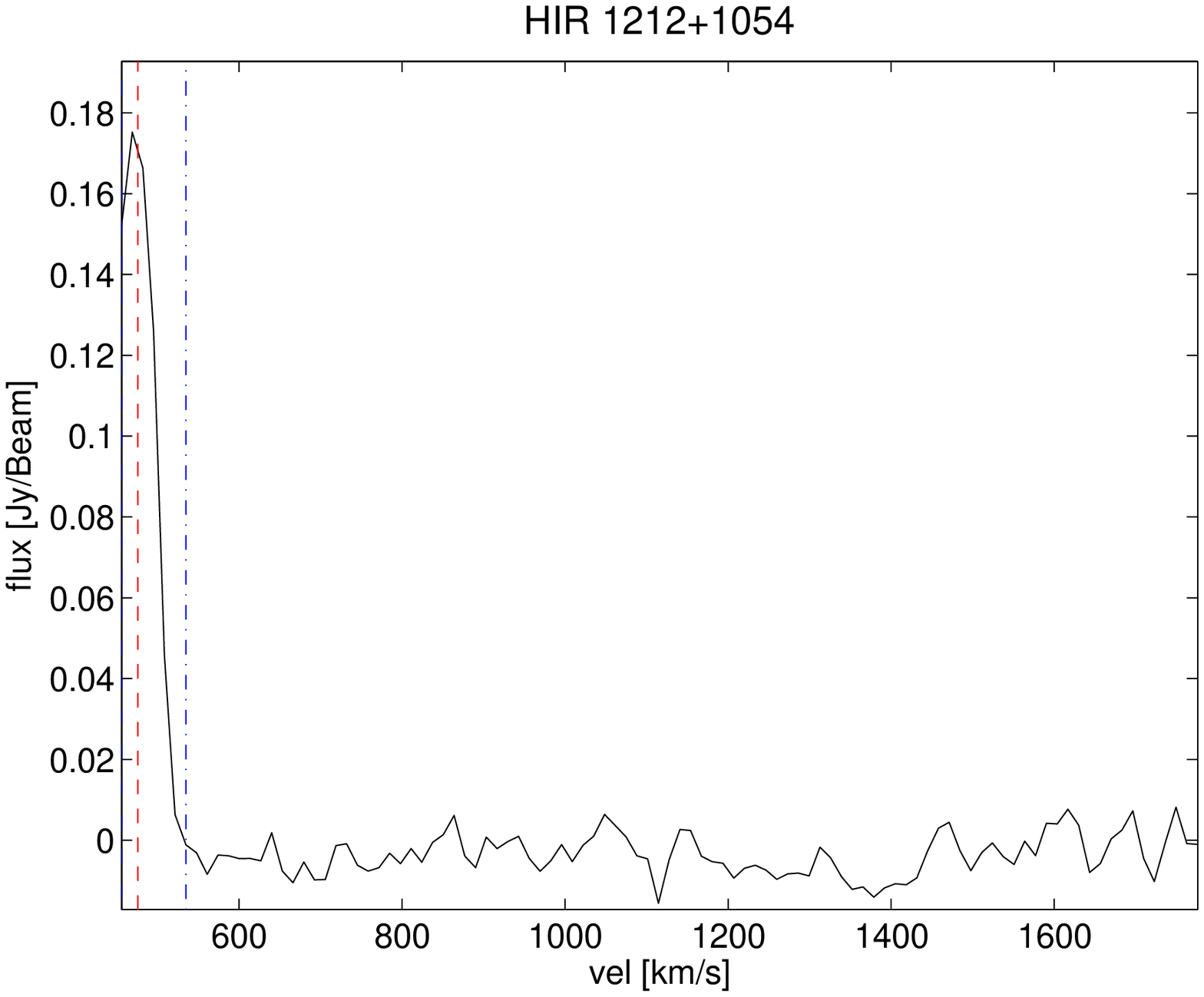}
 \includegraphics[width=0.3\textwidth]{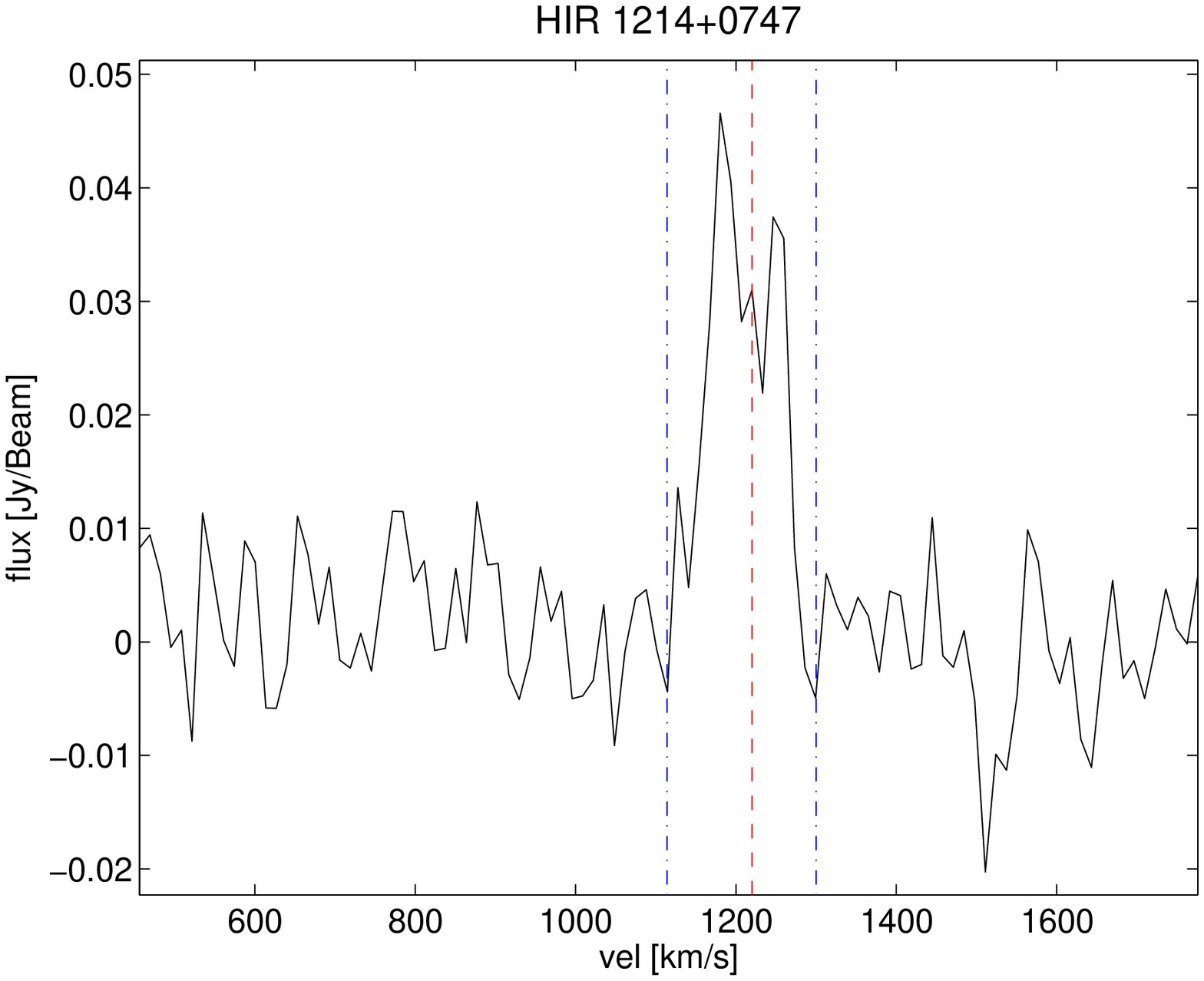}
 \includegraphics[width=0.3\textwidth]{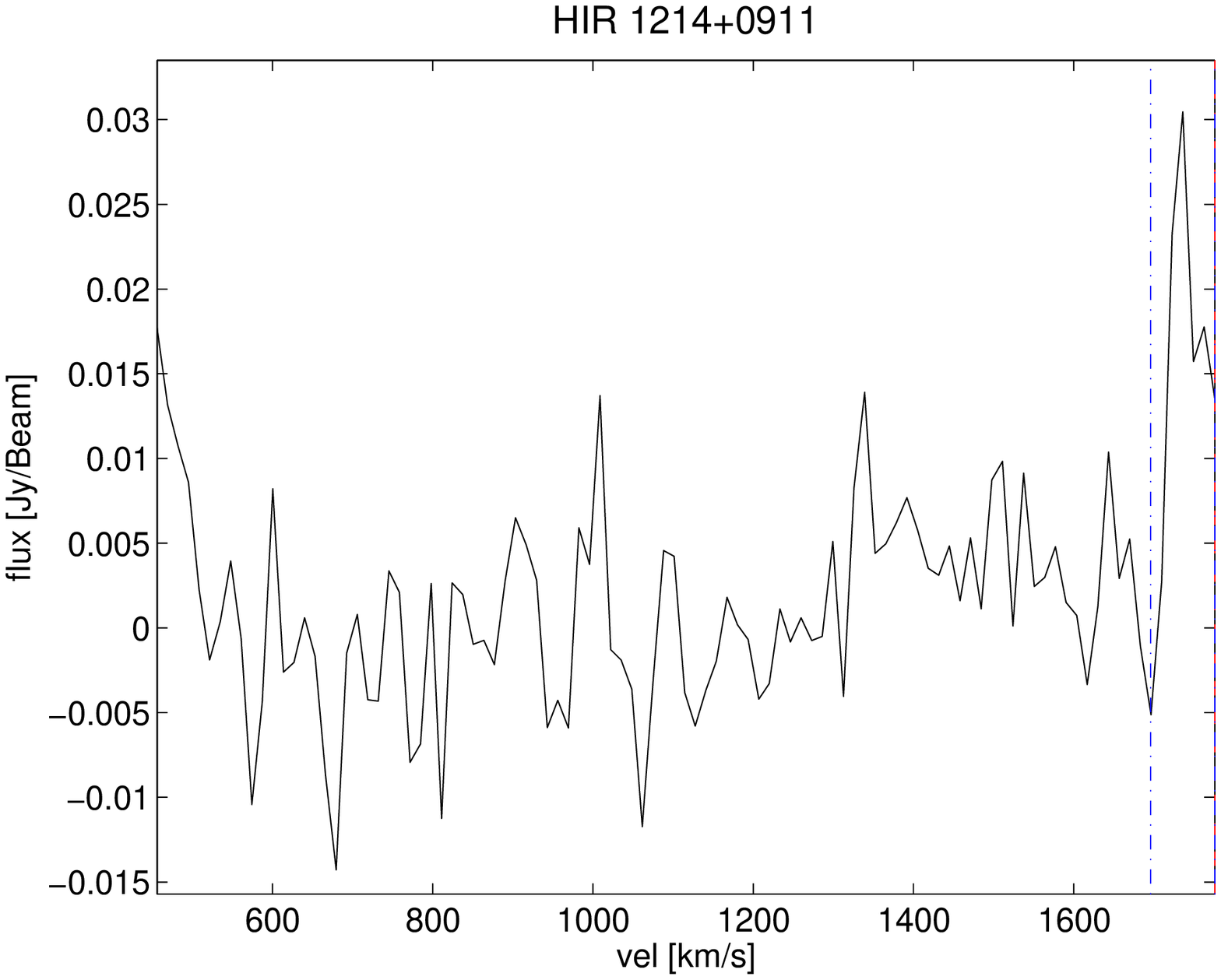}
 \includegraphics[width=0.3\textwidth]{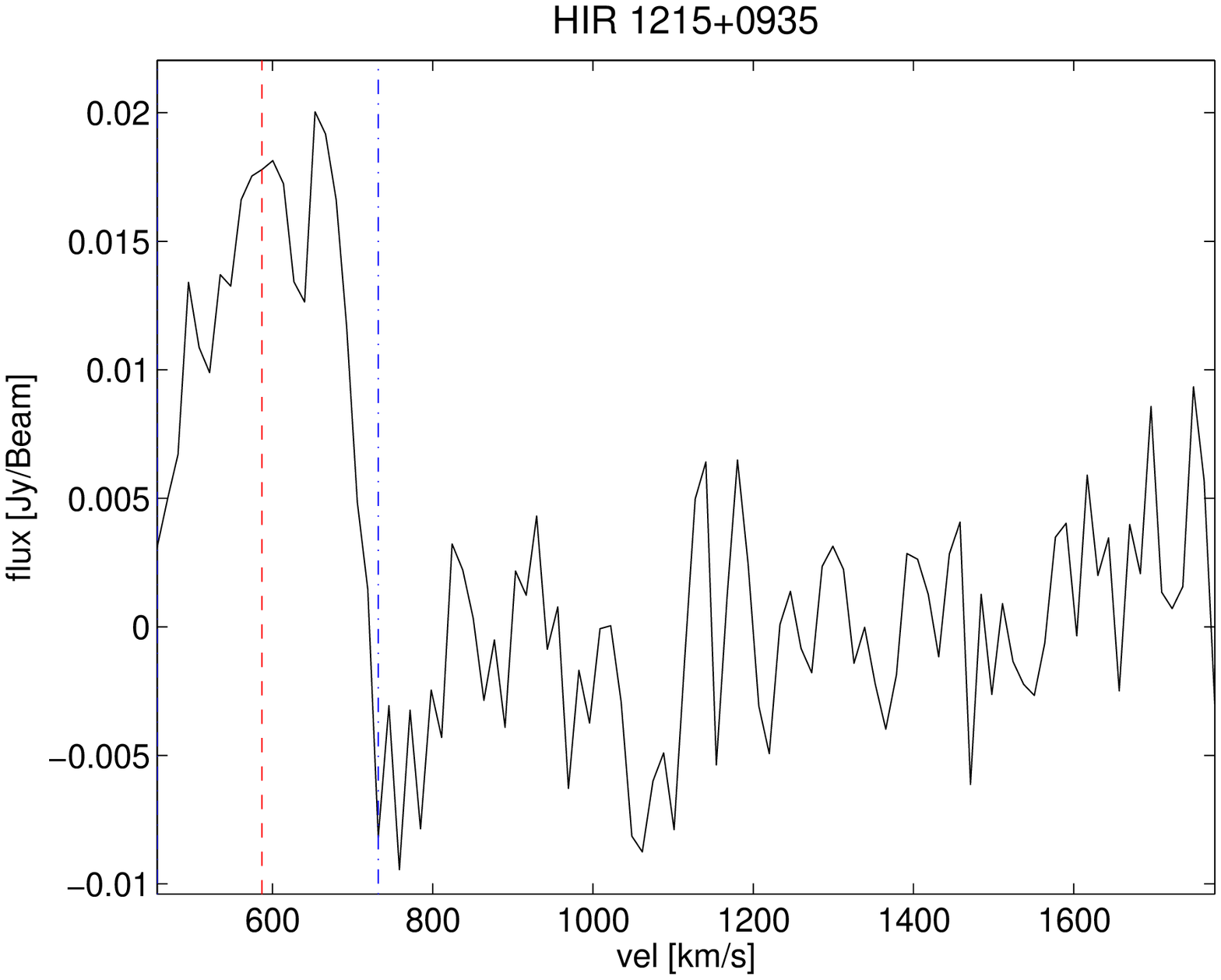}
 \includegraphics[width=0.3\textwidth]{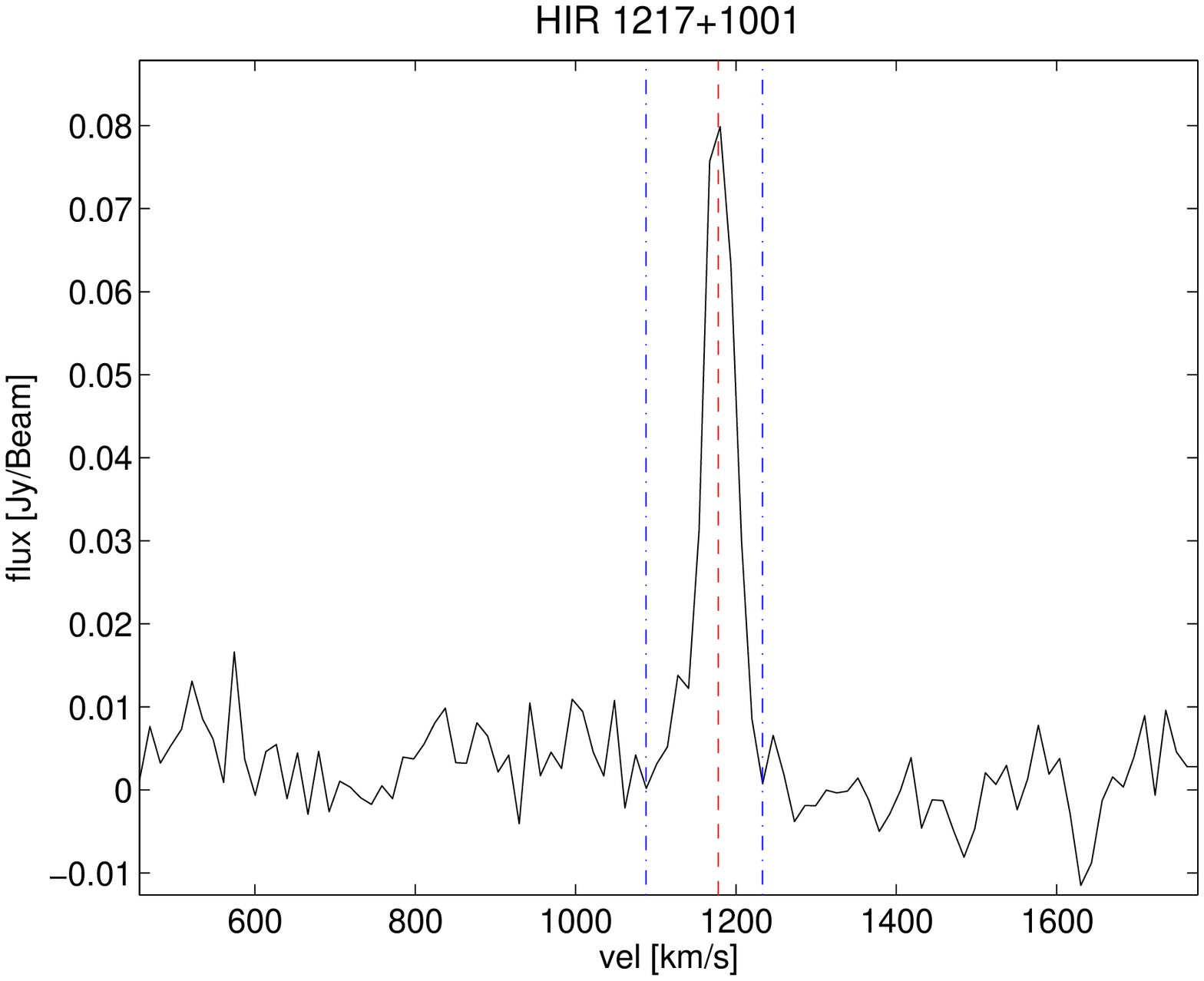}
 \includegraphics[width=0.3\textwidth]{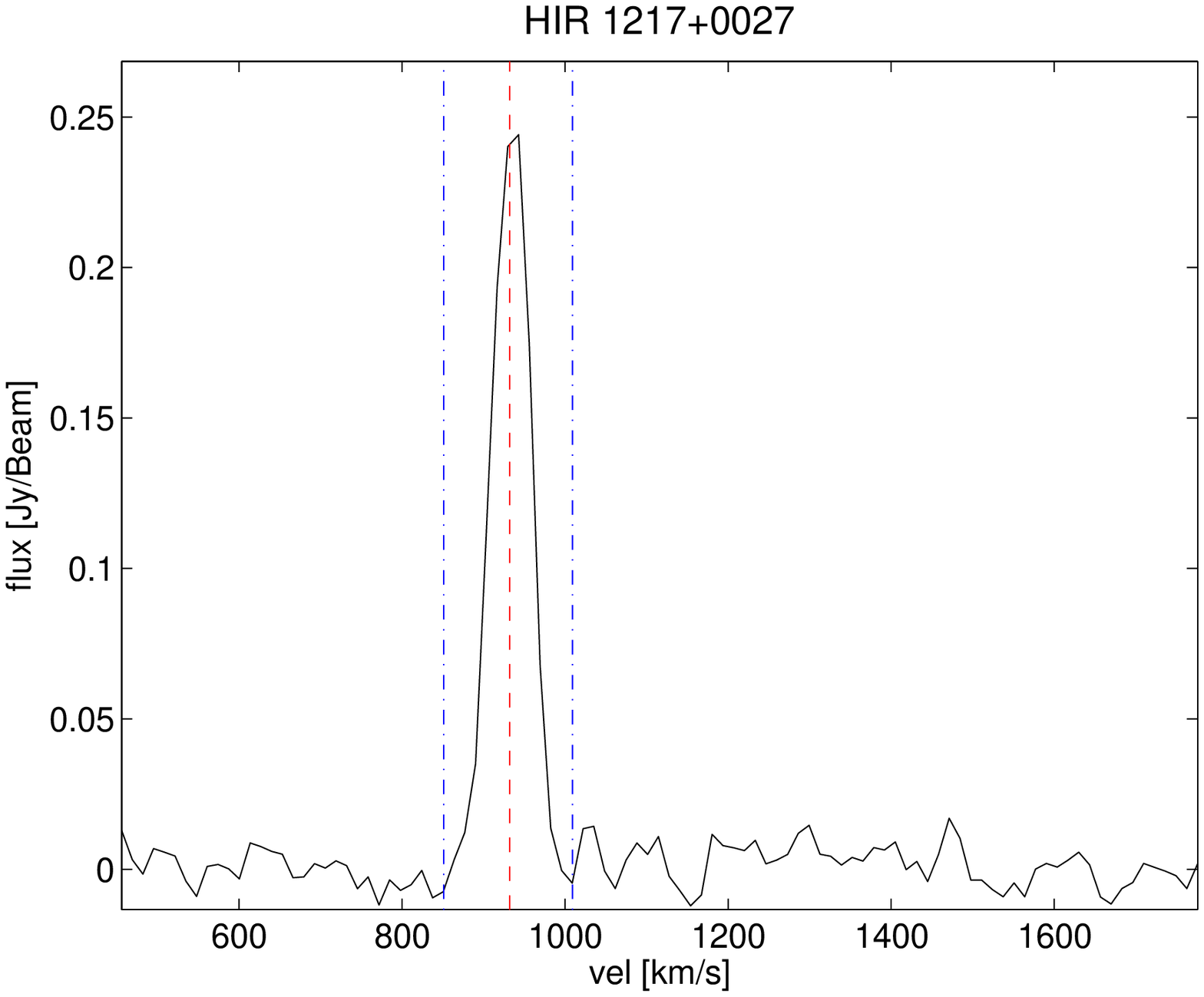}
 \includegraphics[width=0.3\textwidth]{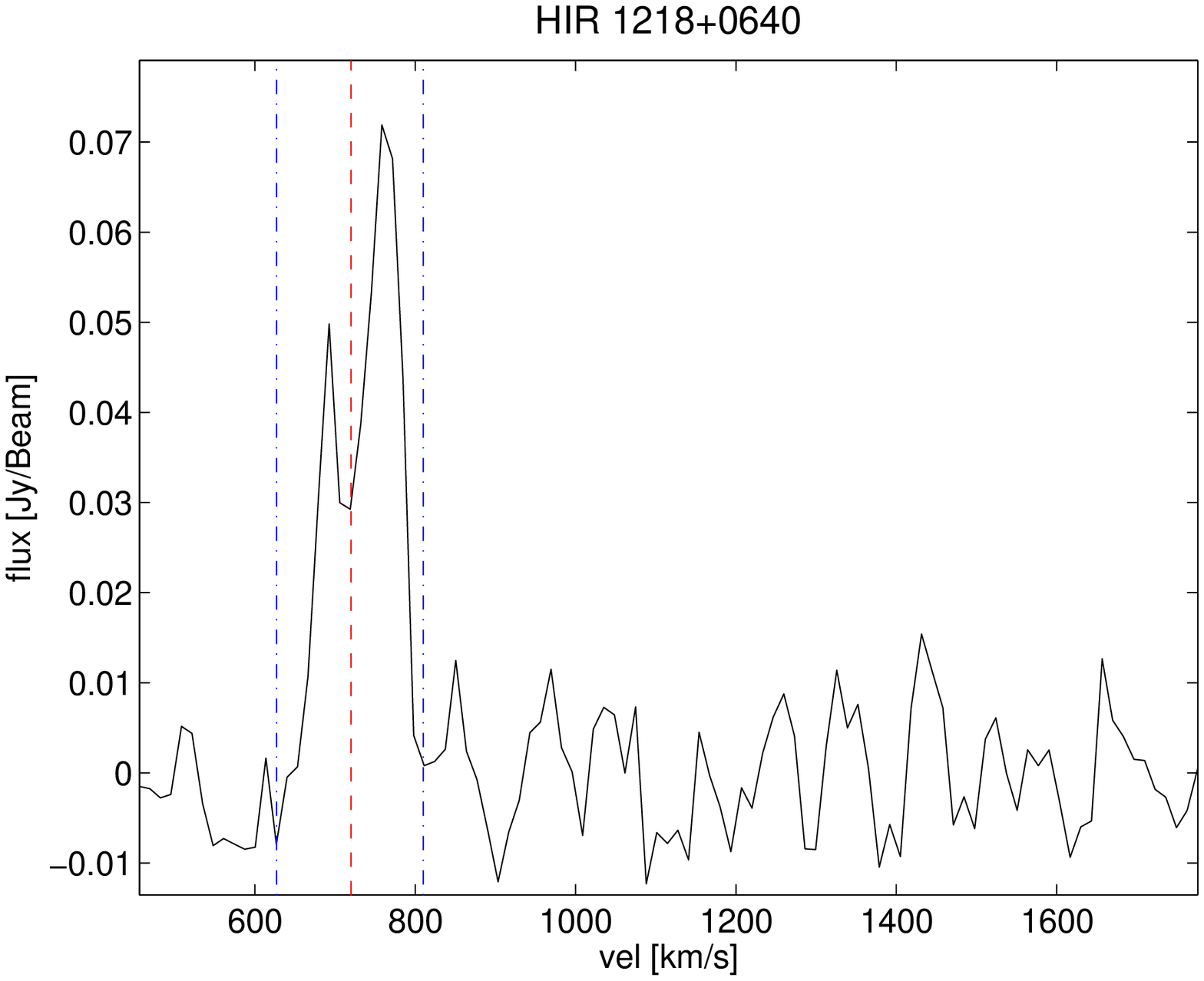}
 \includegraphics[width=0.3\textwidth]{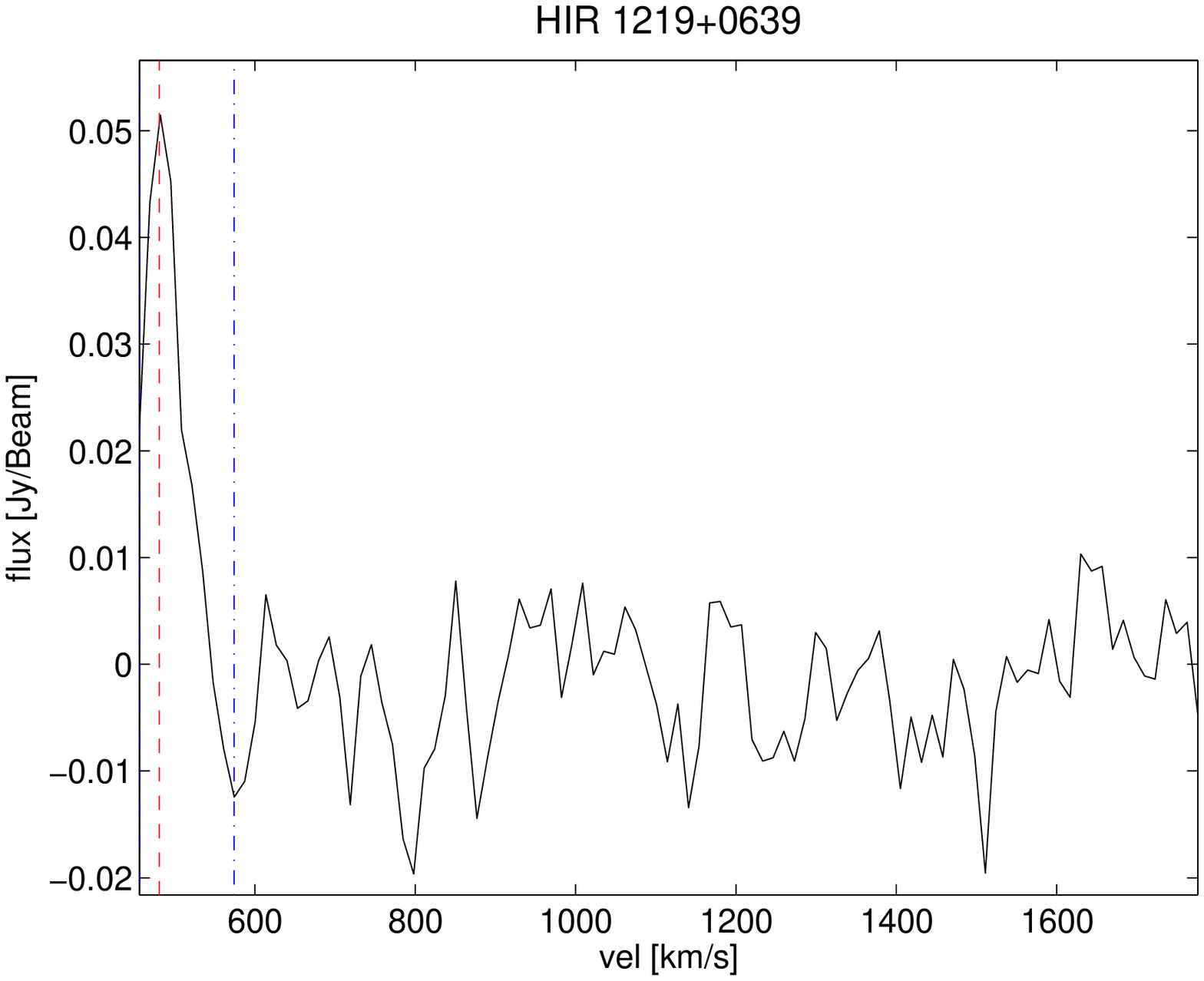}
 \includegraphics[width=0.3\textwidth]{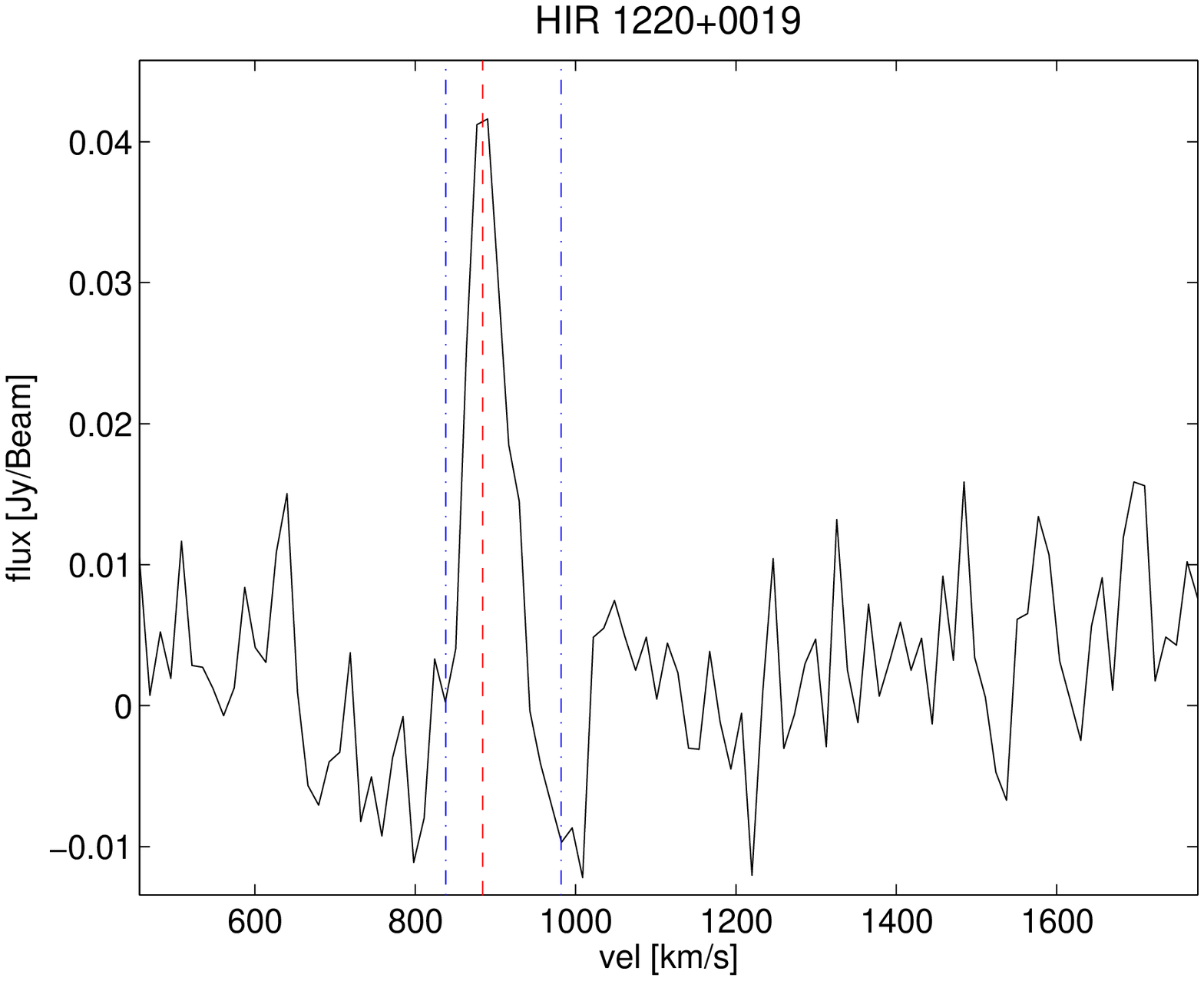}
 \includegraphics[width=0.3\textwidth]{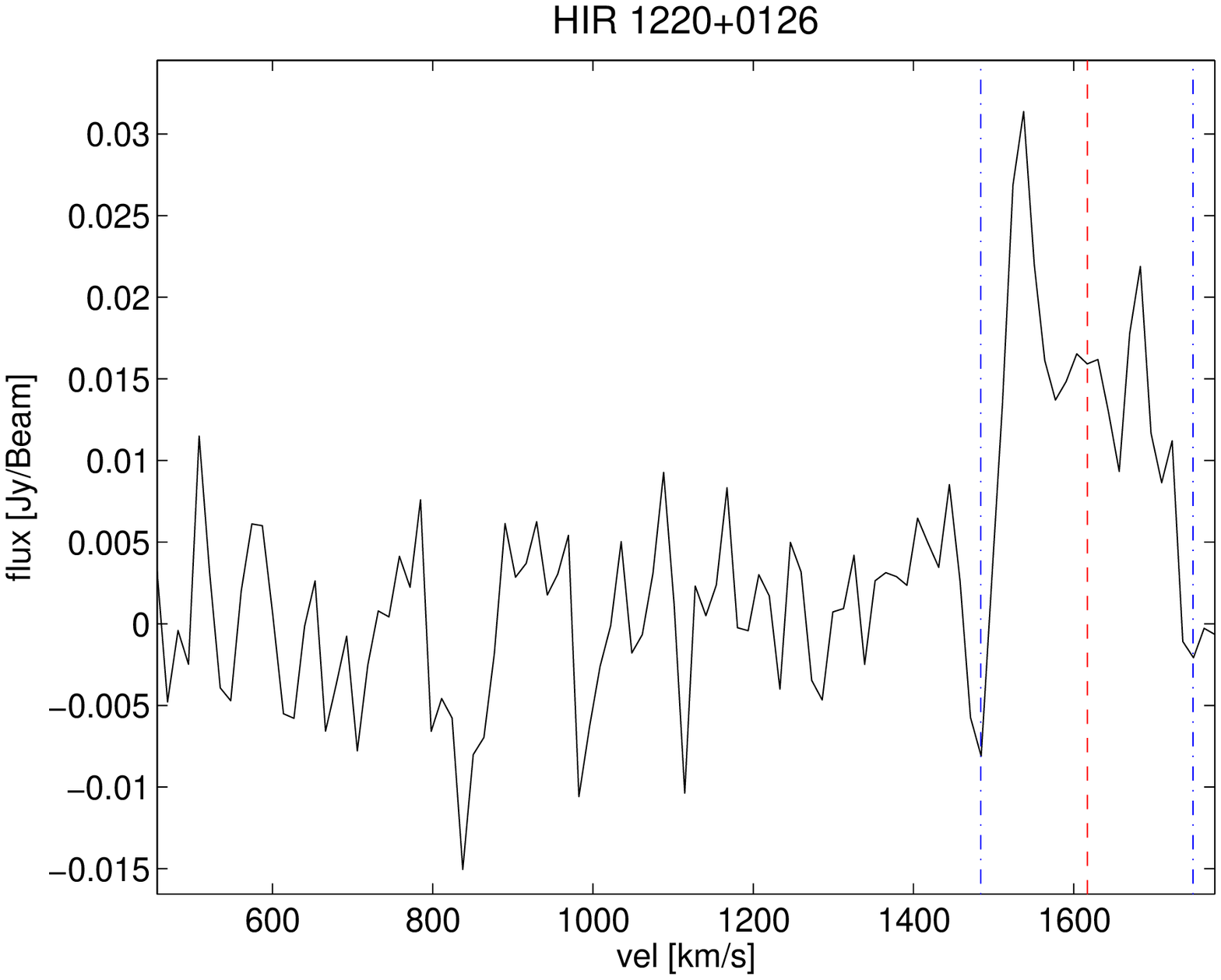}
 \includegraphics[width=0.3\textwidth]{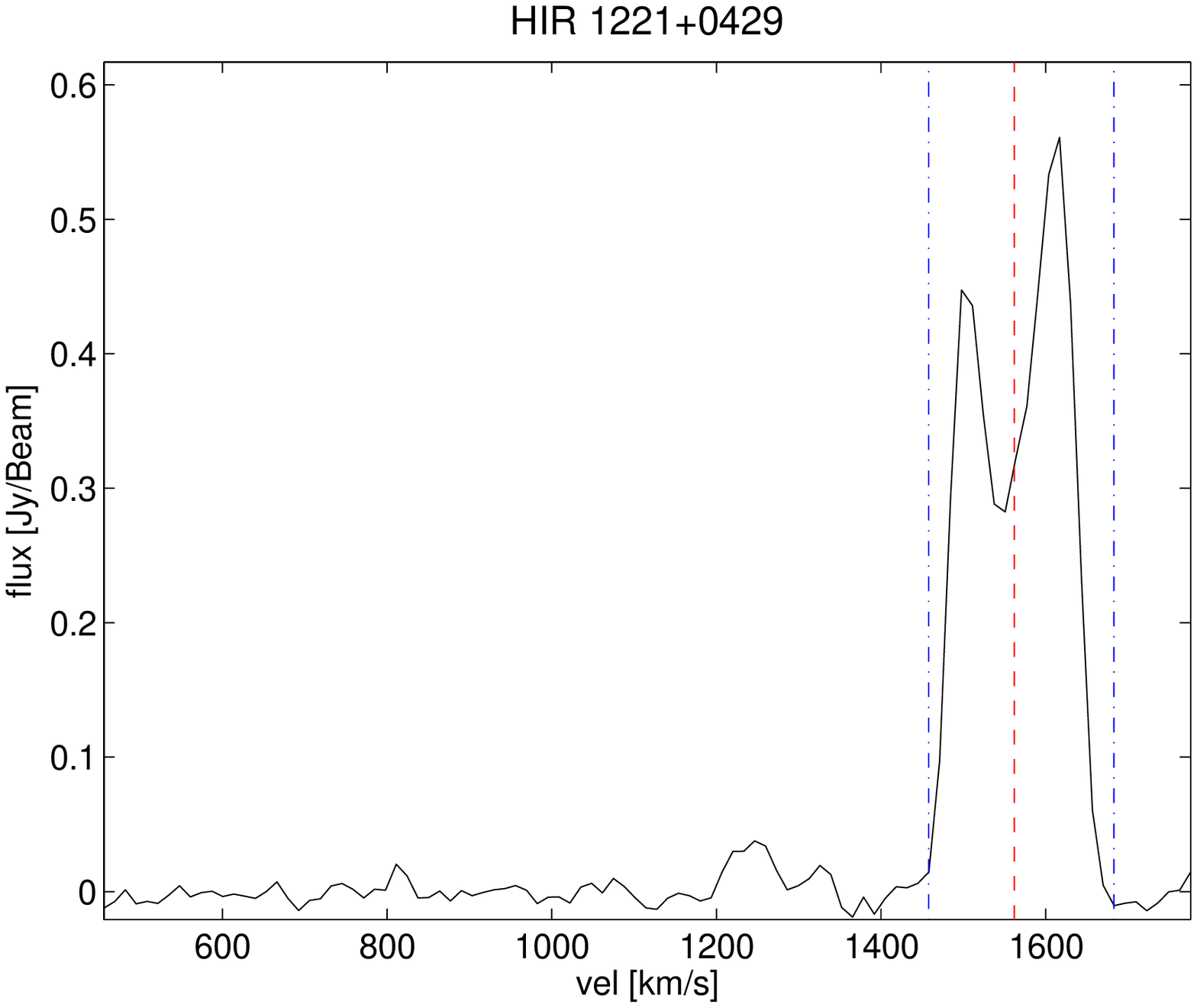}
 \includegraphics[width=0.3\textwidth]{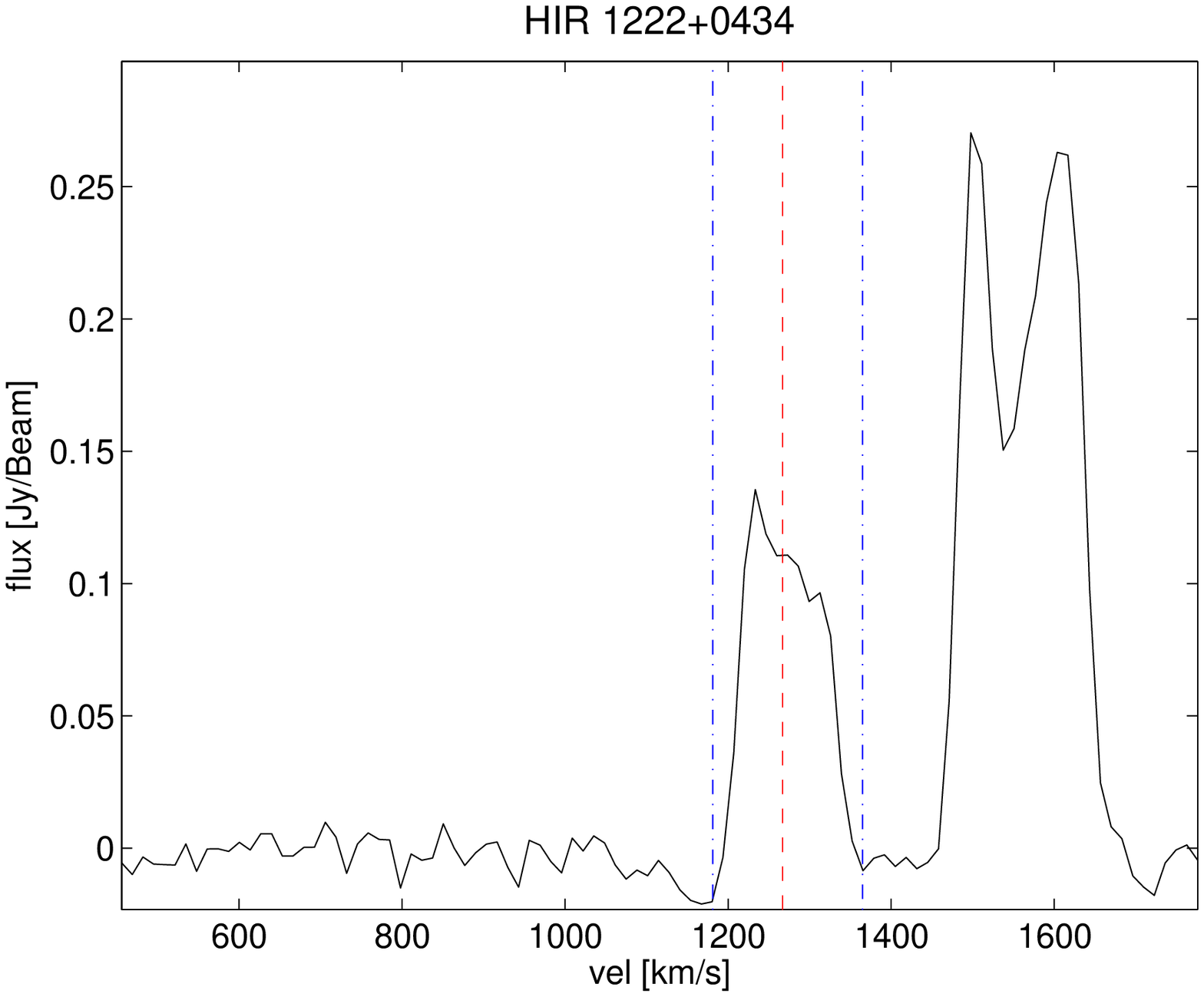}

 \end{center}                                                         
{\bf Fig~\ref{all_spectra}.} (continued)                              
                                                                      
\end{figure*}

\begin{figure*}
  \begin{center}

 \includegraphics[width=0.3\textwidth]{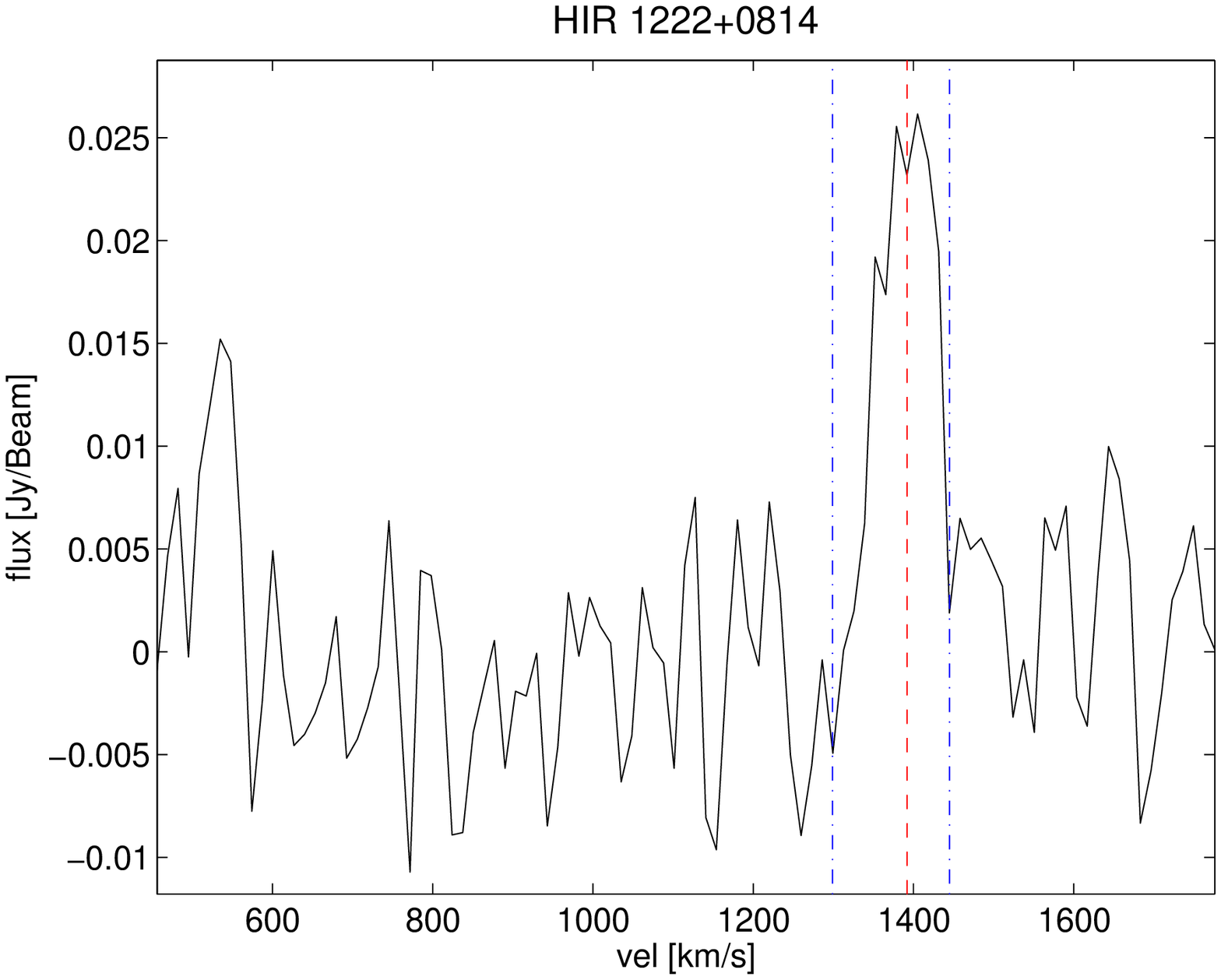}
 \includegraphics[width=0.3\textwidth]{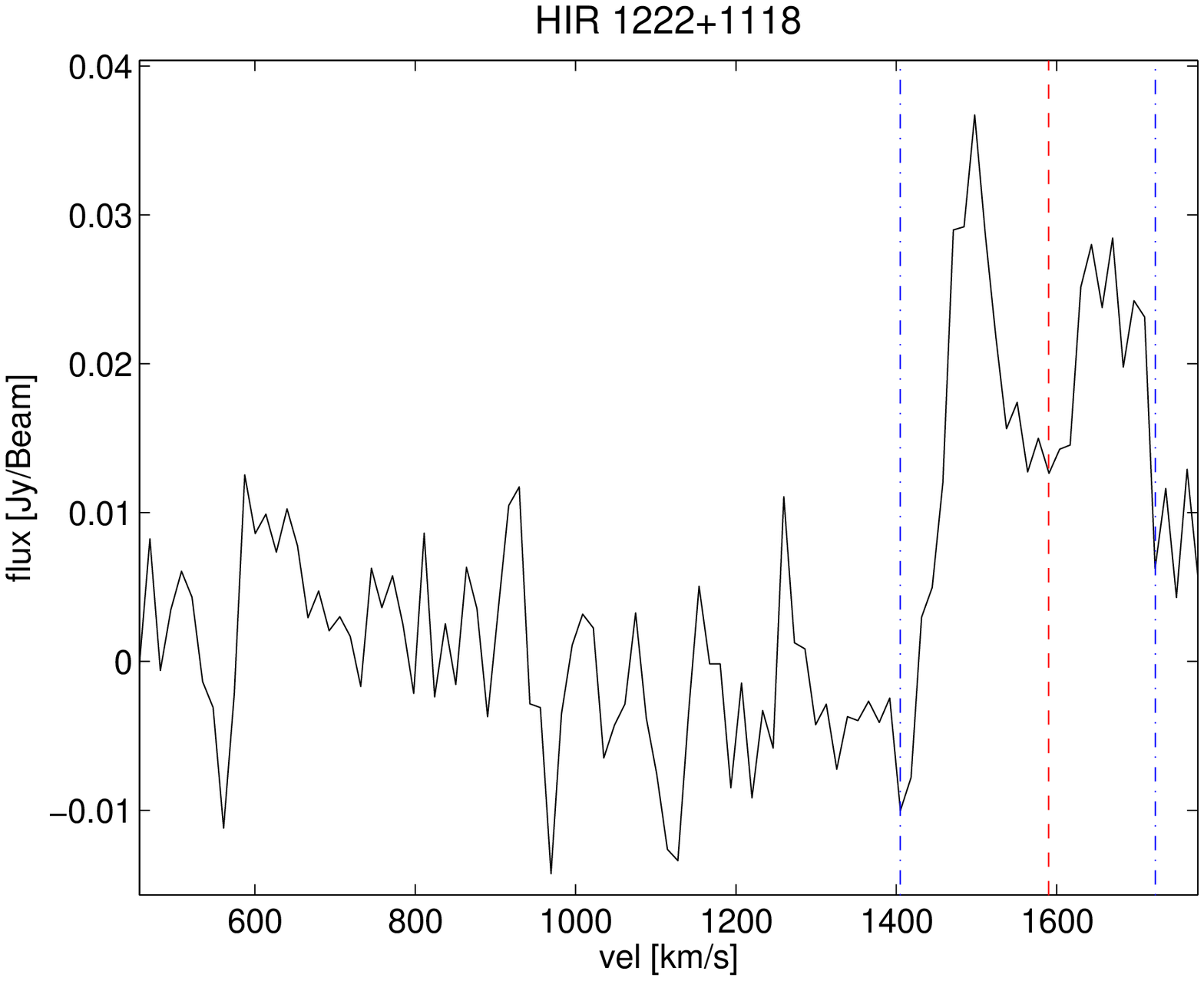}
 \includegraphics[width=0.3\textwidth]{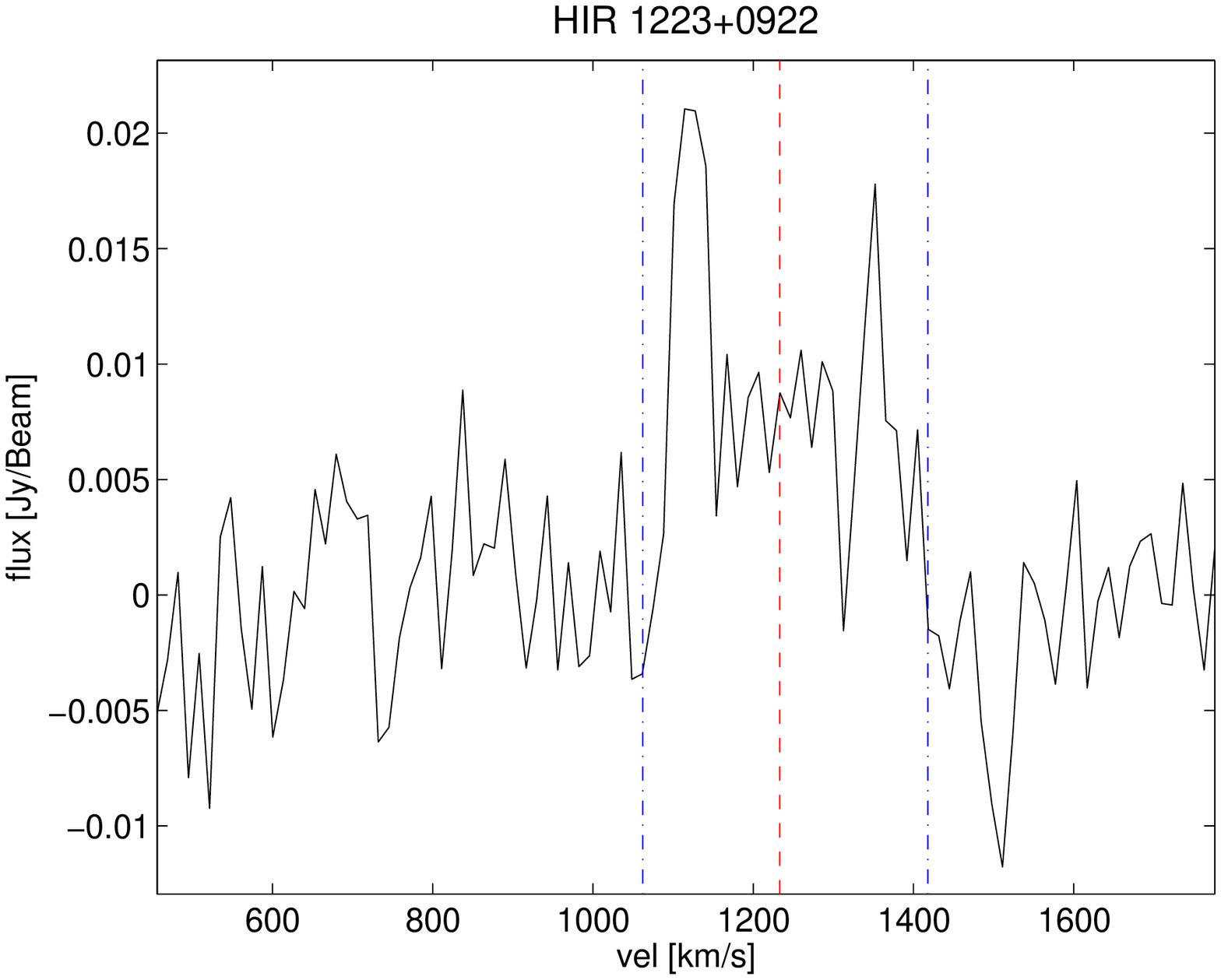}
 \includegraphics[width=0.3\textwidth]{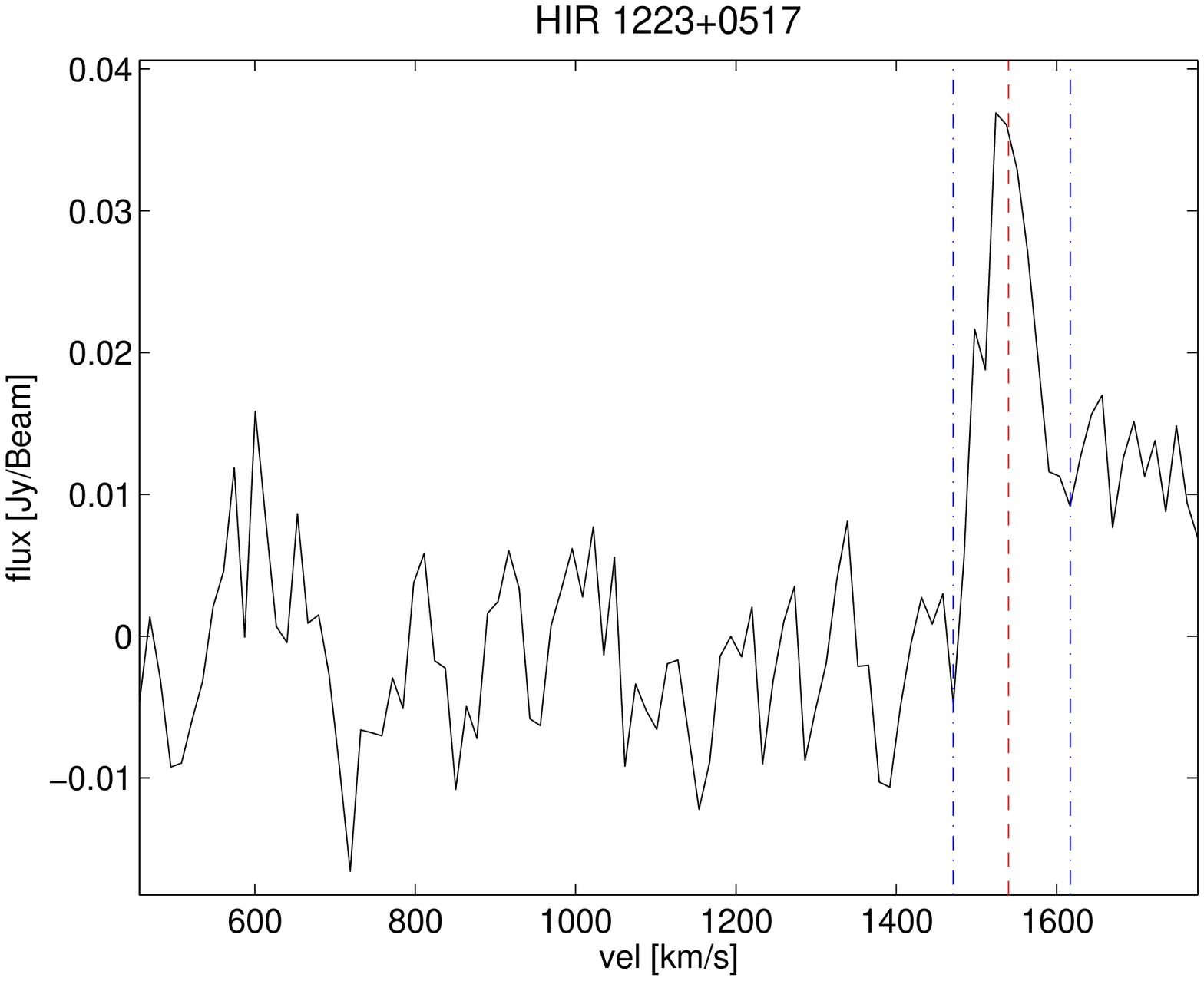}
 \includegraphics[width=0.3\textwidth]{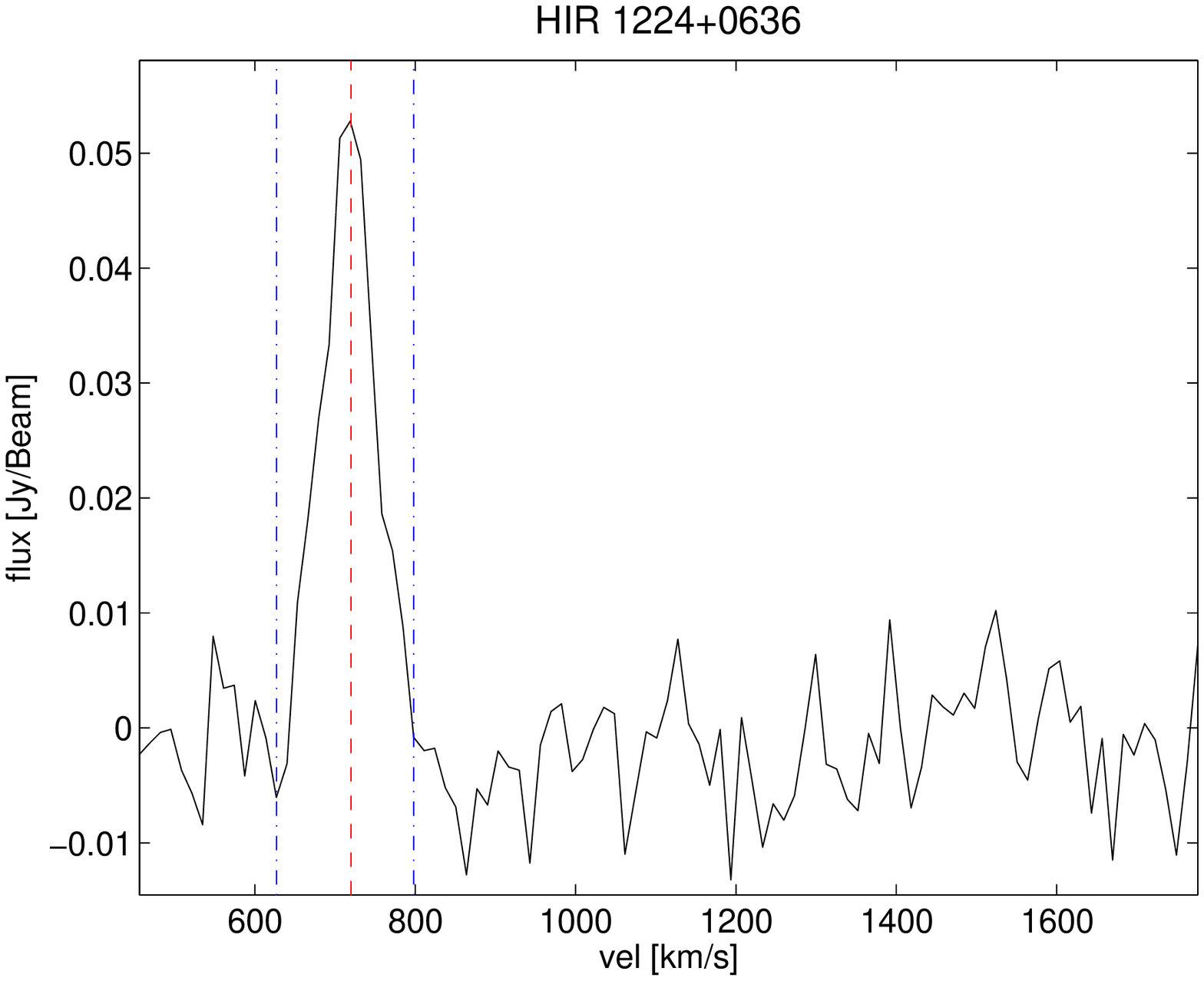}
 \includegraphics[width=0.3\textwidth]{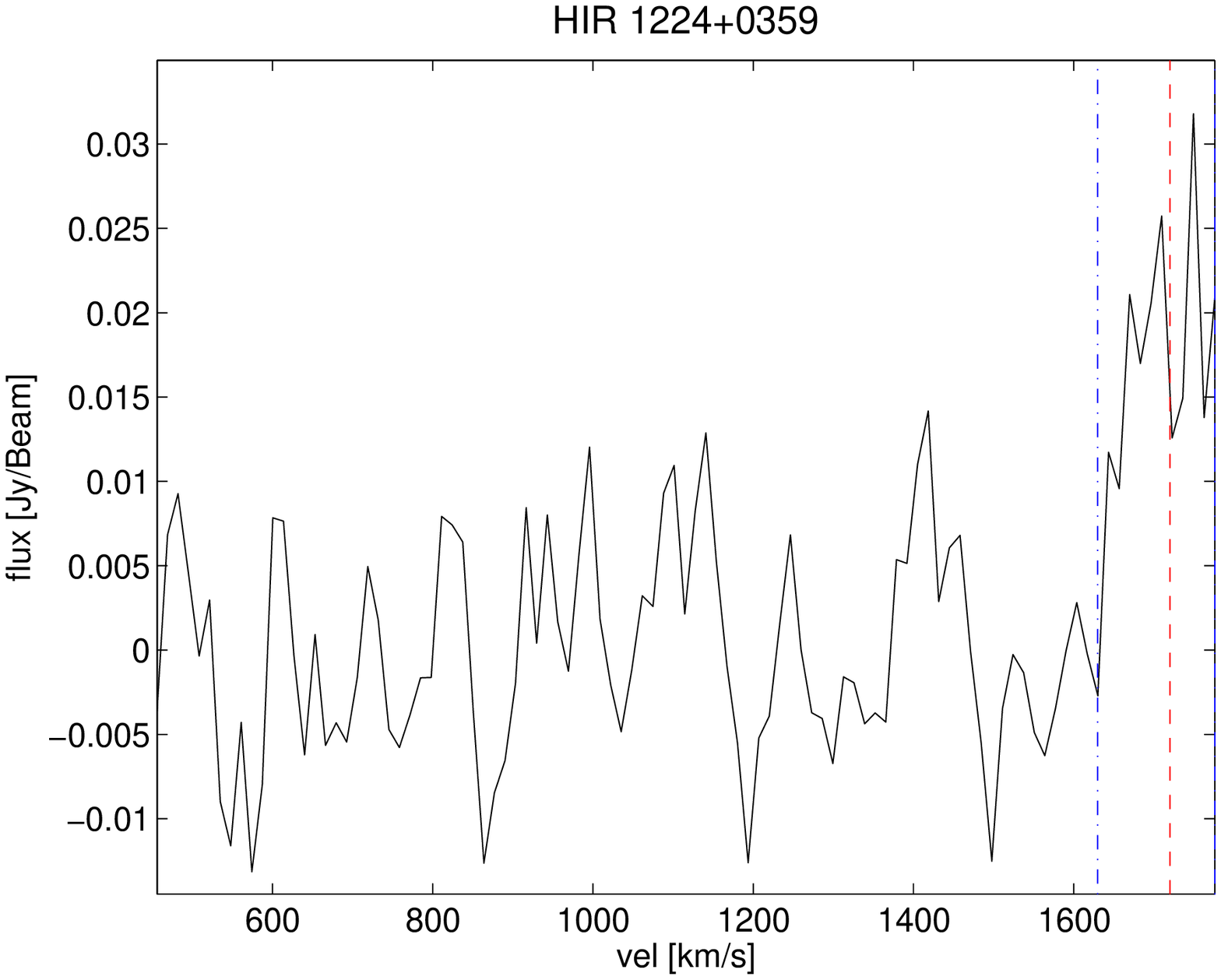}
 \includegraphics[width=0.3\textwidth]{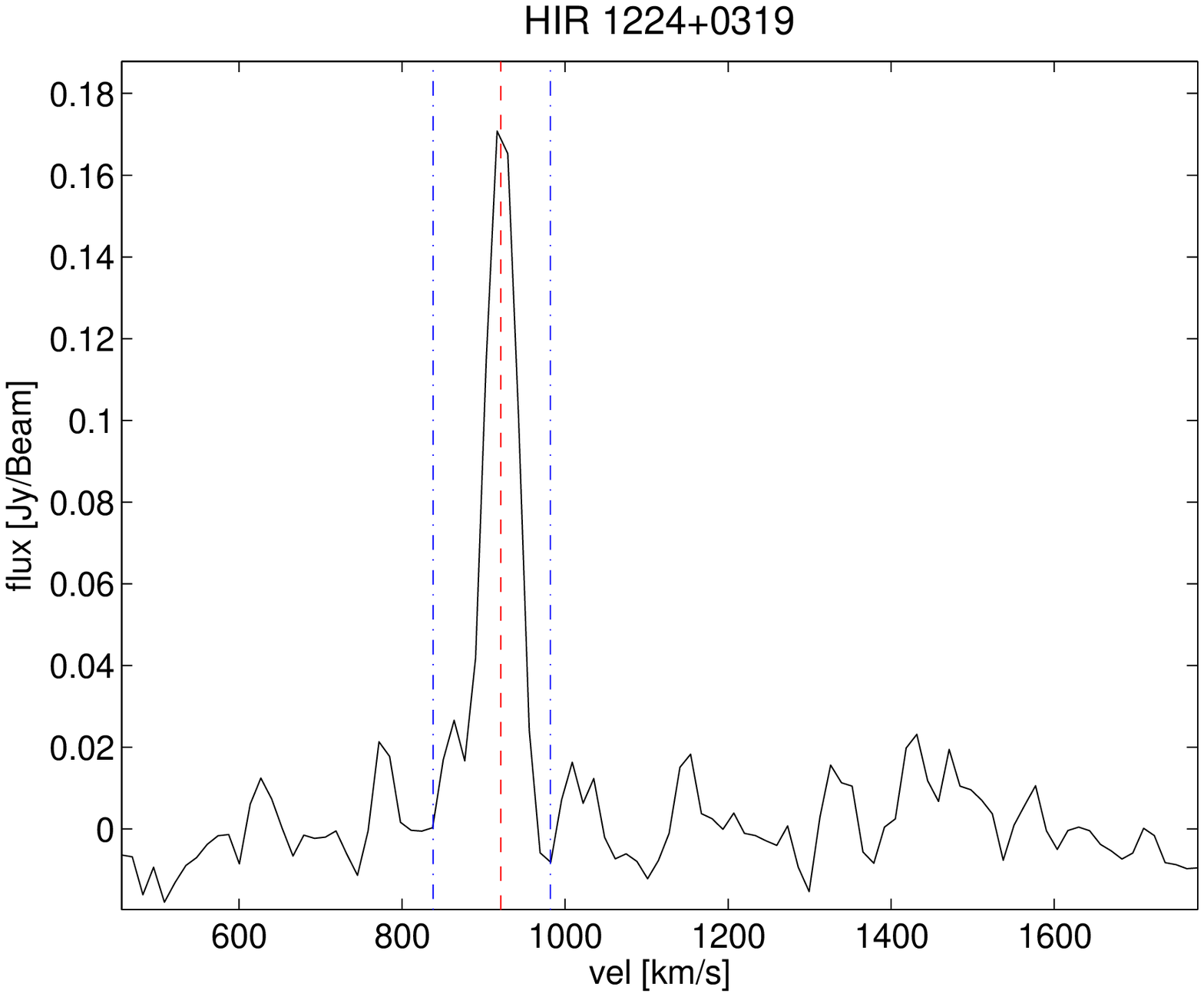}
 \includegraphics[width=0.3\textwidth]{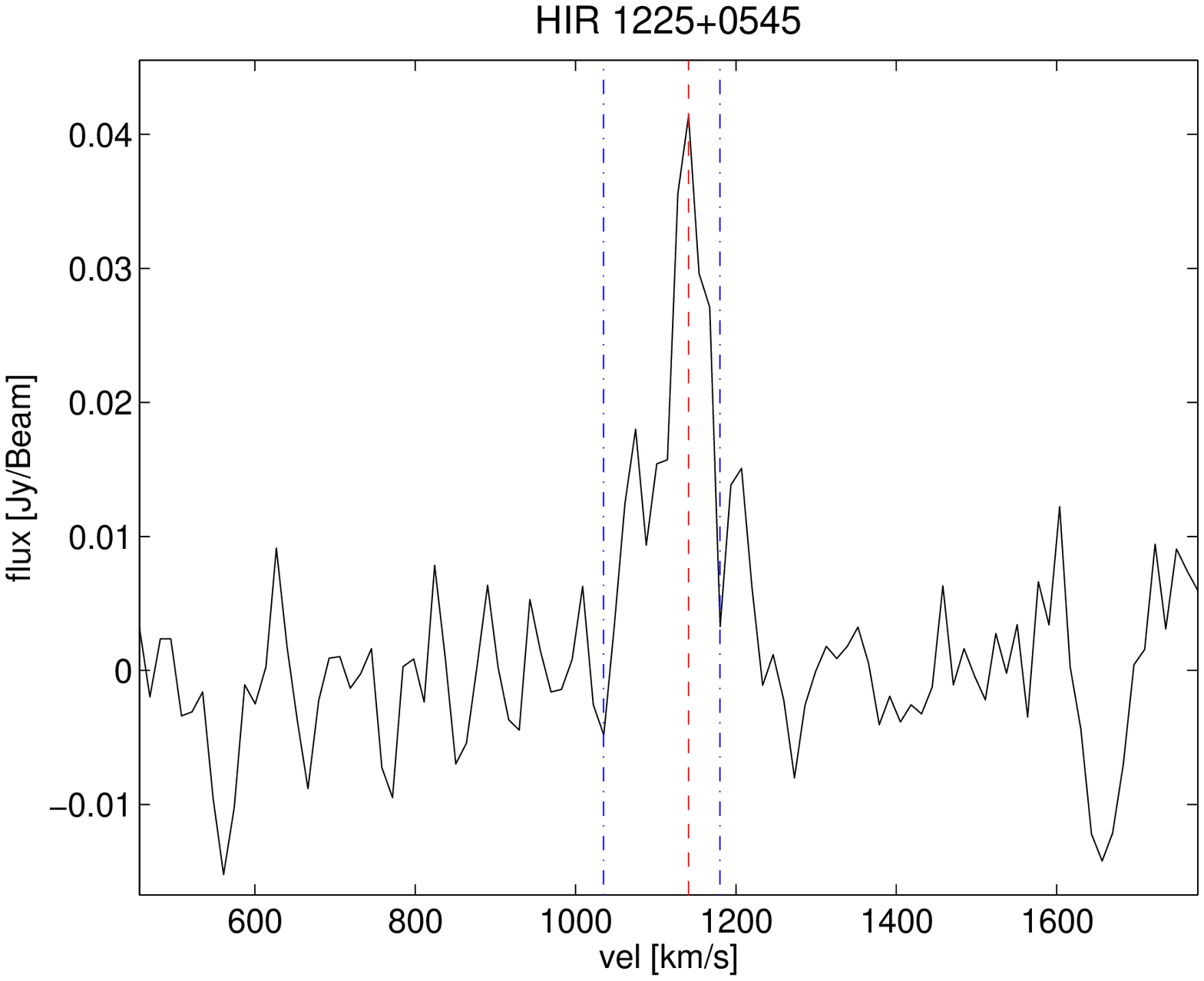}
 \includegraphics[width=0.3\textwidth]{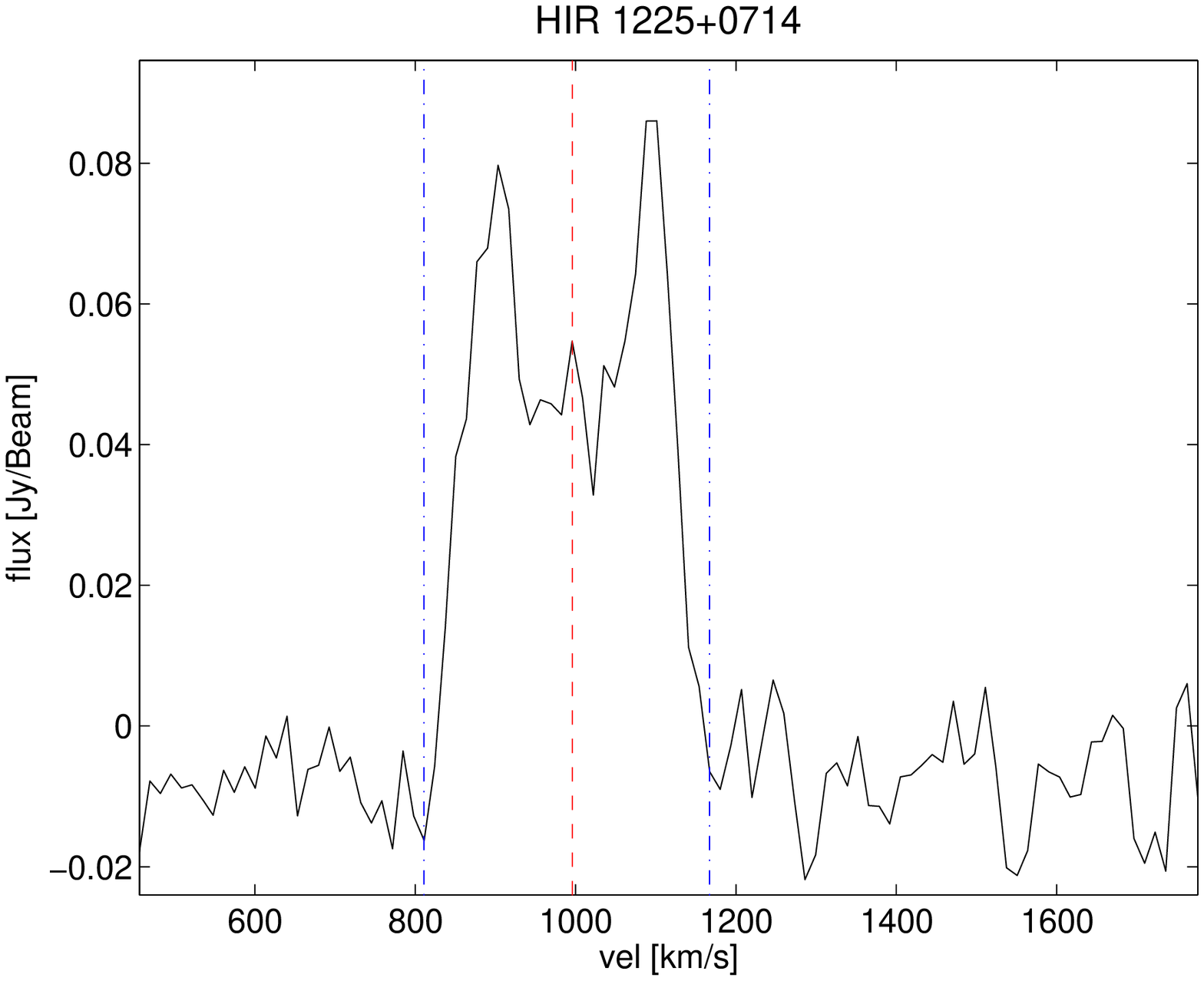}
 \includegraphics[width=0.3\textwidth]{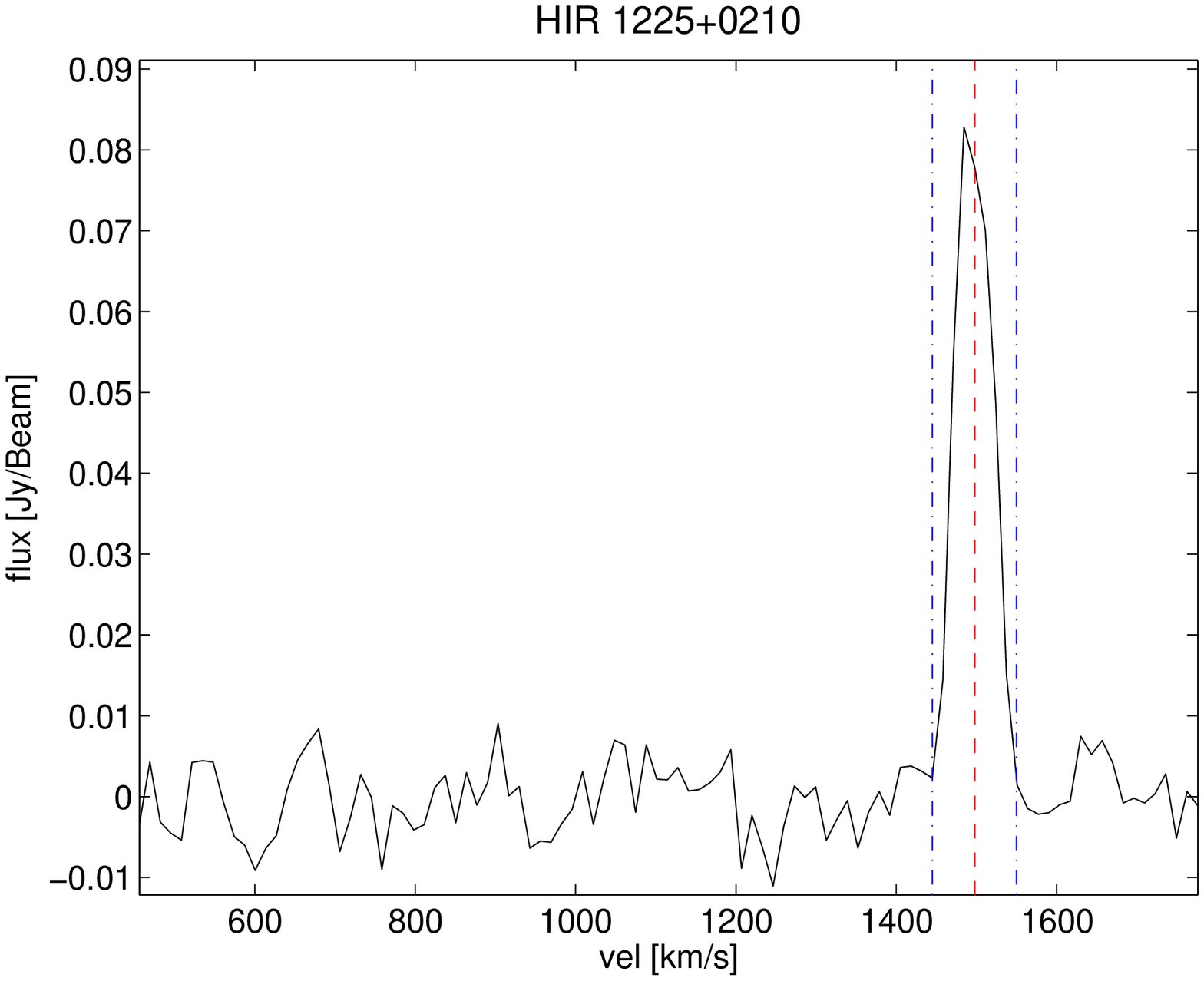}
 \includegraphics[width=0.3\textwidth]{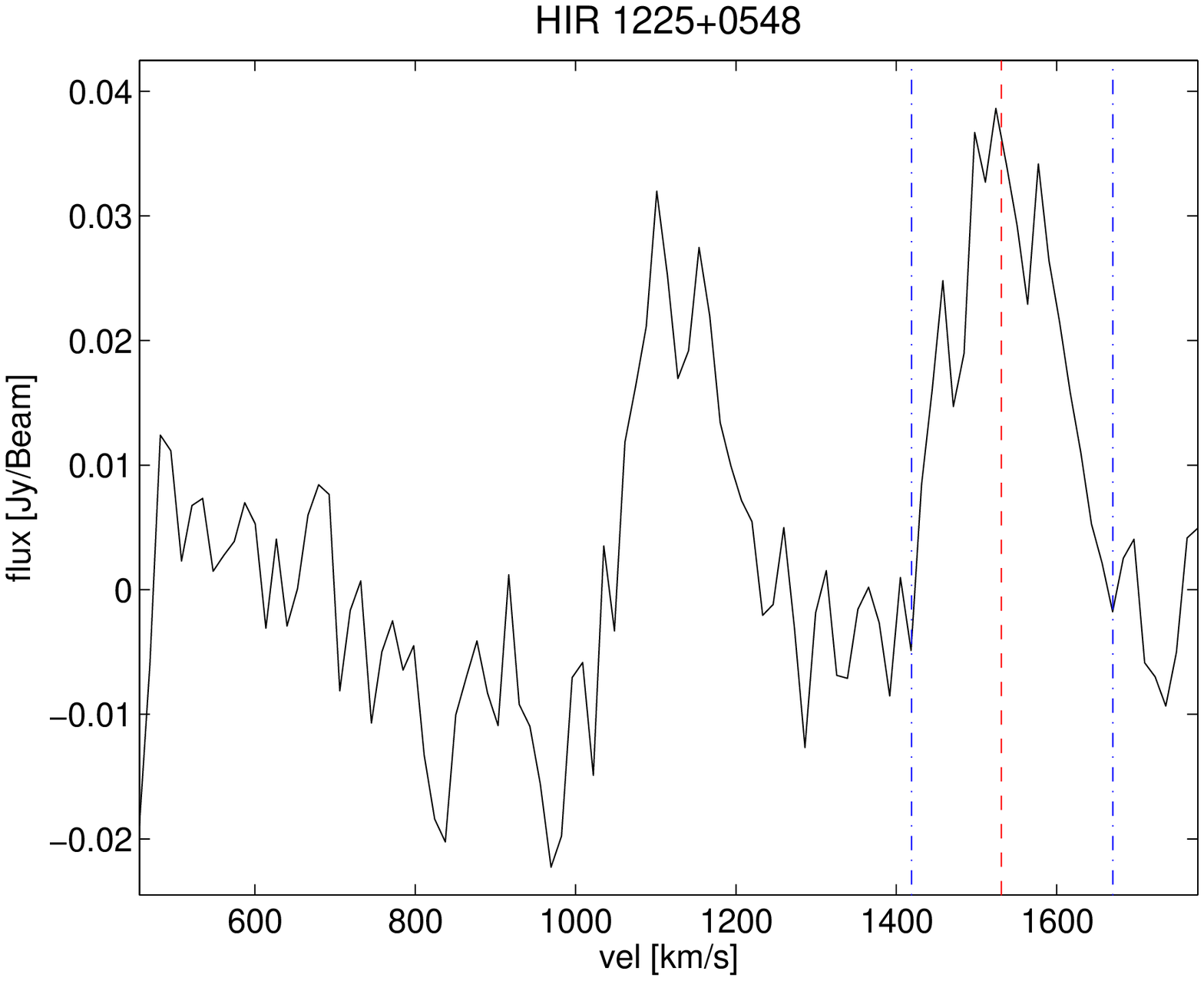}
 \includegraphics[width=0.3\textwidth]{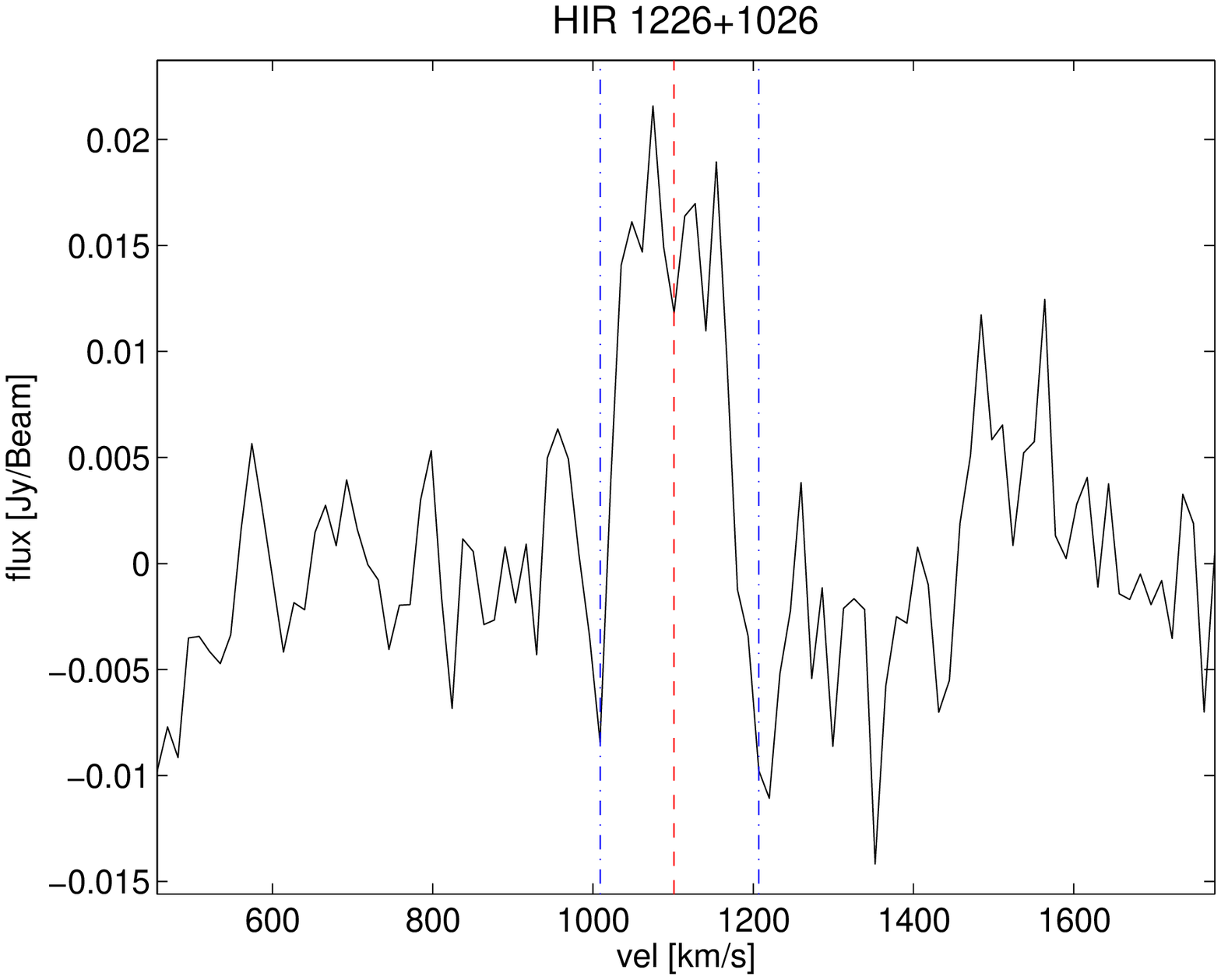}
 \includegraphics[width=0.3\textwidth]{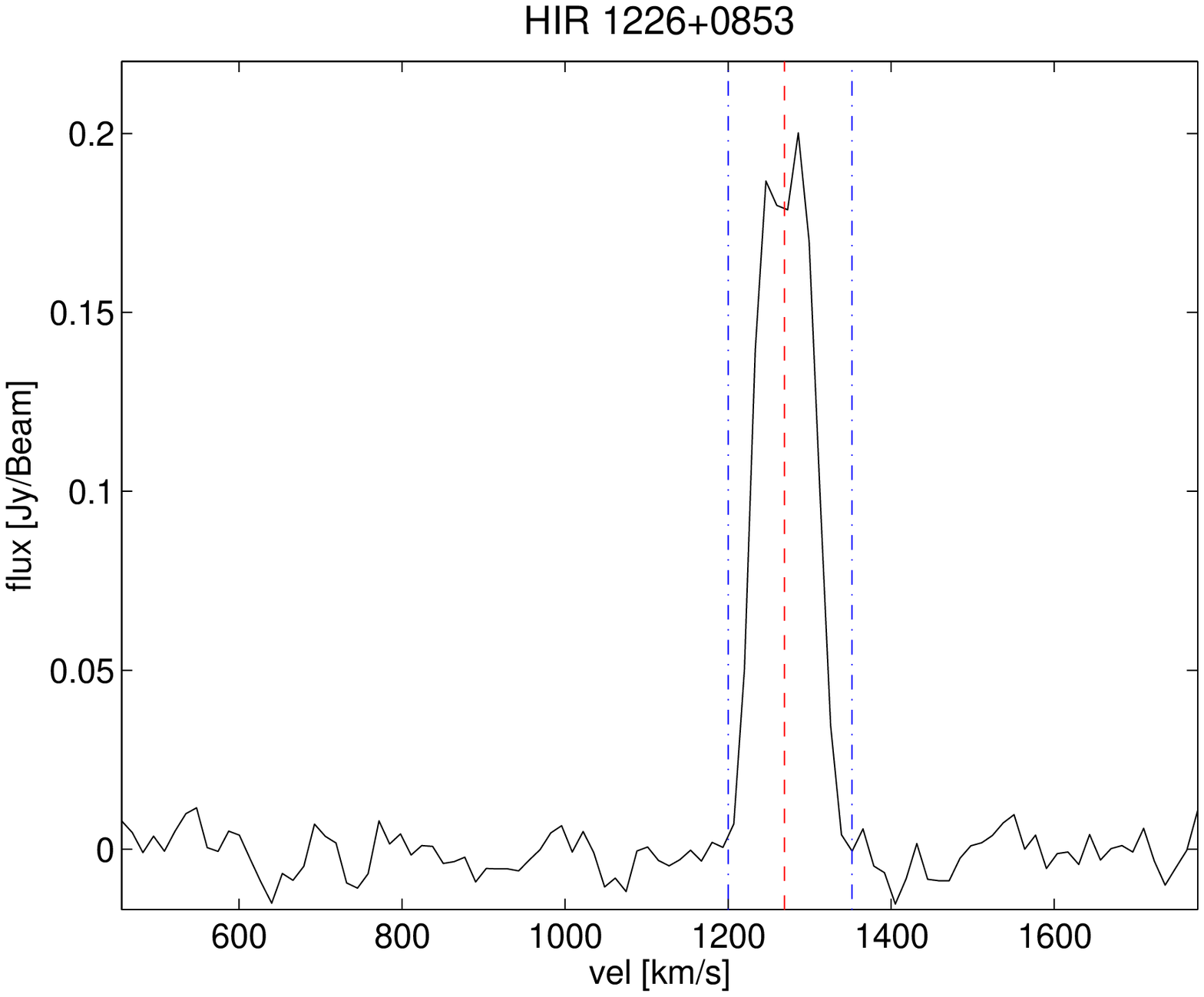}
 \includegraphics[width=0.3\textwidth]{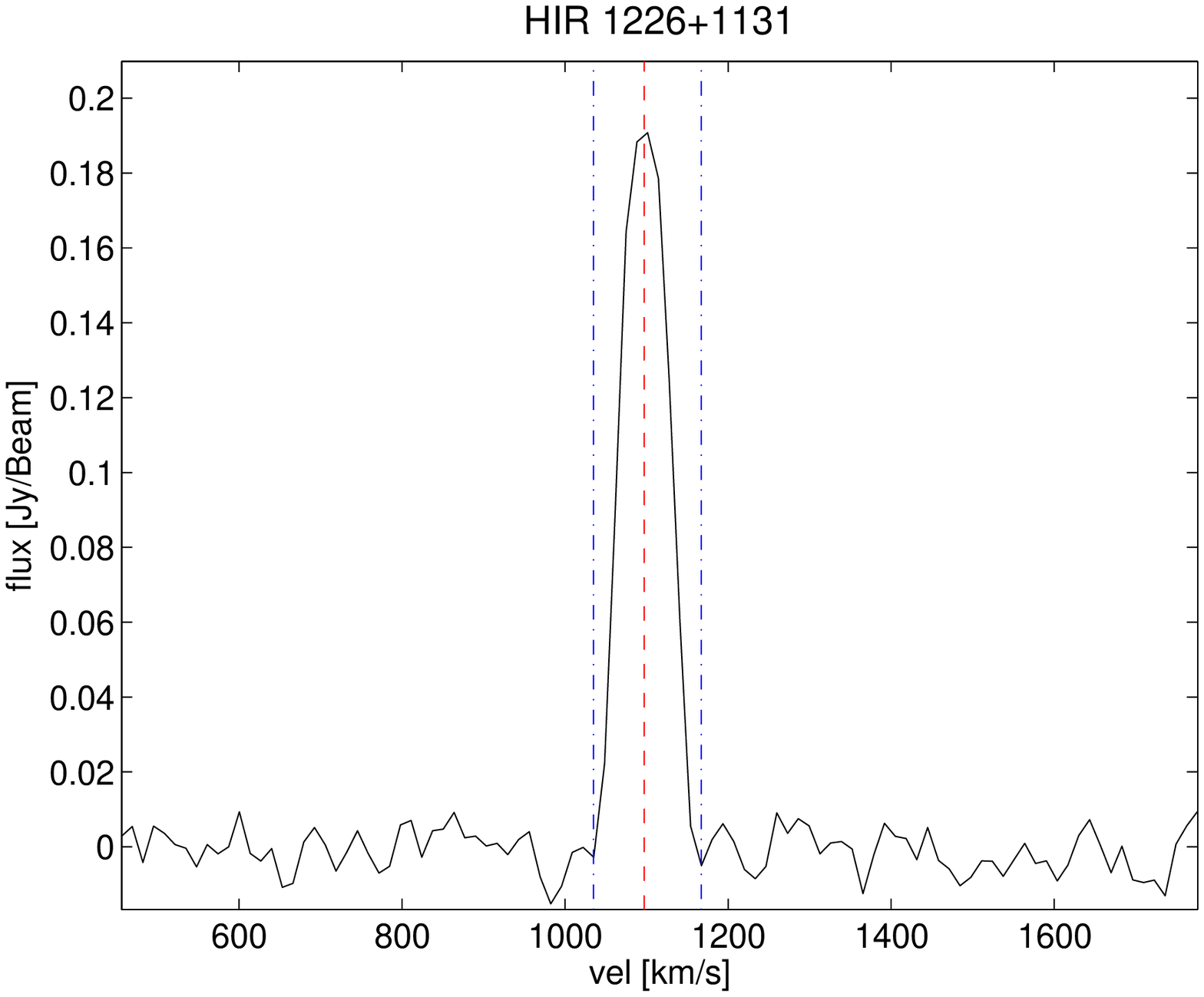}
 \includegraphics[width=0.3\textwidth]{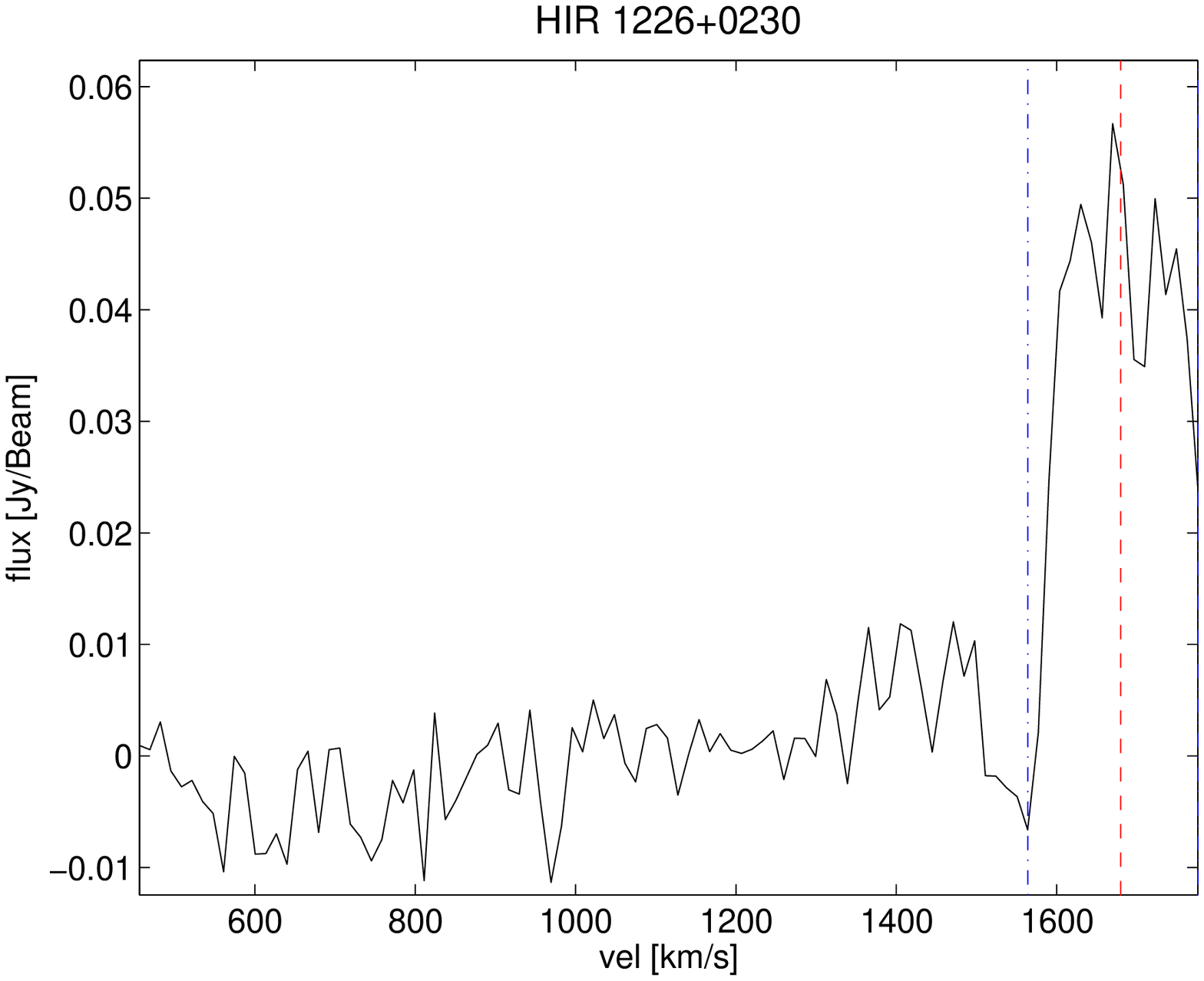}

 \end{center}                                                         
{\bf Fig~\ref{all_spectra}.} (continued)                              
                                                                      
\end{figure*}

\begin{figure*}
  \begin{center}

 \includegraphics[width=0.3\textwidth]{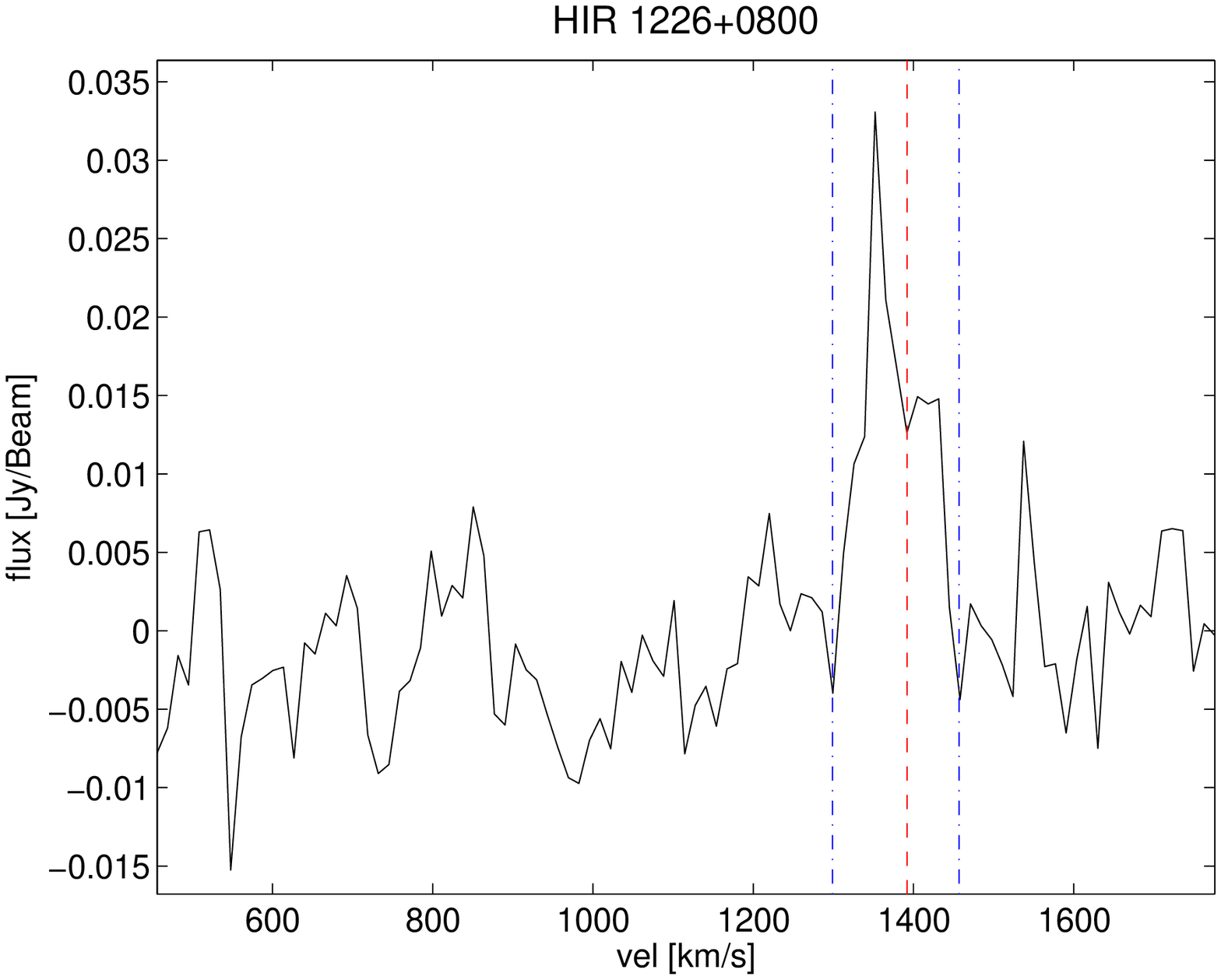}
 \includegraphics[width=0.3\textwidth]{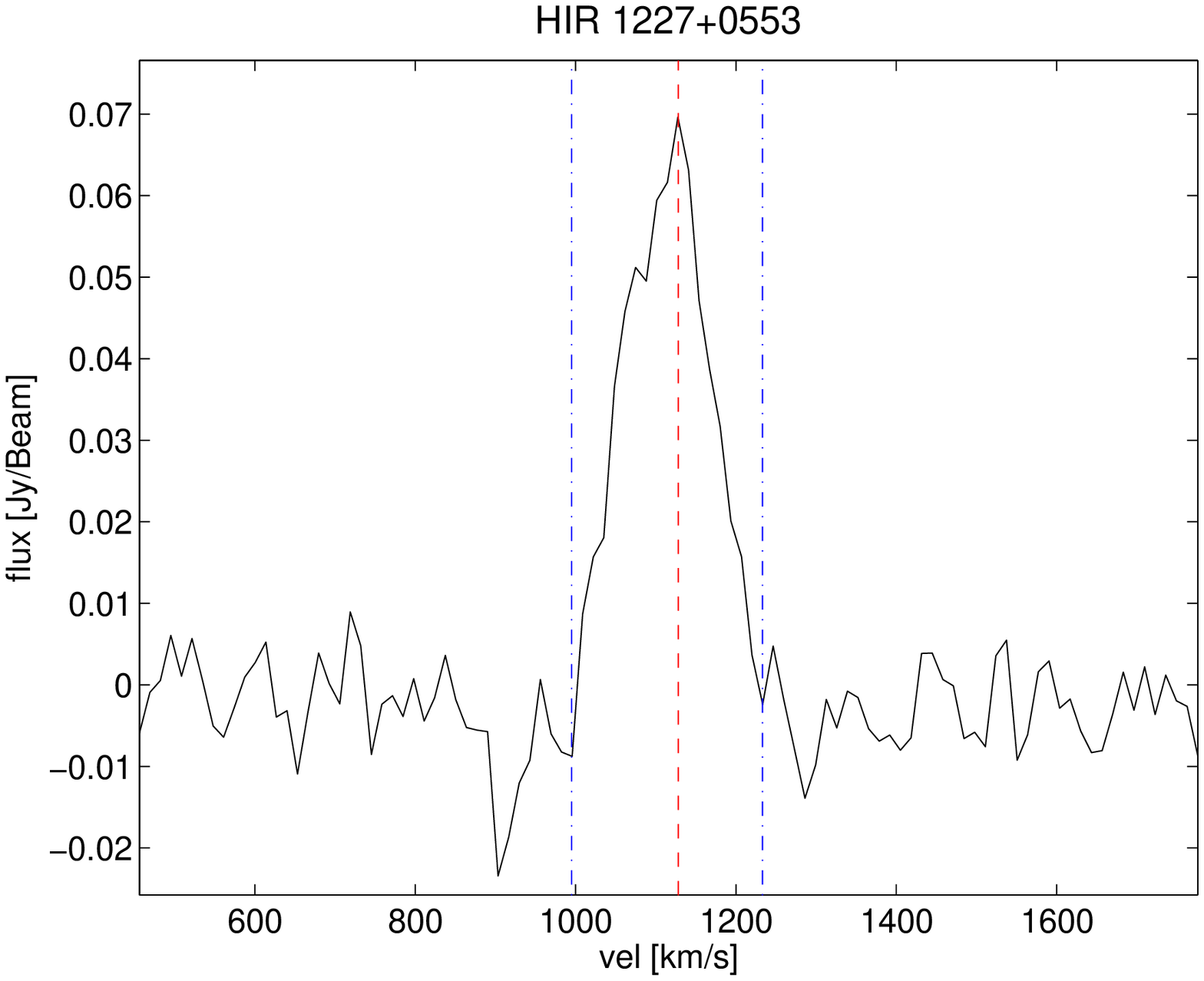}
 \includegraphics[width=0.3\textwidth]{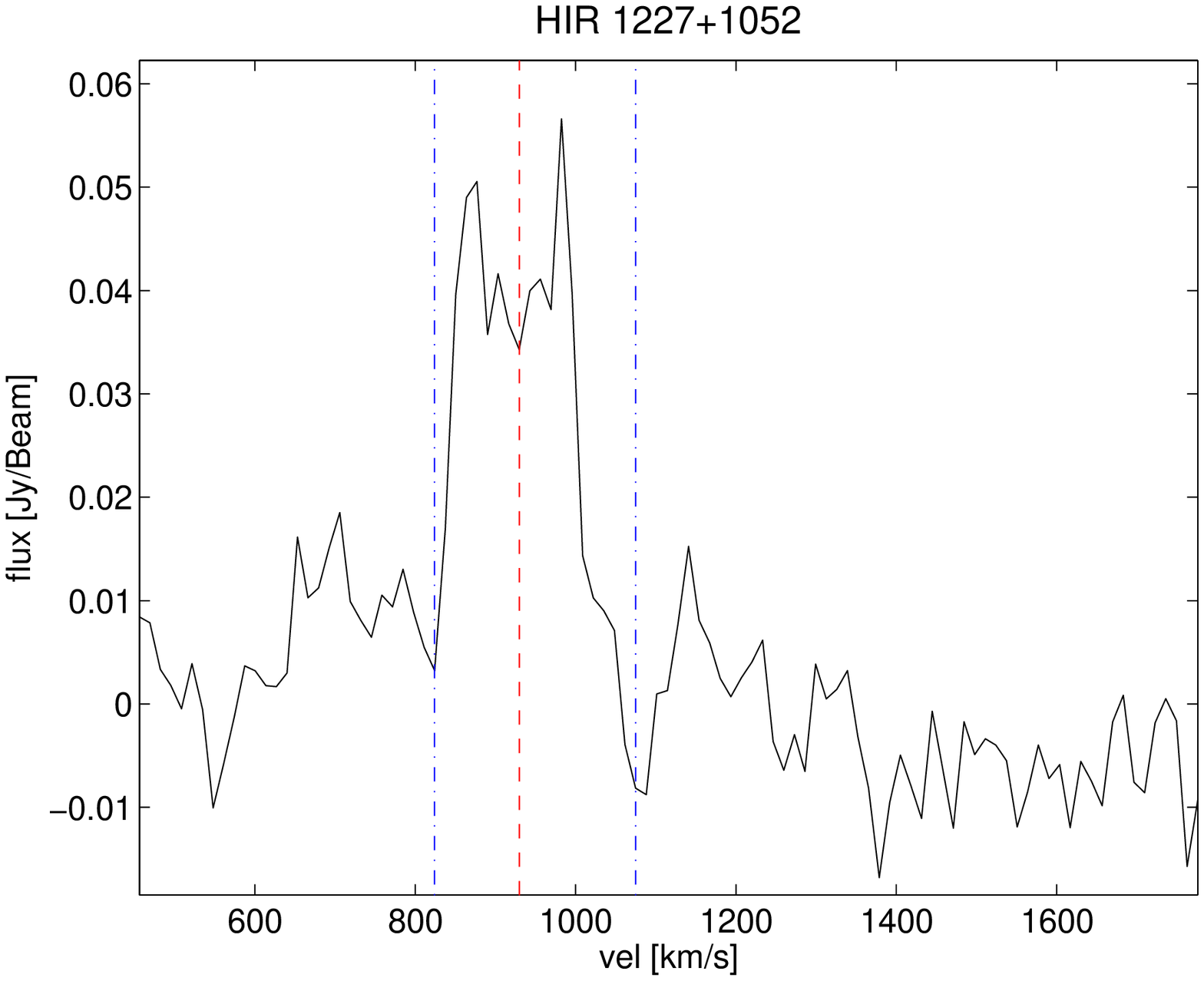}
 \includegraphics[width=0.3\textwidth]{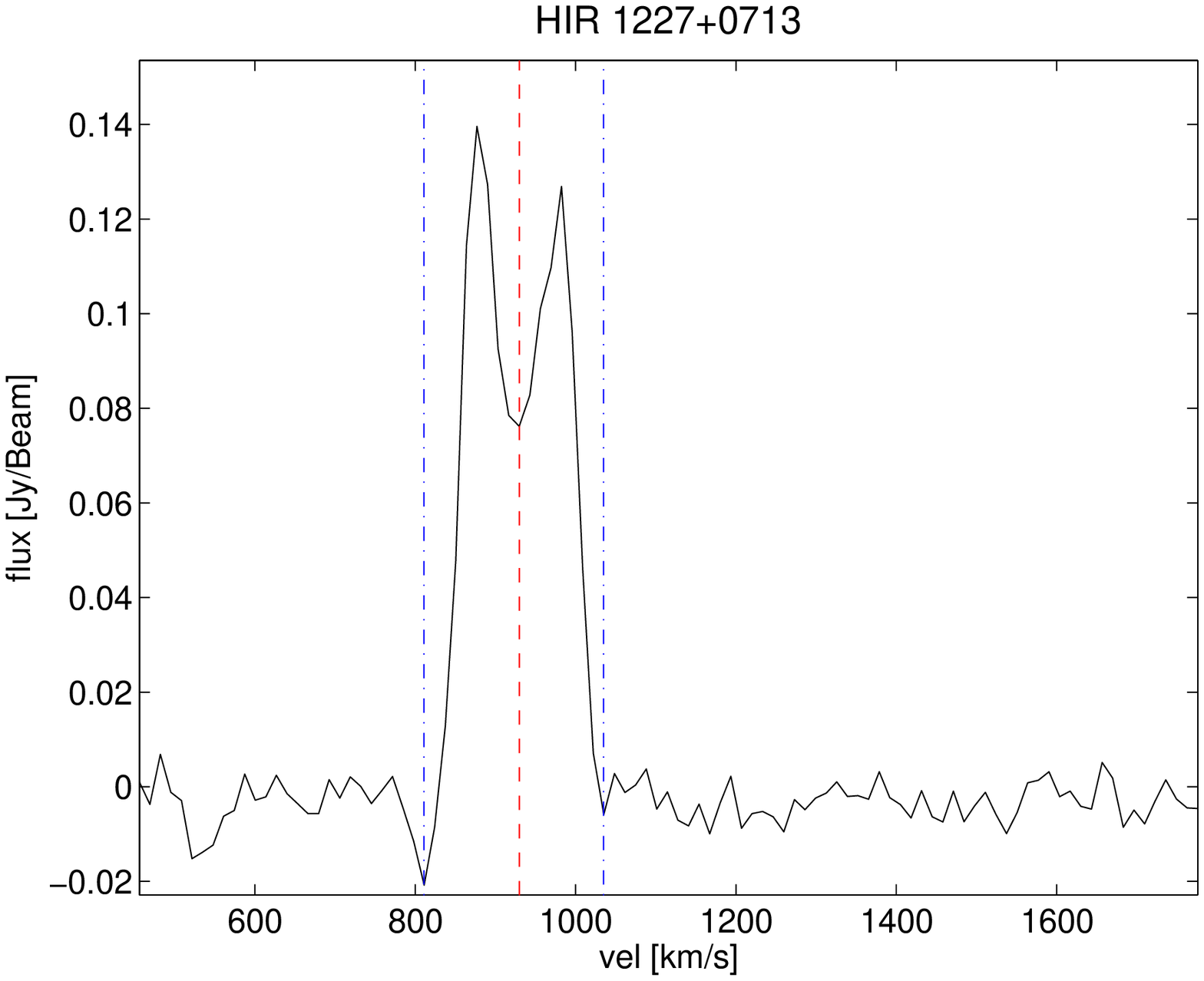}
 \includegraphics[width=0.3\textwidth]{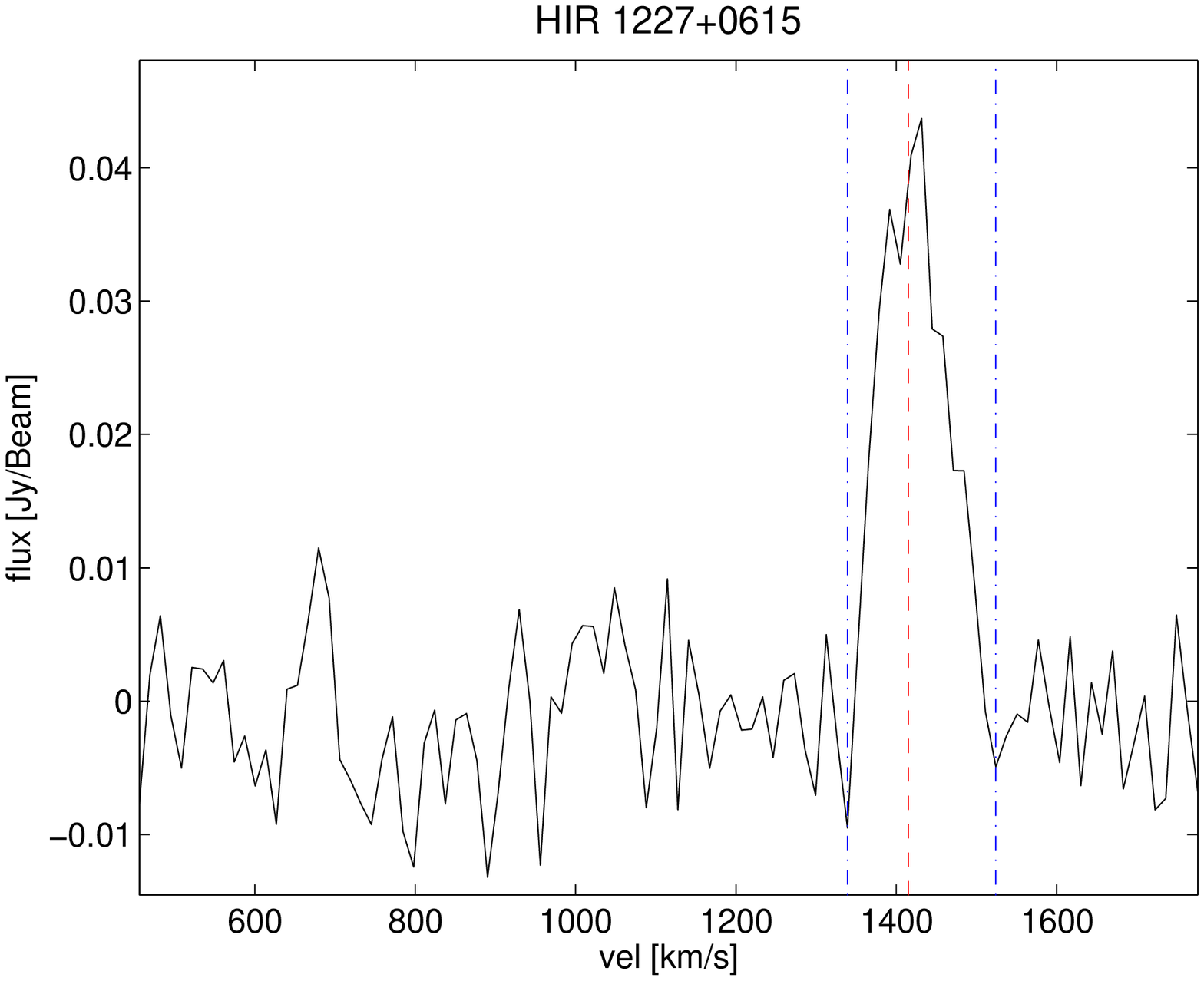}
 \includegraphics[width=0.3\textwidth]{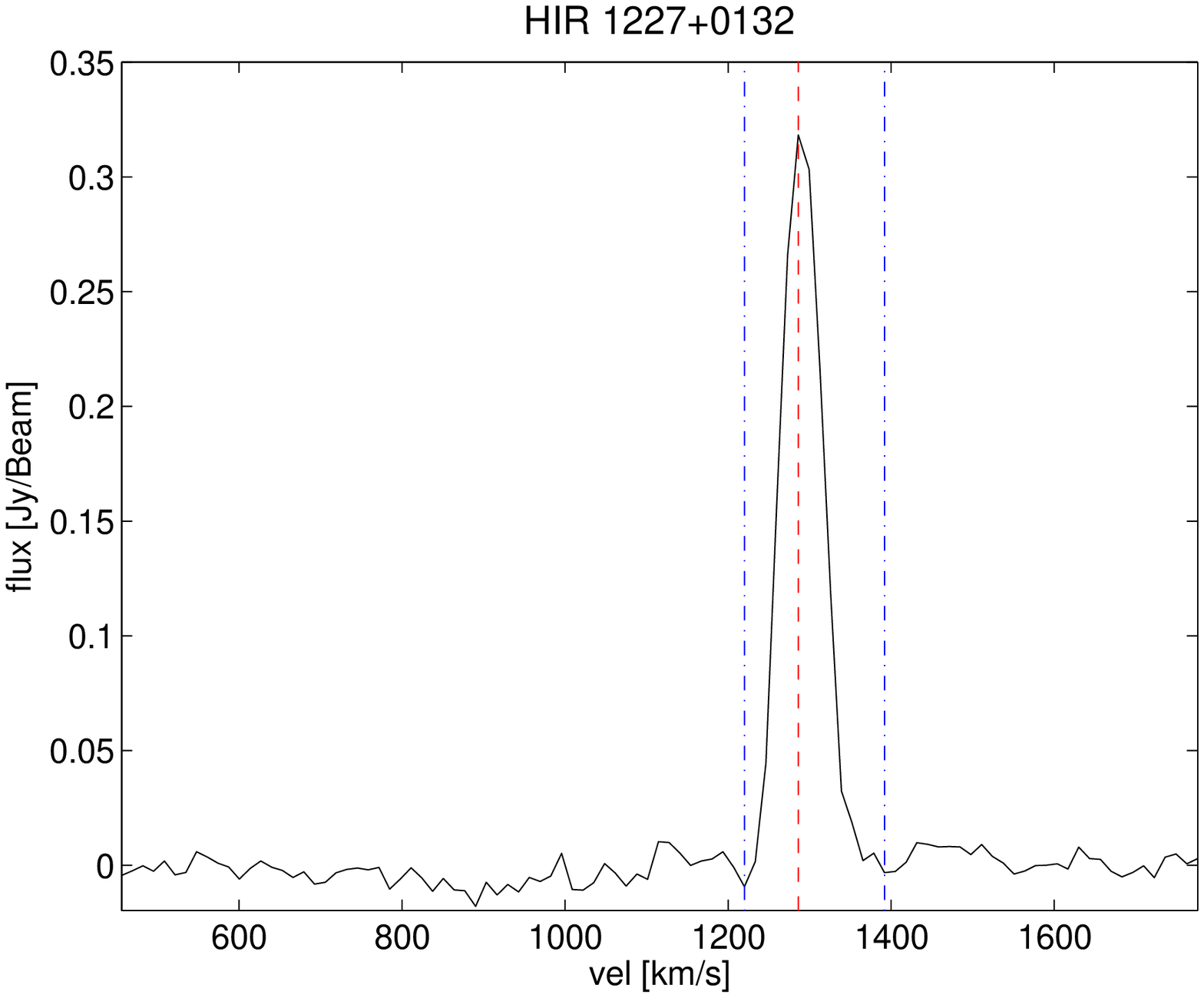}
 \includegraphics[width=0.3\textwidth]{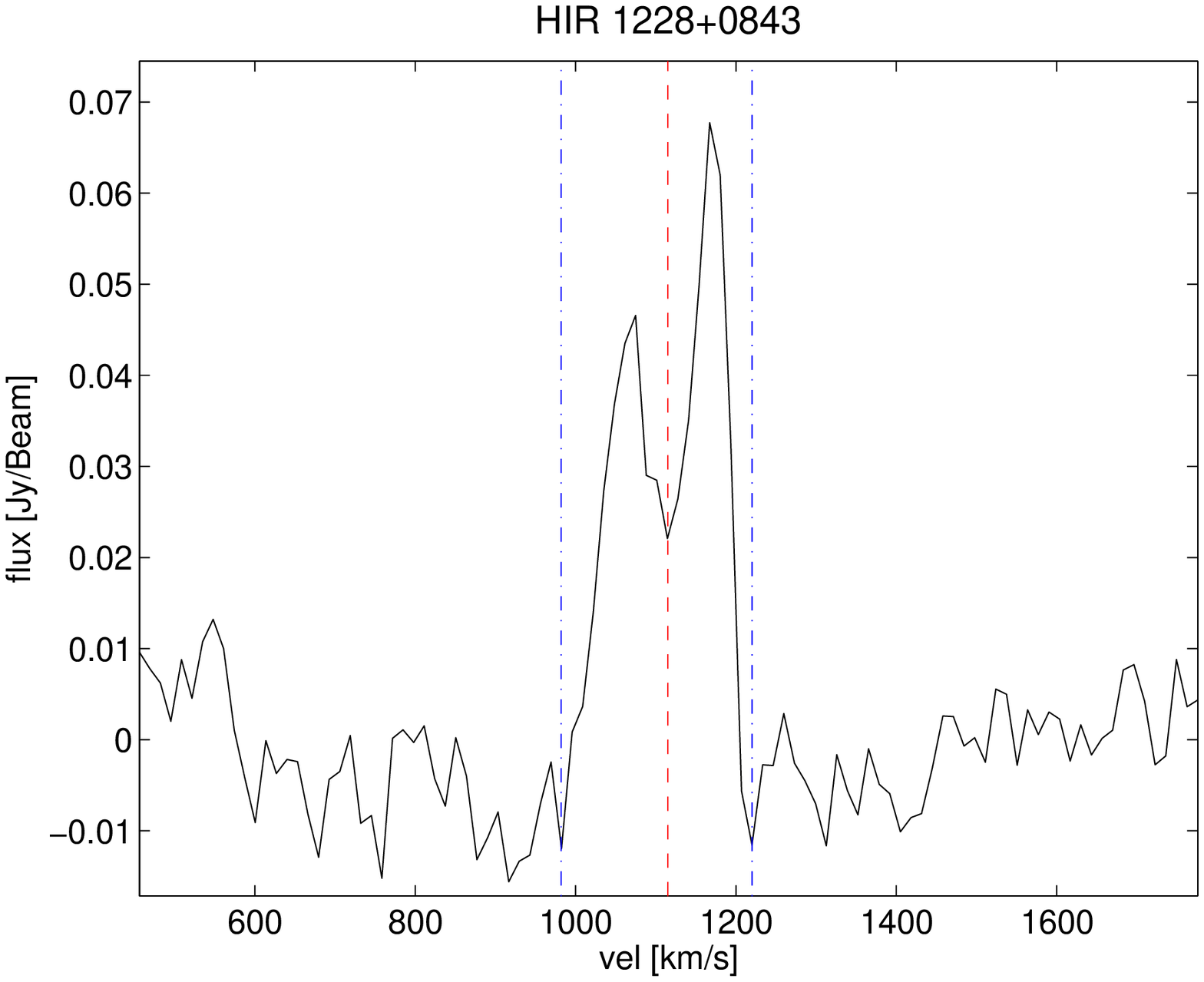}
 \includegraphics[width=0.3\textwidth]{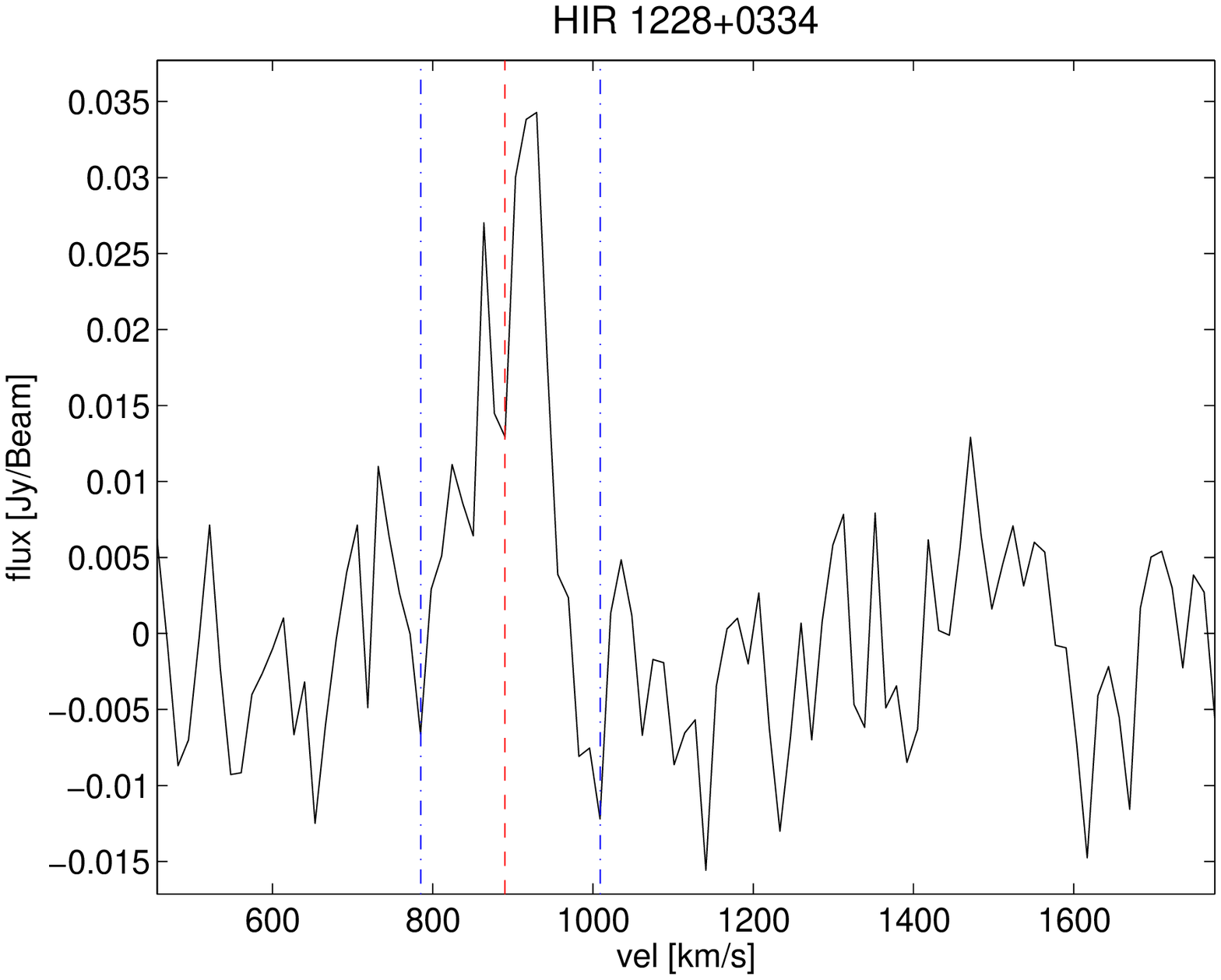}
 \includegraphics[width=0.3\textwidth]{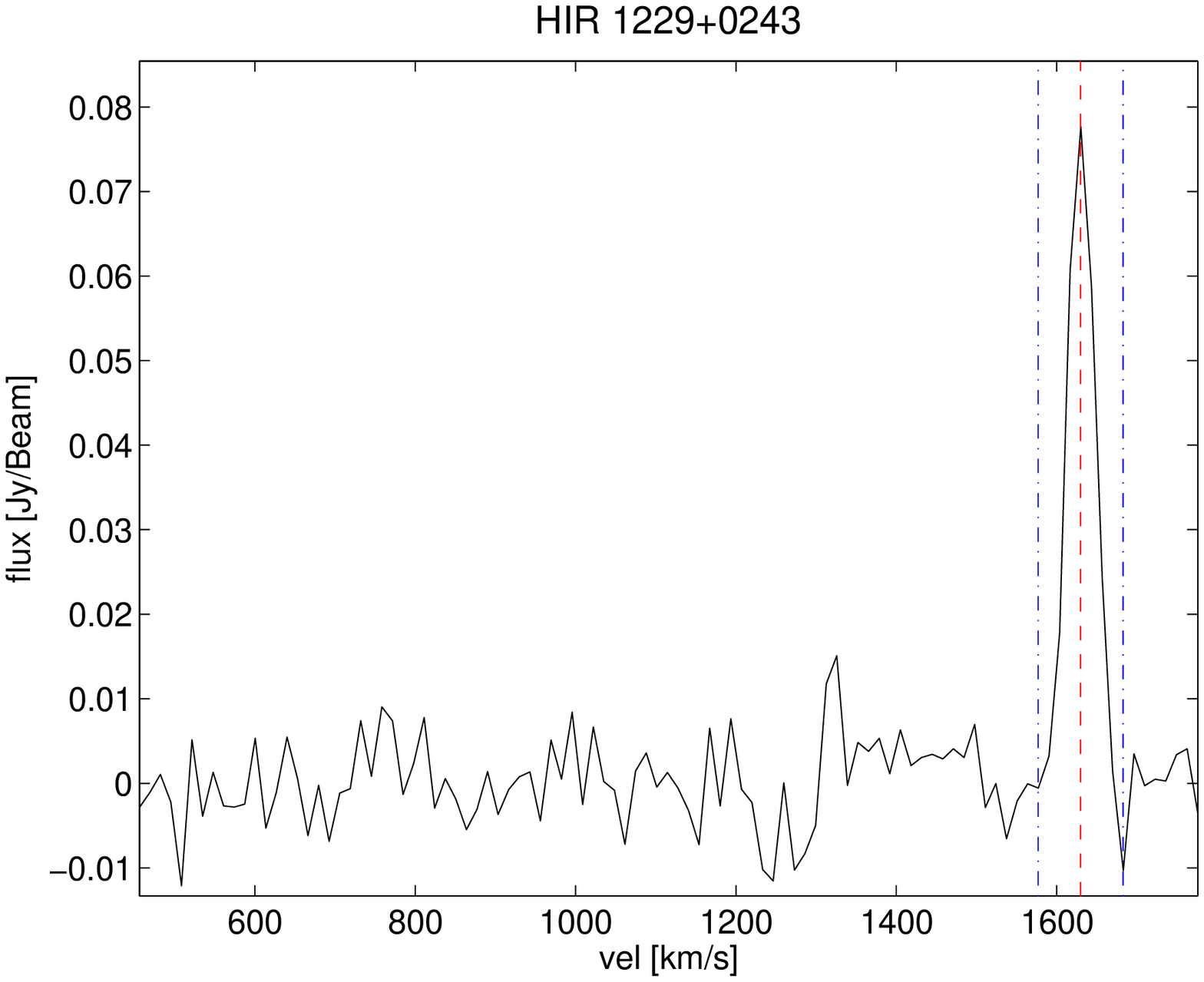}
 \includegraphics[width=0.3\textwidth]{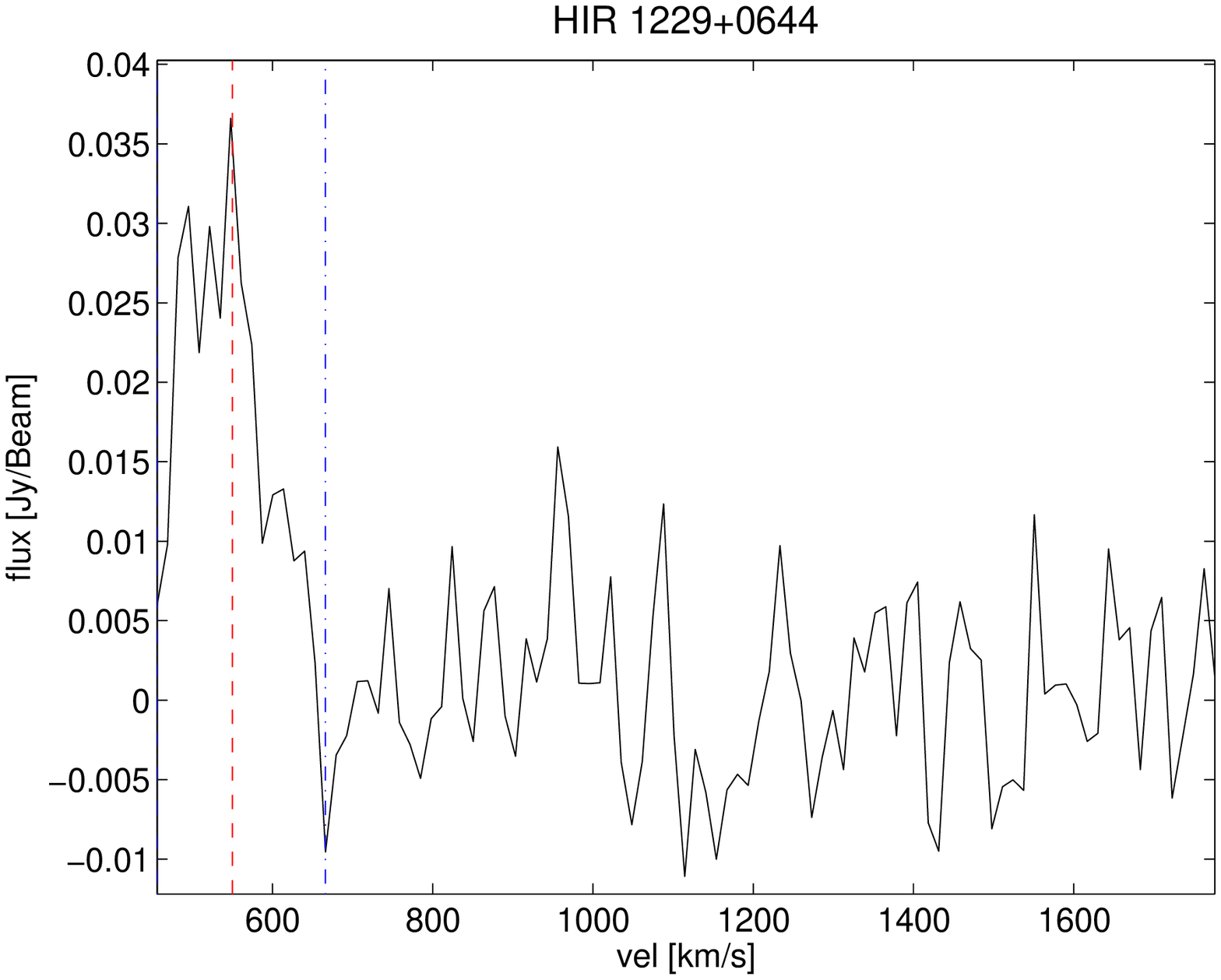}
 \includegraphics[width=0.3\textwidth]{1230+0013_spec.eps}
 \includegraphics[width=0.3\textwidth]{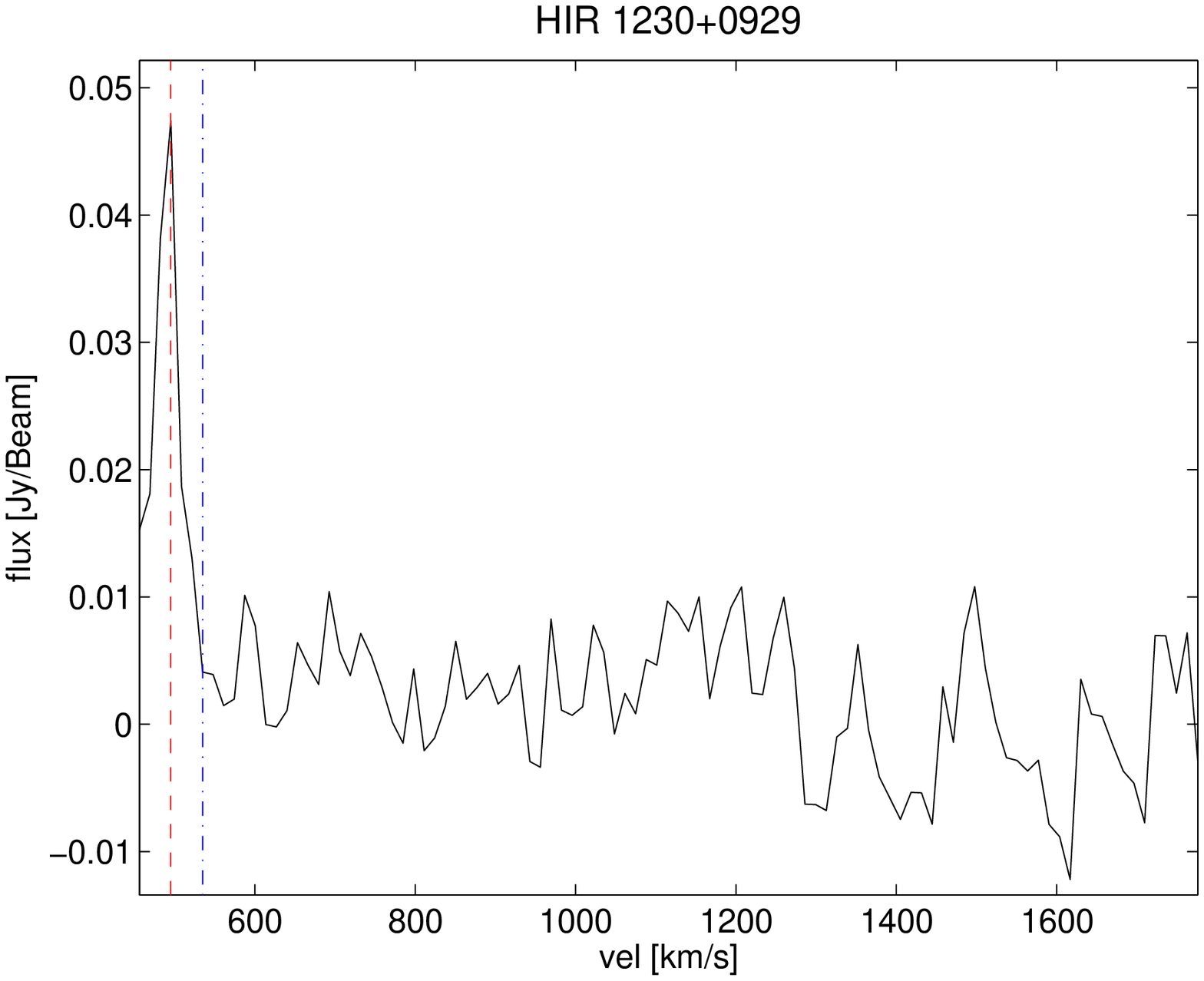}
 \includegraphics[width=0.3\textwidth]{1231+0145_spec.eps}
 \includegraphics[width=0.3\textwidth]{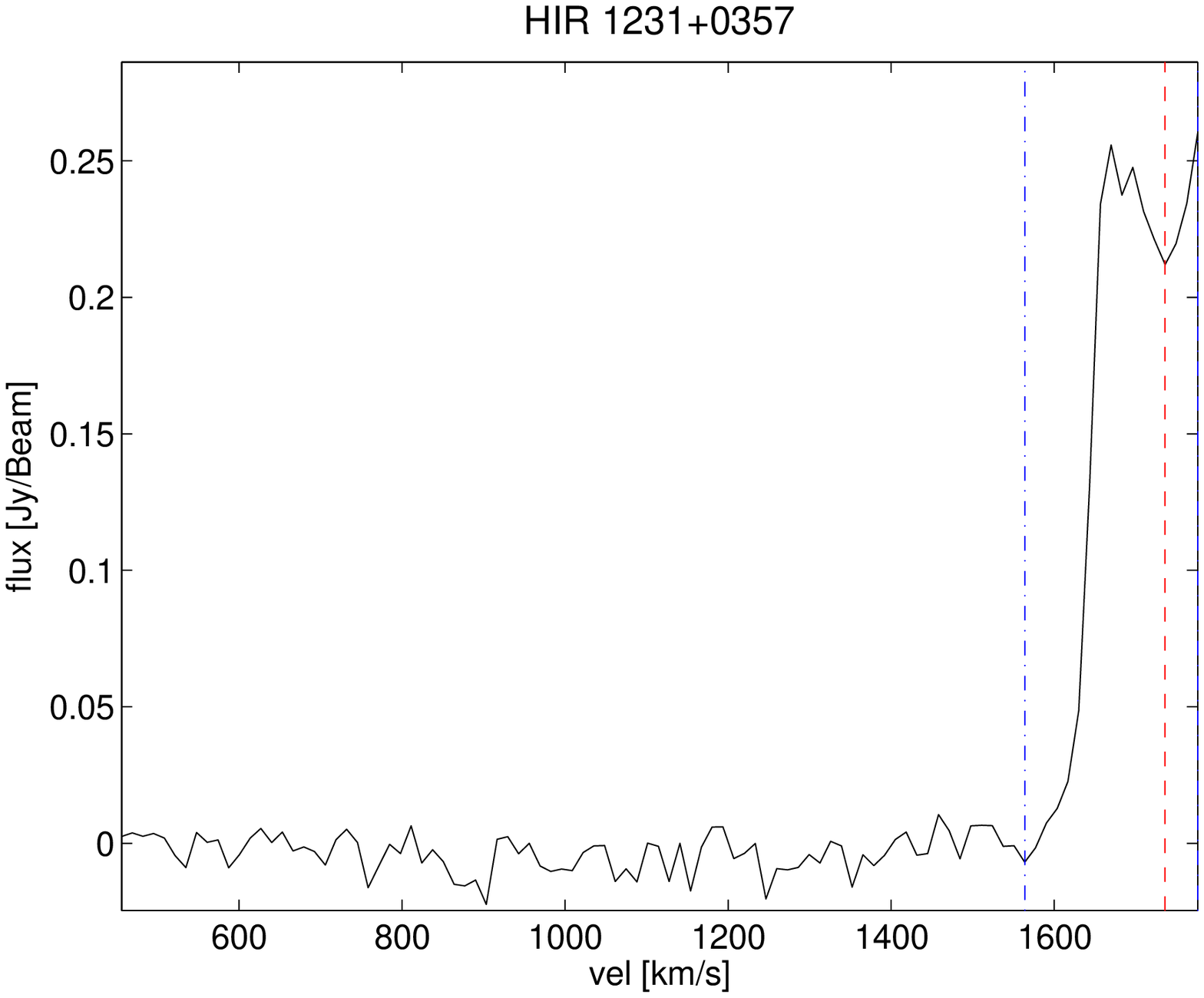}
 \includegraphics[width=0.3\textwidth]{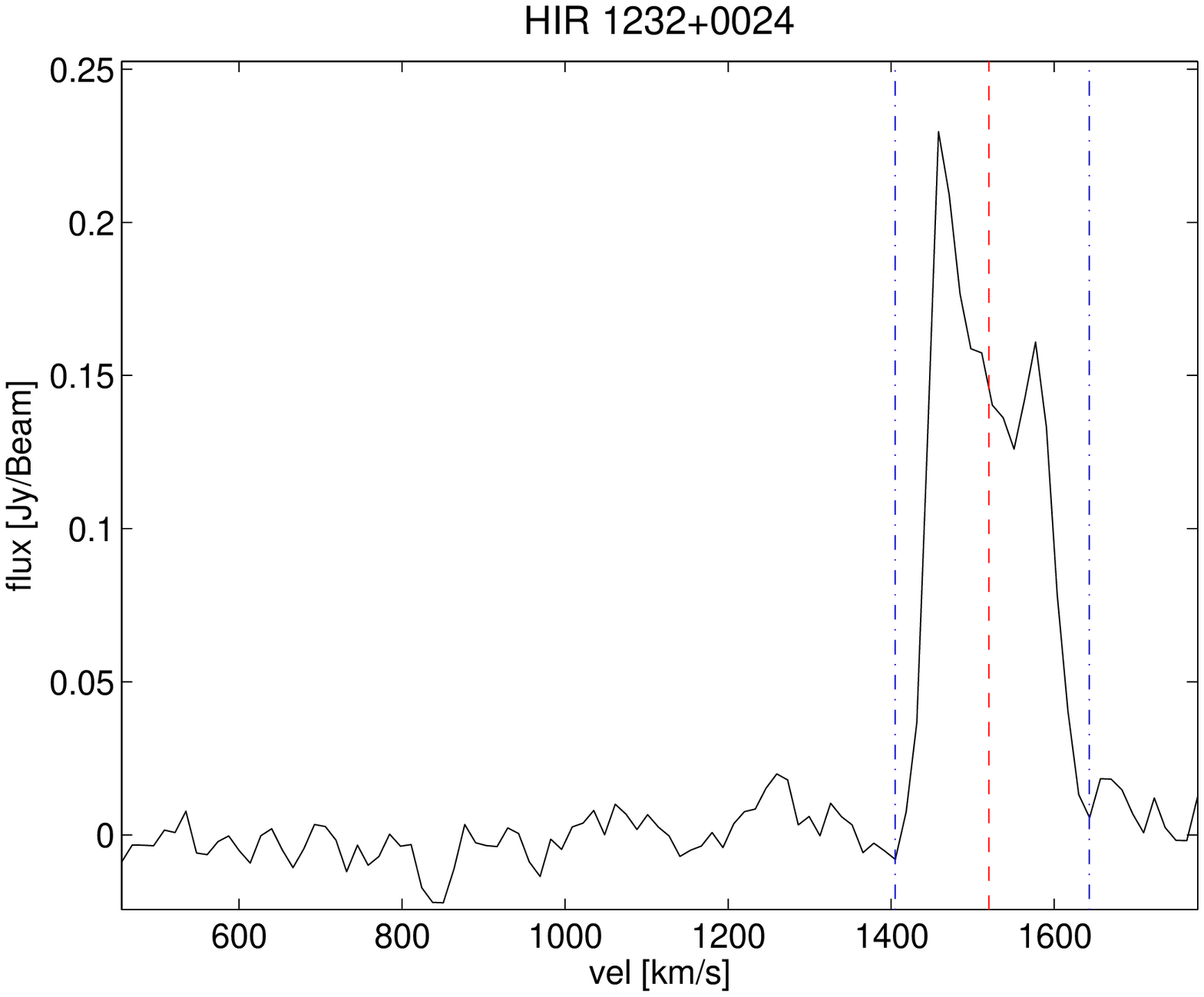}
 
 \end{center}                                                         
{\bf Fig~\ref{all_spectra}.} (continued)                              
                                                                      
\end{figure*}

\begin{figure*}
  \begin{center}

 \includegraphics[width=0.3\textwidth]{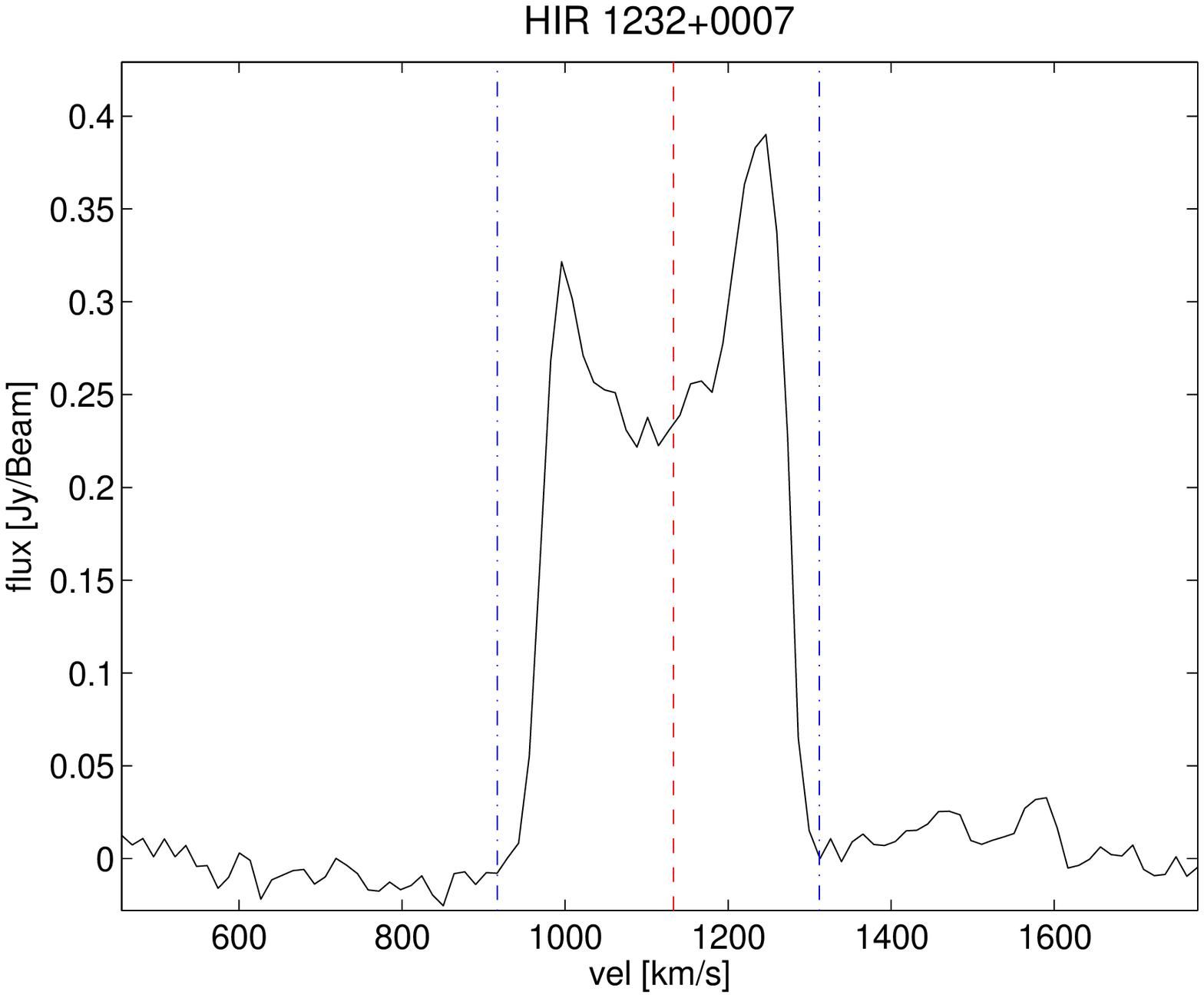}
 \includegraphics[width=0.3\textwidth]{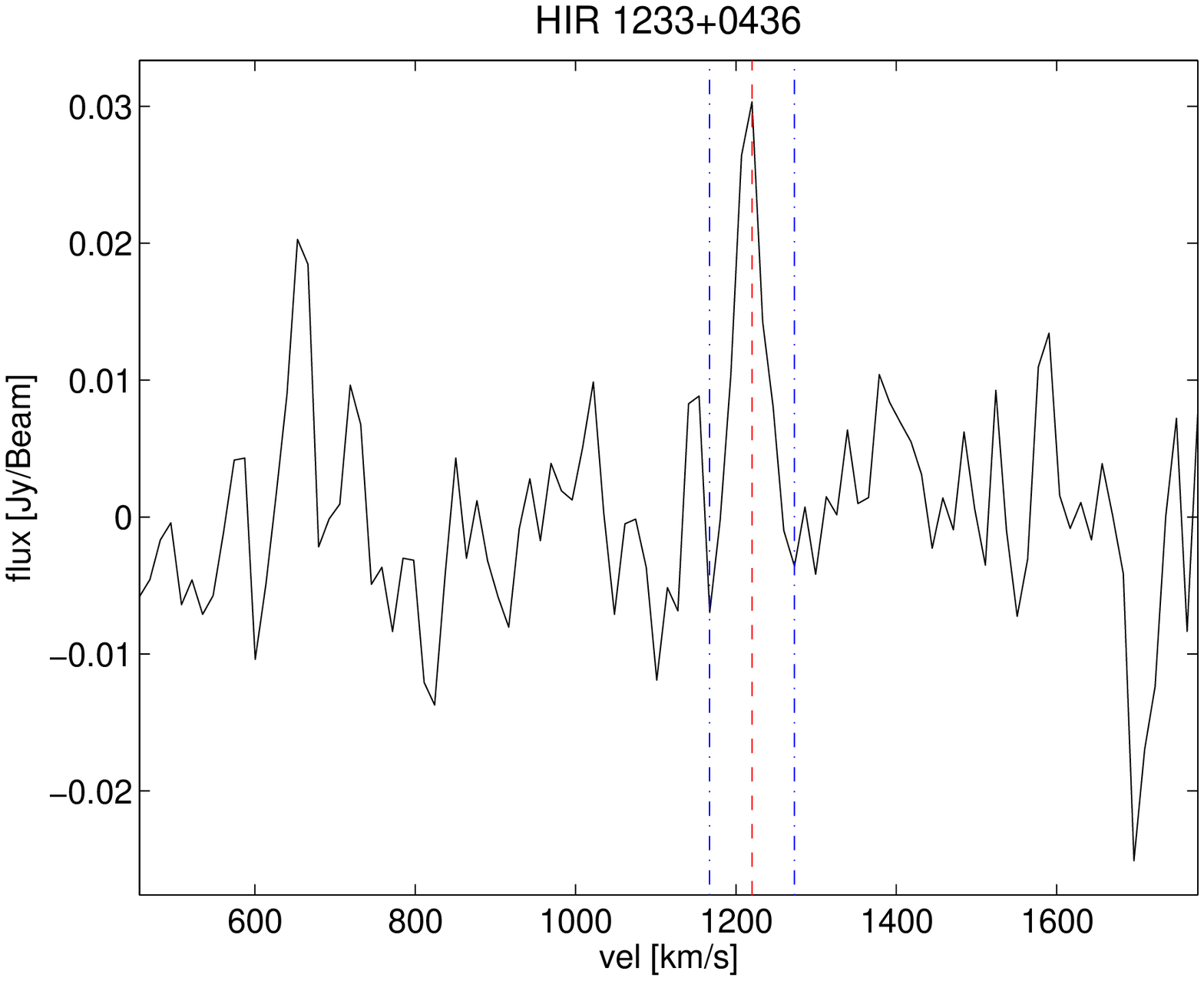}
 \includegraphics[width=0.3\textwidth]{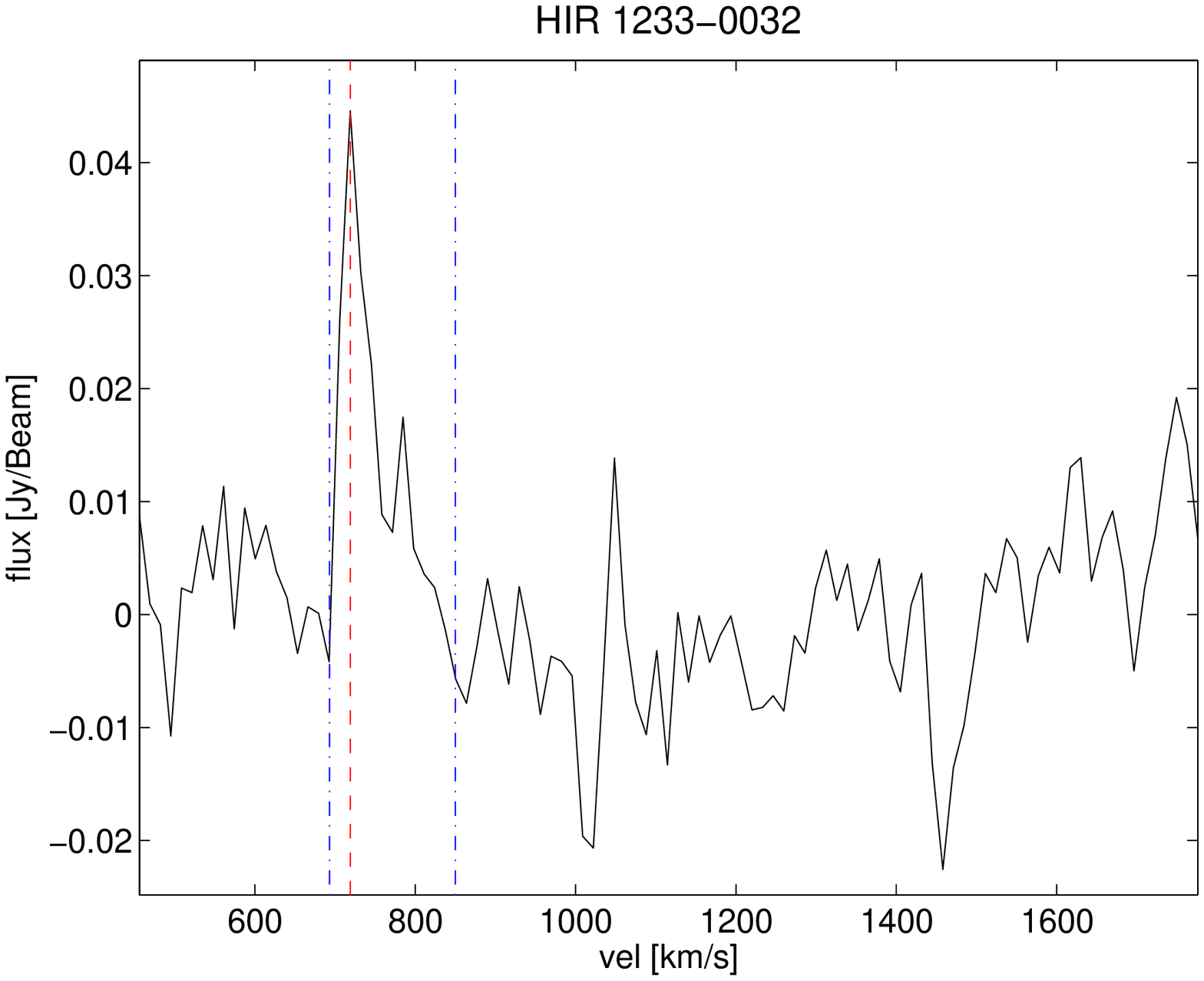}
 \includegraphics[width=0.3\textwidth]{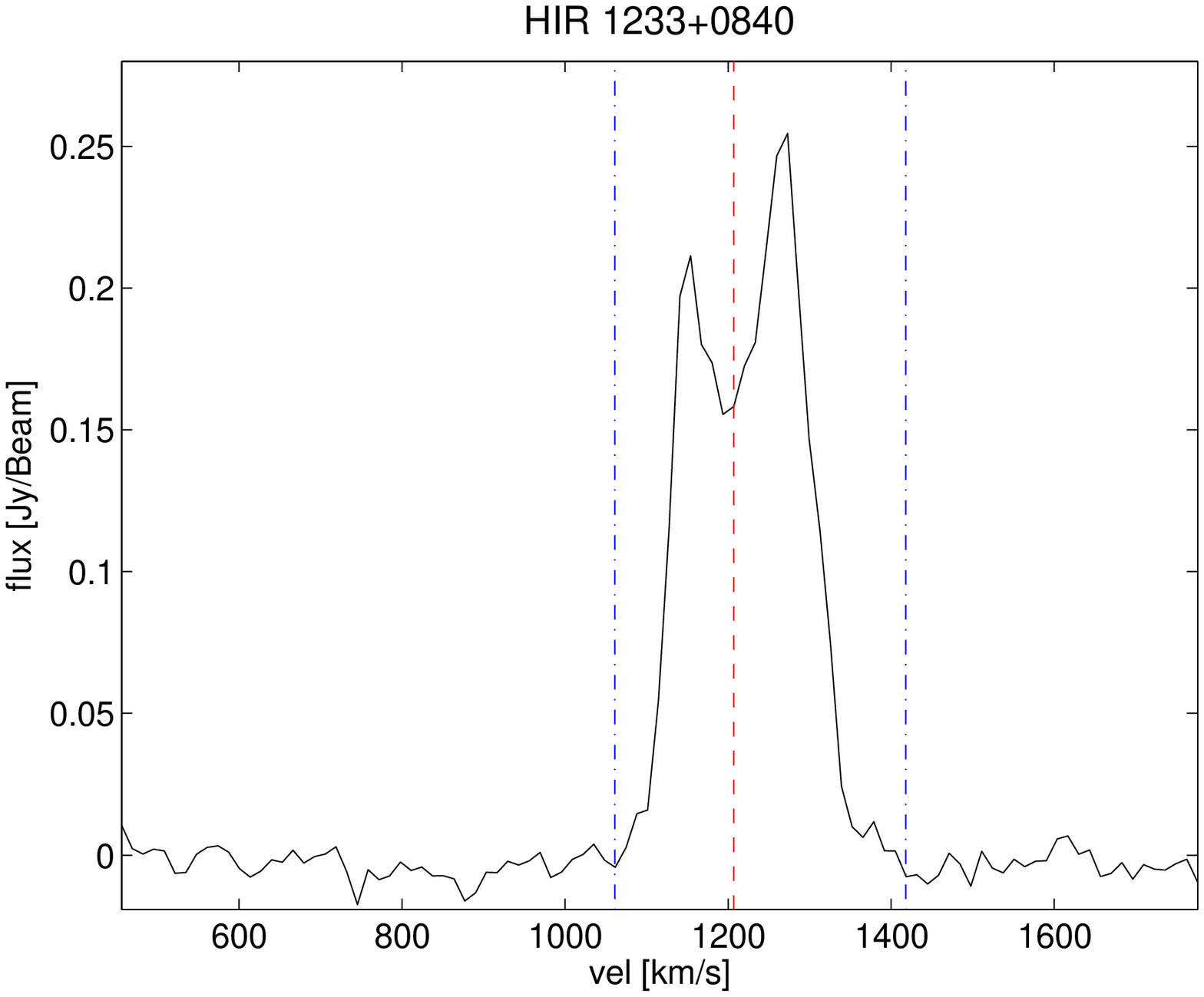}
 \includegraphics[width=0.3\textwidth]{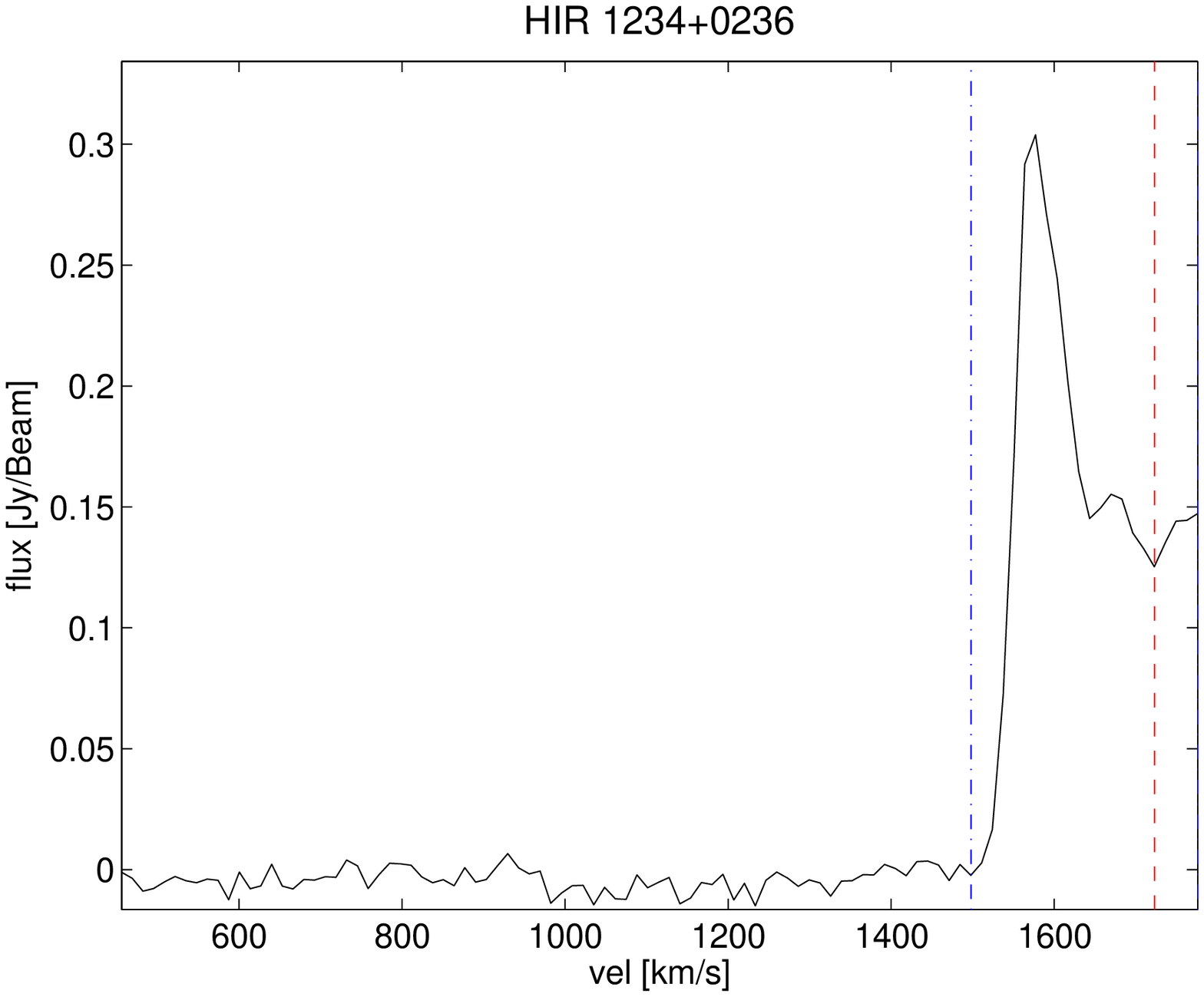}
 \includegraphics[width=0.3\textwidth]{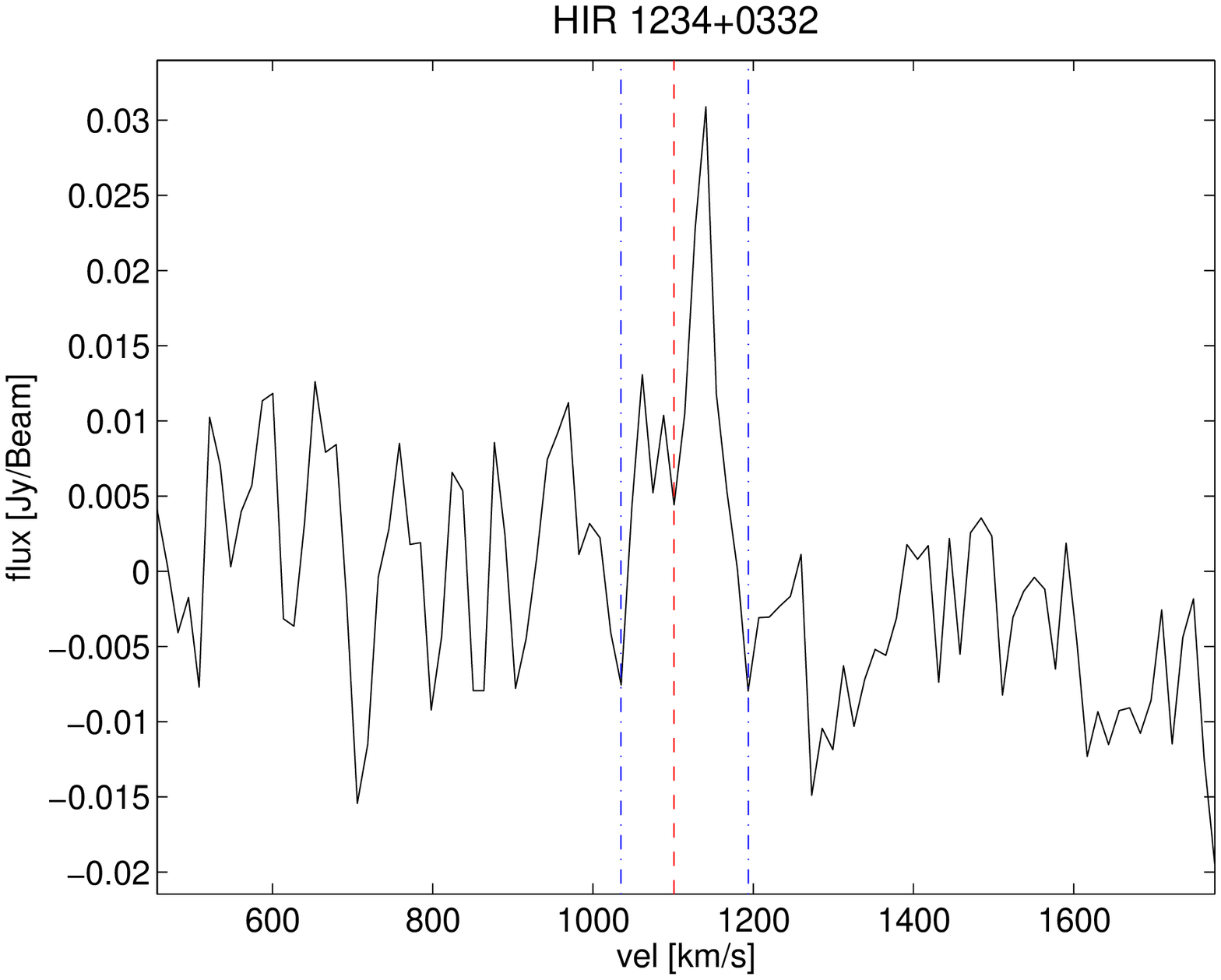}
 \includegraphics[width=0.3\textwidth]{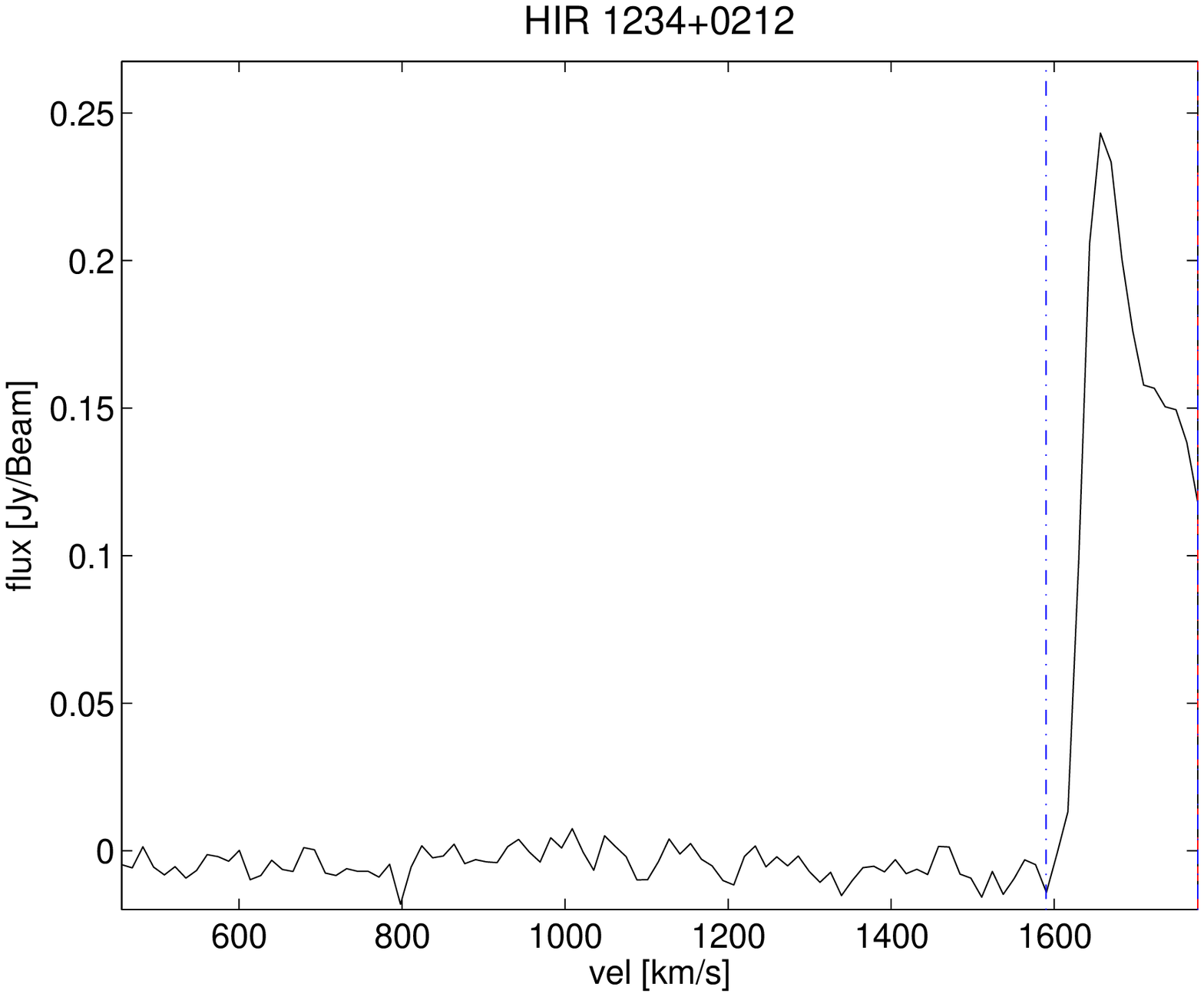}
 \includegraphics[width=0.3\textwidth]{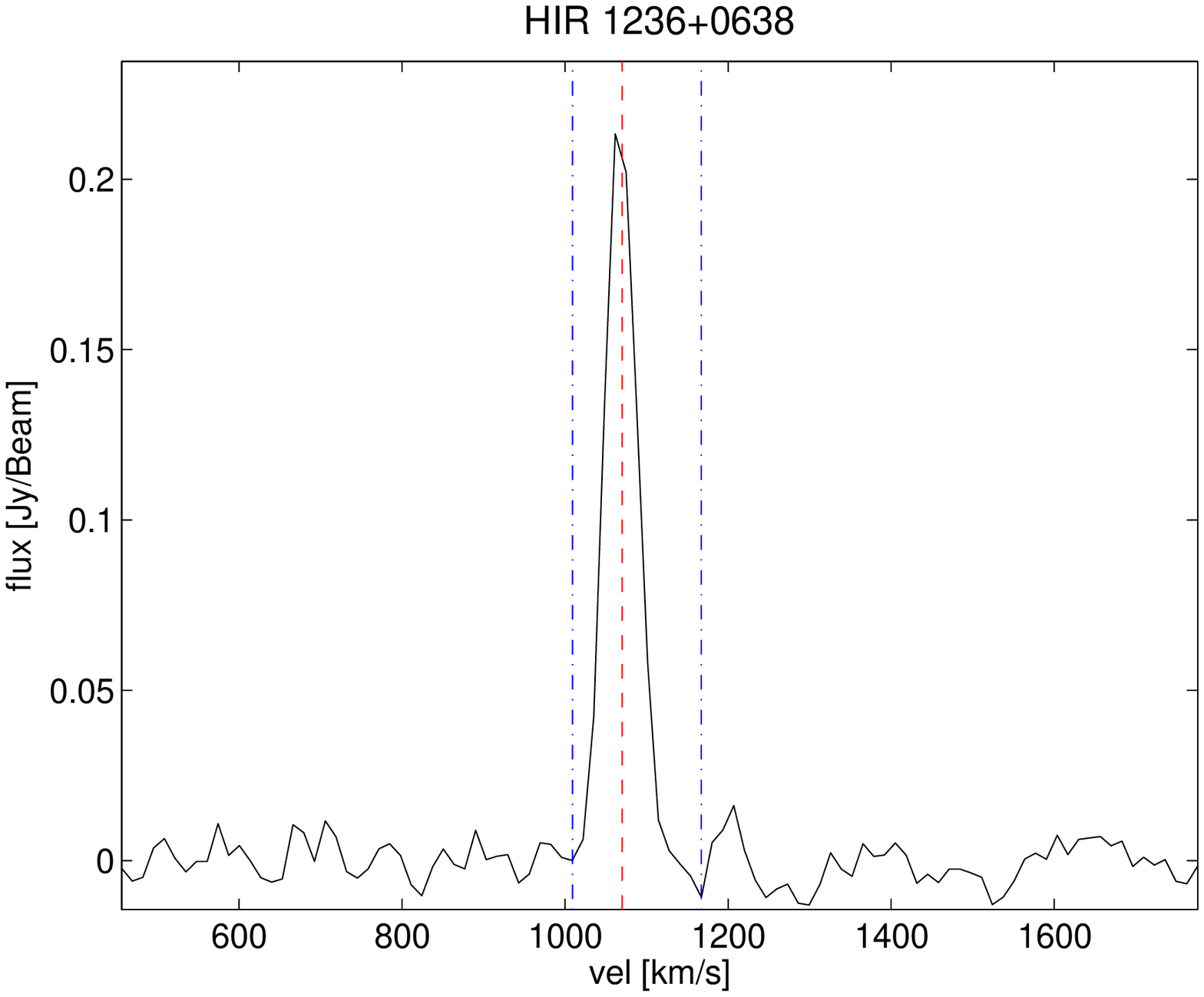}
 \includegraphics[width=0.3\textwidth]{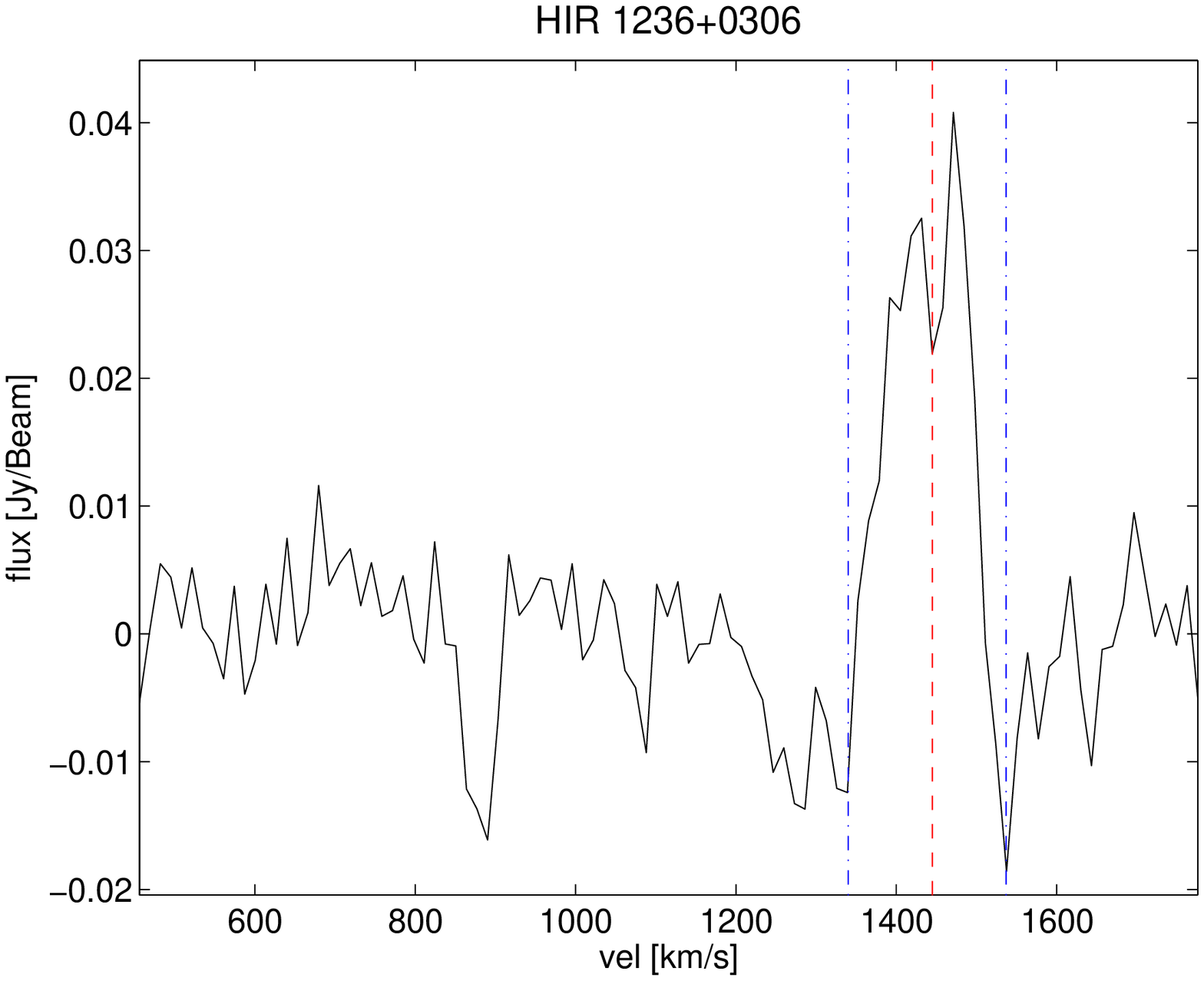}
 \includegraphics[width=0.3\textwidth]{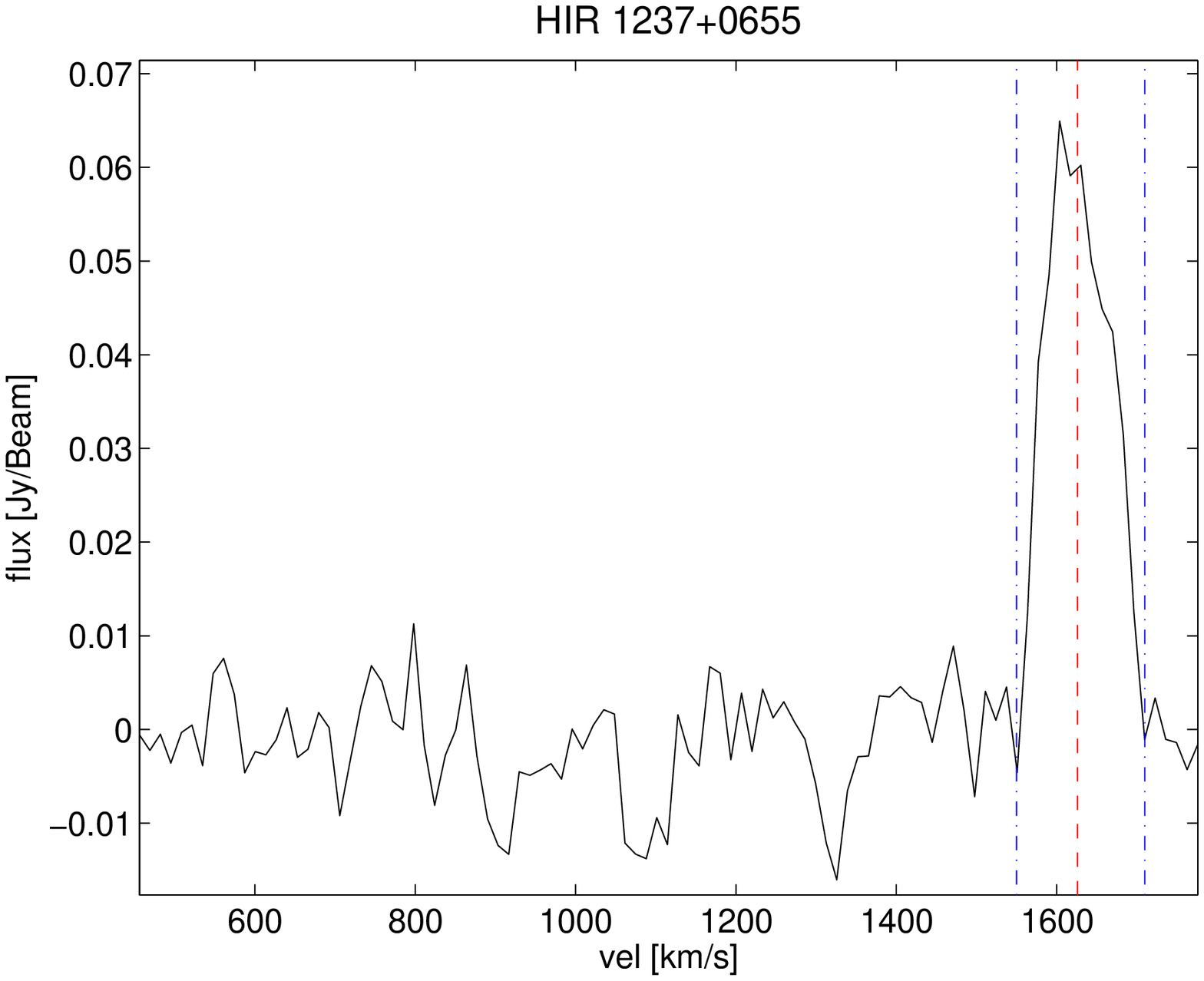}
 \includegraphics[width=0.3\textwidth]{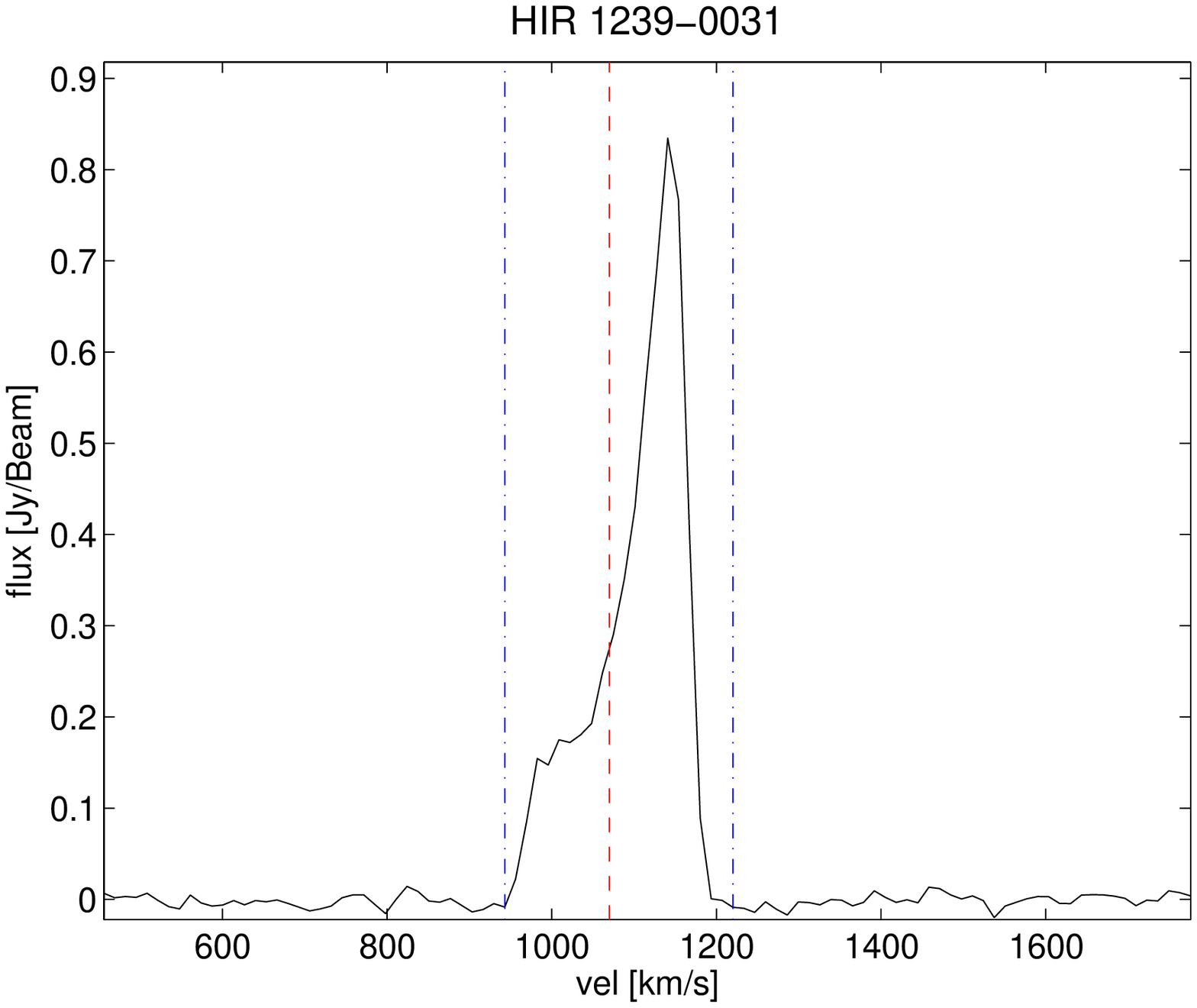}
 \includegraphics[width=0.3\textwidth]{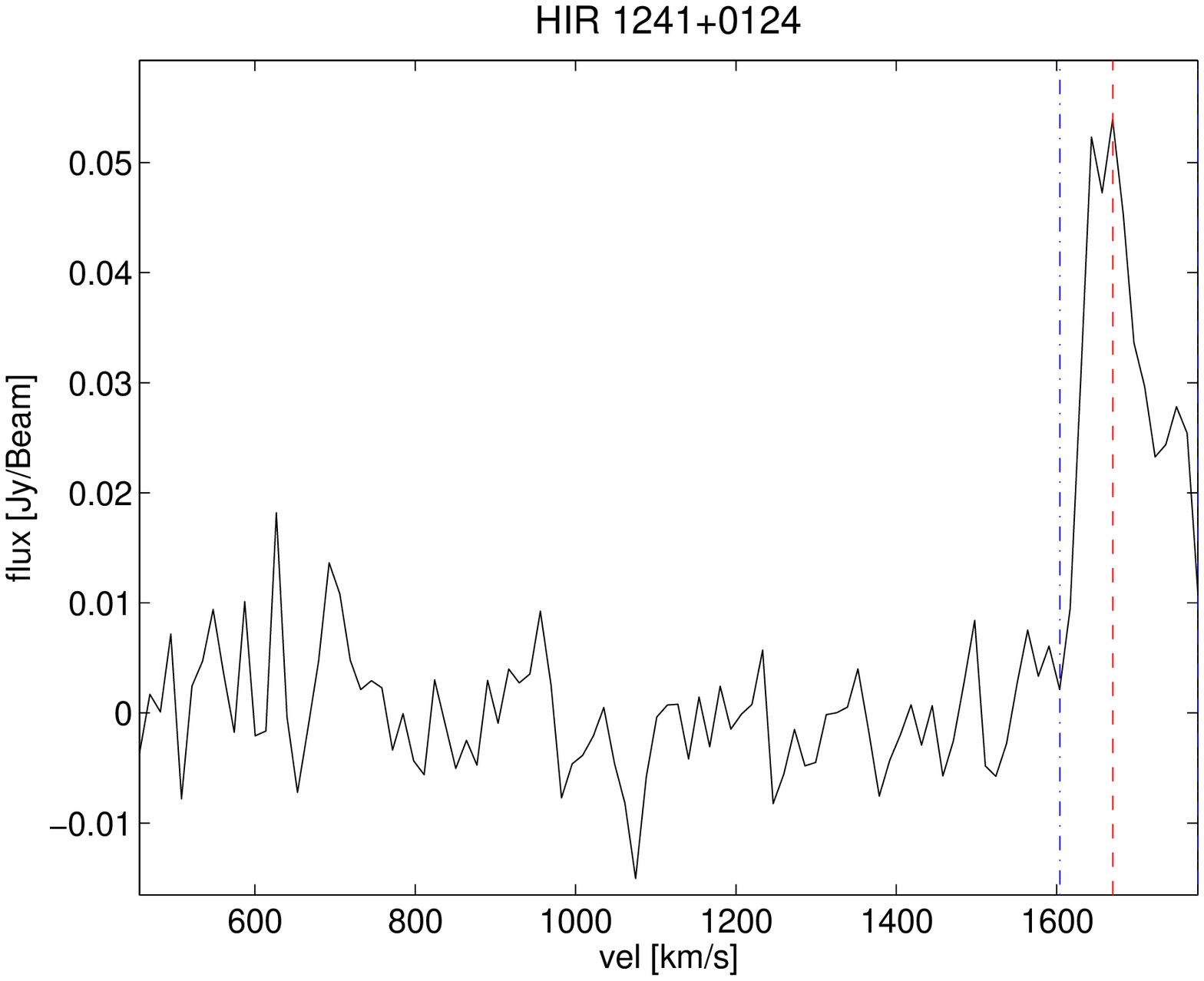}
 \includegraphics[width=0.3\textwidth]{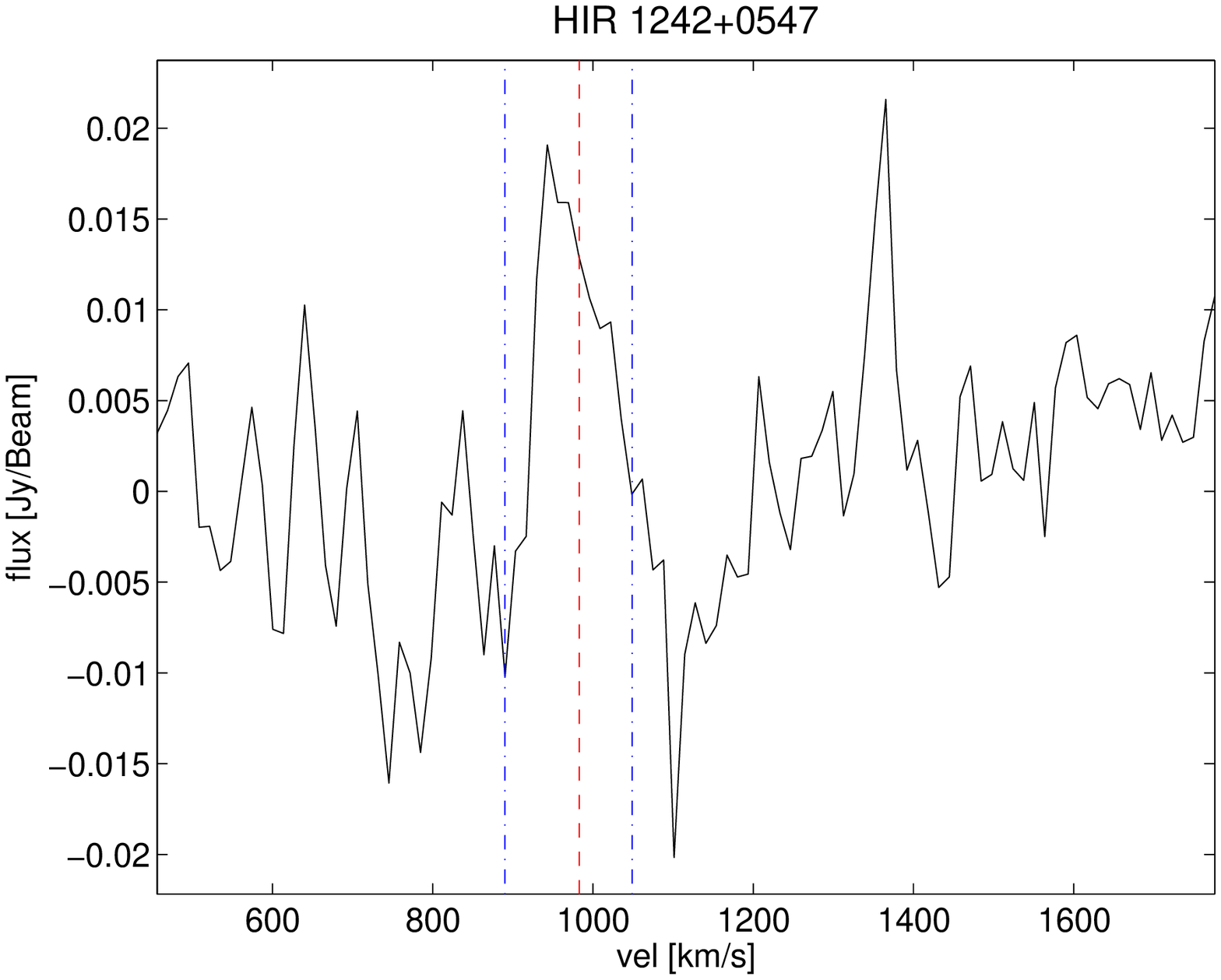}
 \includegraphics[width=0.3\textwidth]{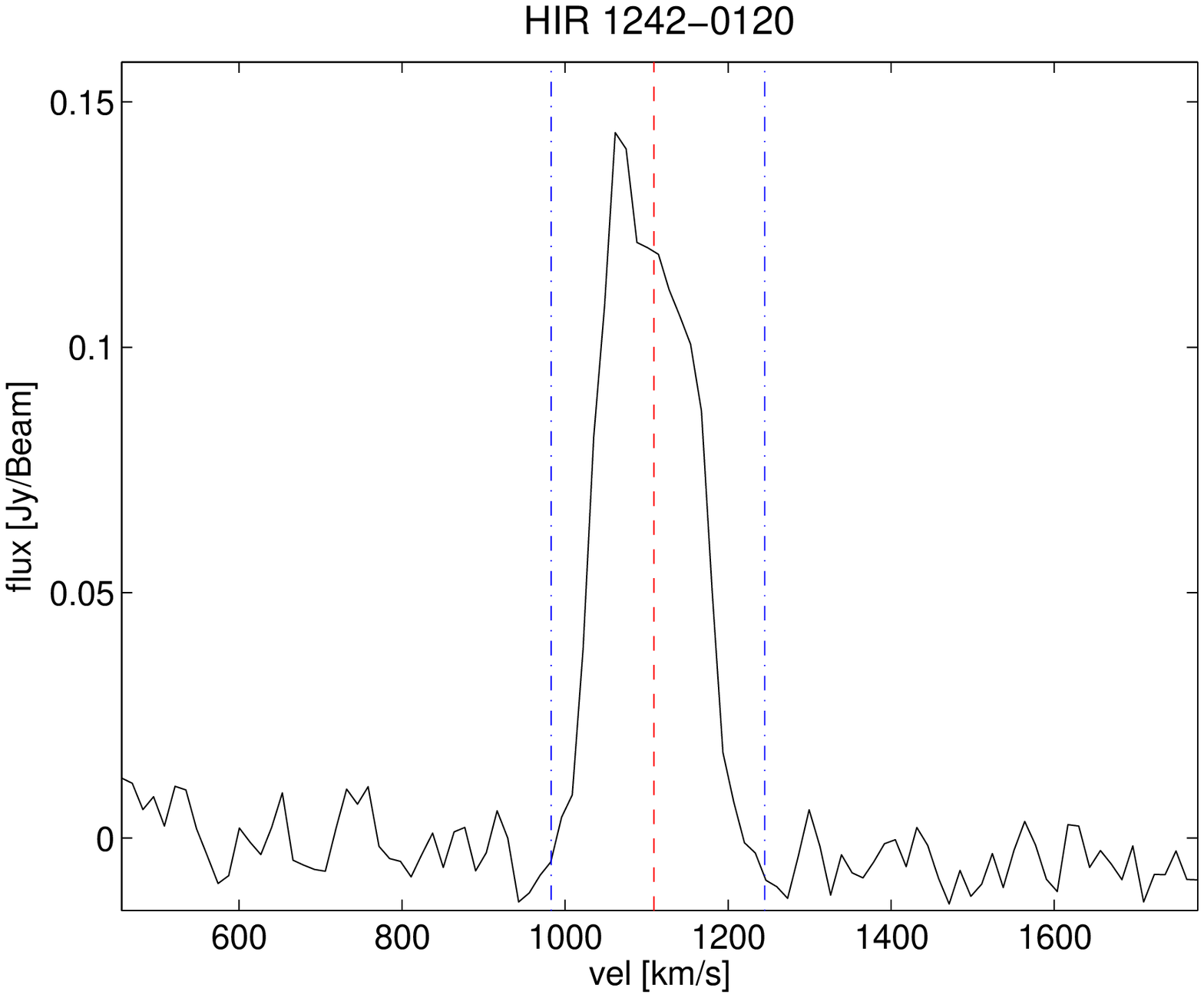}
 \includegraphics[width=0.3\textwidth]{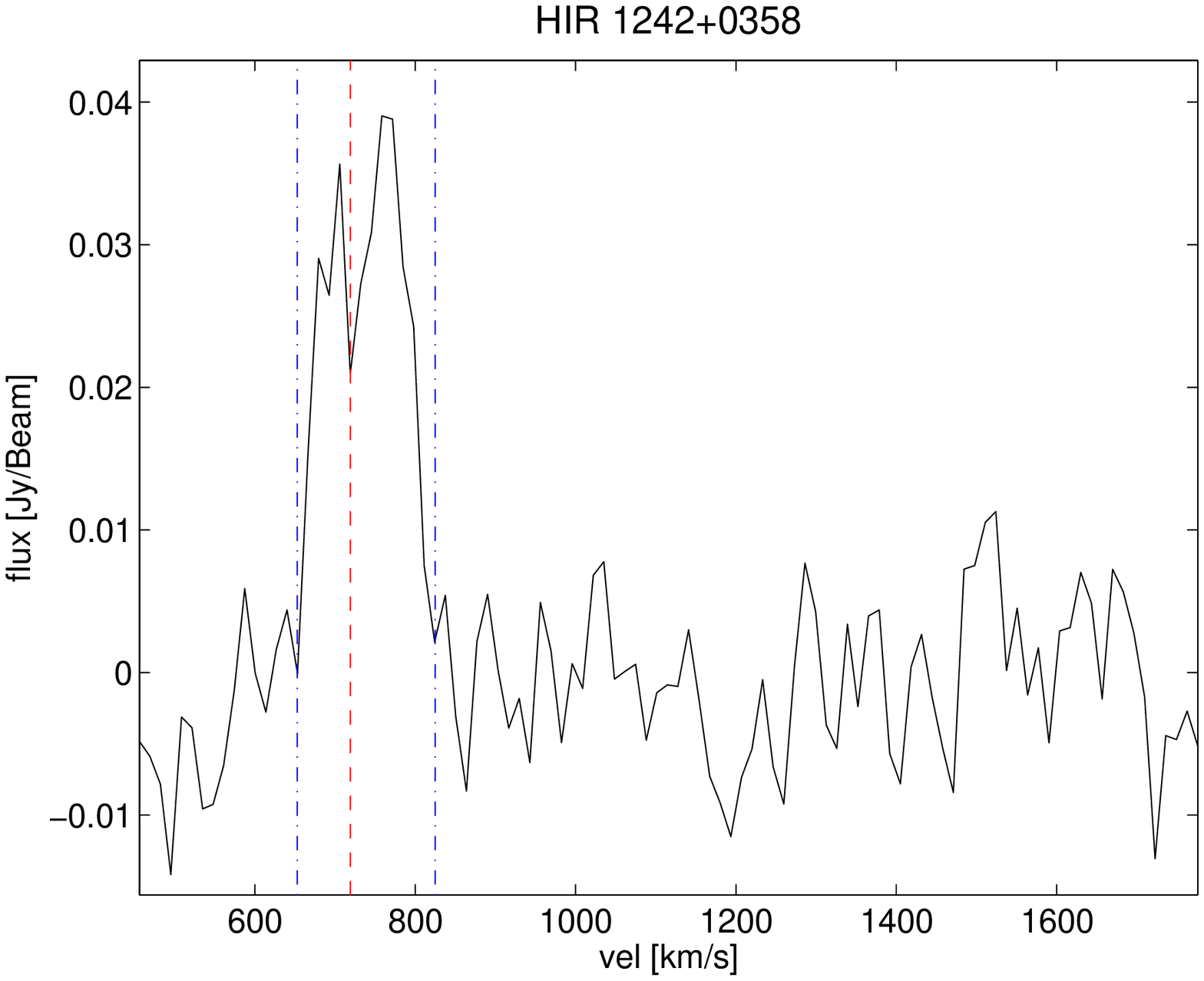}

 \end{center}                                                         
{\bf Fig~\ref{all_spectra}.} (continued)                              
                                                                      
\end{figure*}

\begin{figure*}
  \begin{center}
 \includegraphics[width=0.3\textwidth]{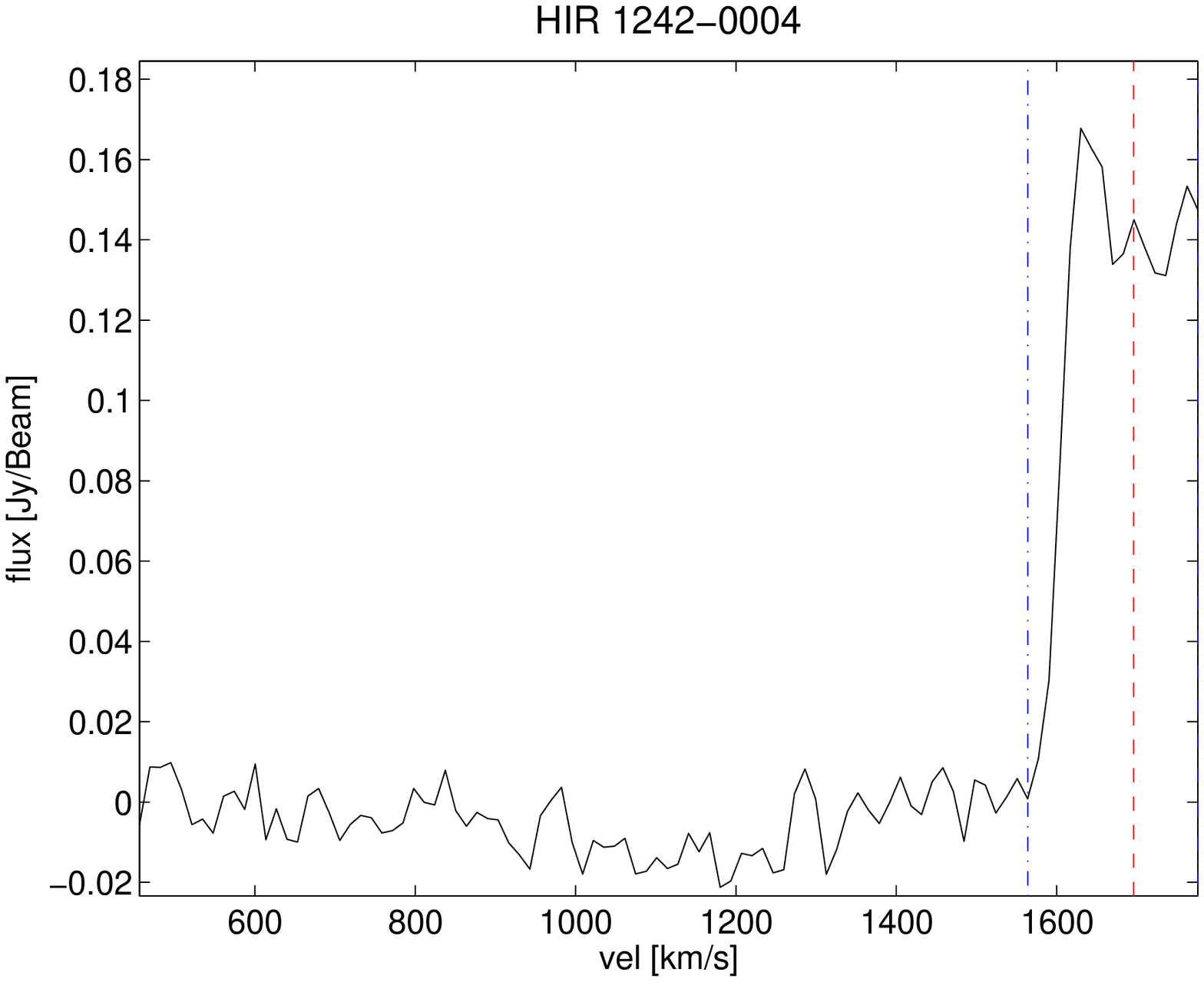}
 \includegraphics[width=0.3\textwidth]{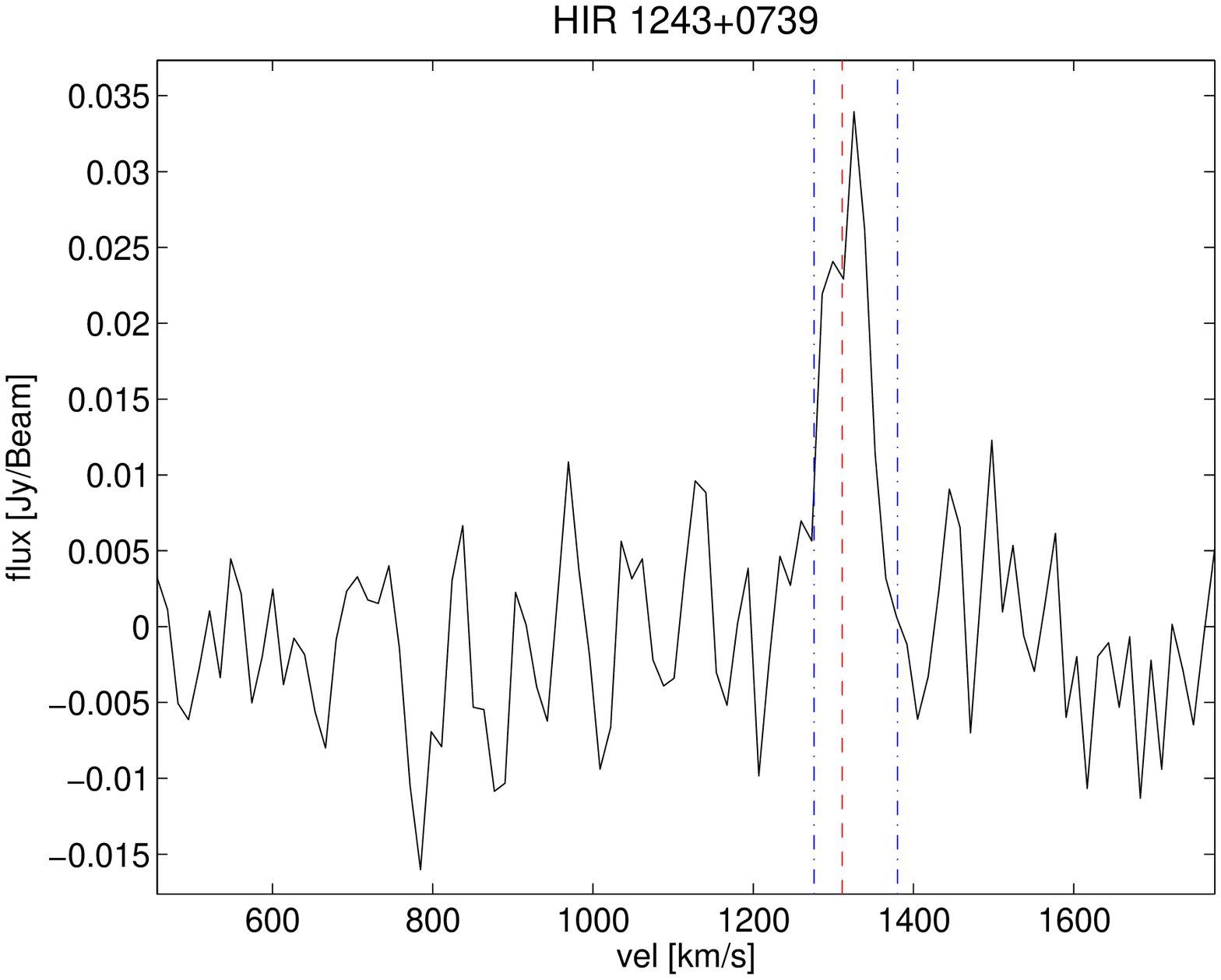}
 \includegraphics[width=0.3\textwidth]{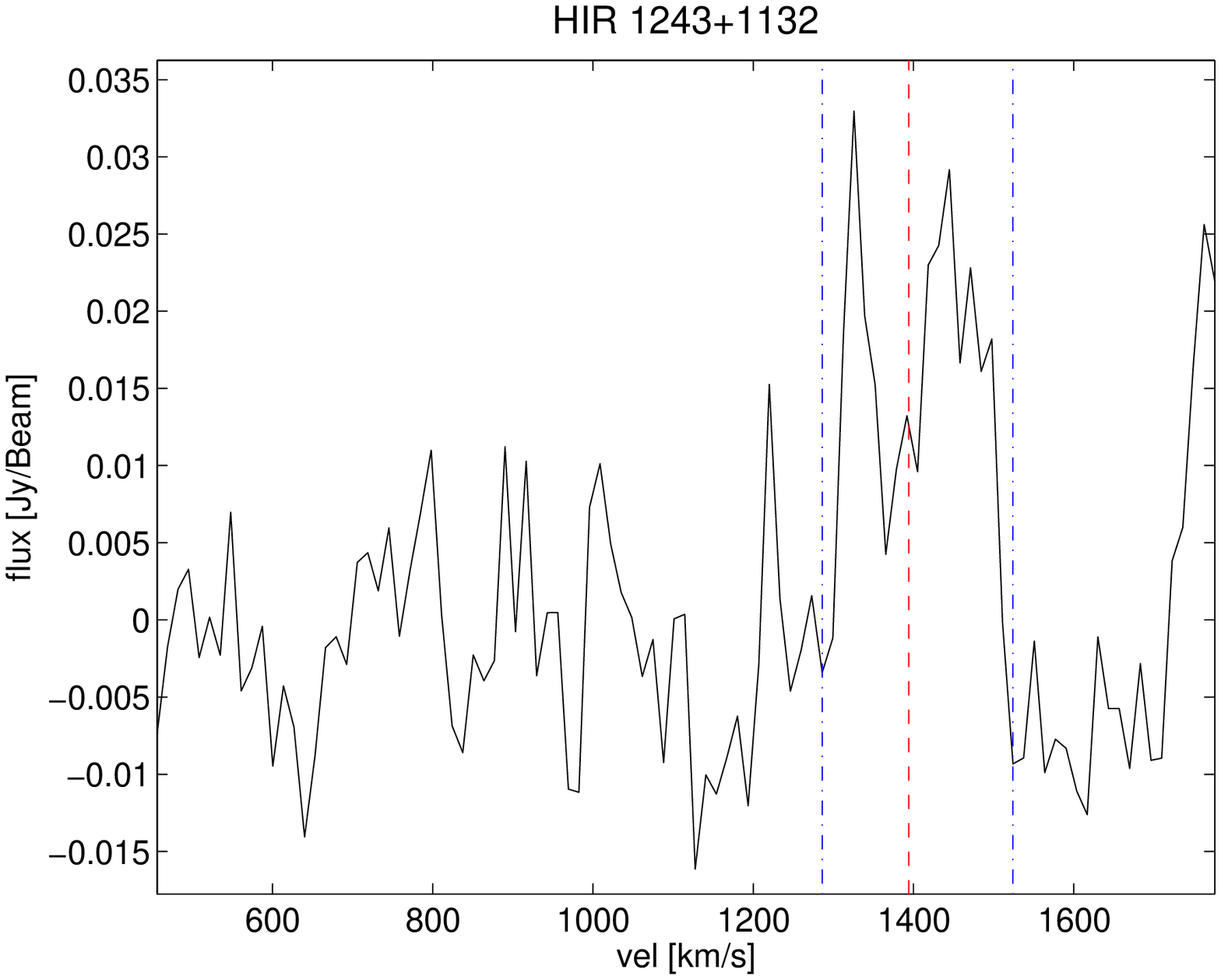}
 \includegraphics[width=0.3\textwidth]{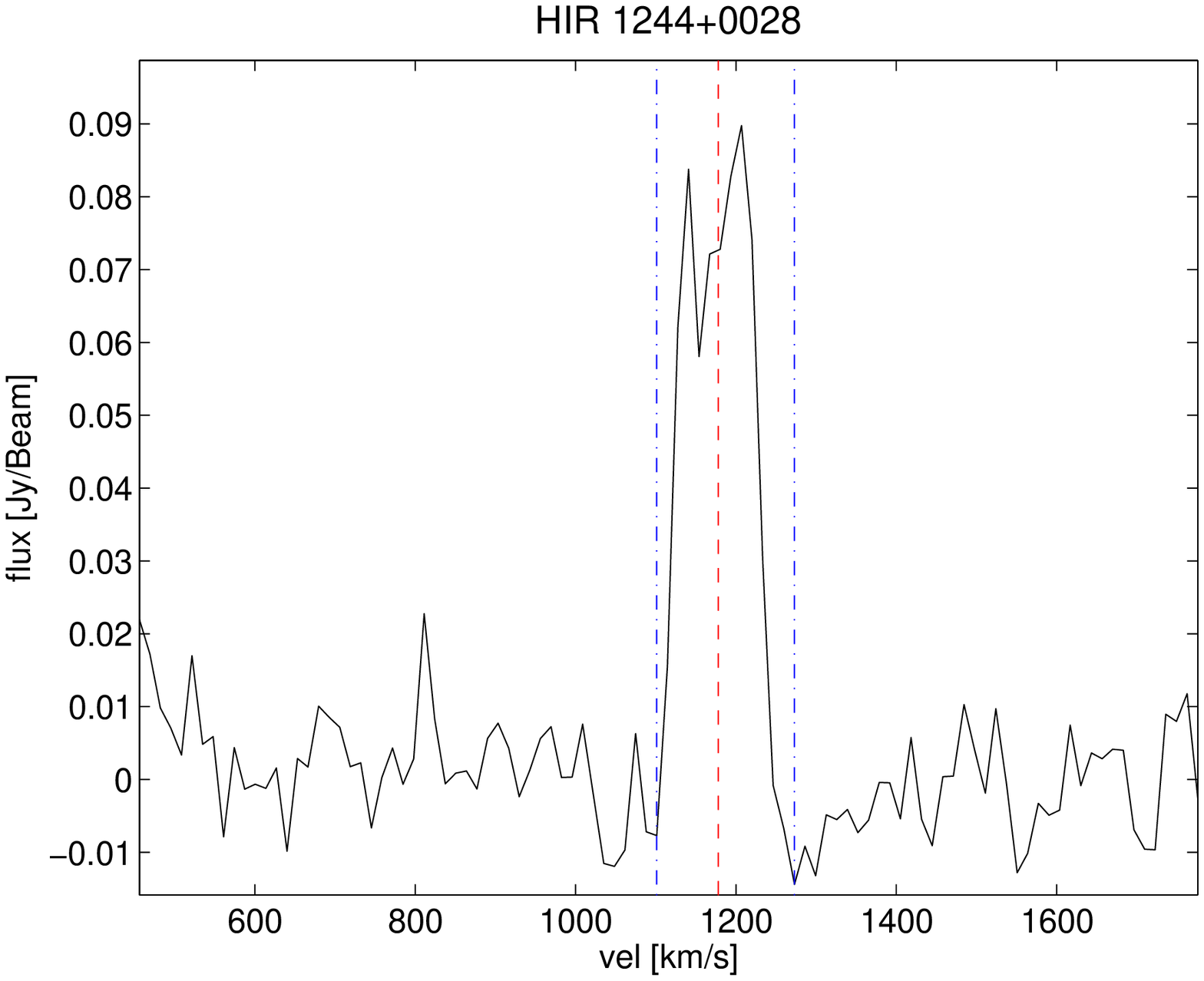}
 \includegraphics[width=0.3\textwidth]{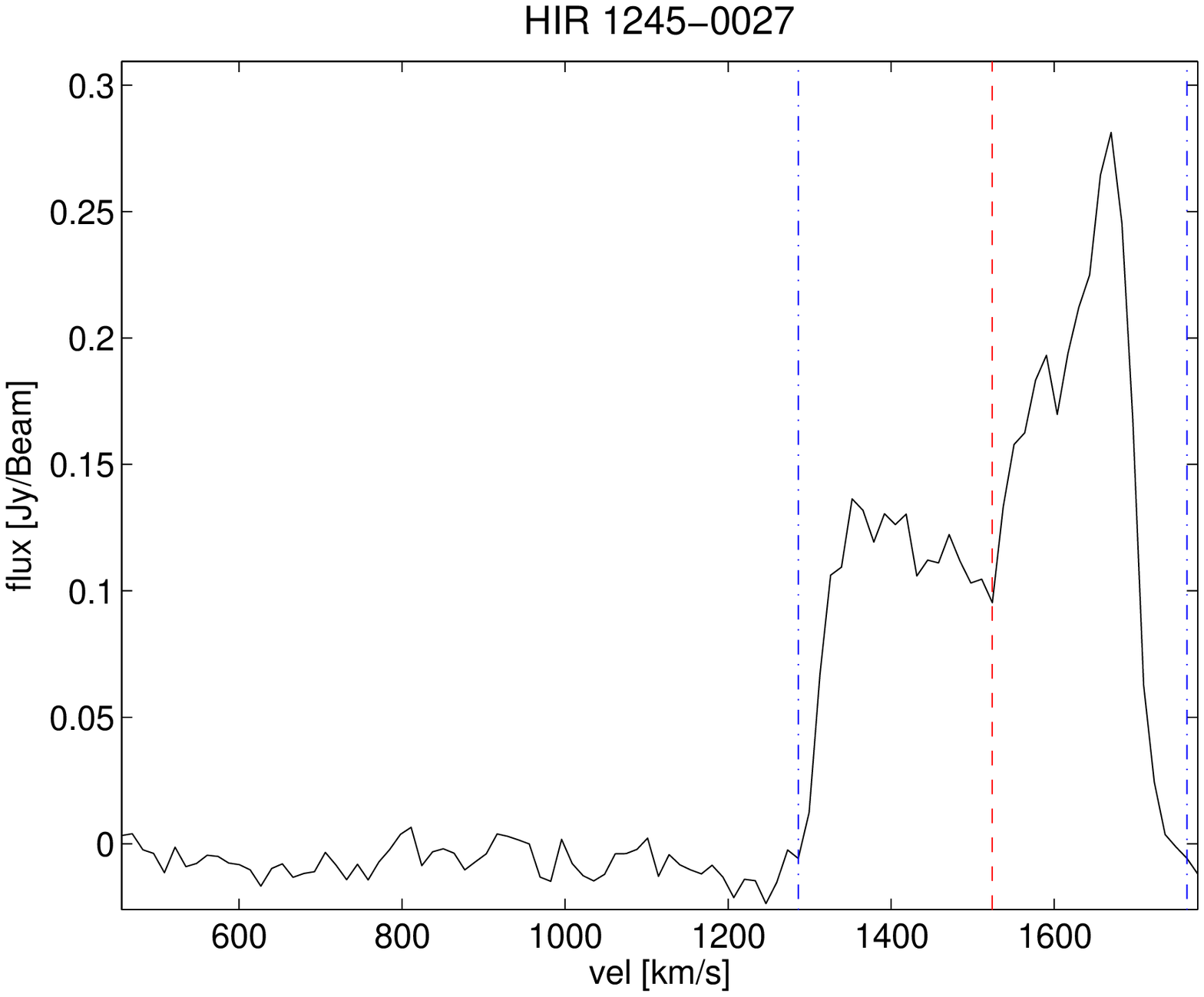}
 \includegraphics[width=0.3\textwidth]{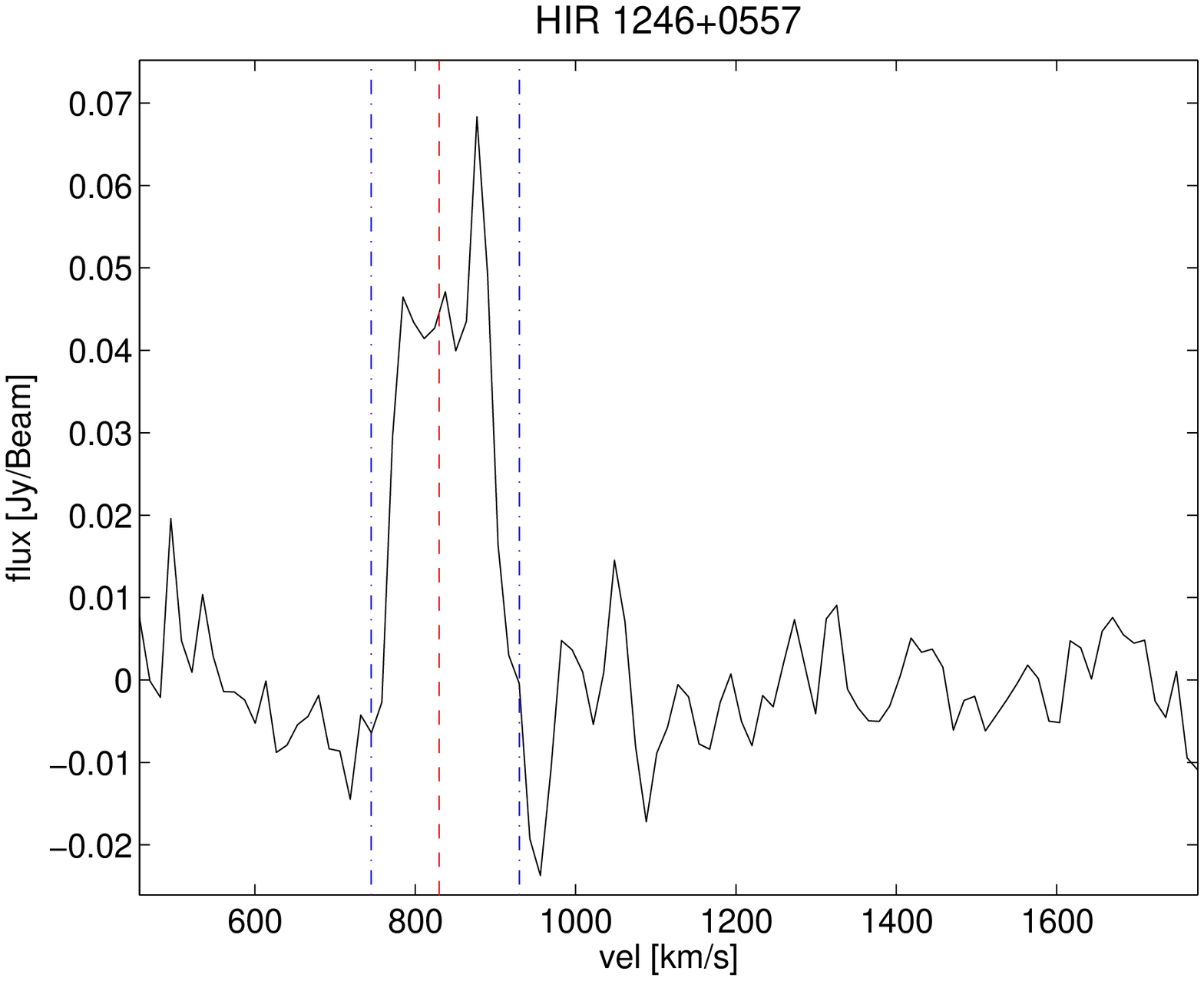}
 \includegraphics[width=0.3\textwidth]{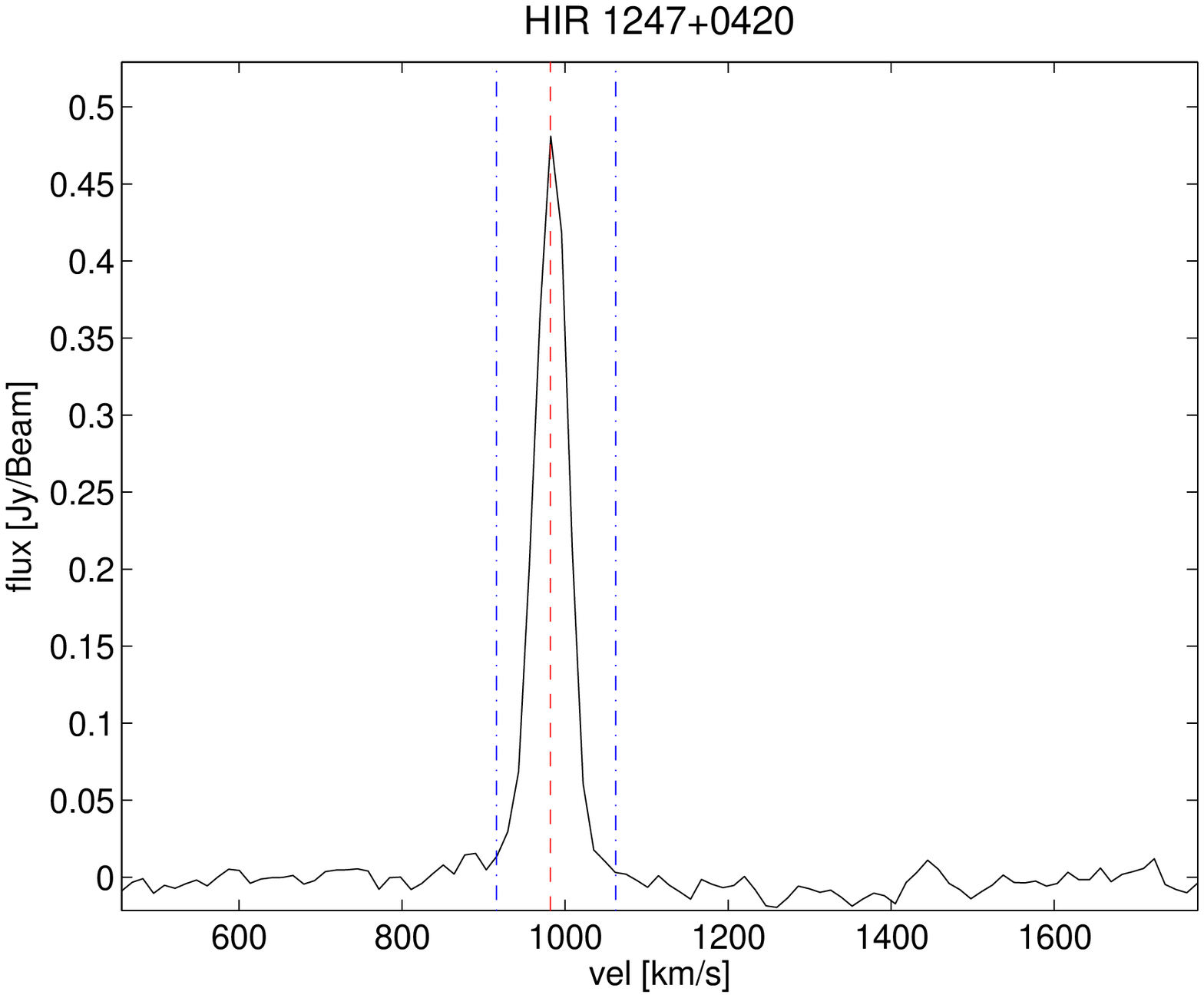}
 \includegraphics[width=0.3\textwidth]{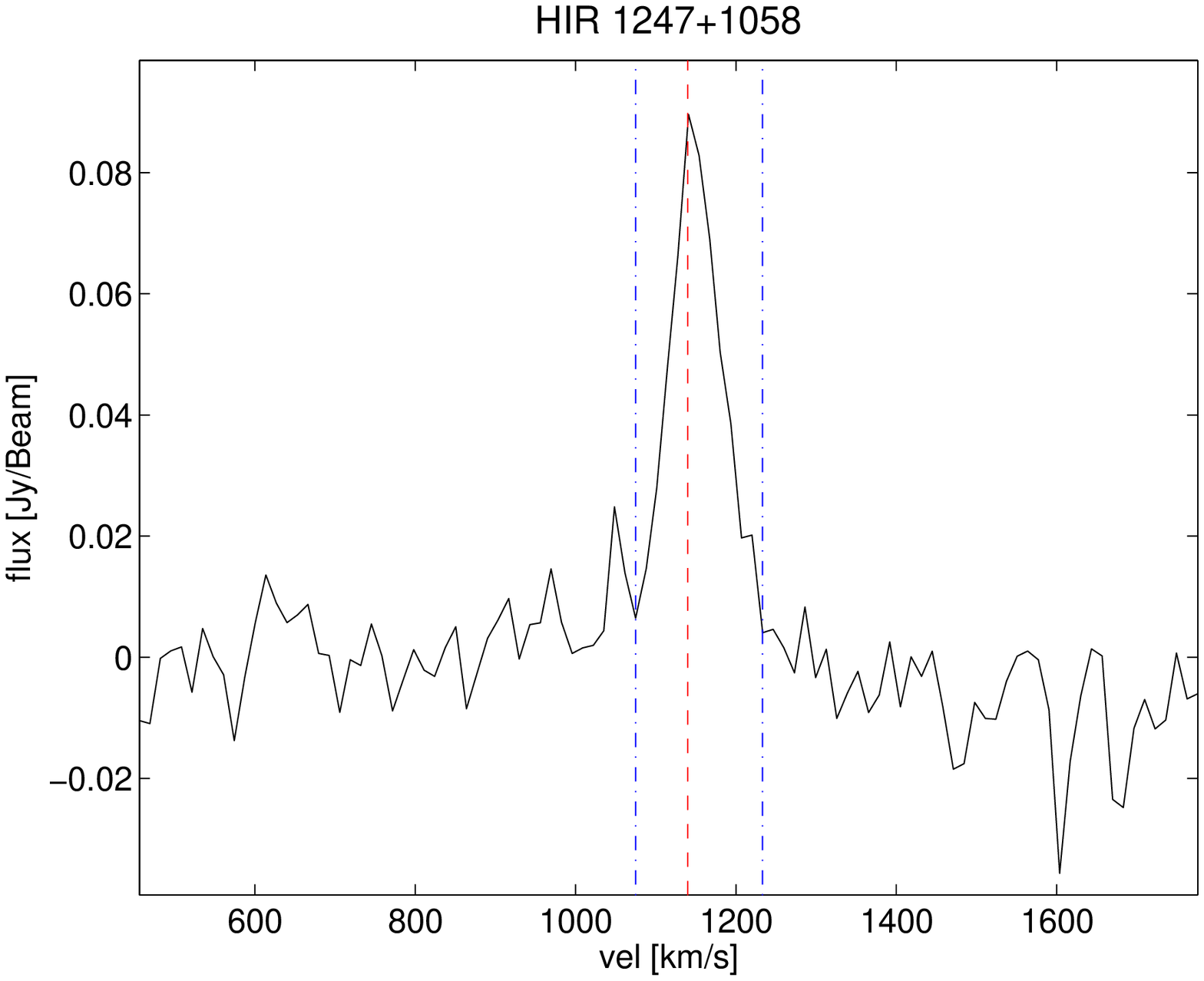}
 \includegraphics[width=0.3\textwidth]{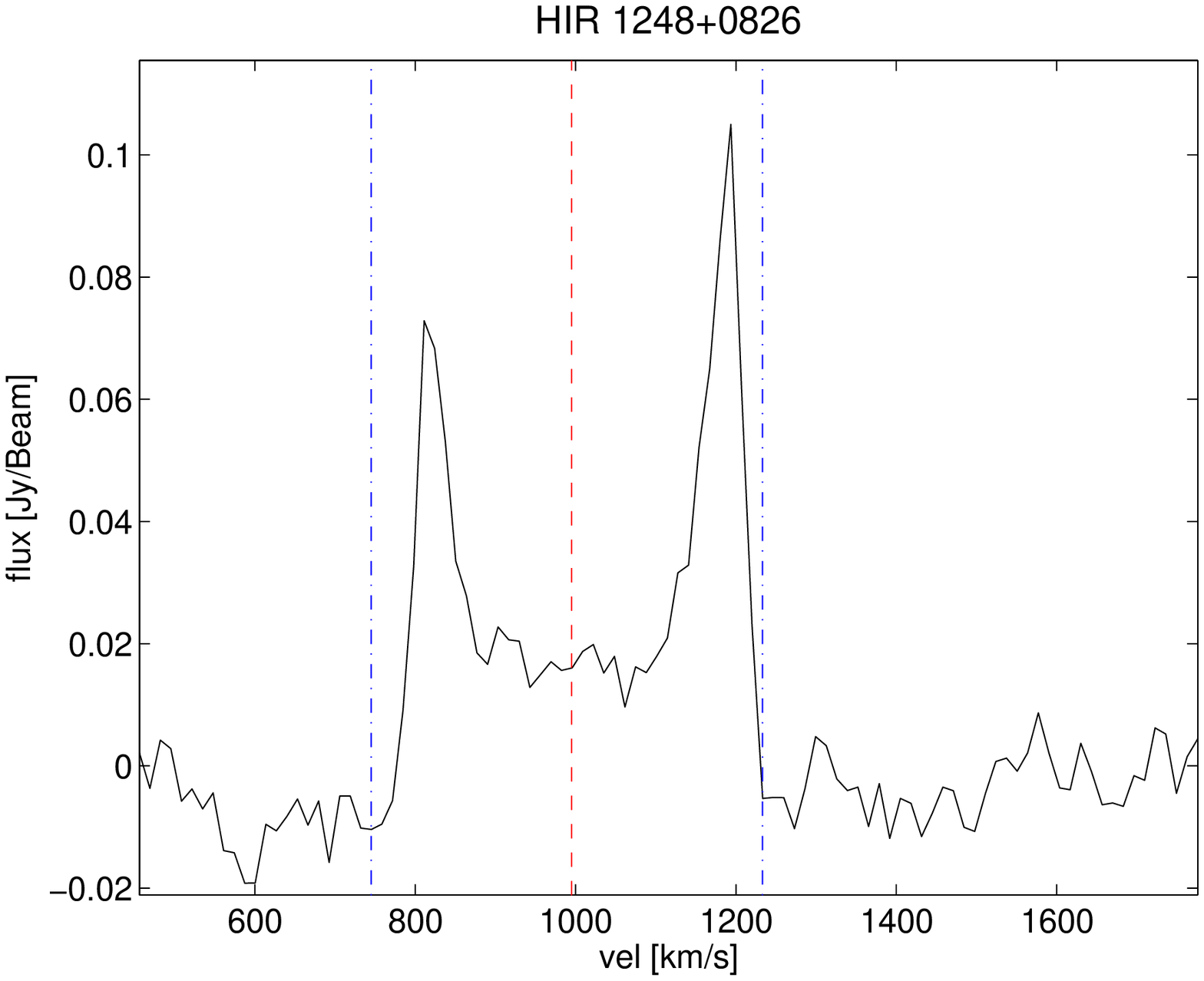}
 \includegraphics[width=0.3\textwidth]{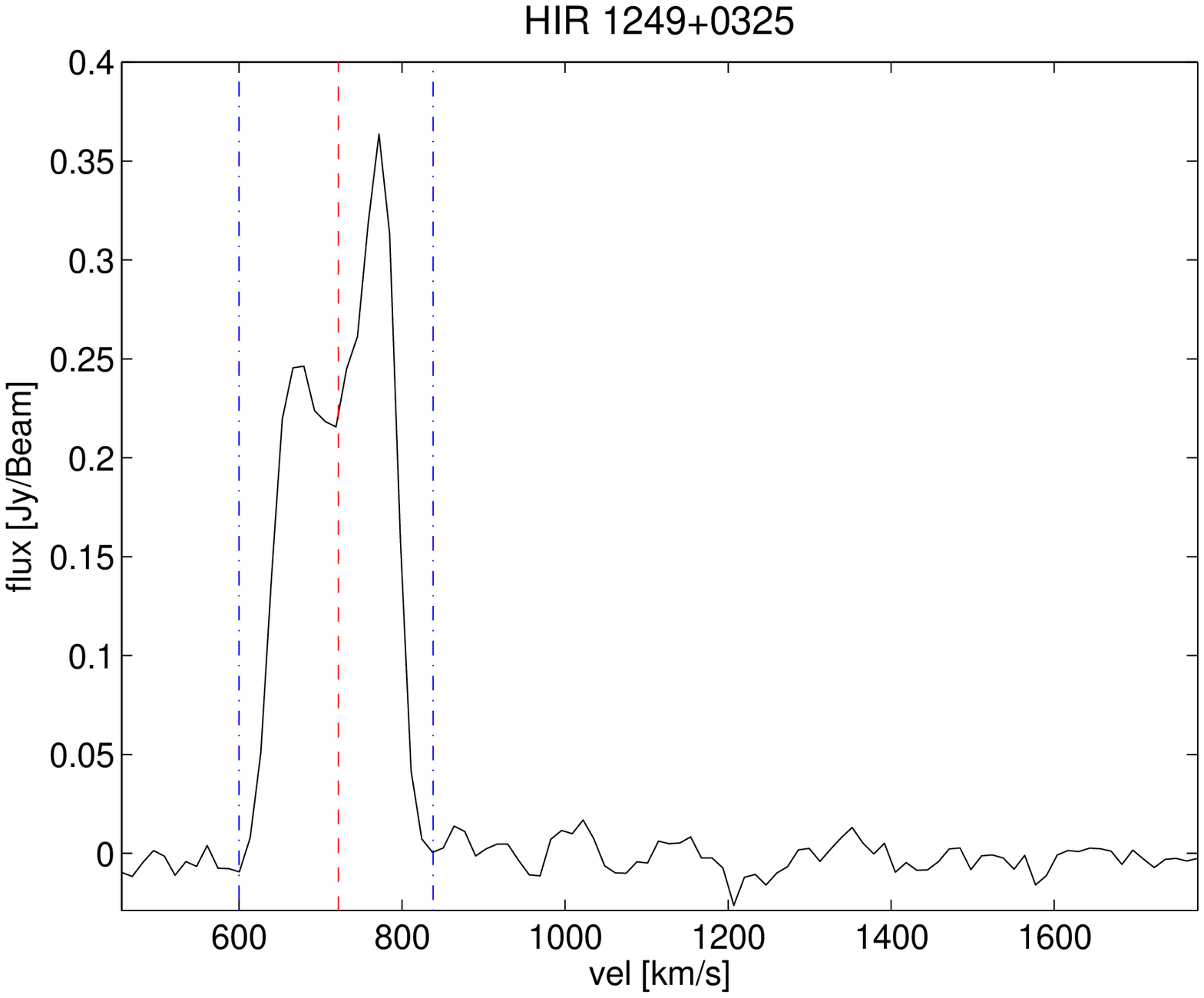}
 \includegraphics[width=0.3\textwidth]{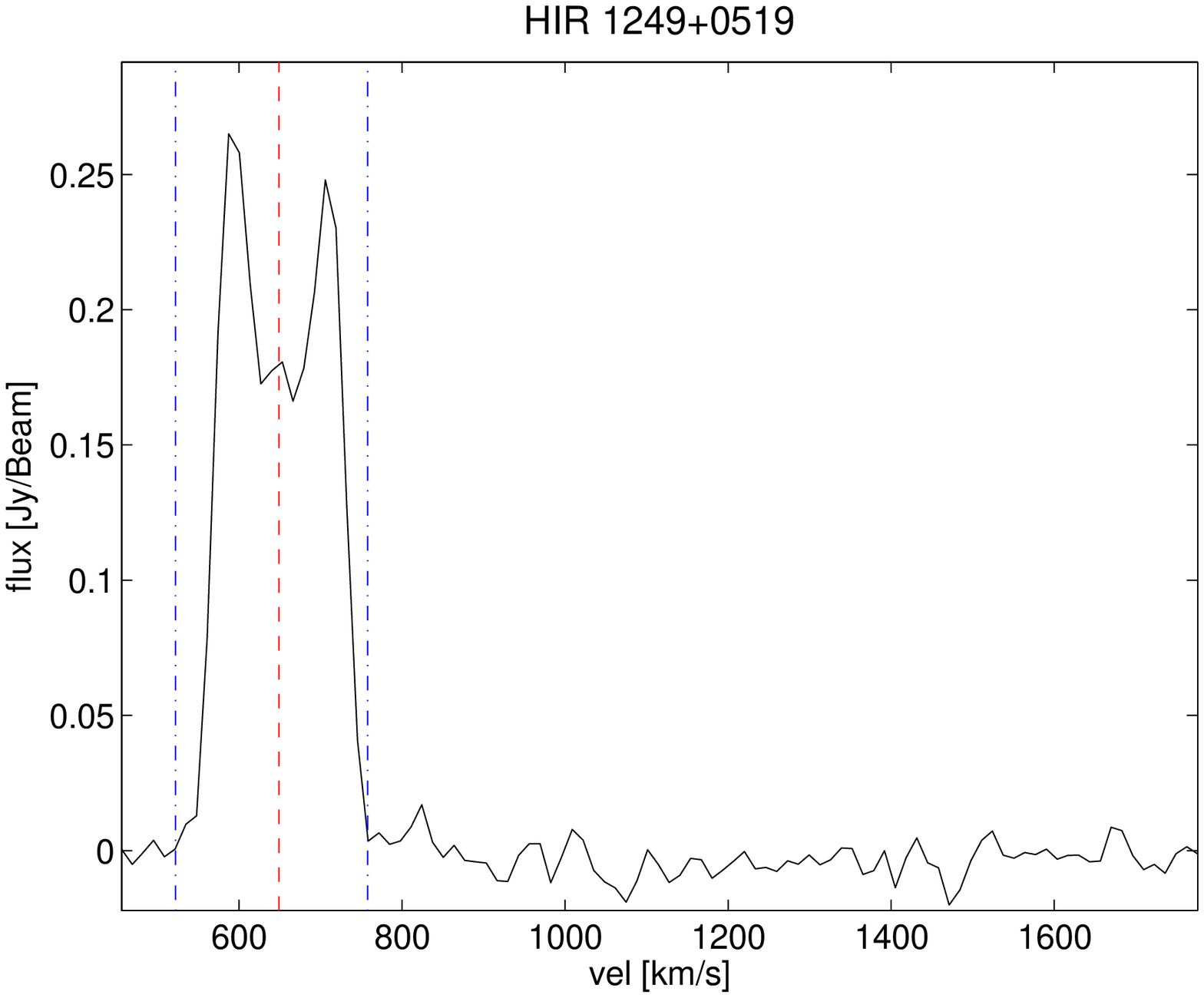}
 \includegraphics[width=0.3\textwidth]{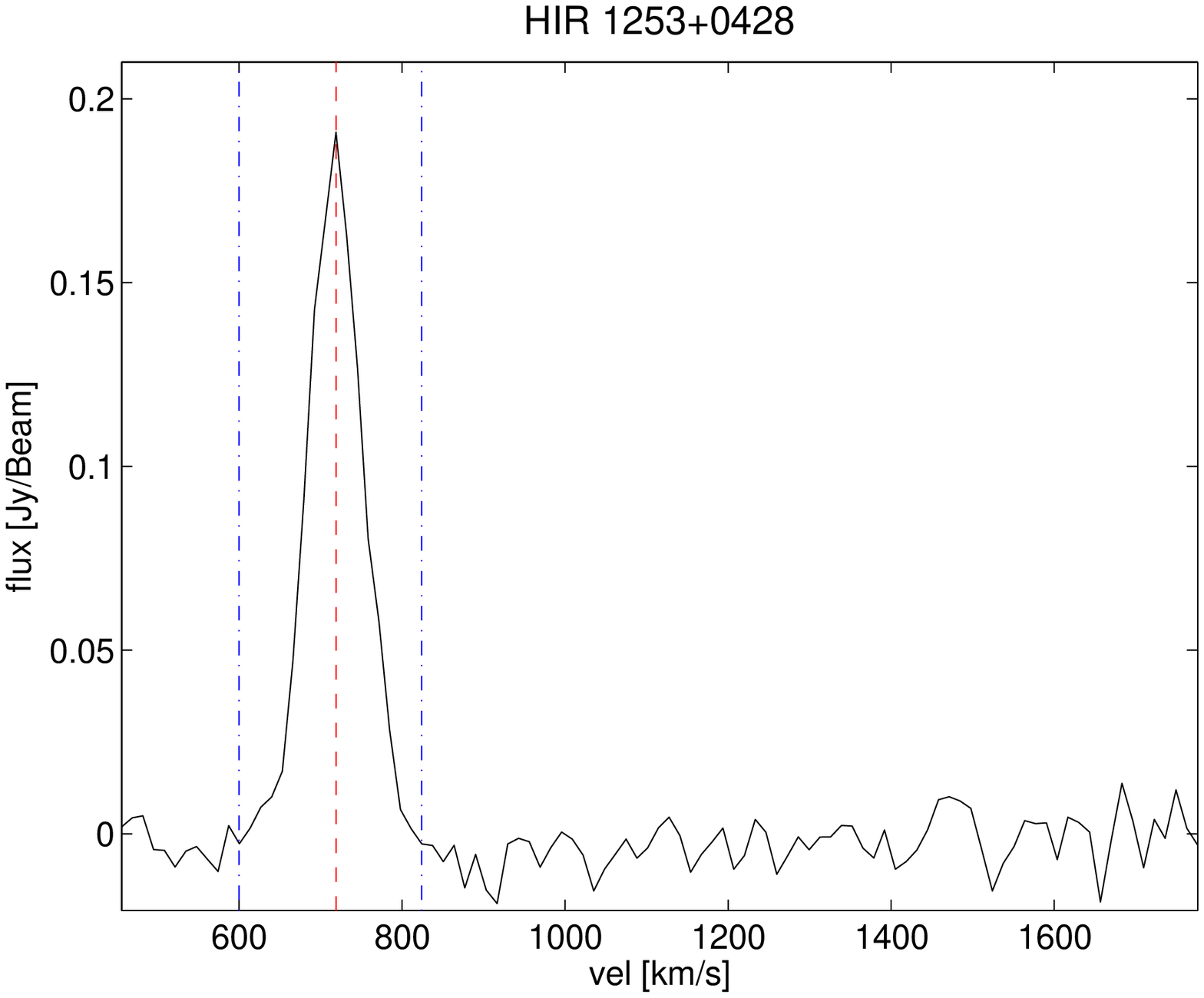}
 \includegraphics[width=0.3\textwidth]{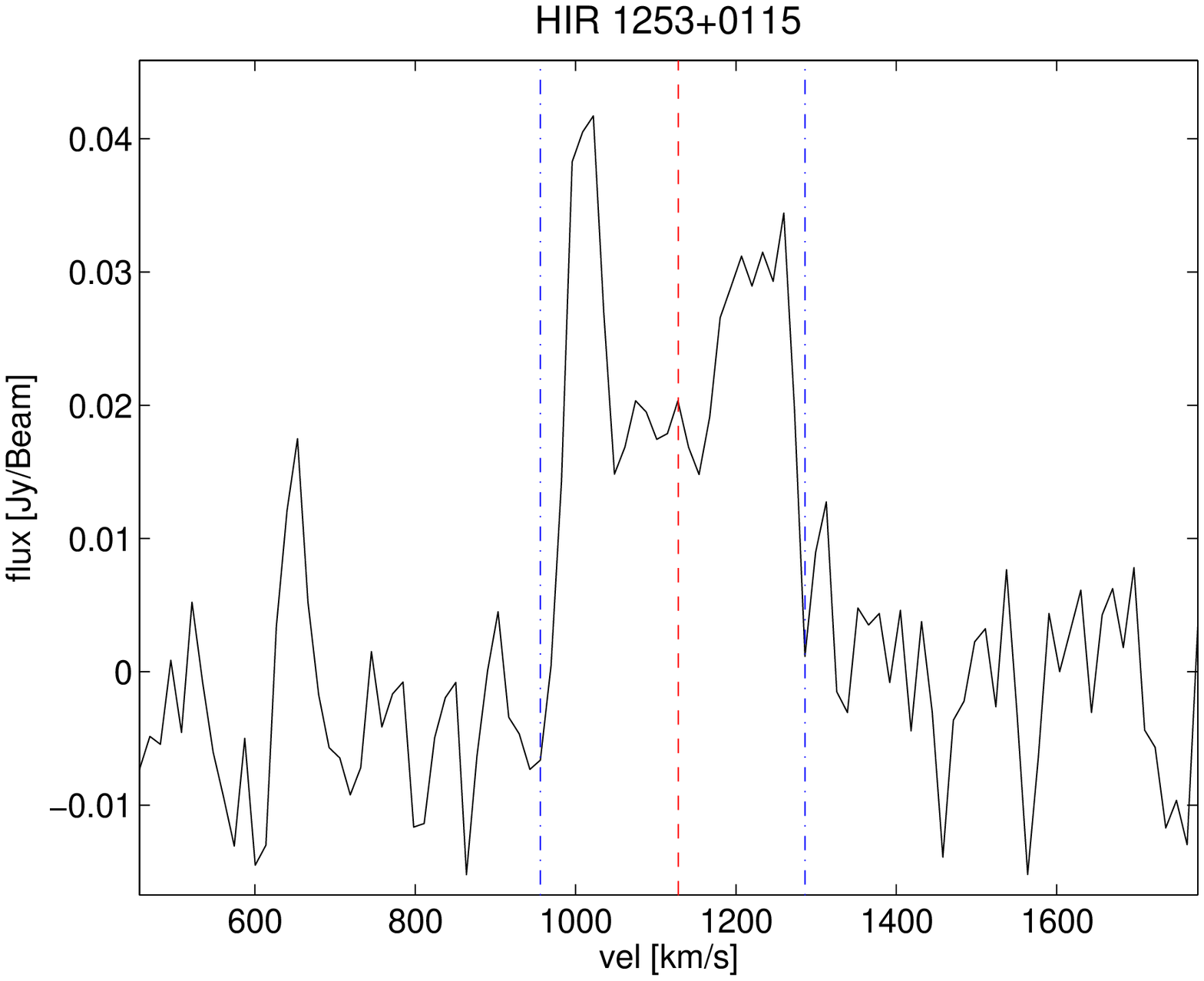}
 \includegraphics[width=0.3\textwidth]{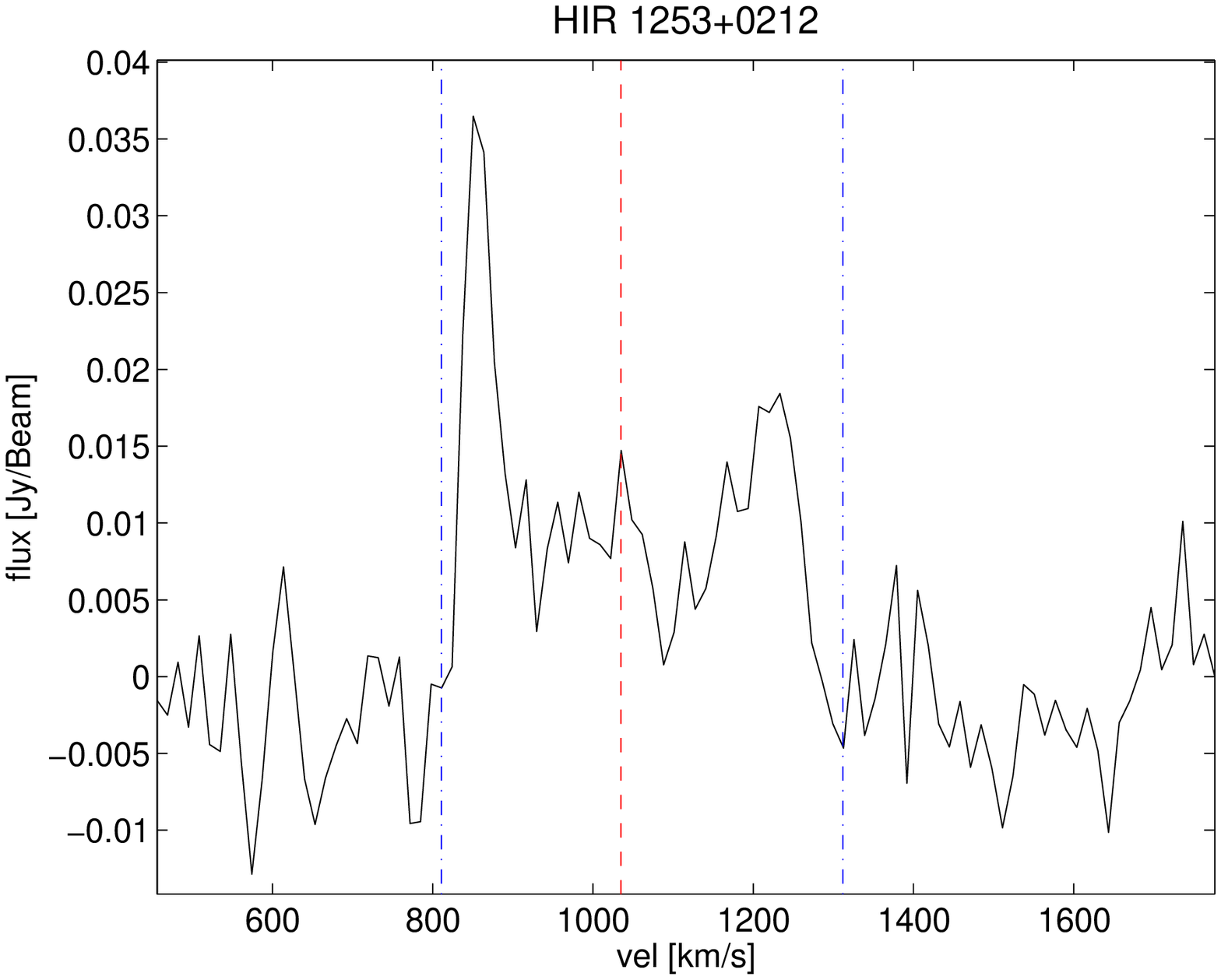}
 \includegraphics[width=0.3\textwidth]{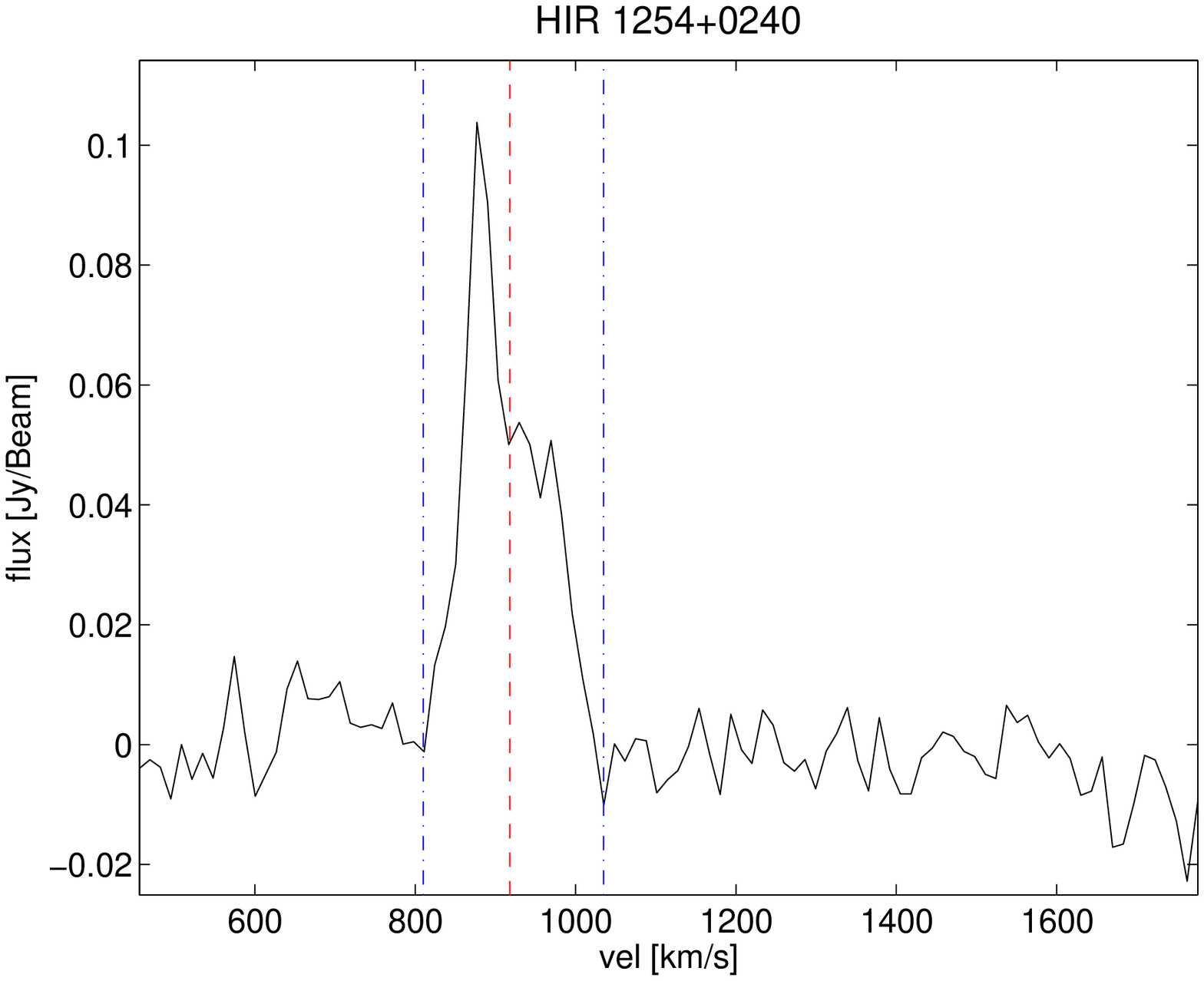}

 \end{center}                                                         
{\bf Fig~\ref{all_spectra}.} (continued)                              
                                                                      
\end{figure*}

\begin{figure*}
  \begin{center}
 \includegraphics[width=0.3\textwidth]{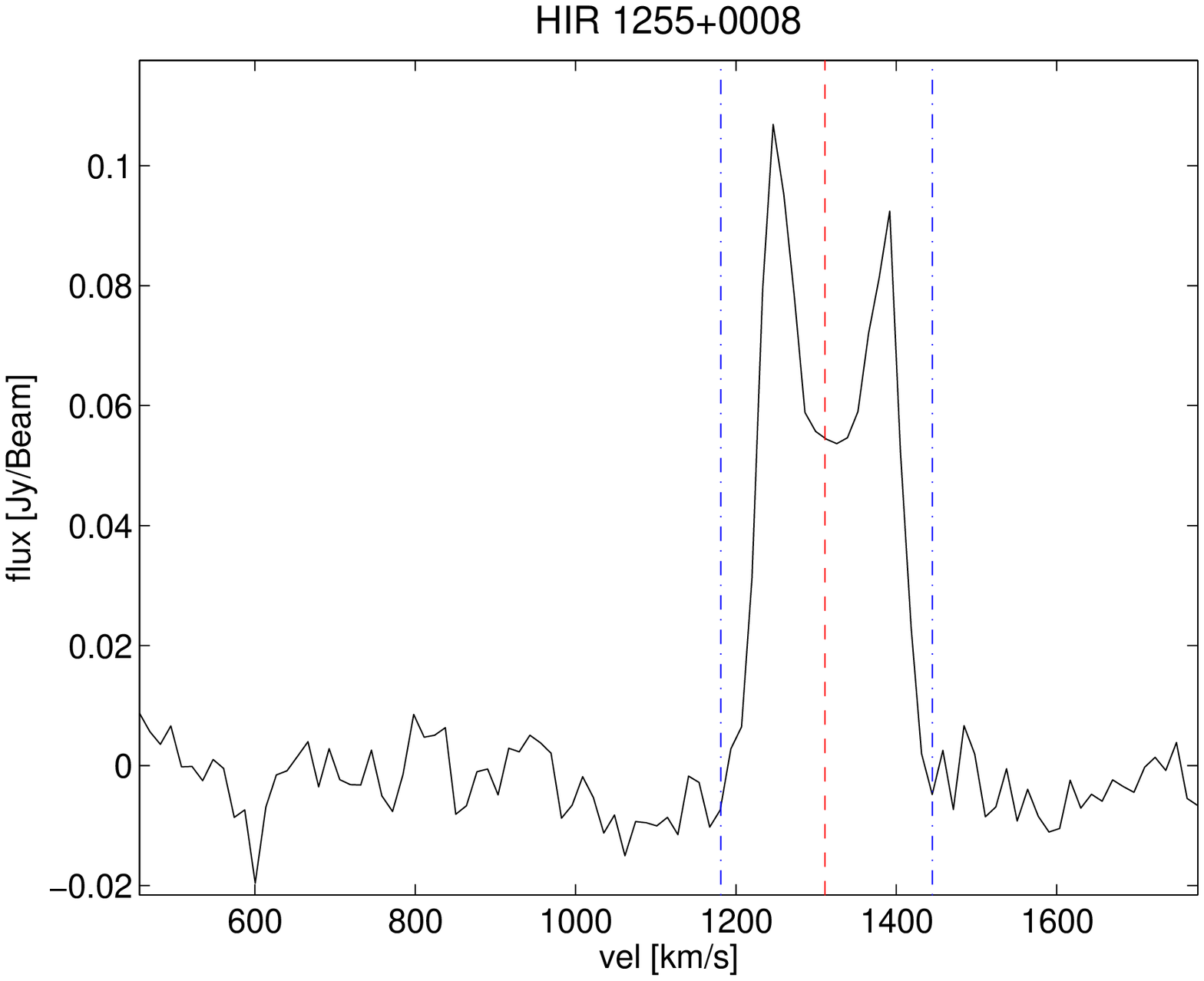}
 \includegraphics[width=0.3\textwidth]{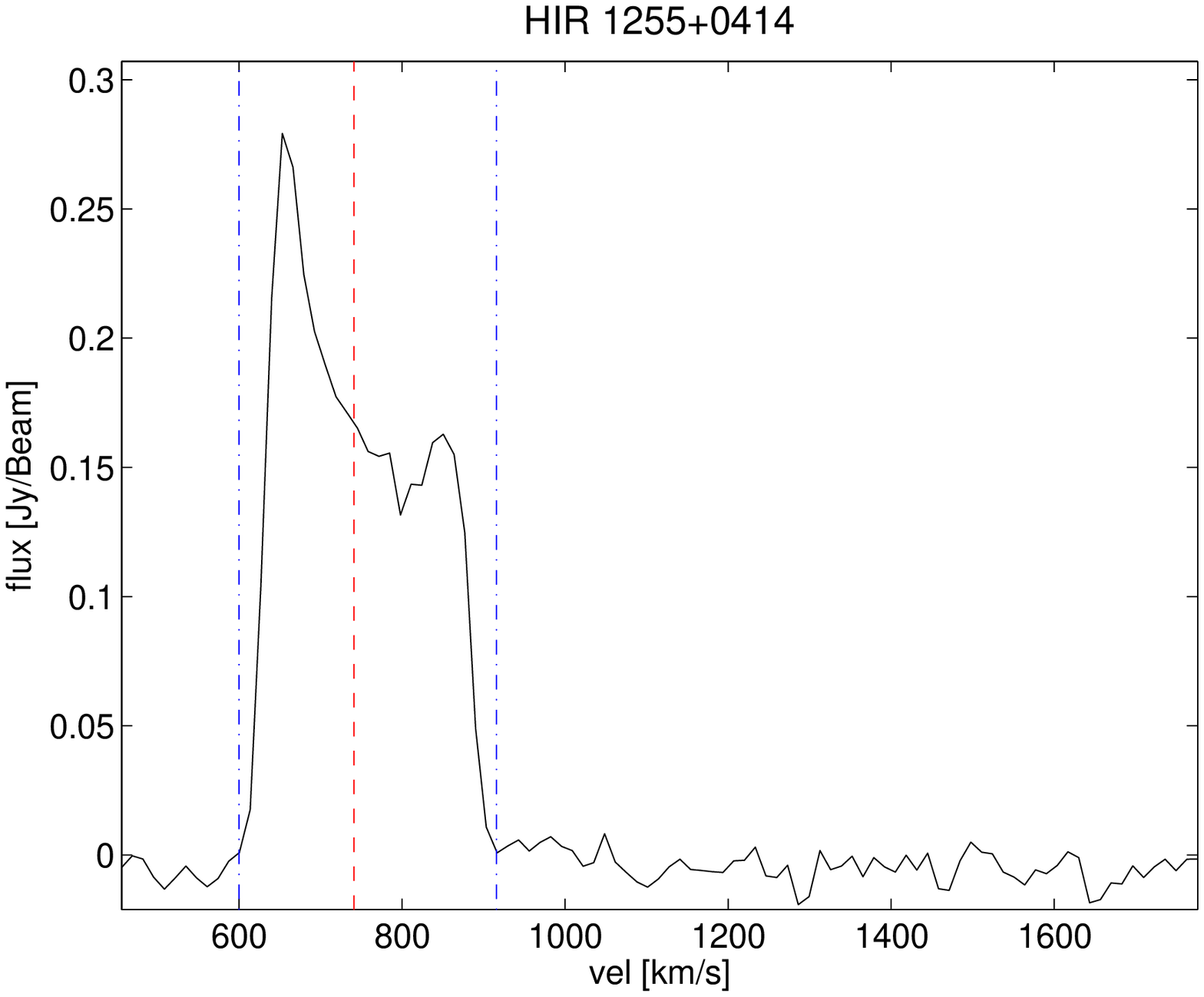}
 \includegraphics[width=0.3\textwidth]{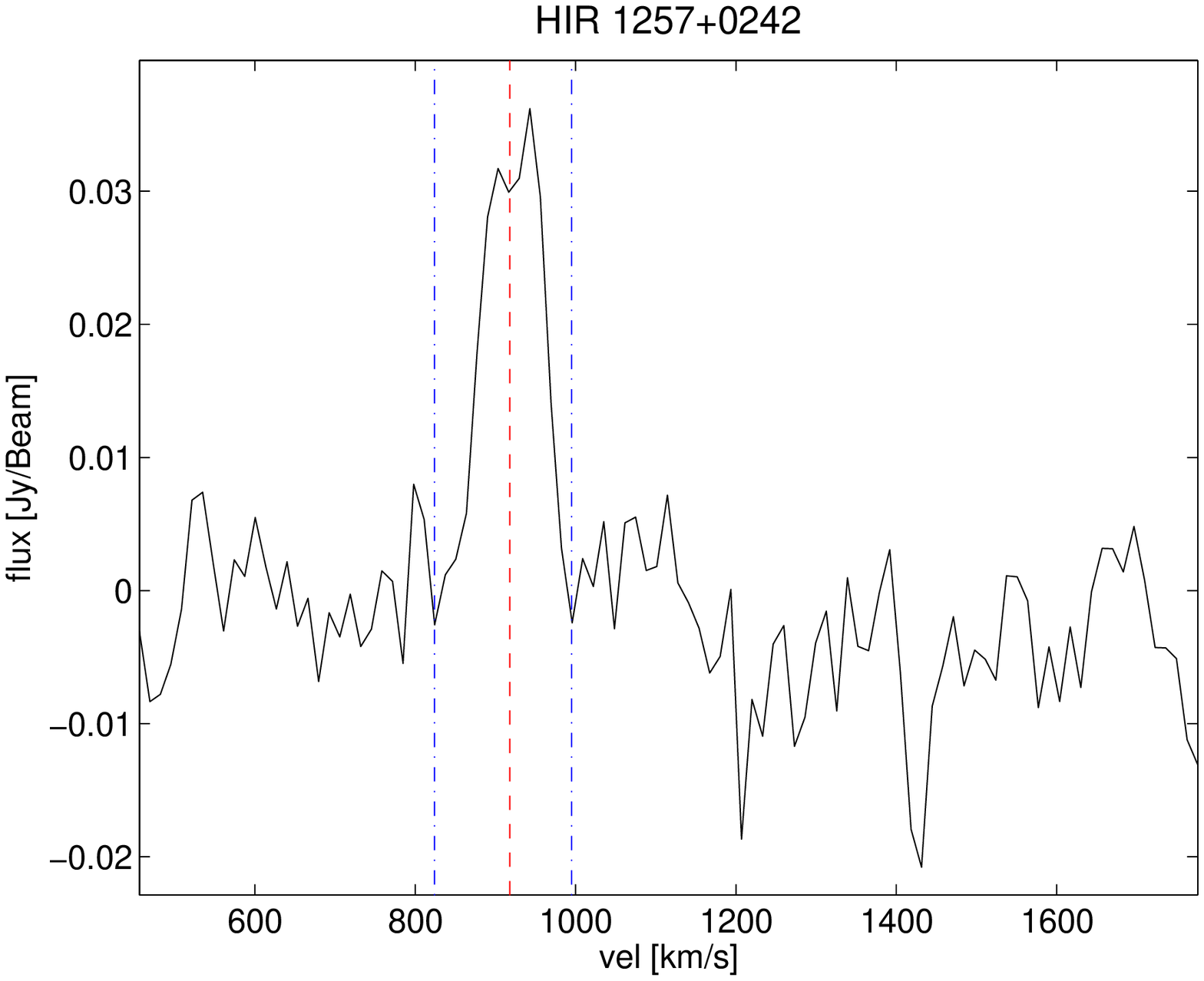}
 \includegraphics[width=0.3\textwidth]{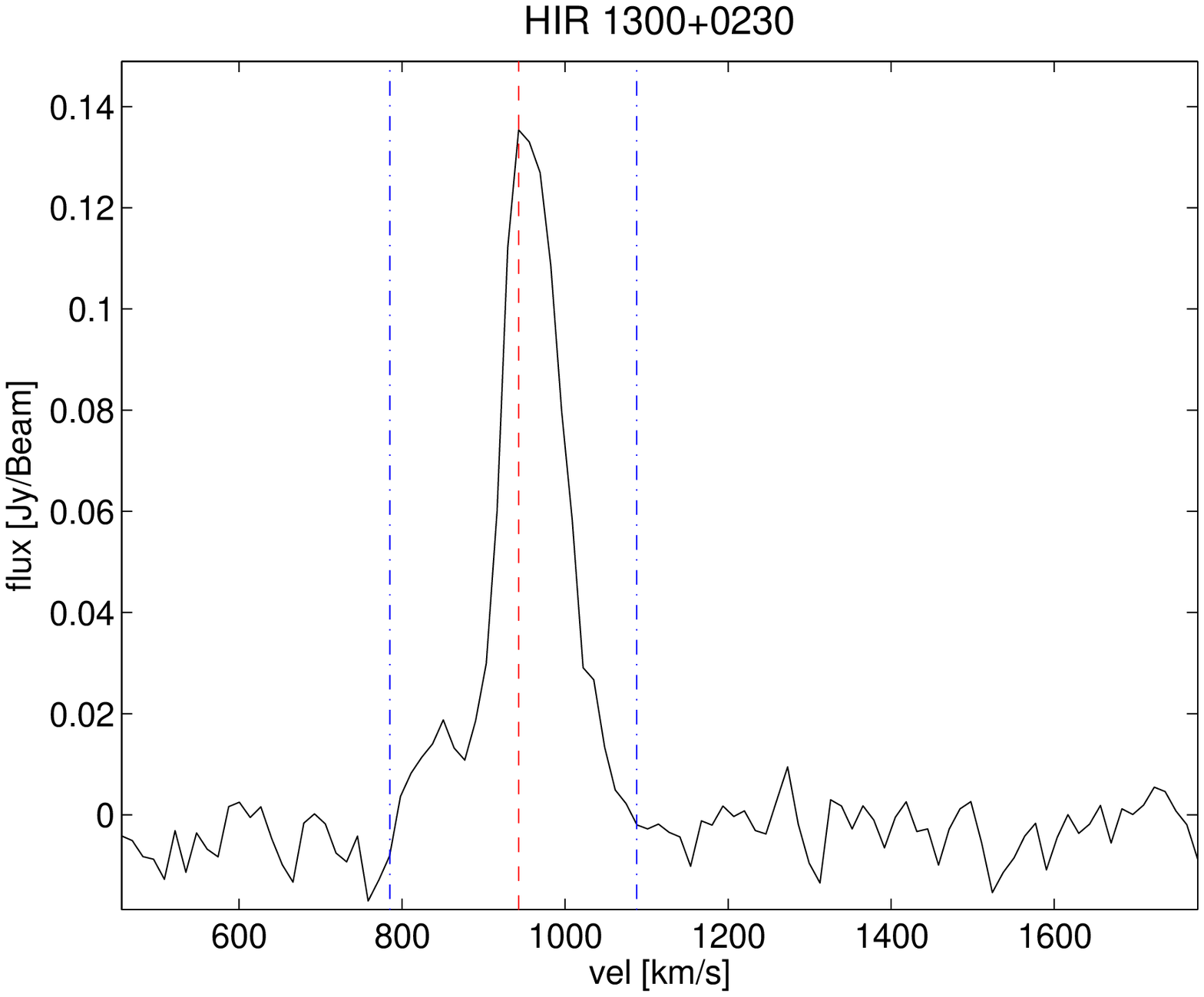}
 \includegraphics[width=0.3\textwidth]{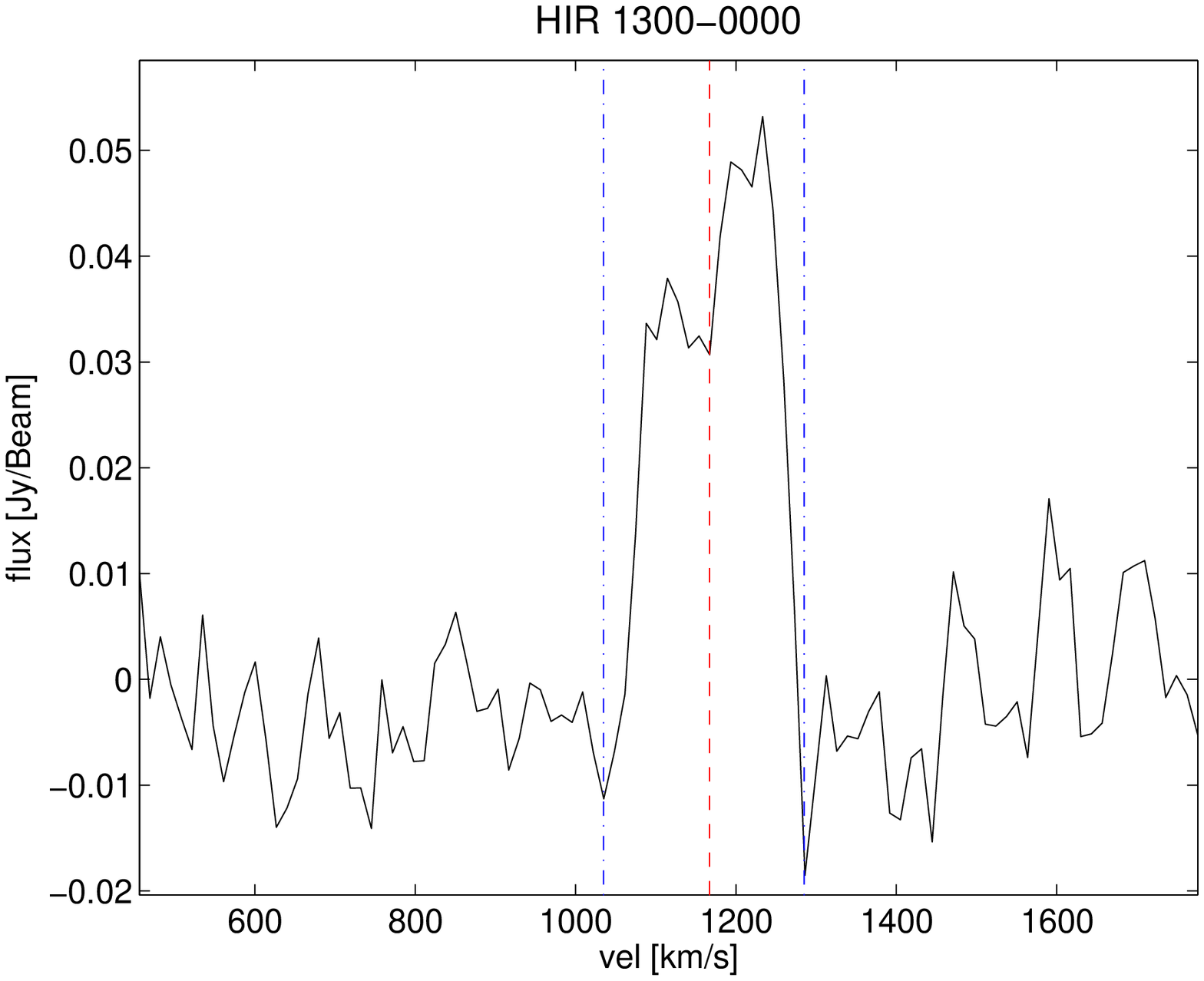}
 \includegraphics[width=0.3\textwidth]{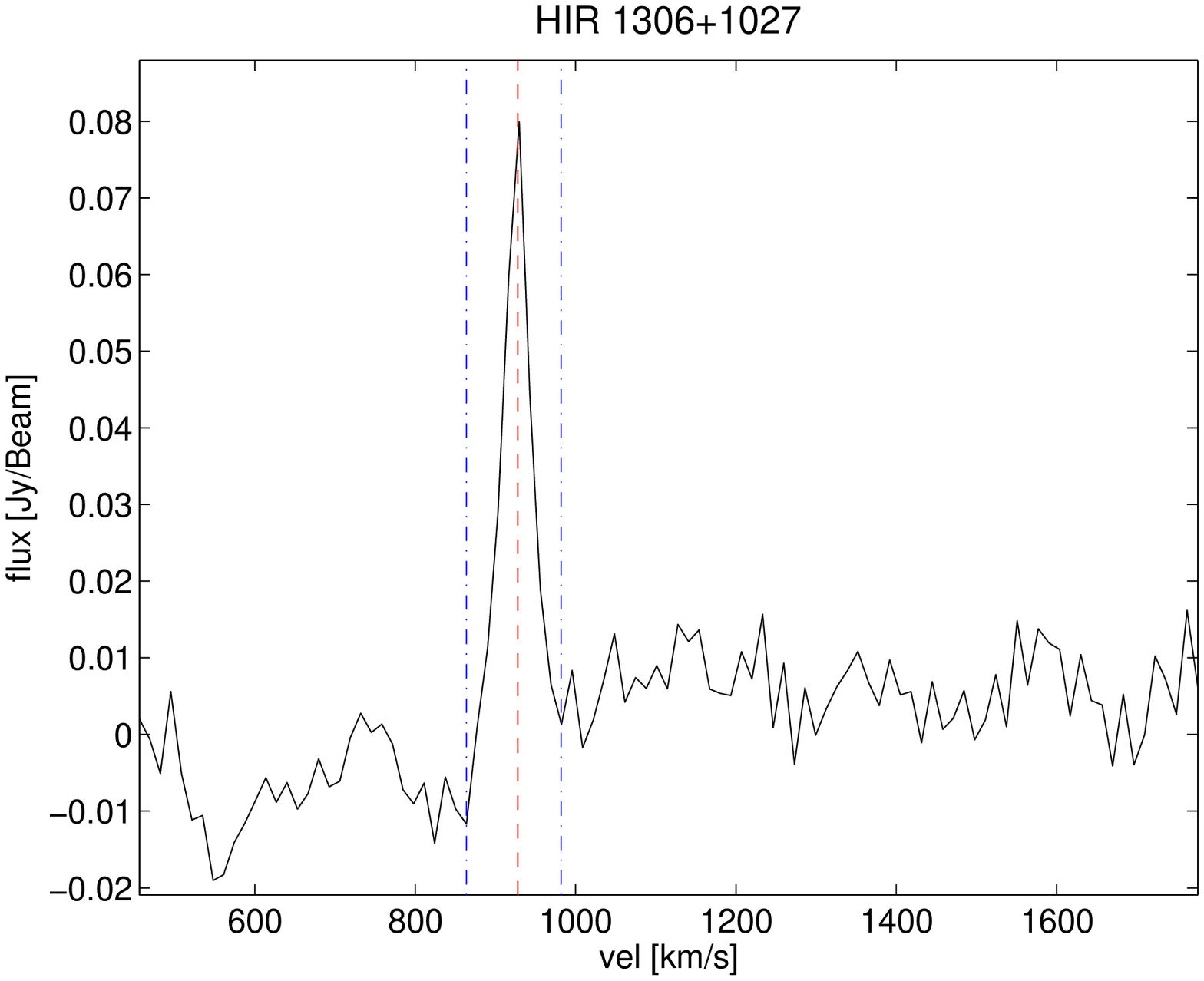}
 \includegraphics[width=0.3\textwidth]{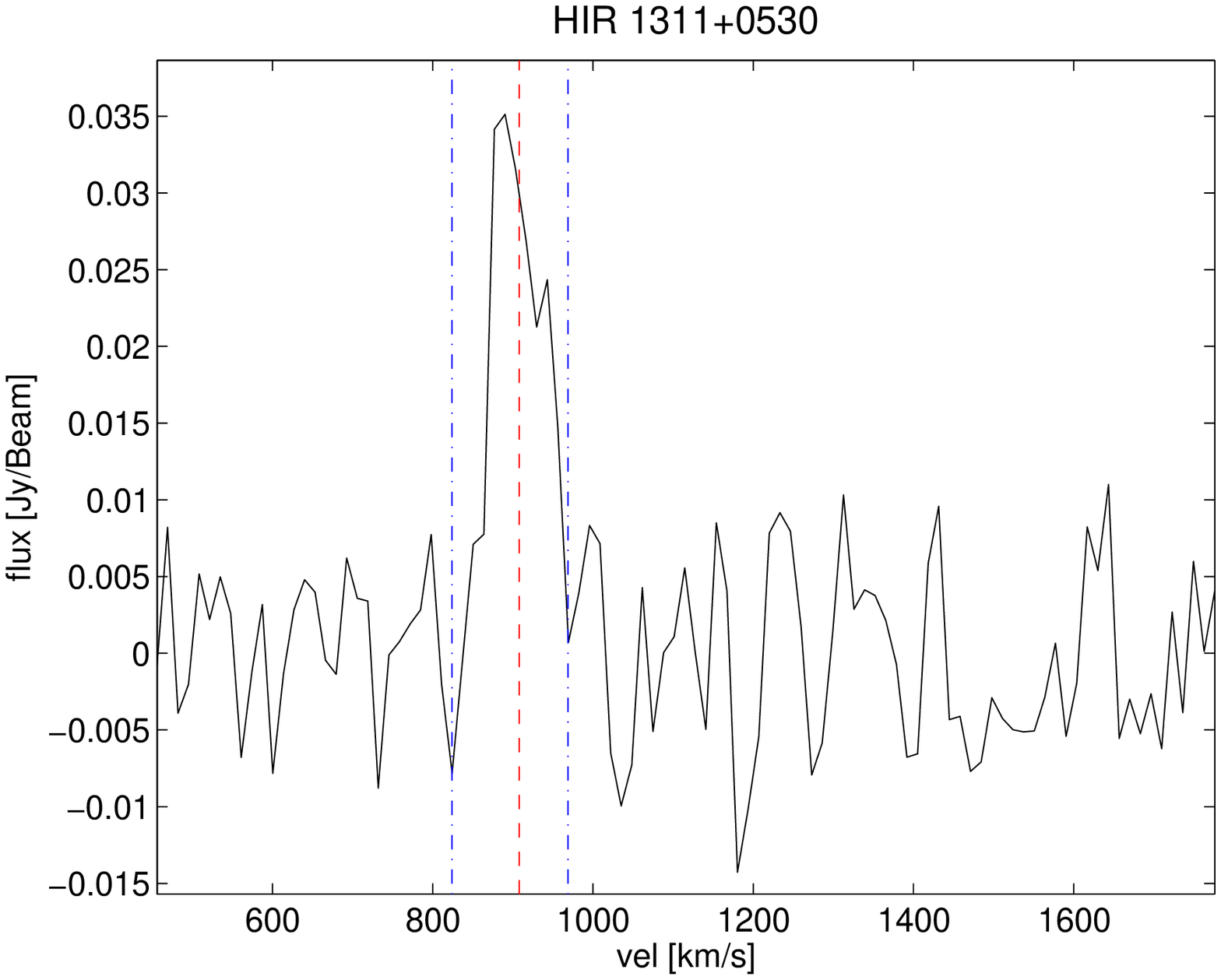}
 \includegraphics[width=0.3\textwidth]{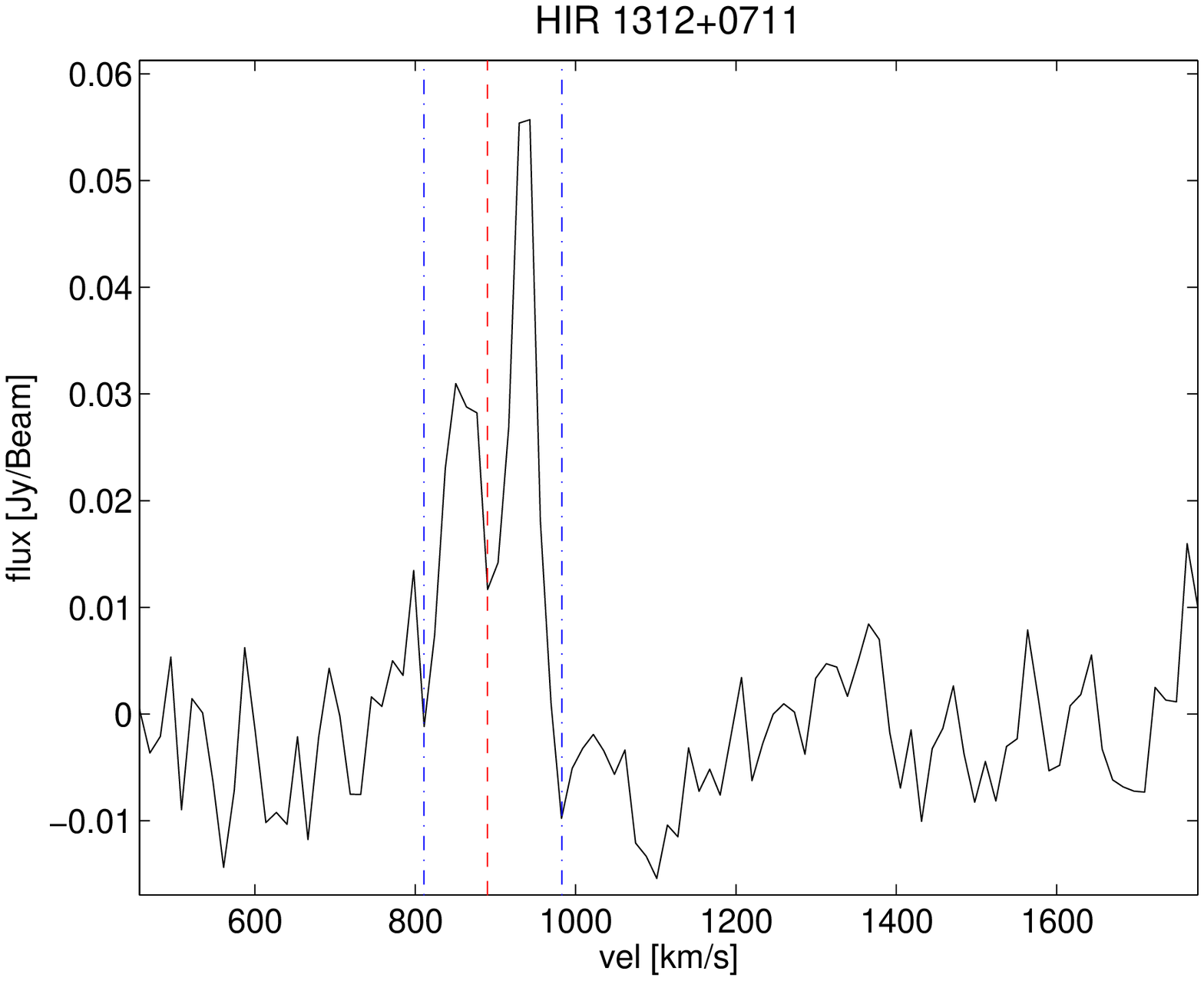}
 \includegraphics[width=0.3\textwidth]{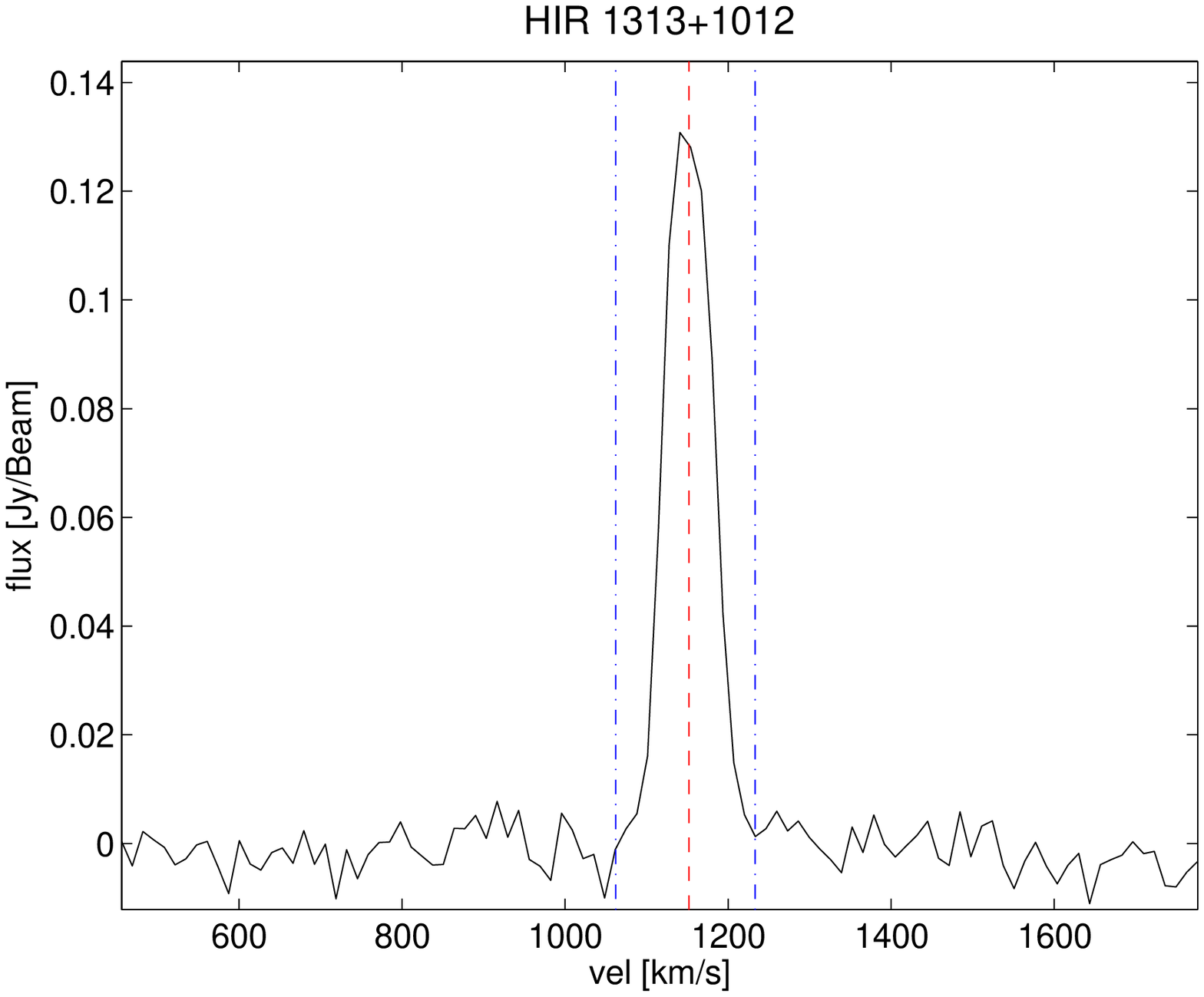}
 \includegraphics[width=0.3\textwidth]{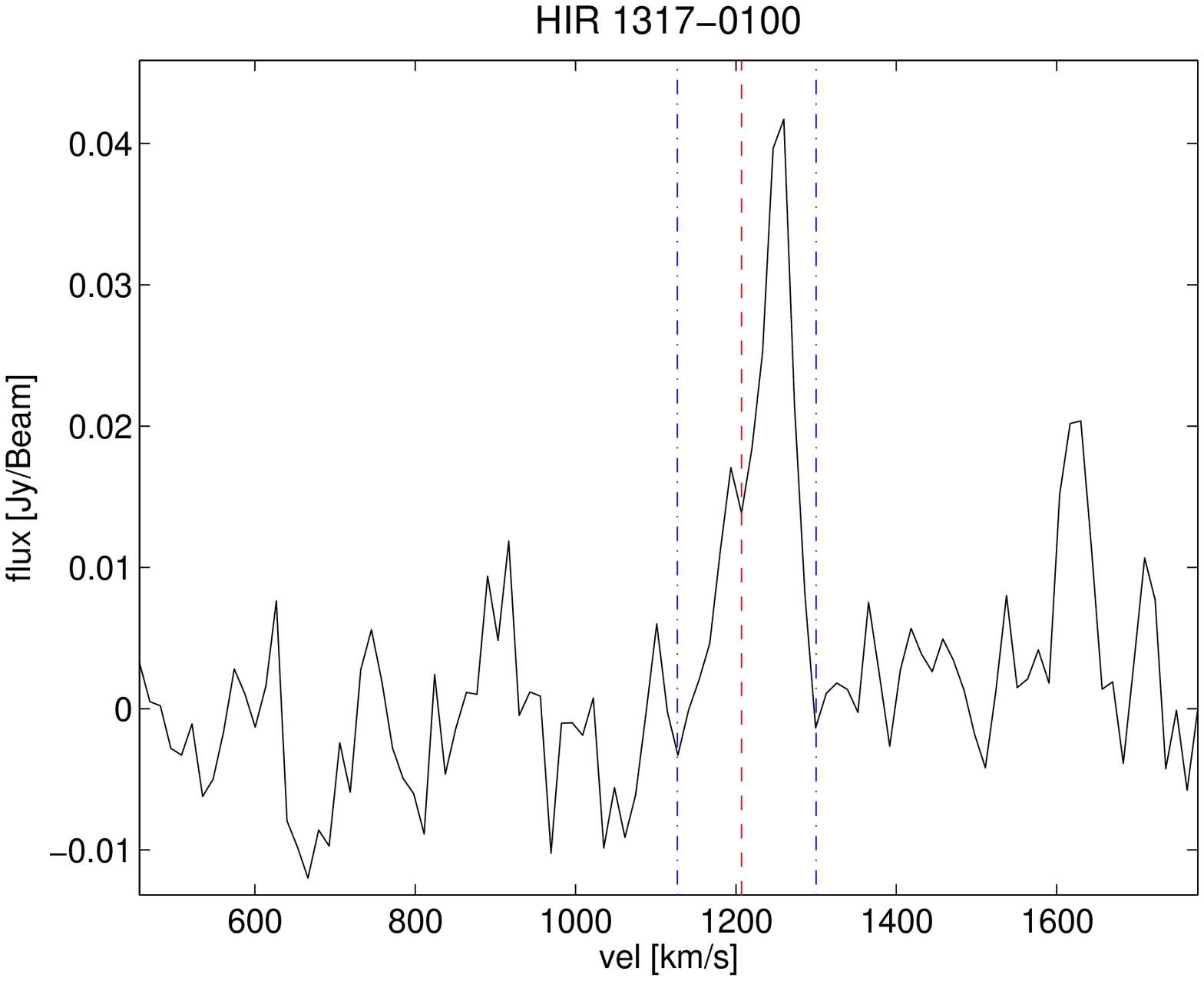}
 \includegraphics[width=0.3\textwidth]{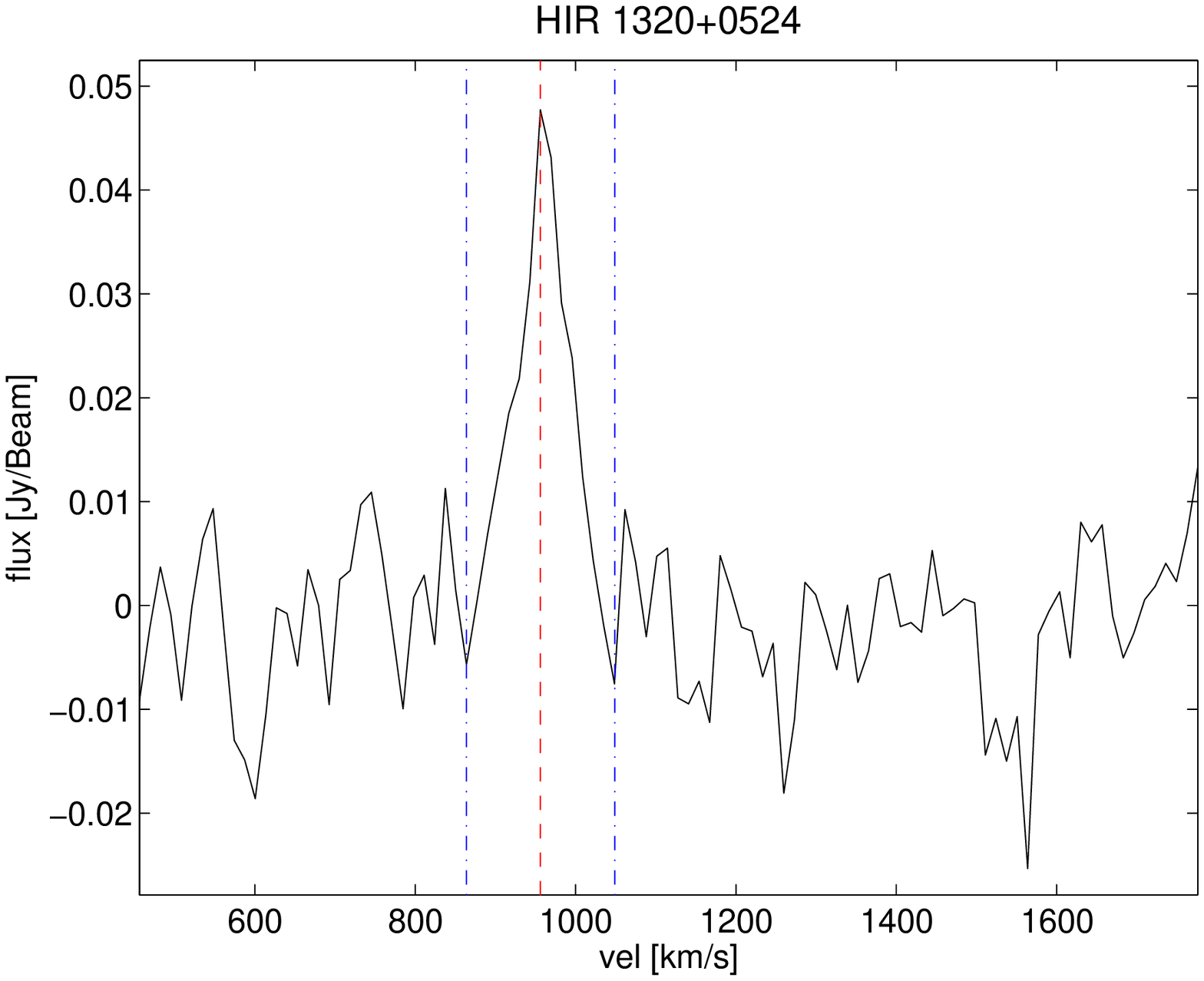}
 \includegraphics[width=0.3\textwidth]{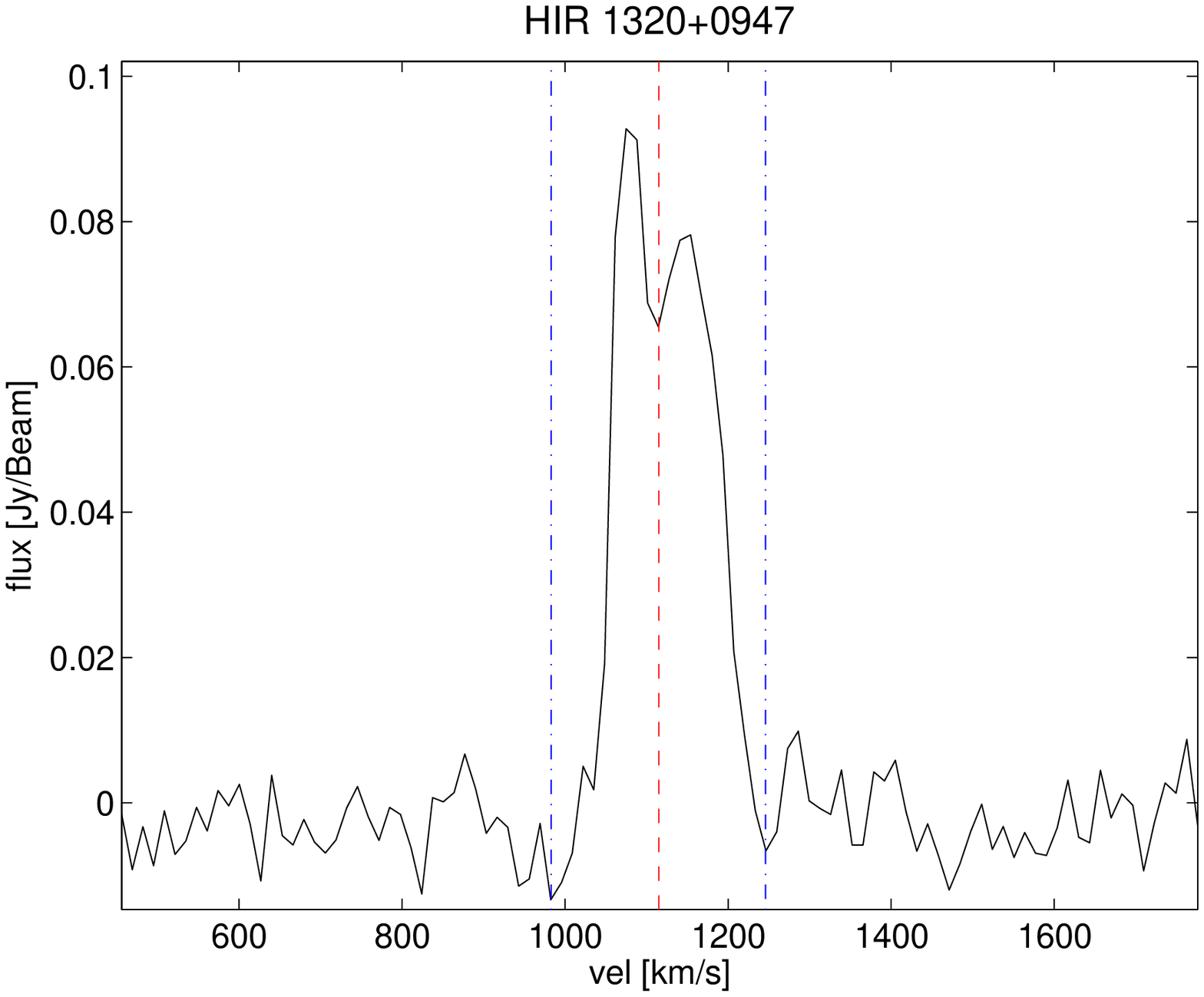}
 \includegraphics[width=0.3\textwidth]{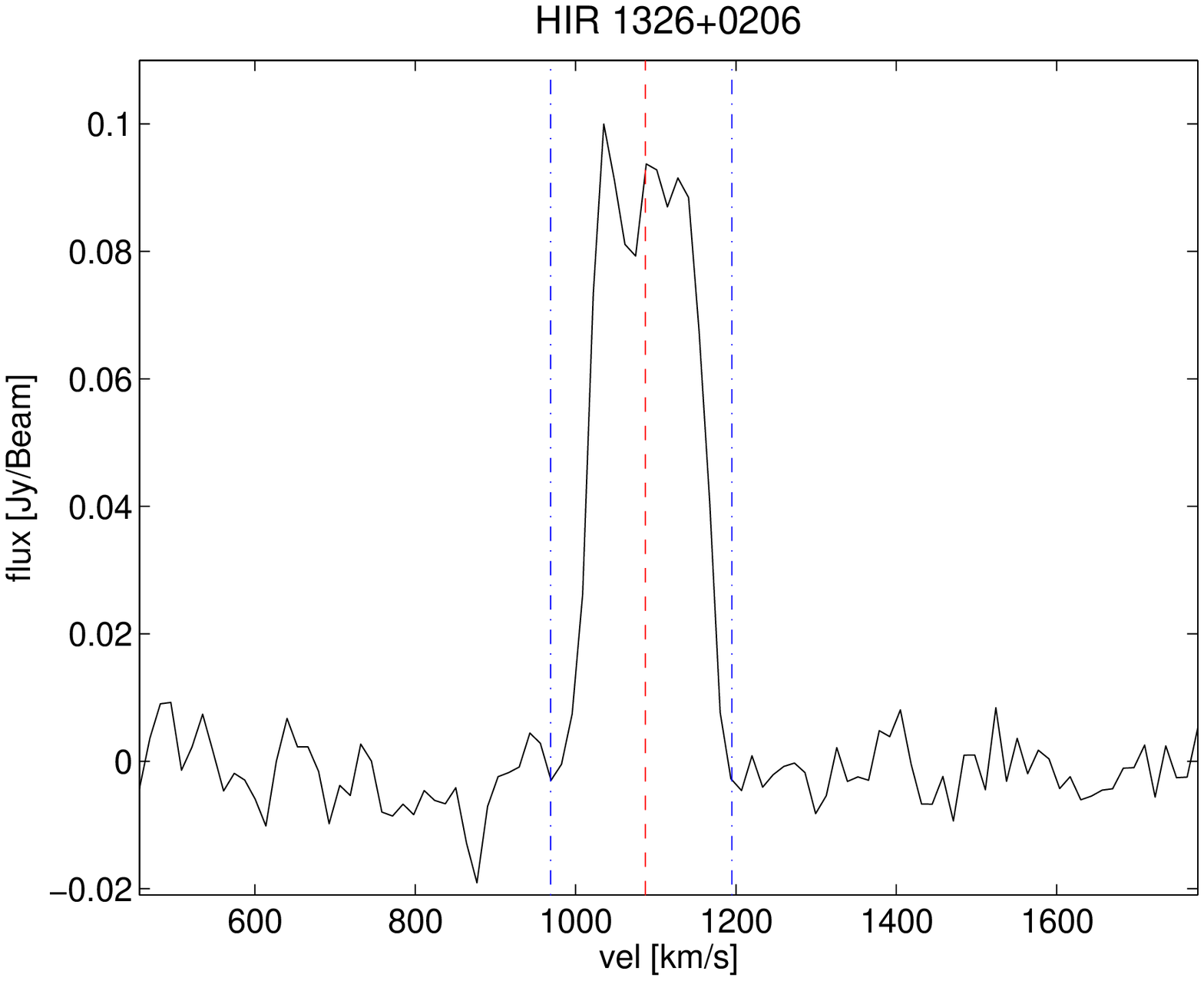}
 \includegraphics[width=0.3\textwidth]{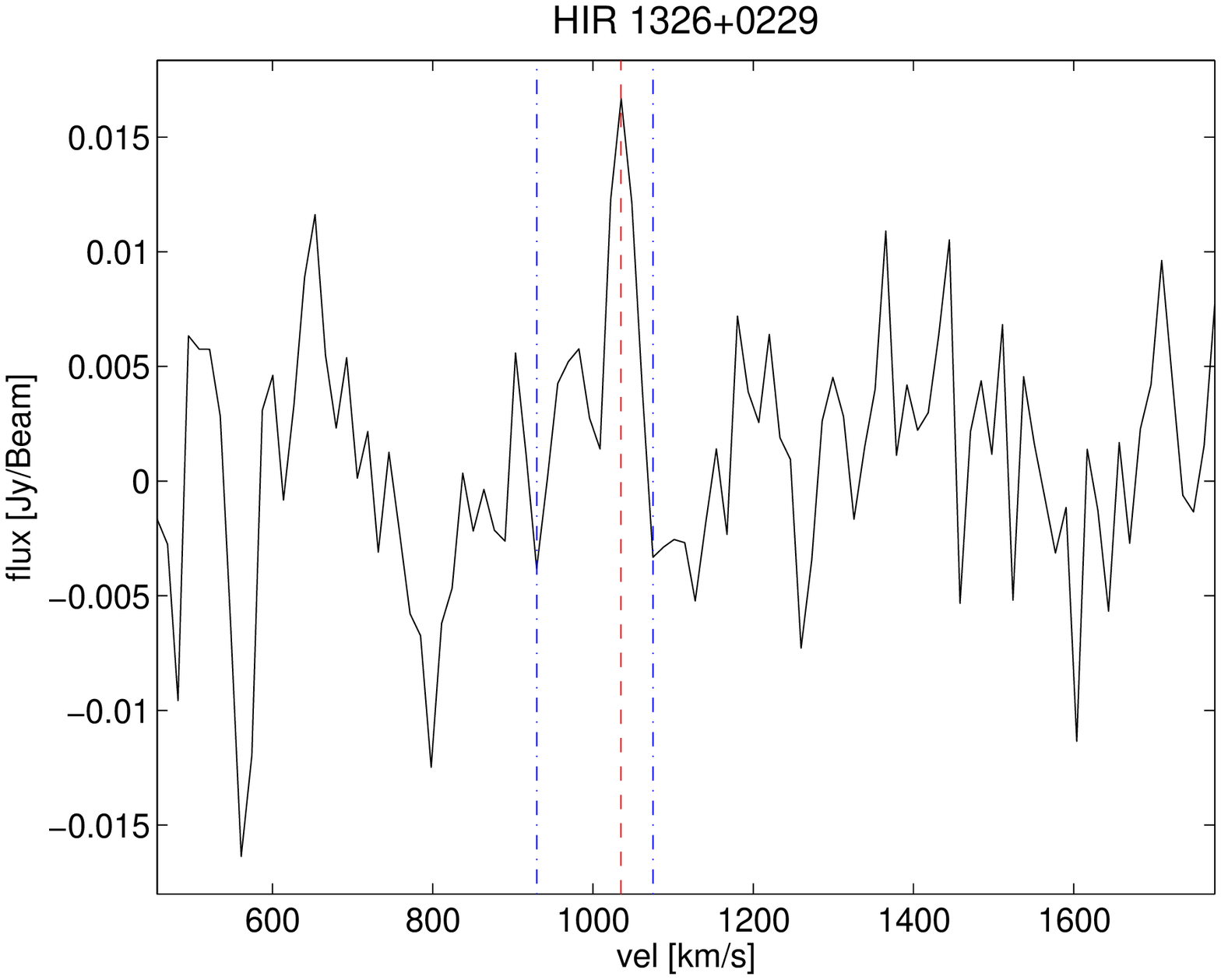}
 \includegraphics[width=0.3\textwidth]{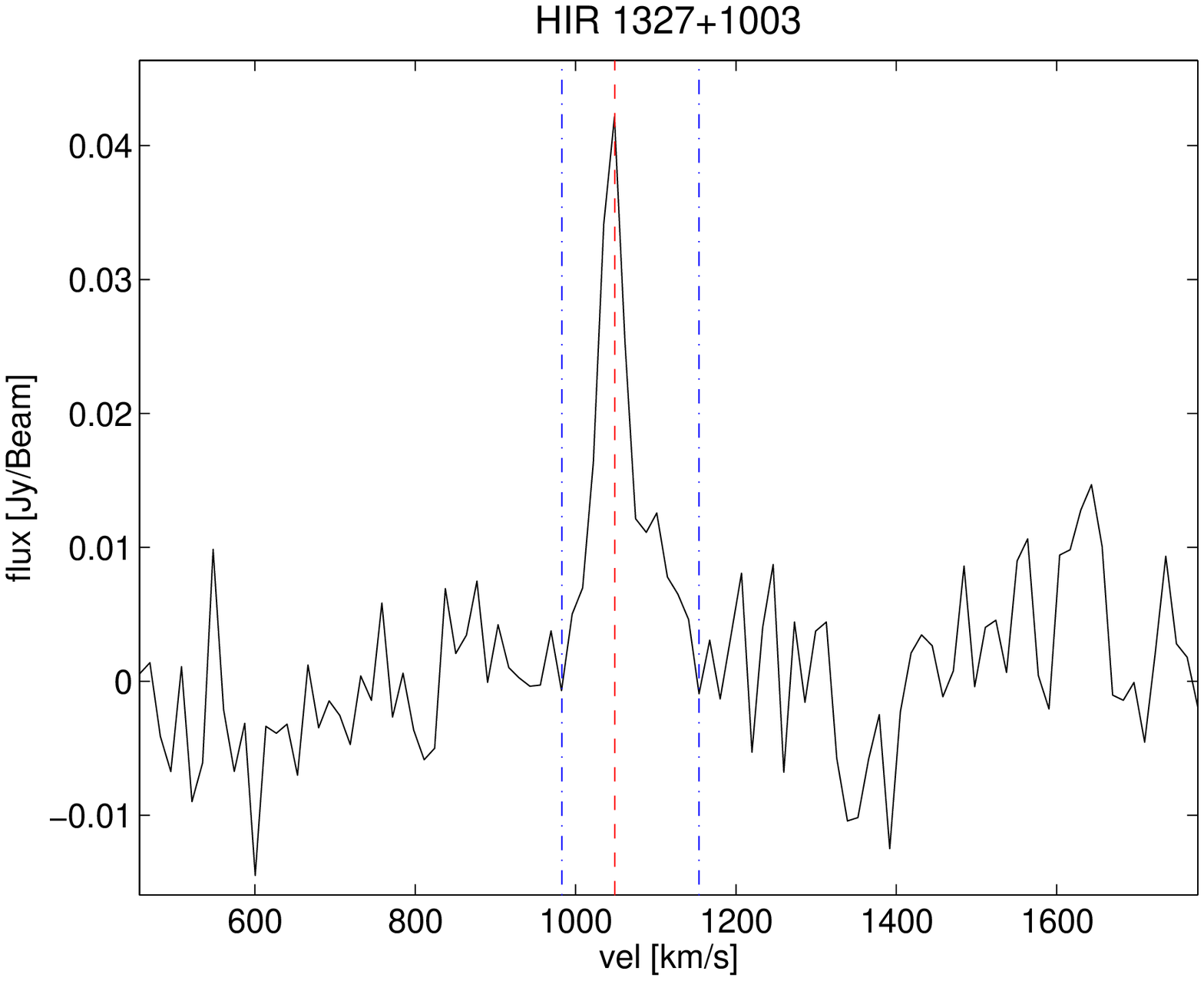}

 \end{center}                                                         
{\bf Fig~\ref{all_spectra}.} (continued)                              
                                                                      
\end{figure*}

\begin{figure*}
  \begin{center}
  
 \includegraphics[width=0.3\textwidth]{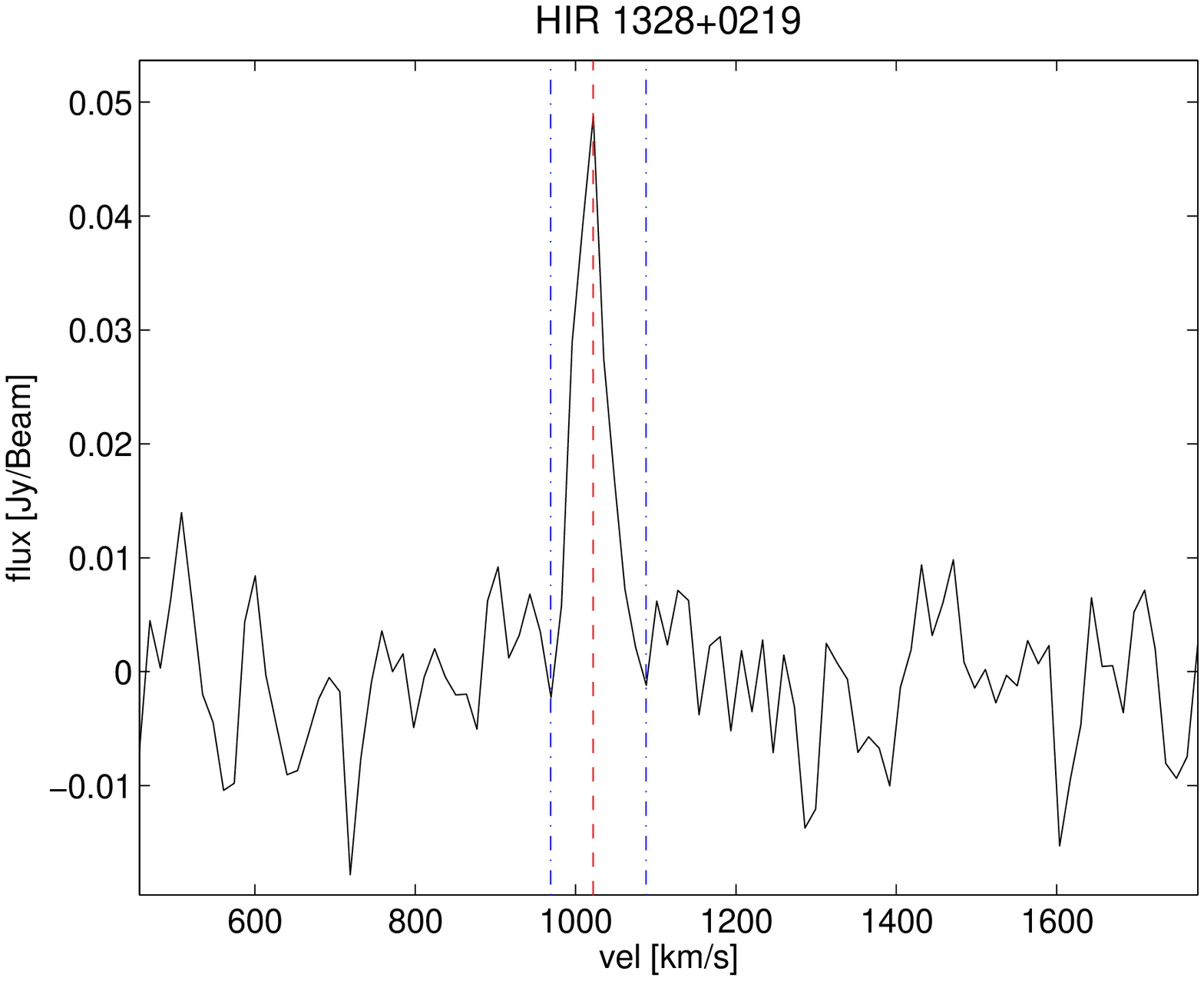}
 \includegraphics[width=0.3\textwidth]{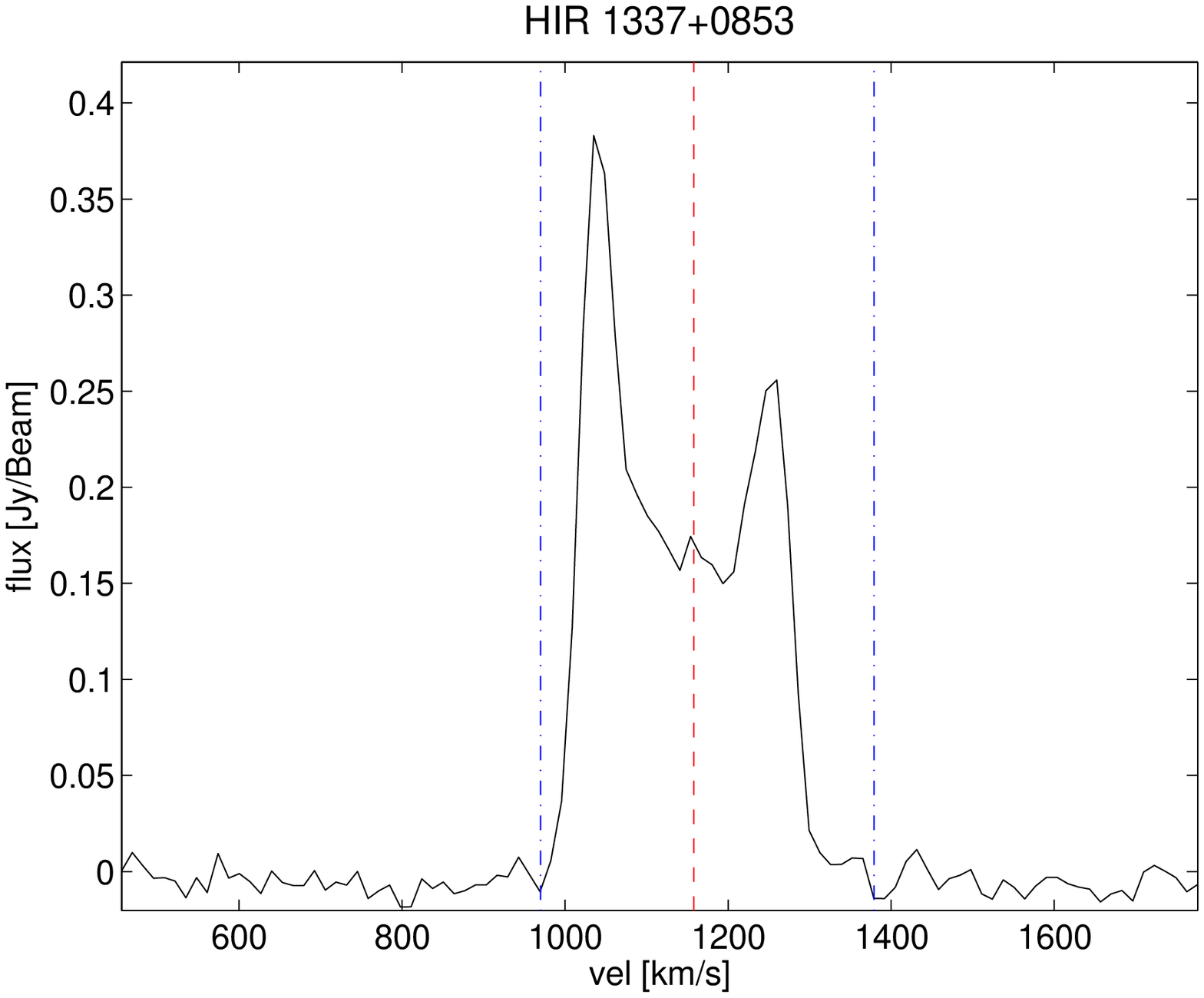}
 \includegraphics[width=0.3\textwidth]{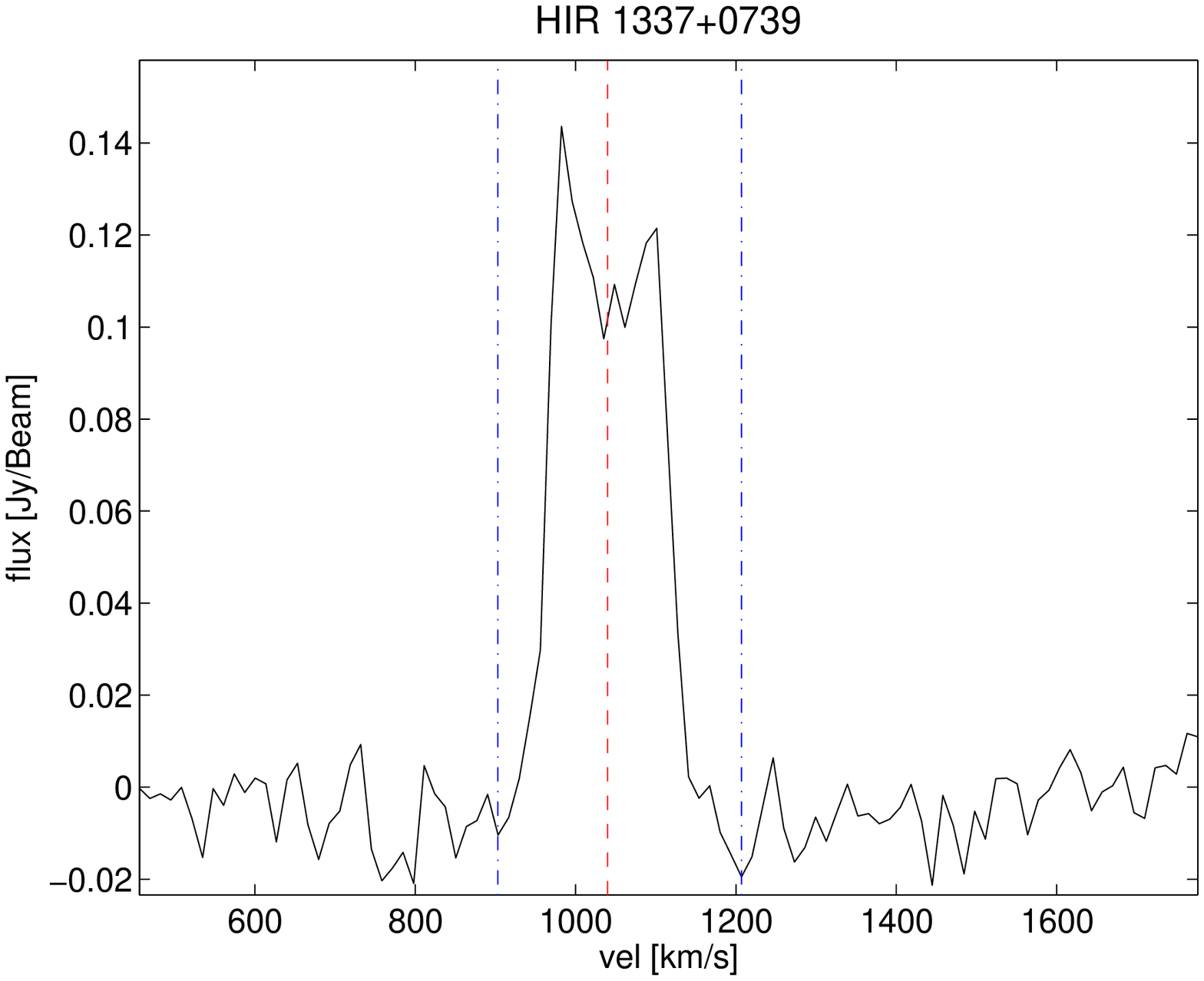}
 \includegraphics[width=0.3\textwidth]{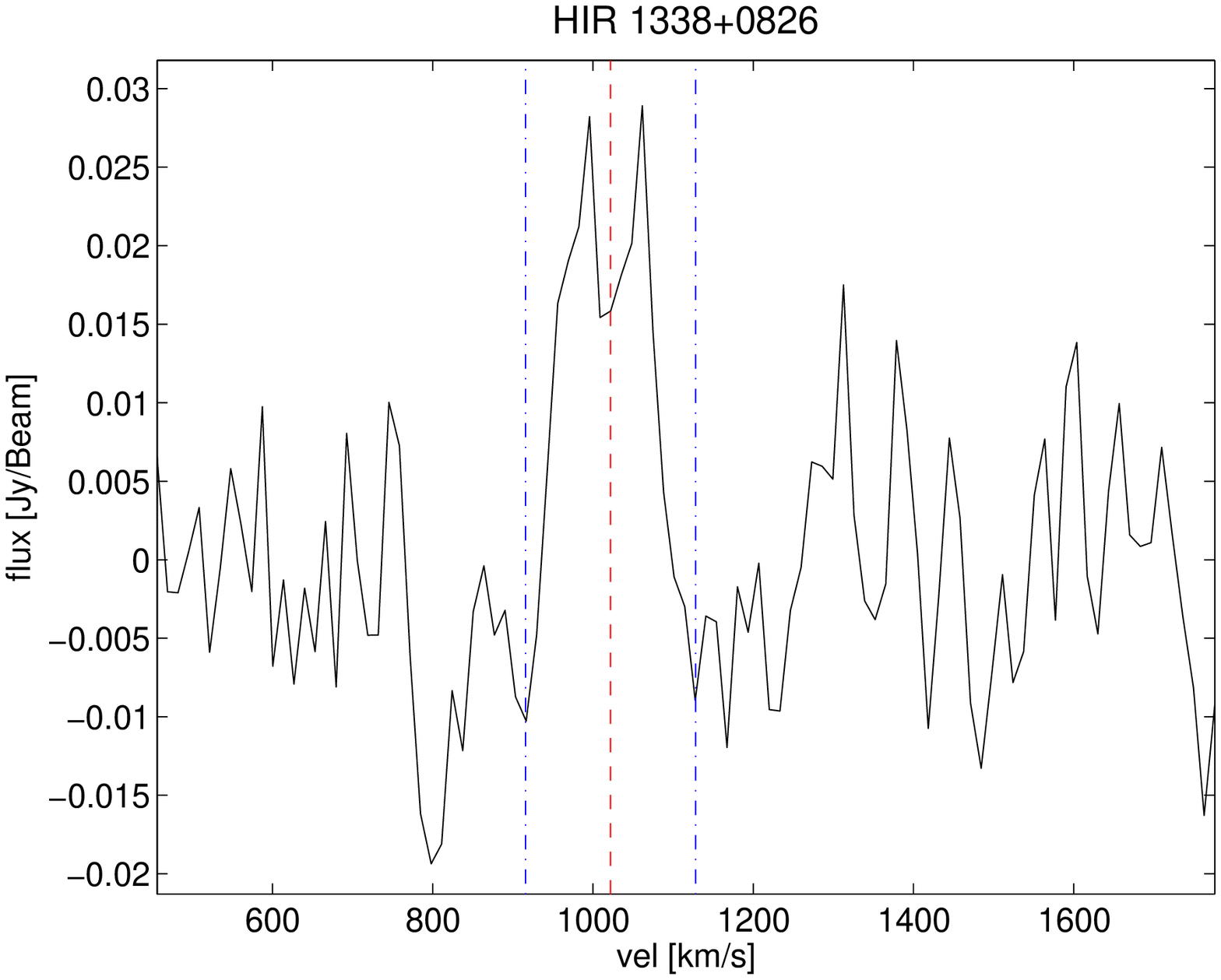}
 \includegraphics[width=0.3\textwidth]{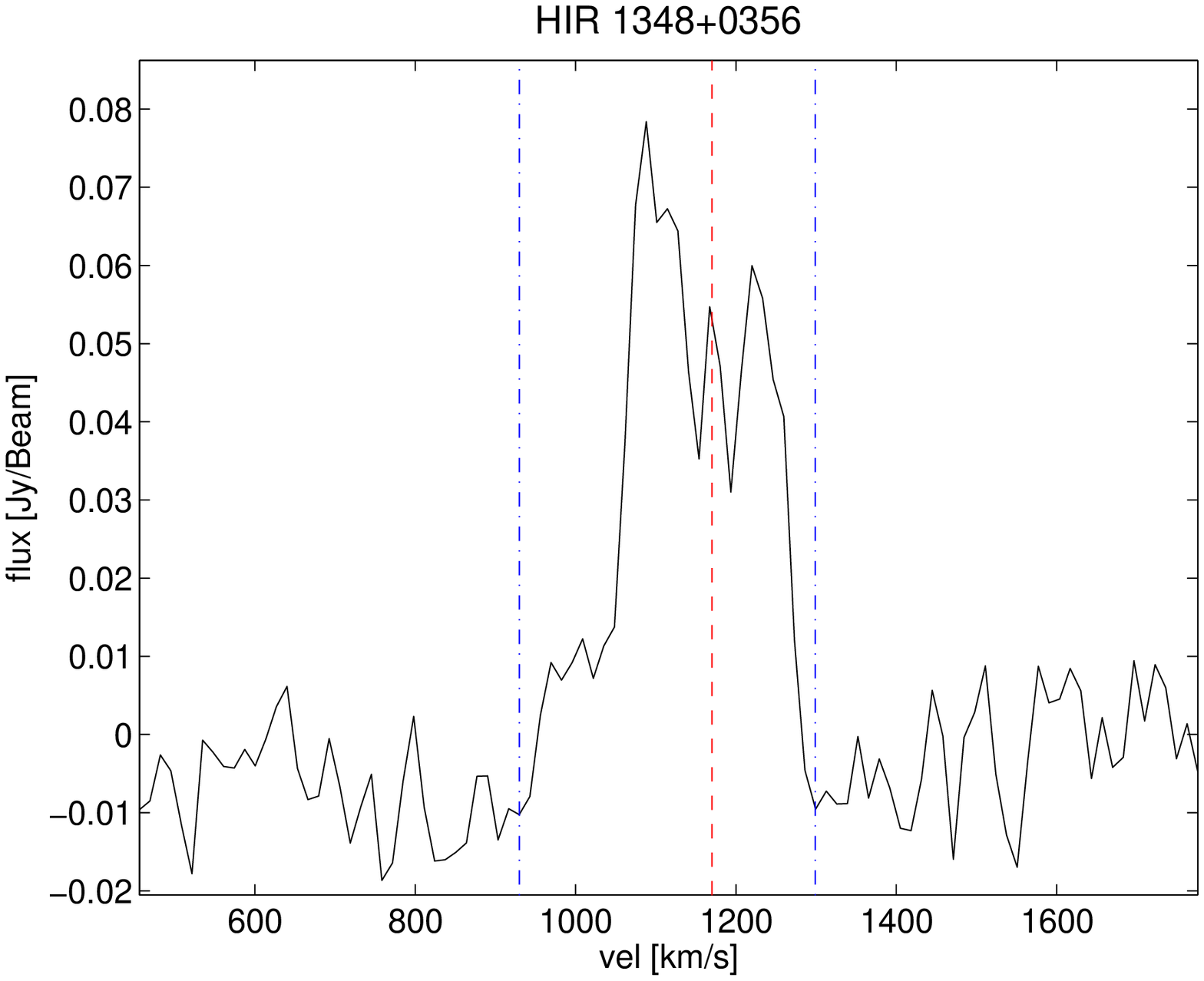}
 \includegraphics[width=0.3\textwidth]{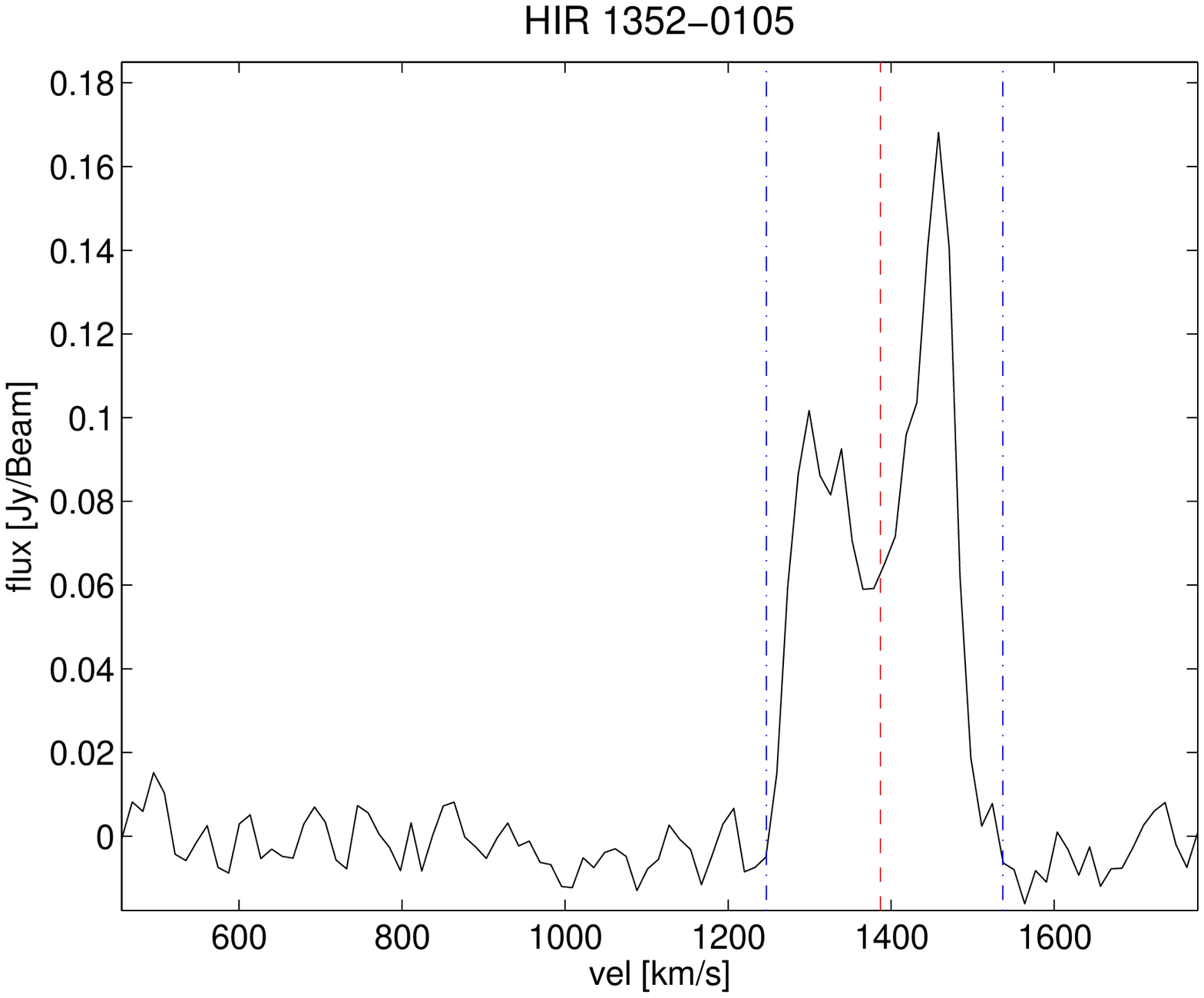}
 \includegraphics[width=0.3\textwidth]{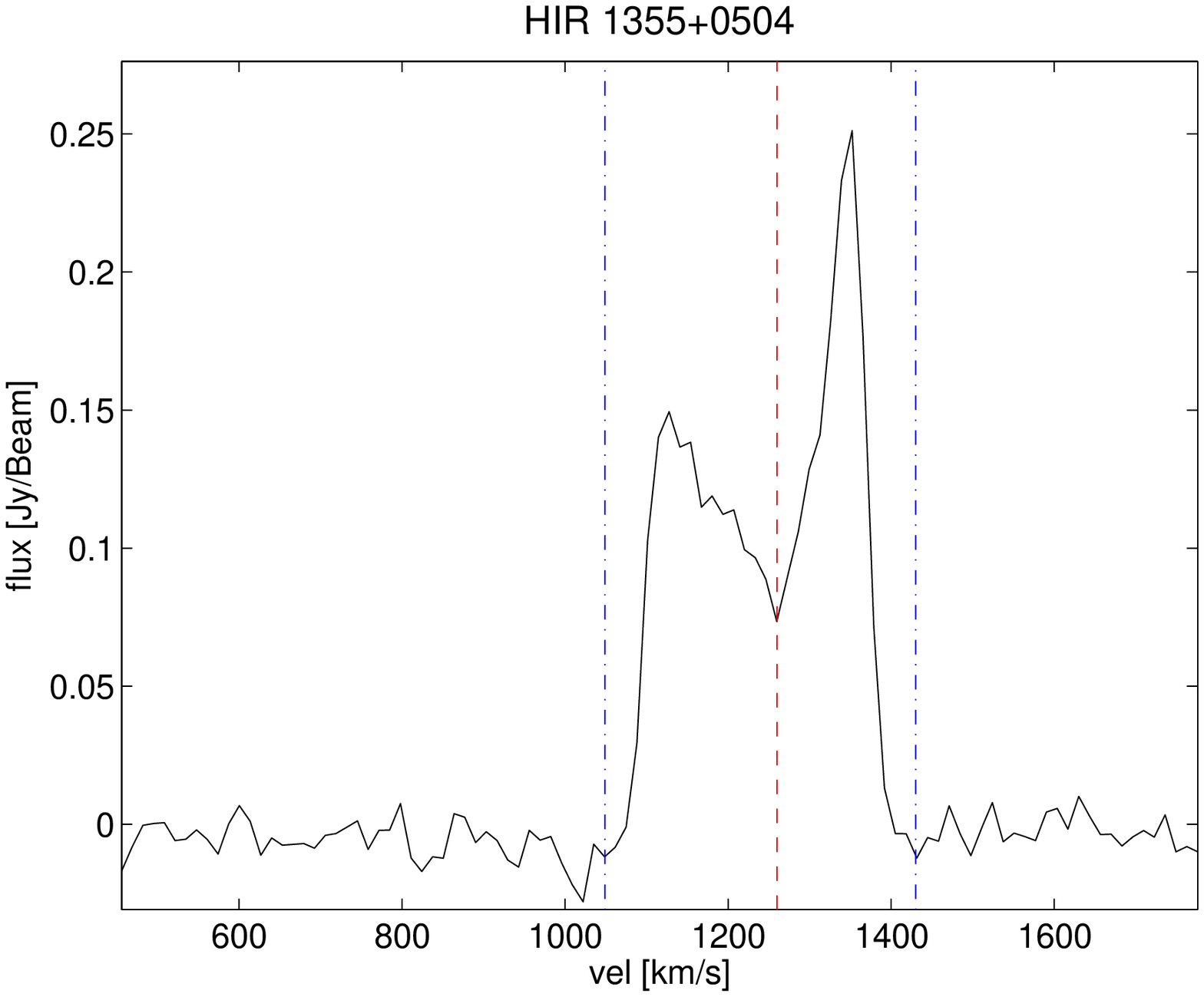}
 \includegraphics[width=0.3\textwidth]{1401+0247_spec.eps}
 \includegraphics[width=0.3\textwidth]{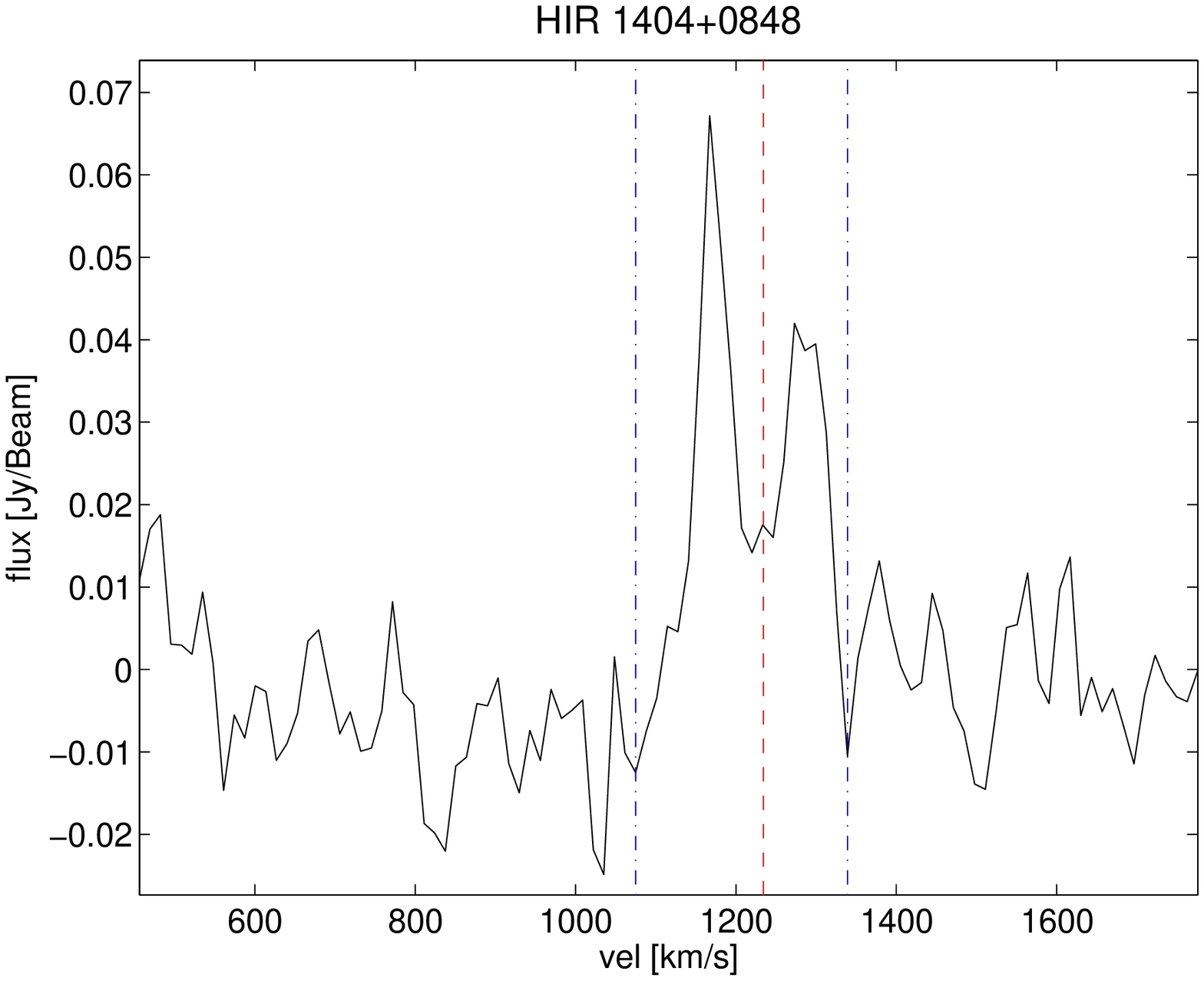}
 \includegraphics[width=0.3\textwidth]{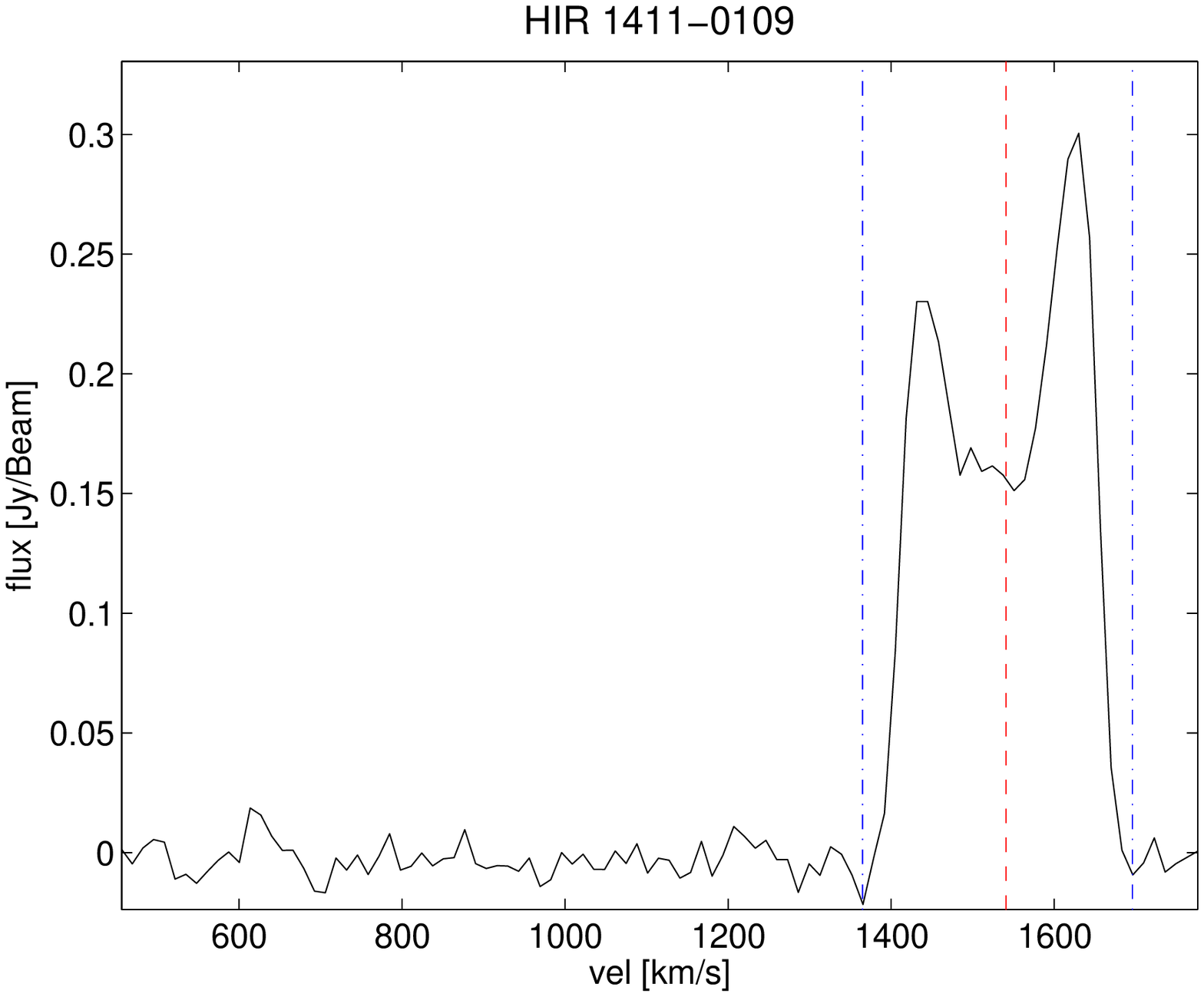}
 \includegraphics[width=0.3\textwidth]{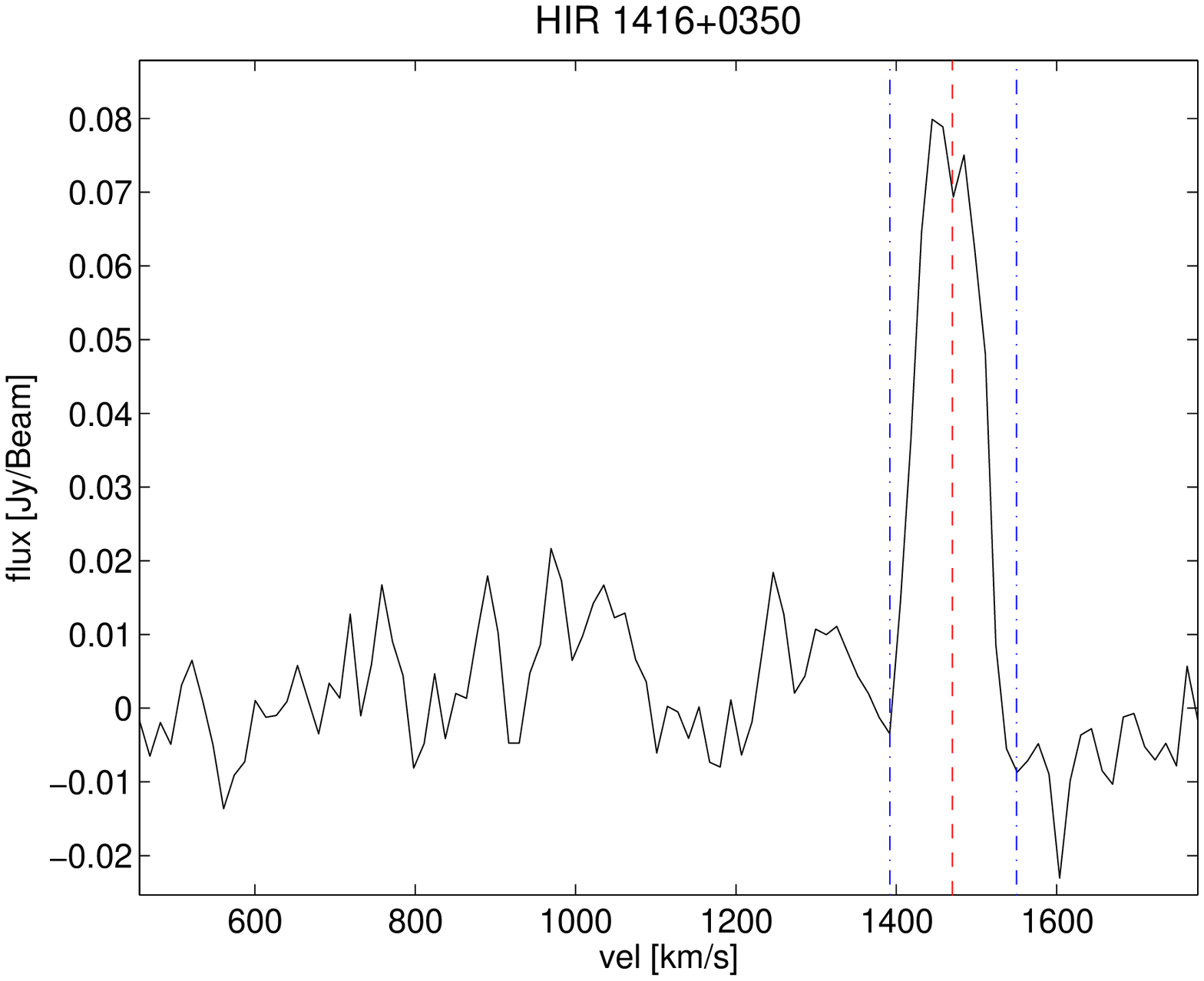}
 \includegraphics[width=0.3\textwidth]{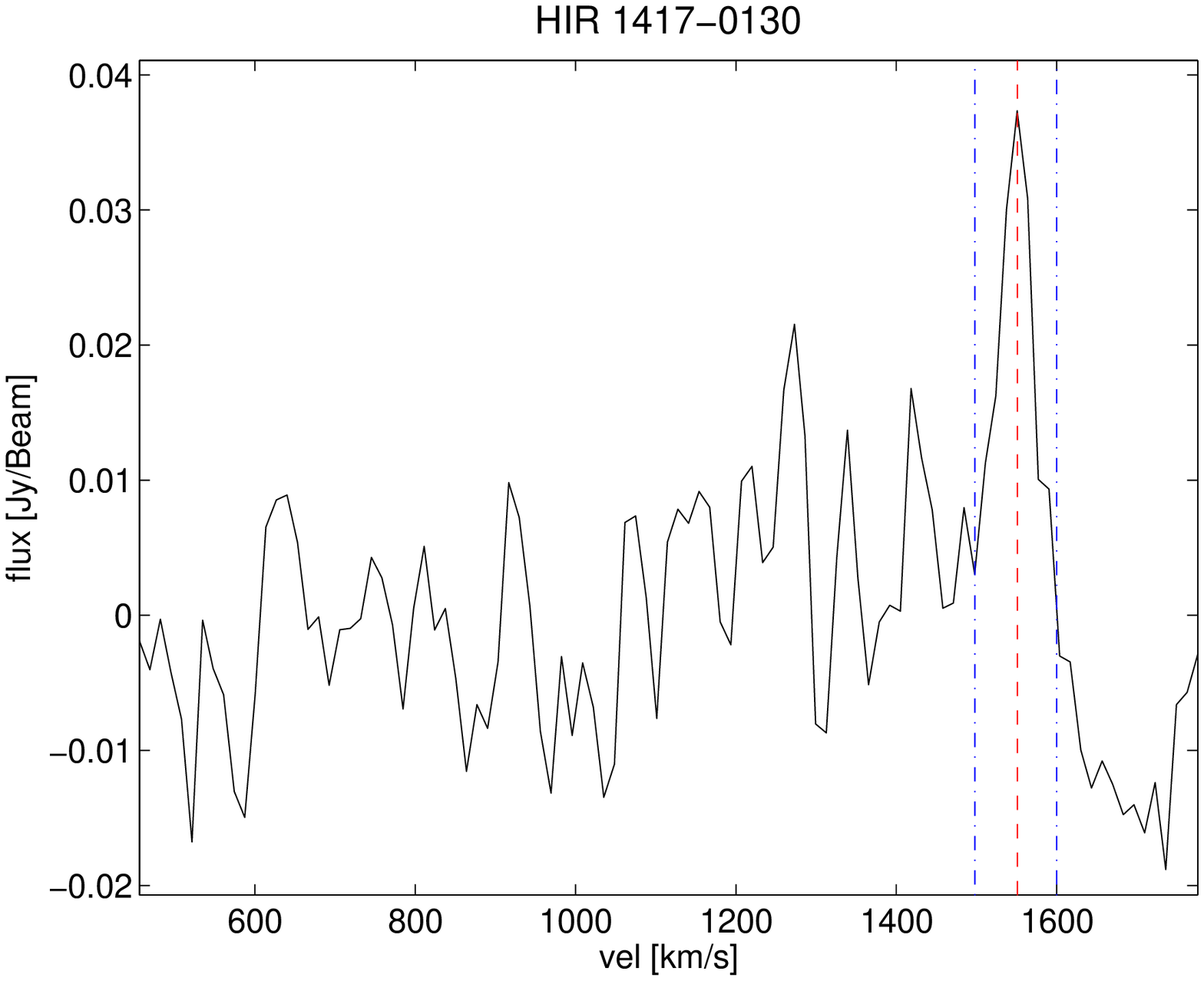}
 \includegraphics[width=0.3\textwidth]{1417+0651_spec.eps}                                                   
 \includegraphics[width=0.3\textwidth]{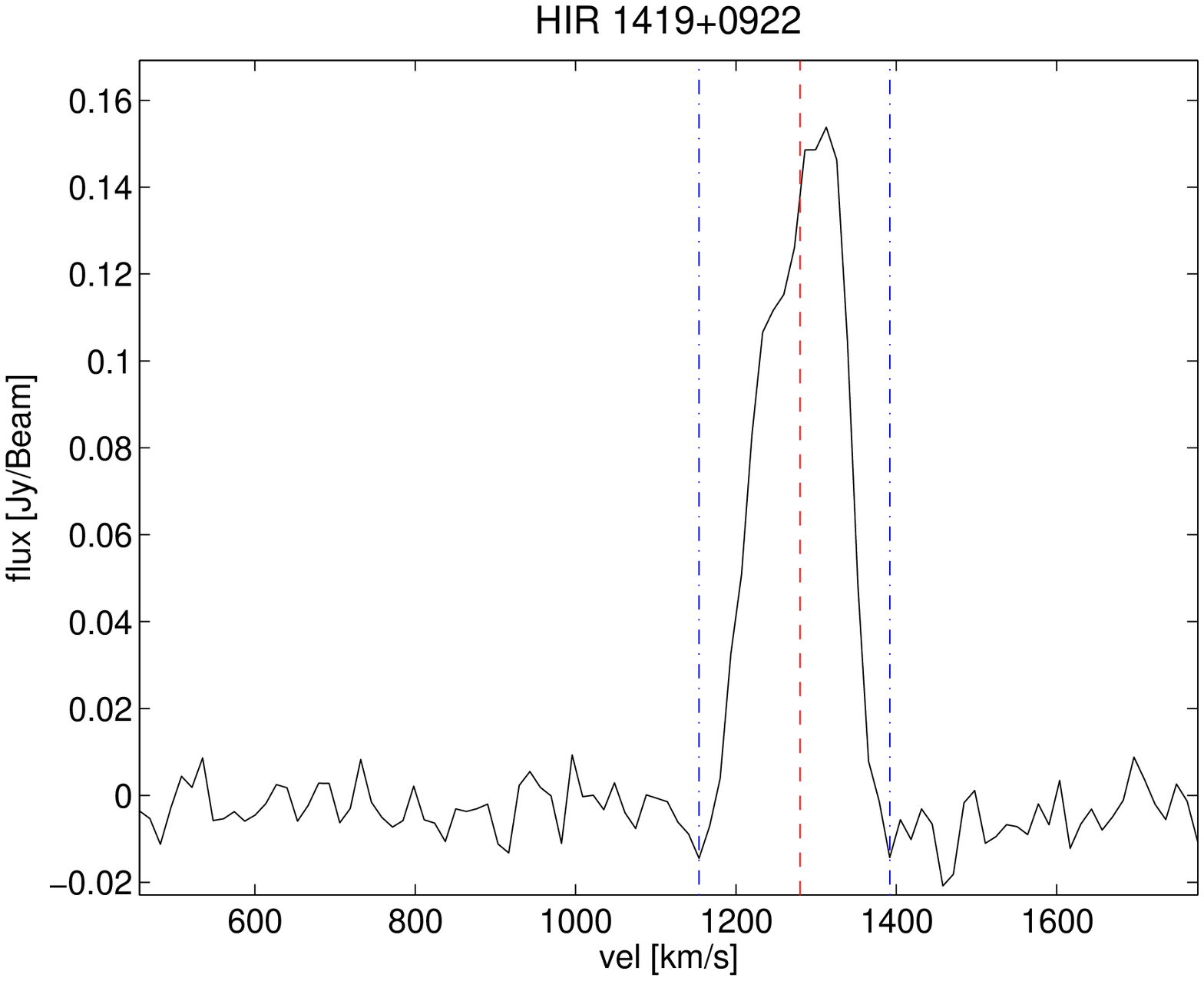}
 \includegraphics[width=0.3\textwidth]{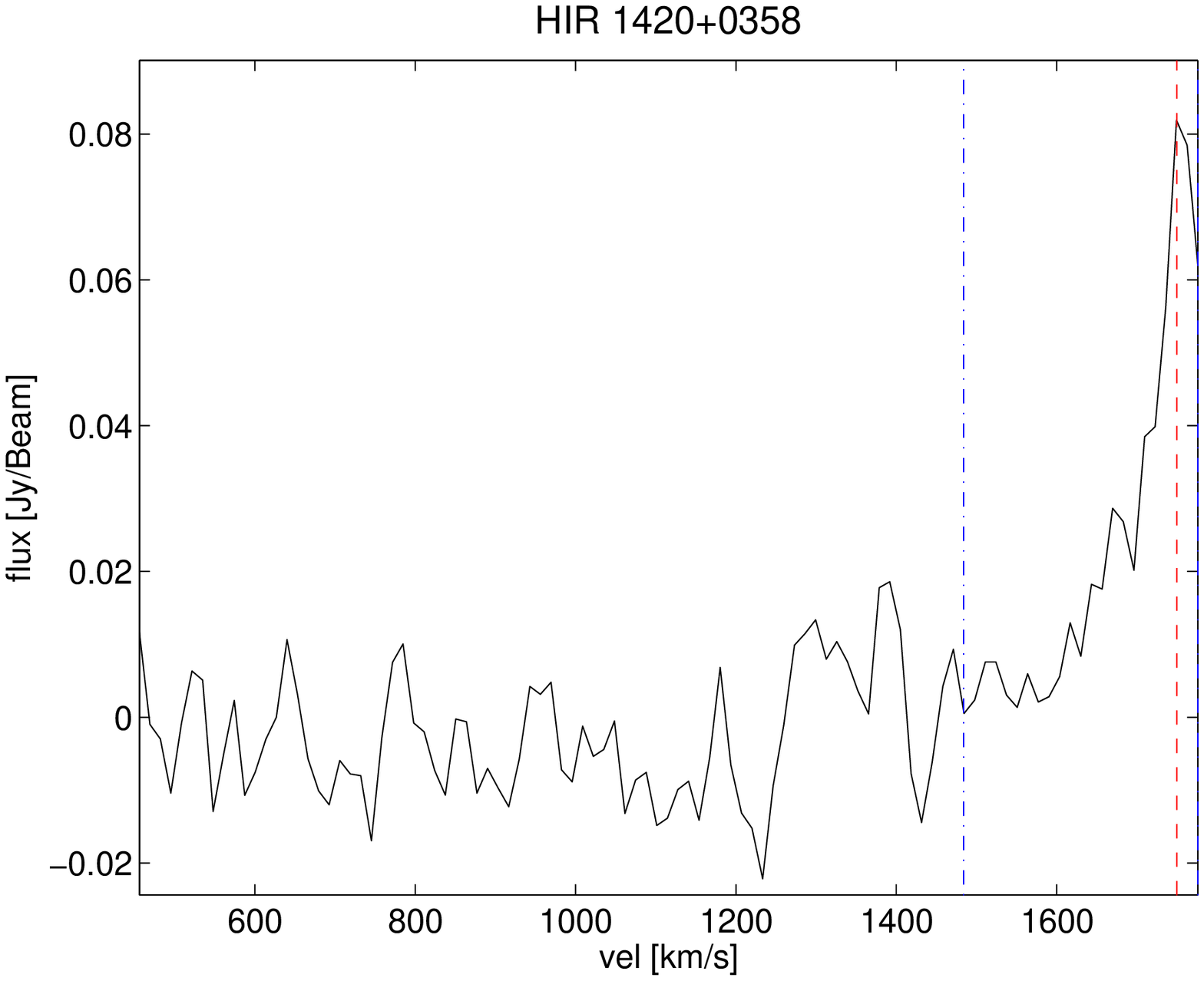}

 \end{center}                                                         
{\bf Fig~\ref{all_spectra}.} (continued)                              
                                                                      
\end{figure*}

\begin{figure*}
  \begin{center}

 \includegraphics[width=0.3\textwidth]{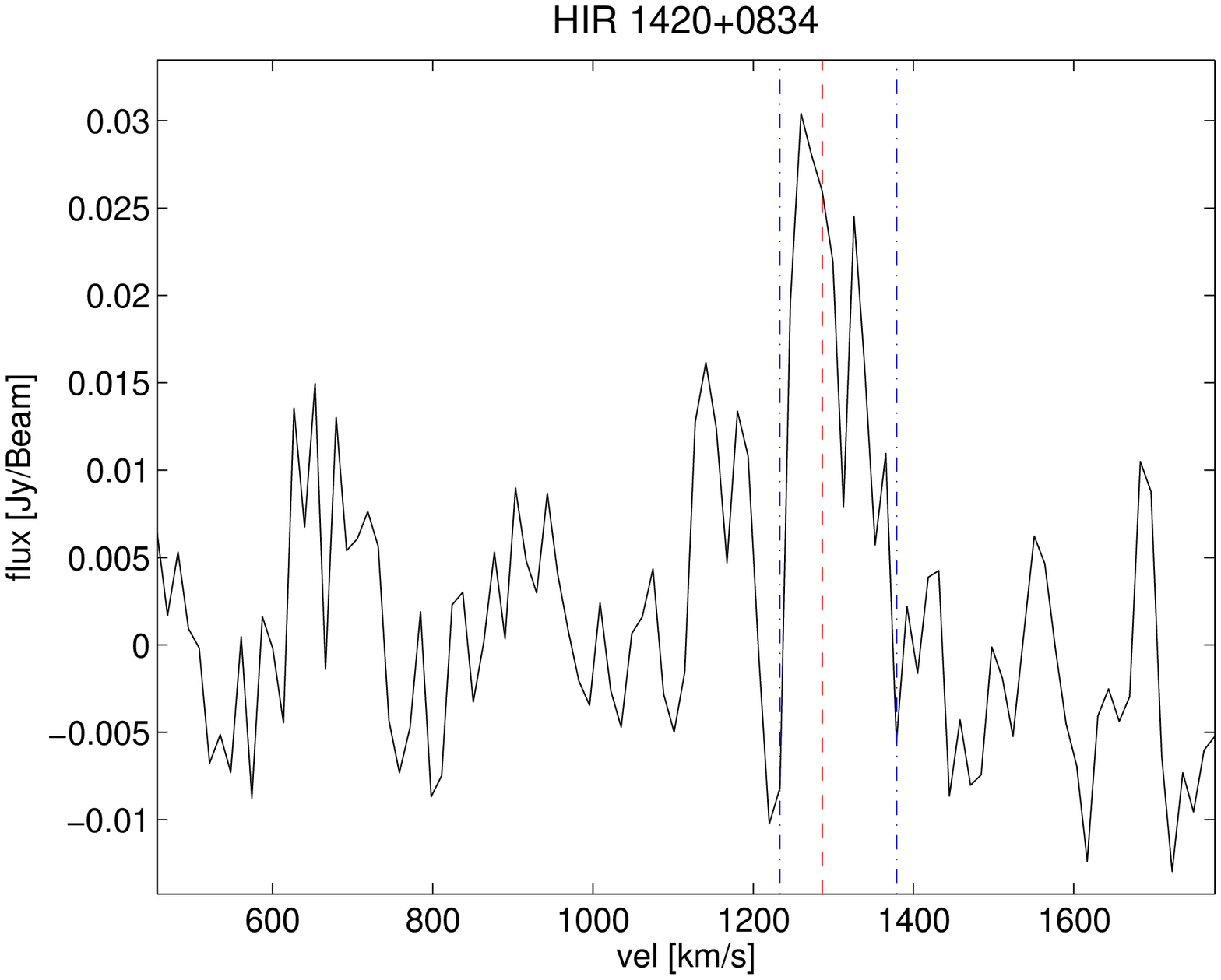}
 \includegraphics[width=0.3\textwidth]{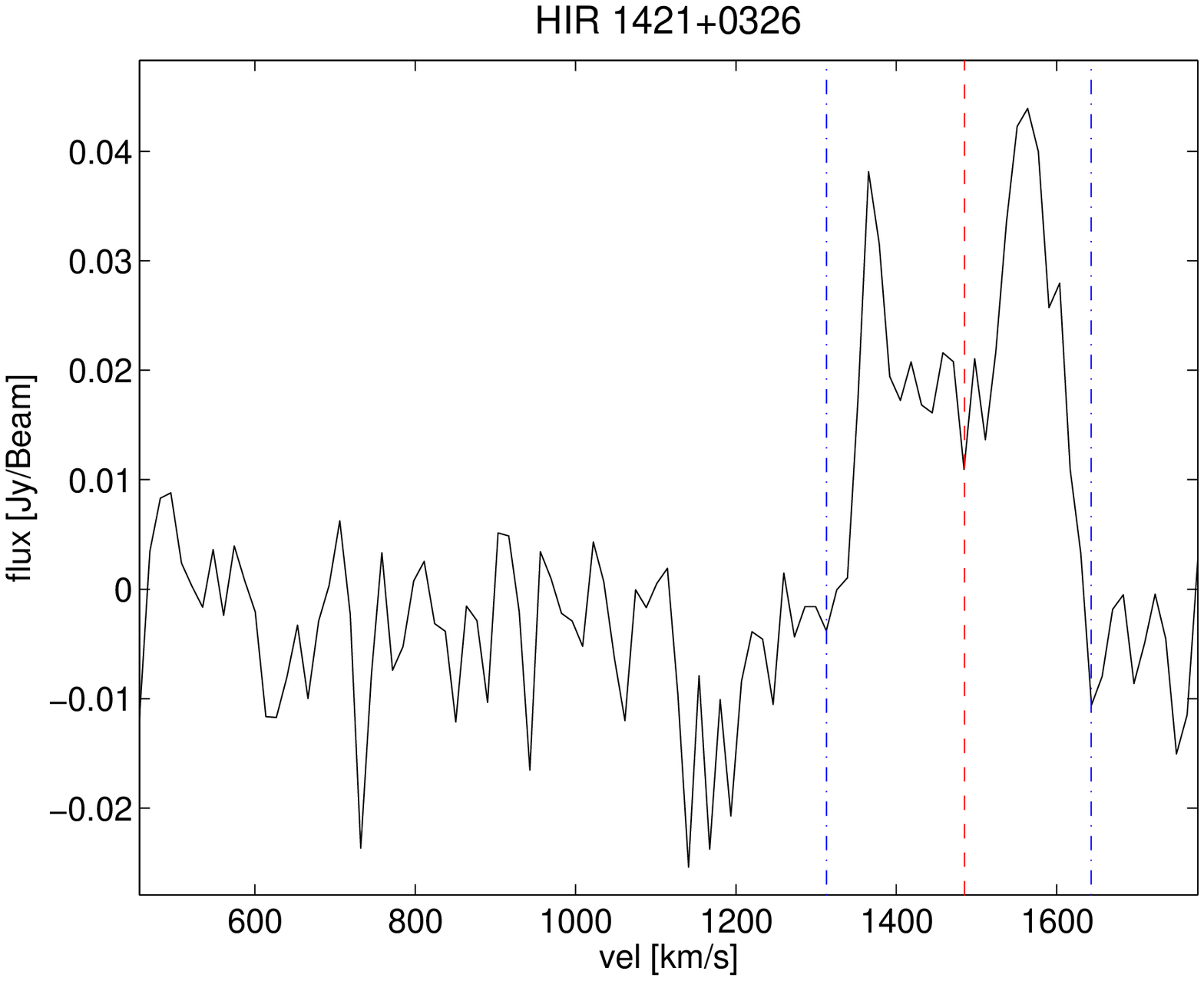}
 \includegraphics[width=0.3\textwidth]{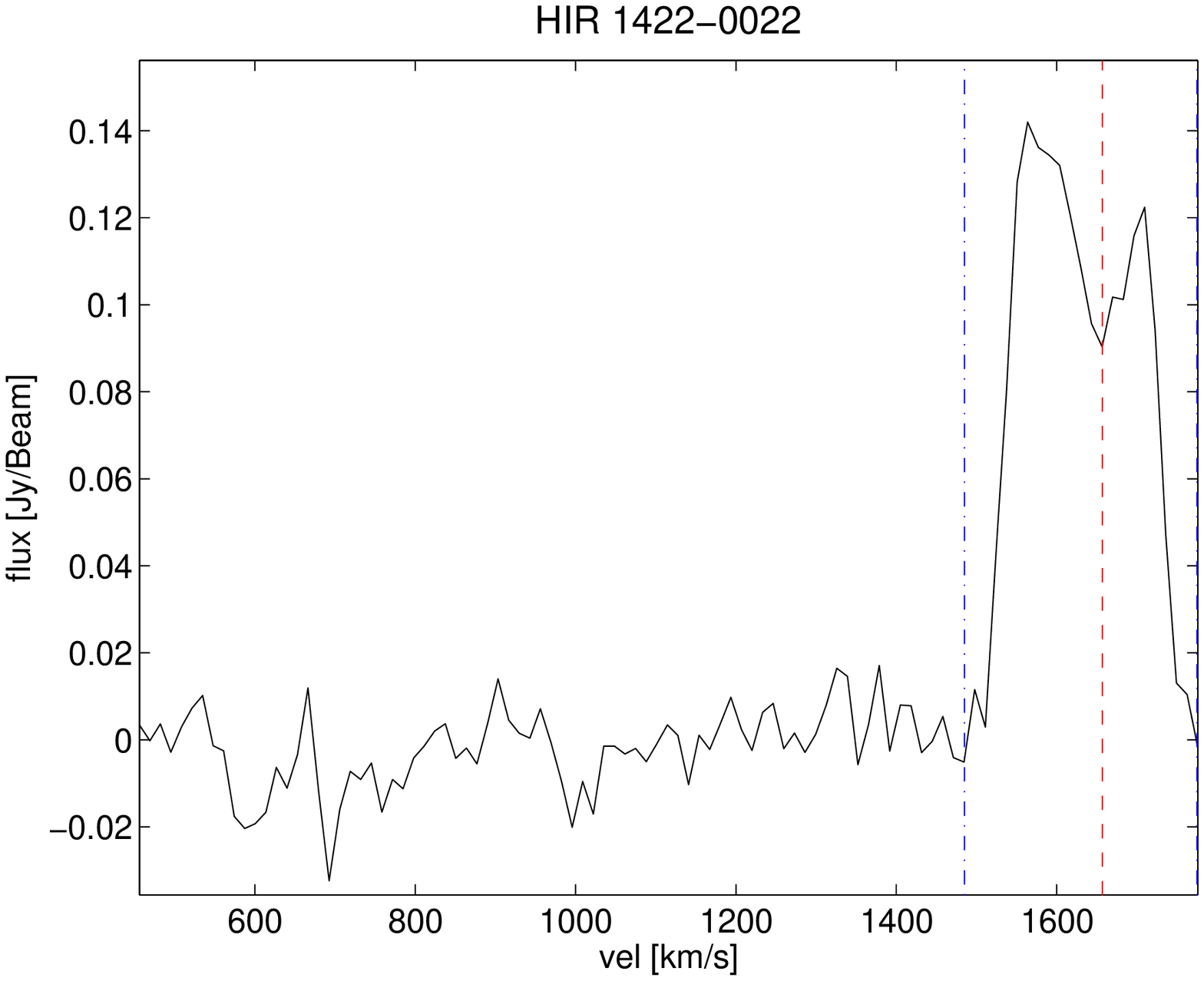}
 \includegraphics[width=0.3\textwidth]{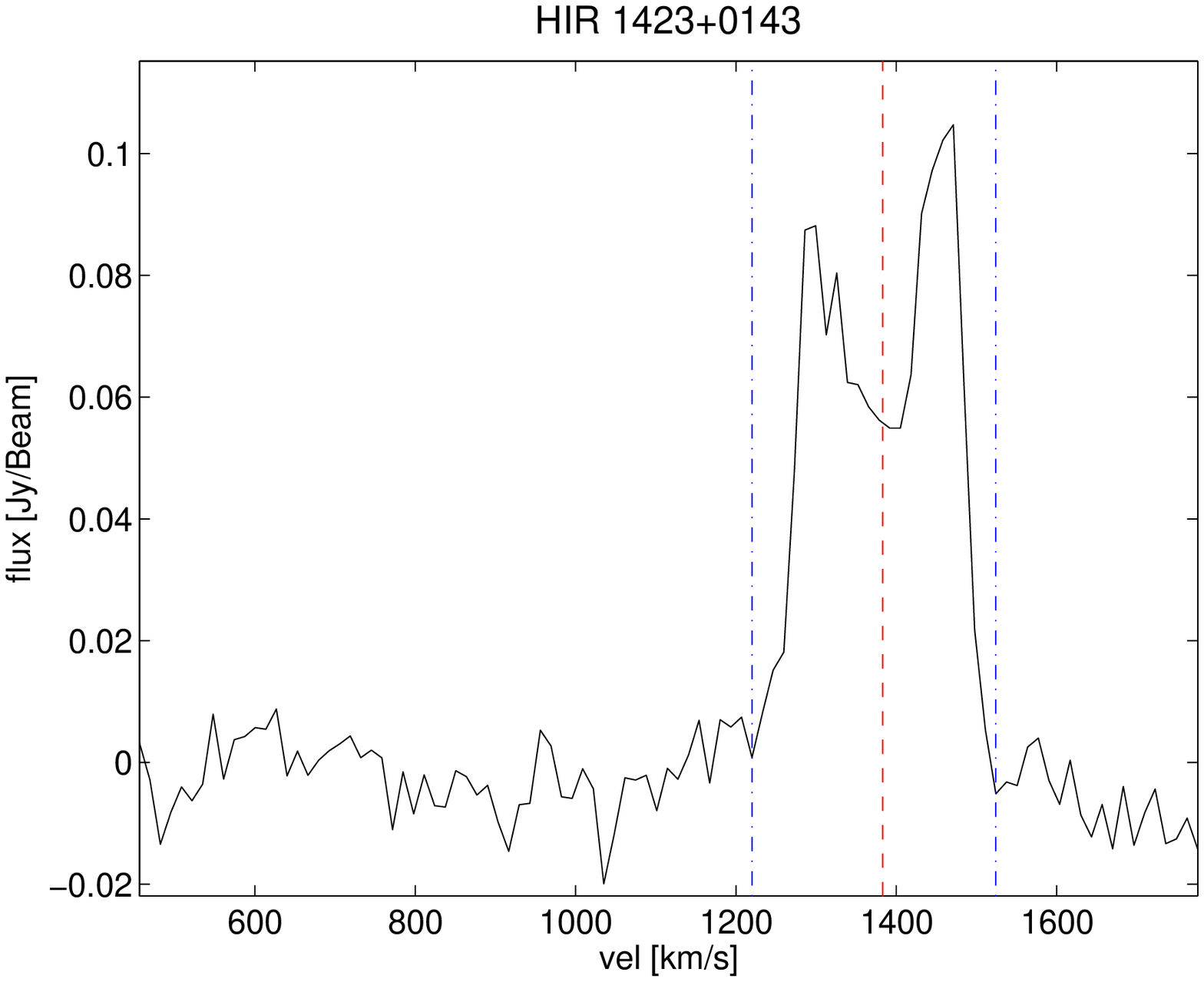}
 \includegraphics[width=0.3\textwidth]{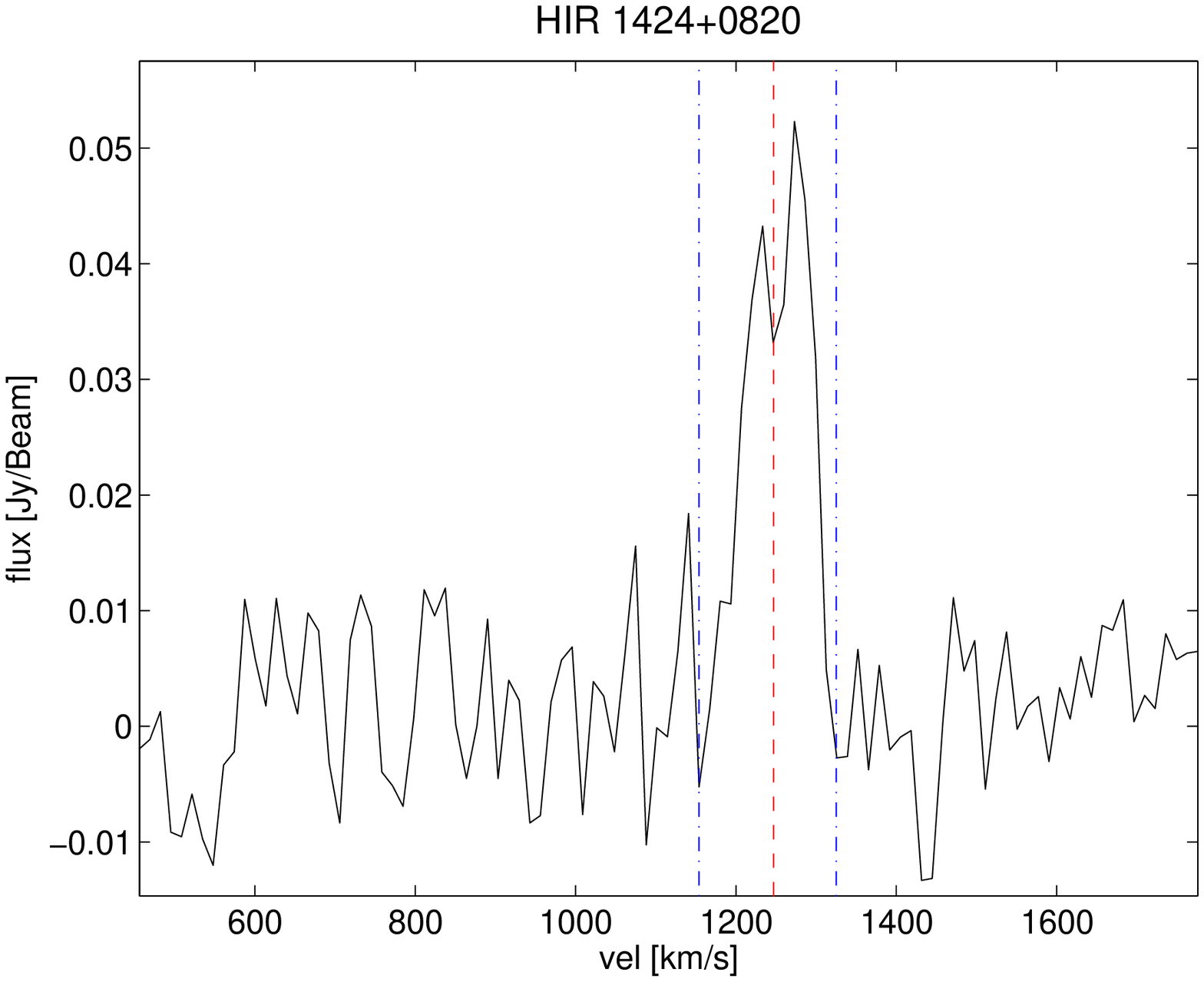}
 \includegraphics[width=0.3\textwidth]{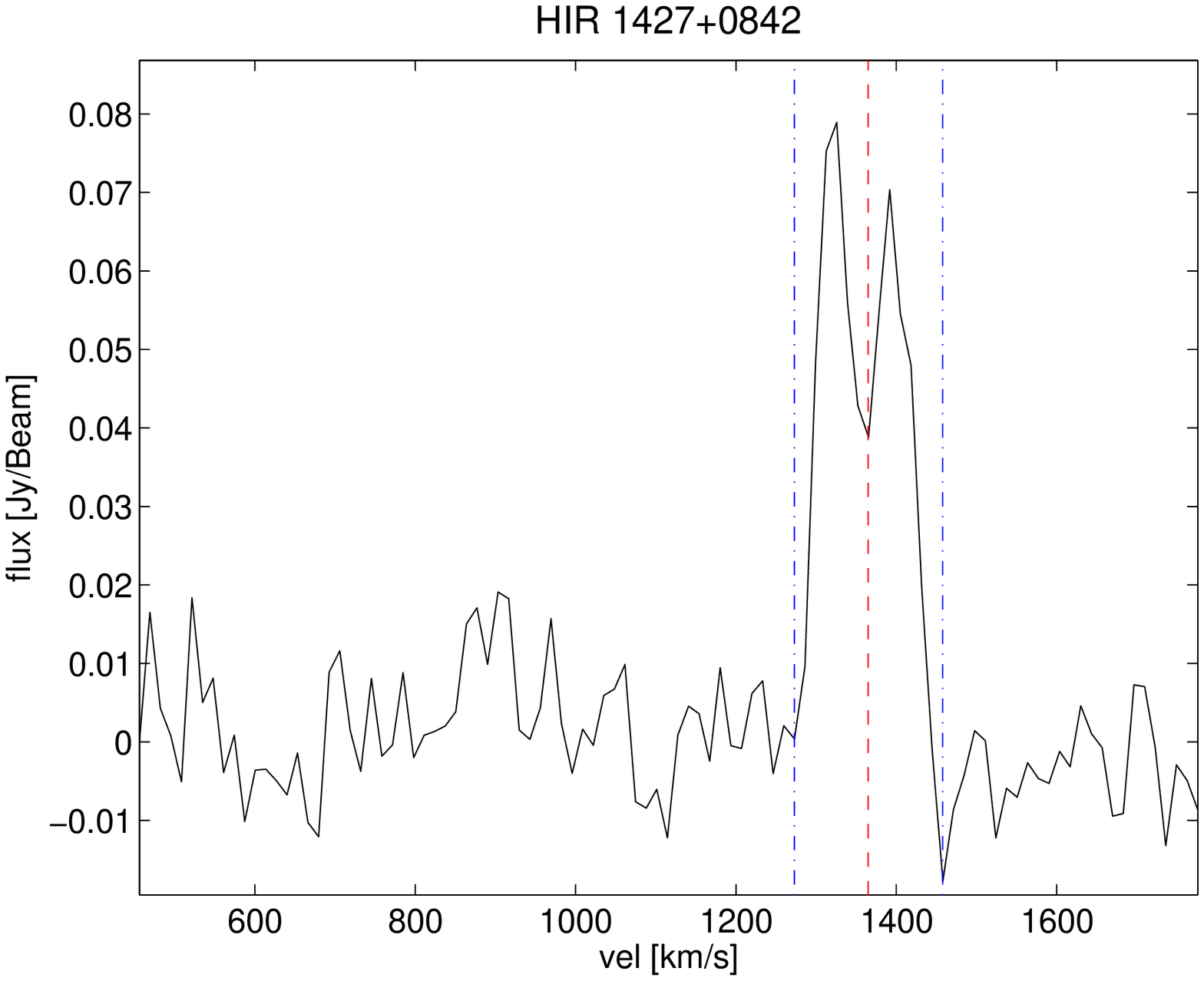}
 \includegraphics[width=0.3\textwidth]{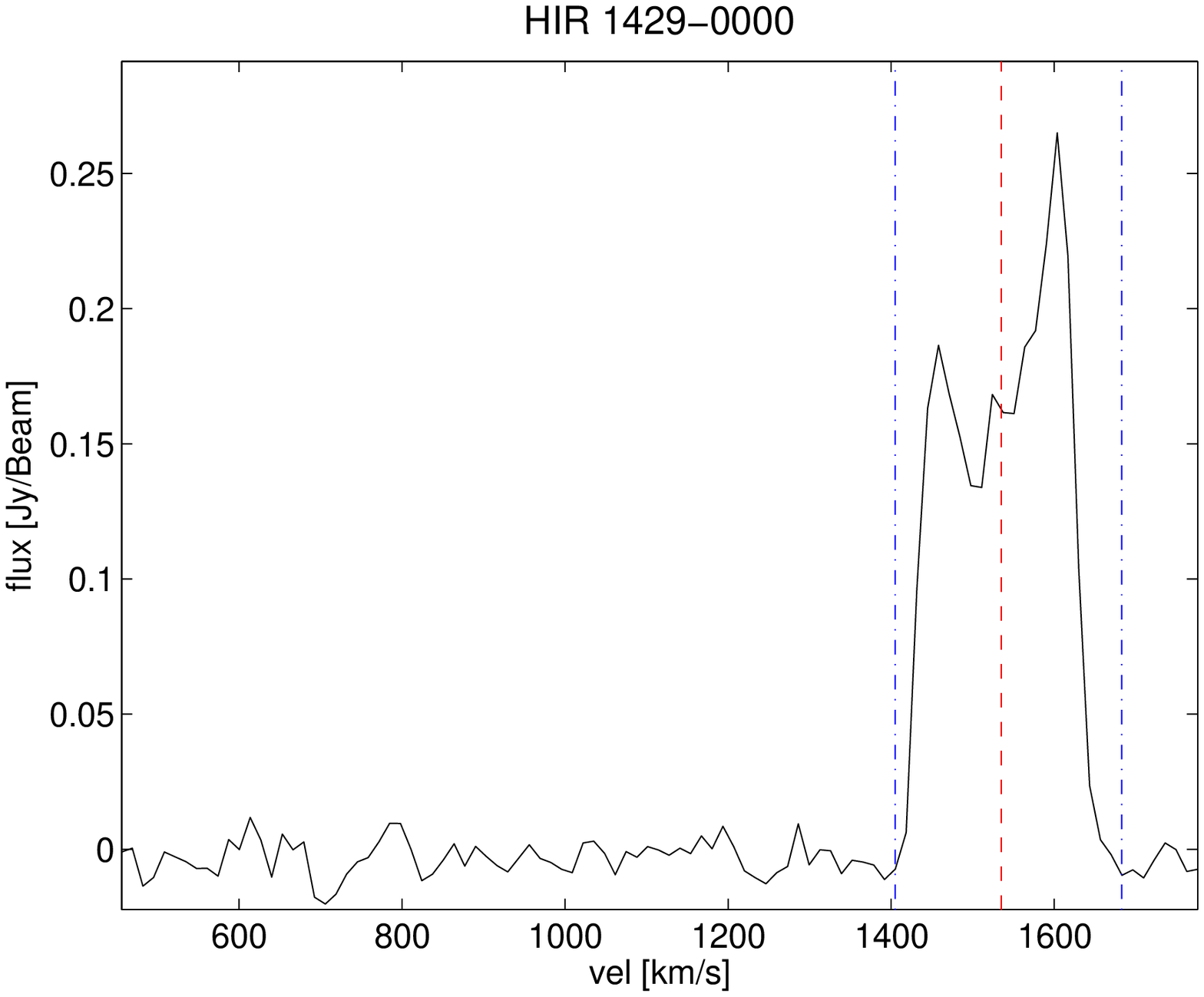}
 \includegraphics[width=0.3\textwidth]{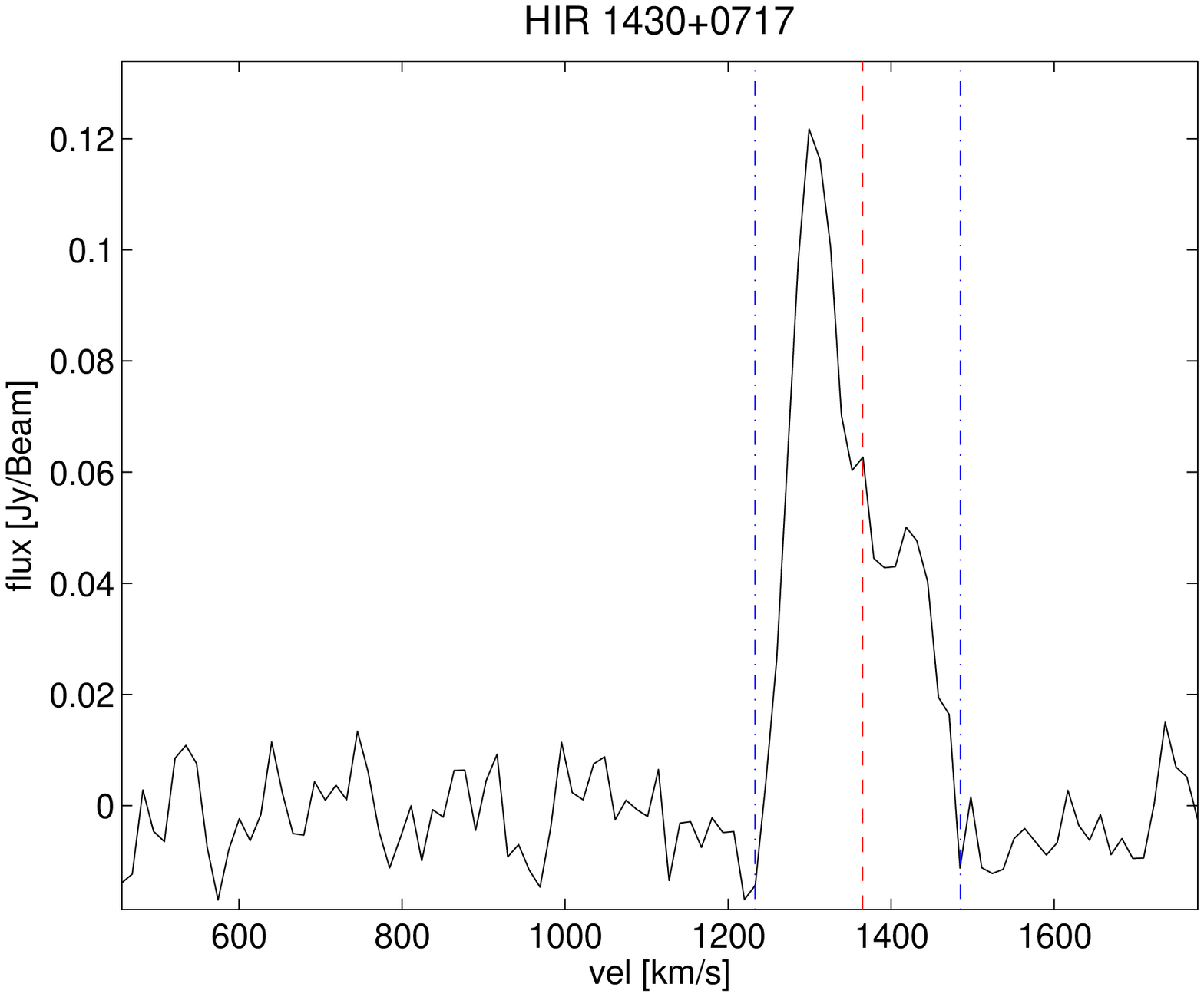}
 \includegraphics[width=0.3\textwidth]{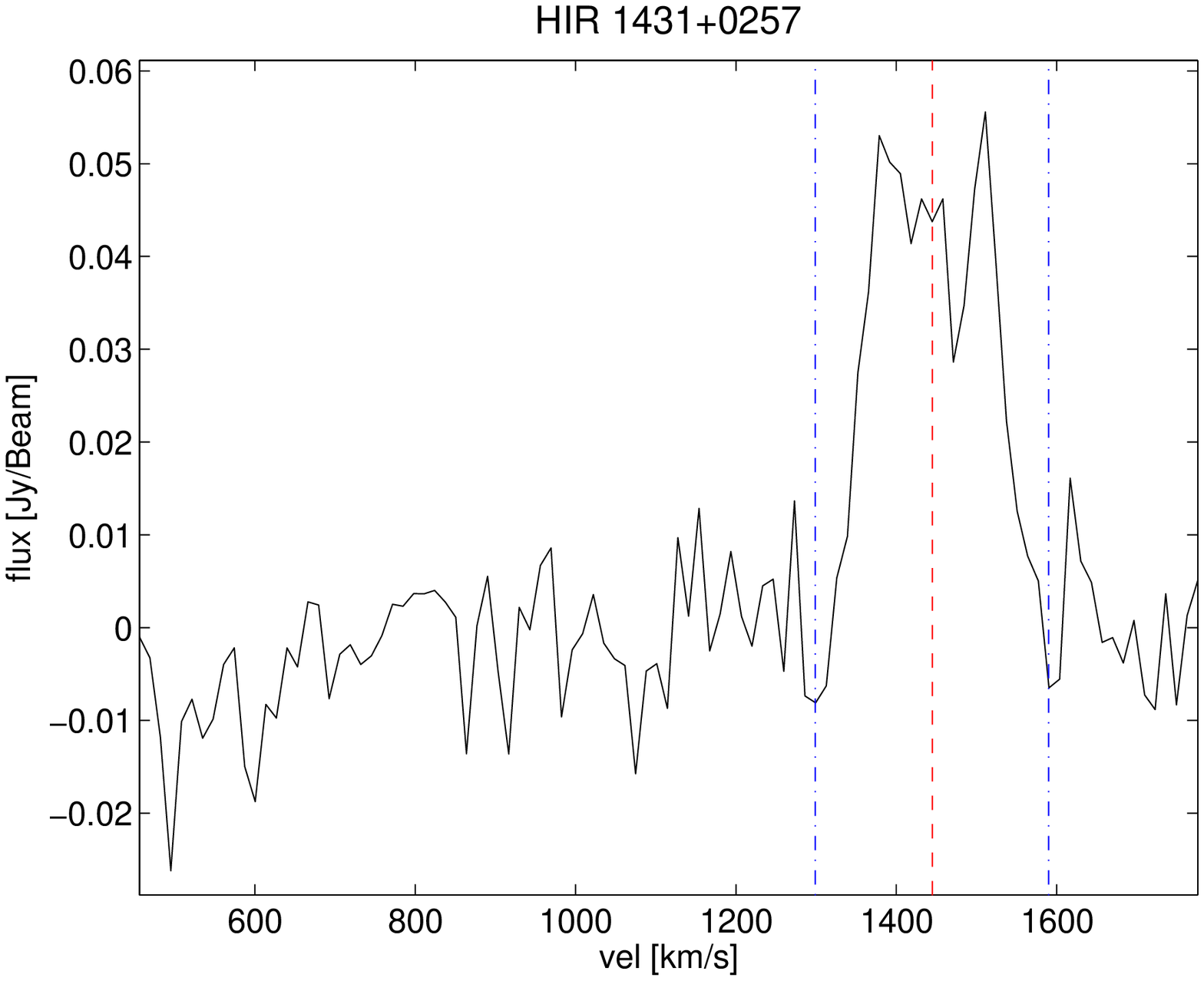}
 \includegraphics[width=0.3\textwidth]{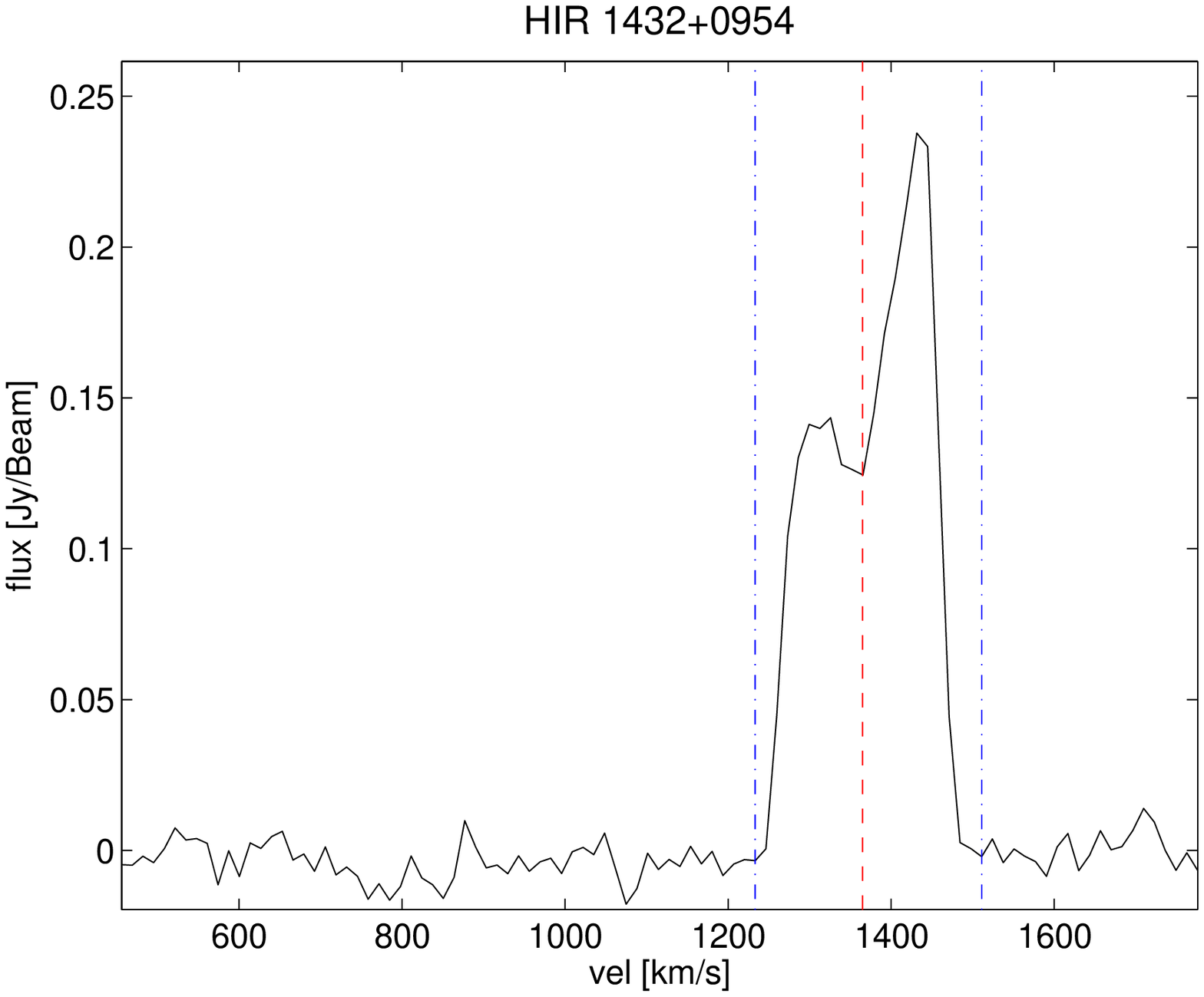}
 \includegraphics[width=0.3\textwidth]{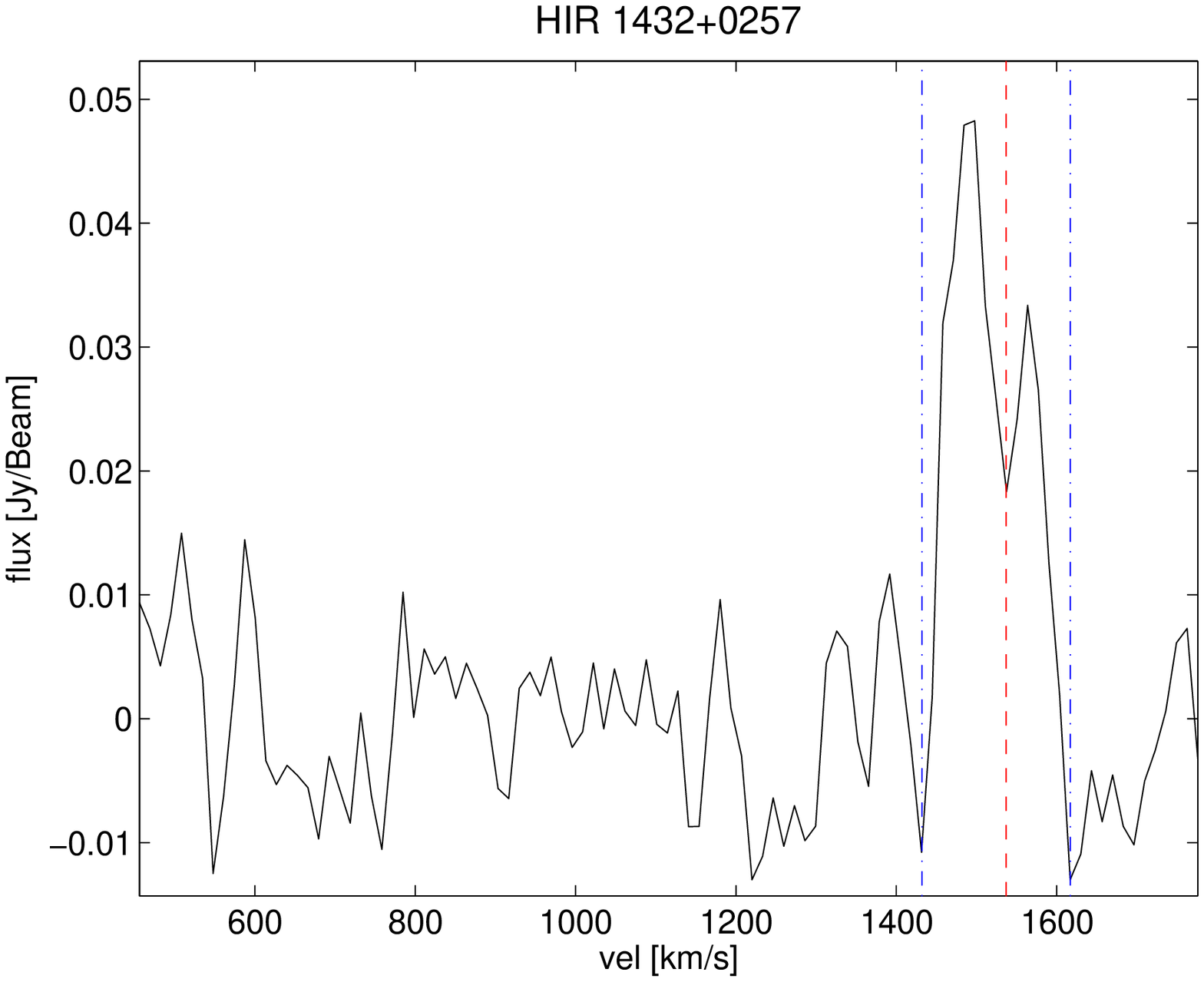}
 \includegraphics[width=0.3\textwidth]{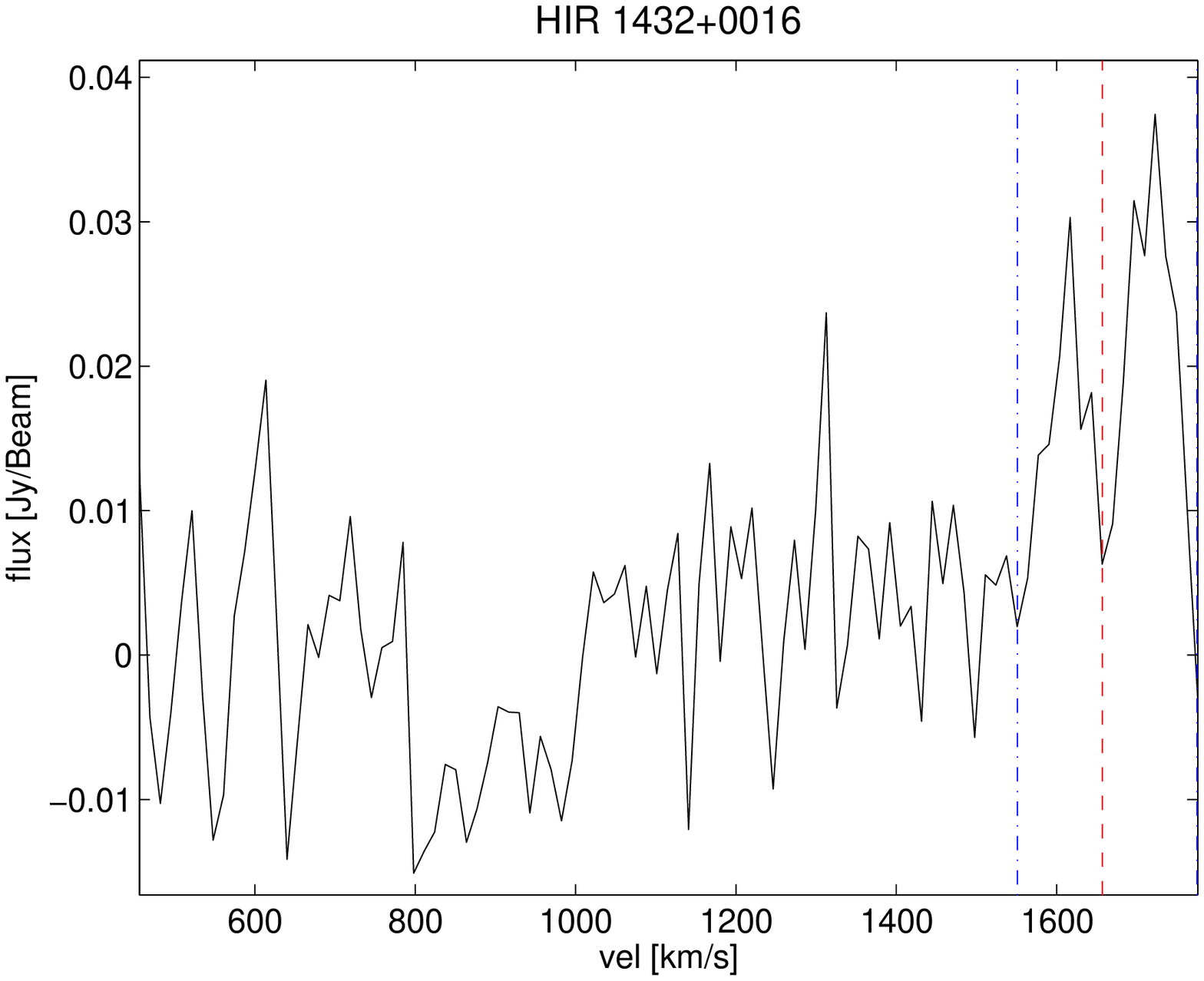}
 \includegraphics[width=0.3\textwidth]{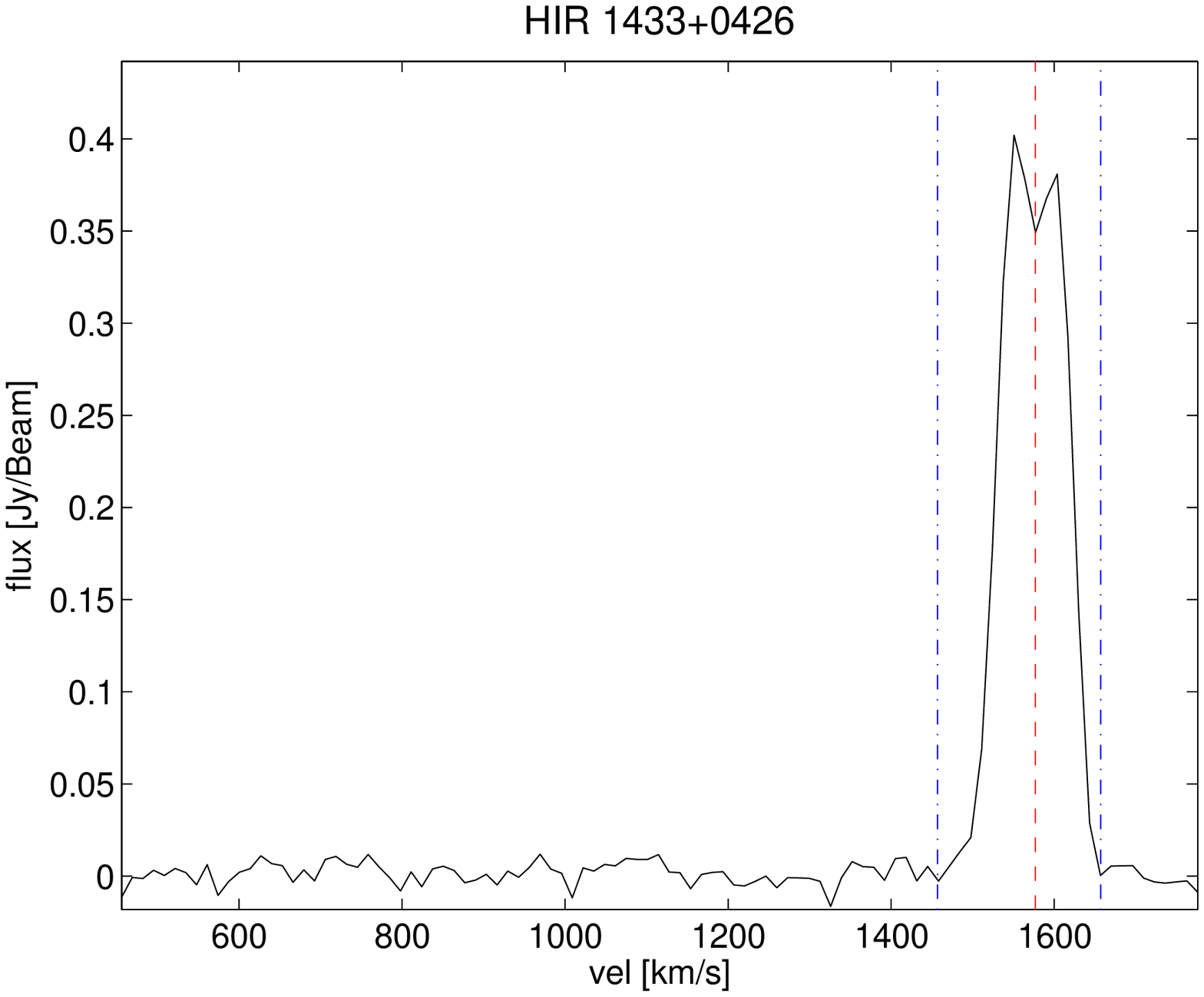}
 \includegraphics[width=0.3\textwidth]{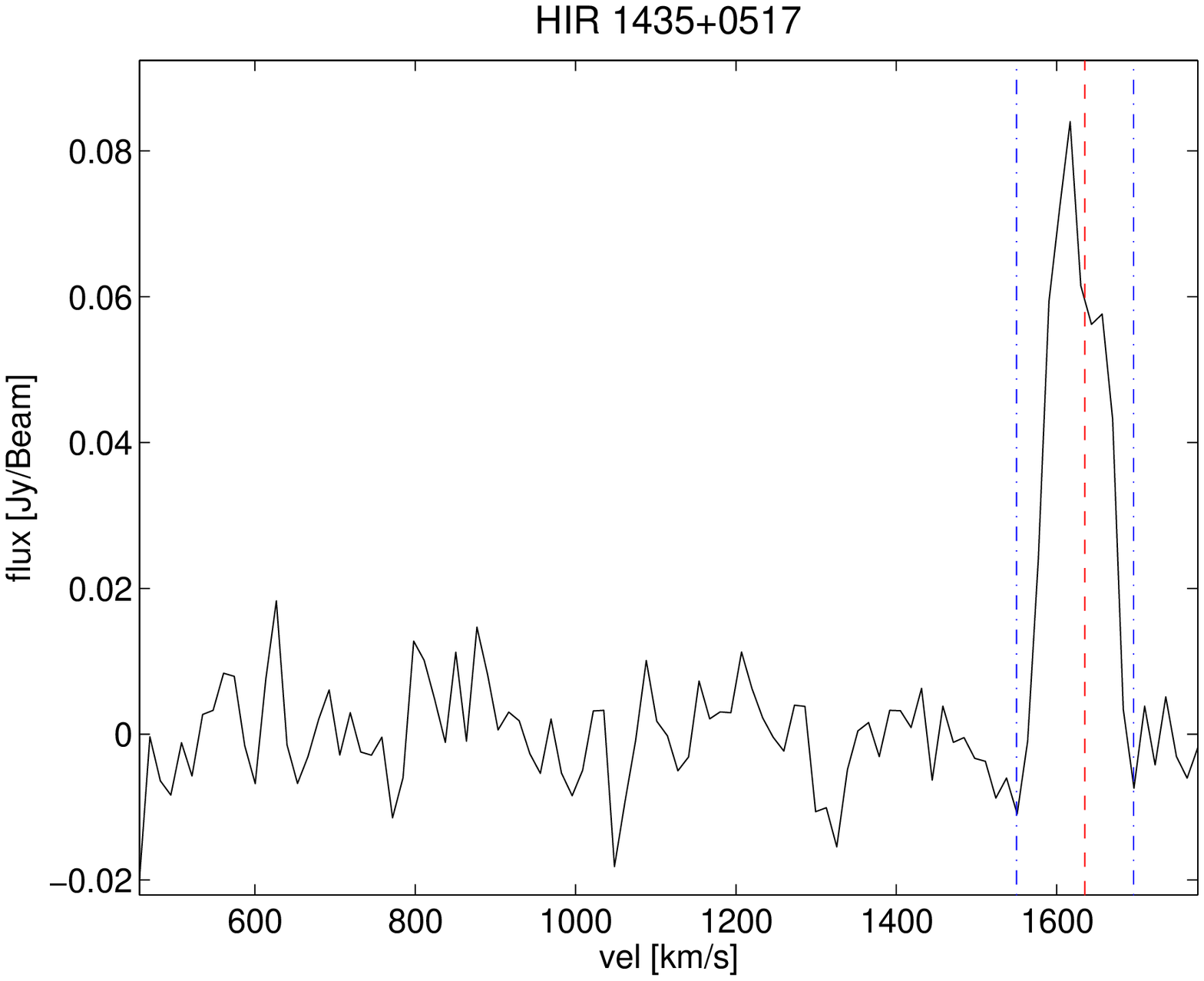}
 \includegraphics[width=0.3\textwidth]{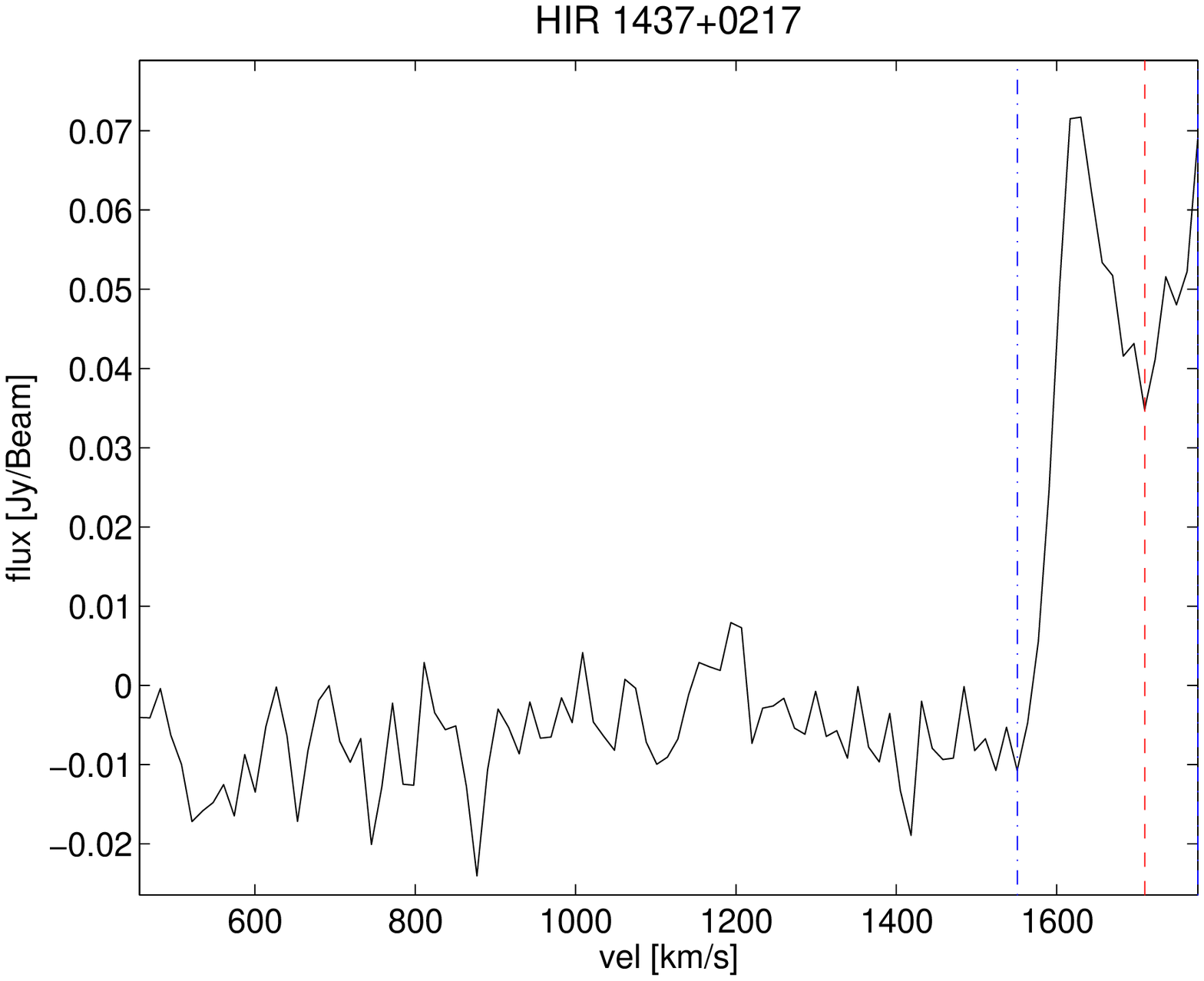}

 \end{center}                                                         
{\bf Fig~\ref{all_spectra}.} (continued)                              
                                                                      
\end{figure*}

\begin{figure*}
  \begin{center}

 \includegraphics[width=0.3\textwidth]{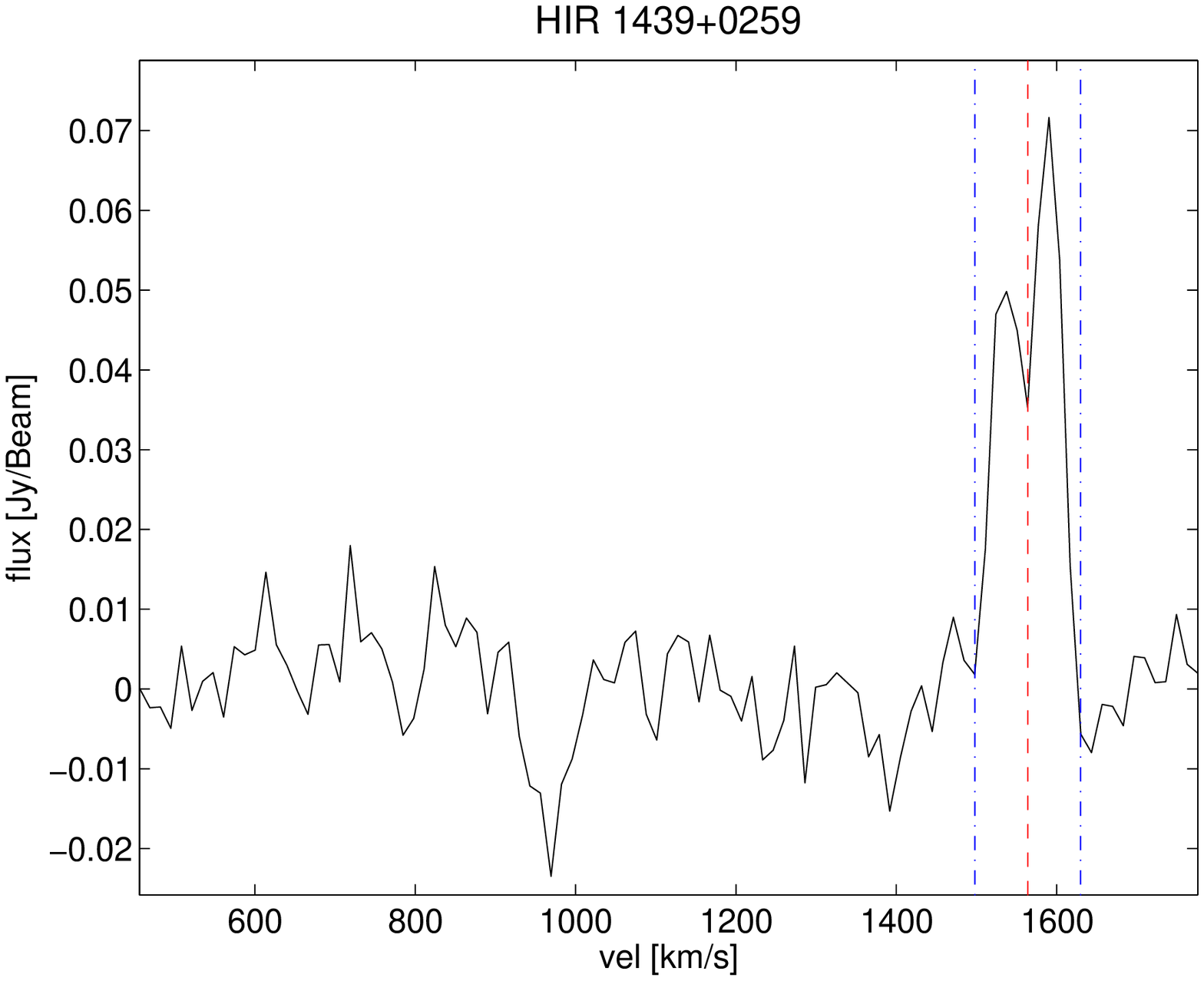}
 \includegraphics[width=0.3\textwidth]{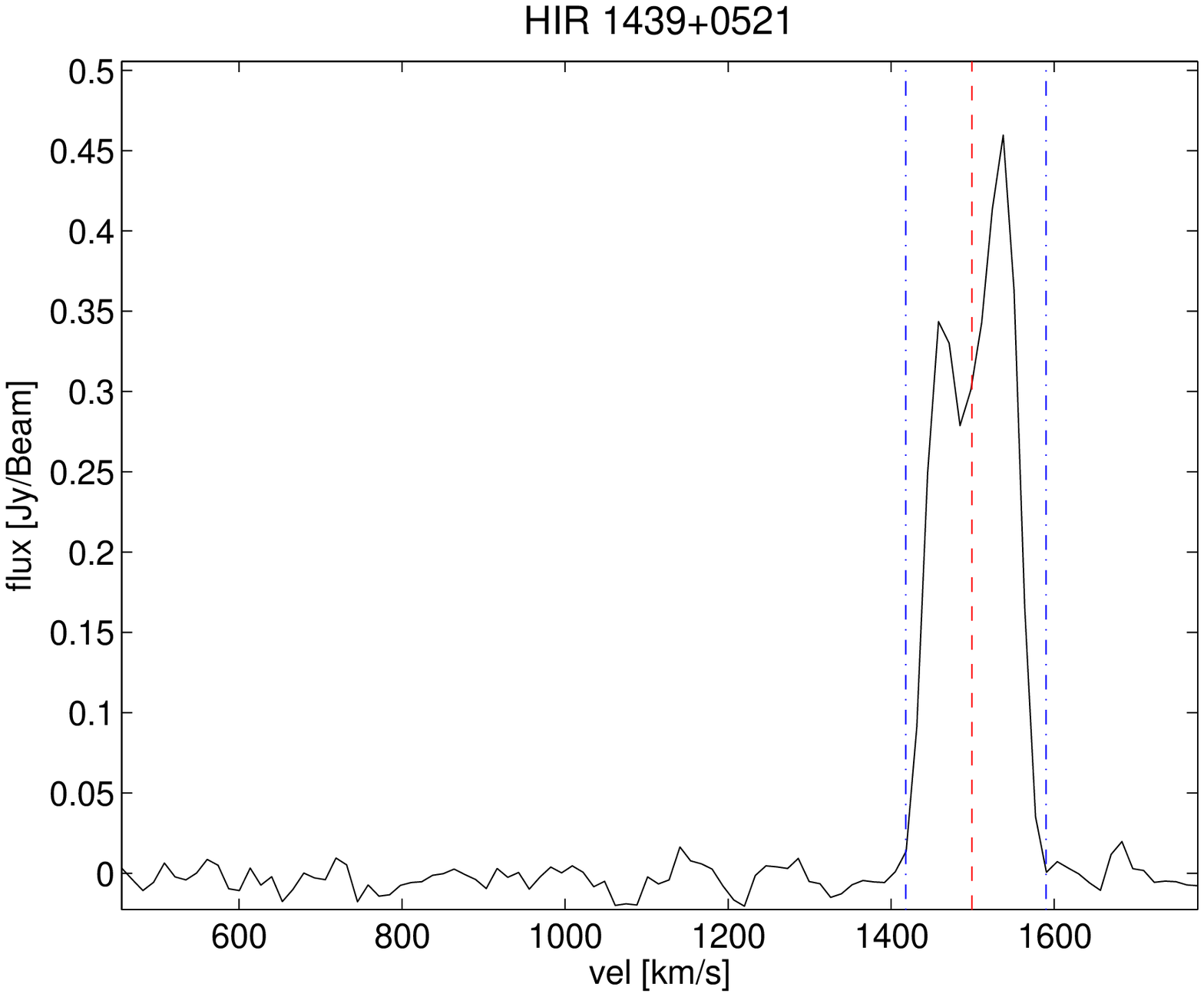}
 \includegraphics[width=0.3\textwidth]{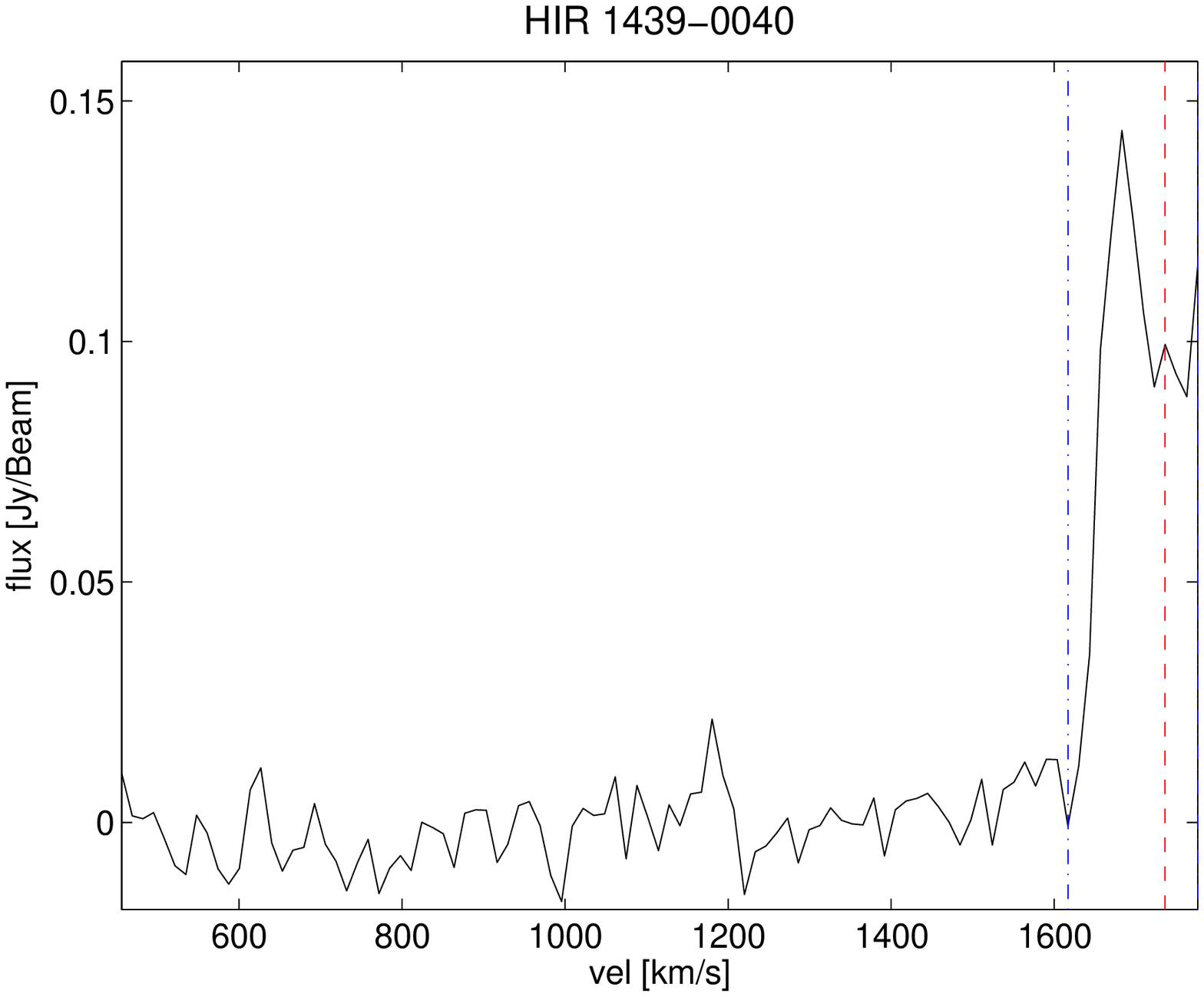}
 \includegraphics[width=0.3\textwidth]{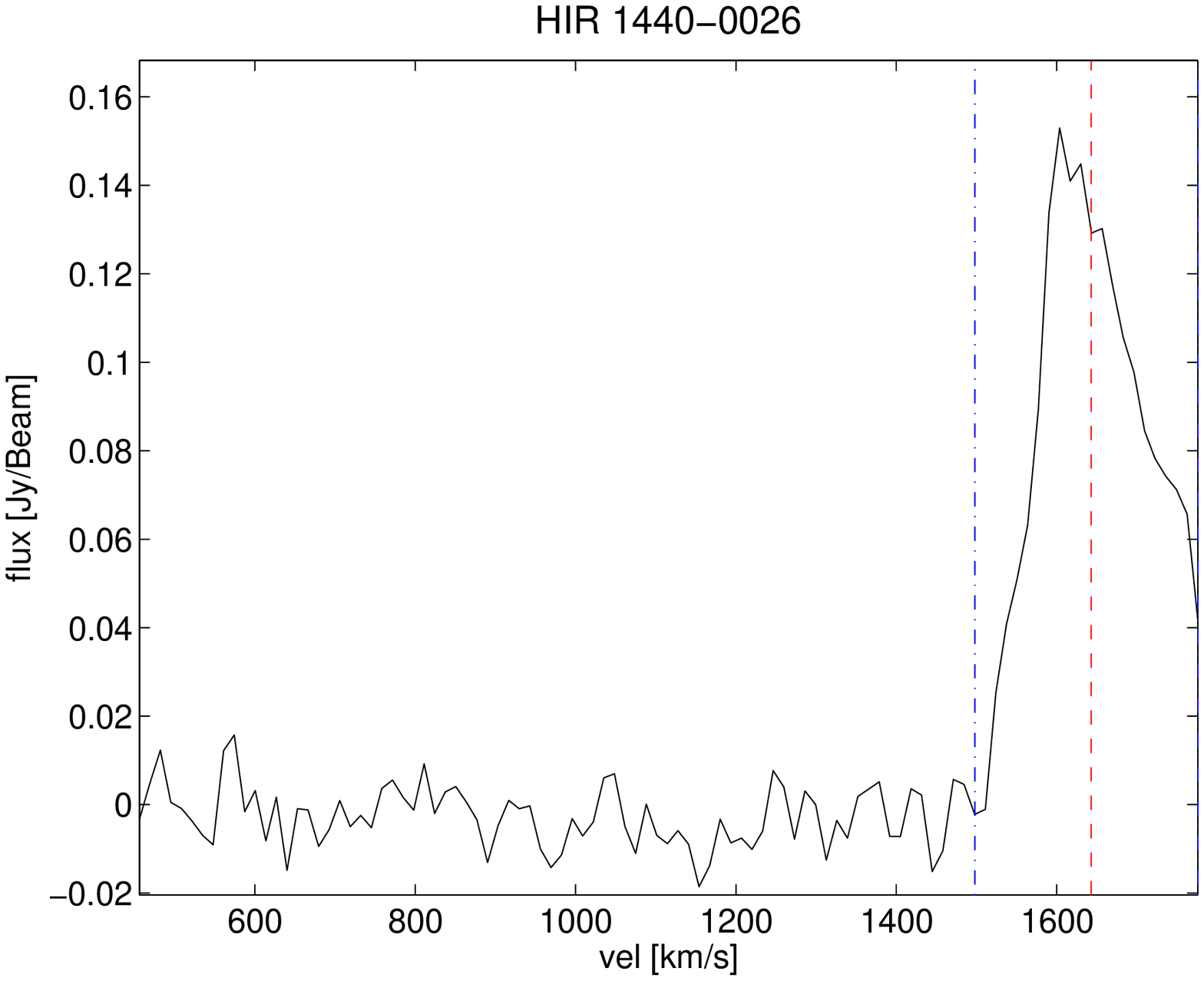}
 \includegraphics[width=0.3\textwidth]{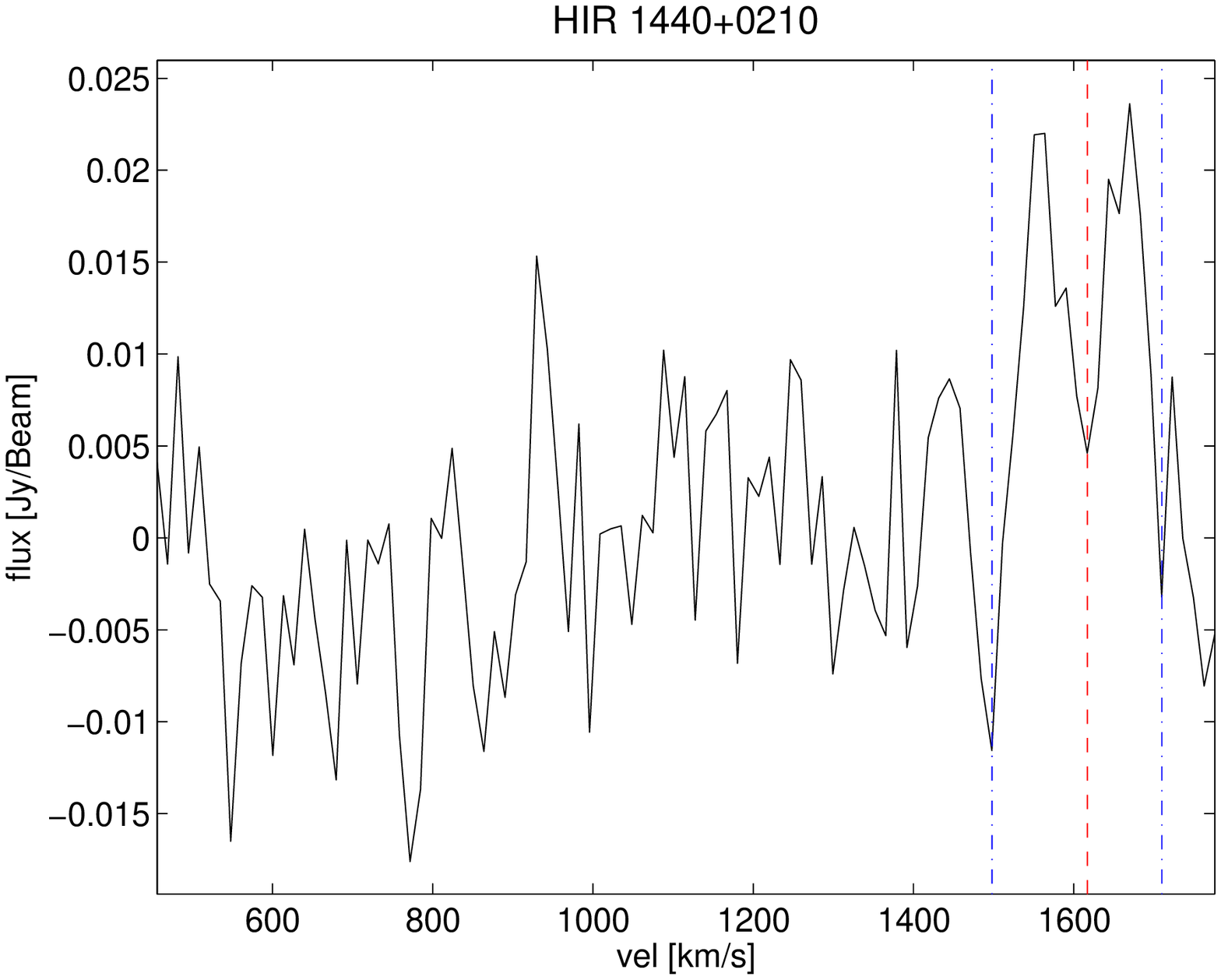}
 \includegraphics[width=0.3\textwidth]{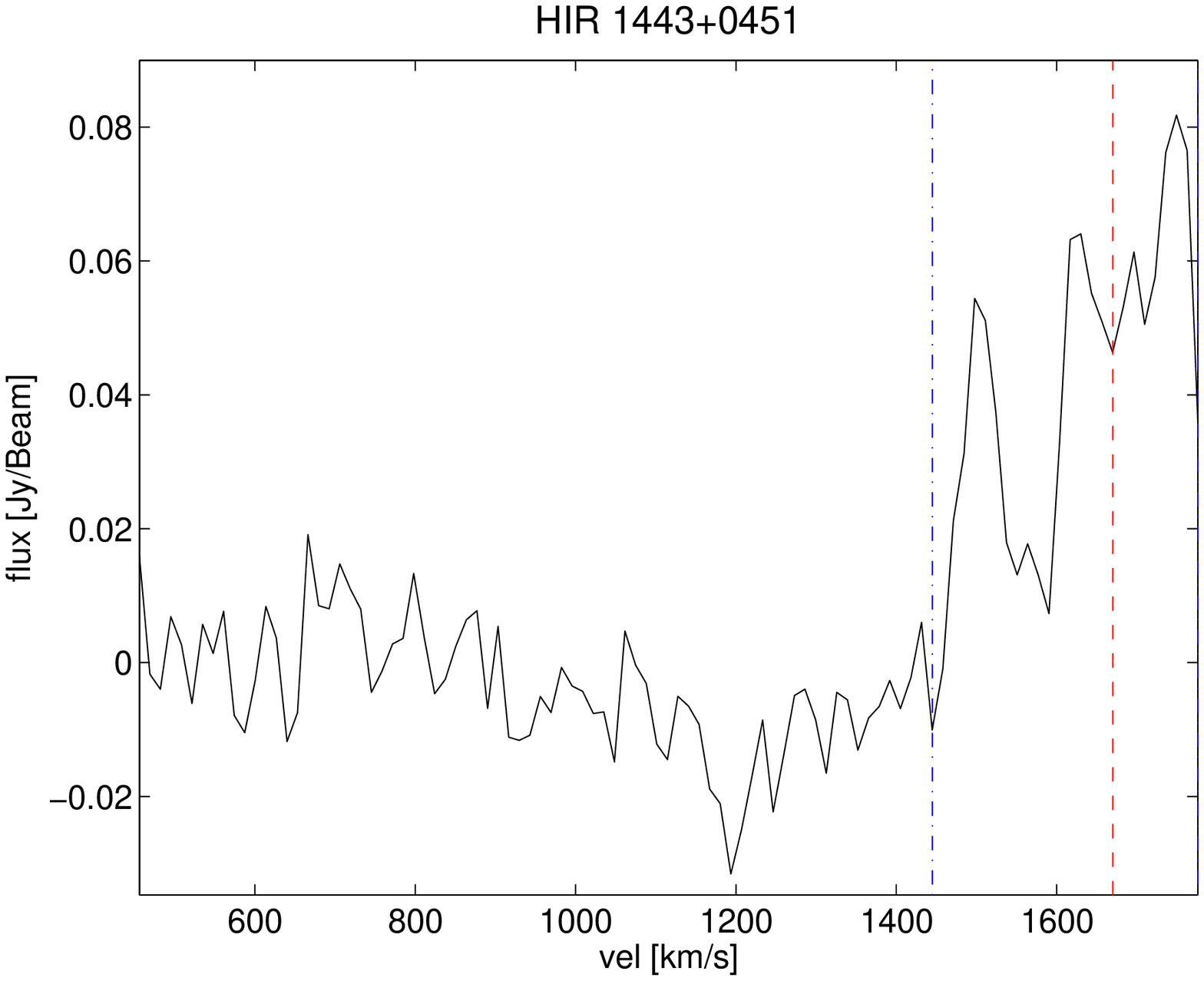}
 \includegraphics[width=0.3\textwidth]{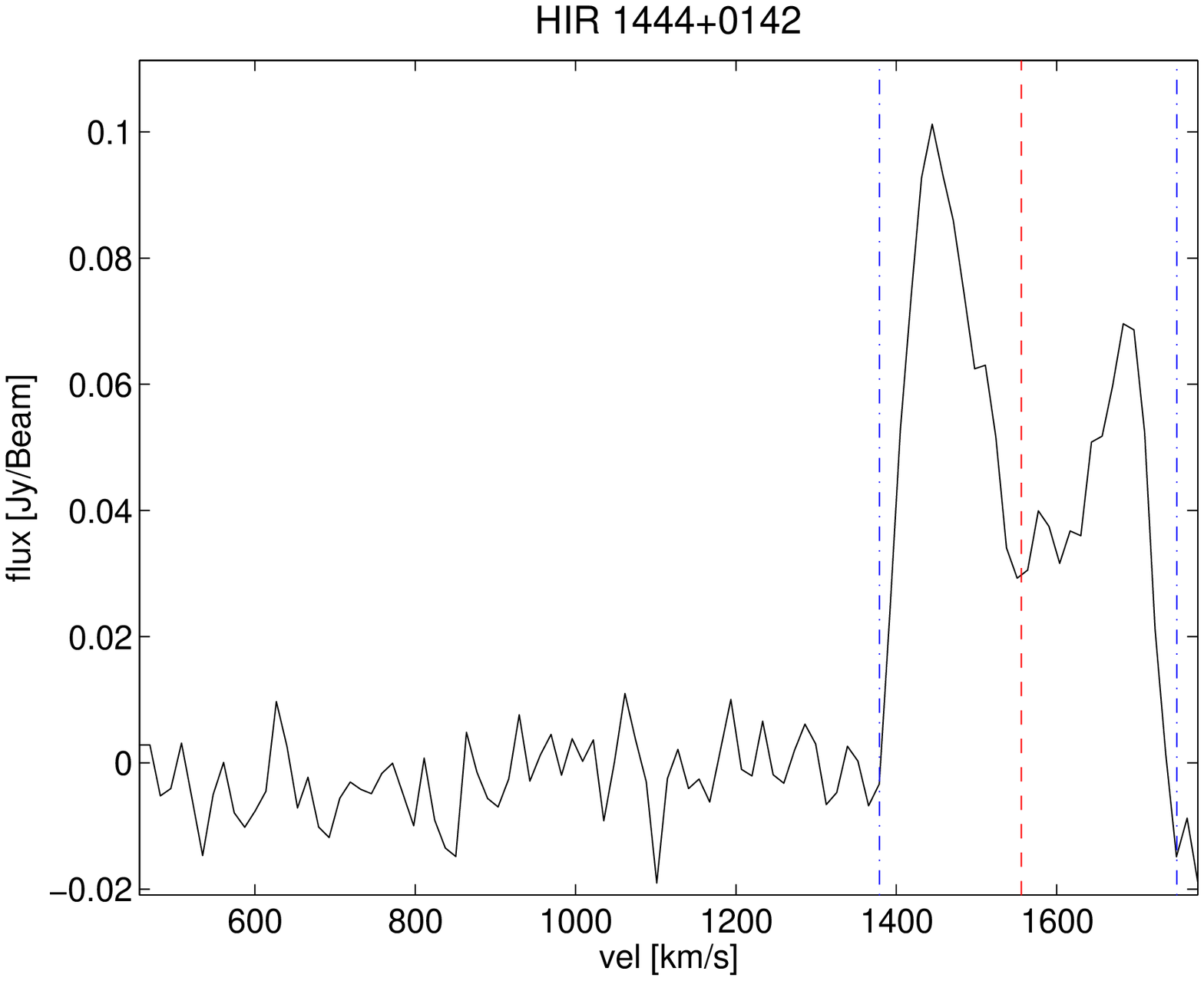}
 \includegraphics[width=0.3\textwidth]{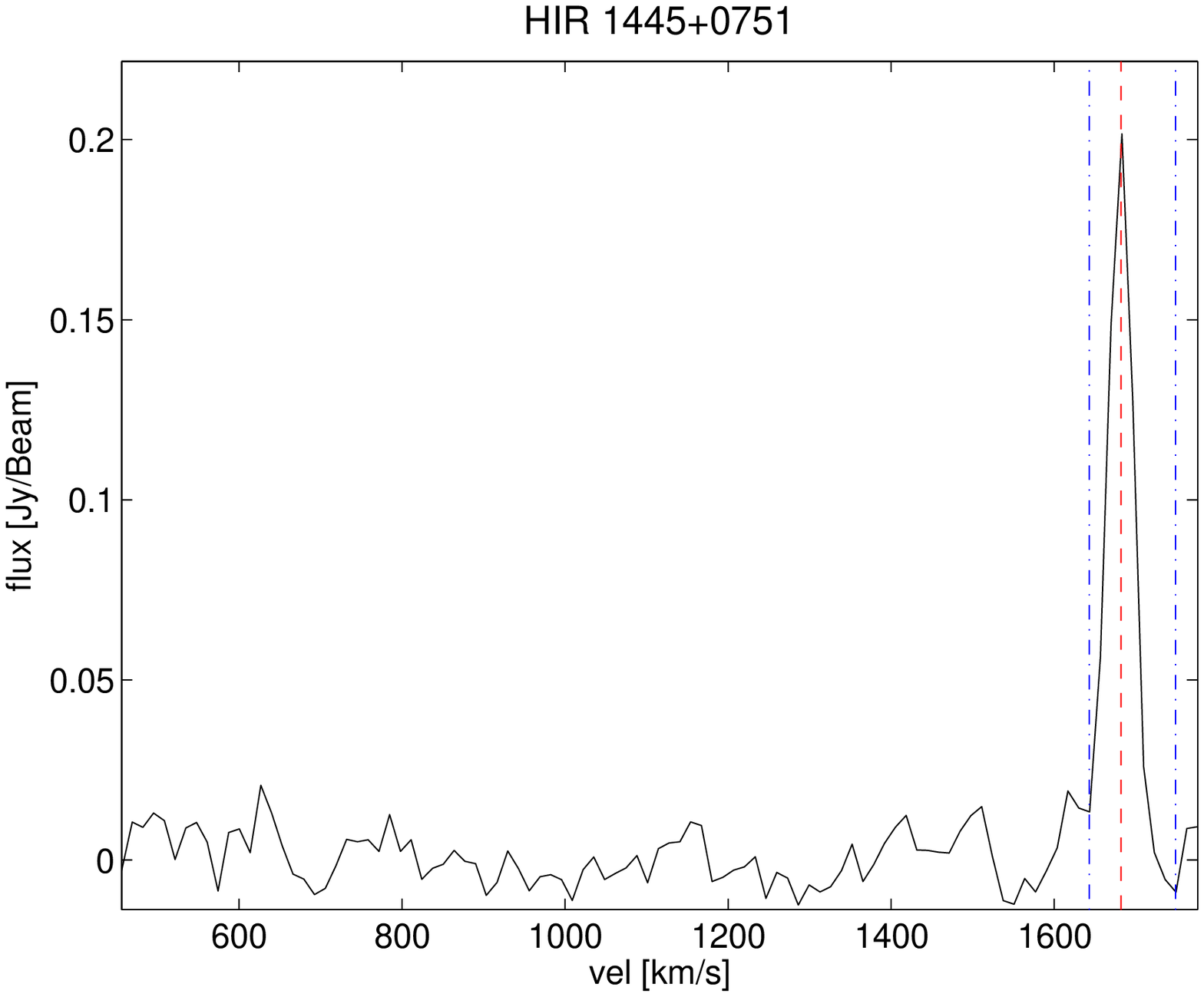}
 \includegraphics[width=0.3\textwidth]{1446+1011_spec.eps}
 \includegraphics[width=0.3\textwidth]{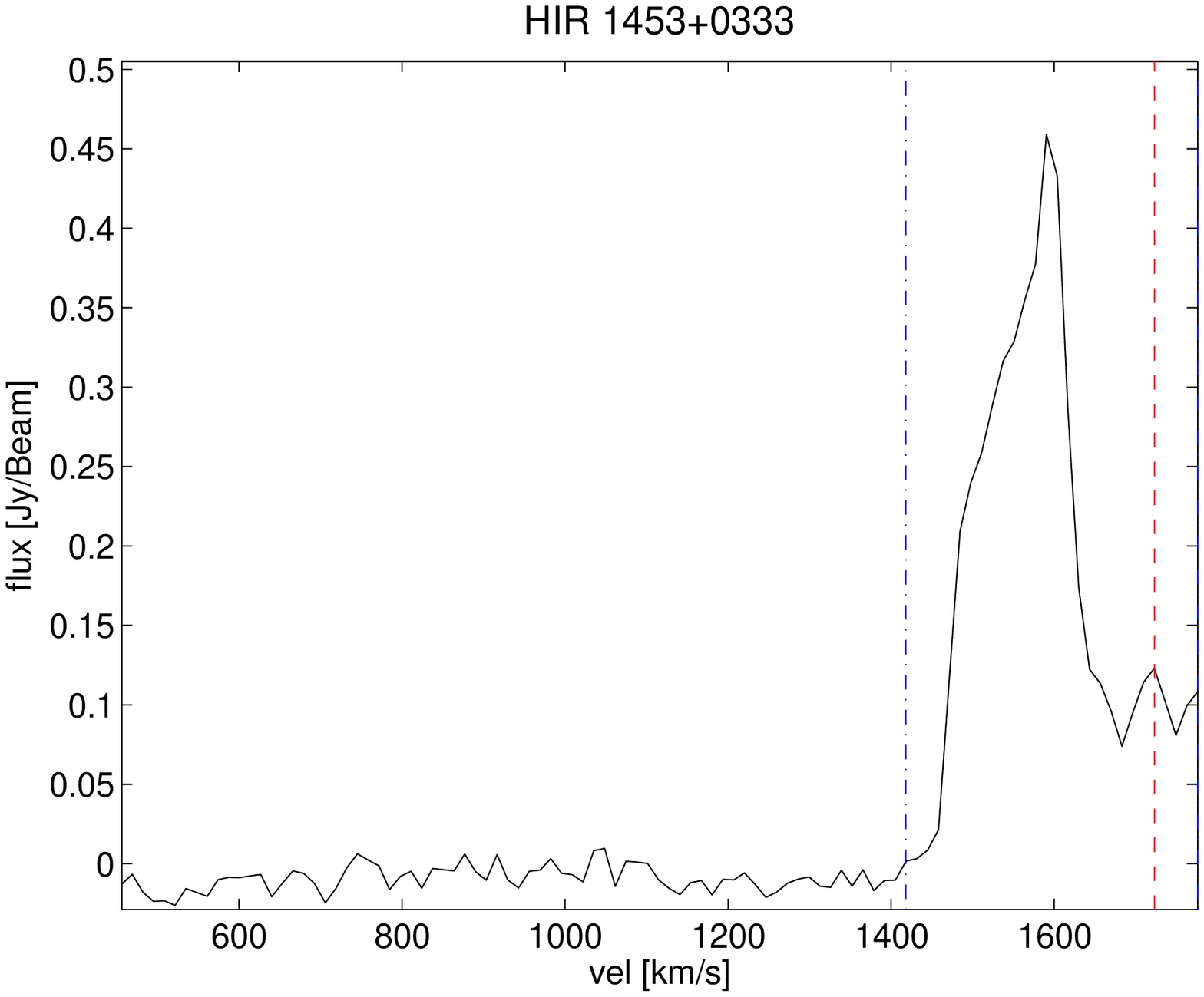}
 \includegraphics[width=0.3\textwidth]{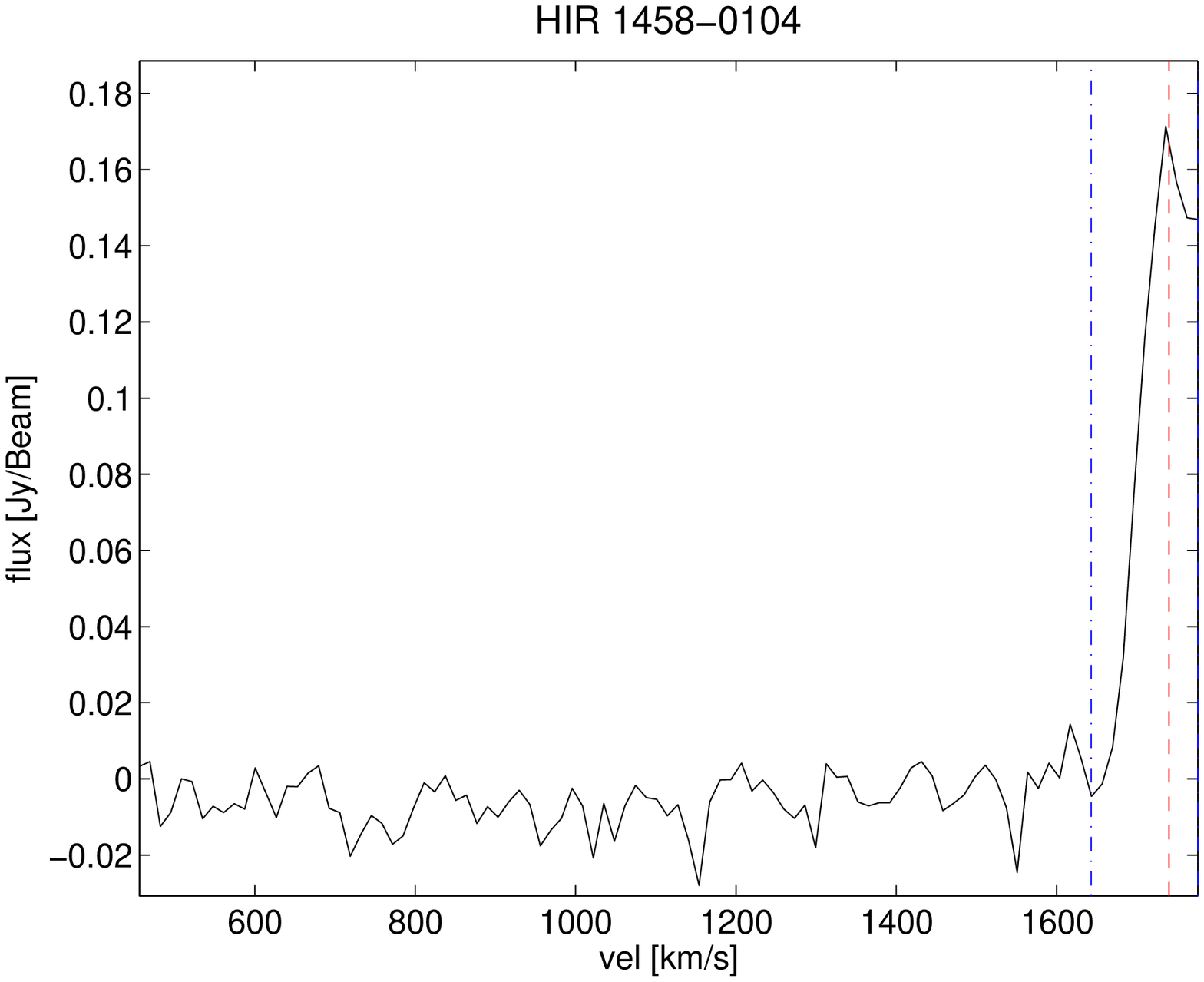}
 \includegraphics[width=0.3\textwidth]{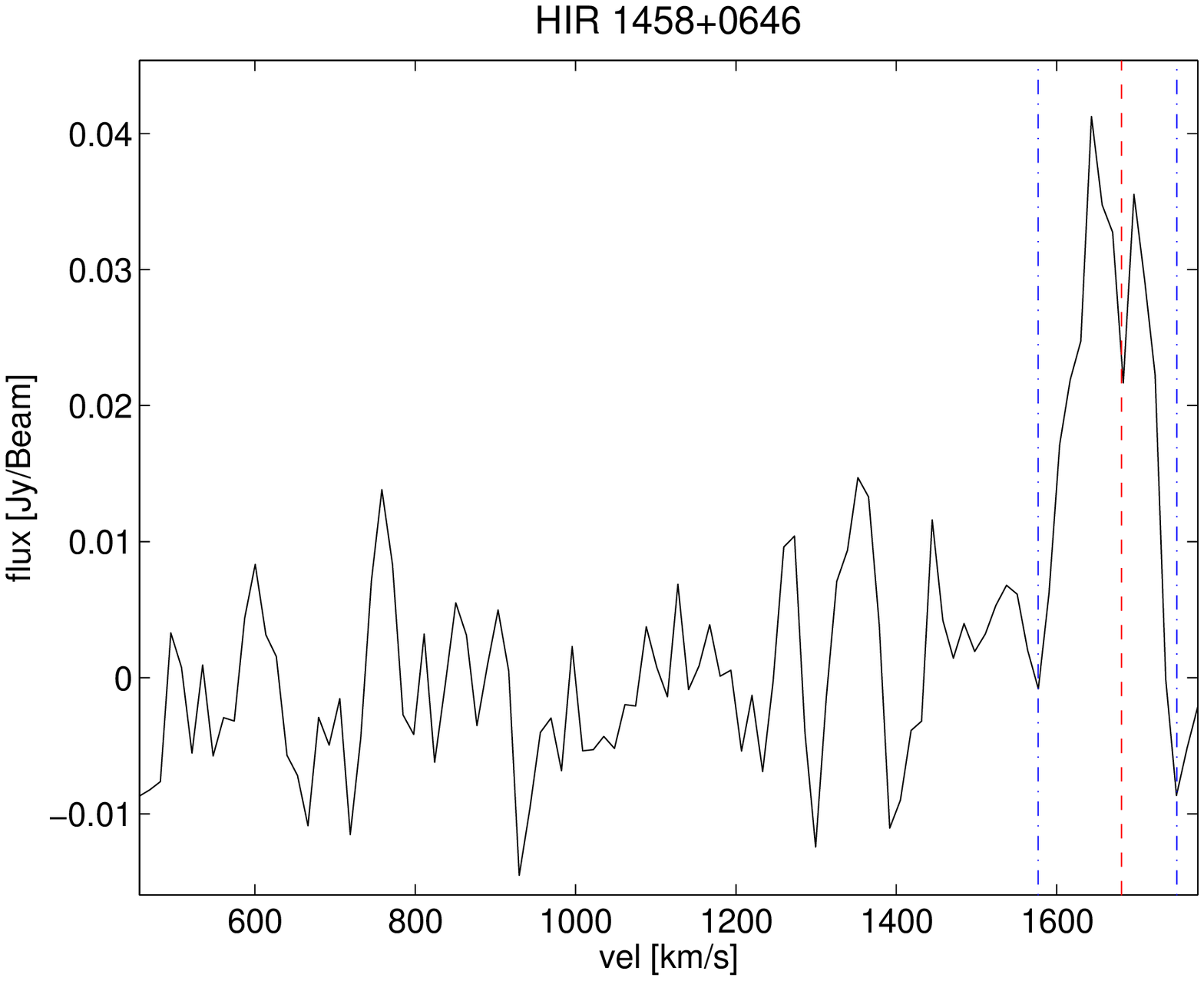}
 \includegraphics[width=0.3\textwidth]{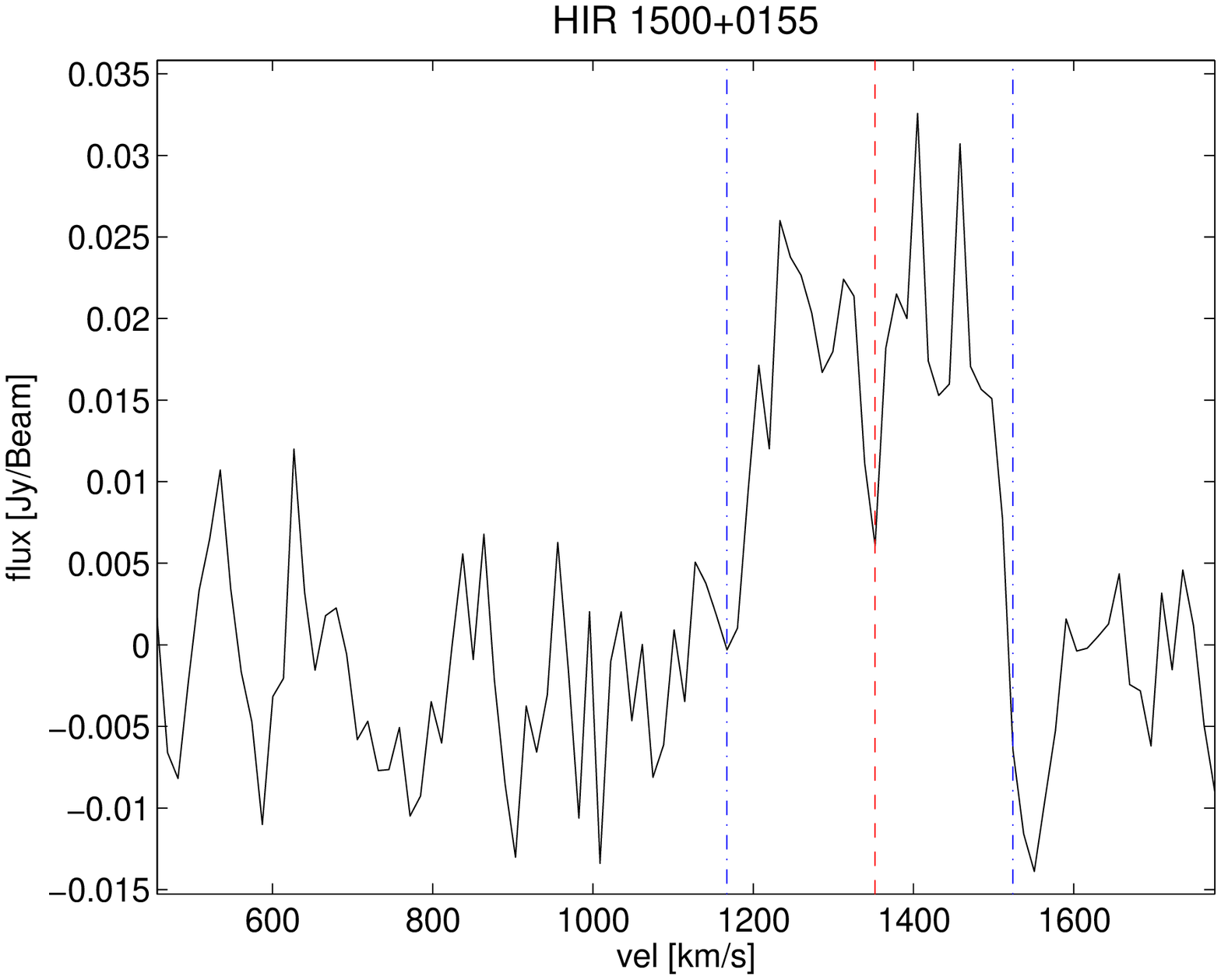}
 \includegraphics[width=0.3\textwidth]{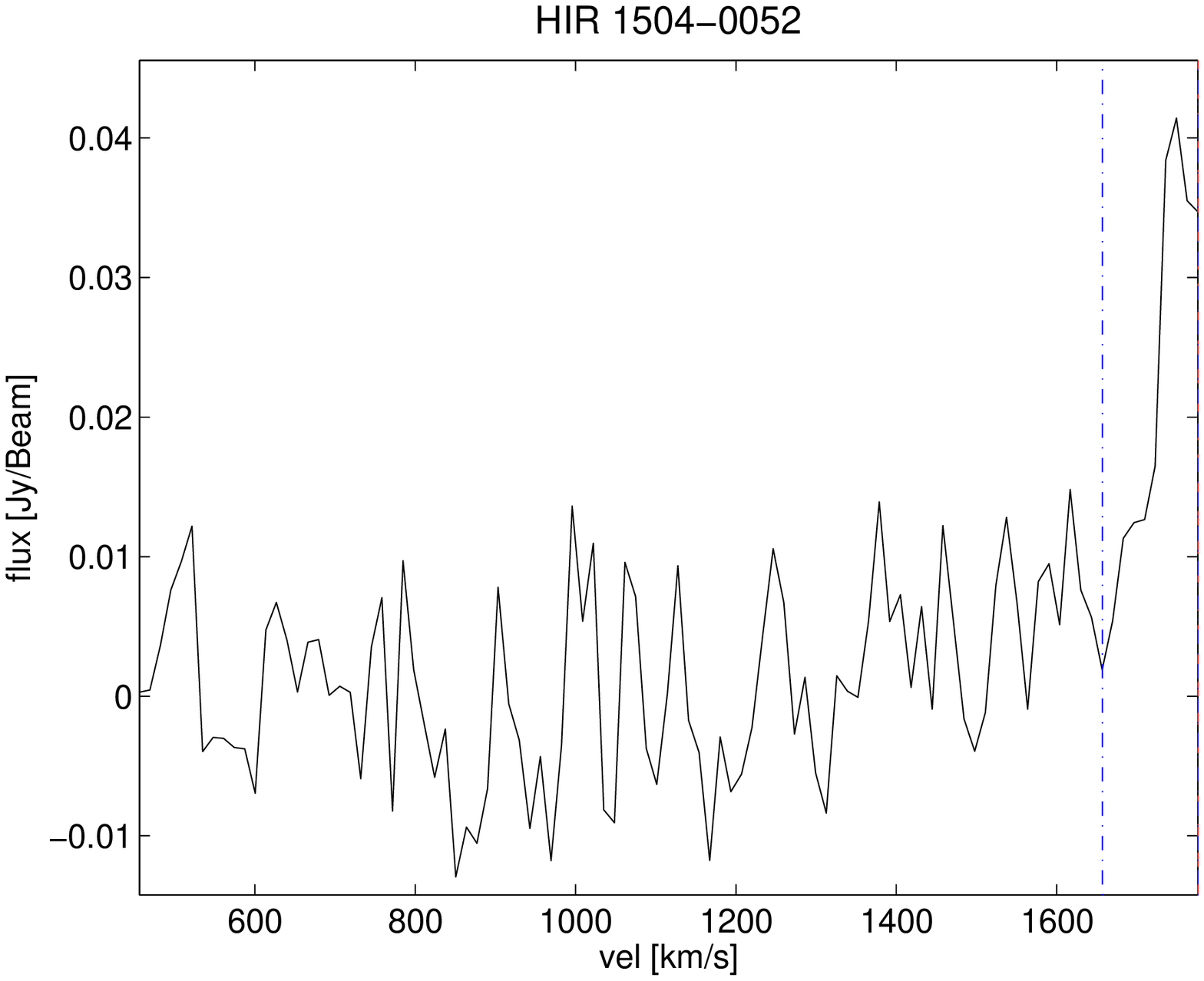}
 \includegraphics[width=0.3\textwidth]{1515+0603_spec.eps}

 \end{center}                                                         
{\bf Fig~\ref{all_spectra}.} (continued)                              
                                                                      
\end{figure*}

\begin{figure*}
  \begin{center}
  
 \includegraphics[width=0.3\textwidth]{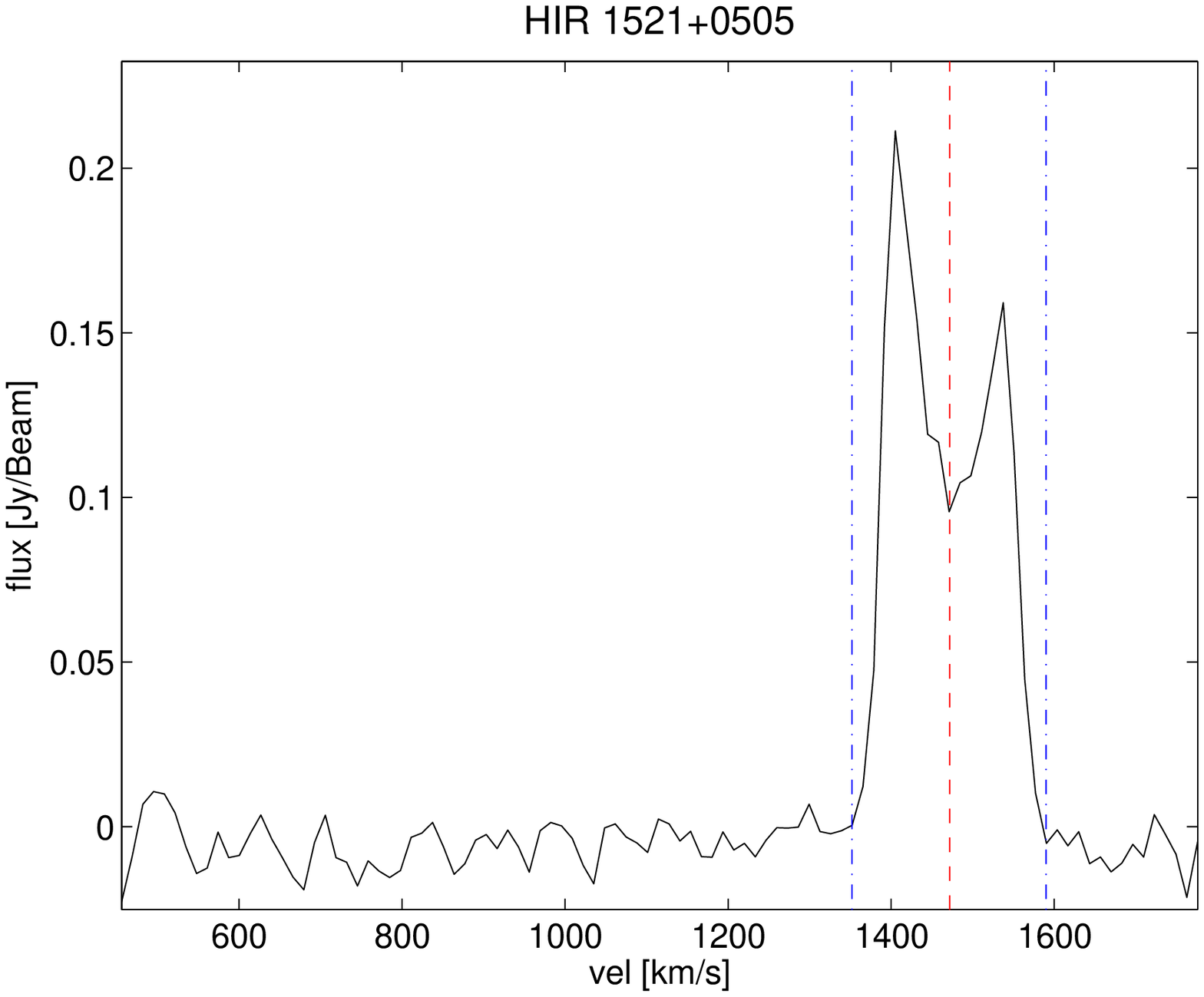}
 \includegraphics[width=0.3\textwidth]{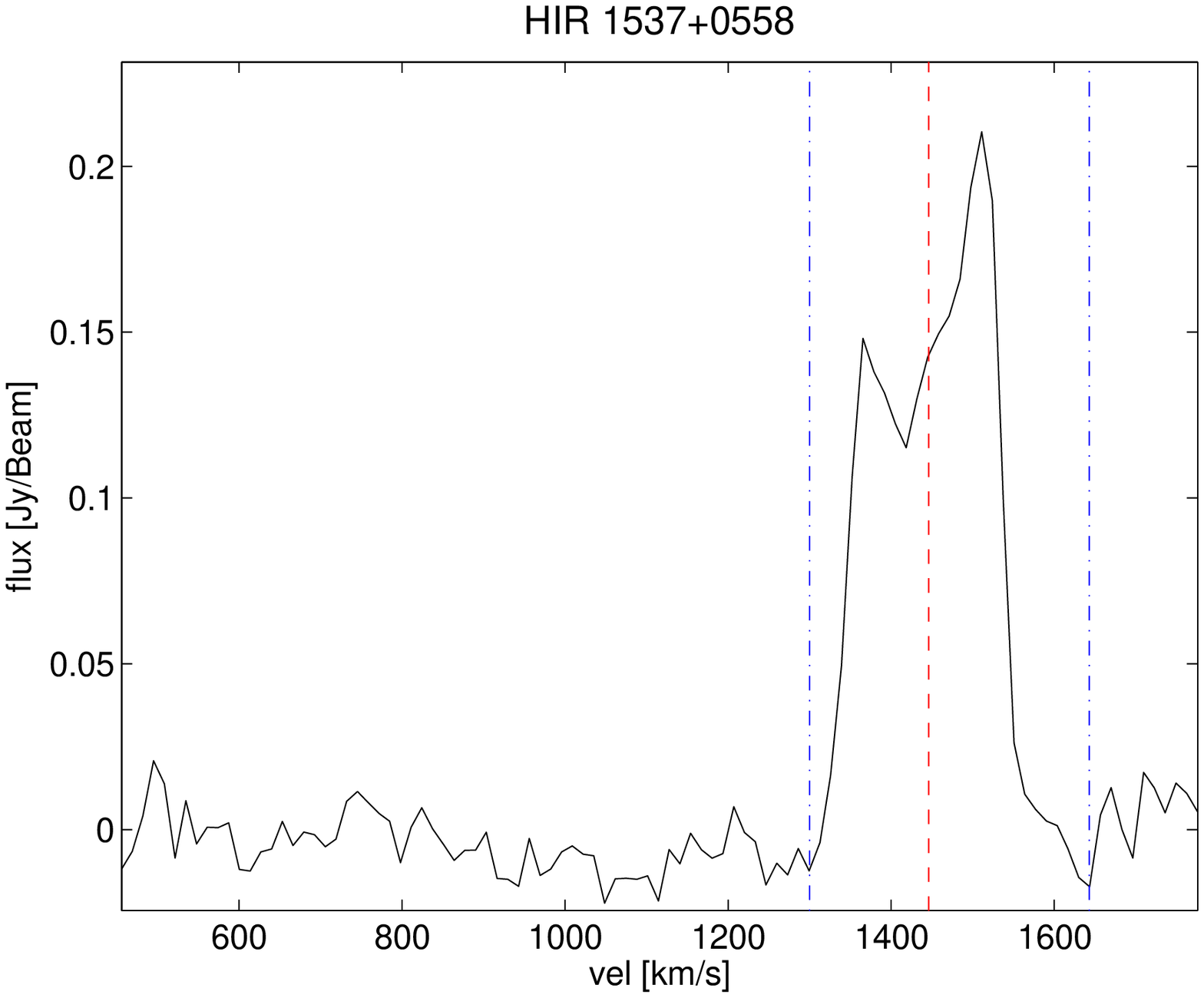}
 \includegraphics[width=0.3\textwidth]{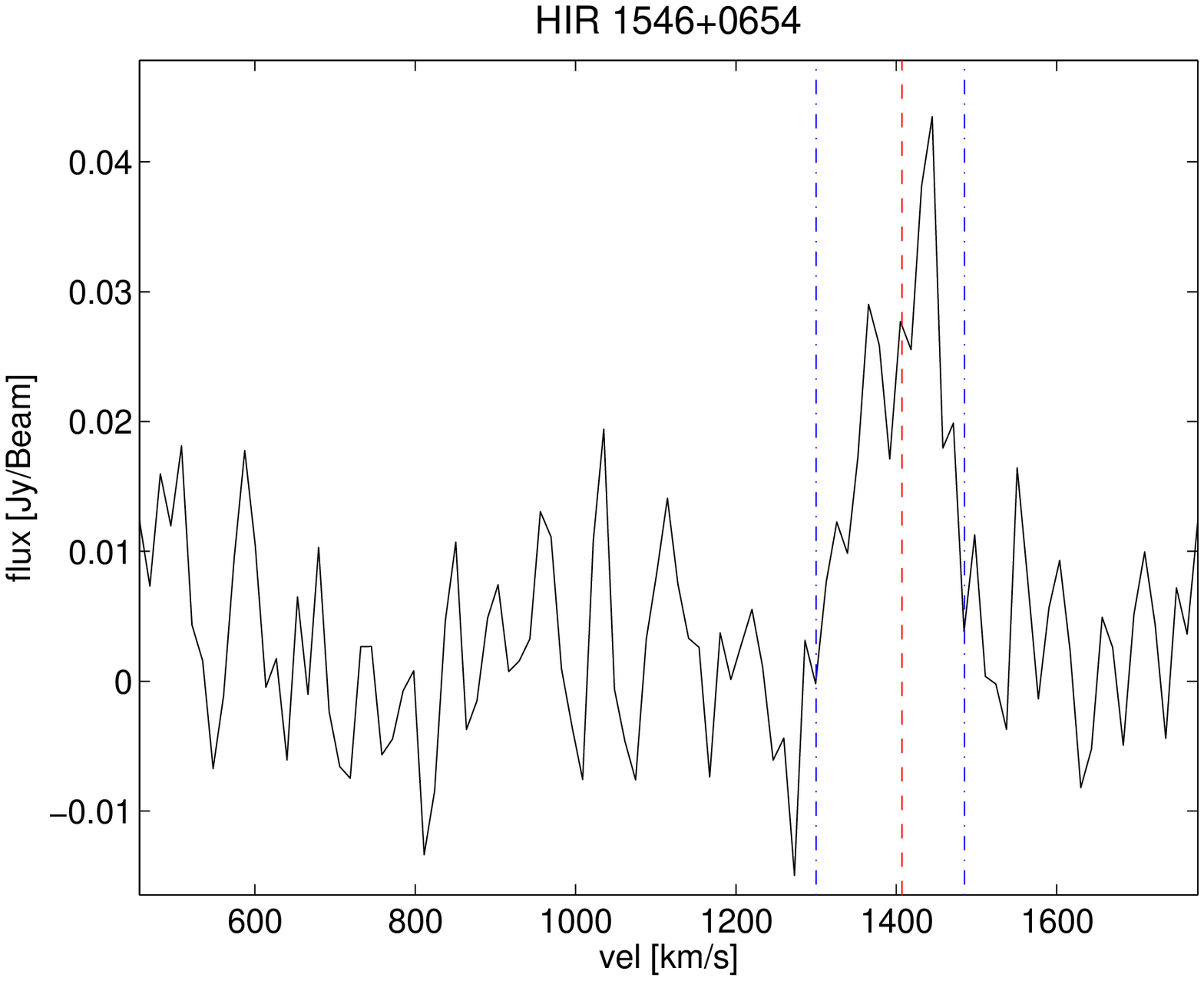}
 \includegraphics[width=0.3\textwidth]{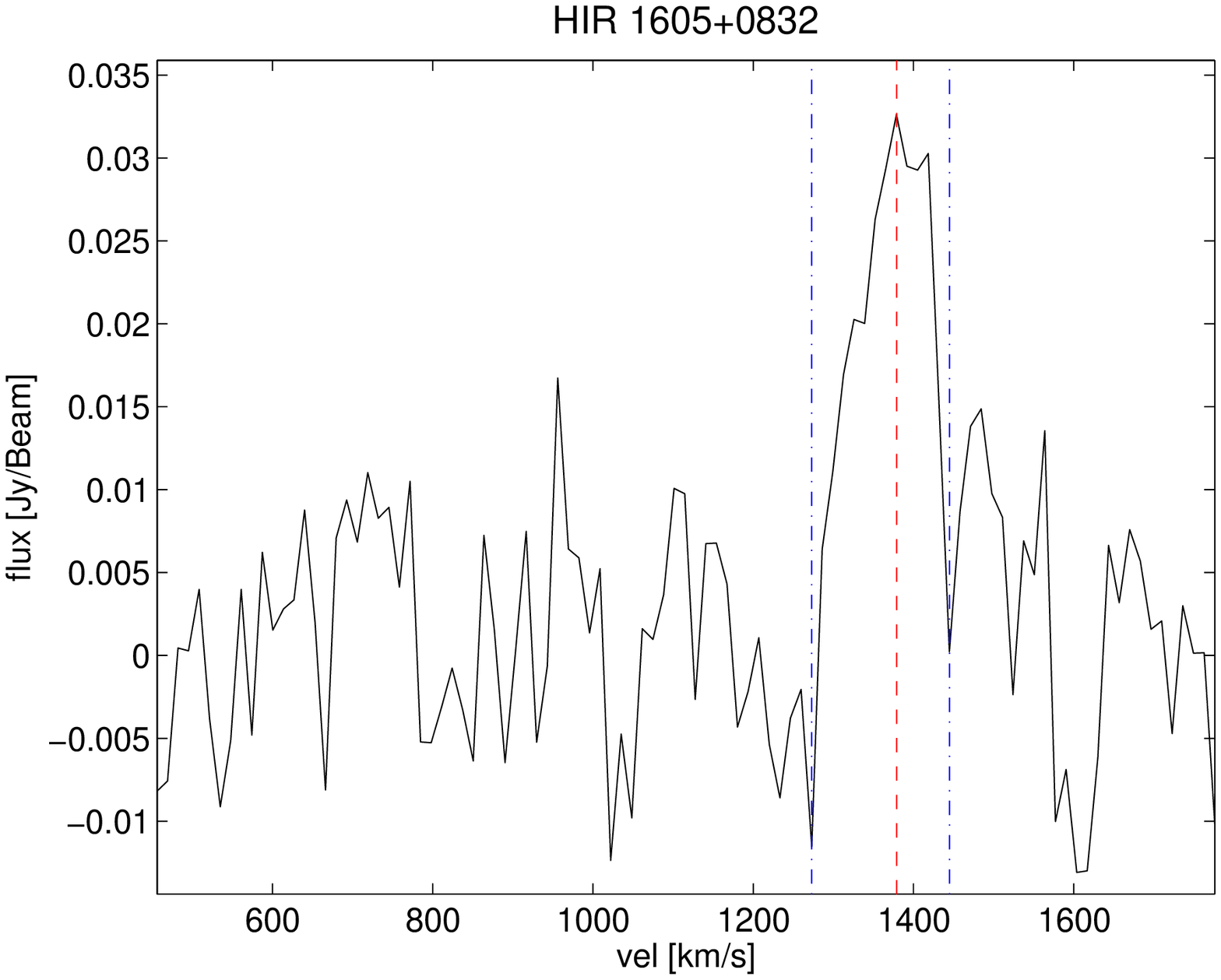}
 \includegraphics[width=0.3\textwidth]{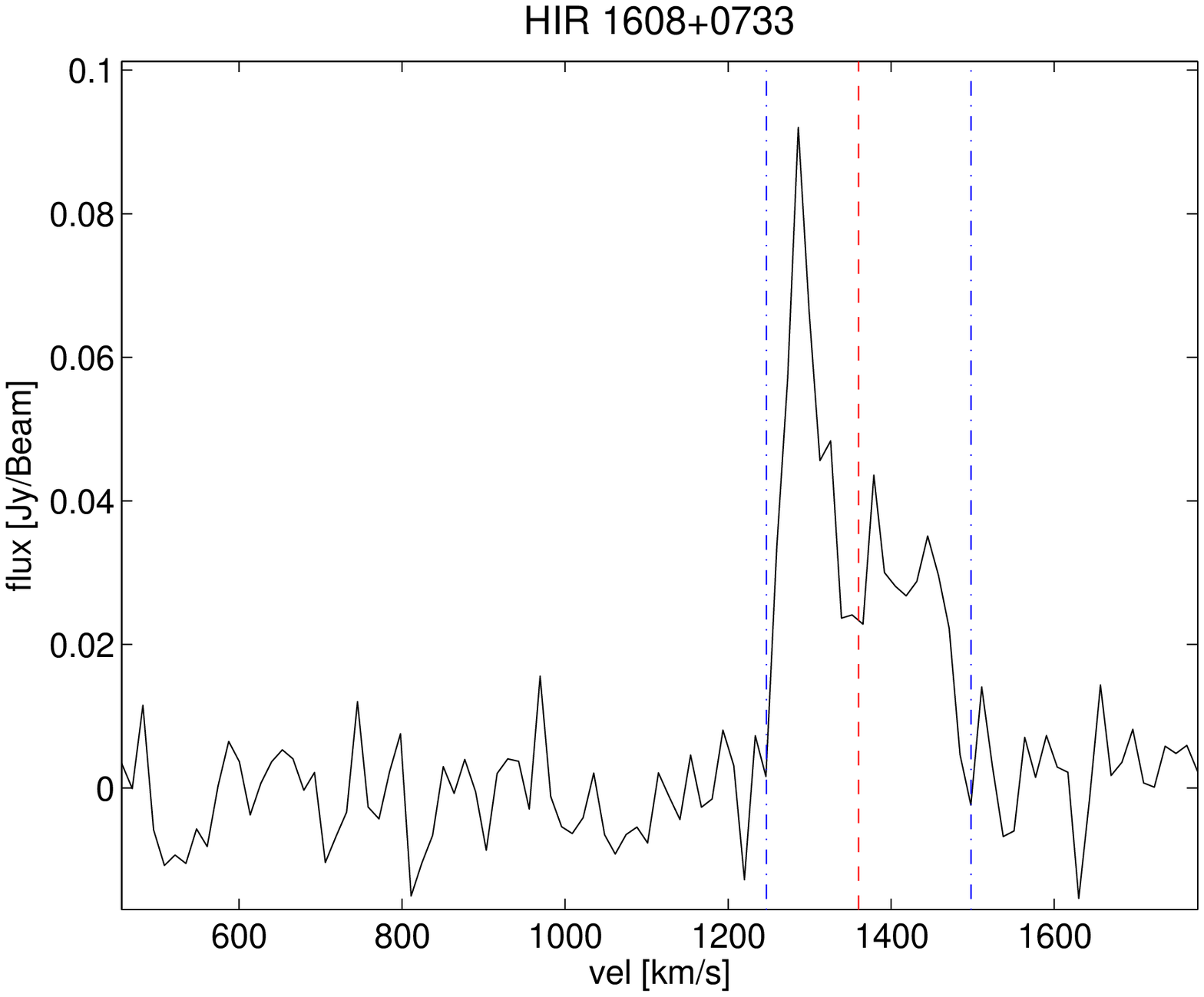}
 \includegraphics[width=0.3\textwidth]{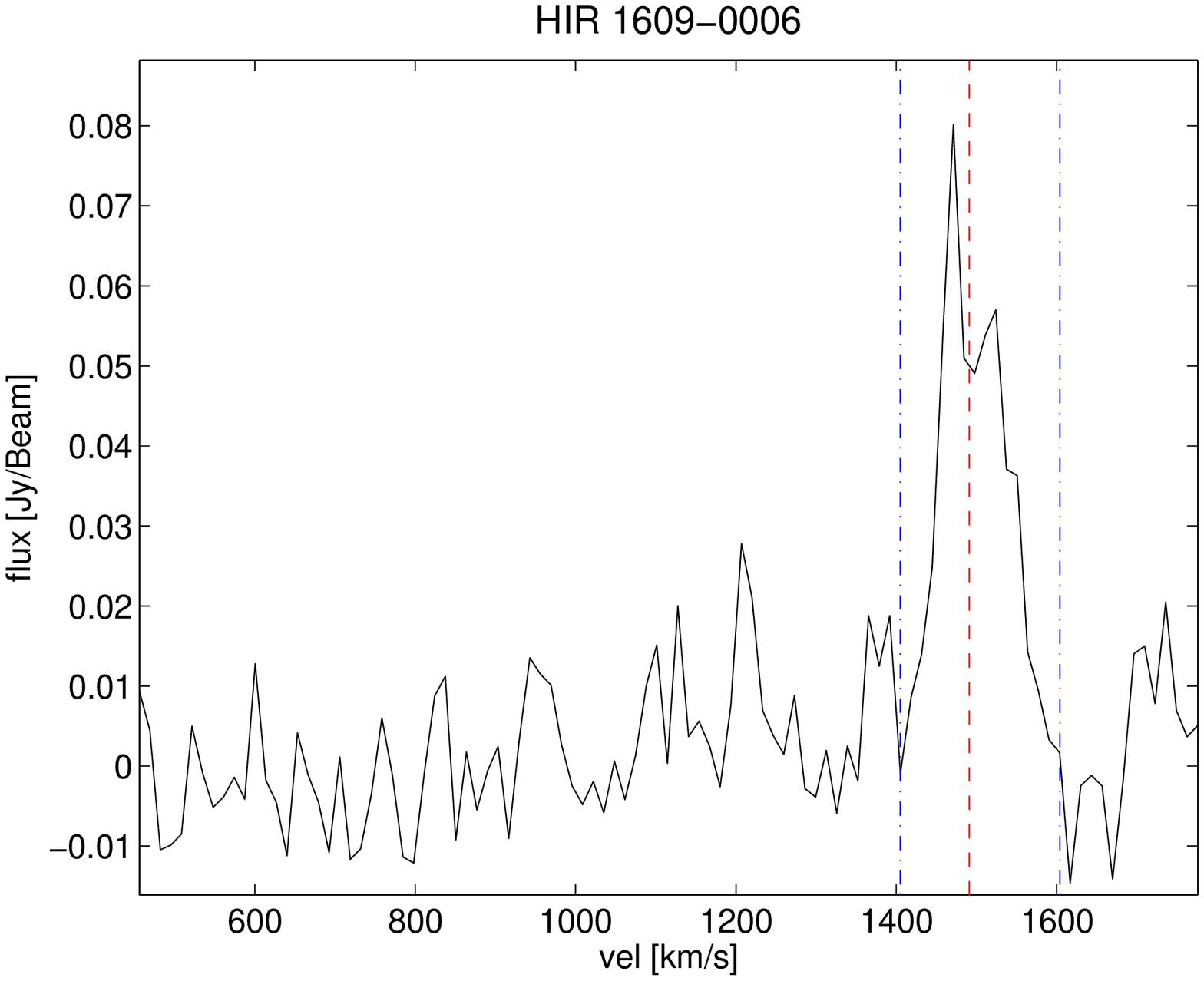}
 \includegraphics[width=0.3\textwidth]{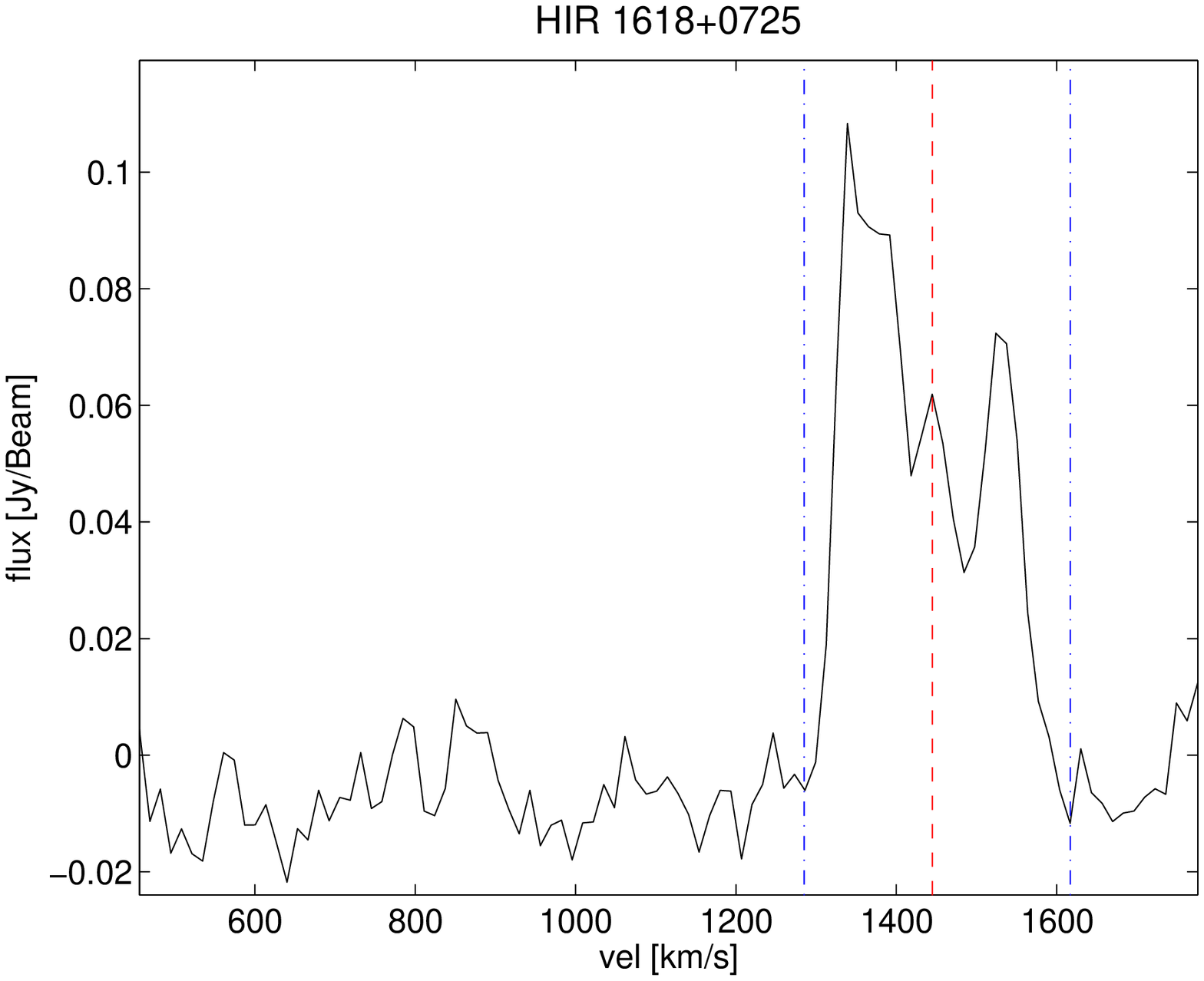}
 \includegraphics[width=0.3\textwidth]{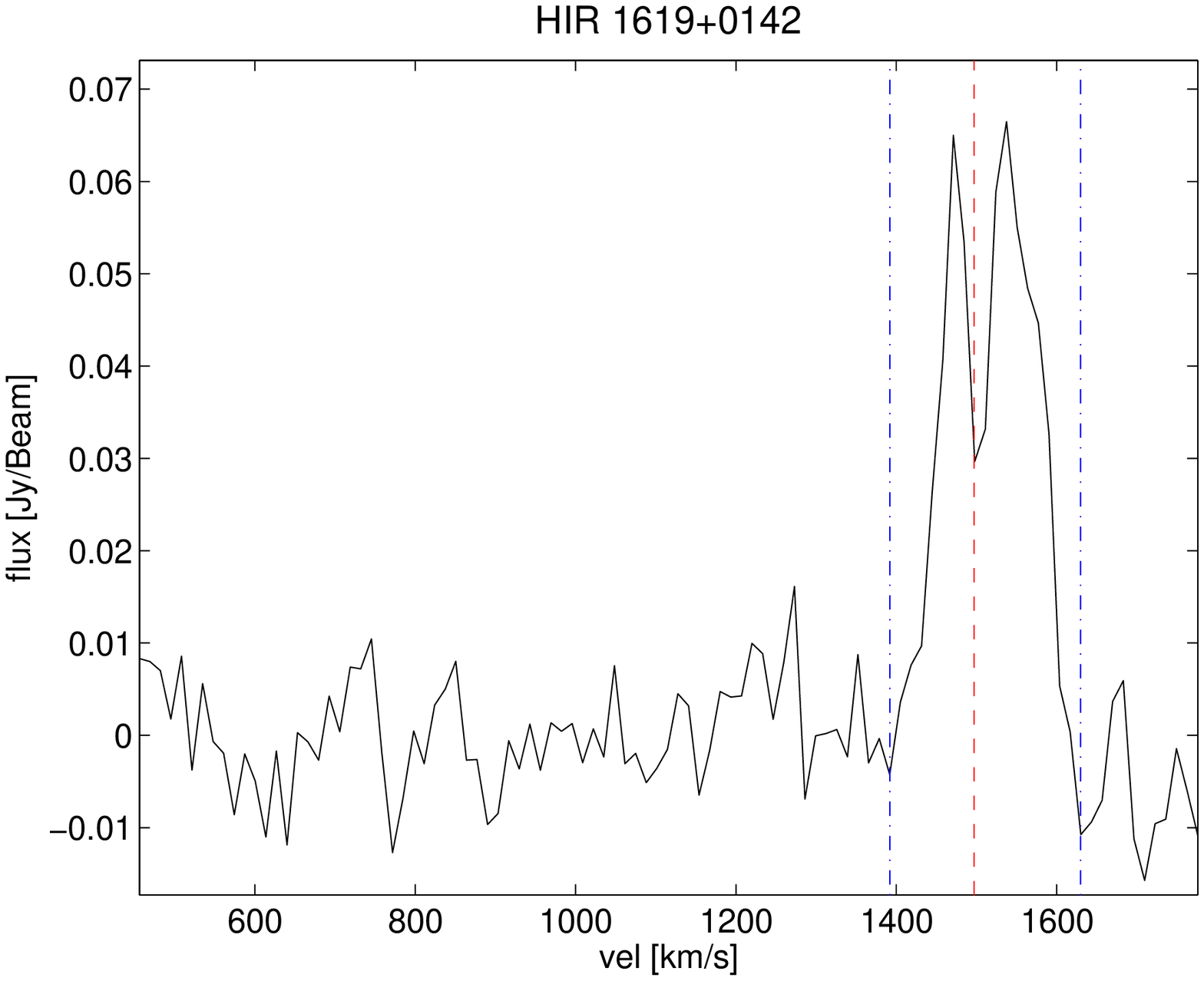}
 \includegraphics[width=0.3\textwidth]{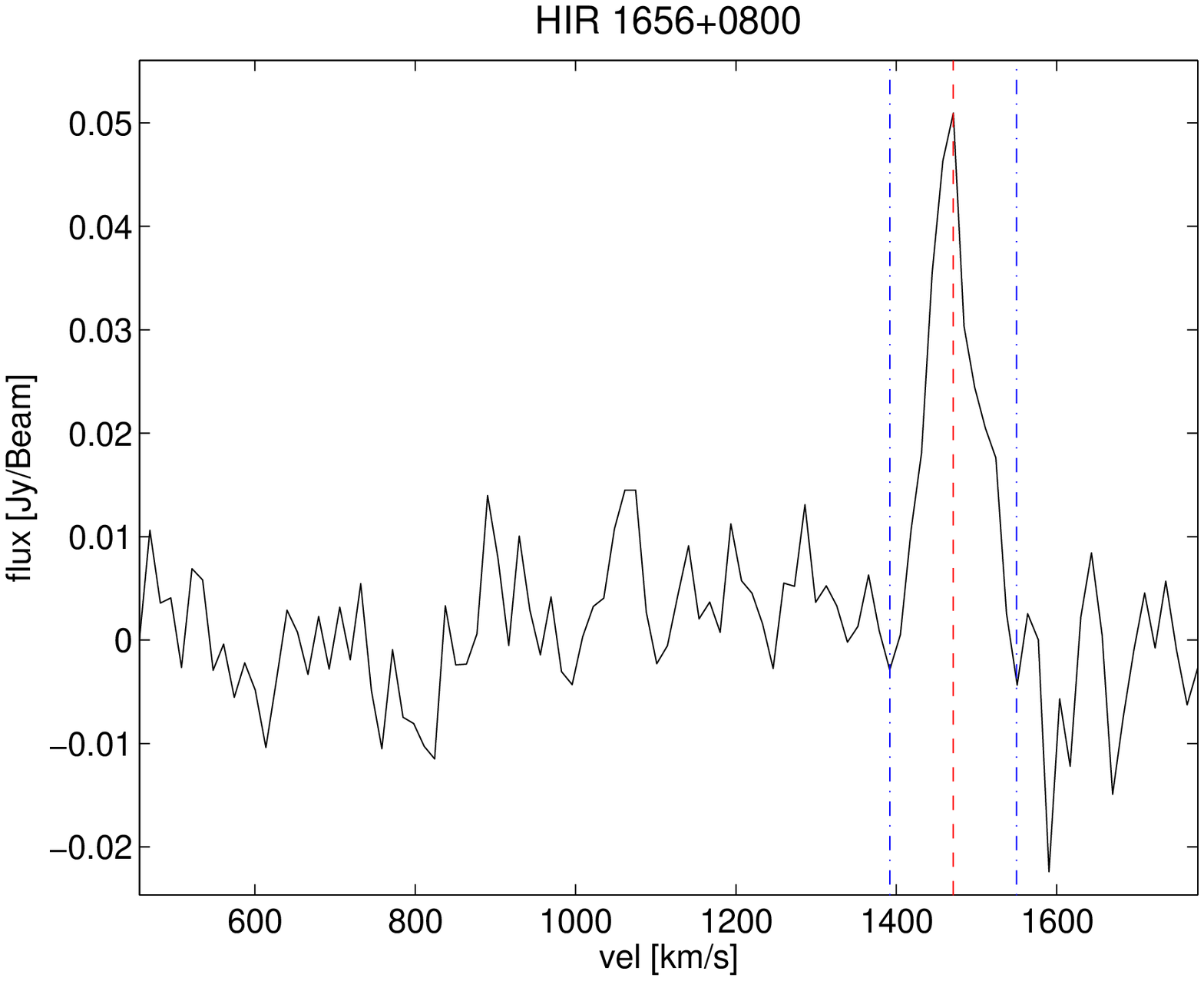}
 \includegraphics[width=0.3\textwidth]{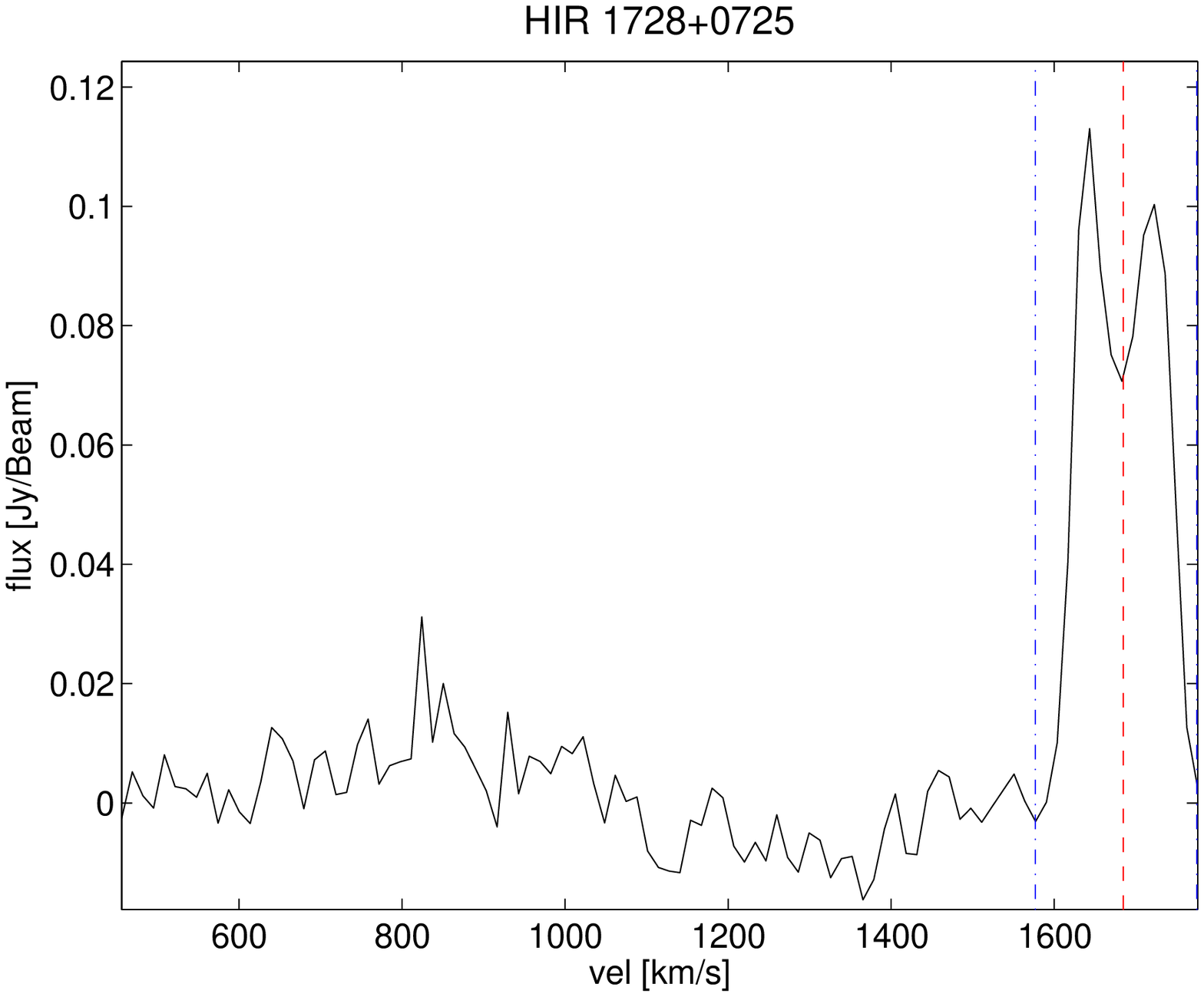}
 \includegraphics[width=0.3\textwidth]{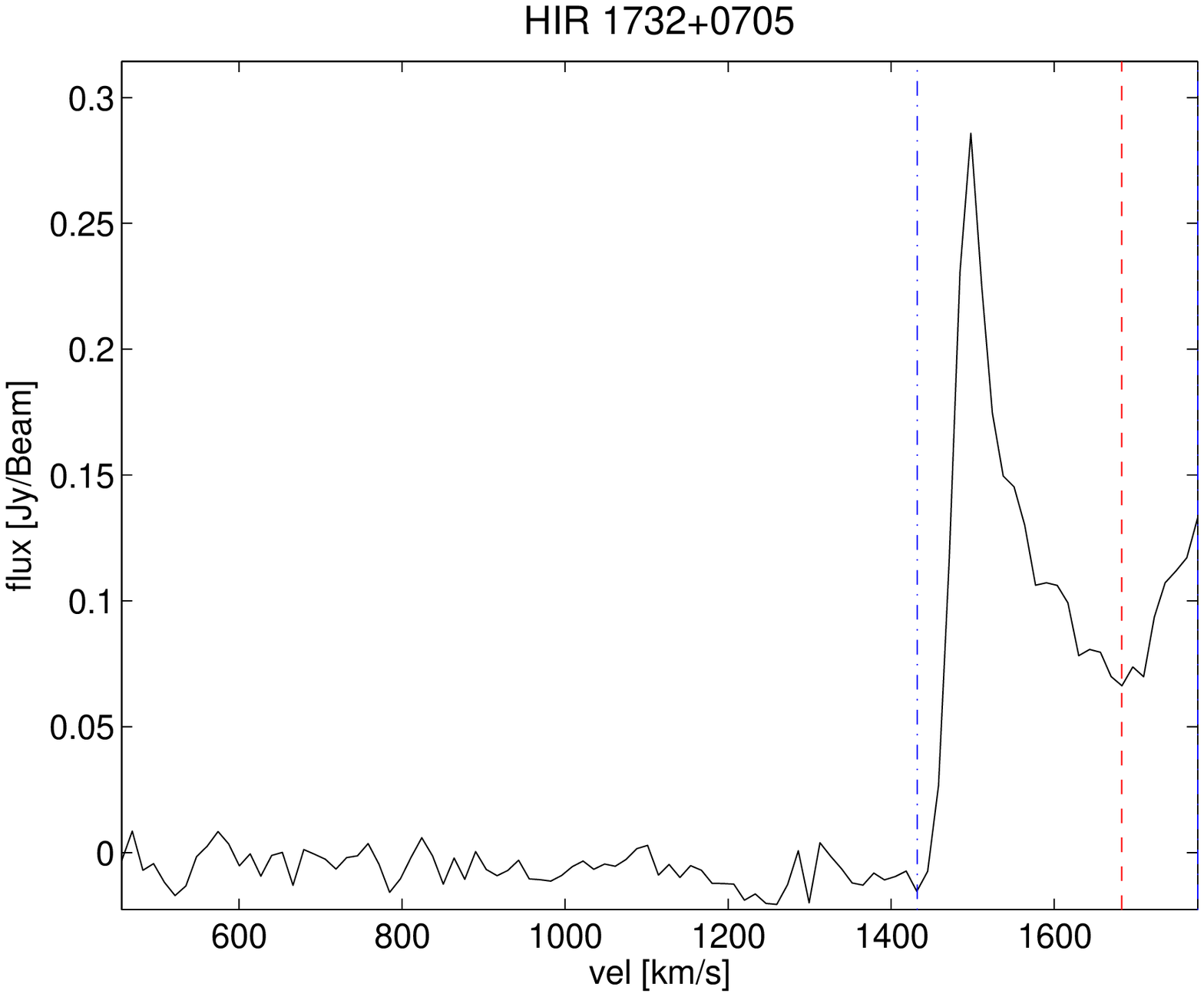}

 \end{center}                                                         
{\bf Fig~\ref{all_spectra}.} (continued)                              
                                                                      
\end{figure*}

\end{appendix}                                                        

\end{document}